\documentclass[12pt,twoside,brazil,english]{report}
\usepackage{lmodern}

\usepackage[T1]{fontenc}
\usepackage[utf8]{inputenc}
\usepackage[a4paper]{geometry}
\usepackage{fancyhdr}
\usepackage{babel}
\usepackage{float}
\usepackage{units}
\usepackage{amsmath}
\usepackage{amssymb}
\usepackage{stmaryrd}
\usepackage{graphicx}
\usepackage{setspace}
\usepackage{wasysym}
\usepackage[unicode=true,pdfusetitle,
 bookmarks=true,bookmarksnumbered=false,bookmarksopen=false,
 breaklinks=false,pdfborder={0 0 1},backref=false,colorlinks=false]
 {hyperref}
\usepackage{breakurl}

\makeatletter

\providecommand{\tabularnewline}{\\}

\usepackage{slashed}

\usepackage{nextpage}
\newcommand\myclearpage{\cleartooddpage[\thispagestyle{empty}]}

\usepackage[fit]{truncate}
\usepackage{simplewick}

\@ifundefined{showcaptionsetup}{}{%
 \PassOptionsToPackage{caption=false}{subfig}}
\usepackage{subfig}
\makeatother

\begin{document}
\begin{center}
\textbf{\large{}Andrés Arturo Gómez Quinto}
\par\end{center}{\large \par}

\vspace{2cm}

\begin{doublespace}
\begin{center}
\textbf{\emph{\Large{}Effective Superpotential and the Renormalization
Group Equation in a Supersymmetric Chern-Simons-Matter Model in the
Superfield Formalism}}
\par\end{center}{\Large \par}
\end{doublespace}

\begin{center}
\vspace{0.25cm}
\par\end{center}

\begin{center}
\includegraphics[scale=0.2]{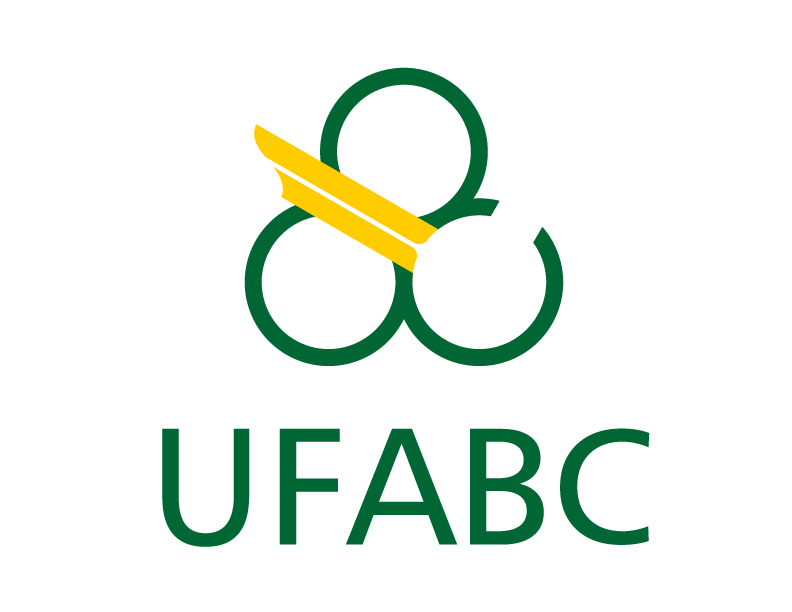}
\par\end{center}

\begin{center}
\vspace{0.25cm}
\par\end{center}

\selectlanguage{brazil}%
\begin{flushright}
\begin{minipage}[t]{0.6\columnwidth}%
Thesis presented to the Graduate Program in Physics of the Universidade Federal do ABC
(UFABC), as a partial requirement to obtain the title of 
PhD in Physics. %
\end{minipage}
\par\end{flushright}

\begin{center}
\vspace{1.5cm}
Advisor: Prof. Dr. Alysson Fábio Ferrari
\par\end{center}

\begin{center}
\vspace{1.75cm}
Santo Andre - SP
\par\end{center}

\begin{center}
2016
\par\end{center}

\selectlanguage{english}%
\begin{center}
{\large{}\thispagestyle{empty}}
\par\end{center}{\large \par}

\begin{flushright}
\begin{minipage}[t]{0.9\columnwidth}%
\begin{flushright}
\vspace{20cm}
\begin{minipage}[b]{0.45\columnwidth}%
\begin{flushright}
\emph{Este trabajo esta dedicado a mi esposa, Indira, a mis viejos,
Juan y Cenia y a mis hermanos, Laura y Juan Carlos.}
\par\end{flushright}%
\end{minipage}
\par\end{flushright}%
\end{minipage} 
\par\end{flushright}

\begin{center}
{\large{}\thispagestyle{empty}}
\par\end{center}{\large \par}

\begin{center}
\begin{minipage}[t]{0.9\columnwidth}%
\begin{center}
\textbf{\large{}\rule[0.5ex]{1\columnwidth}{1pt}}
\par\end{center}{\large \par}
\begin{center}
\textbf{\Large{}Acknowledgments}
\par\end{center}{\Large \par}
\vspace{0.5cm}

\begin{onehalfspace}
First, I would like to thank Prof. Alysson Ferrari for his guidance,
help, patience, for many conversations and friendship. 
\end{onehalfspace}

I also thank 
\begin{itemize}
\item \emph{Coordenação de Aperfeiçoamento de Pessoal de Nivel Superior}
(CAPES) for the financial support, 
\item the professors of the graduate program in Physics at \emph{Universidade
Federal do ABC} (UFABC), who were responsible for a large part of
my scientific training, with classes, discussions, seminars, etc.,
\item all the university staff of UFABC,
\item my colleagues and friends in the group: Carlos Palechor and Luiz Borges
for their discussions and contributions in my studies,
\item the friends I made here: Marcela, Andrés, Diana, Camilo, Julian, Liliana,
John, Dona Ana, Don Salu, Marcio and those who probably I forgot to
mention (apologize me!),
\item my old friends, with whom I am still sharing great moments: Melissa,
Carlos, Fernando, Robinson, Mawin, Jeiner, Cynthia, Rodolfo, and those
who probably I forgot to mention (apologize me!), thank you for your
friendship, there is no happiness without friends specially when you
are far from the family. 
\end{itemize}
I would also give a special thanks to Prof. André Lehum, for the immense
contribution of our discussions. 

Finally, and so very important for me, I would like to thank my family:
my wife Indira Vargas; my parents, Cenia Quinto and Juan Gomez; my
sister, Laura and my brother Juan Carlos, my immense thanks for the
support in these years of study.

\rule[0.5ex]{1\columnwidth}{1pt}%
\end{minipage}
\par\end{center}

\vspace{2cm}

\begin{center}
{\large{}\thispagestyle{empty}}
\par\end{center}{\large \par}

\begin{center}
\textbf{\large{}\rule[0.5ex]{1\columnwidth}{1pt}}
\par\end{center}{\large \par}

\begin{center}
\textbf{\Large{}Abstract}
\par\end{center}{\Large \par}

\vspace{0.5cm}

In this thesis we study the Dynamical Symmetry Breaking (DSB) mechanism
in a supersymmetric Chern-Simons theory in $\left(2+1\right)$ dimensions
coupled to $N$ matter superfields in the superfield formalism. For
this purpose, we developed a mechanism to calculate the effective
superpotencial $K_{\mathrm{eff}}\left(\sigma_{\mathrm{cl}},\alpha\right)$,
where $\sigma_{\mathrm{cl}}$ is a background superfield, and $\alpha$
a gauge-fixing parameter that is introduced in the quantization process.
The possible dependence of the effective potential on the gauge parameter
have been studied in the context of quantum field theory, and it can
have nontrivial consequences to the study of DSB in gauge theories.
Therefore, we did not assume from the start the gauge independence
of the effective potential in our model, as it is customary in the
literature. We developed the formalism of the Nielsen identities in
the superfield language, which is the appropriate formalism to study
DSB when the effective potential is gauge dependent. We also discuss
how to calculate the effective superpotential via the Renormalization
Group Equation (RGE) from the knowledge of the renormalization group
functions of the theory, i.e., $\beta$ functions and anomalous dimensions
$\gamma$. We perform a detailed calculation of these functions at
two loops, finding that these do not depend on $\alpha$, and therefore,
by using the RGE, we calculate the effective superpotencial $K_{\mathrm{eff}}$,
showing that it is also independent of $\alpha$. Then we discuss
the improvement of the calculation of $K_{\mathrm{eff}}$ by summing
up leading logarithms, and we compare this improvement with the one
obtained in the non supersymmetric version of the model. Finally,
we study the DSB finding that it is operational for all reasonable
values of the free parameters, while the improvement obtained from
the RGE in general only produces a small quantitative correction in
the results, instead of the more dramatic qualitative change found
in non supersymmetric models.

\vspace{0.5cm}

\begin{center}
{\large{}\thispagestyle{empty}}
\par\end{center}{\large \par}

\tableofcontents{}

\listoffigures

\listoftables

\begin{center}
{\large{}\thispagestyle{empty}}
\par\end{center}{\large \par}

\begin{center}
\myclearpage
\par\end{center}

\global\long\def\LL{{\rm LL}}
 \global\long\def\NLL{{\rm NLL}}
 \global\long\def\NNLL{{\rm N2LL}}
\global\long\def\eff{{\rm eff}}

\chapter{Introduction}

In the 1970s supersymmetry was discovered in a series of articles
on string theory. Ramond\,\cite{Ramond1971} proposed an equation
for free fermions based on a dual structure theory for bosons, and
in the same period Neveu and Schwarz\,\cite{neveu:1971rx,Neveu1971}
proposed the inclusion of particles with quantum number of interacting
pions in the Ramond model. Another example of a precursor paper to
what would come to be called supersymmetry was the symmetry of a two-dimensional
field theory discussed by Gervais and Sakita\,\cite{Gervais1971}.
On the other hand, almost simultaneously, the Soviet researchers Gol'fand
and Likhtman\,\cite{Golfand1971} extended the algebra of the Poincare
group to a superalgebra and from this built a supersymmetric theory
in four-dimensional space-time. These papers, although seminal, were
generally ignored until much later. Regardless, Volkov and Akulov\,\cite{Volkov1972}
discovered what today might be called spontaneous breaking of supersymmetry,
but they used their formalism to identify the Goldstone fermion associated
with supersymmetry breaking with the neutrino, one idea that was unsuccessful.
In a widely known paper, Wess and Zumino\,\cite{wess:1973kz} related
the bosonic and fermionic states in a quantum field theory in four
dimensions, describing the combination of states with different spin
(integer and half-integer) as a supermultiplet. The Wess-Zumino model
in $\left(3+1\right)$ dimensions is described by the Lagrangian,
\begin{align}
\mathcal{L}= & -\frac{1}{2}\partial_{\mu}A\partial^{\mu}A-\frac{1}{2}\partial_{\mu}B\partial^{\mu}B-\frac{1}{2}\bar{\psi}\gamma^{\mu}\partial_{\mu}\psi+\frac{1}{2}\left(F^{2}+G^{2}\right)\nonumber \\
 & +m\left[FA+GB-\frac{1}{2}\bar{\psi}\psi\right]+g\left[F\left(A^{2}+B^{2}\right)+2GAB-\bar{\psi}\left(A+i\gamma_{5}B\right)\psi\right],\label{eq:Lang-WZ-4dim}
\end{align}
where $A$ and $B$ are the scalar and pseudo-scalar (bosonic) fields
respectively, $\psi$ is a Majorana (fermionic) field, and $F$ and
$G$ are auxiliary scalar and pseudo-scalar fields, respectively.
The auxiliary fields are non-dynamical, being determined by their
algebraic equations of motion,
\begin{eqnarray}
F=-mA-g\left(A^{2}+B^{2}\right), &  & G=-mB-2gAB.
\end{eqnarray}

The Lagrangian in Eq.\,(\ref{eq:Lang-WZ-4dim}) is invariant under
the following infinitesimal supersymmetric transformations:\begin{subequations}
\begin{align}
\delta A= & \bar{\xi}\psi,\\
\delta B= & -i\bar{\xi}\gamma_{5}\psi,\\
\delta\psi= & \partial_{\mu}\left(A+i\gamma_{5}B\right)\gamma^{\mu}\xi+\left(F-i\gamma_{5}G\right)\xi,\\
\delta F= & \bar{\xi}\gamma^{\mu}\partial_{\mu}\psi,\\
\delta G= & -i\bar{\xi}\gamma_{5}\gamma^{\mu}\partial_{\mu}\psi,
\end{align}
\end{subequations}where $\xi$ is a Grassmannian parameter that acts
as an infinitesimal transformation parameter. The supersymmetry transformation
is generated by a Grassmannian generator $Q$ that satisfies
\[
\left\{ Q,Q\right\} =2P,
\]
with $P$ being the generator of space-time translations.

In essence, the supersymmetry is an extension of the symmetry of space-time
(Galilean, Poincare and conformal symmetries) by fermionic generators,
which works as a theoretical method which naturally unifies fermions
and bosons. Therefore, it is natural that supersymmetry is part of
the most modern theoretical physics approaches to find an unified
theory of all fundamental interactions. A fact that suggests why supersymmetry
is important for the unification of interactions is as follows: the
supersymmetry (SUSY) offers a possible way to prevent the no-go theorem
of Coleman and Mandula\,\cite{Coleman1967}, establishing that the
Lie group of all symmetries of a given quantum field theory is composed
of a direct product of the Poincare group with an internal symmetry
group, or, in other words, the internal symmetry transformations always
commute with the Poincare transformations. The hypotheses of this
theorem are very general, consisting of the axioms of quantum field
theory and the assumption that all the symmetries can be described
in terms of Lie groups. One way around this theorem was found by Haag
and \L opusza\'{n}sk\,\cite{Haag1975}, making the assumption that
the infinitesimal generators of symmetry obey a superalgebra, or supersymmetric
Lie algebra. The superalgebra is a generalization of the notion of
Lie algebra, where some generators are fermionic, which means that
some commutation rules are replaced by anti-commutators. 

Besides this formal motivation, supersymmetry is also a possible solution
to hierarchy problems in particle physics. In the Standard Model (SM)
of particles, which is composed of the electroweak theory together
with Quantum Chronodynamics (QCD), the electroweak scale ($m_{EW}\sim10^{3}GeV$)
acquires huge quantum corrections at the Planck scale ($M_{Pl}\sim10^{18}GeV$),
creating a hierarchy problem between these scales, that must be accounted
for with some extraordinary fine tuning. SUSY solves this problem,
since in a supersymmetric version of the SM the hierarchy between
theses scales is attained in a natural manner, without fine tuning.
Supersymmetry can also help us to understand Grand Unified Theories
(GUTs)\,\cite{Weinberg1967,Weinberg1973,Pati1973,Georgi1974,Nath1975},
where the gauge group of the SM is embedded in a larger group, but
in general the gauge couplings do not unify properly at a given scale.
However, in the supersymmetric version of the SM called Minimal Supersymmetric
Standard Model (MSSM), the gauge couplings do unify nicely, provided
the scale of supersymmetry breaking is about $1TeV$. 

If supersymmetry has anything to do with nature, then it must be a
local symmetry. When SUSY is imposed as a local symmetry, Einstein's
theory of general relativity is included automatically, and the result
is said to be a theory of supergravity\,\cite{Nath1975,Freedman1976,Deser1976}.
In the context of string theory, which attempts to be a consistent
quantum gravity theory, the supersymmetry is an essential ingredient,
because all realistic models of string theory are built from superstrings\,\cite{gswbook:1988,gswbook:1988v2,weinbergbookv3}. 

After this mainly historical review of supersymmetry, we will discuss
some concepts developed in Quantum Field Theory (QFT), since we will
be interested in understanding how the effective superpotential work
in the context of supersymmetric models. 
\begin{figure}
\centering{}\includegraphics{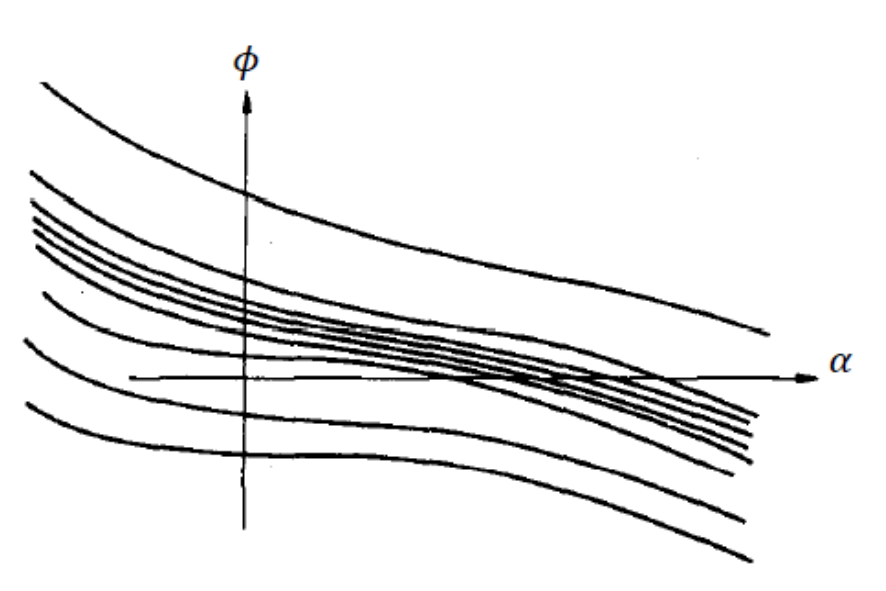}\caption{\label{fig:Characteristic-Nielsen}Characteristic of the partial differential
equation, Eq.\,(\ref{eq:NielsenIdentity-1}), where the effective
potential is constant in the plane $\phi-\alpha$\,\cite{Nielsen:1975fs}. }
\end{figure}

In QFT the effective potential is used to calculate physically meaningful
quantities, such as the masses for physical particles, and therefore
its possible gauge dependence is an important question that has been
studied in the literature for quite some time. Given that physical
observables cannot depend on the choice of gauge-fixing parameter,
it is essential to understand how to extract gauge independent information
from perturbative calculations of the effective action in gauge theories\,\cite{Jackiw:1974cv,Dolan1974,Frere1975}.
A very robust formalism to address this question was developed by
Nielsen, Kugo and Fukuda\,\cite{Nielsen:1975fs,Fukuda:1975di}, providing
identities that encode the behavior of the effective action under
changes of the gauge-fixing parameter. The so-called Nielsen identities
imply that the gauge dependence of the effective action is compensated
by a nonlocal field redefinition. For the effective potential, for
example, the Nielsen identity reads
\begin{align}
\left(\alpha\,\frac{\partial}{\partial\alpha}+C\left(\phi;\alpha\right)\frac{\partial}{\partial\phi}\right)V_{\eff}\left(\phi;\alpha\right) & =0\thinspace,\label{eq:NielsenIdentity-1}
\end{align}
where $\alpha$ is the gauge-fixing parameter, $\phi$ is the vacuum
expectation value of the scalar field, $V_{\eff}$ is the quantum
effective potential, and $C\left(\phi;\alpha\right)$ is a functional
that can be calculated in terms of Feynman diagrams. Equation\,(\ref{eq:NielsenIdentity-1})
defines characteristic curves in the $\phi-\alpha$ plane, such as
the ones represented in Figure\,\ref{fig:Characteristic-Nielsen},
over which the effective potential is constant. This mean that changes
in the gauge-fixing parameter are compensated by changes in the vacuum
expectation value. A consequence of this relation is that physical
quantities defined at extrema of the effective potential become gauge
independent\,\cite{Nielsen:1975fs,Aitchison:1983ns,Johnston1985}.
We find in the literature many examples of the application of the
Nielsen identities: in condensed matter physics, QCD, Quantum Electrodynamics
(QED), the SM, and Aharony-Bergman-Jafferis-Maldacena (ABJM) theory,
to name a few\,\cite{DoNascimento1987,Breckenridge1995,Gambino2000,Iguri2001,Gerhold2003,Lewandowski2013,Upadhyay2016}.

The Dynamical Symmetry Breaking (DSB) studied by Coleman and Weinberg\,\cite{Coleman:1973jx}
constitutes a very appealing scenario in QFT, where quantum corrections
are entirely responsible for the appearance of nontrivial minima of
the effective potential. When a given Lagrangian $\hat{\mathcal{L}}$
depends on the fields $\varphi$ and dimensionless coupling constants,
and there is no mass parameter in the classical Lagrangian, we say
that the model is scale invariant or conformally invariant. The essence
of the Coleman-Weinberg mechanism is to naturally generate mass via
quantum corrections, starting from a given conformal model. In the
case of massless scalar electrodynamics, for example, the classical
potential $\frac{\lambda}{4!}\varphi_{c}^{4}$ receives quantum corrections
and becomes
\begin{align}
V_{\mathrm{eff}} & =\frac{\lambda}{4!}\phi^{4}+\left(\frac{5}{1152\pi^{2}}\,\lambda^{2}+\frac{3}{64\pi^{2}}\,e^{2}\right)\phi^{4}\left(\ln\left[\frac{\phi^{2}}{M^{2}}\right]-\frac{25}{6}\right),\label{eq:CW-potential}
\end{align}
where $\lambda$ is the scalar coupling constant, $e$ is the gauge
coupling constant and $M$ is the renormalization scale. This corrected
potential has a minima at $\phi\neq0$, signaling the presence of
symmetry breaking in $\lambda=\frac{33}{8\pi^{2}}\,e^{4}$. In the
case of a classically scale invariant model, all mass scales are generated
by these quantum corrections and are fixed as functions of the symmetry
breaking scale. This scenario would be particularly interesting in
the SM, but earlier calculations pointed to a dead end\,\cite{Sher:1988mj}.
The one loop effective potential for the SM reads\,\cite{Elias:2003xp}
\begin{align}
V_{\eff} & =\frac{\lambda}{4!}\phi^{4}+\phi^{4}\left[\frac{12\lambda^{2}-3g_{t}^{3}}{64\pi^{2}}+\frac{3\left(3g_{2}^{4}+2g_{2}^{2}g^{\prime2}+g^{\prime4}\right)}{1024\pi^{2}}\right]\left(\ln\left[\frac{\phi^{2}}{M^{2}}\right]-\frac{25}{6}\right),
\end{align}
where $g_{2}$, $g^{\prime}$ and $g_{t}$ are the $SU\left(2\right)$,
the $U\left(1\right)$ and the top quark Yukawa coupling constants,
respectively. The large value of $g_{t}^{2}\sim1\gg g_{2}^{2},g^{\prime2}$
makes $\lambda\sim{\cal O}\left(g_{t}^{2}\right)\sim1$, so that subsequent
leading-logarithm terms such as $\sim\lambda^{3}\phi^{4}\ln^{2}\left[\frac{\phi^{2}}{M^{2}}\right]$
would be too large to ignore, rendering the one-loop approximation
useless to study a possible DSB. 

However, it has been shown that this conclusion, based on the effective
potential calculated up to one loop level, could be modified substantially
by using the Renormalization Group Equation (RGE)\,\cite{elias:2003zm,Elias:2003xp,Chishtie:2005hr,Elias2005,Meissner:2006zh,Meissner:2007xv,Meissner:2008gj,Meissner:2008uw,Quinto2014}
to sum up infinite subsets of higher loop contributions to the effective
potential. Indeed, in\,\cite{Elias:2003xp} the authors were the
first to use the RGE to sum up all leading-logarithm terms, finding
an improved version of effective potential. More than a quantitative
correction over the one loop result, this improvement lead to a new
scenario, where DSB was operational, in the sense that a consistent
perturbative minimum of the effective potential were found at $\phi\neq0$.
More recent calculations were able to include corrections up to five
loops in the effective potential\,\cite{Chishtie:2010ni,Steele:2012av},
bringing the prediction for the Higgs mass relatively close to the
experimental value indicated by the Large Hadron Collider (LHC) (for
other works regarding conformal symmetry in the SM, see for example\,\cite{Englert:2013gz,Chun:2013soa}).

Besides being a viable ingredient to the phenomenology of SM, DSB
also occurs in other contexts, such as lower dimensional theories.
Particularly interesting are models involving the Chern-Simons (CS)
term in $\left(2+1\right)$ dimensions\,\cite{Deser:1981wh,Schonfeld1981},
which exhibit fascinating properties such as massive gauge fields,
exotic statistics, and fractional spin, relevant qualities for the
study of the quantum Hall effect\,\cite{Pr.Gir}. Supersymmetric
CS theories have also been studied for quite some time\,\cite{Avdeev:1991za,ruizruiz:1997jq,lehum:2007nf,Ferrari:2010ex,Lehum:2010tt}.
In the context of superstring field theory and supergravity, they
have recently attracted much attention due to their relation to M2-branes\,\cite{Gaiotto:2007qi}:
the superconformal field theory describing multiple M2-branes is dual
to the $D=11$ supergravity on $AdS_{4}\times S^{7}$, which has $\mathcal{N}=8$
supersymmetry. However, the on-shell degrees of freedom of this theory
are exhausted by bosons and physical fermions making its gauge sector
to have no on-shell degrees of freedom. These requirements are satisfied
by a CS-matter theory called Bagger\textendash Lambert\textendash Gustavsson
(BLG) theory\,\cite{Gustavsson:2008dy,Bagger:2007vi,Bagger:2007jr,Bandres:2008ry,Antonyan:2008jf},
which describes two M2-branes. Relaxing the requirement of manifest
$\mathcal{N}=8$ supersymmetry, this approach can be generalized to
$\mathcal{N}=6$ CS-matter theory with the gauge group $U_{k}(N)\times U_{-k}(N)$
($k$ and $-k$ are CS levels)\,\cite{Aharony:2008ug,Naghdi:2011ex},
known as ABJM theory. The quantization of such model was thoroughly
studied in\,\cite{Faizal:2011en,Faizal:2014dca,Upadhyay:2014oda,Faizal:2012dj,Queiruga:2015fzn,Akerblom:2009gx,Bianchi:2009rf,Bianchi:2009ja,Bianchi:2010cx}.
Also, detailed calculations of the effective superpotential within
$\mathcal{N}=2$ superfield theories in $\left(2+1\right)$ dimension
have been reported in\,\cite{Buchbinder:2012zd,Buchbinder:2015swa}. 

The basic renormalization properties of CS models have been studied
in\,\cite{Chen:1990sc,Avdeev:1991za,Chen:1992ee,ruizruiz:1996yf,Tan:1997ew,Alves:1999hw,dias:2003pw}.
We shall be particularly interested in models with scale invariance
at the classical level, that is, with a pure CS field coupled to massless
scalars and fermions with Yukawa quartic interactions, and scalar
$\varphi^{6}$ self-interactions. In these models, the one-loop corrections
calculated using the dimensional reduction scheme\,\cite{Siegel:1979wq}
are rather trivial, since no singularities appear, and no DSB happens
either; at the two loop level, however, one finds renormalizable divergences.
Also, the two loops effective potential $V_{\eff}$ exhibits a nontrivial
minimum, signalizing the appearance of DSB. Due to the nontrivial
$\beta$ and $\gamma$ functions at two loops level, one may obtain
an improvement in the calculation of $V_{\eff}$ by imposing the RGE,
\begin{align}
\left[\mu\frac{\partial}{\partial\mu}+\beta_{x_{i}}\frac{\partial}{\partial x_{i}}-\gamma_{\varphi}\phi\frac{\partial}{\partial\phi}\right]V_{\eff}\left(\phi;\mu,x_{i},L\right) & =0\,,\label{eq:RGE1}
\end{align}
where $x$ denotes collectively the coupling constants of the theory,
$\mu$ is the mass scale introduced by the regularization, $\gamma_{\varphi}$
is the anomalous dimension of scalar field $\varphi$, 
\begin{align}
L & =\ln\left[\frac{\phi^{2}}{\mu}\right],\label{eq:defL}
\end{align}
and $\phi$ is the vacuum expectation value of $\varphi$. This improved
effective potential was calculated in\,\cite{Dias:2010it}, and it
was shown to imply in considerable changes in the properties of DSB
in this model, thus providing another context where the consideration
of the RGE is essential to a proper analysis of the phase structure
of the model. 

The supersymmetric CS model in the superfield formalism was considered
by Avdeev \emph{et.al.}\,\cite{Avdeev:1991za,Avdeev:1992jt}, which
computed the renormalization group functions in the Abelian and non-Abelian
cases using a particular gauge-fixing parameter in the $R_{\xi}$
gauge, assuming as true the gauge invariance. However, one must keep
in mind that, on general grounds, renormalization group functions
can depend on the choice of the gauge-fixing parameter\,\cite{Collins_Book,Fazio2001,Bell2013,DiLuzio2014,Bell2015}.
More recently, a perturbative study of a supersymmetric CS model in
$\left(2+1\right)$ dimensions coupled to Higgs field were presented
in\,\cite{lehum:2007nf}, showing the spontaneous breaking of the
gauge symmetry $U\left(1\right)$ by the first quantum corrections
of the effective action. The dynamical breaking of gauge symmetry
in a scale invariant supersymmetric CS model in $(2+1)$ dimensions
was studied in\,\cite{Ferrari:2010ex}, where the masses of the gauge
and matter superfields are dynamically generated by quantum corrections,
at two loops. A similar study in an abelian supersymmetric CS model
in $(2+1)$ dimensions with $\mathcal{N}=1,\,2$ supersymmetry in
the large $N$ limit, with arbitrary gauge-fixing parameter, was presented
in\,\cite{Lehum:2010tt}. 

In this work, we will consider supersymmetric models containing the
CS field with arbitrary gauge-fixing parameter in $\left(2+1\right)$
dimension in the superfield formalism. Our objective is to verify
whether the consideration of the RGE also induces considerable modifications
in the scenario of DSB in these models. We also want to settle the
question of a possible gauge dependence of the effective superpotential.
In order to do that, we start with the study of the Nielsen identity
in the context of supersymmetric CS theory, working in the superfield
formalism. We find the general Becchi-Rouet-Stora-Tyutin (BRST) transformations
associated to this theory, then use this result to obtain the Nielsen
identity for the effective superpotential $V_{\eff}^{S}$. The complete
development of the Nielsen formalism in the superfield language is
the first result of this thesis. We then calculate the renormalization
group functions with arbitrary gauge-fixing parameter, and we show
explicitly that these are gauge independent, a result we will show
can provide information on the gauge independence of the effective
potential. These results are reported in\,\cite{Quinto2016a}

The direct application of the Nielsen identity for superfield models
in $\left(2+1\right)$ space-time dimensions is complicated by the
difficulty in calculating the complete effective superpotential $V_{\eff}^{S}$
in the superfield language\,\cite{Ferrari:2009zx}. Because of that,
we will focus on the part of the effective superpotential which does
not depend on supercovariant derivatives of the background scalar
superfield, $K_{\eff}$. We will show that $K_{\eff}$ can be calculated
from the renormalization group functions of the theory, together with
the RGE. It follows then, from the gauge invariance of the renormalization
group functions, that $K_{\eff}$ also does not depend on the gauge
choice. We briefly discuss how to extend these results to calculate
the complete $V_{\eff}^{S}$. 

Then, we calculate the effective superpotential of a supersymmetric
CS theory coupled to $N$ matter superfields, up to two loops. To
this end, we use the RGE and the $\beta$ and $\gamma$ function calculated
in\,\cite{Quinto2016a}. With this result at hand, we discuss how
we can reorganize the expansion of the effective superpotential in
terms of Leading Logs (LL), Next-to-Leading Logs (NLL) contributions,
and so on, in a way that allows us to calculate coefficients arising
from higher orders corrections, thus improving the two loops evaluation
of the effective superpotential. We are able to find an improved effective
superpotential, which will be used to study DSB in our model. We show
that, contrary to what happens in the non-supersymmetric case\,\cite{Dias:2010it},
here the RGE improvement in general leads only to a slight modification
in the DSB scenario. Most of these results are presented in\,\cite{Quinto2016}.

In this thesis, all calculations are done in the superfield language\,\cite{gates:1983nr,buchbinder:1998qv},
in which the supersymmetry is manifest in all stages of the calculations.
The thesis is organized as follows: in Chapter\,\ref{chap:Supersymmetry-in-(2+1)},
we present the conventions and some definitions about supersymmetry
in $\left(2+1\right)$ dimensions to be used here. In Chapter\,\ref{chap:Nielsen-Identity-for},
we study the BRST transformations and the Nielsen identity for the
effective superpotential $V_{\eff}^{S}\left(\alpha,\sigma_{\mathrm{cl}}\right)$.
In Chapter\,\ref{chap:Renormalization-Group-Functions}, we calculate
the renormalization group functions, $\beta$ and $\gamma$ functions,
up to two loops, with arbitrary gauge-fixing parameter. In Chapter\,\ref{chap:RGE-Impro-DBS},
we calculate the effective superpotential $K_{\eff}\left(\alpha,\sigma_{\mathrm{cl}}\right)$
and its improvement using the RGE. From this result, we study the
DSB properties. Chapter\,\ref{chap:Conclusions-and-perspectives}
presents our conclusions and perspectives for the future. In Appendix\,\ref{chap:One-Loop-Correction}
we give an explicit example on how to calculate a one loop superdiagram.
Appendix\,\ref{chap:Two-loops-corrections} contains all the details
about the calculation of superdiagrams involved in the two loops evaluation
of the renormalization group functions. Finally, in Appendix\,\ref{chap:Useful-integrals}
we present the set of Feynman integrals used in this work.
\begin{center}
\myclearpage
\par\end{center}

\chapter{\label{chap:Supersymmetry-in-(2+1)}Supersymmetry in $(2+1)$ Dimensions}

In this chapter we review the notations and conventions necessary
to work with supersymmetry in $\left(2+1\right)$ dimensions, providing
an elementary introduction to the concept and the different types
of superfields that will be used here and the D-algebra that is important
in the evaluation of superdiagrams. 

\section{Conventions\label{subsec:Convention-index}}

We show the notations and conventions to be used in this thesis, following\,\cite{gates:1983nr}.
In the $\left(2+1\right)$ dimensional space-time with metric $\eta_{mn}=\mbox{diag}\left(-,+,+\right)$,
where $m,n=0,1,2$, the Lorentz group is $\mbox{SL}\left(2,\mathbb{R}\right)$
instead of $\mbox{SL}\left(2,\mathbb{C}\right)$ as in a $\left(3+1\right)$
dimensional space-time, and the fundamental representation of the
Lorentz group operate on two real components spinors $\psi^{\alpha}=\left(\psi^{1},\psi^{2}\right)$\footnote{Which happens to be a Majorana spinor.}.
All the spinor used here belong to a Grassmann algebra, meaning they
anti-commute,
\begin{align}
\psi^{\alpha}\psi^{\beta} & =-\psi^{\beta}\psi^{\alpha},\label{eq:C0}
\end{align}
therefore $\psi^{\alpha}\psi^{\alpha}=0$ and, as $\alpha=1,\,2$,
we have $\psi^{\alpha_{1}}\psi^{\alpha_{2}}...\psi^{\alpha_{n}}=0$,
$\forall\,n>2$. Greek letters $\alpha,\beta,\mu...\,$ will always
be used to label the components of a spinor, and a vector will be
represented by a bi-spinor, 
\begin{align}
x^{\alpha\beta} & =x^{m}\left(\gamma_{m}\right)^{\alpha\beta},\label{eq:C0-1}
\end{align}
where the $\gamma$ matrices are,
\begin{eqnarray}
\left(\gamma^{0}\right)_{\,\,\beta}^{\alpha}=-i\sigma^{2}, & \left(\gamma^{1}\right)_{\,\,\beta}^{\alpha}=\sigma^{1}, & \left(\gamma^{2}\right)_{\,\,\beta}^{\alpha}=\sigma^{3}\,,\label{eq:C1}
\end{eqnarray}
$\sigma^{i}$ being the Pauli matrices, and
\begin{equation}
\left\{ \left(\gamma_{m}\right)_{\,\beta}^{\alpha},\left(\gamma_{n}\right)_{\,\alpha}^{\sigma}\right\} =2\,\eta_{mn}\,\delta_{\,\beta}^{\sigma}\thinspace.\label{eq:C2}
\end{equation}
The $\gamma$ matrices with both lower indices $\left(\gamma^{m}\right)_{\alpha\beta}=\left(-\mbox{1},-\sigma^{3},\sigma^{1}\right)$
are symmetrical, therefore all vectors are represented by symmetric
bi-spinors.

The indices of the spinors are raised or lowered by contraction with
the anti-symmetric symbol, 
\begin{gather}
C_{\alpha\beta}=-C_{\beta\alpha}=-C^{\alpha\beta}=\begin{pmatrix}0 & -i\\
i & 0
\end{pmatrix},\label{eq:C3}
\end{gather}
with the property
\begin{align}
C_{\alpha\beta}C^{\gamma\delta} & \equiv\delta_{\alpha}^{\gamma}\delta_{\beta}^{\delta}-\delta_{\beta}^{\gamma}\delta_{\alpha}^{\delta}\,.\label{eq:C4}
\end{align}
The contraction of the spinor indices is defined by,
\begin{eqnarray}
\psi^{\alpha}=C^{\alpha\beta}\psi_{\beta}, &  & \psi_{\alpha}=\psi^{\beta}C_{\beta\alpha},\label{eq:C5}
\end{eqnarray}
and the following combination is a scalar,
\begin{align}
\psi^{2} & =\frac{1}{2}\psi^{\alpha}\psi_{\alpha}.\label{eq:C6}
\end{align}
Due to the anti-symmetry of $C_{\alpha\beta}$, the order in which
the indices are contracted is important, and Eq.\,(\ref{eq:C5})
define the order to be used in this work. It is important to say that,
due to the fact that $C_{\alpha\beta}$ is purely imaginary, $\psi^{\alpha}$
is real, whereas $\psi_{\alpha}$ is imaginary, and therefore $\psi^{2}$
is Hermitian.

\section{Supersymmetry and superfields}

\subsection{Superfields}

The superspace is an extension of the usual space-time that include
anti-commutative coordinates. Superfields are functions defined over
this space, with supercoordinates $z=\left(x^{\mu\nu},\theta^{\alpha}\right)=\left(x,\theta\right)$,
which have the property $z^{\dagger}=z$. We define the following
derivatives,
\begin{alignat}{1}
\partial_{\mu}\theta^{\nu} & \equiv\delta_{\mu}^{\,\,\nu}\,,\label{eq:S1}\\
\partial_{\mu\nu}x^{\sigma\tau} & \equiv\frac{1}{2}\left(\delta_{\mu}^{\sigma}\delta_{\nu}^{\tau}+\delta_{\nu}^{\sigma}\delta_{\mu}^{\tau}\right)\,,\label{eq:S2}
\end{alignat}
where Eq.\,(\ref{eq:S1}) define the derivation with respect to a
spinor coordinate and\,(\ref{eq:S2}) the derivation with rspect
to a space-time coordinate. The momentum operator satisfies
\begin{align}
\left(i\,\partial_{\mu\nu}\right)^{\dagger} & =i\,\partial_{\mu\nu}\,.\label{eq:S3}
\end{align}

The integration over an anti-commutative variable $\theta_{\alpha}$
is defined by\begin{subequations}\label{eq:S4}
\begin{align}
\int d\theta_{\alpha} & =\partial_{\alpha}\,,\\
\int d\theta_{\alpha}\theta^{\beta} & =\delta_{\alpha}^{\,\beta}\,,\\
\int d^{2}\theta\,\theta^{2} & =-1\,.
\end{align}
\end{subequations}We can define a function with the usual properties
of the Dirac delta for Grassmannian variables; if we compare the previous
equation with $\int d^{2}\theta\,\delta^{2}\left(\theta\right)=1$,
we can define
\begin{gather}
\delta^{2}\left(\theta\right)=-\theta^{2}=-\frac{1}{2}\theta^{\alpha}\theta_{\alpha}\,,\label{eq:S6}
\end{gather}
or, in general 
\begin{gather}
\delta^{2}\left(\theta-\theta'\right)=\underset{j=1}{\overset{2}{\Pi}}\left(\theta_{j}-\theta'_{j}\right)\,.\label{eq:S6-1}
\end{gather}

Superfields are functions defined over the superspace, for example:
$\Phi\left(z\right),$ $\Phi_{\alpha}\left(z\right)...$ A superfield
without a Lorentz index $\Phi\left(z\right)$ is called a scalar superfield,
a superfield with one index $\Phi_{\alpha}\left(z\right)$ is a spinor
superfield, and so on. Only these two types of superfields will be
relevant for our study. The transformations of these functions under
the action of the generators of the Poincare group, $P_{\mu\nu}$
(translations) and $M_{\alpha\beta}$ (rotations), are the usual from
field theory, taking into account that the bosonic coordinates $x^{\mu\nu}$
transform as a vector, and the fermionic coordinates $\theta^{\alpha}$
transforms as spinors.

The superalgebra or graded Poincare algebra is obtained by the inclusion
of a new supersymmetric spinor generator $Q_{\alpha}$, which satisfies
the following supersymmetric algebra or superalgebra:\begin{subequations}
\begin{align}
\left[P_{\mu\nu},P_{\rho\sigma}\right] & =0\thinspace,\label{eq:S7}\\
\left\{ Q_{\mu},Q_{\nu}\right\}  & =2P_{\mu\nu}\thinspace,\label{eq:S8}\\
\left[Q_{\mu},P_{\nu\rho}\right] & =0\thinspace.\label{eq:S9}
\end{align}
\end{subequations}This superalgebra is realized as differential operators
acting on superfields $\Phi\left(z\right),$ $\Phi_{\alpha}\left(z\right),\thinspace\ldots$
as follows\begin{subequations}
\begin{align}
Q_{\mu} & =i\left(\partial_{\mu}-i\theta^{\nu}\partial_{\nu\mu}\right)\,,\label{eq:S10}\\
P_{\mu\nu} & =i\,\partial_{\mu\nu}\,.\label{eq:S11}
\end{align}
\end{subequations}

The supersymmetric transformation acting on a scalar superfield is
\begin{align}
\Phi\left(x^{\mu\nu},\theta^{\nu}\right) & =\exp\left(i\left[\xi^{\lambda\rho}P_{\lambda\rho}+\eta^{\lambda}Q_{\lambda}\right]\right)\Phi\left(x'^{\mu\nu},\theta'^{\nu}\right)\,,\label{eq:S11-1}
\end{align}
where the infinitesimal generator $\xi^{\lambda\rho}P_{\lambda\rho}+\eta^{\lambda}Q_{\lambda}$
generates the following transformation on the supercoordinates, \begin{subequations}
\begin{align}
x'^{\mu\nu} & =x^{\mu\nu}+\xi^{\mu\nu}-\frac{1}{2}i\left(\theta^{\mu}\eta^{\nu}-\theta^{\nu}\eta^{\mu}\right)\,,\label{eq:S12}\\
\theta'^{\mu} & =\theta^{\mu}+\eta^{\mu}\,,\label{eq:S13}
\end{align}
\end{subequations}with $\xi^{\lambda\rho}$ and $\eta^{\lambda}$
being real constants parameters, $\xi$ a vector and $\eta$ an anti-commutative
spinor. These parameters generate a translation on space-time and
Grassmann coordinates.

\subsection{D-algebras\label{subsec:D-=0000E1lgebras}}

Equations \,(\ref{eq:S10}) - (\ref{eq:S13}) provide a representation
of the superalgebra given in Eqs.\,(\ref{eq:S7}) - (\ref{eq:S9}).
We notice that due to Eqs.\,(\ref{eq:S7}), (\ref{eq:S8}), (\ref{eq:S10})
and\,(\ref{eq:S11}), not only $\Phi$ and functions of $\Phi$,
but also their space-time derivatives $\partial_{\mu\nu}\Phi$, ...,
lead to a supersymmetric representation. Nevertheless, this is not
the case for the spinor derivative $\partial_{\mu}$ because it does
not commute with the supersymmetric spinor generator $Q_{\mu}$. This
motivates the introduction of the supercovariant derivatives, 
\begin{align}
D_{\alpha} & =\partial_{\alpha}-i\,\theta^{\beta}\partial_{\beta\alpha}\,,\label{eq:S14-0}
\end{align}
that anticommute with the generators of the supersymmetry. We will
provide a brief summary on the properties of these supercovariant
derivatives, since they are essential in defining many relevant supersymmetric
actions.

The supercovariant derivative anticommute with the supersymmetric
generators,
\begin{equation}
\left\{ D_{\mu},Q_{\nu}\right\} =0,\label{eq:S14}
\end{equation}
and commute with the translation generators,
\begin{equation}
\left[D_{\mu},P_{\beta\gamma}\right]=0.\label{eq:S16}
\end{equation}
Finally, from Eq.\,(\ref{eq:S14-0}), we can show that, 
\begin{align}
\left\{ D_{\mu},\,D_{\nu}\right\}  & =2i\,\partial_{\mu\nu}=2P_{\mu\nu}\,.\label{eq:S15}
\end{align}
Next, multiplying $D_{\alpha}D_{\beta}\left(C^{\alpha\beta}C_{\mu\nu}\right)$
and using Eq.\,(\ref{eq:C4}), we find
\begin{align}
\left[D_{\mu},D_{\nu}\right] & =-2\,C_{\mu\nu}D^{2}\,.\label{eq:S17}
\end{align}

We will now deduce some identities that will be of much help for the
manipulation of supercovariant derivatives. We start by adding Eqs.\,(\ref{eq:S15})
and\,(\ref{eq:S17}) to obtain 
\begin{align}
D_{\mu}D_{\nu} & =i\partial_{\mu\nu}-C_{\mu\nu}D^{2}.\label{eq:S18}
\end{align}
If we multiply by $D^{\nu}$ in the left hand side of Eq.\,(\ref{eq:S15}),
using Eq.\,(\ref{eq:S18}) we get,
\begin{align}
D^{\nu}D_{\mu}D_{\nu} & =0\,,\label{eq:S19}
\end{align}
and using Eqs.\,(\ref{eq:S15}) - (\ref{eq:S19}) we obtain 
\begin{align}
\left(D^{2}\right)^{2} & =\square\thinspace,\label{eq:S20}
\end{align}
where it was used the identity
\begin{align}
\partial^{\alpha\beta}\partial_{\nu\beta} & =\delta_{\nu}^{\alpha}\square\,,\label{eq:S21}
\end{align}
and $\square=\nicefrac{1}{2}\thinspace\partial^{\alpha\beta}\partial_{\alpha\beta}$
is called the d'Alembertian. Finally, if we multiply $D^{\nu}$ in
the left hand side of Eq.\,(\ref{eq:S15}), using Eq.\,(\ref{eq:S18})
and\,(\ref{eq:S19}), we have
\begin{gather}
D^{2}D_{\mu}=-D_{\mu}D^{2}=i\,\partial_{\mu\beta}D^{\beta}.\label{eq:S22}
\end{gather}

Supercovariant derivatives satisfy the Leibniz rules and can be integrated
by parts when they appear inside of an integral $d^{3}x\,d^{2}\theta$,
since they are composed of derivatives in $x$ and $\theta$. Finally,
from Eq.\,(\ref{eq:S6}) and the $\left(D^{\alpha},D^{2}...\right)$,
we can prove the following rules that will be used for the calculation
of superdiagrams,
\begin{alignat}{1}
\delta^{2}\left(\theta_{n}-\theta_{m}\right)\delta^{2}\left(\theta_{m}-\theta_{n}\right) & =0,\nonumber \\
\delta^{2}\left(\theta_{n}-\theta_{m}\right)D_{\alpha}\delta^{2}\left(\theta_{m}-\theta_{n}\right) & =0,\nonumber \\
\delta^{2}\left(\theta_{n}-\theta_{m}\right)D^{2}\delta^{2}\left(\theta_{m}-\theta_{n}\right) & =\delta^{2}\left(\theta_{n}-\theta_{m}\right).\label{eq:S23}
\end{alignat}

\subsection{Components by expansion and projections}

Superfields can be expanded in a power series with respect to the
Grassmannian coordinates $\theta$, and since these have the property
that their square is zero, the power series have a finite number of
terms. For example, we consider a scalar superfield written in terms
of a power series in $\theta$,
\begin{gather}
\Phi\left(x,\theta\right)=\varphi\left(x\right)+\theta^{\alpha}\Psi_{\alpha}\left(x\right)-\theta^{2}E\left(x\right)\,,\label{eq:24}
\end{gather}
where $\varphi\left(x\right)$, $\Psi_{\alpha}\left(x\right)$ and
$E\left(x\right)$ are a scalar field, a Dirac field and an auxiliary
scalar field, respectively. They are the component fields of the superfield
$\Phi\left(x^{\mu\nu},\theta^{\mu}\right)$, as obtained by power
expansion in $\theta$. A more efficient way tothem comes from observing
that the components in Eq.\,(\ref{eq:24}) can also be defined by
projection:\begin{subequations} 
\begin{align}
\varphi\left(x\right) & =\Phi\left(x,\theta\right)|,\\
\Psi_{\alpha}\left(x\right) & =D_{\alpha}\Phi\left(x,\theta\right)|,\\
E\left(x\right) & =D^{2}\Phi\left(x,\theta\right)|.\label{eq:S32}
\end{align}
\end{subequations}where $|$ means that after the differentiation,
just the $\theta$ independent part is kept. Finally, it is convenient
to notice that 
\begin{gather}
\int d^{3}x\,d^{2}\theta\,\Phi\left(x,\theta\right)=\int d^{3}x\,\partial^{2}\Phi\left(x,\theta\right)=\int d^{3}x\,D^{2}\Phi\left(x,\theta\right)|\thinspace.\label{eq:23-1}
\end{gather}

\section{Scalar multiplet}

We consider a complex scalar superfield\begin{subequations}\label{eq:Define-P1-P2}
\begin{align}
\Phi\left(x,\theta\right) & =\frac{1}{\sqrt{2}}\left[\Phi_{1}\left(x,\theta\right)+i\Phi_{2}\left(x,\theta\right)\right],\\
\overline{\Phi}\left(x,\theta\right) & =\frac{1}{\sqrt{2}}\left[\Phi_{1}\left(x,\theta\right)-i\Phi_{2}\left(x,\theta\right)\right],
\end{align}
\end{subequations}where $\Phi_{1}\left(x,\theta\right)$ and $\Phi_{2}\left(x,\theta\right)$
are real scalar superfields. On dimensional grounds, we postulate
the free Lagrangian 
\begin{gather}
\mathcal{S}_{\mathrm{K}}=-\frac{1}{2}\,\int d^{3}x\,d^{2}\theta\left(D^{\alpha}\overline{\Phi}\,D_{\alpha}\Phi\right),\label{eq:S33}
\end{gather}
 which is invariant under constant phase transformations,\begin{subequations}\label{eq:S43}
\begin{align}
\Phi & \rightarrow\Phi'=e^{i\mathcal{K}}\,\Phi,\\
\overline{\Phi} & \rightarrow\overline{\Phi}'=\overline{\Phi}\,e^{-i\mathcal{K}}.
\end{align}
\end{subequations}where $\mathcal{K}$ is a constant scalar superfield.
In the case of a gauge theory, we will have to consider $\mathcal{K}$
as a general, non-constant scalar superfield $\mathcal{K}\left(x,\theta\right)$,
i.e.,
\begin{gather}
\mathcal{K}\left(x,\theta\right)=\omega\left(x\right)+\theta^{\mu}\sigma_{\mu}\left(x\right)-\theta^{2}\tau\left(x\right)\thinspace,\label{eq:S44}
\end{gather}
where $w\left(x\right)$, $\tau\left(x\right)$ and $\sigma_{\mu}\left(x\right)$
can be defined by projection as\begin{subequations}\label{eq:S45}
\begin{align}
\omega\left(x\right) & =\mathcal{K}|,\\
\sigma_{\alpha}\left(x\right) & =D_{\alpha}\mathcal{K}|,\\
\tau\left(x\right) & =D^{2}\mathcal{K}|.
\end{align}
\end{subequations}

From Eq.\,(\ref{eq:S8}), we can see that $\theta$ has mass dimension
$-\frac{1}{2}$. Therefore, we can say that $\left[\Phi\left(x,\theta\right)\right]=M^{\frac{1}{2}}$
and then $\left[\Psi_{\mu}\left(x\right)\right]=M^{1}$. Thus $\left[\varphi\left(x\right)\right]=M^{\frac{1}{2}}$
and $\left[E\left(x\right)\right]=M^{\frac{3}{2}}$, this imply that
$E\left(x\right)$ will not be a dynamical field, so it is called
an auxiliary field; the same applies to the $\tau$ field in\,(\ref{eq:S44}).
These dimensional considerations are also the motivation for the definition
of the free massless kinetic action in Eq.\,(\ref{eq:S33}).

The action in terms of components fields can be obtained by power
expansion and integration in $\theta$. Notwithstanding, we prefer
to use the alternative procedure, based in projection, Eq.\,(\ref{eq:S32}),
which consist in integrating by parts Eq.\,(\ref{eq:S33}),
\begin{align}
\mathcal{S}_{\mathrm{K}}= & -\frac{1}{2}\underset{\mbox{ surface integral}}{\underbrace{\int d^{3}x\,d^{2}\theta D^{\alpha}\left(\overline{\Phi}D_{\alpha}\Phi\right)}}+\frac{1}{2}\int d^{3}x\,d^{2}\theta\,\overline{\Phi}D^{\alpha}D_{\alpha}\Phi=\int d^{3}x\,d^{2}\theta\,\overline{\Phi}D^{2}\Phi,
\end{align}
next, using the identity\,(\ref{eq:23-1}),
\begin{align*}
\mathcal{S}_{\mathrm{K}} & =\int d^{3}xD^{2}\left(\overline{\Phi}D^{2}\Phi\right)|=\int d^{3}x\left[D^{2}\overline{\Phi}D^{2}\Phi+\overline{\Phi}\left(D^{2}\right)^{2}\Phi+D^{\alpha}\overline{\Phi}D_{\alpha}D^{2}\Phi\right]|,
\end{align*}
and finally, using the projections defined in\,(\ref{eq:S32}), we
obtain 
\begin{align}
\mathcal{S}_{\mathrm{K}} & =\frac{1}{2}\int d^{3}x\left[\overline{E}\left(x\right)E\left(x\right)+\overline{\varphi}\left(x\right)\,\square\,\varphi\left(x\right)+\overline{\Psi}^{\alpha}\left(x\right)\,i\,\partial_{\alpha}^{\,\beta}\Psi_{\beta}\left(x\right)\right],\label{eq:S34}
\end{align}
where the auxiliary fields $E\left(x\right)$ and $\overline{E}\left(x\right)$
can be eliminated by means of their equation of motion ($E=0$ and
$\overline{E}=0$ in the massless case). Mass and interaction terms
can be included in Eq.\,(\ref{eq:S33}): 
\begin{align}
\mathcal{S}_{\mathrm{I}} & =\int d^{2}\theta d^{3}x\,\left(m\bar{\Phi}\Phi+V\left(\Phi,\bar{\Phi}\right)\right)\thinspace,
\end{align}
where $V\left(\Phi,\bar{\Phi}\right)$ is a function involving monomials
of three or more powers of the scalar superfield.

\section{\label{sec:Multipleto-vetorial}Vector multiplet}

A spinor gauge superfield $\Gamma_{\alpha}\left(x,\theta\right)$
can be used to describe a massless gauge vector field, together with
its fermionic superpartner.

If the supercovariant spinor derivative $D_{\alpha}$ is substituted
by the gauge supercovariant derivative $\nabla_{\alpha}$,
\begin{gather}
D_{\alpha}\rightarrow\nabla_{\alpha}=D_{\alpha}-ig\,\Gamma_{\alpha}\,,\label{eq:S46}
\end{gather}
and using Eq.\,(\ref{eq:S33}), we obtain
\begin{gather}
\mathcal{S}=-\frac{1}{2}\int d^{3}x\,d^{2}\theta\left(\overline{\nabla_{\alpha}\Phi}\nabla^{\alpha}\Phi\right)\,,\label{eq:S46-1}
\end{gather}
where this Lagrangian describes the coupling of a spinor superfield
with a complex scalar superfield, and it is invariant over local phase
transformations of the form Eq.\,(\ref{eq:S43}).

We can expand the gauge superfield $\Gamma_{\alpha}\left(x,\theta\right)$
in power series of $\theta$,
\begin{gather}
\Gamma_{\alpha}\left(x,\theta\right)=\chi_{\alpha}\left(x\right)-\theta_{\alpha}R\left(x\right)+i\theta^{\beta}A_{\beta\alpha}\left(x\right)-\theta^{2}\Lambda_{\alpha}\left(x\right),\label{eq:S47}
\end{gather}
where\begin{subequations}
\begin{align}
\chi_{\alpha}\left(x\right) & =\left.\Gamma_{\alpha}\left(x,\theta\right)\right|,\\
R\left(x\right) & =\frac{1}{2}\left.D^{\alpha}\Gamma_{\alpha}\left(x,\theta\right)\right|,\\
A_{\alpha\beta}\left(x\right) & =-\frac{i}{2}\left.\left[D_{\alpha}\Gamma_{\beta}\left(x,\theta\right)+D_{\beta}\Gamma_{\alpha}\left(x,\theta\right)\right]\right|,\\
\Lambda_{\alpha}\left(x\right) & =2\rho_{\alpha}\left(x\right)+i\partial_{\beta\alpha}\chi^{\beta}\left(x\right)=\left.D^{2}\Gamma_{\alpha}\left(x,\theta\right)\right|,
\end{align}
\end{subequations}where $\chi_{\alpha}\left(x\right)$ and $R\left(x\right)$
are auxiliary fields and $A_{\alpha\beta}$ is the vector potential
written as a bi-spinor. The mass dimensions of these fields can be
shown to be $\left[\Gamma_{\alpha}\left(x,\theta\right)\right]=M^{\frac{1}{2}}$
, $\left[\chi_{\alpha}\left(x\right)\right]=M^{\frac{1}{2}}$, $\left[R\left(x\right)\right]=M^{1}$
and $\left[A_{\alpha\beta}\left(x\right)\right]=M^{1}$. 

Introducing the so-called field strength superfield,
\begin{gather}
W_{\alpha}\left(x,\theta\right)\equiv\frac{1}{2}D^{\beta}D_{\alpha}\Gamma_{\beta}\left(x,\theta\right)=\frac{1}{2}D_{\alpha}D^{\beta}\Gamma_{\beta}\left(x,\theta\right)+D^{2}\Gamma_{\alpha}\left(x,\theta\right)\,,\label{eq:S51}
\end{gather}
we can check that $W_{\alpha}$ is gauge invariant and satisfies a
Bianchi identity\footnote{In usual Abelian gauge field theory, the field strength tensor $F_{\mu\nu}$
satisfy the Bianchi identity because they are written in terms of
the vector potential as $F_{\mu\nu}=\partial_{\mu}A_{\nu}-\partial_{\nu}A_{\mu}\,$;
these are identities that do not bring dynamic information. For gauge
theories described by supercovariant derivatives, the Bianchi identities
are just the Jacobi identities,
\begin{align*}
[\nabla_{\alpha}\,,[\nabla_{\beta}\,,\nabla_{\gamma}\}\} & =0\,,
\end{align*}
where $[\,\,\}$ is the graduated anti-symmetrization symbol, similar
to the usual anti-symmetrization symbol but with an extra factor of
$(-1)$ for each interchange between fermion indices.},
\begin{equation}
D^{\alpha}W_{\alpha}=0\thinspace.
\end{equation}
We expand the superfield $W_{\alpha}\left(x,\theta\right)$ in power
series in $\theta$,
\begin{gather}
W_{\alpha}\left(x,\theta\right)=\rho_{\alpha}\left(x\right)+\theta^{\beta}f_{\beta\alpha}\left(x\right)-\theta^{2}\Theta\left(x\right),\label{eq:S52}
\end{gather}
where\begin{subequations}
\begin{align}
\rho_{\alpha}\left(x\right) & =\left.W_{\alpha}\left(x,\theta\right)\right|,\\
f_{\beta\alpha}\left(x\right) & =\frac{1}{2}\left[\partial_{\beta\rho}A_{\,\,\alpha}^{\rho}+\partial_{\alpha\rho}A_{\,\,\beta}^{\rho}\right]=\left.D_{\beta}W_{\alpha}\left(x,\theta\right)\right|,\\
\Theta\left(x\right) & =i\partial_{\alpha}^{\,\,\beta}\rho_{\beta}\left(x\right)=\left.D^{2}W_{\alpha}\left(x,\theta\right)\right|,
\end{align}
\end{subequations}with $\rho_{\alpha}\left(x\right)$ being the photino
field, and $f_{\alpha\beta}\left(x\right)$ \textbackslash{}the bi-spinor
Maxwell field strength. We can verify that $\left[W_{\alpha}\left(x,\theta\right)\right]=M^{\frac{3}{2}}$,
$\left[\rho_{\alpha}\left(x\right)\right]=M^{\frac{3}{2}}$ and $\left[f_{\alpha\beta}\left(x\right)\right]=M^{2}$
are their mass dimensions.

We can also define a Lagrangian involving the vector superfield. In
this thesis, we will be interested in studying the Abelian supersymmetric
CS action defined as
\begin{align}
\mathcal{S}_{\mathrm{CS}} & =-\frac{1}{2}\int d^{3}x\,d^{2}\theta\,\Gamma^{\alpha}W_{\alpha},\label{eq:S57-1}
\end{align}
or, in terms of components fields,
\begin{align}
\mathcal{S}_{\mathrm{CS}} & =-\int d^{3}x\left(A^{\alpha\beta}\left(x\right)f_{\alpha\beta}\left(x\right)-\rho^{\alpha}\left(x\right)\rho_{\alpha}\left(x\right)\right).\label{eq:S57-2}
\end{align}
To rewrite Eq.\,(\ref{eq:S46-1}), describing the interaction between
the gauge and the scalar superfields, in terms of component fields,
the more efficient way is to define the components of $\Phi$ via
projection,
\begin{align}
\varphi\left(x\right) & =\Phi\left(x,\theta\right)|\nonumber \\
\Psi_{\alpha}\left(x\right) & =\nabla_{\alpha}\Phi\left(x,\theta\right)|\nonumber \\
F\left(x\right) & =\nabla^{2}\Phi\left(x,\theta\right)|,\label{eq:S60}
\end{align}
and use that 
\begin{gather}
\int d^{3}x\,d^{2}\theta=\int d^{3}x\,D^{2}|=\int d^{3}x\,\nabla^{2}|\,.\label{eq:S62}
\end{gather}
Therefore, Eq.\,(\ref{eq:S46-1}) can be written as follows,
\begin{align}
\mathcal{S} & =\int d^{3}x\,\nabla^{2}\left[\overline{\Phi}\nabla^{2}\Phi\right]|\nonumber \\
 & =\int d^{3}x\left[\nabla^{2}\overline{\Phi}\nabla^{2}\Phi+\frac{1}{2}\nabla^{\alpha}\overline{\Phi}\nabla_{\alpha}\nabla^{2}\Phi+\frac{1}{2}\,C_{\alpha\beta}\nabla^{\beta}\overline{\Phi}\,C^{\alpha\rho}\,\nabla_{\rho}\nabla^{2}\Phi+\overline{\Phi}\left(\nabla^{2}\right)^{2}\Phi\right]|\nonumber \\
 & =\int d^{3}x\left[\overline{F}\left(x\right)F\left(x\right)+i\overline{\Psi}^{\alpha}\left(x\right)\left(\partial_{\alpha}^{\beta}-ig\,A_{\alpha}^{\beta}\left(x\right)\right)\Psi_{\beta}\left(x\right)+i\overline{\Psi}^{\alpha}\left(x\right)\Lambda_{\alpha}\left(x\right)\varphi\left(x\right)\right.\nonumber \\
 & \left.-i\overline{\varphi}\left(x\right)\Lambda^{\alpha}\left(x\right)\Psi_{\alpha}\left(x\right)+\overline{\varphi}\left(x\right)\left(\partial_{\alpha\beta}-ig\,A_{\alpha\beta}\left(x\right)\right)^{2}\varphi\left(x\right)\right]\,,\label{eq:S63}
\end{align}
where we used the algebra of supercovariant derivatives and the property,
\begin{align}
\left(\nabla^{2}\right)^{2} & =\square-iW^{\alpha}\nabla_{\alpha}\,.\label{eq:S64}
\end{align}

\begin{center}
\myclearpage
\par\end{center}

\chapter{\label{chap:Nielsen-Identity-for}Nielsen Identity}

In this chapter we study the Nielsen identity for the supersymmetric
CS matter model in the superfield formalism. The Nielsen identity
is essential to understand the gauge invariance of the symmetry breaking
mechanism, and it is calculated by using the BRST invariance of the
model. We will also discuss the technical difficulties in applying
this identity to the complete effective superpotential $V_{\eff}^{S}$.

\section{\label{sec:Building-an-invariant-Lagrangian}Building a Lagrangian
invariant under BRST transformations}

We will investigate the BRST transformations in $\mathcal{N}=1$ supersymmetric
CS-Higgs model in $\left(2+1\right)$ dimensions
\begin{align}
\mathcal{S}_{\mathrm{CS}} & =\int d^{5}z\mathcal{L}_{\mathrm{CSM}}\left(z\right)=\int d^{5}z\left\{ -\frac{1}{2}\Gamma^{\alpha}W_{\alpha}-\frac{1}{2}\overline{\nabla^{\alpha}\Phi}\nabla_{\alpha}\Phi-m\,\overline{\Phi}\Phi+\frac{\lambda}{4}\left(\overline{\Phi}\Phi\right)^{2}\right\} ,\label{eq:actionCSM}
\end{align}
where $m$ is a mass parameter. In this chapter, we work with a nonvanishing
mass parameter $m$ (Higgs model), instead of a theory with conformal
invariance at the classical level (Coleman-Weinberg\,\cite{Coleman:1973jx}
model) for the sake of generality, and also because the $m=0$ case
would be more complicated to analyze in the context of the Nielsen
identities\,\cite{Kang1974,Nielsen:1975fs,Aitchison:1983ns}. 

The action in Eq.\,(\ref{eq:actionCSM}) is invariant under the gauge
transformations
\begin{eqnarray}
\delta\Phi=ig\mathcal{K}\Phi,\: & \delta\overline{\Phi}=-ig\mathcal{K}\overline{\Phi},\: & \delta\Gamma_{\alpha}=D_{\alpha}\mathcal{K},\label{eq:GaugeTrans}
\end{eqnarray}
where $\mathcal{K}$ is a scalar superfield. From the gauge transformations
we can obtain a set of BRST transformations\,\cite{Becchi1974,Ryder1996,SrednickiTQC}.
First we define $\mathcal{K}=\varepsilon\,C$, where $C$ is a ghost
superfields and $\epsilon$ is a constant infinitesimal parameter,
both being Grassmannian (so that $\varepsilon^{2}=0$). Therefore,
Eq.\,(\ref{eq:GaugeTrans}) can be rewritten as
\begin{eqnarray}
\delta_{B}\Phi=i\varepsilon gC\Phi,\: & \delta_{B}\overline{\Phi}=-i\varepsilon gC\overline{\Phi},\: & \delta_{B}\Gamma_{\alpha}=-\varepsilon D_{\alpha}C.\label{eq:BRSTTrans-1}
\end{eqnarray}
The BRST transformations for ghosts fields $C$ and $\overline{C}$
are defined as usual, i.e., in the Abelian model, 
\begin{align}
\delta_{B}C & =0,\label{eq:BRSTTrans-2}
\end{align}
since $\delta_{B}C$ would depend on the structure constants of the
gauge group, and
\begin{align}
\delta_{B}\overline{C} & =\epsilon B\left(z\right),\label{eq:BRSTTrans-3}
\end{align}
where $B\left(z\right)$ is a scalar superfield, known as the Nakanishi-Lautrup
auxiliary field in quantum field theory. Also, it is known that all
BRST transformations are nilpotent, i.e.,
\begin{equation}
\delta_{B}^{2}C=\delta_{B}^{2}\overline{C}=\delta_{B}^{2}\Gamma_{\alpha}=\delta_{B}^{2}\Phi=\delta_{B}^{2}\overline{\Phi}=0\,,
\end{equation}
which implies that $\delta_{B}B\left(z\right)=0.$ 

We define the total Lagrangian as
\begin{align}
\mathcal{L}_{t} & =\mathcal{L}_{\mathrm{CSM}}+\delta_{B}\mathcal{O},\label{eq:Lagran-t}
\end{align}
which is invariant by the BRST transformations, Eqs.\,(\ref{eq:BRSTTrans-1})
- (\ref{eq:BRSTTrans-3}). We choose the operator 
\begin{align*}
\mathcal{O}\left(z\right) & =\overline{C}\left(z\right)\left(-\alpha\,\frac{1}{4}B\left(z\right)+\frac{1}{2}\,F\left(z\right)\right),
\end{align*}
where $F\left(z\right)$ and $\alpha$ are the gauge fixing function
and a gauge-fixing parameter, respectively. Applying the BRST transformation
on the operator $\mathcal{O}\left(z\right)$, we have
\begin{align}
\delta_{B}\mathcal{O} & =-\varepsilon\,\frac{\alpha}{4}B^{2}+\varepsilon\frac{1}{2}B\,F+\frac{1}{2}\overline{C}\left(\delta_{B}F\right).\label{eq:Delta-O}
\end{align}
There is no derivatives acting on the scalar superfield $B\left(z\right)$,
which appears only quadratically and linearly in $\delta_{B}\mathcal{O}$.
Therefore, we can perform the path integral over it, the result being
equivalent to solving the classical equation of motion with respect
to $B$,
\begin{align}
\frac{\partial\delta_{B}\mathcal{O}}{\partial B} & =-\varepsilon\,\frac{\alpha}{2}\,B+\varepsilon\,\frac{1}{2}\,F=0,\label{eq:Cond-Delta-O}
\end{align}
and from this, we find
\begin{align}
B & =\alpha\,F,\label{eq:Valor-B}
\end{align}
which leads to
\begin{align}
\mathcal{O}\left(z\right) & =\frac{1}{4}\,\overline{C}\left(z\right)\,F\left(z\right),\label{eq:Operador-O}
\end{align}
and therefore
\begin{align}
\delta_{B}\mathcal{O} & =\varepsilon\,\frac{1}{4\alpha}\,F^{2}+\frac{1}{2}\,\overline{C}\left(\delta_{B}F\right).\label{eq:DeltaBRST-O}
\end{align}
The total Lagrangian, Eq.\,(\ref{eq:Lagran-t}) becomes
\begin{align}
\mathcal{L}_{t} & =\mathcal{L}_{\mathrm{CSM}}+\frac{1}{4\alpha}\,F^{2}+\frac{1}{2}\,\overline{C}\,\frac{\delta F}{\delta\mathcal{K}}\,C\,.\label{eq:Lagran-Total}
\end{align}
By rewriting the scalar superfields as in Eq.\,(\ref{eq:Define-P1-P2}),
we obtain
\begin{align}
\mathcal{L}_{t} & =\frac{1}{2}\,\Gamma_{\alpha}W^{\alpha}+\frac{1}{2}\left(\Phi_{1}D^{2}\Phi_{1}+\Phi_{2}D^{2}\Phi_{2}\right)-\frac{1}{2}m\left(\Phi_{1}^{2}+\Phi_{2}^{2}\right)-\frac{1}{4}g^{2}C^{\alpha\beta}\Gamma_{\beta}\Gamma_{\alpha}\left(\Phi_{1}^{2}+\Phi_{2}^{2}\right)\nonumber \\
 & +\frac{1}{2}\,g\left[\Phi_{1}D^{\alpha}\Phi_{2}-\Phi_{2}D^{\alpha}\Phi_{1}\right]\Gamma_{\alpha}-\frac{\lambda}{16}\,\left(\Phi_{1}^{2}+\Phi_{2}^{2}\right)^{2}+\frac{1}{4\alpha}\,F^{2}+\frac{1}{2}\,\overline{C}\,\frac{\delta F}{\delta\mathcal{K}}\,C.\label{eq:Lagran-Total-1}
\end{align}

We will use a supersymmetric generalization of the $R_{\xi}$ defined
in usual QFT\,\cite{SrednickiTQC,Ferrari:2010ex,lehum:2007nf}, defined
by the gauge fixing function
\begin{align}
F & =D^{\alpha}\Gamma_{\alpha}+d\,g\,\Phi_{2},\label{eq:Fixing-Gauge}
\end{align}
where $d$ is an arbitrary parameter that can be chosen to eliminate
the mixing between the $\Phi_{2}$ and $\Gamma_{\alpha}$, for example\,\cite{Ferrari:2010ex,lehum:2007nf}.
We will leave the value of $d$ unspecified, in which case in general
one would need to consider mixed propagators to evaluate quantum corrections\,\cite{Nielsen:1975fs,Aitchison:1983ns,Johnston1985,DoNascimento1987}.
With this choice of gauge fixing, we end up with the Lagrangian
\begin{align}
\mathcal{L}_{t} & =\frac{1}{4}\Gamma_{\alpha}W^{\alpha}+\frac{1}{2}\left(\Phi_{1}\left[D^{2}-m\right]\Phi_{1}+\Phi_{2}\left[D^{2}-m\right]\Phi_{2}\right)+\frac{1}{2}\,g\left[\Phi_{1}D^{\alpha}\Phi_{2}-\Phi_{2}D^{\alpha}\Phi_{1}\right]\Gamma_{\alpha}\nonumber \\
 & -\frac{1}{4}\,g^{2}C^{\alpha\beta}\Gamma_{\beta}\Gamma_{\alpha}\left(\Phi_{1}^{2}+\Phi_{2}^{2}\right)+\frac{\lambda}{16}\left(\Phi_{1}^{2}+\Phi_{2}^{2}\right)^{2}+\frac{1}{4\alpha}\,F^{2}+\overline{C}\,D^{2}\,C+\frac{1}{2}\,d\,g^{2}\,\Phi_{1}\overline{C}C,\label{eq:Lagran-Total-2}
\end{align}
which is invariant under the BRST transformations\begin{subequations}
\begin{align}
\delta_{B}\Gamma_{\alpha} & =-\varepsilon D_{\alpha}C,\\
\delta_{B}\Phi_{1} & =-\varepsilon g\Phi_{2}C,\\
\delta_{B}\Phi_{2} & =\varepsilon g\Phi_{1}C,\\
\delta_{B}\overline{C} & =-\varepsilon\frac{1}{\alpha}\,F,\\
\delta_{B}C & =0\,.
\end{align}
\end{subequations}Finally, we add to the Lagrangian the source terms
\begin{equation}
\mathcal{L}_{\mathrm{source}}=J^{\mu}\Gamma_{\mu}+\overline{\eta}C+\overline{C}\eta+f_{1}\Phi_{1}+f_{2}\Phi_{2}-gK_{1}\Phi_{2}C+gK_{2}\Phi_{1}C+h\mathcal{O}\thinspace,\label{eq:Lagran-Fonte}
\end{equation}
where $J^{\mu},\,\overline{\eta},\,\eta,\,f_{1}$, and $f_{2}$ are
the sources for the basic superfields, while $K_{1},\,K_{2}$ and
$h$ are sources for the composite operators.

\section{\label{sec:Calculating-the-Nielsen-Identity}Obtaining the Nielsen
identity}

Our starting point is the generating functional,
\begin{align}
Z\left[J_{a}\right] & =e^{iW\left[J_{a}\right]}=N\int\mathcal{D}\phi_{a}e^{iS},\label{eq:Func-Geradora-1}
\end{align}
or
\begin{align}
W\left[J_{a}\right] & =-i\,\ln Z\left[J_{a}\right],\label{eq:Func-Geradora-2}
\end{align}
where $J_{a}$ represent all the sources and $\mathcal{D}\phi_{a}$
the path integral over all superfields present in the action 
\begin{equation}
S=\int d^{5}z\,\left(\mathcal{L}_{t}+\mathcal{L}_{\mathrm{source}}\right).
\end{equation}

Applying the BRST transformation $\phi_{a}\rightarrow\phi_{a}+\delta_{B}\phi_{a}$
on $W\left[J_{a}\right]$, we observe that the generating functional
does not change since the functional integration covers all the possible
field configurations; in other words, the functional only depends
on the sources and not on the superfields. Therefore, this BRST transformation
leaves $\mathcal{L}_{t}$ and the functional measure invariant. This
implies that
\begin{align}
0 & =\frac{1}{Z\left[J_{a}\right]}N\int\mathcal{D}\phi_{a}e^{iS}\int d^{5}z\left(\delta_{B}\mathcal{L}_{\mathrm{source}}\right)\,,\label{eq:BRST-Fonte-1}
\end{align}
where
\begin{align}
\delta_{B}\mathcal{L}_{\mathrm{source}} & =\varepsilon J^{\mu}D_{\mu}C-\varepsilon\frac{1}{\alpha}\,F\eta-\varepsilon gf_{1}\Phi_{2}C+\varepsilon gf_{2}\Phi_{1}C+h\varepsilon\mathcal{\tilde{O}}\,,\label{eq:BRST-Fonte-2}
\end{align}
and
\begin{equation}
\delta_{B}\mathcal{O}=\varepsilon\,\mathcal{\tilde{O}}\,.\label{eq:Def-O}
\end{equation}

We also quote the useful relations, familiar from standard QFT,
\begin{align}
\frac{\delta W\left[J_{a}\right]}{\delta J_{\mu}} & =\frac{1}{Z\left[J_{a}\right]}N\int\mathcal{D}\phi_{a}e^{iS}\Gamma_{\mu}=\left\langle \Gamma_{\mu}\right\rangle \,,\label{eq:VMeio-Gamma}
\end{align}
as well as \begin{subequations}\label{VMeio-SUperfields}
\begin{eqnarray}
\frac{\delta W\left[J_{a}\right]}{\delta\eta}=-\overline{C}\,, &  & \frac{\delta W\left[J_{a}\right]}{\delta\overline{\eta}}=C\,,\\
\frac{\delta W\left[J_{a}\right]}{\delta f_{1}}=\Phi_{1}\,, &  & \frac{\delta W\left[J_{a}\right]}{\delta f_{2}}=\Phi_{2}\,,\\
\frac{\delta W\left[J_{a}\right]}{\delta K_{1}}=-g\,\Phi_{2}C\,, &  & \frac{\delta W\left[J_{a}\right]}{\delta K_{2}}=g\,\Phi_{1}C\,.
\end{eqnarray}
\end{subequations}Substituting Eqs.\,(\ref{eq:VMeio-Gamma}) and\,(\ref{VMeio-SUperfields})
into Eq.\,(\ref{eq:BRST-Fonte-1}), we obtain
\begin{align}
 & -\frac{1}{Z\left[J_{a}\right]}N\int\mathcal{D}\phi_{a}e^{iS}\int d^{5}z\left(h\mathcal{\mathcal{\tilde{O}}}\right)=\nonumber \\
 & \int d^{5}z\left(J^{\mu}D_{\mu}\frac{\delta W\left[J_{a}\right]}{\delta\overline{\eta}}-\frac{1}{\alpha}\left(D^{\mu}\frac{\delta W\left[J_{a}\right]}{\delta J_{\mu}}+d\,g\frac{\delta W\left[J_{a}\right]}{\delta f_{2}}\right)\eta\right.\nonumber \\
 & \left.+f_{1}\frac{\delta W\left[J_{a}\right]}{\delta K_{1}}+f_{2}\frac{\delta W\left[J_{a}\right]}{\delta K_{2}}\right)\,.\label{eq:BRST-Fonte-3}
\end{align}

The quantum effective action is defined by means of a partial Legendre
transformation,
\begin{align}
\Gamma\left[\phi_{a}\right] & =W\left[J_{a}\right]-\int d^{5}z\left(J^{\mu}\Gamma_{\mu}+\overline{\eta}C+\overline{C}\eta+f_{1}\Phi_{1}+f_{2}\Phi_{2}\right)\,,\label{eq:Funcional-efetiva}
\end{align}
where the sources $h$ and $K_{i}$ are not Legendre transformed.
Besides the usual relations,\begin{subequations}\label{Eq:Valor-Fontes}
\begin{eqnarray}
\frac{\delta\Gamma\left[\phi_{a}\right]}{\delta\Gamma^{\mu}}=J^{\mu}\,, &  & \frac{\delta\Gamma\left[\phi_{a}\right]}{\delta\overline{C}}=\eta\,,\\
\frac{\delta\Gamma\left[\phi_{a}\right]}{\delta C}=-\overline{\eta}\,, &  & \frac{\delta\Gamma\left[\phi_{a}\right]}{\delta\Phi_{1}}=-f_{1}\,,\\
\frac{\delta\Gamma\left[\phi_{a}\right]}{\delta\Phi_{2}}=-f_{2}\,.
\end{eqnarray}
\end{subequations}one can also prove that\,\cite{Aitchison:1983ns},\begin{subequations}\label{Eq:Relations}
\begin{align}
\frac{\delta\Gamma\left[\phi_{a}\right]}{\delta h} & =\frac{\delta W\left[J_{a}\right]}{\delta h}\,,\\
\frac{\delta\Gamma\left[\phi_{a}\right]}{\delta K_{1}} & =\frac{\delta W\left[J_{a}\right]}{\delta K_{i}}\,,\\
\frac{\partial\Gamma\left[\phi_{a}\right]}{\partial\alpha} & =\frac{\partial W\left[J_{a}\right]}{\partial\alpha}\,.
\end{align}
\end{subequations}

Now, substituting Eqs.\,(\ref{eq:VMeio-Gamma}), (\ref{VMeio-SUperfields}),
(\ref{Eq:Valor-Fontes}) and\,(\ref{Eq:Relations}) into Eq.\,(\ref{eq:BRST-Fonte-3}),
we find
\begin{align}
-\frac{1}{Z\left[J_{a}\right]}N\int\mathcal{D}\phi_{a}e^{iS}\int d^{5}z\left(h\mathcal{\tilde{O}}\right) & =\int d^{5}z\left(\frac{\delta\Gamma\left[\phi_{a}\right]}{\delta\Gamma^{\mu}}D_{\mu}C-\frac{1}{\alpha}\,F\frac{\delta\Gamma\left[\phi_{a}\right]}{\delta\overline{C}}\right.\nonumber \\
 & \left.-\frac{\delta\Gamma\left[\phi_{a}\right]}{\delta\Phi_{1}}\frac{\delta\Gamma\left[\phi_{a}\right]}{\delta K_{1}}-\frac{\delta\Gamma\left[\phi_{a}\right]}{\delta\Phi_{2}}\frac{\delta\Gamma\left[\phi_{a}\right]}{\delta K_{2}}\right)\,,\label{eq:BRST-Fonte-4}
\end{align}
which after functional differentiation with respect to $h$ and taking
$h=0$, reduces to
\begin{align}
-\frac{1}{Z\left[J_{a}\right]}N\int\mathcal{D}\phi_{a}e^{iS}\mathcal{\mathcal{\tilde{O}}}= & \int d^{5}z\left(\frac{\delta\Gamma\left[\mathcal{O}\left(z\right)\right]}{\delta\Gamma^{\mu}}D_{\mu}C-\frac{1}{\alpha}\,F\frac{\delta\Gamma\left[\mathcal{O}\left(z\right)\right]}{\delta\overline{C}}\right.\nonumber \\
 & -\frac{\delta\Gamma\left[\mathcal{O}\left(z\right)\right]}{\delta\Phi_{1}}\frac{\delta\Gamma\left[J_{a}\right]}{\delta K_{1}}-\frac{\delta\Gamma\left[\phi_{a}\right]}{\delta\Phi_{1}}\frac{\delta\Gamma\left[\mathcal{O}\left(z\right)\right]}{\delta K_{1}}\nonumber \\
 & \left.-\frac{\delta\Gamma\left[\mathcal{O}\left(z\right)\right]}{\delta\Phi_{2}}\frac{\delta\Gamma\left[\phi_{a}\right]}{\delta K_{2}}-\frac{\delta\Gamma\left[\phi_{a}\right]}{\delta\Phi_{2}}\frac{\delta\Gamma\left[\mathcal{O}\left(z\right)\right]}{\delta K_{2}}\right)\,,\label{eq:BRST-Fonte-5}
\end{align}
where
\begin{align}
\delta\Gamma\left[\mathcal{O}\left(z\right)\right] & =\left.\frac{\delta\Gamma\left[\phi_{a}\right]}{\delta h}\right|_{h=0}\,.\label{eq:BRST-Fonte-5-Complemento}
\end{align}

The operator $\tilde{\mathcal{O}}$ in Eq.\,(\ref{eq:BRST-Fonte-5}),
in the class of supersymmetric $R_{\xi}$ gauge we are considering,
is given explicitly by
\begin{align}
\mathcal{\tilde{O}} & =\frac{1}{4\alpha}\,F^{2}+\overline{C}D^{2}C+\frac{1}{2}d\,g^{2}\,\Phi_{1}\overline{C}C\,.\label{eq:Operador-O-bar}
\end{align}
By using the equation of motion $\delta S/\delta\overline{C}=0,$
we have
\begin{align}
-\overline{C}\eta & =\overline{C}D^{2}C+\frac{1}{2}d\,g^{2}\,\Phi_{1}\overline{C}C\,,\label{eq:Cond-EM}
\end{align}
which, substituted into Eq.\,(\ref{eq:Operador-O-bar}), leads to
\begin{align}
\tilde{\mathcal{O}} & =\frac{1}{4\alpha}\,F^{2}-\overline{C}\eta\,.\label{eq:Operador-O-bar-1}
\end{align}
By differentiation of Eq.\,(\ref{eq:Func-Geradora-2}) with respect
to the gauge-fixing parameter $\alpha$, considering Eq.\,(\ref{eq:Lagran-Total-2}),
one obtains that
\begin{align}
\alpha\frac{\partial W\left[J_{a}\right]}{\partial\alpha} & =-\frac{1}{Z\left[J_{a}\right]}N\int\mathcal{D}\phi_{a}\int d^{5}z\,\frac{1}{4\alpha}\,F^{2}e^{iS}\,,\label{eq:Parte-1-Operador}
\end{align}
and proceeding similarly, one can show that
\begin{align}
\frac{\delta W\left[J_{a}\right]}{\delta\eta\left(r\right)} & =-\frac{1}{Z\left[J_{a}\right]}N\int\mathcal{D}\phi_{a}e^{iS}\int d^{5}z\delta^{5}\left(r-z\right)\overline{C}\left(z\right)\,,
\end{align}
then, after functional integration,
\begin{align}
-\int d^{5}r\,\eta\left(r\right)\frac{\delta W\left[J_{a}\right]}{\delta\eta\left(r\right)} & =\frac{1}{Z\left[J_{a}\right]}N\int\mathcal{D}\phi_{a}e^{iS}\int d^{5}r\,\eta\left(r\right)\overline{C}\left(r\right)\,.\label{eq:Parte-2-Operador}
\end{align}

Theses relations, Eqs.\,(\ref{eq:Parte-1-Operador}) and\,(\ref{eq:Parte-2-Operador})
together with Eqs.\,(\ref{eq:Operador-O-bar-1}) in\,(\ref{eq:BRST-Fonte-5}),
lead us to
\begin{align}
 & \alpha\frac{\partial W\left[J_{a}\right]}{\partial\alpha}-\int d^{5}z\,\eta\left(z\right)\frac{\delta W\left[J_{a}\right]}{\delta\eta\left(z\right)}\nonumber \\
 & =\int d^{5}r\int d^{5}z\left(\frac{\delta\Gamma\left[\mathcal{O}\left(z\right)\right]}{\delta\Gamma^{\mu}}D_{\mu}C-\frac{1}{\alpha}\,F\frac{\delta\Gamma\left[\mathcal{O}\left(z\right)\right]}{\delta\overline{C}}-\frac{\delta\Gamma\left[\mathcal{O}\left(z\right)\right]}{\delta\Phi_{1}}\frac{\delta\Gamma\left[\phi_{a}\right]}{\delta K_{1}}\right.\nonumber \\
 & \left.-\frac{\delta\Gamma\left[\phi_{a}\right]}{\delta\Phi_{1}}\frac{\delta\Gamma\left[\mathcal{O}\left(z\right)\right]}{\delta K_{1}}-\frac{\delta\Gamma\left[\mathcal{O}\left(z\right)\right]}{\delta\Phi_{2}}\frac{\delta\Gamma\left[\phi_{a}\right]}{\delta K_{2}}-\frac{\delta\Gamma\left[\phi_{a}\right]}{\delta\Phi_{2}}\frac{\delta\Gamma\left[\mathcal{O}\left(z\right)\right]}{\delta K_{2}}\right)\,,
\end{align}
then, using Eqs.\,(\ref{Eq:Relations}), (\ref{Eq:Valor-Fontes})
and\,(\ref{VMeio-SUperfields}) on the left hand side, we obtain
\begin{align}
\alpha\frac{\partial\Gamma\left[\phi_{a}\right]}{\partial\alpha}+\int d^{5}z\,\frac{\delta\Gamma\left[\phi_{a}\right]}{\delta\overline{C}}\,\overline{C} & =\int d^{5}r\int d^{5}z\left(\frac{\delta\Gamma\left[\mathcal{O}\left(z\right)\right]}{\delta\Gamma^{\mu}}D_{\mu}C-\frac{1}{\alpha}\,F\frac{\delta\Gamma\left[\mathcal{O}\left(z\right)\right]}{\delta\overline{C}}\right.\nonumber \\
 & -\frac{\delta\Gamma\left[\mathcal{O}\left(z\right)\right]}{\delta\Phi_{1}}\frac{\delta\Gamma\left[\phi_{a}\right]}{\delta K_{1}}-\frac{\delta\Gamma\left[\phi_{a}\right]}{\delta\Phi_{1}}\frac{\delta\Gamma\left[\mathcal{O}\left(z\right)\right]}{\delta K_{1}}-\frac{\delta\Gamma\left[\mathcal{O}\left(z\right)\right]}{\delta\Phi_{2}}\frac{\delta\Gamma\left[\phi_{a}\right]}{\delta K_{2}}\nonumber \\
 & \left.-\frac{\delta\Gamma\left[\phi_{a}\right]}{\delta\Phi_{2}}\frac{\delta\Gamma\left[\mathcal{O}\left(z\right)\right]}{\delta K_{2}}\right)\,.\label{eq:BRST-Fonte-final}
\end{align}
This expression is the base for obtaining the Nielsen identity. By
taking $\overline{C}=C=\Phi_{2}=\Gamma_{\alpha}=0$, and using the
ghost number conservation in Eq.\,(\ref{eq:BRST-Fonte-final}), we
arrive to a simple result:
\begin{align}
\alpha\frac{\partial\Gamma\left[\phi_{a}\right]}{\partial\alpha} & =-\int d^{5}r\int d^{5}z\left(\frac{\delta\Gamma\left[\phi_{a}\right]}{\delta\Phi_{1}}\frac{\delta\Gamma\left[\mathcal{O}\left(z\right)\right]}{\delta K_{1}}+\frac{\delta\Gamma\left[\phi_{a}\right]}{\delta\Phi_{2}}\frac{\delta\Gamma\left[\mathcal{O}\left(z\right)\right]}{\delta K_{2}}\right)\,.\label{eq:Reduce-Nielsen}
\end{align}

The effective superpotential is obtained by setting $\Phi_{1}=\sigma_{\mathrm{cl}}$
in the effective action, 
\begin{equation}
\left.\Gamma\left[\Phi_{1},\alpha\right]\right|_{\Phi_{1}=\sigma_{\mathrm{cl}}}=V_{\eff}^{S}\left(\sigma_{\mathrm{cl}},\,\alpha\right)\thinspace,
\end{equation}
where
\begin{equation}
\sigma_{\mathrm{cl}}=\sigma_{1}-\theta^{2}\sigma_{2}\,,\label{eq:background-superfield}
\end{equation}
is the space-time constant expectation value of the scalar superfield,
called of \emph{background superfield}, where $\sigma_{1}$ is a scalar
field and $\sigma_{2}$ is an auxiliary scalar field. The superfield
$\sigma_{\mathrm{cl}}$ has the same properties of a scalar superfield,
except there is not a spinor component because its presence would
imply a Lorentz violation. Consdering Eq.\,(\ref{eq:Reduce-Nielsen})
for the case of the effective superpotential, we have
\begin{align}
\left[\alpha\frac{\partial}{\partial\alpha}+C^{S}\left(\sigma_{\mathrm{cl}},\alpha\right)\frac{\partial}{\partial\sigma_{\mathrm{cl}}}\right]V_{\mathrm{eff}}^{S}\left(\sigma_{\mathrm{cl}},\,\alpha\right) & =0\thinspace,\label{eq:Identidade-Nielsen-V}
\end{align}
where 
\begin{align}
C^{S}\left(\sigma_{\mathrm{cl}},\alpha\right) & =\int d^{5}z\left.\frac{\delta^{2}\Gamma\left[\mathcal{O}\left(z\right)\right]}{\delta K_{1}\left(0\right)\delta h\left(y\right)}\right|_{K_{1}=h=0}\thinspace,\label{eq:C-Phi-alpha}
\end{align}
which is the Nielsen identity for the effective superpotential.

\section{\label{sec:On-the-gauge}On the gauge (in)dependence of the Effective
Superpotential}

The effective superpotential in three space-time dimensions has the
general form
\begin{align}
V_{\eff}^{S}\left(\sigma_{\mathrm{cl}},\,\alpha\right) & =-\int d^{5}z\,\left[K_{\eff}\left(\sigma_{\mathrm{cl}},\,\alpha\right)+\mathcal{F}\left(D_{\alpha}\sigma_{\mathrm{cl}},\,D^{\alpha}\sigma_{\mathrm{cl}},\,D^{2}\sigma_{\mathrm{cl}},\,\sigma_{\mathrm{cl}},\alpha\right)\right]\,,\label{eq:SuperPotencialCompleto}
\end{align}
where we made explicit the potential gauge dependence. Here, $K_{\eff}$
is the part of the effective superpotential that does not depend on
derivatives of the background classical superfield $\sigma_{\mathrm{cl}}$,
similar to the Kälerian effective superpotential defined in four-dimensional
models\,\cite{gates:1983nr,buchbinder:1998qv}. In the context of
dynamical gauge symmetry breaking, it is enough to consider only the
$K_{\eff}$\,\cite{Ferrari:2010ex,Quinto2016,lehum:2007nf,Lehum:2010tt,Queiruga:2015fzn},
while a study of a possible supersymmetry breaking would involve also
the knowledge of ${\cal F}$\,\cite{Gallegos:2011ux,Lehum:2013rpa}.

An explicit perturbative evaluation of $V_{\eff}^{S}$ starting from
Eq.\,(\ref{eq:Lagran-Total-2}), within the superfield formalism,
is in general quite difficult. The root of this problem is the fact
that the classical superfield $\sigma_{\mathrm{cl}}$ is space-time
constant, but its supercovariant derivatives do not vanish, $D_{\alpha}\sigma_{\mathrm{cl}}\neq0$.
This, for example, complicates the calculation of the free superpropagators
of the model, since powers of $\sigma_{\mathrm{cl}}$ appear in the
quadratic operators that have to be inverted. This leads to the appearance
of non-covariant superpropagators, as discussed in\,\cite{Gallegos:2011ag,Gallegos:2011ux}.
In these works, the effective superpotential of SCSM and SQED models
was calculated up to two loops. Part of these calculations was performed
in an arbitrary gauge, but the final results for the effective superpotential
were only obtained in the Landau gauge. Another perspective on the
difficulties of evaluating the full effective superpotential in the
superfield formalism, using heat kernel techniques, can be found in\,\cite{Ferrari:2009zx}. 

One possibility to obtain a full computation of $V_{\eff}^{S}$ within
the superfield formalism would involve the use of the RGE. This approach
will be used to calculate $K_{\eff}$ in the massless limit in Chapter\,\ref{chap:RGE-Impro-DBS}.
Essentially, one may consider $K_{\eff}$ as a function of the single
mass scale $\sigma_{1}$, and this constrains the form of the radiative
corrections to $K_{\eff}$, so that a simple ansatz can be made, which
inserted in the RGE provides a set of recursive equations from which
the coefficients of the leading logs contributions to $K_{\eff}$
can be found. In principle this technique could be extended for the
full effective superpotential $V_{\eff}^{S}$, but in this case more
complicated, multiscale techniques would be needed\,\cite{Ford1994,Ford1997},
since $V_{\eff}^{S}$ should be considered as a function of both $\sigma_{1}$
and $\sigma_{2}$. 

The derivation of the effective superpotential from the renormalization
group functions, by means of the RGE, may allow one to infer from
the gauge (in)dependence of the beta functions and anomalous dimensions
the gauge (in)dependence of the effective superpotential itself. This
is something we can do for the $K_{\eff}$, since we establish in
Chapter\,\ref{chap:RGE-Impro-DBS} that it can be calculated from
the renormalization group functions without any ambiguity. Therefore,
in the next Chapter, we will present a detailed computation of the
beta and gamma functions in an arbitrary gauge, showing that the result
is indeed gauge independent. By this reasoning, we can conclude that
$K_{\eff}$ does not depend on the gauge parameter. That means, when
only $K_{\eff}$ is considered, the Nielsen identity\,(\ref{eq:Identidade-Nielsen-V})
is trivially satisfied with $C^{S}=0$. 

These results may suggest the gauge independence of the whole effective
superpotential $V_{\eff}^{S}$, but without explicitly establishing
that the RGE fixes the form of $\mathcal{F}$ in some approximation,
without ambiguities, from the renormalization group functions (which
we know are gauge independent), we believe this is still an open question.
The discussion of\,\cite{Gallegos:2011ag,Gallegos:2011ux} is not
conclusive in this regard, since most of the results are presented
in a specific gauge, but some gauge dependence was found in the effective
superpotential of the supersymmetric QED model. It is also not simple
to use the Nielsen identity itself to investigate this point since,
as discussed in\,\cite{Nielsen:1975fs,Aitchison:1983ns}, the calculation
of $C^{S}$ in the massless case is complicated by the fact that different
loop orders contribute to $C^{S}$ in a given order in the coupling
constants. 
\begin{center}
\myclearpage
\par\end{center}

\chapter{\label{chap:Renormalization-Group-Functions}Renormalization Group
Functions}

In this chapter we calculate the renormalization group functions of
the model for arbitrary gauge-fixing parameter, finding them to be
independent of the gauge choice. This result will be used to argue
that $K_{\eff}$ also does not depend on the gauge-fixing parameter. 

\section{\label{sec:Calculation-of-beta-function}Gauge invariance of the
renormalization group functions}

In this section, we consider the $m=0$ version of Eq.\,(\ref{eq:actionCSM}),
generalized to exhibit a global $SU\left(N\right)$ symmetry,
\begin{align}
\mathcal{S}_{\mathrm{CS}} & =\int d^{5}z\left\{ -\frac{1}{2}\Gamma^{\alpha}W_{\alpha}-\frac{1}{2}\overline{\nabla^{\alpha}\Phi_{a}}\nabla_{\alpha}\Phi_{a}+\frac{\lambda}{4}\left(\overline{\Phi_{a}}\Phi_{a}\right)^{2}\right\} ,\label{eq:M1}
\end{align}
where $a=1,\,...\,N$. We follow the basic conventions for three-dimensional
supersymmetry described in Chapter\,\ref{chap:Supersymmetry-in-(2+1)}.

In the quantization process, we introduce the gauge-fixing action,
\begin{align}
\mathcal{S}_{\mathrm{GF}} & =\frac{1}{4\,\alpha}\int d^{5}z\left(D^{\alpha}\Gamma_{\alpha}\right)^{2}\thinspace,\label{eq:GF}
\end{align}
but differently from what is done in the literature\,\cite{Avdeev:1991za,lehum:2007nf,Lehum:2010tt},
we will perform all calculations with an arbitrary gauge-fixing parameter.
Then, from Eqs.\,(\ref{eq:M1}) and\,(\ref{eq:GF}) we have
\begin{align}
\mathcal{S} & =\int d^{5}z\left\{ \frac{1}{4}\Gamma_{\alpha}\left[D^{\beta}D^{\alpha}+\frac{1}{\alpha}\,D^{\alpha}D^{\beta}\right]\Gamma_{\beta}-\frac{1}{2}\,g^{2}C_{\beta\alpha}\Gamma^{\alpha}\Gamma^{\beta}\overline{\Phi}_{a}\Phi_{a}\right.\nonumber \\
 & \left.+\overline{\Phi}_{a}D^{2}\Phi_{a}-\frac{1}{2}\,i\,g\left[\Gamma^{\alpha}\overline{\Phi}_{a}D_{\alpha}\Phi_{a}-\left(D^{\alpha}\overline{\Phi}_{a}\right)\Gamma_{\alpha}\Phi_{a}\right]+\frac{\lambda}{4}\left(\overline{\Phi}_{a}\Phi_{a}\right)^{2}+\mathcal{L}_{ct}\right\} ,\label{eq:M3}
\end{align}
where the counterterm Lagrangian is
\begin{align}
\mathcal{L}_{ct} & =\frac{1}{4}\left(Z_{\Gamma}-1\right)\Gamma_{\alpha}D^{\beta}D^{\alpha}\Gamma_{\beta}+\frac{1}{2}\left(Z_{\Phi}-1\right)\overline{\nabla^{\alpha}\Phi_{a}}\nabla_{\alpha}\Phi_{a}+\frac{\lambda}{4}\,Z_{\lambda}\left(\overline{\Phi_{a}}\Phi_{a}\right)^{2},\label{eq:M3CT}
\end{align}
$Z_{\Gamma}$, $Z_{\Phi}$, and $Z_{\lambda}$ being the counterterms
needed to make the renormalized quantities finite in each order of
perturbation theory. 

\section{\label{sec:Feynman-rules}Feynman rules}

In this section we present the explicit derivation of the Feynman
rules in our model.

\subsection{Scalar propagator}

From Eq.\,(\ref{eq:M3}) we can find the scalar propagator. For this
purpose, we start with the generating functional
\begin{align}
Z\left[J_{a},\overline{J}_{a}\right] & =\int\mathcal{D}\Phi_{a}\,\mathcal{D}\overline{\Phi}_{a}\exp\left[i\int d^{5}z\left(\overline{\Phi}_{a}D^{2}\Phi_{a}+\overline{J}_{a}\Phi_{a}+J_{a}\overline{\Phi}_{a}\right)\right]\,,\label{eq:PS1}
\end{align}
where $J_{a}$ and $\overline{J}_{a}$ are the sources associated
to the scalar superfields $\Phi_{a}$ and $\overline{\Phi}_{a}$,
respectively. Then, the two-point Green's function in position space
is
\begin{align}
\left\langle 0\left|\mbox{T}\overline{\Phi}_{i}\left(z_{1}\right)\Phi_{j}\left(z_{2}\right)\right|0\right\rangle  & =\left.\frac{1}{Z_{0}}\left(\frac{1}{i}\frac{\delta}{\delta J_{i}\left(z_{1}\right)}\right)\left(-\frac{1}{i}\frac{\delta}{\delta\overline{J_{j}}\left(z_{2}\right)}\right)Z\left[J_{a},\overline{J}_{a}\right]\right|_{J=\overline{J}=0}.\label{eq:PS2}
\end{align}
By shifting the superfields $\Phi_{a}$ and $\overline{\Phi}_{a}$
\begin{align}
\Phi_{a}\rightarrow & \Phi_{a}-\frac{J_{a}}{D^{2}}\,,\label{eq:PS3}\\
\overline{\Phi}_{a}\rightarrow & \overline{\Phi}_{a}-\frac{\overline{J}_{a}}{D^{2}}\,,\label{eq:PS4}
\end{align}
and substituting in\,(\ref{eq:PS1}), we find
\begin{align}
Z\left[J_{a},\overline{J}_{a}\right] & =\int\mathcal{D}\Phi_{a}\,\mathcal{D}\overline{\Phi}_{a}\exp\left[i\int d^{5}z\left(\overline{\Phi}_{a}D^{2}\Phi_{a}-\frac{\overline{J}_{a}\,J_{a}}{D^{2}}\right)\right]\nonumber \\
 & =Z_{0}\int\mathcal{D}\Phi_{a}\,\mathcal{D}\overline{\Phi}_{a}\exp\left[-i\int d^{5}z\,d^{5}z'\,\overline{J}_{a}\left(z\right)\triangle\left(z,z'\right)J_{a}\left(z'\right)\right]\,,\label{eq:PS5}
\end{align}
where 
\begin{equation}
\triangle\left(z,z'\right)=\frac{1}{D^{2}}\delta^{5}\left(z-z'\right)=\frac{1}{D^{2}}\delta^{3}\left(x-x'\right)\delta^{2}\left(\theta-\theta'\right).\label{eq:PS6}
\end{equation}

If we substitute Eq.\,(\ref{eq:PS5}) into Eq.\,(\ref{eq:PS2})
and perform the functional derivations, we obtain 
\begin{align}
\left\langle 0\left|\mbox{T}\overline{\Phi}_{i}\left(z_{1}\right)\Phi_{j}\left(z_{2}\right)\right|0\right\rangle  & =-\frac{i\,\delta_{ij}}{D^{2}}\delta^{5}\left(z_{1}-z_{2}\right)=-\frac{i\,\delta_{ij}D^{2}}{\boxempty}\delta^{5}\left(z_{1}-z_{2}\right)\,.\label{eq:PS7}
\end{align}
We know that $G_{ij}^{\left(2\right)}\left(z_{1},z_{2}\right)=\left\langle 0\left|\mbox{T}\overline{\Phi}_{i}\left(z_{1}\right)\Phi_{j}\left(z_{2}\right)\right|0\right\rangle $,
so by using the Fourier transformation in Eq.\,(\ref{eq:PS7}), we
have
\begin{align}
G_{ij}^{\left(2\right)}\left(k_{1},k_{2}\right) & =\int d^{3}x_{1}\,d^{3}x_{2}\,\exp\left[-i\,\left(k_{1}x_{1}+k_{2}x_{2}\right)\right]\tilde{G}^{\left(2\right)}\left(z_{1},z_{2}\right)\nonumber \\
 & =\left(2\pi\right)^{3}\delta^{2}\left(k_{1}+k_{2}\right)\left(\frac{i\,\delta_{ij}\,D^{2}}{k_{1}}\delta^{2}\left(\theta_{1}-\theta_{2}\right)\right)\,,\label{eq:PS8}
\end{align}
where it was used that $\left(D^{2}\right)^{2}=-k^{2}$ (see Eq.\,(\ref{eq:S20})).
Therefore, we find the scalar propagator in momentum space to be
\begin{align}
\contraction{}{\overline{\Phi}}{_{i}(k,\theta_{1}){}}{\Phi}\nomathglue{\overline{\Phi}_{i}\left(k,\theta_{1}\right){}\Phi_{j}\left(-k,\theta_{2}\right)}=\left\langle 0\left|\mbox{T}\overline{\Phi}_{i}\left(k,\theta_{1}\right)\Phi_{j}\left(-k,\theta_{2}\right)\right|0\right\rangle  & =i\delta_{ij}\frac{D^{2}}{k^{2}}\delta^{2}\left(\theta_{1}-\theta_{2}\right).\label{eq:ProgatorScalar}
\end{align}

\subsection{Gauge propagator}

We calculate the gauge propagator associated to the gauge superfield
$\Gamma_{\alpha}$, see Eq.\,(\ref{eq:M3}). The relevant part of
the generating functional is
\begin{equation}
Z\left[J^{\alpha}\right]=\int\mathcal{D}\Gamma\exp\left[\frac{1}{2}\,i\int d^{5}z\left(\Gamma_{\alpha}P^{\alpha\beta}\Gamma_{\beta}+2\,J^{\rho}\Gamma_{\rho}\right)\right],\label{eq:PG1}
\end{equation}
where $J^{\alpha}$ is the source of the gauge superfield, which behaves
as a spinor, and 
\begin{align}
P^{\alpha\beta} & =\frac{1}{2}D^{\beta}D^{\alpha}+\frac{1}{2\alpha}D^{\alpha}D^{\beta}\,.\label{eq:PG2}
\end{align}
We shift the gauge superfield in Eq.\,(\ref{eq:PG1}) as,
\begin{align}
\Gamma_{\rho} & \rightarrow\Gamma_{\rho}+Q_{\rho\theta}J^{\theta}\,,\label{eq:PG3}
\end{align}
where
\begin{align}
P^{\alpha\beta}Q_{\beta\theta} & =\delta_{\,\,\theta}^{\alpha}\,,\label{eq:PG4}
\end{align}
\begin{align}
Q_{\beta\theta} & =b_{1}D_{\theta}D_{\beta}+b_{2}D_{\beta}D_{\theta}\,,\label{eq:Form-Q}
\end{align}
and $b_{1}$ and $b_{2}$ are constants to be determined. We obtain
\begin{align}
Z\left[J_{\alpha}\right] & =\int\mathcal{D}\Gamma\,\exp\left[\frac{1}{2}i\,\int d^{5}z\left(\Gamma_{\alpha}P^{\alpha\beta}\Gamma_{\beta}+J_{\alpha}Q^{\alpha\beta}J_{\beta}\right)\right]\nonumber \\
 & =Z_{0}\int\mathcal{D}\Gamma\,\exp\left[\frac{1}{2}\int d^{5}zd^{5}z'\,J_{\alpha}\left(z\right)\Delta_{z\,z'}^{\alpha\beta}J_{\beta}\left(z'\right)\right],\,\label{eq:PG5}
\end{align}
where
\begin{equation}
\Delta_{z\,z'}^{\alpha\beta}=i\,Q^{\alpha\beta}\,\delta^{5}\left(z-z'\right)\,.\label{eq:PG6}
\end{equation}
The two-point Green's function in position space is
\begin{align}
\contraction{}{\Gamma}{_{\beta}\left(z_{1}\right){}}{\Gamma}\nomathglue{\Gamma_{\beta}\left(z_{1}\right){}\Gamma_{\theta}\left(z_{2}\right)} & =\left.\frac{1}{Z_{0}}\left(\frac{1}{i}\frac{\delta}{\delta J_{\beta}\left(z_{1}\right)}\right)\left(\frac{1}{i}\frac{\delta}{\delta J_{\theta}\left(z_{2}\right)}\right)Z\left[J_{\alpha}\right]\right|_{J_{\beta}=J_{\theta}=0}\nonumber \\
 & =i\,Q_{\beta\theta}\,\delta^{5}\left(z_{1}-z_{2}\right)\,.\label{eq:PG7}
\end{align}
To calculate the explicit form of $Q$, we use Eq.\,(\ref{eq:PG4})
and the identities shown in Section\,\ref{subsec:D-=0000E1lgebras},
which lead to
\begin{align}
\begin{cases}
\delta_{\,\theta}^{\alpha}\, & =-2\,b_{1}\,\delta_{\,\theta}^{\alpha}\,\boxempty\,,\\
0 & =b_{1}\,D^{2}D^{\alpha}D_{\theta}-b_{2}\,D^{2}D^{\alpha}D_{\theta}\,,
\end{cases}
\end{align}
from this system of equations, we find 
\begin{align}
b_{2} & =\alpha\,b_{1}=-\frac{\alpha}{2\,\boxempty}\,.
\end{align}
This fixes the form of the gauge superpropagator to be
\begin{align}
\contraction{}{\Gamma}{_{\beta}\left(z_{1}\right){}}{\Gamma}\nomathglue{\Gamma_{\beta}\left(z_{1}\right){}\Gamma_{\theta}\left(z_{2}\right)} & =-\frac{1}{2\,\boxempty}\,i\left(D_{\theta}D_{\beta}+\alpha\,D_{\beta}D_{\theta}\right)\delta^{5}\left(z_{1}-z_{2}\right)\,,\label{eq:PG12}
\end{align}
or, in momentum space, 
\begin{align}
\contraction{}{\Gamma}{_{\beta}\left(k,\theta_{1}\right){}}{\Gamma}\nomathglue{\Gamma_{\beta}\left(k,\theta_{1}\right){}\Gamma_{\theta}\left(-k,\theta_{2}\right)} & =\frac{i}{2k^{2}}\left(D_{\theta}D_{\beta}+\alpha\,D_{\beta}D_{\theta}\right)\delta^{2}\left(\theta_{1}-\theta_{2}\right)\,,\label{eq:PropagatorGauge}
\end{align}
where 
\begin{equation}
D_{\rho}D_{\beta}+\alpha\,D_{\beta}D_{\rho}=b\left(\alpha\right)\,k_{\rho\beta}+a\left(\alpha\right)\,C_{\beta\alpha}\,D^{2}\thinspace,
\end{equation}
and 
\begin{equation}
a\left(\alpha\right)=1-\alpha\thinspace;\thinspace b\left(\alpha\right)\equiv1+\alpha\thinspace.\label{eq:defAB}
\end{equation}

\subsection{Vertices}

To compute the vertices of the theory, we use the Wick's theorem.
For this purpose, we start with the correlation functions, which are
vacuum expectation values of time-ordered products of a finite number
of free superfield operators,
\begin{equation}
\left\langle 0\right|\mbox{T}\Phi\left(z_{1}\right)...\Phi\left(z_{n}\right)\left|0\right\rangle \,.\label{eq:CFunction}
\end{equation}
As an example, for $n=2$ Eq.\,(\ref{eq:CFunction}) represents the
scalar propagator,
\begin{equation}
\contraction{}{\Phi}{\left(z_{1}\right)}{\Phi}\nomathglue{\Phi\left(z_{1}\right){}\Phi\left(z_{2}\right)}=\left\langle 0\left|\mbox{T}\Phi\left(z_{1}\right)\Phi\left(z_{2}\right)\right|0\right\rangle \,.\label{eq:FP}
\end{equation}
Wick's theorem allows us to turn any expression of the form Eq.\,(\ref{eq:CFunction})
into a sum of products of Feynman propagators. This theorem can be
extended to gauge superfields, taking care of the signals appearing
due to the anti-commuting properties of spinors.
\begin{center}
\begin{figure}
\centering{}\includegraphics[scale=0.6]{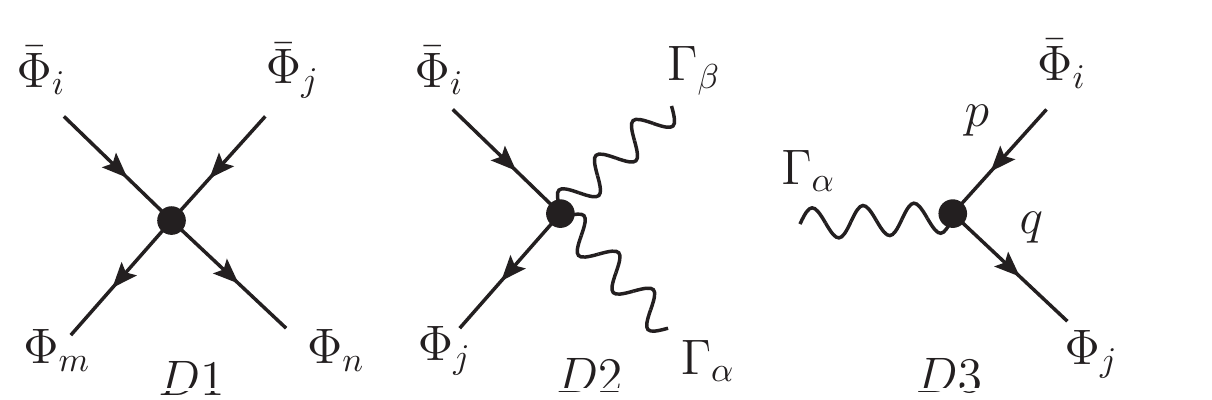}\caption{\label{fig:The-vertices-Diagrams-1}Elementary vertices of the model,
where a continuous line is associated to a scalar superfield, and
a wavy line to the gauge superfield. }
\end{figure}
\par\end{center}

If we consider n-point Green's functions,
\begin{align}
G^{\left(n\right)}\left(z_{1},...,z_{n}\right) & =\frac{\left\langle 0\right|\mbox{T}\Phi\left(z_{1}\right)\Gamma_{\alpha}\left(z_{2}\right)...\Phi\left(z_{n}\right)\Gamma_{\beta}\left(z_{m}\right)\exp\left[i\int d^{5}z\,\mathcal{L}_{\mathrm{int}}\left(\Phi\left(z\right),\Gamma_{\alpha}\left(z\right)\right)\right]\left|0\right\rangle }{\left\langle 0\right|\mbox{T }\exp\left[i\int d^{5}z\,\mathcal{L}_{\mathrm{int}}\left(\Phi\left(z\right),\Gamma_{\alpha}\left(z\right)\right)\right]\left|0\right\rangle }\,,\label{eq:GFunction-n}
\end{align}
where the interaction Lagrangian $\mathcal{L}_{\mathrm{int}}$ is
a function of the complex scalar $\Phi_{a}\left(z\right)$ and the
gauge superfields $\Gamma_{\alpha}$$\left(z\right)$, it is possible
to find the elementary vertices of the theory. . We start with the
four-point Green's function associated to the complex scalar superfield,
for this we only need to use the numerator in\,(\ref{eq:GFunction-n})
and\,(\ref{eq:M1}),
\begin{align}
 & \left\langle 0\right|\mbox{T }\overline{\Phi}_{i}\left(x\right)\Phi_{j}\left(y\right)\overline{\Phi}_{k}\left(w\right)\Phi_{l}\left(s\right)\left[1+i\left(\frac{\lambda}{4}\right)\int d^{5}z\left(\overline{\Phi}_{a}\left(z\right)\Phi_{a}\left(z\right)\right)^{2}+...\right]\left|0\right\rangle \,,
\end{align}
after expanding the exponential function. In our case, we need only
the order $\mathcal{O}\left(\lambda\right)$ term, then choosing only
the second term of the previous equation, we have
\[
i\left(\frac{\lambda}{4}\right)\left\langle 0\right|\mbox{T}\overline{\Phi}_{i}\left(x\right)\Phi_{j}\left(y\right)\overline{\Phi}_{k}\left(w\right)\Phi_{l}\left(s\right)\int d^{5}z\overline{\Phi}_{a}\left(z\right)\Phi_{a}\left(z\right)\overline{\Phi}_{b}\left(z\right)\Phi_{b}\left(z\right)\left|0\right\rangle 
\]
\vfill{}
\begin{eqnarray*}
=i\left(\frac{\lambda}{4}\right)\int d^{5}z\left(\contraction{}{\overline{\Phi}}{_{i}\left(x\right)\Phi_{j}\left(y\right)\overline{\Phi}_{k}\left(w\right)\Phi_{l}\left(s\right)\overline{\Phi}_{a}\left(z\right)}{\Phi}
\contraction[2ex]{\overline{\Phi}_{i}\left(x\right)}{\Phi}{_{j}\left(y\right)\overline{\Phi}_{k}\left(w\right)\Phi_{l}\left(s\right)}{\overline{\Phi}}
\bcontraction{\overline{\Phi}_{i}\left(x\right)\Phi_{j}\left(y\right)}{\overline{\Phi}}{_{k}\left(w\right)\Phi_{l}\left(s\right)\overline{\Phi}_{a}\left(z\right)\Phi_{a}\left(z\right)\overline{\Phi}_{b}\left(z\right)}{\Phi}
\bcontraction[2ex]{\overline{\Phi}_{i}\left(x\right)\Phi_{j}\left(y\right)\overline{\Phi}_{k}\left(w\right)}{\Phi}{_{l}\left(s\right)\overline{\Phi}_{a}\left(z\right)\Phi_{a}\left(z\right)}{\overline{\Phi}}
\overline{\Phi}_{i}\left(x\right)\Phi_{j}\left(y\right)\overline{\Phi}_{k}\left(w\right)\Phi_{l}\left(s\right)\overline{\Phi}_{a}\left(z\right)\Phi_{a}\left(z\right)\overline{\Phi}_{b}\left(z\right)\Phi_{b}\left(z\right)\right.\\
+\contraction{}{\overline{\Phi}}{_{i}\left(x\right)\Phi_{j}\left(y\right)\overline{\Phi}_{k}\left(w\right)\Phi_{l}\left(s\right)\overline{\Phi}_{a}\left(z\right)}{\Phi}
\contraction[2ex]{\overline{\Phi}_{i}\left(x\right)}{\Phi}{_{j}\left(y\right)\overline{\Phi}_{k}\left(w\right)\Phi_{l}\left(s\right)\overline{\Phi}_{a}\left(z\right)\Phi_{a}\left(z\right)}{\overline{\Phi}}
\bcontraction{\overline{\Phi}_{i}\left(x\right)\Phi_{j}\left(y\right)}{\overline{\Phi}}{_{k}\left(w\right)\Phi_{l}\left(s\right)\overline{\Phi}_{a}\left(z\right)\Phi_{a}\left(z\right)\overline{\Phi}_{b}\left(z\right)}{\Phi}
\bcontraction[2ex]{\overline{\Phi}_{i}\left(x\right)\Phi_{j}\left(y\right)\overline{\Phi}_{k}\left(w\right)}{\Phi}{_{l}\left(s\right)}{\overline{\Phi}}
\overline{\Phi}_{i}\left(x\right)\Phi_{j}\left(y\right)\overline{\Phi}_{k}\left(w\right)\Phi_{l}\left(s\right)\overline{\Phi}_{a}\left(z\right)\Phi_{a}\left(z\right)\overline{\Phi}_{b}\left(z\right)\Phi_{b}\left(z\right)\\
+\contraction{}{\overline{\Phi}}{_{i}\left(x\right)\Phi_{j}\left(y\right)\overline{\Phi}_{k}\left(w\right)\Phi_{l}\left(s\right)\overline{\Phi}_{a}\left(z\right)\Phi_{a}\left(z\right)\overline{\Phi}_{b}\left(z\right)}{\Phi}
\contraction[2ex]{\overline{\Phi}_{i}\left(x\right)}{\Phi}{_{j}\left(y\right)\overline{\Phi}_{k}\left(w\right)\Phi_{l}\left(s\right)}{\overline{\Phi}}
\bcontraction{\overline{\Phi}_{i}\left(x\right)\Phi_{j}\left(y\right)}{\overline{\Phi}}{_{k}\left(w\right)\Phi_{l}\left(s\right)\overline{\Phi}_{a}\left(z\right)}{\Phi}
\bcontraction[2ex]{\overline{\Phi}_{i}\left(x\right)\Phi_{j}\left(y\right)\overline{\Phi}_{k}\left(w\right)}{\Phi}{_{l}\left(s\right)\overline{\Phi}_{a}\left(z\right)\Phi_{a}\left(z\right)}{\overline{\Phi}}
\overline{\Phi}_{i}\left(x\right)\Phi_{j}\left(y\right)\overline{\Phi}_{k}\left(w\right)\Phi_{l}\left(s\right)\overline{\Phi}_{a}\left(z\right)\Phi_{a}\left(z\right)\overline{\Phi}_{b}\left(z\right)\Phi_{b}\left(z\right)\\
+\left.\contraction{}{\overline{\Phi}}{_{i}\left(x\right)\Phi_{j}\left(y\right)\overline{\Phi}_{k}\left(w\right)\Phi_{l}\left(s\right)\overline{\Phi}_{a}\left(z\right)\Phi_{a}\left(z\right)\overline{\Phi}_{b}\left(z\right)}{\Phi}
\contraction[2ex]{\overline{\Phi}_{i}\left(x\right)}{\Phi}{_{j}\left(y\right)\overline{\Phi}_{k}\left(w\right)\Phi_{l}\left(s\right)\overline{\Phi}_{a}\left(z\right)\Phi_{a}\left(z\right)}{\overline{\Phi}}
\bcontraction{\overline{\Phi}_{i}\left(x\right)\Phi_{j}\left(y\right)}{\overline{\Phi}}{_{k}\left(w\right)\Phi_{l}\left(s\right)\overline{\Phi}_{a}\left(z\right)}{\Phi}
\bcontraction[2ex]{\overline{\Phi}_{i}\left(x\right)\Phi_{j}\left(y\right)\overline{\Phi}_{k}\left(w\right)}{\Phi}{_{l}\left(s\right)}{\overline{\Phi}}
\overline{\Phi}_{i}\left(x\right)\Phi_{j}\left(y\right)\overline{\Phi}_{k}\left(w\right)\Phi_{l}\left(s\right)\overline{\Phi}_{a}\left(z\right)\Phi_{a}\left(z\right)\overline{\Phi}_{b}\left(z\right)\Phi_{b}\left(z\right)\right)
\end{eqnarray*}
\begin{align*}
= & i\,\frac{\lambda}{2}\left(\delta_{ij}\delta_{kl}+\delta_{il}\delta_{jk}\right)\int d^{5}z\,\Delta\left(x,y,w,s-z\right)
\end{align*}
where $\Delta$ was introduced to denote the product of the scalar
propagators. From this expression, we can find in momentum space the
four-point vertex function, at tree level,
\begin{align}
i\,V_{\left(\overline{\Phi}_{a}\Phi_{a}\right)^{2}} & =\frac{1}{2}\,i\,\lambda\left(\delta_{in}\delta_{jm}+\delta_{im}\delta_{jn}\right)\int d^{2}\theta,\label{eq:Vertex4PointsScalar}
\end{align}
where the corresponding Feynman diagram is represented by $D1$ in
the Figure\,\ref{fig:The-vertices-Diagrams-1}.

Next, we do the same procedure described before for the vertices involving
the gauge superfield, taking care of the signs generated by the interchange
of gauge superfields $\Gamma_{\alpha}\left(z\right)$. We start with
\[
\left\langle 0\right|\mbox{T }\Gamma^{\beta}\left(w\right)\Gamma_{\beta}\left(s\right)\overline{\Phi}_{i}\left(x\right)\Phi_{j}\left(y\right)\left[1+i\left(-\frac{1}{2}g^{2}\right)\int d^{5}z\Gamma^{\alpha}\left(z\right)\Gamma_{\alpha}\left(z\right)\overline{\Phi}_{a}\left(z\right)\Phi_{a}\left(z\right)+...\right]\left|0\right\rangle \,,
\]
considering the second term on the expansion, we have
\[
\left\langle 0\right|\mbox{T }\Gamma_{\sigma}\left(w\right)\Gamma_{\lambda}\left(s\right)\overline{\Phi}_{i}\left(x\right)\Phi_{j}\left(y\right)\left[-\frac{1}{2}\,i\,g^{2}C^{\alpha\beta}\int d^{5}z\Gamma_{\beta}\left(z\right)\Gamma_{\alpha}\left(z\right)\overline{\Phi}_{a}\left(z\right)\Phi_{a}\left(z\right)\right]\left|0\right\rangle 
\]
\begin{eqnarray*}
=-\frac{1}{2}\, i\, g^{2}\,C^{\alpha\beta}\int d^{5}z\left(\contraction{}{\Gamma}{_{\sigma}\left(w\right)\Gamma_{\beta}\left(s\right)\overline{\Phi}_{i}\left(x\right)\Phi_{j}\left(y\right)}{\Gamma}
\contraction[2ex]{\Gamma^{\beta}\left(w\right)}{\Gamma}{_{\beta}\left(s\right)\overline{\Phi}_{i}\left(x\right)\Phi_{j}\left(y\right)\Gamma^{\alpha}\left(z\right)}{\Gamma}
\bcontraction{\Gamma^{\beta}\left(w\right)\Gamma_{\beta}\left(s\right)}{\overline{\Phi}}{_{i}\left(x\right)\Phi_{j}\left(y\right)\Gamma^{\alpha}\left(z\right)\Gamma_{\alpha}\left(z\right)\overline{\Phi}_{a}\left(z\right)}{\Phi}
\bcontraction[2ex]{\Gamma^{\beta}\left(w\right)\Gamma_{\beta}\left(s\right)\overline{\Phi}_{i}\left(x\right)}{\Phi}{_{j}\left(y\right)\Gamma^{\alpha}\left(z\right)\Gamma_{\alpha}\left(z\right)}{\overline{\Phi}}
\Gamma_{\sigma}\left(w\right)\Gamma_{\lambda}\left(s\right)\overline{\Phi}_{i}\left(x\right)\Phi_{j}\left(y\right)\Gamma_{\beta}\left(z\right)\Gamma_{\alpha}\left(z\right)\overline{\Phi}_{a}\left(z\right)\Phi_{a}\left(z\right)\right.\\
+\left.\contraction{}{\Gamma}{^{\beta}\left(w\right)\Gamma_{\beta}\left(s\right)\overline{\Phi}_{i}\left(x\right)\Phi_{j}\left(y\right)\Gamma^{\alpha}\left(z\right)}{\Gamma}
\contraction[2ex]{\Gamma^{\beta}\left(w\right)}{\Gamma}{_{\beta}\left(s\right)\overline{\Phi}_{i}\left(x\right)\Phi_{j}\left(y\right)}{\Gamma}
\bcontraction{\Gamma^{\beta}\left(w\right)\Gamma_{\beta}\left(s\right)}{\overline{\Phi}}{_{i}\left(x\right)\Phi_{j}\left(y\right)\Gamma^{\alpha}\left(z\right)\Gamma_{\alpha}\left(z\right)\overline{\Phi}_{a}\left(z\right)}{\Phi}
\bcontraction[2ex]{\Gamma^{\beta}\left(w\right)\Gamma_{\beta}\left(s\right)\overline{\Phi}_{i}\left(x\right)}{\Phi}{_{j}\left(y\right)\Gamma^{\alpha}\left(z\right)\Gamma_{\alpha}\left(z\right)}{\overline{\Phi}}
\Gamma_{\sigma}\left(w\right)\Gamma_{\lambda}\left(s\right)\overline{\Phi}_{i}\left(x\right)\Phi_{j}\left(y\right)\Gamma_{\beta}\left(z\right)\Gamma_{\alpha}\left(z\right)\overline{\Phi}_{a}\left(z\right)\Phi_{a}\left(z\right)\right)
\end{eqnarray*}
\begin{align*}
= & \,i\,g^{2}\,C^{\alpha\beta}\int d^{5}z\contraction{}{\Gamma}{_{\sigma}\left(w\right)}{\Gamma}\Gamma_{\sigma}\left(w\right)\Gamma_{\beta}\left(z\right)\contraction{}{\Gamma}{_{\lambda}\left(s\right)}{\Gamma}\Gamma_{\lambda}\left(s\right)\Gamma_{\alpha}\left(z\right)\contraction{}{\overline{\Phi}}{_{i}\left(x\right)}{\Phi}\overline{\Phi}_{i}\left(x\right)\Phi_{a}\left(z\right)\contraction{}{\overline{\Phi}}{_{a}\left(z\right)}{\Phi}\overline{\Phi}_{a}\left(z\right)\Phi_{j}\left(y\right)
\end{align*}
\vfill{}
and in momentum space, we find
\begin{align}
i\,V_{\overline{\Phi}_{a}\Phi_{a}\Gamma_{\beta}\Gamma_{\alpha}} & =i\,g^{2}\,\delta_{ij}\,C^{\alpha\beta}\int d^{2}\theta,\label{eq:Vertex4PointGauge-Scalar}
\end{align}
where the corresponding Feynman diagram is represented by $D2$ in
the Figure\,\ref{fig:The-vertices-Diagrams-1}.

Finally, performing a similar procedure, we find
\begin{align}
i\,V_{\Phi_{a}D^{\alpha}\overline{\Phi}_{a}\Gamma_{\alpha}-\Phi_{a}D^{\alpha}\Phi_{a}\Gamma_{\alpha}} & =\frac{1}{2}\,\delta_{ij}\,g\,\left[D^{\alpha}\left(p\right)-D^{\alpha}\left(-q\right)\right]\int d^{2}\theta,\label{eq:Vertex3PointGauge-Scalar}
\end{align}
where the corresponding Feynman diagram is represented by $D3$ in
the Figure\,\ref{fig:The-vertices-Diagrams-1}.

\section{Quantum corrections up two loops and renormalization group functions}

In this section we calculate the quantum corrections to the vertices
functions, up two loops, using the Feynman rules derived in Section\,\ref{sec:Feynman-rules};
from these, we will be able to calculate the $\beta$ and $\gamma$
functions of our model. The use of regularization by dimensional reduction
means that all super-algebra manipulations are performed in three
dimensions, while momentum integrals are calculated at dimension $d=3-\epsilon$
(see in Appendix\,\ref{chap:Useful-integrals} more details about
the momentum integrals). The use of this regularization scheme guarantees
that the one loop corrections are finite, see an example given in
Appendix\,\ref{chap:One-Loop-Correction}. The full set of topologies
was generated using the \textsc{Mathematica }package\textsc{ FeynArts\,\cite{Hahn2001}},
and the final diagrams drawn with the program\textsc{ JaxoDraw\,\cite{Binosi:2008ig}.}
All algebraic manipulations of supercovariant derivatives needed for
the evaluation of the superdiagrams were performed with the \textsc{Mathematica}
package \textsc{SusyMath}\,\cite{ferrarisusymath}; explicit details
about our calculations are presented in the Appendix\,\ref{chap:Two-loops-corrections}. 
\begin{center}
\begin{figure}
\centering{}\includegraphics[scale=0.8]{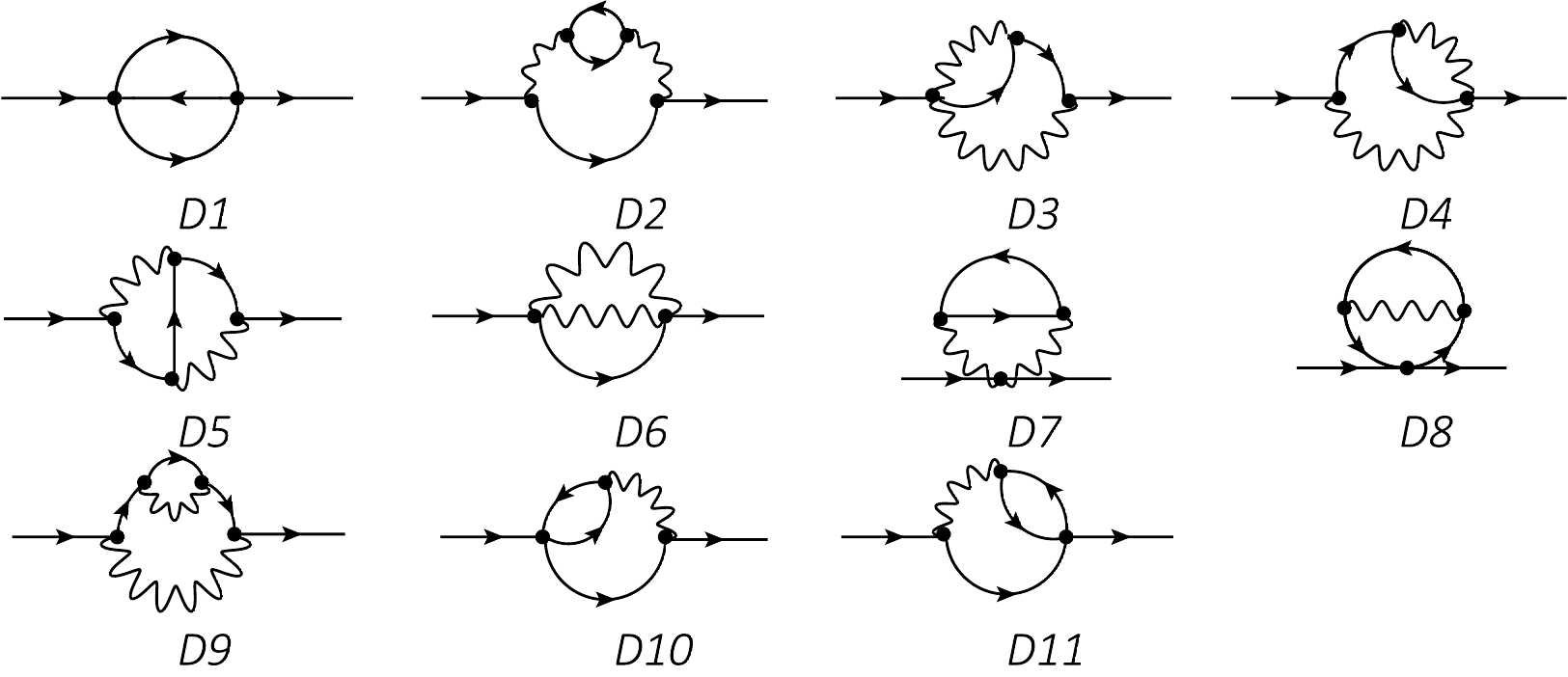}\caption{\label{fig:The-contribution-to-Two-Points-scalar}Two loops diagrams
contributing to the two-point vertex function of the scalar superfield
$\Phi$. }
\end{figure}
\par\end{center}

\begin{center}
\begin{table}
\centering{}%
\begin{tabular}{lccclccclcc}
 &  &  &  &  &  &  &  &  &  & \tabularnewline
\hline 
\hline 
$D1$ &  & $-2\left(N+1\right)\lambda^{2}$ &  & $D2$ &  & $\left(a+b\right)^{2}g^{4}N$ &  & $D3$ &  & $a\left(b-3\,a\right)g^{4}$\tabularnewline
$D4$ &  & $a\left(b-3\,a\right)\,g^{4}$ &  & $D5$ &  & $\frac{1}{2}\left(3\,a^{2}+2\,a\,b+3\,b^{2}\right)g^{4}$ &  & $D6$ &  & $2\,a^{2}\,g^{4}$\tabularnewline
$D7$ &  & $0$ &  & $D8$ &  & $0$ &  & $D9$ &  & $4\,a^{2}\,g^{4}$\tabularnewline
$D10$ &  & $0$ &  & $D11$ &  & $0$ &  &  &  & \tabularnewline
\hline 
\hline 
 &  &  &  &  &  &  &  &  &  & \tabularnewline
\end{tabular}\caption{\label{tab:Results-of-PhiPhi Diagrams}Divergent contributions from
each diagram presented in Figure\,\ref{fig:The-contribution-to-Two-Points-scalar},
with the common factor $\frac{1}{8}\left(\frac{i}{32\pi^{2}\epsilon}\right)\int\frac{d^{3}p}{\left(2\pi\right)^{3}}\,d^{2}\theta\,\overline{\Phi}_{i}\left(p,\theta\right)D^{2}\Phi_{i}\left(-p,\theta\right)$
omitted.}
\end{table}
\par\end{center}

We start by calculating the two-point vertex functions associated
to scalar and gauge superfields. For the case of the scalar superfield,
the diagrams that contribute are represented in Figure\,\ref{fig:The-contribution-to-Two-Points-scalar},
and the corresponding results are given in Table\,\ref{tab:Results-of-PhiPhi Diagrams}.
For the divergent part of the two points vertex function, at two loops,
we can write
\begin{align}
S_{\bar{\Phi}\Phi}^{2\,\mathrm{loop}} & =\frac{i}{4\left(32\pi^{2}\epsilon\right)}\left[-\left(N+1\right)\lambda^{2}+\frac{1}{4}\left(a+b\right)^{2}\left(2\,N+3\right)g^{4}\right]\nonumber \\
 & \times\int\frac{d^{3}p}{\left(2\pi\right)^{3}}\,d^{2}\theta\,\overline{\Phi}_{i}\left(p,\theta\right)D^{2}\Phi_{i}\left(-p,\theta\right)\,,\label{eq:PhiPhi contribution}
\end{align}
which fixes the value of the $Z_{\Phi}$ counterterm as
\begin{align}
Z_{\Phi} & =1+\frac{1}{4\left(32\pi^{2}\epsilon\right)}\left[-\left(N+1\right)\lambda^{2}+\frac{1}{4}\left(a+b\right)^{2}\left(2\,N+3\right)g^{4}\right].\label{eq:Ct ZPhi}
\end{align}
Remembering Eq.\,(\ref{eq:defAB}), we see that $Z_{\Phi}$ depends
only on the gauge independent combination $\left(a+b\right)$. From
this, it follows that the anomalous dimension $\gamma_{\Phi}$ will
also be gauge independent.
\begin{center}
\begin{figure}
\centering{}\includegraphics[scale=0.7]{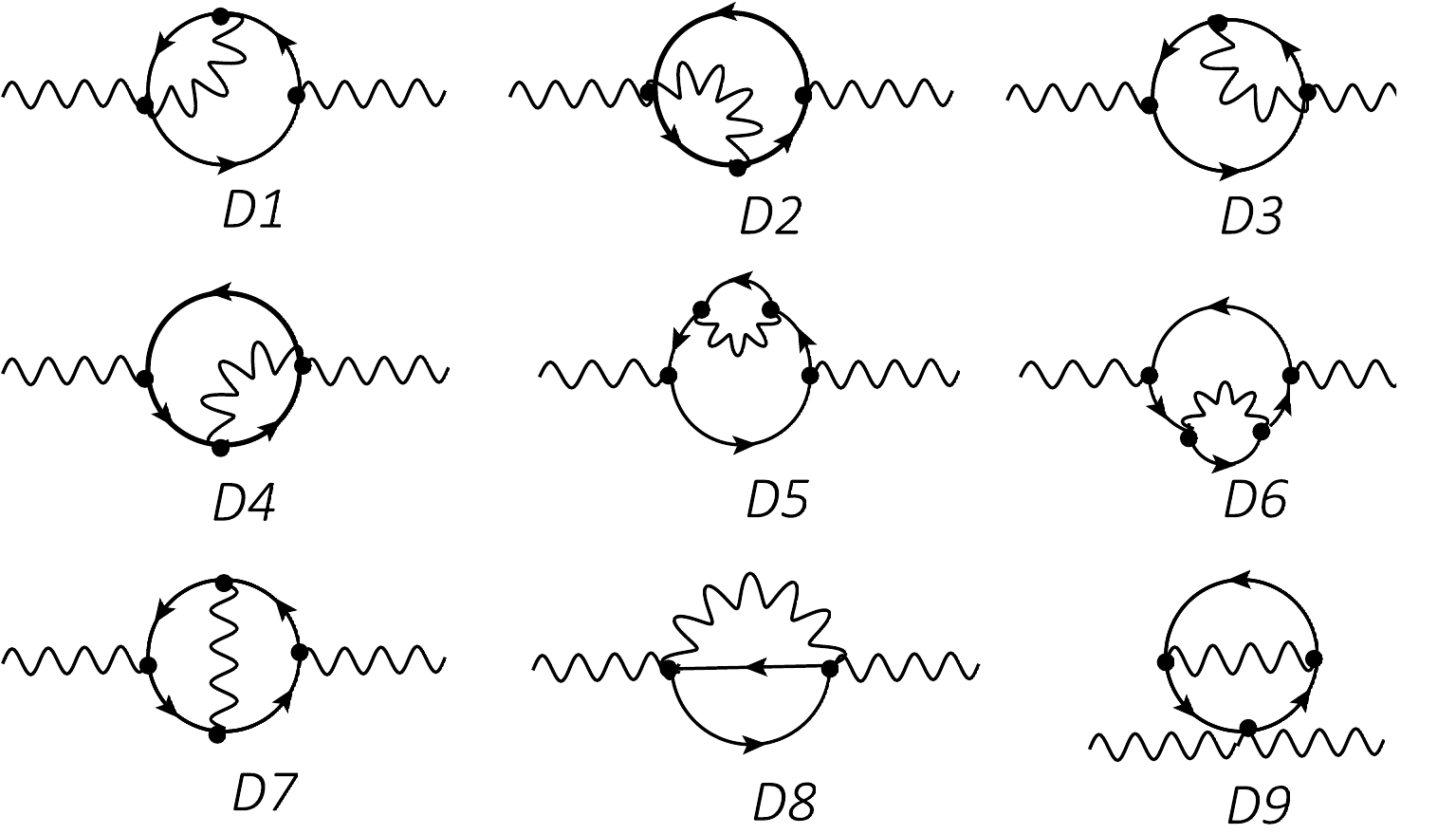}\caption{\label{fig:The-contribution-to-Two-Points-gauge}Two loop diagrams
contributing to the two-point vertex function of the gauge superfield
$\Gamma_{\alpha}$. }
\end{figure}
\par\end{center}

\begin{center}
\begin{table}
\centering{}%
\begin{tabular}{lcccccc}
 &  &  &  &  &  & \tabularnewline
\hline 
\hline 
$D1$ &  & $\left(3\,a-b\right)\left(-p_{\alpha\beta}+3\,C_{\beta\alpha}\,D^{2}\right)$ &  & $D2$ &  & $\left(3\,a-b\right)\left(-p_{\alpha\beta}+3\,C_{\beta\alpha}\,D^{2}\right)$\tabularnewline
$D3$ &  & $\left(3\,a-b\right)\left(-p_{\alpha\beta}+3\,C_{\beta\alpha}\,D^{2}\right)$ &  & $D4$ &  & $\left(3\,a-b\right)\left(-p_{\alpha\beta}+3\,C_{\beta\alpha}\,D^{2}\right)$\tabularnewline
$D5$ &  & $-2\,a\left(p_{\alpha\beta}+3\,C_{\beta\alpha}\,D^{2}\right)$ &  & $D6$ &  & $-2\,a\left(p_{\alpha\beta}\,+3\,C_{\beta\alpha}\,D^{2}\right)$\tabularnewline
$D7$ &  & $4\left(\left(4\,a+b\right)p_{\alpha\beta}+3\,b\,C_{\beta\alpha}\,D^{2}\right)$ &  & $D8$ &  & $-b\,p_{\alpha\beta}-3\,a\,C_{\beta\alpha}\,D^{2}$\tabularnewline
$D9$ &  & $0$ &  &  &  & \tabularnewline
\hline 
\hline 
 &  &  &  &  &  & \tabularnewline
\end{tabular}\caption{\label{tab:Result-of-GammaGamma}Divergent contributions from each
diagram in Figure\,\ref{fig:The-contribution-to-Two-Points-gauge},
omitting the common factor $\frac{1}{8}\left(\frac{N}{192\pi^{2}\epsilon}\right)i\,g^{4}\int\frac{d^{3}p}{\left(2\pi\right)^{3}}\,d^{2}\theta\,\Gamma^{\alpha}\left(p,\theta\right)\Gamma^{\beta}\left(-p,\theta\right)$.}
\end{table}
\par\end{center}

The next step is to compute, up to two loops, the two-point vertex
function of the gauge superfield $\Gamma_{\alpha}$. The diagrams
involved are represented in Figure\,\ref{fig:The-contribution-to-Two-Points-gauge},
with the respective divergent contributions given in Table\,\ref{tab:Result-of-GammaGamma}.
We verify that, for any gauge choice, all divergences cancel among
the diagrams in Figure\,\ref{fig:The-contribution-to-Two-Points-gauge}.
As a consequence, no infinite wave function renormalization of the
CS superfield is needed, and therefore the anomalous dimension $\gamma_{\Gamma}$
vanishes ($Z_{\Gamma}=1$). This result extends for the massless matter
case according to the Coleman-Hill theorem\,\cite{Coleman:1985zi},
which states that in Abelian theories there are no quantum correction
above one loop order for the CS coefficient, and it was also verified
in previous calculations performed in a specific gauge\,\cite{Avdeev:1991za},
as well as in the non-supersymmetric version of the model\,\cite{dias:2003pw}.

Finally, the evaluation of the divergent part of the four-point vertex
function associated to the scalar superfield $\Phi$, up to two loops,
involve all diagrams in Figure\,\ref{fig:The-contribution-to-Four-Points-scalar},
these $41$ topologies being equivalent to $371$ diagrams, see Appendix\,\ref{sec:Four-points-vertex}
for more details. The results are given in Table\,\ref{tab:Result-of-Four-Point},
and lead to
\begin{align}
S_{\left(\bar{\Phi}\Phi\right)^{2}}^{2\,\mathrm{loop}} & =\frac{i}{4\left(32\pi^{2}\epsilon\right)}\left[-\left(5\,N+11\right)\lambda^{3}-\left(a+b\right)\lambda^{2}\,g^{2}+\frac{1}{4}\left(a+b\right)^{2}\left(2\,N+5\right)\lambda\,g^{4}\right.\nonumber \\
 & \left.+\frac{1}{4}\left(a+b\right)^{3}\left(N+3\right)g^{6}\right]\left(\delta_{im}\,\delta_{nj}+\delta_{jm}\,\delta_{ni}\right)\int_{\theta}\overline{\Phi}_{i}\Phi_{m}\Phi_{n}\overline{\Phi}_{j}\thinspace,\label{eq:PhiPhiPhiPhi contribution}
\end{align}
where $\int_{\theta}\overline{\Phi}_{i}\Phi_{m}\Phi_{n}\overline{\Phi}_{j}\equiv\int d^{2}\theta\overline{\Phi}_{i}\left(0,\theta\right)\Phi_{m}\left(0,\theta\right)\Phi_{n}\left(0,\theta\right)\overline{\Phi}_{j}\left(0,\theta\right)$.
Here we chose all the external momenta to be zero, since the tree-level
vertex factor
\begin{equation}
\frac{1}{2}\,i\,\lambda Z_{\lambda}\left(\delta_{in}\delta_{jm}+\delta_{im}\delta_{jn}\right)\int d^{2}\theta,
\end{equation}
does not depend of external momenta. Equation\,(\ref{eq:PhiPhiPhiPhi contribution})
implies that
\begin{align}
Z_{\lambda} & =1+\frac{1}{2\left(32\pi^{2}\epsilon\right)}\left[\left(5\,N+11\right)\lambda^{2}+\left(a+b\right)\lambda\,g^{2}-\frac{1}{4}\left(a+b\right)^{2}\left(2\,N+5\right)g^{4}\right.\nonumber \\
 & \left.-\frac{1}{4}\left(a+b\right)^{3}\left(N+3\right)\lambda^{-1}g^{6}\right].\label{eq:Ct ZLambda}
\end{align}
Again, this results turns out to be independent of the gauge parameter
$\alpha$.

In conclusion, all the renormalization constants $Z_{\Phi}$, $Z_{\Gamma}$,
and $Z_{\lambda}$ are independent of the choice of the gauge-fixing
parameter. This same property will follow for the renormalization
group functions that are calculated from these constants.
\begin{center}
\begin{figure}
\centering{}\includegraphics[scale=0.9]{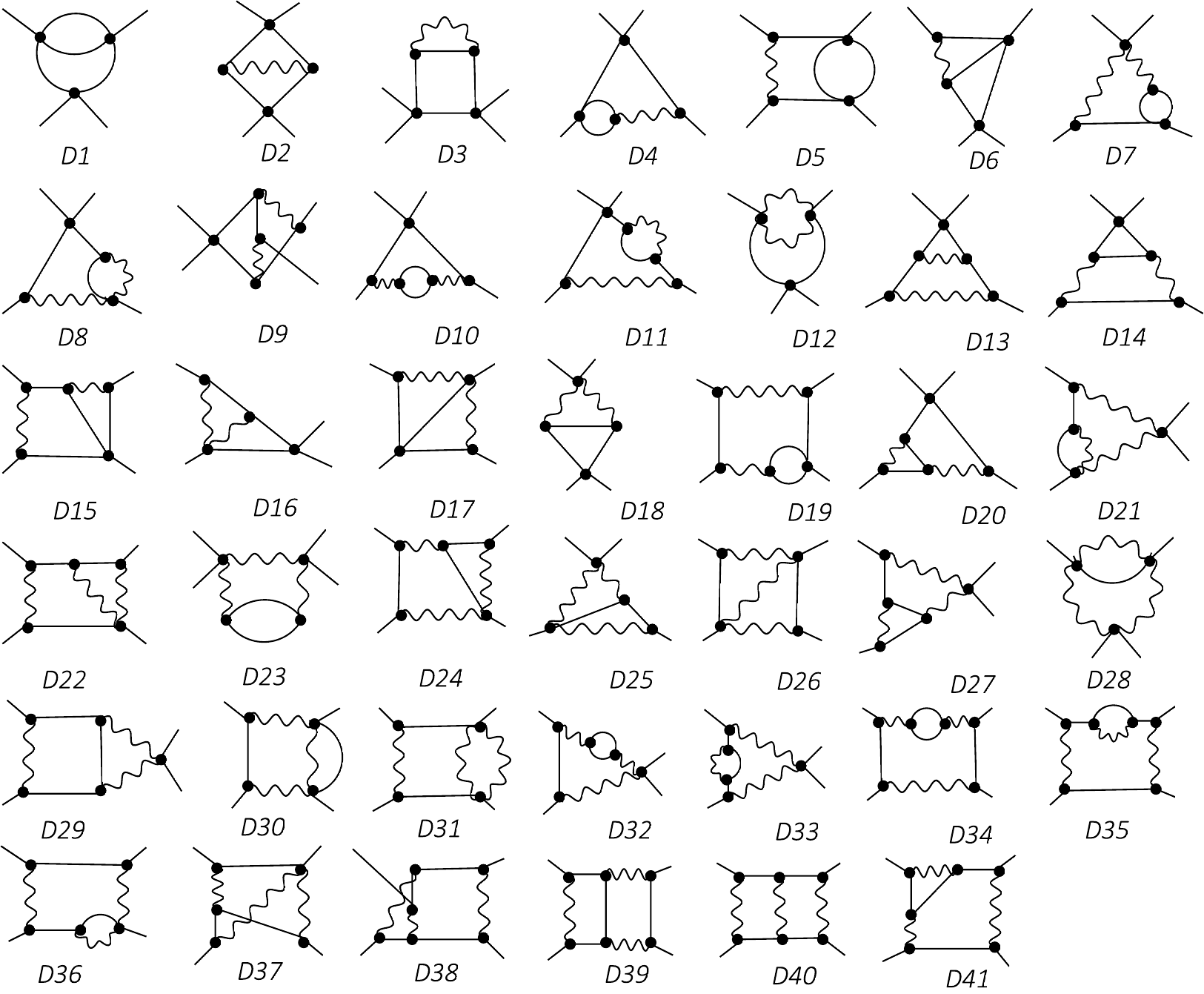}\caption{\label{fig:The-contribution-to-Four-Points-scalar}The complete set
of two loops diagrams contributing to the four-point scalar vertex
function.}
\end{figure}
\par\end{center}

\begin{center}
\begin{table}
\centering{}%
\begin{tabular}{lccclcc}
 &  &  &  &  &  & \tabularnewline
\hline 
\hline 
$D1$ &  & $-\frac{1}{4}\left(5N+11\right)\lambda^{3}$ &  & $D2$ &  & $\frac{1}{4}\,a\left(N+2\right)\lambda^{2}\,g^{2}$\tabularnewline
$D3$ &  & $-\frac{1}{4}\,a\left(N+4\right)\lambda^{2}\,g^{2}$ &  & $D4$ &  & $0$\tabularnewline
$D5$ &  & $-\frac{1}{4}\left(a-b\right)\lambda^{2}\,g^{2}$ &  & $D6$ &  & $\frac{1}{2}\left(a-b\right)\lambda^{2}\,g^{2}$\tabularnewline
$D7$ &  & $0$ &  & $D8$ &  & $\frac{1}{8}\left(b^{2}-4\,a\,b+3\,a^{2}\right)\lambda\,g^{4}$\tabularnewline
$D9$ &  & $\frac{3}{16}\left(a-b\right)^{2}\lambda\,g^{4}$ &  & $D10$ &  & $0$\tabularnewline
$D11$ &  & $\frac{1}{4}\,a\left(b-a\right)\lambda\,g^{4}$ &  & $D12$ &  & $\frac{3}{4}\,a^{2}\lambda\,g^{4}$\tabularnewline
$D13$ &  & $\frac{3}{4}\,a\,\left(a-b\right)\lambda\,g^{4}$ &  & $D14$ &  & $0$\tabularnewline
$D15$ &  & $-\frac{1}{4}\left(a-b\right)^{2}\lambda\,g^{4}$ &  & $D16$ &  & $\frac{3}{2}\,a\,\left(b-a\right)\lambda\,g^{4}$\tabularnewline
$D16$ &  & $\frac{3}{2}\,a\,\left(b-a\right)\lambda\,g^{4}$ &  & $D17$ &  & $\frac{1}{4}\left(a-b\right)^{2}\lambda\,g^{4}$\tabularnewline
$D18$ &  & $\frac{1}{8}\left(a+b\right)^{2}\left(1+N\right)\lambda\,g^{4}$ &  & $D19$ &  & $0$\tabularnewline
$D20$ &  & $-\frac{1}{8}\left(a-b\right)^{2}\lambda\,g^{4}$ &  & $D21$ &  & $-\frac{1}{8}\left(a-b\right)^{2}\left(3\,a-b\right)g^{6}$\tabularnewline
$D22$ &  & $0$ &  & $D23$ &  & $\frac{1}{16}\left(a+b\right)^{3}N\,g^{6}$\tabularnewline
$D24$ &  & $\frac{1}{8}\left(a-b\right)^{3}g^{6}$ &  & $D25$ &  & $-\frac{1}{4}\left(a-b\right)\left(a^{2}+b^{2}\right)g^{6}$\tabularnewline
$D26$ &  & $0$ &  & $D27$ &  & $\frac{1}{8}\left(a-b\right)^{3}g^{6}$\tabularnewline
$D28$ &  & $\frac{1}{2}\,a\,\left(a^{2}+2\,b^{2}\right)g^{6}$ &  & $D29$ &  & $\frac{1}{16}\left(a-b\right)\left(a+b\right)^{2}g^{6}$\tabularnewline
$D30$ &  & $-\frac{1}{8}\left(a-b\right)^{2}\left(2\,a-b\right)g^{6}$ &  & $D31$ &  & $0$\tabularnewline
$D32$ &  & $0$ &  & $D33$ &  & $\frac{1}{8}\,a\left(a-b\right)^{2}g^{6}$\tabularnewline
$D34$ &  & $0$ &  & $D35$ &  & $-\frac{1}{8}\,a\left(a-b\right)^{2}g^{6}$\tabularnewline
$D36$ &  & $\frac{1}{8}\left(a-b\right)^{2}\left(3\,a-b\right)g^{6}$ &  & $D37$ &  & $0$\tabularnewline
$D38$ &  & $0$ &  & $D39$ &  & $0$\tabularnewline
$D40$ &  & $0$ &  & $D41$ &  & $\frac{1}{8}\left(b-a\right)^{3}g^{6}$\tabularnewline
\hline 
\hline 
 &  &  &  &  &  & \tabularnewline
\end{tabular}\caption{\label{tab:Result-of-Four-Point}Divergent contributions arising from
the diagrams presented in Figure\,\ref{fig:The-contribution-to-Four-Points-scalar};
all contributions include the common factor $\frac{i}{32\pi^{2}\epsilon}\left(\delta_{im}\,\delta_{nj}+\delta_{jm}\,\delta_{ni}\right)\int d^{2}\theta\,\Phi_{n}\Phi_{m}\overline{\Phi}_{i}\overline{\Phi}_{j}$,
where all external momenta were set to zero.}
\end{table}
\par\end{center}

The explicit relations between bare and renormalized quantities are
given by\begin{subequations}\label{eq:bare-relation}
\begin{align}
\Gamma_{0}^{\alpha} & =Z_{\Gamma}^{\frac{1}{2}}\Gamma^{\alpha},\\
\Phi_{0} & =Z_{\Phi}^{\frac{1}{2}}\Phi,\\
\alpha_{0} & =\alpha\,Z_{\Gamma},\\
g_{0} & =\mu^{\frac{\epsilon}{2}}g\,Z_{\Gamma}^{-\frac{1}{2}},\\
\lambda_{0} & =\mu^{\epsilon}\lambda\,Z_{\lambda}Z_{\Phi}^{-2},
\end{align}
\end{subequations}where $\mu$ is a mass parameter introduced to
keep $g$ and $\lambda$ dimensionless. From these, we can obtain
the renormalization group functions,\begin{subequations}\label{eq:RGE-relations}
\begin{align}
\gamma_{\Gamma} & \equiv-\frac{\mu}{\Gamma}\frac{d}{d\mu}\Gamma=\frac{\mu}{2\,Z_{\Gamma}}\frac{d}{d\mu}Z_{\Gamma},\\
\gamma_{\Phi} & =\frac{\mu}{2\,Z_{\Phi}}\frac{d}{d\mu}Z_{\Phi},\\
\beta_{\alpha} & \equiv\mu\frac{d}{d\mu}\alpha=-2\,\alpha\,\gamma_{\Gamma},\\
\beta_{g} & =g\,\gamma_{\Gamma},\\
\beta_{\lambda} & =\epsilon\,\frac{\lambda^{2}}{Z_{\lambda}}\frac{\partial}{\partial\lambda}Z_{\lambda}+\frac{\epsilon}{2}\frac{\lambda\,g}{Z_{\lambda}}\frac{\partial}{\partial g}Z_{\lambda}+4\lambda\gamma_{\Phi}.
\end{align}
\end{subequations}From these definitions, and the results presented
in this section, we obtain\begin{subequations}\label{eq:RGEfunctions}
\begin{align}
\beta_{\lambda} & =c_{3}\,\lambda^{3}+c_{2}\,\lambda^{2}\,y+c_{1}\,\lambda\,y^{2}+c_{0}\,y^{3}\thinspace,\\
\gamma_{\Gamma}=\beta_{\alpha}=\beta_{y} & =0\thinspace,\\
\gamma_{\Phi} & =d_{2}\,\lambda^{2}+d_{0}\,y^{2}\thinspace,
\end{align}
\end{subequations}in terms of the redefined gauge coupling constant
\begin{equation}
y=g^{2}\thinspace.
\end{equation}
The numerical coefficients present in\,(\ref{eq:RGEfunctions}) are
given by
\begin{eqnarray}
c_{3}=\frac{3}{16\pi^{2}}\left(N+2\right), &  & c_{2}=\frac{1}{16\pi^{2}},\nonumber \\
c_{1}=-\frac{2}{16\pi^{2}}\left(N+2\right), &  & c_{0}=-\frac{1}{16\pi^{2}}\left(N+3\right),\nonumber \\
d_{2}=\frac{1}{4\left(32\pi^{2}\right)}\left(N+1\right), &  & d_{1}=0,\nonumber \\
d_{0}=-\frac{1}{4\left(32\pi^{2}\right)}\left(2N+3\right).\label{eq:Coefficient-Beta}
\end{eqnarray}

As stated before, these functions are independent of the gauge-fixing
parameter. Our results agree with\,\cite{Avdeev:1991za}, except
for overall numerical factors that arise due to their different definition
of the scale $\mu$. We stress however that in\,\cite{Avdeev:1991za}
the authors considered as true the gauge independence of these functions,
a fact that has been checked explicitly here.
\begin{center}
\myclearpage
\par\end{center}

\chapter{\label{chap:RGE-Impro-DBS}Renormalization Group Improvement and
Dynamical Symmetry Breaking}

In this chapter, we investigate the consequences of the RGE in the
determination of the effective superpotential and the study of DSB
in an $\mathcal{N}=1$ supersymmetric theory including an Abelian
CS superfield coupled to $N$ scalar superfields in $(2+1)$ dimensional
space-time, as already detailed in Chapter\,\ref{chap:Renormalization-Group-Functions}.
The classical Lagrangian presents scale invariance, which is broken
by radiative corrections to the effective superpotential. We calculate
the effective superpotential up to two-loops by using the RGE and
the beta functions and anomalous dimensions found in Chapter\,\ref{chap:Renormalization-Group-Functions}.
We then show how the RGE can be used to improve this calculation,
by summing up properly defined series of leading logs (LL), next-to-leading
logs (NLL) contributions, and so on... We conclude that even if the
RGE improvement procedure can indeed be applied in a supersymmetric
model, the effects of the consideration of the RGE are not so dramatic
as it happens in the non-supersymmetric case. 

\section{\label{sec:Model}Calculation of the Effective Superpotential}

The main object we shall be interested in studying is the three-dimensional
effective superpotential. To define this object, we consider a shift
in the $N$-th component of $\Phi_{a}$ in\,(\ref{eq:M1}),
\begin{align}
\Phi_{\mathrm{N}} & =\Phi_{\mathrm{N}}^{q}+\sigma_{\mathrm{cl}},\label{eq:M2}
\end{align}
where $\sigma_{\mathrm{cl}}$ was defined in Eq.\,(\ref{eq:background-superfield}).
On general grounds, the effective superpotential is given by Eq.\,(\ref{eq:SuperPotencialCompleto}).

As discussed in\,\cite{Ferrari:2010ex}, the knowledge of $K_{\mathrm{eff}}\left(\sigma_{\mathrm{cl}},\alpha\right)$
in\,(\ref{eq:SuperPotencialCompleto}) is sufficient for investigating
the dynamical breaking of gauge symmetry, and consequential generation
of a mass scale in the model. For simplicity, in this work we will
restrict ourselves to the calculation of $K_{\mathrm{eff}}\left(\sigma_{\mathrm{cl}},\alpha\right)$,
which we shall call the\emph{ effective superpotential} from now on.
The effective superpotential $K_{\mathrm{eff}}\left(\sigma_{\mathrm{cl}},\alpha\right)$
is particularly well suited for the approach that we develop in this
thesis, since we will be able to calculate it by using a simple ansatz,
from the renormalization group functions for the model\,(\ref{eq:M1})
found in Chapter\,\ref{chap:Renormalization-Group-Functions}, see
Eq.\,(\ref{eq:RGEfunctions}).

We shall use for $K_{\mathrm{eff}}\left(\sigma_{\mathrm{cl}},\alpha\right)$
the ansatz
\begin{equation}
K_{\mathrm{eff}}\left(\sigma_{\mathrm{cl}},\alpha\right)=-\frac{1}{4}\sigma_{\mathrm{cl}}^{4}\,S_{\mathrm{eff}}\left(\sigma_{\mathrm{cl}},\lambda,y,\alpha,L\right)\thinspace,\label{eq:KeffAnsatz1}
\end{equation}
where
\begin{equation}
S_{\mathrm{eff}}\left(\sigma_{\mathrm{cl}},\lambda,y,\alpha,L\right)=A\left(y,\lambda,\alpha\right)+B\left(y,\lambda,\alpha\right)L+C\left(y,\lambda,\alpha\right)L^{2}+\cdots\thinspace,\label{eq:KeffAnsatz2}
\end{equation}
and $A,\thinspace B,\thinspace C,\thinspace\ldots$ are defined as
series in powers of the coupling constants $y$ and $\lambda$, and
$L=\ln\left[\frac{\sigma_{\mathrm{cl}}^{2}}{\mu}\right]$. We will
eventually adopt a shorthand notation where $x$ denotes any of the
two couplings in our model, so that a monomial like $y^{n}\lambda^{m}$
will be written as $x^{m+n}$. These ansatz comes from the conformal
invariance at tree-level, leading to the fact that we can have only
one type of logarithm appearing in the quantum corrections. Comparison
with the tree level action\,(\ref{eq:M1}) shows us that
\begin{equation}
A\left(y,\lambda,\alpha\right)=\lambda+{\cal O}\left(x^{2}\right)\thinspace.\label{eq:A}
\end{equation}
The value of $A\left(y,\lambda,\alpha\right)$ will be fixed by the
CW normalization of the effective superpotential, 
\begin{equation}
\frac{1}{4!}\frac{d^{4}K_{\mathrm{eff}}\left(\sigma_{\mathrm{cl}},\alpha\right)}{d^{4}\sigma_{\mathrm{cl}}}=\frac{\lambda}{4}\thinspace,\label{eq:CWcondition}
\end{equation}
so only the $L$ dependent pieces of $K_{\mathrm{eff}}\left(\sigma_{\mathrm{cl}},\alpha\right)$,
involving $B,\thinspace C,\thinspace\ldots$, have to be calculated.

The other ingredient that we will need is the RGE. To obtain this
equation, we start with 
\begin{align}
K_{\mathrm{eff}}^{0}\left(\sigma_{\mathrm{cl}};\lambda_{0},y_{0},\alpha_{0}\right) & =K_{\mathrm{eff}}\left(\sigma_{\mathrm{cl}};\lambda\left(\mu\right),y\left(\mu\right),\alpha\left(\mu\right),\mu,L\right)\,,\label{eq:K0}
\end{align}
where $K^{0}$ is independent of the arbitrary mass scale $\mu$.
By deriving Eq,\,(\ref{eq:K0}) with respect to $\mu$ 
\begin{align}
0 & =\left(\mu\frac{\partial}{\partial\mu}+\mu\frac{d\lambda}{d\mu}\frac{\partial}{\partial\lambda}+\mu\frac{dy}{d\mu}\frac{\partial}{\partial y}+\mu\frac{d\alpha}{d\mu}\frac{\partial}{\partial\alpha}+\mu\frac{d\sigma_{\mathrm{cl}}}{d\mu}\frac{\partial}{\partial\sigma_{\mathrm{cl}}}\right)K_{\mathrm{eff}}\left(\sigma_{\mathrm{cl}};\lambda,y,\alpha,\mu,L\right)\,,\label{eq:RGE-K-1}
\end{align}
and using Eqs.\,(\ref{eq:RGE-relations}), we have 
\begin{align}
0 & =\left(\mu\frac{\partial}{\partial\mu}+\beta_{\lambda}\frac{\partial}{\partial\lambda}+\beta_{y}\frac{\partial}{\partial y}+\beta_{\alpha}\frac{\partial}{\partial\alpha}-\gamma_{\Phi}\sigma_{\mathrm{cl}}\frac{\partial}{\partial\sigma_{\mathrm{cl}}}\right)K_{\mathrm{eff}}\left(\sigma_{\mathrm{cl}};\lambda,y,\alpha,\mu,L\right)\,,\label{eq:RGE-Keff}
\end{align}
where we used that $\gamma_{\Phi}=\gamma_{\sigma}$\footnote{Using the shift of superfield $\Phi$, Eq.\,(\ref{eq:M2}) and\,(\ref{eq:bare-relation})
we have
\begin{align}
\Phi_{0} & =Z_{\Phi}^{\frac{1}{2}}\left(\Phi+\sigma_{\mathrm{cl}}\right)\,,
\end{align}
by deriving with respect to $\mu$, we find
\begin{align}
\Phi\left(\gamma_{\Phi}-\frac{1}{2}\frac{\mu}{Z_{\Phi}}\frac{d}{d\mu}Z_{\Phi}\right)+\sigma_{\mathrm{cl}}\left(\gamma_{\sigma}-\frac{1}{2}\frac{\mu}{Z_{\Phi}}\frac{d}{d\mu}Z_{\Phi}\right) & =0\,,
\end{align}
and from this, we have $\gamma_{\Phi}=\gamma_{\sigma}.$ }. From Eq,\,(\ref{eq:defL}), it follows that 
\begin{equation}
\partial_{L}=\frac{1}{2}\sigma_{\mathrm{cl}}\partial_{\sigma_{\mathrm{cl}}}=-\mu\partial_{\mu}\thinspace,
\end{equation}
where it was used the notation $\partial_{t}\equiv\frac{\partial}{\partial t}$,
and inserting\,(\ref{eq:KeffAnsatz1}) into\,(\ref{eq:RGE-Keff}),
we obtain an alternative form for the RGE,
\begin{equation}
\left[-\left(1+2\gamma_{\Phi}\right)\partial_{L}+\beta_{\lambda}\,\partial_{\lambda}-4\,\gamma_{\Phi}\right]S_{\mathrm{eff}}\left(\sigma_{\mathrm{cl}},\lambda,y,\alpha,L\right)=0\thinspace,\label{eq:RGE2}
\end{equation}
where it was used that $\beta_{y}=\beta_{\alpha}=0,$ which will be
used hereafter. 

Inserting the ansatz\,(\ref{eq:KeffAnsatz2}) in\,(\ref{eq:RGE2}),
and separating the resulting expression by orders of $L$, we obtain
a series of equations, of which we quote the first two:
\begin{equation}
-\left(1+2\gamma_{\Phi}\right)B\left(y,\lambda,\alpha\right)+\beta_{\lambda}\,\partial_{\lambda}\,A\left(y,\lambda,\alpha\right)-4\,\gamma_{\Phi}\,A\left(y,\lambda,\alpha\right)=0\thinspace,\label{eq:orderL0}
\end{equation}
and
\begin{equation}
-2\left(1+2\gamma_{\Phi}\right)C\left(y,\lambda,\alpha\right)+\beta_{\lambda}\,\partial_{\lambda}\,B\left(y,\lambda,\alpha\right)-4\,\gamma_{\Phi}\,B\left(y,\lambda,\alpha\right)=0\thinspace.\label{eq:orderL1}
\end{equation}
We now consider that all functions appearing in these equations are
defined as series in powers of the couplings $x$, writing Eq.\,(\ref{eq:orderL0})
as
\begin{multline}
-\left(B^{\left(1\right)}+B^{\left(2\right)}+B^{\left(3\right)}+\cdots\right)-2\left(\gamma_{\Phi}^{\left(2\right)}+\gamma_{\Phi}^{\left(3\right)}+\cdots\right)\left(B^{\left(1\right)}+B^{\left(2\right)}+B^{\left(3\right)}+\cdots\right)\\
+\left(\beta_{\lambda}^{\left(3\right)}+\beta_{\lambda}^{\left(4\right)}+\cdots\right)\left(\partial_{\lambda}A^{\left(1\right)}+\partial_{\lambda}A^{\left(2\right)}+\cdots\right)-4\left(\gamma_{\Phi}^{\left(2\right)}+\gamma_{\Phi}^{\left(3\right)}+\cdots\right)\left(A^{\left(1\right)}+A^{\left(2\right)}+\cdots\right)=0\thinspace,\label{eq:orderL0powers}
\end{multline}
where the numbers in the superscripts denote the power of $x$ of
each term. Since all terms of the previous equation start at order
$x^{3}$, except the first, we conclude that $B^{\left(1\right)}=B^{\left(2\right)}=0$,
and obtain the relation
\begin{equation}
B^{\left(3\right)}=\beta_{\lambda}^{\left(3\right)}-4\,\lambda\,\gamma_{\Phi}^{\left(2\right)}\thinspace,
\end{equation}
after using Eq.\,(\ref{eq:A}). This last equation fixes the coefficients
of $B^{\left(3\right)}$ in terms of the (known) coefficients of $\beta_{\lambda}^{\left(3\right)}$
and $\gamma_{\Phi}^{\left(2\right)}$, in the following form,
\begin{equation}
B^{\left(3\right)}=b_{3}\,\lambda^{3}+b_{2}\,\lambda^{2}\,y+b_{1}\,\lambda\,y^{2}+b_{0}\,y^{3}\thinspace,\label{eq:B3}
\end{equation}
where\begin{subequations}\label{eq:B3coefs}
\begin{align}
b_{0} & =c_{0},\\
b_{1} & =c_{1}-4\,d_{0}\,,\\
b_{2} & =c_{2}\,,\\
b_{3} & =c_{3}-4\,d_{2}.
\end{align}
\end{subequations}We observe that $B^{\left(3\right)}$ do not depend
on the gauge-fixing parameter, inheriting this property from the renormalization
group function. The corrections of order $x^{3}L$, which we have
found for $S_{\eff}$, could be obtained by a two-loop calculation
of the effective superpotential, using supergraph methods. Since we
do not know the coefficients of $\beta_{\lambda}^{\left(4\right)}$
and $\gamma_{\Phi}^{\left(3\right)}$, which would appear from higher
loop corrections, we cannot use Eq.\,(\ref{eq:orderL0}) to calculate
further coefficients of $B$ or $A$. 

Now looking at Eq.\,(\ref{eq:orderL1}) expanded in powers of the
couplings,
\begin{align}
-2\left(C^{\left(1\right)}+C^{\left(2\right)}+C^{\left(3\right)}+C^{\left(4\right)}+C^{\left(5\right)}+\cdots\right)+\left(\beta_{\lambda}^{\left(3\right)}+\beta_{\lambda}^{\left(4\right)}+\cdots\right)\partial_{\lambda}\,B^{\left(3\right)}\nonumber \\
-4\left(\gamma_{\Phi}^{\left(2\right)}+\gamma_{\Phi}^{\left(3\right)}+\cdots\right)\left(C^{\left(1\right)}+C^{\left(2\right)}+C^{\left(3\right)}+\cdots\right)-4\left(\gamma_{\Phi}^{\left(2\right)}+\gamma_{\Phi}^{\left(3\right)}+\cdots\right)B^{\left(3\right)} & =0,
\end{align}
one may conclude that $C\left(\lambda,y,\alpha\right)$ starts at
order $x^{5}$, and obtain the relation,
\begin{equation}
C^{\left(5\right)}=\frac{1}{2}\,\beta_{\lambda}^{\left(3\right)}\,\partial_{\lambda}\,B^{\left(3\right)}-2\,\gamma_{\Phi}^{\left(2\right)}\,B^{\left(3\right)}\thinspace,\label{eq:C5-1}
\end{equation}
from which the coefficients of the form $x^{5}L^{2}$ of $S_{\eff}$
are calculated from known coefficients of the beta functions, anomalous
dimension, and $B^{\left(3\right)}$. The end result is as follows,
\begin{multline}
C^{\left(5\right)}=\lambda^{5}\left(\frac{3}{2}\,c_{3}\,b_{3}-2\,d_{2}\,b_{3}\right)+\lambda^{4}\,y\left(c_{3}\,b_{2}+\frac{3}{2}\,c_{2}\,b_{3}-2\,d_{2}\,b_{2}\right)\\
+\lambda^{3}\,y^{2}\left(\frac{1}{2}\,c_{3}\,b_{1}+c_{2}\,b_{2}+\frac{3}{2}\,c_{1}\,b_{3}-2d_{0}\,b_{3}-2\,d_{2}\,b_{1}\right)\\
+\lambda^{2}\,y^{3}\left(\frac{1}{2}\,c_{2}\,b_{1}+c_{1}\,b_{2}+\frac{3}{2}\,c_{0}\,b_{3}-2d_{0}\,b_{2}-2\,d_{2}\,b_{0}\right)\\
+\lambda\,y^{4}\left(\frac{1}{2}\,c_{1}\,b_{1}+c_{0}\,b_{2}-2\,d_{0}\,b_{1}\right)+y^{5}\left(\frac{1}{2}\,c_{0}\,b_{1}-2\,d_{0}\,b_{0}\right)\thinspace.\label{eq:C5explicit}
\end{multline}
As before, we see that $C^{\left(5\right)}$ do not depend on the
gauge-fixing parameter. As a result, the RGE allows us to calculate
terms of order $x^{5}L^{2}$ which, in our model, would appear only
at four loops in an explicit evaluation of $K_{\eff}$. The Eq.\,(\ref{eq:orderL1})
does not provide us with order $x^{6}L^{2},\thinspace x^{7}L^{2},\thinspace\ldots$
terms, since we do not have knowledge of higher orders coefficients
of $\beta_{\lambda}$ and $\gamma_{\Phi}$. 

At this point, it is clear that one could go on calculating order
$x^{7}L^{3},\thinspace x^{9}L^{4},\thinspace\ldots$ terms from Eqs.\,(\ref{eq:KeffAnsatz2})
and\,(\ref{eq:RGE2}), obtaining contributions to the effective superpotential
arising from higher loop orders, based only on the information we
have from the two loop calculation of $\beta_{\lambda}$ and $\gamma_{\Phi}$.
Therefore, we can conclude that the effective superpotential does
not depend on the gauge-fixing parameter. In the next section, we
will give an explanation of this pattern of coefficients we are able
to calculate, interpreting it as a leading logs summation of the effective
superpotential.

\section{\label{sec:Summing-up-Leading}RGE improvement and DSB: A short review
of the four dimensional case}

We now review the procedure for the RGE improvement of the effective
potential that was applied to the Standard Model in\,\cite{elias:2003zm,Chishtie:2005hr,Chishtie:2010ni,Steele:2012av}
and to the non supersymmetric version of the model studied in this
work in\,\cite{Dias:2010it}. In doing so, we will be able to pinpoint
the differences we find in the supersymmetric three dimensional case,
still recognizing that the procedure outlined in the previous section
is essentially the same used in these works. 

Consider a scale invariant $\varphi^{4}$ model in $\left(3+1\right)$
dimensions, coupled to other fermionic or gauge fields via a set of
couplings denoted collectively by $x$. The effective potential ${\cal V}_{\eff}\left(\phi;\mu,x,{\cal L}\right)$
satisfy the RGE
\begin{equation}
\left[\mu\frac{\partial}{\partial\mu}+\beta_{x}\frac{\partial}{\partial x}-\gamma_{\varphi}\phi\frac{\partial}{\partial\phi}\right]{\cal V}_{\eff}\left(\phi;\mu,x,{\cal L}\right)=0\,,\label{eq:calRGE}
\end{equation}
where now 
\begin{equation}
{\cal L}=\ln\left[\frac{\phi^{2}}{\mu^{2}}\right]\thinspace.\label{eq:calLogCW}
\end{equation}
As before, we can rewrite this in a more convenient fashion by defining
\begin{equation}
{\cal V}_{\eff}\left(\phi;\mu,x,{\cal L}\right)=\phi^{4}{\cal S}_{\eff}\left(\mu,x,{\cal L}\right)\thinspace,\label{eq:defcalS}
\end{equation}
so that Eq.\,(\ref{eq:defcalS}) implies 
\begin{equation}
\left[-\left(2+2\gamma_{\varphi}\right)\frac{\partial}{\partial{\cal L}}+\beta_{x}\frac{\partial}{\partial x}-4\gamma_{\varphi}\right]{\cal S}_{\eff}\left(\mu,x,{\cal L}\right)=0\,.\label{eq:calRGE2}
\end{equation}

The central point of the general approach to RGE improvement discussed
in the aforementioned references is to reorganize the contributions
to ${\cal S}_{\eff}\left(\mu,x,{\cal L}\right)$ arising from different
loop orders according to the difference between the aggregate power
of the couplings $x$ and the logs ${\cal L}$, that is to say,
\begin{equation}
{\cal S}_{\eff}\left(x,{\cal L}\right)={\cal S}_{\eff}^{\LL}\left(x,{\cal L}\right)+{\cal S}_{\eff}^{\NLL}\left(x,{\cal L}\right)+\cdots\,,\label{eq:calSLL}
\end{equation}
where ${\cal S}_{\eff}^{\LL}$ contains the leading logs contributions,
\begin{equation}
{\cal S}_{\eff}^{\LL}\left(x,{\cal L}\right)=\sum_{n\geq1}{\cal C}_{n}^{\LL}x^{n}{\cal L}^{n-1}\,,\label{eq:defSLL}
\end{equation}
${\cal S}_{\eff}^{\NLL}$ contains the next to leading logs terms,
\begin{equation}
{\cal S}_{\eff}^{\NLL}\left(x,{\cal L}\right)=\sum_{n\geq2}{\cal C}_{n}^{\NLL}x^{n}{\cal L}^{n-2}\,,\label{eq:defSNLL}
\end{equation}
and so on. Insertion of the ansatz\,(\ref{eq:calSLL}) into the RGE\,(\ref{eq:calRGE2})
gives a set of coupled differential equations, of which we quote the
first two,
\begin{equation}
\left[-2\frac{\partial}{\partial{\cal L}}+\beta_{x}^{\left(2\right)}\frac{\partial}{\partial x}\right]{\cal S}_{\eff}^{\LL}\left(x,{\cal L}\right)=0\,,\label{eq:eqs1}
\end{equation}
and
\begin{equation}
\left[-2\frac{\partial}{\partial{\cal L}}+\beta_{x}^{\left(2\right)}\frac{\partial}{\partial x}\right]{\cal S}_{\eff}^{\NLL}\left(x,{\cal L}\right)+\left[\beta_{x}^{\left(3\right)}\frac{\partial}{\partial x}-4\gamma_{\varphi}^{\left(2\right)}\right]{\cal S}_{\eff}^{\LL}\left(x,{\cal L}\right)=0\,.\label{eq:eqs2}
\end{equation}
Equation\,(\ref{eq:eqs1}) results in a first order difference equation
for ${\cal C}_{n}^{\LL}$, so the knowledge of the initial coefficient
${\cal C}_{1}^{\LL}$ and the order $x^{2}$ contribution to the beta
function from loop calculations allows one to calculate all ${\cal C}_{n}^{\LL}$,
therefore summing up all the leading logs contributions to the effective
potential. This summation was the key to making the DSB scenario viable
in the scale invariant Standard Model as shown in\,\cite{elias:2003zm}.
One does not need to stop at this point, however, since Eq.\,(\ref{eq:eqs2})
can also be used to sum up the next to leading logs, after ${\cal S}_{\eff}^{\LL}$
was calculated, provided one knows the first coefficient ${\cal C}_{2}^{\NLL}$
of the series, as well as $\beta_{x}^{\left(3\right)}$ and $\gamma_{\varphi}^{\left(2\right)}$.
That means one can sum up sequentially several subseries of coefficients
contributing to the effective potential, until exhausting the perturbative
information encoded in $\beta_{x}$, $\gamma_{\varphi}$, and the
$V_{\eff}$ calculated up to a certain loop order. This is a systematic
procedure to extract the maximum amount of information concerning
the effective potential from a perturbative calculation. 

One important technical detail is that the renormalization group functions
are usually calculated in the Minimal Subtraction (MS) renormalization
scheme, and they need to be adapted to the procedure outlined in this
section, as it was first pointed out in\,\cite{Ford:1991hw}. For
simplicity, let us consider the case of a theory with a single coupling
$x$. In the MS scheme, divergent integrals (in four space-time dimension)
appear with a factor
\begin{equation}
\tilde{{\cal L}}=\ln\left[\frac{x\,\phi^{2}}{2\,\mu^{2}}\right]\thinspace,\label{eq:calLogMS}
\end{equation}
while in the so-called CW scheme, the effective potential depends
on a log of the form\,(\ref{eq:calLogCW}). Both schemes can be related
by a redefinition of the mass scale $\mu$, 
\begin{equation}
\mu_{\mathrm{MS}}^{2}=f\left(x\right)\mu_{\mathrm{CW}}^{2}\thinspace,
\end{equation}
which can be shown to imply the following relation between the beta
functions in both schemes,
\begin{alignat}{1}
\beta_{\mathrm{CW}} & =\beta_{\mathrm{MS}}\left(1-\frac{1}{2}\beta_{\mathrm{MS}}\,\partial_{x}\ln f\right)^{-1}\thinspace.\label{eq:relationbetas}
\end{alignat}
In four space-time dimensions, divergences usually start at one loop,
generating order $x^{2}$ contributions to $\beta_{MS}$, therefore,
\begin{align}
\beta_{\mathrm{CW}} & =\left(\beta_{\mathrm{MS}}^{\left(2\right)}+\beta_{\mathrm{MS}}^{\left(3\right)}+\cdots\right)\left(1+{\cal O}\left(x\right)\right)\nonumber \\
 & =\beta_{\mathrm{MS}}^{\left(2\right)}+{\cal O}\left(x^{3}\right)\thinspace.\label{eq:relCWMS}
\end{align}
The conclusion is that at one loop level, both beta functions can
be used interchangeably, but if calculations are done at two loops
or more, one has to adapt the MS functions to be used in the calculation
of the CW effective potential. The same reasoning concerning the beta
functions can be applied to the anomalous dimension, with similar
conclusions. 

To gain further insight into this problem, we present the following
argument: in the MS and CW schemes, the effective potential would
be calculated at one loop level in the forms
\begin{equation}
V_{\mathrm{MS}}=\phi^{4}\left(\tilde{A}\left(x\right)+\tilde{B}\left(x\right)\tilde{{\cal L}}\right)\thinspace,\label{eq:VMS1}
\end{equation}
and
\begin{equation}
V_{\mathrm{CW}}=\phi^{4}\left(A\left(x\right)+B\left(x\right){\cal L}\right)\thinspace.\label{eq:VCW1}
\end{equation}
From Eqs.\,(\ref{eq:calLogCW}) and\,(\ref{eq:calLogMS}), we have
\begin{equation}
\tilde{{\cal L}}={\cal L}+\ln\left[\frac{x}{2}\right]\thinspace,\label{eq:relLogs}
\end{equation}
and therefore one can rewrite $V_{\mathrm{MS}}$ in a form compatible
with the CW scheme as follows,
\begin{equation}
V_{\mathrm{MS}}=\phi^{4}\left[\left(\tilde{A}\left(x\right)+\ln\frac{x}{2}\tilde{B}\left(x\right)\right)+\tilde{B}\left(x\right){\cal L}\right]\thinspace.
\end{equation}
Since the value of $A\left(x\right)$ is immaterial in the CW potential,
being fixed by the CW condition\,(\ref{eq:CWcondition}), we conclude
that $\tilde{B}\left(x\right)=B\left(x\right)$ and that both $V_{\mathrm{MS}}$
and $V_{\mathrm{CW}}$ end up giving identical results at one loop.
At two loops, however, $V_{\mathrm{MS}}$ contains a term of the form
$\tilde{C}\left(x\right)\tilde{{\cal L}}^{2}$, so after employing\,(\ref{eq:relLogs}),
one would find a difference in the relevant term proportional to ${\cal L}$,
meaning both potentials are not equivalent at two loops. The net result
is that, at the two loop level, the RGE can be used to relate renormalization
group functions and the effective potential in the CW and the MS scheme,
but not interchangeably.

\section{\label{sec:RGE-Improvement-in3d}RGE improvement in the three dimensional
supersymmetric case}

Now we discuss how to adapt the procedure outlined in Section\,\ref{sec:Summing-up-Leading}
to our model. First of all, we consider the problem of interchangeability
of MS and CW renormalization group functions when using the RGE to
calculate the effective potential. In the supersymmetric three-dimensional
model considered by us, divergences only start at two loops, and the
beta functions start at order $x^{3}$. This means that instead of
Eq.\,(\ref{eq:relCWMS}) we have
\begin{align}
\beta_{\mathrm{CW}} & =\left(\beta_{\mathrm{MS}}^{\left(3\right)}+\beta_{\mathrm{MS}}^{\left(4\right)}+\cdots\right)\left(1+{\cal O}\left(x\right)\right)\nonumber \\
 & =\beta_{\mathrm{MS}}^{\left(3\right)}+{\cal O}\left(x^{4}\right)\thinspace.
\end{align}
Also, $K_{\eff}$ acquires a term proportional to $\tilde{{\cal L}}^{2}$
only at loop orders greater than two. This means we are safe to use
interchangeably functions calculated in the MS and the CW scheme,
as we have done in Section\,\ref{sec:Model}.

It is still not clear that the series of terms we calculated in section\,\ref{sec:Model},
of orders $x^{2n+1}L^{n}$, have any relation to the leading logs
summation described in Section\,\ref{sec:Summing-up-Leading}. Indeed,
by repeating the steps outlined in the start of that section for our
model, the fact that $\beta_{\lambda}^{\left(2\right)}=0$ together
with Eq.\,(\ref{eq:eqs1}) would imply ${\cal C}_{n}^{LL}=0$ for
$n>1$, which in its turn would also trivialize Eq.\,(\ref{eq:eqs2}).
The conclusion would be that the RGE does not allows us to calculate
any new contribution for the effective superpotential.

Actually, this apparent problem is a consequence of the particular
pattern of divergences that appear in our model, whenever we use dimensional
regularization to evaluate Feynman integrals. In four dimensional
non supersymmetric theories, divergences in general occur at any loop
order $n$, the leading logs being of the order $x^{n+1}L^{n}$. In
three dimensional supersymmetric models, divergences start only at
two loops, and are of the order $x^{3}L$. At three loops, the only
divergences arise from two loops subdiagrams, of the order $x^{4}L$.
At four loops, we find again superficially divergent diagrams, of
order $x^{5}L^{2}$, while five loops diagrams contain at most four
and two loops divergent subdiagrams, of order $x^{6}L^{2}$ and $x^{6}L$.
This pattern suggests that new superficial divergences appear only
at even loops, and are of the order $x^{2n+1}L^{n}$, and these terms
should be identified as ``leading logs'' in our case, despite the
fact that the difference between the power of coupling constants and
logs is not the same for these terms. Careful consideration of this
divergence pattern suggests for supersymmetric three dimensional models
the definition,
\begin{equation}
S_{\eff}\left(x,L\right)=S_{\eff}^{\LL}\left(x,L\right)+S_{\eff}^{\NLL}\left(x,L\right)+\cdots\,,\label{eq:ansatz3D}
\end{equation}
where leading logs contributions are of the form
\begin{equation}
S_{\eff}^{\LL}\left(x,L\right)=\sum_{n\geq0}C_{n}^{\LL}x^{2n+1}L^{n}\thinspace,\label{eq:LL3d}
\end{equation}
next to leading logs are given by
\begin{equation}
S_{\eff}^{\NLL}\left(x,L\right)=\sum_{n\geq0}\left(C_{n}^{\NLL}x^{2n+2}L^{n}+D_{n}^{\NLL}x^{2n+3}L^{n}\right)\thinspace,\label{eq:NLL3d}
\end{equation}
and so on. Using Eqs.\,(\ref{eq:LL3d}), (\ref{eq:NLL3d}) and the
RGE\,(\ref{eq:RGE2}), we obtain\begin{subequations}\label{eq:relations-a}
\begin{align}
\partial_{L}S_{\eff}^{\LL} & =\sum_{n\geq0}\left(n+1\right)C_{n+1}^{\LL}x^{2n+3}L^{n}\,,\\
C_{\gamma_{\Phi}^{\left(2\right)}}x^{2}\partial_{L}S_{\eff}^{\LL} & =C_{\gamma_{\Phi}^{\left(2\right)}}\sum_{n\geq0}\left(n+1\right)C_{n}^{\LL}x^{2n+5}L^{n}\,,\\
C_{\beta_{\lambda}^{\left(3\right)}}x^{3}\partial_{x}S_{\eff}^{\LL} & =C_{\beta_{\lambda}^{\left(3\right)}}\sum_{n\geq0}\left(2n+1\right)C_{n}^{\LL}x^{2n+3}L^{n}\,,\\
C_{\gamma_{\Phi}^{\left(2\right)}}x^{2}S_{\eff}^{\LL} & =C_{\gamma_{\Phi}^{\left(2\right)}}\sum_{n\geq0}C_{n}^{\LL}x^{2n+3}L^{n}\,,\\
\partial_{L}S_{\eff}^{\NLL} & =\sum_{n\geq0}\left(n+1\right)\left(C_{n+1}^{\NLL}x^{2n+4}L^{n}+D_{n+1}^{\NLL}x^{2n+5}L^{n}\right)\,,\\
C_{\gamma_{\Phi}^{\left(2\right)}}x^{2}\partial_{L}S_{\eff}^{\NLL} & =C_{\gamma_{\Phi}^{\left(2\right)}}\sum_{n\geq0}\left(n+1\right)\left(C_{n+1}^{\NLL}x^{2n+6}L^{n}+D_{n+1}^{\NLL}x^{2n+7}L^{n}\right)\,,\\
C_{\beta_{\lambda}^{\left(3\right)}}x^{3}\partial_{x}S_{\eff}^{\NLL} & =C_{\beta_{\lambda}^{\left(3\right)}}\sum_{n\geq0}\left(\left(2n+2\right)C_{n}^{\NLL}x^{2n+4}L^{n}+\left(2n+3\right)D_{n}^{\NLL}x^{2n+5}L^{n}\right)\,,\\
C_{\gamma_{\Phi}^{\left(2\right)}}x^{2}S_{\eff}^{\NLL} & =C_{\gamma_{\Phi}^{\left(2\right)}}\sum_{n\geq0}\left(C_{n}^{\NLL}x^{2n+4}L^{n}+D_{n}^{\NLL}x^{2n+5}L^{n}\right)\,,
\end{align}
\end{subequations}where $C_{\beta_{\lambda}^{\left(3\right)}}$ and $C_{\gamma_{\Phi}^{\left(2\right)}}$ are the coefficients of the renormalization group functions. Inserting Eq.\,(\ref{eq:relations-a}) into the
RGE\,(\ref{eq:RGE2}) gives us
\begin{multline}
\sum_{n\geq0}\left[\left(-\left(n+1\right)C_{n+1}^{\LL}+\left(2n+1\right)C_{\beta_{\lambda}^{\left(3\right)}}C_{n}^{\LL}-4C_{\gamma_{\Phi}^{\left(2\right)}}C_{n}^{\LL}\right)x^{2n+3}L^{n}+{\cal O}\left(x^{2n+5}L^{n}\right)\right]=0\thinspace,\label{eq:eqs3}
\end{multline}
which, very much like Eq.\,(\ref{eq:eqs1}), provides a first order
difference equation for $C_{n}^{\LL}$, now involving the order $x^{3}$
terms in the beta function, as well as the order $x^{2}$ terms of
the anomalous dimension. From this equation, the whole series of leading
logs coefficients,
\begin{align}
C_{n+1}^{\LL} & =\left(\frac{2n+1}{n+1}\,C_{\beta_{\lambda}^{\left(3\right)}}-\frac{4}{n+1}\,C_{\gamma_{\Phi}^{\left(2\right)}}\right)C_{n}^{\LL}
\end{align}
may be (in principle) determined, and so $S_{\eff}^{\LL}\left(x,L\right)$
is obtained from the two loop information we have at hand. Looking
at other coefficients of the sum in Eq.\,(\ref{eq:eqs3}) would provide
equations for the calculation of next to leading log contributions,
and so on. The result is that the leading logs summation procedure
can be applied to three dimensional supersymmetric models, yet with
nontrivial modifications, taking into account the peculiar divergence
structure of such models.

To actually apply this technique to our model, one has to generalize
the equations in the last paragraph to the case of two couplings,
which involves dealing with double sums of the form
\begin{equation}
S_{\eff}^{\LL}\left(\lambda,y,L\right)=\sum_{n\geq0\,}\sum_{\,0\leq\ell\leq2n+1}C_{\ell,\,2n+1-\ell}^{\LL}\,\lambda^{\ell}\,y^{2n+1-\ell}L^{n}\thinspace.
\end{equation}
We will only need the first, third and fourth terms from the RGE\,(\ref{eq:RGE2}),
since the second, involving $\gamma_{\Phi}\partial_{L}$, can be shown
to appear only at the next-to-leading log level. According to the
coefficients in\,(\ref{eq:Coefficient-Beta}), we obtain the initial
values for $S_{\eff}^{\LL}$ ,
\begin{align}
C_{01}^{\LL}=0,\,\, & C_{10}^{\LL}=1,\label{eq:CLL-coefficient}\\
C_{30}^{\LL}=b_{3},\,\, & C_{21}^{\LL}=b_{2}\,,\,C_{12}^{\LL}=b_{1}\,,\,C_{03}^{\LL}=b_{0}\,.
\end{align}
Substituting\,(\ref{eq:RGEfunctions}) into $S_{\eff}^{\LL}$ we
obtain
\begin{align}
\partial_{L}S_{\eff}^{\LL} & =\sum_{n\geq0\,}\sum_{\,0\leq\ell\leq2n+3}\left(n+1\right)C_{l,\,2n+3-\ell}^{\LL}\,\lambda^{\ell}\,y^{2n+3-\ell}L^{n}\thinspace,\\
\gamma_{\Phi}^{\left(2\right)}\partial_{L}S_{\eff}^{\LL} & =\sum_{n\geq0\,}\sum_{\,0\leq\ell\leq2n+3}\left(n+1\right)\left(d_{2}\,C_{l,\,2n+3-\ell}^{\LL}\,\lambda^{\ell+2}\,y^{2n+3-\ell}\right.\nonumber \\
 & \left.+d_{0}\,C_{l,\,2n+5-\ell}^{\LL}\,\lambda^{\ell}\,y^{2n+5-\ell}\right)L^{n}\thinspace,\\
\beta_{\lambda}^{\left(3\right)}\partial_{\lambda}S_{\eff}^{\LL} & =\sum_{n\geq0\,}\sum_{\,m\leq\ell\leq2n+m\,\,}\sum_{0\leq m\leq3}\left(l-m+1\right)c_{m}\,C_{l-m+1,\,2n-\ell+m}^{\LL}\,\lambda^{l}\,y^{2n+3-l}L^{n}\thinspace,\\
\gamma_{\Phi}^{\left(2\right)}S_{\eff}^{\LL} & =\sum_{n\geq0\,}\sum_{\,0\leq\ell\leq2n+1}\left(d_{2}\,C_{l,\,2n+1-\ell}^{\LL}\,\lambda^{\ell+2}\,y^{2n+1-\ell}\right.\nonumber \\
 & \left.+d_{0}\,C_{l,\,2n+1-\ell}^{\LL}\,\lambda^{\ell}\,y^{2n+3-\ell}\right)L^{n}\thinspace.
\end{align}
To calculate the first terms of the leading logs series, we expand
explicitly these sums for $n=0$ and $n=1$,
\begin{align}
\partial_{L}S_{\eff}^{\LL} & =C_{30}^{\LL}\,\lambda^{3}+C_{2,1}^{\LL}\,\lambda^{2}\,y+C_{1\text{2}}^{\LL}\,\lambda\,y^{2}+C_{03}^{\LL}y^{3}+2\,C_{50}^{\LL}\,\lambda^{5}\,L+2\,C_{41}^{\LL}\,\lambda^{4}\,y\,L,\nonumber \\
 & +2\,C_{32}^{\LL}\,\lambda^{3}\,y^{2}\,L+2\,C_{23}^{\LL}\,\lambda^{2}\,y^{3}\,L+2\,C_{14}^{\LL}\,\lambda\,y^{4}\,L+2\,C_{05}^{\LL}\,y^{5}\,L+\ldots\,,\label{eq:Expansion-1}
\end{align}
\begin{align}
\gamma_{\Phi}^{\left(2\right)}\partial_{L}S_{\eff}^{\LL} & =d_{2}\,C_{12}^{\LL}\,\lambda^{3}\,y^{2}+d_{0}\,C_{14}^{\LL}\,\lambda\,y^{4}+d_{2}\,C_{21}^{\LL}\,\lambda^{4}\,y+d_{0}\,C_{23}^{\LL}\,\lambda^{2}\,y^{3}\nonumber \\
 & +d_{2}\,C_{30}^{\LL}\,\lambda^{5}+d_{0}\,C_{32}^{\LL}\,\lambda^{3}\,y^{2}+2\left(d_{2}\,C_{05}^{\LL}\,\lambda^{2}\,y^{5}+d_{0}\,C_{07}^{\LL}\,y^{7}\right)L\nonumber \\
 & +2\left(d_{2}\,C_{14}^{\LL}\,\lambda^{3}\,y^{4}+d_{0}\,C_{16}^{\LL}\,\lambda\,y^{6}\right)L+2\left(d_{2}\,C_{23}^{\LL}\,\lambda^{4}\,y^{3}+d_{0}\,C_{25}^{\LL}\,\lambda^{2}\,y^{5}\right)L\thinspace,\nonumber \\
 & +2\left(d_{2}\,C_{32}^{\LL}\,\lambda^{5}\,y^{2}+d_{0}\,C_{34}^{\LL}\,\lambda^{3}\,y^{4}\right)L+2\left(d_{2}\,C_{41}^{\LL}\,\lambda^{6}\,y+d_{0}\,C_{43}^{\LL}\,\lambda^{4}\,y^{3}\right)L\thinspace,\nonumber \\
 & +2\left(d_{2}\,C_{41}^{\LL}\,\lambda^{6}\,y+d_{0}\,C_{43}^{\LL}\,\lambda^{4}\,y^{3}\right)L+\ldots\thinspace,\label{eq:Expansion-2}
\end{align}
\begin{align}
\beta_{\lambda}^{\left(3\right)}\partial_{\lambda}S_{\eff}^{\LL} & =C_{10}^{\LL}\left(c_{3}\,\lambda^{3}+c_{2}\,\lambda^{2}\,y+c_{1}\,\lambda\,y^{2}+c_{0}\,y^{3}\right)+3\,c_{3}\,C_{30}^{\LL}\,\lambda^{5}\,L\nonumber \\
 & +\left(3\,c_{2}\,C_{30}^{\LL}+2\,c_{3}\,C_{21}^{\LL}\right)\lambda^{4}y\,L+\left(3\,c_{1}\,C_{30}^{\LL}+2\,c_{2}\,C_{21}^{\LL}+c_{3}\,C_{12}^{\LL}\right)\lambda^{3}\,y^{2}\,L\nonumber \\
 & +\left(3\,c_{0}\,C_{30}^{\LL}+2\,c_{1}\,C_{21}^{\LL}+c_{2}\,C_{12}^{\LL}\right)\lambda^{2}\,y^{3}\,L+\left(2\,c_{0}\,C_{21}^{\LL}+c_{1}\,C_{12}^{\LL}\right)\lambda\,y^{4}\,L\nonumber \\
 & +c_{0}\,C_{12}^{\LL}\,y^{5}\,L+\ldots\,,\label{eq:Expansion-3}
\end{align}
\begin{align}
\gamma_{\Phi}^{\left(2\right)}S_{\eff}^{\LL} & =d_{2}\left(C_{10}^{\LL}\,\lambda^{3}+C_{01}^{\LL}\,\lambda^{2}\,y\right)+d_{0}\left(C_{10}^{\LL}\,\lambda\,y^{2}+C_{01}^{\LL}\,y^{3}\right)+d_{2}\,C_{30}^{\LL}\,\lambda^{5}L\nonumber \\
 & +d_{2}\,C_{21}^{\LL}\,\lambda^{4}\,y\,L+\left(d_{2}\,C_{12}^{\LL}+d_{0}\,C_{30}^{\LL}\right)\lambda^{3}\,y^{2}\,L+\left(d_{2}\,C_{03}^{\LL}+d_{0}\,C_{21}^{\LL}\right)\lambda^{2}\,y^{3}\,L\nonumber \\
 & +d_{0}\,C_{12}^{\LL}\lambda\,y^{4}\,L+d_{0}\,C_{03}^{\LL}\,y^{5}\,L+\ldots\,.\label{eq:Expansion-4}
\end{align}
 Substituting these results in Eq.\,(\ref{eq:RGE2}), at order
$L^{0}$ we have the relations 
\begin{align}
\begin{cases}
\left(-C_{30}^{\LL}+C_{10}^{\LL}\,c_{3}-4\,d_{2}\,C_{10}^{\LL}\right)\lambda^{3} & =0\\
\left(-C_{2,1}^{\LL}+C_{10}^{\LL}\,c_{2}-4\,d_{2}\,C_{01}^{\LL}\right)\lambda^{2}\,y & =0\\
\left(-C_{12}^{\LL}+C_{10}^{\LL}\,c_{1}-4\,d_{0}\,C_{10}^{\LL}\right)\lambda\,y^{2} & =0\\
\left(-C_{03}^{\LL}+C_{10}^{\LL}\,c_{0}-4\,d_{0}\,C_{01}^{\LL}\right)y^{3} & =0
\end{cases}
\end{align}
From this system of equations and\,(\ref{eq:CLL-coefficient}), we
find the same result obtained in\,(\ref{eq:B3coefs}). If we continue
with the order $L^{1}$, we reproduce the results obtained in\,(\ref{eq:C5-1}).

To systematically calculate higher order coefficients, we developed
a \textsc{Mathematica} code to calculate the coefficients $C^{\LL}$
up to an arbitrary (finite) order. With this code we could calculate
corrections to $S_{\eff}^{\LL}$ up to the order $x^{41}L^{20}$ in
a few seconds, for example. This result will be used, in the next
section, to study the modifications introduced by the leading logs
summation in the DSB in our model. The code as well as the explicit
results are available as a Supplementary Material in\,\cite{Quinto2016}.

\section{\label{sec:DYNAMICAL-BREAKING-OF}Dynamical Breaking of Symmetry }

In this section we study the dynamical breaking of the conformal symmetry
that occurs in the theory we are considering, based on the improved
effective superpotential that was obtained in the previous section
by summing up leading logs contributions. More explicitly, we consider,
\begin{align}
K_{\eff}^{I}\left(\sigma_{\mathrm{cl}}\right) & =-\frac{1}{4}\sigma_{\mathrm{cl}}^{4}\left[S_{\eff}^{\LL}\left(\lambda,y,L\right)+\rho\right],\label{eq:KalherianInprove}
\end{align}
$\rho$ being a finite renormalization constant. The constant $\rho$
is fixed using the CW normalization condition\,(\ref{eq:CWcondition}).
Requiring that the $K_{\eff}^{I}\left(\sigma_{\mathrm{cl}}\right)$
has a minimum at $\sigma_{\mathrm{cl}}^{2}=\mu$ means imposing that
\begin{align}
\left.\frac{d}{d\sigma}K_{\eff}^{I}\left(\sigma_{\mathrm{cl}}\right)\right|_{\sigma_{\mathrm{cl}}^{2}=\mu} & =0\thinspace,\label{eq:minimo}
\end{align}
which can be used to determine the value of $\lambda$ as a function
of the free parameters $y$ and $N$. 

Upon explicit calculation, Eq.\,(\ref{eq:minimo}) turns out to be
a polynomial equation in $\lambda$, and among its solutions we look
for those which are real and positive, and correspond to a minimum
of the potential, i.e., 
\begin{align}
M_{\Sigma}=\left.\frac{d^{2}}{d\sigma^{2}}K_{\eff}^{I}\left(\sigma_{\mathrm{cl}}\right)\right|_{\sigma_{\mathrm{cl}}^{2}=\mu} & >0\thinspace.
\end{align}

This procedure was implemented in a \textsc{Mathematica} program,
and we can verify whether DSB is operational for any given value of
$y\leq1$ and $N$, both considering the unimproved (which is the
effective superpotential calculated at two loops, i.e., including
only the $B^{\left(3\right)}$ term in\,(\ref{eq:KeffAnsatz2}))
and the improved effective superpotential (the one including all higher
order leading logs corrections, calculated by the procure described
in the last section). This way, we can verify whether the effect of
the RGE improvement is so dramatic as it has found to be in other
cases. 

In general, we found DSB being operational for most reasonable value
of its free parameters. For the case of the unimproved effective superpotential,
this is true for any $y\leq1$ and $N$. In the case of the improved
effective superpotential, Figure\,\ref{fig:DBS} displays a plot
of the allowed regions in the $y-N$ parameter space. We notice that
the region of the parameter space where DSB is operational is slightly
smaller for the improved superpotential compared to the unimproved
case. The same pattern were found in the non supersymmetric version
of this model\,\cite{Dias:2010it}, but in that case this difference
was much more relevant. It is also noteworthy that for some values
of $y$ and $N$, one may find actually two viable solutions for $\lambda$
as a function of the free parameters. For larger values of $N$, this
multiplicity of solutions dominates the parameter space of the model,
as can be seen in Fig.\,\ref{fig:DBS-b}. 

Choosing for example the values $y=0.5$ and $N=1$, we find using
the improved superpotential calculated including corrections up to
order $x^{41}L^{20}$ that $\lambda^{I}=0.0118182863005$. To compare,
by choosing the same values of $y$ and $N$, but using the unimproved
two-loop effective superpotential, we find $\lambda=0.0118485063225$.
The difference between the two values being only of order $0.255\%$,
we say that for these values of $y$ and $N$, the improvement of
the effective superpotential by means of the summation of the leading
logs contributions provides only a small quantitative change on the
parameters of the DSB. We verify this happens for most of the parameter
space of the model.

In Figure\,\ref{fig:a}, we show the behavior of $\lambda^{I}$ and
$\lambda$ for a range of values of $y\in\left[0,1\right]$, with
specific values of $N$, to wit $N=1,\,10,\,50,\,100$. It becomes
evident that the difference between the DSB described by the improved
and the unimproved superpotential is significant only for larger values
of $N$, and even so, for not so small values of $y$. In particular,
for large $N$, such as $N=10\times10^{3},\,10\times10^{4},\,10\times10^{5},\,10\times10^{6}$,
the Figure\,\ref{fig:b} show us that in essentially the whole range
of values of $y$ (in the perturbative regime) two solutions can be
found for $\lambda^{I}$.

Finally, the incremental aspect of the RGE improvement in the present
case can also be seen by plotting both the improved and unimproved
effective superpotentials as in Figure\,\ref{fig:UnimproveandImprove},
where only by choosing relatively high values of $y$ and $N$ we
were able to get two graphs that do not superimpose. We note that
the improved superpotential is shallower than the unimproved one,
which is a general feature observed also in four dimensional cases\,\cite{elias:2003zm,Elias:2004hi,Elias2005}.
\begin{center}
\begin{figure}
\centering{}\includegraphics[scale=0.8]{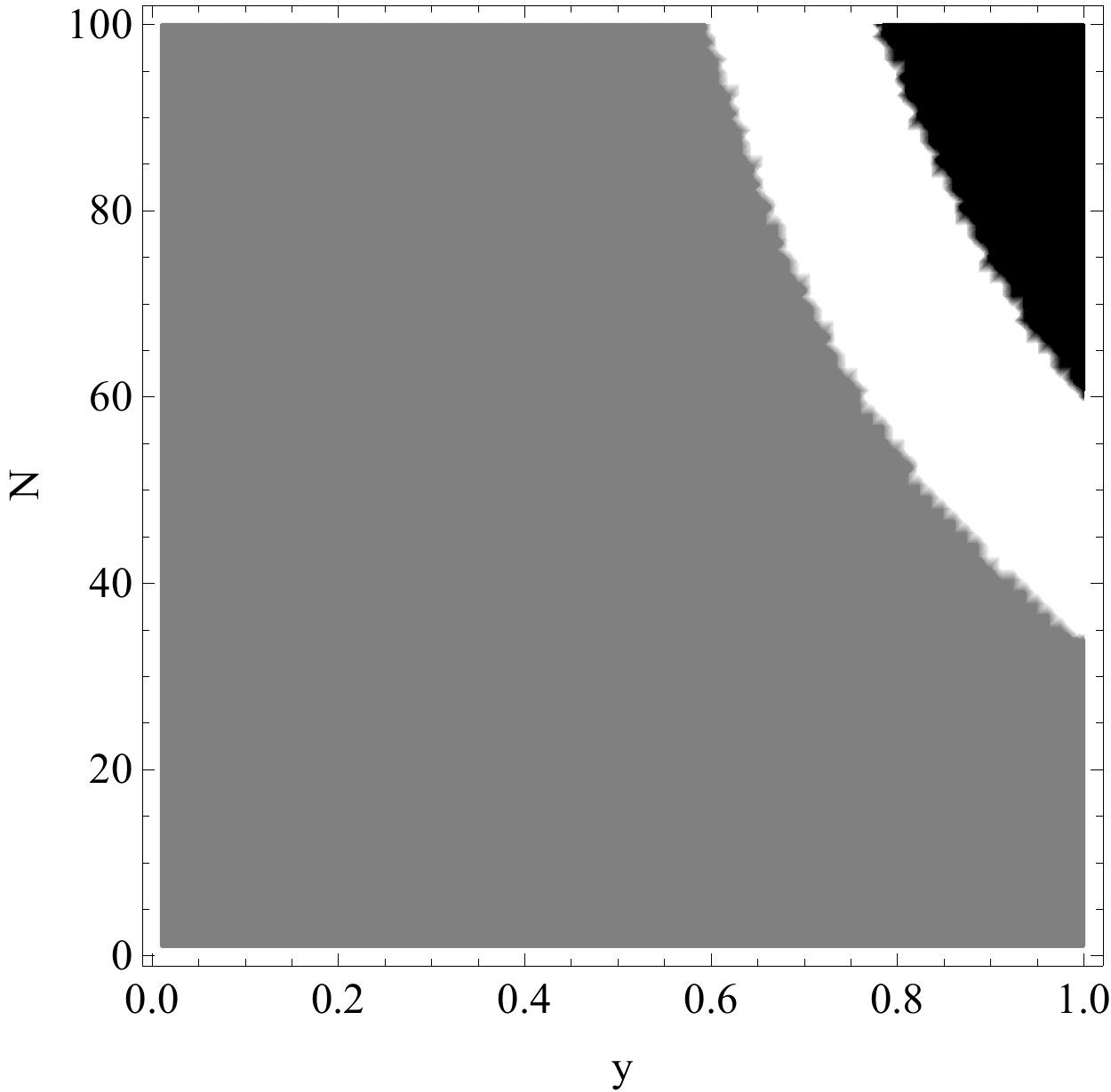}\caption{\label{fig:DBS} Solutions for $\lambda$ in the improved case: one
(gray region), zero (white region) and two (black region)}
\end{figure}
\par\end{center}

\begin{center}
\begin{figure}
\begin{centering}
\subfloat[]{\centering{}\includegraphics[scale=0.8]{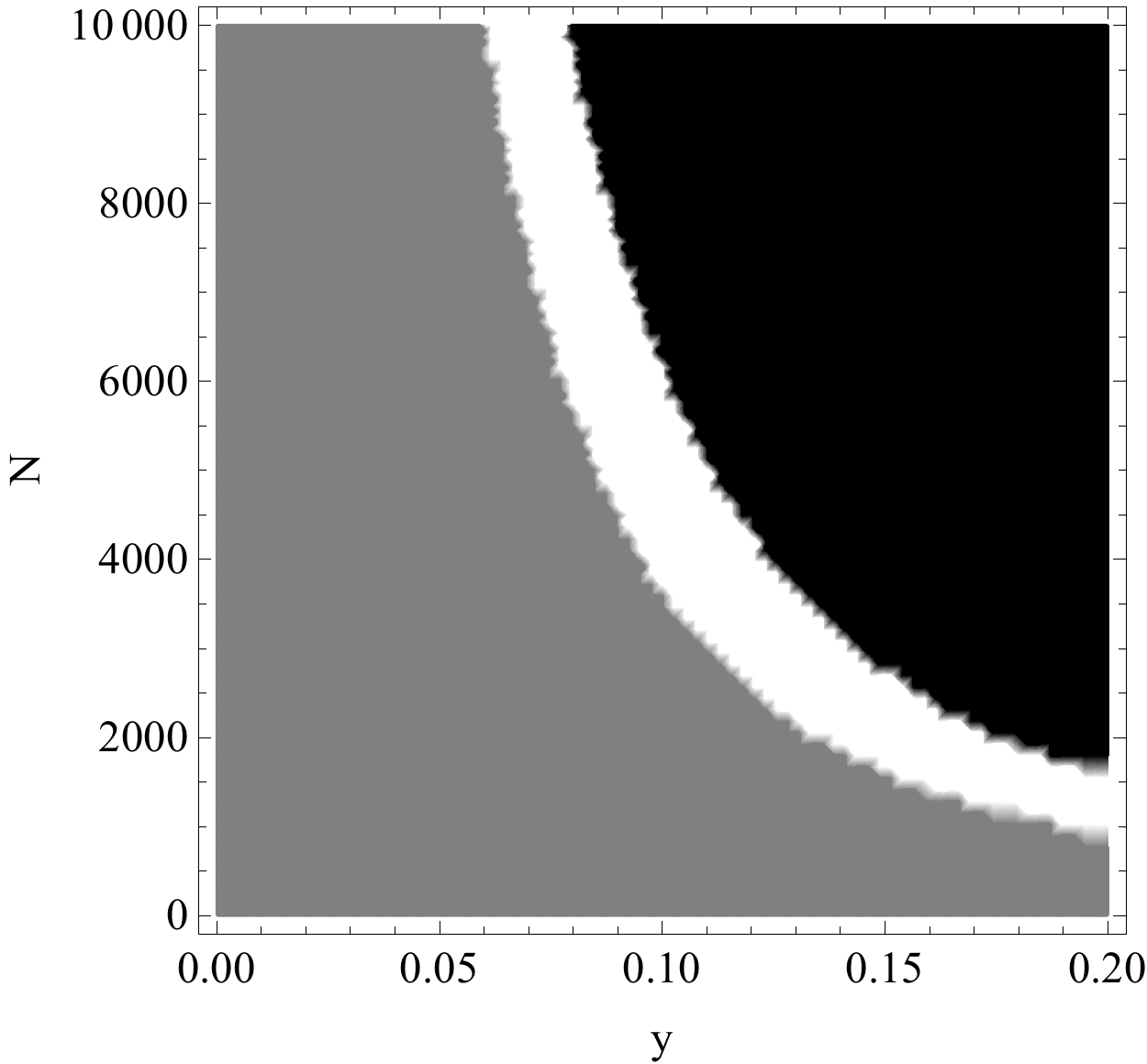}}
\par\end{centering}
\begin{centering}
\subfloat[]{\centering{}\includegraphics[scale=0.8]{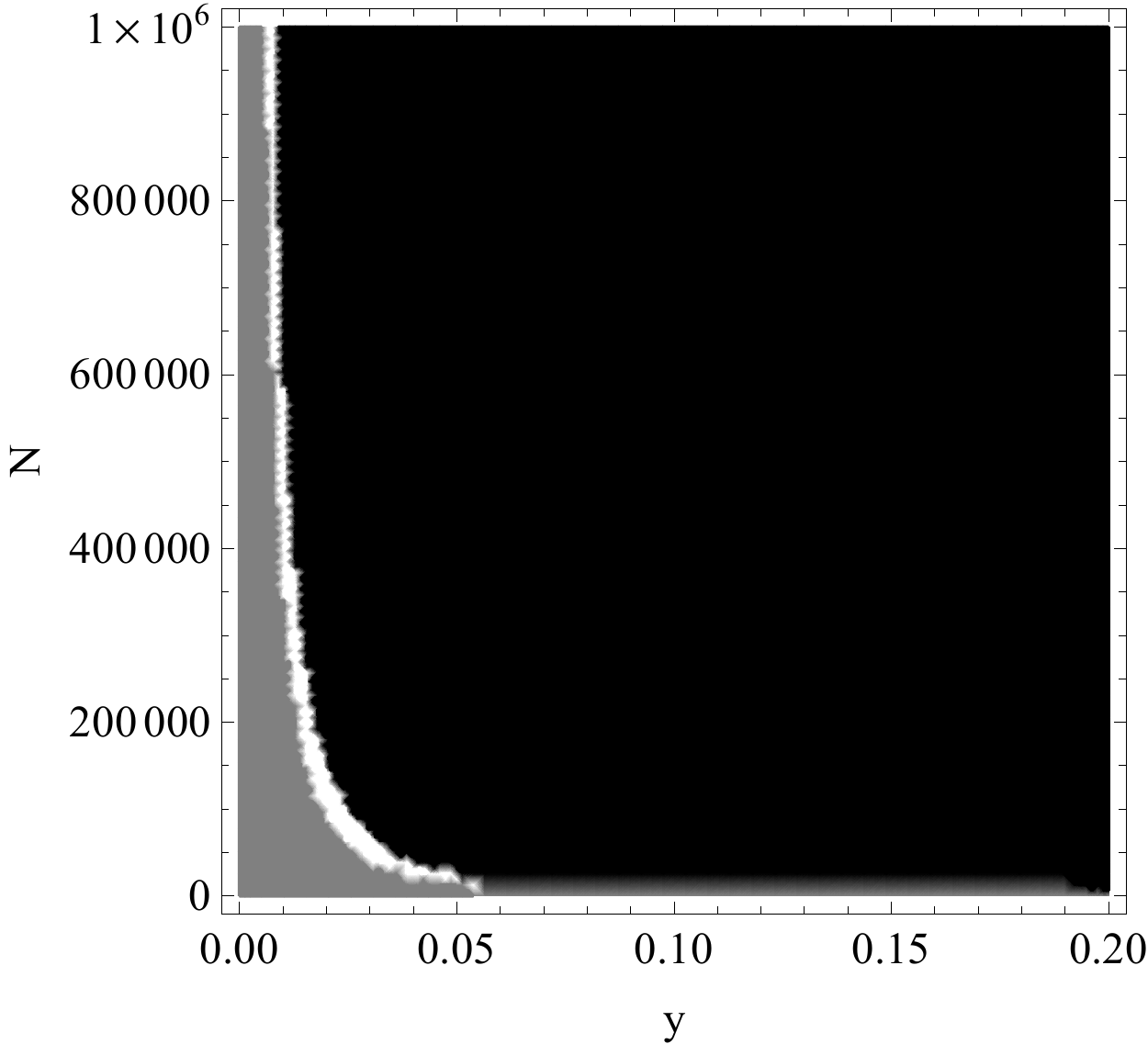}}
\par\end{centering}
\caption{\label{fig:DBS-b}Same as Figure\,(\ref{fig:DBS}), but with larger
$N$. We observe that for larger values of $N$, the region where
we have two solutions for $\lambda$ dominates the parameter space
of the model.}
\end{figure}
\par\end{center}

\begin{center}
\begin{figure}
\begin{centering}
\subfloat[\label{fig:a}]{\begin{centering}
\includegraphics[scale=0.7]{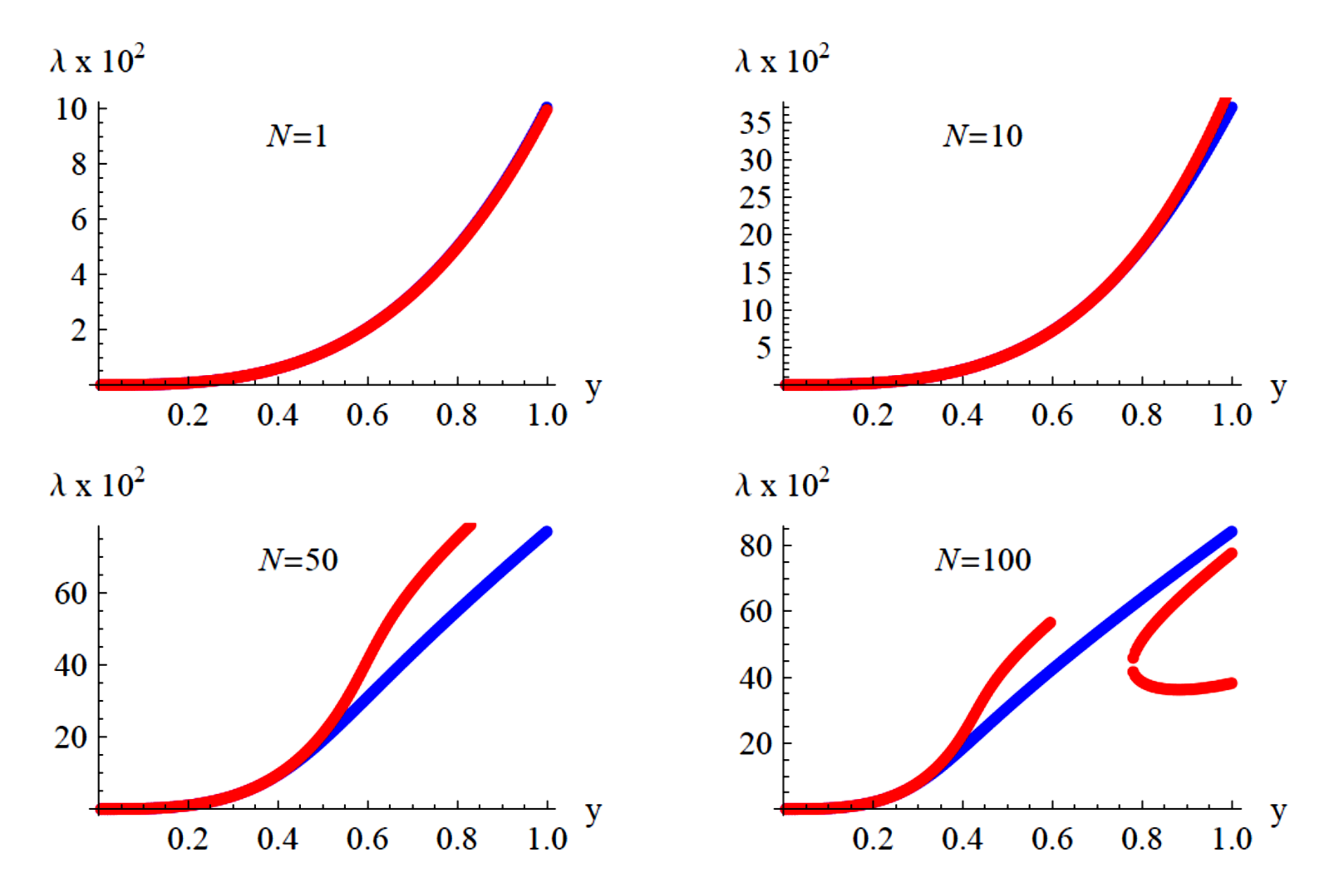}
\par\end{centering}
}
\par\end{centering}
\centering{}\subfloat[\label{fig:b}]{\begin{centering}
\includegraphics[scale=0.7]{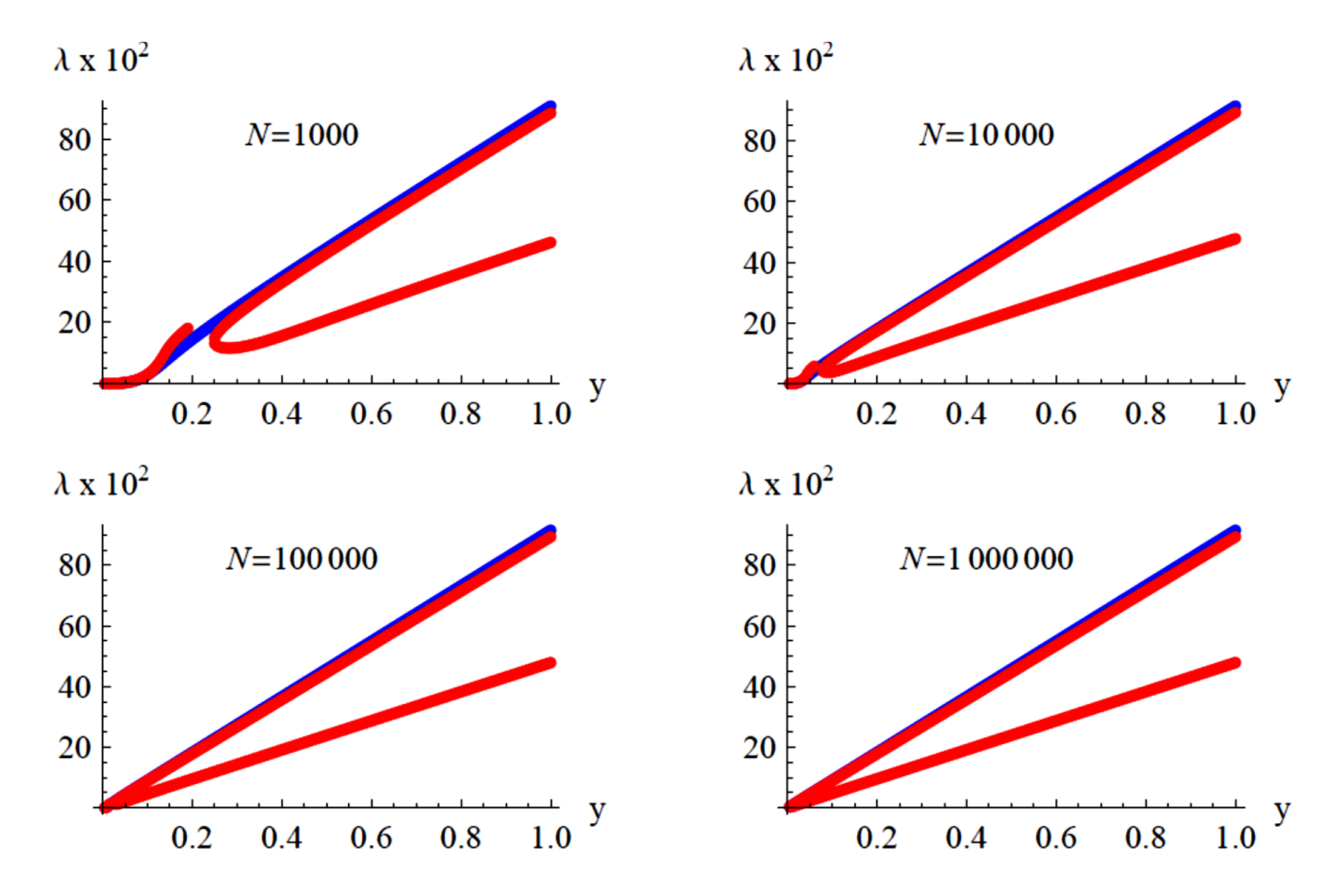}
\par\end{centering}
}\caption{\label{fig:Graph-behaviour}Behavior of the unimproved $\lambda$
(blue line) and improved $\lambda^{I}$ (red line) with the variation
of the coupling constant $y$: a) for $N=1,\,10,\,50,\,100$ and b)
$N=10\times10^{2},\,10\times10^{4},\,10\times10^{5},\,10\times10^{6}$. }
\end{figure}
\begin{figure}
\centering{}\includegraphics[scale=0.7]{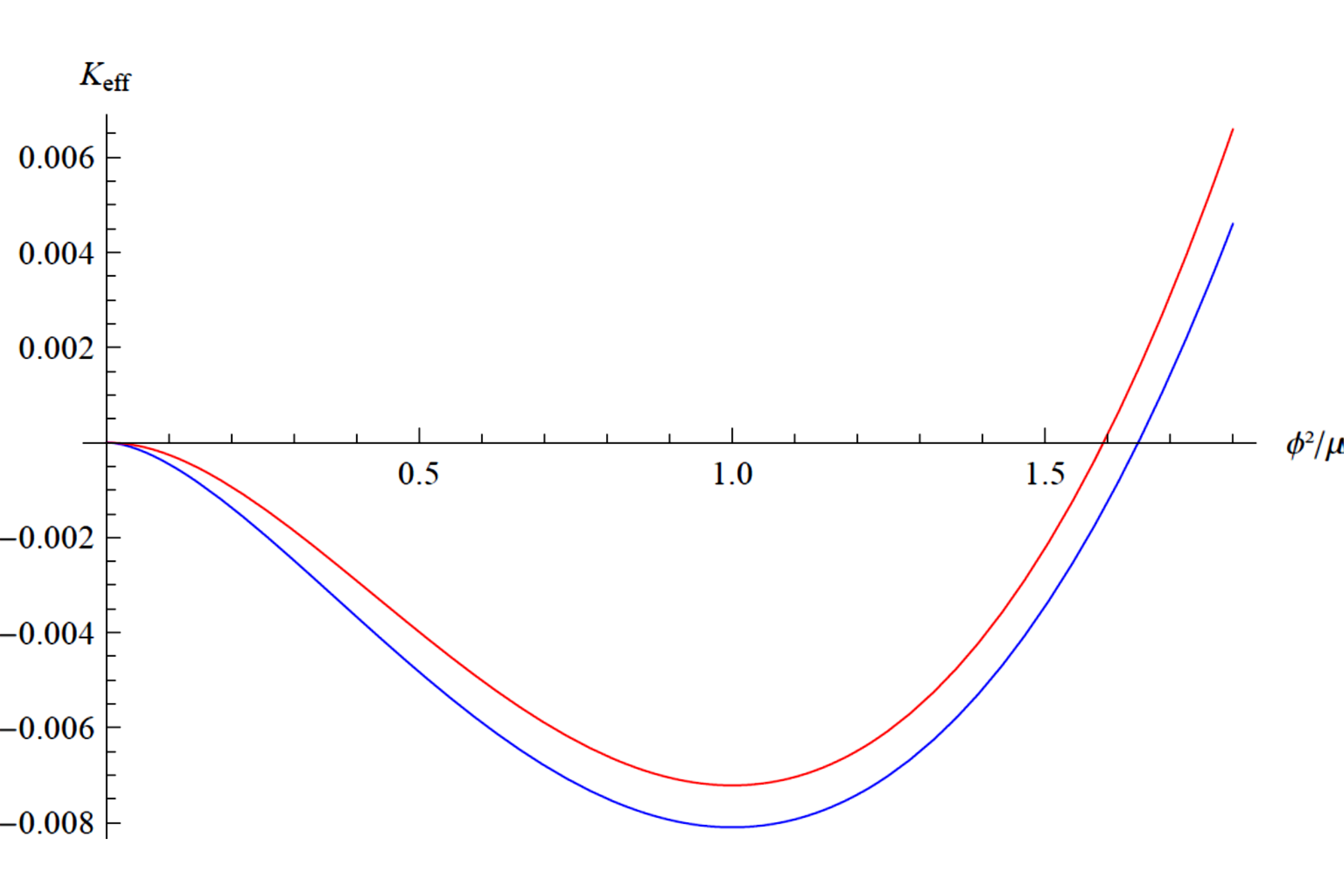}\caption{\label{fig:UnimproveandImprove}Comparison of the unimproved (blue
line) and improved (red line) effective superpotential, for $y=0.5$
and $N=65$. }
\end{figure}
\par\end{center}

\begin{center}
\myclearpage
\par\end{center}

\chapter{\label{chap:Conclusions-and-perspectives}Conclusions and perspectives}

The gauge independence of physical observables is an essential point
to consider when studying gauge theories. In relation to the effective
potential, which plays an essential role in the DSB mechanism in many
relevant models, the Nielsen identity is the key to understand how
the effective potential can depend on the gauge choice, and yet physical
quantities evaluated at its minima can be gauge independent. The DSB
mechanism is central for the formulation of a consistent quantum field
theory of the known elementary interactions, and the possibility that
quantum corrections of a symmetric potential could alone induce such
symmetry breaking is a rather interesting one, not only for its mathematical
elegance, but also for physical reasons. Recently, for example, a
DSB mechanism in a scale-invariant version of the Standard Model is
being discussed as a viable mechanism for generating a mass for the
Higgs particle compatible with experimental observations. The idea
of using the RGE to improve the calculation of the effective potential,
summing up terms arising from higher loop orders organized as leading
logarithms, next to leading logarithms, and so on, is central to this
approach. We have shown how this program can be applied to a supersymmetric
model in the superfield formalism, which is the main technical result
of this thesis. 

In this thesis we studied the Nielsen identity for a supersymmetric
CS model in the superfield formalism. After deriving the Nielsen identity
in the superfield language, we argue that an explicit calculation
of the complete effective superpotential $V_{\eff}^{S}$, including
any possible gauge dependence, is still a technically difficult task.
As a first step in this direction, we calculated the part of the effective
superpotential that do not depend on supercovariant derivatives of
the background scalar superfield, $K_{\eff}$\,\cite{Ferrari:2009zx,buchbinder:1998qv}.
This calculation was made without fixing the gauge-fixing parameter
$\alpha$, and we were able to verify explicitly that all divergent
corrections to the vertex functions at two loop order are independent
of $\alpha$. From these vertex functions, we calculate the renormalization
group functions. Our results agree with the ones previously found
in the literature\,\cite{Avdeev:1991za,Avdeev:1992jt}, that were
calculated in a specific gauge. We developed a method where the use
of the renormalization group functions together with the RGE allows
us to calculate the improved effective superpotential in the superfield
formalism and showing its independence on the gauge-fixing parameter.
The effective superpotential is used to study the DSB mechanism in
our model. The end result is that DSB is operational for reasonable
values of the free parameters, and the RGE improvement produces in
general only a small quantitative change in the properties of the
model. Only for higher values of the gauge coupling and the number
of matter fields we obtain a significant difference on the properties
of DSB from the RGE improvement.

In this particular model, therefore, the effects of the RGE improvement
were not so dramatic as in its non supersymmetric counterpart, however
the question remains whether the same might happen in different models.
As a future perspective, we will try to extend these results to the
full effective superpotential $V_{\eff}^{S}$. This should be possible
with the results given in this thesis, and the same techniques used
here, generalized to the multiscale case as discussed in\,\cite{Ford1994,Ford1997}.
Also, a generalization of this study for non Abelian models and ABJM
theories would be an interesting endeavor. 

To finish, we would like to briefly comment on another study that
was made by us during the development of this thesis, where we studied
the improvement of the effective potential in the context of the scalar
QED and QCD with a colorless scalar, in order to verify the effect
of a possible non-perturbative infrared fixed points associated to
a conformal phase\,\cite{Quinto2014}. This work represent another
instance of the use of the RGE to gain further physical insights on
the dynamics of four dimensional gauge theory.
\begin{center}
\myclearpage
\par\end{center}

\appendix

\chapter{\label{chap:One-Loop-Correction}One Loop Correction}

\section{Example: Two-point vertex function of gauge superfield}

In this section, for didactic purposes, we present the explicit calculation
of the two-point vertex function at one loop associated to the gauge
superfield, Figure\,\ref{fig:Example-of-one-gauge}. 

First, we need to find the correct sign in front of the, for this
purpose, one may use the Wick's theorem, but we note that for practical
reasons we will use the Wick's contractions only to determine the
correct sign of the amplitude. We start with, 
\[
\frac{1}{2}g\left[D_{\Phi}^{\alpha}-D_{\overline{\Phi}}^{\alpha}\right]_{1}\overline{\Phi}_{1\,a}\Phi_{1\,a}\Gamma_{1\,\alpha}=\left(\frac{1}{2}\,g\right)\,\mathcal{D}_{1}^{\alpha}\overline{\Phi}_{1\,a}\Phi_{1\,a}\Gamma_{1\,\alpha}\,,
\]
this term is the complete trilinear vertex, and we introduced the
notation 
\[
\mathcal{D}^{\alpha}=D_{\Phi}^{\alpha}-D_{\overline{\Phi}}^{\alpha}\thinspace,
\]
to facilitate the manipulations of the contractions, for example,
\begin{align*}
 & \left(\frac{1}{2}\,g\right)^{4}\mathcal{D}_{1}^{\alpha}\overline{\Phi}_{1\,a}\Phi_{1\,a}\Gamma_{1\,\alpha}\mathcal{D}_{2}^{\beta}\overline{\Phi}_{2\,b}\Phi_{2\,b}\Gamma_{2\,\beta}\\
= & -\frac{1}{16}\,g^{4}\,\mathcal{D}_{1}^{\alpha}\mathcal{D}_{2}^{\beta}:\overline{\Phi}_{1\,a}\Phi_{1\,a}\Gamma_{1\,\alpha}:\overline{\Phi}_{2\,b}\Phi_{2\,b}\Gamma_{2\,\beta}
\end{align*}

\begin{center}
$=-\frac{1}{16}\,g^{4}\,\mathcal{D}_{1}^{\alpha}\mathcal{D}_{2}^{\beta}
\contraction{}{\overline{\Phi}}{_{1\,a}\Phi_{1\,a}\Gamma_{1\,\alpha}\overline{\Phi}_{2\,b}}{\Phi}
\contraction[2ex]{\overline{\Phi}_{1\,a}}{\Phi}{_{1\,a}\Gamma_{1\,\alpha}}{\overline{\Phi}}
\bcontraction{\overline{\Phi}_{1\,a}\Phi_{1\,a}}{\Gamma}{_{1\,\alpha}\overline{\Phi}_{2\,b}\Phi_{2\,b}}{\Gamma}
\overline{\Phi}_{1\,a}\Phi_{1\,a}\Gamma_{1\,\alpha}\overline{\Phi}_{2\,b}\Phi_{2\,b}\Gamma_{2\,\beta}
$
\par\end{center}

\begin{center}
$=-\frac{1}{16}\,g^{4}\,\mathcal{D}_{1}^{\alpha}\mathcal{D}_{2}^{\beta}
\contraction{}{\overline{\Phi}}{_{1\,a}}{\Phi}\overline{\Phi}_{1\,a}\Phi_{2\,b}
\contraction{}{\Phi}{_{1\,a}}{\overline{\Phi}}\Phi_{1\,a}\overline{\Phi}_{2\,b}
\contraction{}{\Gamma}{_{1\,\alpha}}{\Gamma}\Gamma_{1\,\alpha}\Gamma_{2\,\beta},
$
\par\end{center}

\begin{center}
\begin{figure}
\centering{}\includegraphics[scale=0.6]{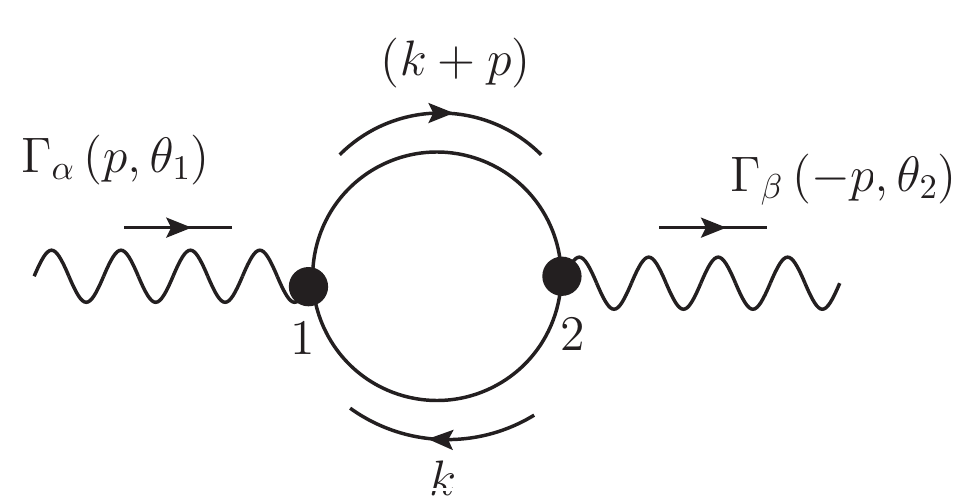}\caption{\label{fig:Example-of-one-gauge}$\mathcal{S}_{\Gamma\Gamma}^{\left(a\right)-1l}$.}
\end{figure}
\par\end{center}

Now, with this procedure, we find the correct sign of the amplitude.
Using the Feynman rules shown in Section\,\ref{sec:Feynman-rules},
we can calculate the diagram $\mathcal{S}_{\Gamma\Gamma}^{\left(a\right)-1l}$,
in the Figure\,\ref{fig:Example-of-one-gauge}, finding
\begin{align}
\mathcal{S}_{\Gamma\Gamma}^{\left(a\right)-1l}= & -\frac{1}{s}\left(\frac{g}{2}\delta_{ij}\mathcal{D}_{1}^{\alpha}\right)\left(\frac{g}{2}\delta_{kl}\mathcal{D}_{2}^{\beta}\right)\int d^{2}\theta_{1}d^{2}\theta_{2}\int\frac{d^{3}p}{\left(2\pi\right)^{3}}\int\frac{d^{D}k}{\left(2\pi\right)^{3}}\,\frac{i\delta_{jk}\,D^{2}}{\left(k+p\right)^{2}}\delta_{12}^{\left(-\left(k+p\right)\right)}\nonumber \\
 & \times\frac{i\delta_{il}\,D^{2}}{\left(k\right)^{2}}\delta_{21}^{\left(-k\right)}\Gamma_{\alpha}\left(p,\theta_{1}\right)\Gamma_{\beta}\left(-p,\theta_{2}\right)\thinspace,\label{eq:D1-one-loop-Gauge}
\end{align}
where $s$ is a symmetric factor, we compute it as, $\frac{1}{s}=\frac{1}{2!}\times2\times1\times1$,
meaning: the $2!$ in the denominator come from the Taylor expansion
because we have two identical vertices, the first numerical factor
``$2$'' come from the external gauge lines (wave line) of the vertices
$1$ and $2$, the second numerical factor ``$1$'' come from the
symmetric factor from the trilinear vertex, and the last numerical
factor come from the combinations of the internal lines that connect
vertex 1 with vertex 2, which in the case amount to a single choice
because the scalar superfield is complex. 

Also, 
\begin{align}
\mathcal{D}_{1}^{\alpha} & \equiv\left[D_{1}^{\alpha}\left(k\right)-D_{1}^{\alpha}\left(-\left(k+p\right)\right)\right],\label{eq:D1-Mathcal-D1-alpha}\\
\mathcal{D}_{2}^{\beta} & \equiv\left[D_{2}^{\beta}\left(k+p\right)-D_{2}^{\beta}\left(-\left(k\right)\right)\right],\label{eq:D1-Mathcal-D2-beta}
\end{align}
which substituted into Eq.\,(\ref{eq:D1-one-loop-Gauge}), leads
to
\begin{align}
\mathcal{S}_{\Gamma\Gamma}^{\left(a\right)-1l}= & -i^{2}\frac{g^{2}}{4}N\int d^{2}\theta_{1}\,\int d^{2}\theta_{2}\int\frac{d^{3}p}{\left(2\pi\right)^{3}}\int\frac{d^{D}k}{\left(2\pi\right)^{3}}\frac{1}{\left(k+p\right)^{2}k^{2}}\times\nonumber \\
 & \left\{ -D_{2}^{\beta}\left(k+p\right)D^{2}\delta_{12}^{\left(-\left(k+p\right)\right)}D_{1}^{\alpha}\left(k\right)D^{2}\delta_{21}^{\left(-k\right)}\Gamma_{\alpha}\left(p,\theta_{1}\right)\Gamma_{\beta}\left(-p,\theta_{2}\right)\right.\nonumber \\
 & -D_{1}^{\alpha}\left(-\left(k+p\right)\right)D_{2}^{\beta}\left(k+p\right)D^{2}\delta_{12}^{\left(-\left(k+p\right)\right)}D^{2}\delta_{21}^{\left(-k\right)}\Gamma_{\alpha}\left(p,\theta_{1}\right)\Gamma_{\beta}\left(-p,\theta_{2}\right)\nonumber \\
 & +D^{2}\delta_{12}^{\left(-\left(k+p\right)\right)}D_{2}^{\beta}\left(-\left(k\right)\right)D_{1}^{\alpha}\left(k\right)D^{2}\delta_{21}^{\left(-k\right)}\Gamma_{\alpha}\left(p,\theta_{1}\right)\Gamma_{\beta}\left(-p,\theta_{2}\right)\nonumber \\
 & \left.+D_{1}^{\alpha}\left(-\left(k+p\right)\right)D^{2}\delta_{12}^{\left(-\left(k+p\right)\right)}D_{2}^{\beta}\left(-\left(k\right)\right)D^{2}\delta_{21}^{\left(-k\right)}\Gamma_{\alpha}\left(p,\theta_{1}\right)\Gamma_{\beta}\left(-p,\theta_{2}\right)\right\} ,\label{eq:D1-one-loop-1}
\end{align}
then, integrating by parts the first term in the second line in Eq.\,(\ref{eq:D1-one-loop-1}),
\begin{align}
 & -D_{2}^{\beta}\left(k+p\right)D^{2}\delta_{12}^{\left(-\left(k+p\right)\right)}D_{1}^{\alpha}\left(k\right)D^{2}\delta_{21}^{\left(-k\right)}\Gamma_{\alpha}\left(p,\theta_{1}\right)\Gamma_{\beta}\left(-p,\theta_{2}\right)=\nonumber \\
 & =-\frac{1}{2}\delta_{12}\left\{ D_{\rho1}D_{1}^{\rho}D_{2}^{\beta}D_{1}^{\alpha}D_{2}^{2}\delta_{21}\Gamma_{\alpha}^{\left(1\right)}\Gamma_{\beta}^{\left(2\right)}+D_{\rho1}D_{2}^{\beta}D_{1}^{\alpha}D_{2}^{2}\delta_{21}D_{1}^{\rho}\Gamma_{\alpha}^{\left(1\right)}\Gamma_{\beta}^{\left(2\right)}\right.\nonumber \\
 & +D_{\rho1}D_{1}^{\rho}D_{1}^{\alpha}D_{2}^{2}\delta_{21}\Gamma_{\alpha}^{\left(1\right)}D_{2}^{\beta}\Gamma_{\beta}^{\left(2\right)}-D_{\rho1}D_{1}^{\alpha}D_{2}^{2}\delta_{21}D_{1}^{\rho}\Gamma_{\alpha}^{\left(1\right)}D_{2}^{\beta}\Gamma_{\beta}^{\left(2\right)}\nonumber \\
 & -D_{1}^{\rho}D_{2}^{\beta}D_{1}^{\alpha}D_{2}^{2}\delta_{21}D_{\rho1}\Gamma_{\alpha}^{\left(1\right)}\Gamma_{\beta}^{\left(2\right)}+D_{2}^{\beta}D_{1}^{\alpha}D_{2}^{2}\delta_{21}D_{\rho1}D_{1}^{\rho}\Gamma_{\alpha}^{\left(1\right)}\Gamma_{\beta}^{\left(2\right)}\nonumber \\
 & \left.+D_{1}^{\rho}D_{1}^{\alpha}D_{2}^{2}\delta_{21}D_{\rho1}\Gamma_{\alpha}^{\left(1\right)}D_{2}^{\beta}\Gamma_{\beta}^{\left(2\right)}+D_{1}^{\alpha}D_{2}^{2}\delta_{21}D_{\rho1}D_{1}^{\rho}\Gamma_{\alpha}^{\left(1\right)}D_{2}^{\beta}\Gamma_{\beta}^{\left(2\right)}\right\} ,
\end{align}
where $\Gamma_{\alpha}\left(p,\theta_{1}\right)=\Gamma_{\alpha}^{\left(1\right)}$.

By using Eq.\,(\ref{eq:S23}), we can see that some terms are zero,
therefore the previous equation is reduced to 
\begin{align}
 & -D_{2}^{\beta}\left(k+p\right)D^{2}\delta_{12}^{\left(-\left(k+p\right)\right)}D_{1}^{\alpha}\left(k\right)D^{2}\delta_{21}^{\left(-k\right)}\Gamma_{\alpha}\left(p,\theta_{1}\right)\Gamma_{\beta}\left(-p,\theta_{2}\right)=\nonumber \\
 & \delta_{12}\left\{ -D_{2}^{\beta}D_{2}^{2}D_{2}^{\alpha}D_{2}^{2}\delta_{21}\Gamma_{\alpha}^{\left(1\right)}\Gamma_{\beta}^{\left(1\right)}-D_{2}^{\beta}D_{2}^{2}D_{2}^{\alpha}\delta_{21}D_{1}^{2}\Gamma_{\alpha}^{\left(1\right)}\Gamma_{\beta}^{\left(1\right)}\right.\nonumber \\
 & \left.+D_{2}^{2}D_{2}^{\alpha}D_{2}^{\rho}\delta_{21}D_{\rho1}\Gamma_{\alpha}^{\left(1\right)}D_{2}^{\beta}\Gamma^{\left(2\right)}\right\} ,\label{eq:D1-one-loop-2}
\end{align}
now, using the D-algebra, Section\,\ref{subsec:D-=0000E1lgebras}
in Eq.\,(\ref{eq:D1-one-loop-2}) and integrating in $\theta_{2}$,
we obtain
\begin{align}
 & -\int d^{2}\theta_{2}D_{2}^{\beta}\left(k+p\right)D^{2}\delta_{12}^{\left(-\left(k+p\right)\right)}D_{1}^{\alpha}\left(k\right)D^{2}\delta_{21}^{\left(-k\right)}\Gamma_{\alpha}\left(p,\theta_{1}\right)\Gamma_{\beta}\left(-p,\theta_{2}\right)\nonumber \\
 & =\left(k^{2}+k^{\alpha\rho}\,p_{\rho}^{\,\beta}\right)\Gamma_{\alpha}\left(p,\theta\right)\Gamma_{\beta}\left(-p,\theta\right).\label{eq:D1-one-loop-3}
\end{align}

Using a similar procedure, we can find the form of the second term
in the third line of Eq.\,(\ref{eq:D1-one-loop-1}),
\begin{align}
 & -\int d^{2}\theta_{2}\,D_{1}^{\alpha}\left(-\left(k+p\right)\right)D_{2}^{\beta}\left(k+p\right)D^{2}\delta_{12}^{\left(-\left(k+p\right)\right)}D^{2}\delta_{21}^{\left(-k\right)}\Gamma_{\alpha}\left(p,\theta_{1}\right)\Gamma_{\beta}\left(-p,\theta_{2}\right)\nonumber \\
 & =\left(k^{2}+p^{2}+2\,k\cdot p\right)\Gamma_{\alpha}\left(p,\theta\right)\Gamma^{\alpha}\left(-p,\theta\right)+\left(k^{\alpha\beta}+p^{\alpha\beta}\right)\Gamma_{\alpha}\left(p,\theta\right)D^{2}\Gamma_{\beta}\left(-p,\theta\right).\label{eq:D1-one-loop-4}
\end{align}
For the third term in the four line of Eq.\,(\ref{eq:D1-one-loop-1}),
\begin{align}
 & \int d^{2}\theta D^{2}\delta_{12}^{\left(-\left(k+p\right)\right)}D_{2}^{\beta}\left(-\left(k\right)\right)D_{1}^{\alpha}\left(k\right)D^{2}\delta_{21}^{\left(-k\right)}\Gamma_{\alpha}\left(p,\theta_{1}\right)\Gamma_{\beta}\left(-p,\theta_{2}\right)\nonumber \\
 & =-\left(k^{2}+k^{\beta\alpha}D^{2}\right)\Gamma_{\alpha}\left(p,\theta\right)\Gamma_{\beta}\left(-p,\theta\right)\,.\label{eq:D1-one-loop-5}
\end{align}
The fourth term in Eq.\,(\ref{eq:D1-one-loop-1}) yields
\begin{align}
 & \int d^{2}\theta D_{1}^{\alpha}\left(-\left(k+p\right)\right)D^{2}\delta_{12}^{\left(-\left(k+p\right)\right)}D_{2}^{\beta}\left(-\left(k\right)\right)D^{2}\delta_{21}^{\left(-k\right)}\Gamma_{\alpha}\left(p,\theta_{1}\right)\Gamma_{\beta}\left(-p,\theta_{2}\right)\nonumber \\
 & =-\left(k^{2}+k^{\beta\rho}p_{\rho}^{\,\alpha}\right)\Gamma_{\alpha}\left(p,\theta\right)\Gamma_{\beta}\left(-p,\theta\right)\,.\label{eq:D1-one-loop-6}
\end{align}
Finally, substituting Eqs.\,(\ref{eq:D1-one-loop-3}) - (\ref{eq:D1-one-loop-6})
into Eq.\,(\ref{eq:D1-one-loop-1}) we find
\begin{align}
\mathcal{S}_{\Gamma\Gamma}^{\left(a\right)-1l}= & \frac{g^{2}}{4}N\int d^{2}\theta\int\frac{d^{3}p}{\left(2\pi\right)^{3}}\int\frac{d^{D}k}{\left(2\pi\right)^{3}}\frac{1}{\left(k+p\right)^{2}k^{2}}\left\{ -\left(4\,k^{2}+p^{2}+2k\cdot p\right)C^{\alpha\beta}\right.\nonumber \\
 & \left.+\left(k^{\alpha\rho}p_{\rho}^{\,\beta}-k^{\beta\rho}p_{\rho}^{\,\alpha}+p^{\beta\alpha}D^{2}\right)\right\} \Gamma_{\alpha}\left(p,\theta\right)\Gamma_{\beta}\left(-p,\theta\right)\,.\label{eq:DT1}
\end{align}
If it use the dimensional regularization this contribution turns out
to be finite.
\begin{center}
\myclearpage
\par\end{center}

\chapter{\label{chap:Two-loops-corrections}Two loops corrections}

Here we give details about the perturbative calculations presented
in Chapter\,\ref{chap:Renormalization-Group-Functions}. The general
procedure was exemplified for an one-loop graph in Appendix\,\ref{chap:One-Loop-Correction}.
For the two-loop diagrams considered here, we use the \textsc{Mathematica
}packages\textsc{ FeynArts\,\cite{Hahn2001}} to perform all the
topologies and the \textsc{Mathematica} package \textsc{SusyMath}\,\cite{ferrarisusymath}
to perform the integration by parts, and then we use a set of integrals
that are given in Appendix\,\ref{chap:Useful-integrals}.

\section{Two-point vertex function to gauge superfield}

We present a summary of the details of the calculation of the two-point
vertex function, up to two loops, for each diagram presented in Figure\,\ref{fig:SGG}. 
\begin{center}
\begin{figure}[!t]
\begin{centering}
\subfloat[]{\centering{}\includegraphics[scale=0.5]{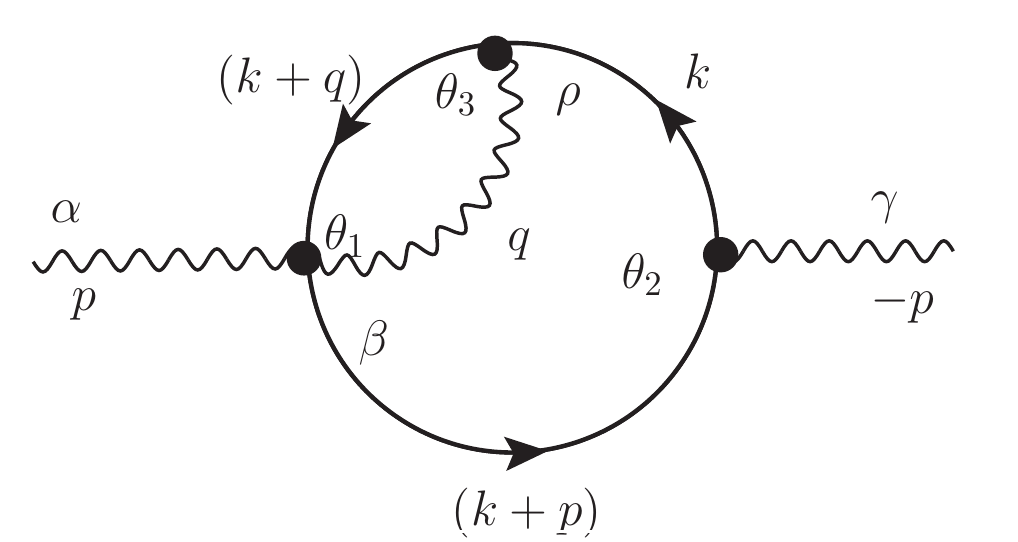}}\subfloat[]{\centering{}\includegraphics[scale=0.5]{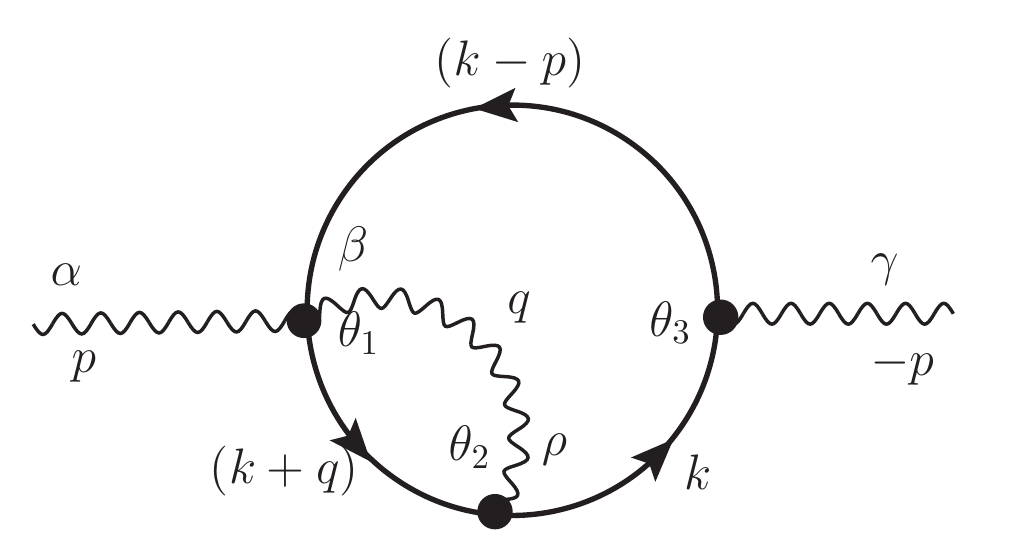}}\subfloat[]{\centering{}\includegraphics[scale=0.5]{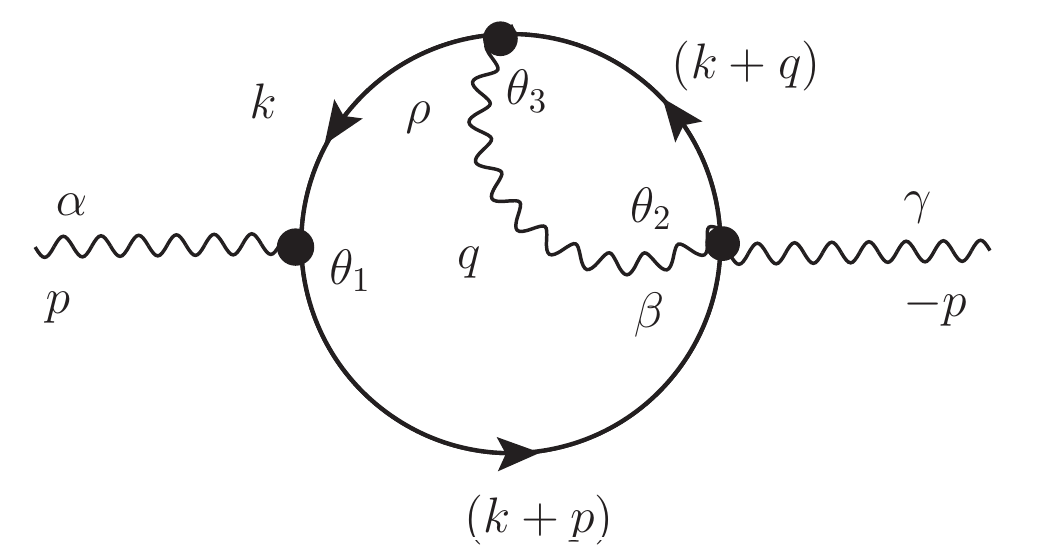}}
\par\end{centering}
\begin{centering}
\subfloat[]{\centering{}\includegraphics[scale=0.5]{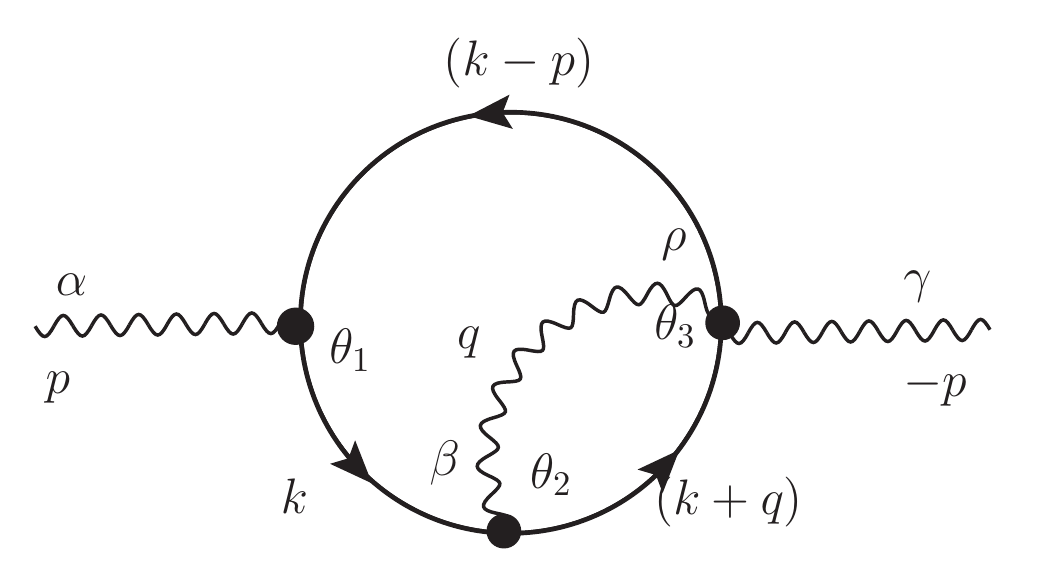}}\subfloat[]{\centering{}\includegraphics[scale=0.5]{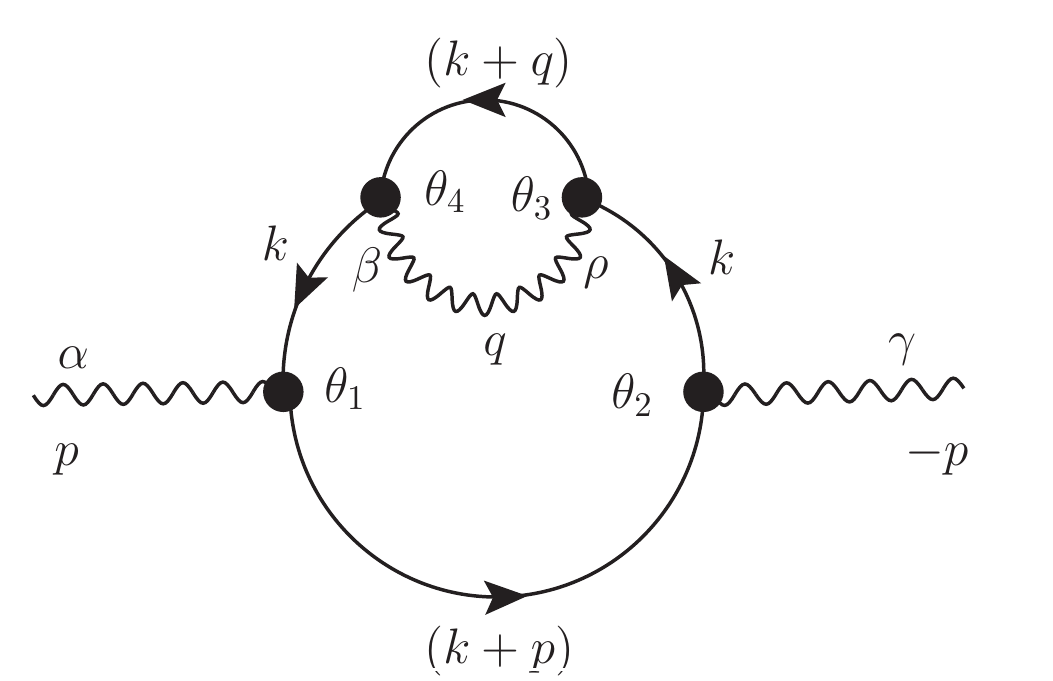}}\subfloat[]{\centering{}\includegraphics[scale=0.5]{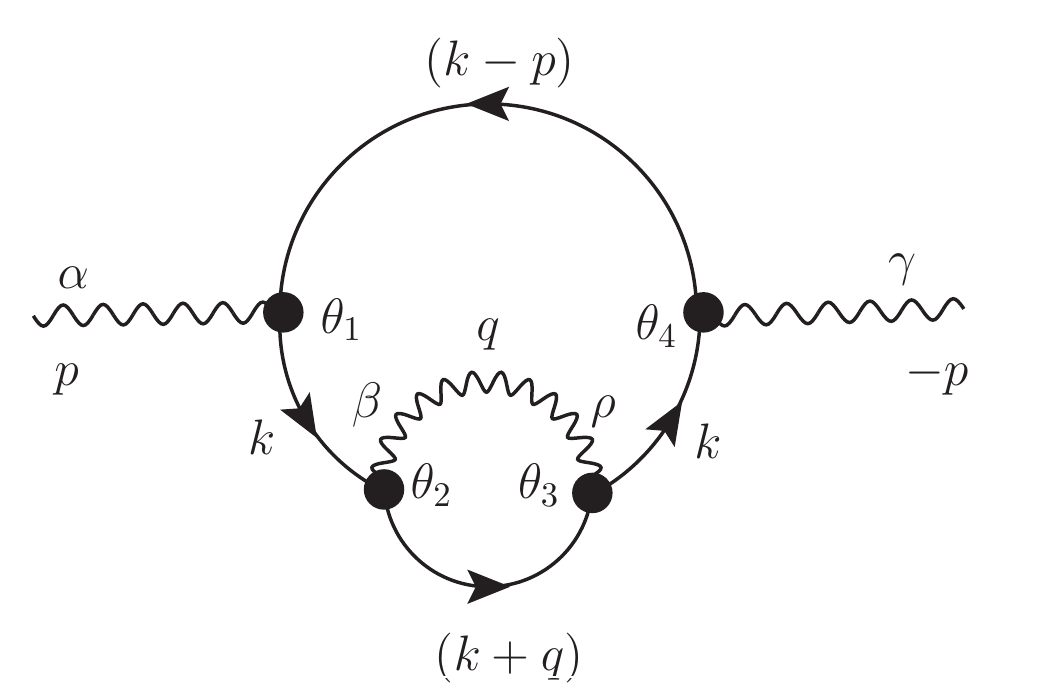}}
\par\end{centering}
\begin{centering}
\subfloat[]{\centering{}\includegraphics[scale=0.5]{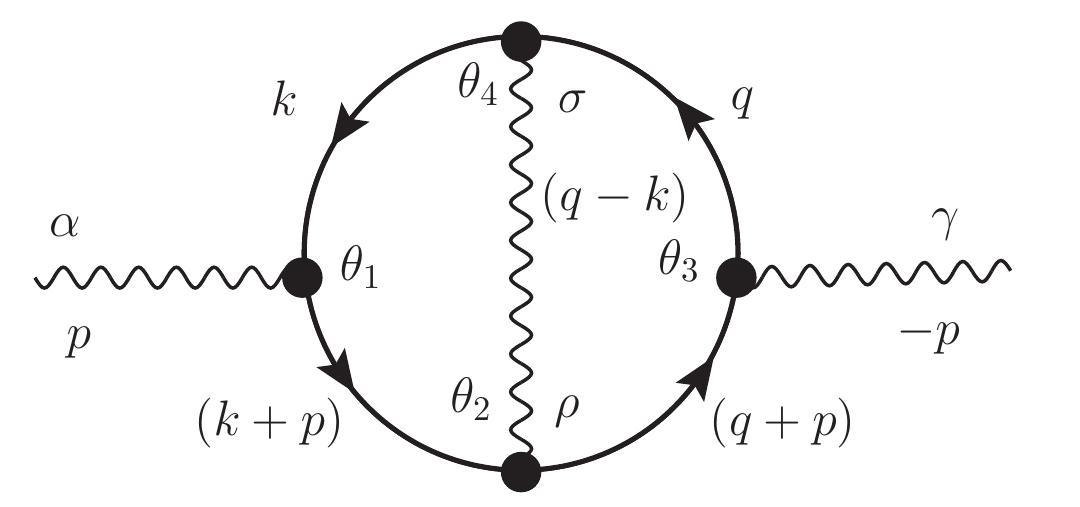}}\subfloat[]{\centering{}\includegraphics[scale=0.5]{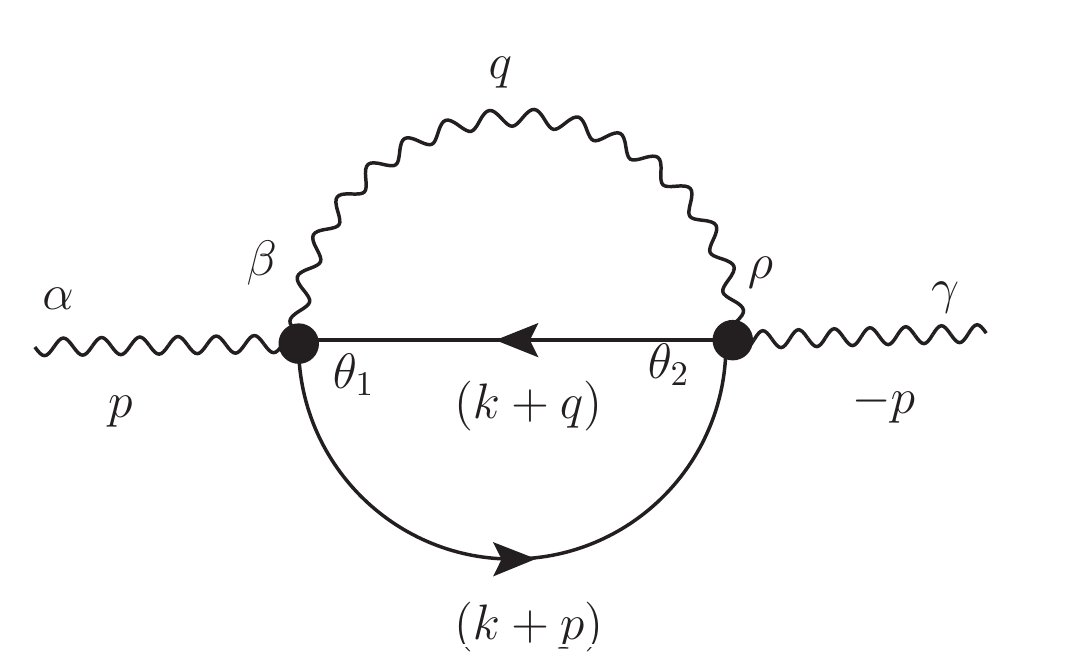}}\subfloat[]{\centering{}\includegraphics[scale=0.5]{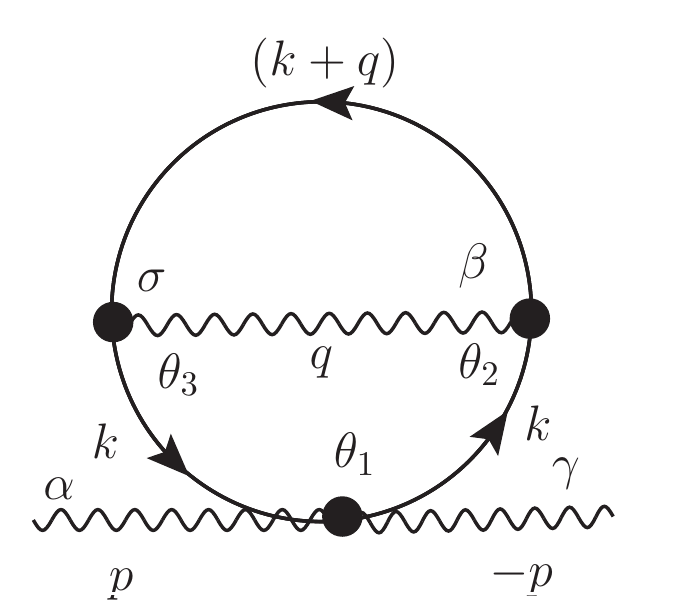}}
\par\end{centering}
\centering{}\caption{\label{fig:SGG}$\mathcal{S}_{\Gamma\Gamma}$}
\end{figure}
\par\end{center}

We start with the first diagram, i.e. $\mathcal{S}_{\Gamma\Gamma}^{\left(a\right)}$,
\begin{align}
\mathcal{S}_{\Gamma\Gamma}^{\left(a\right)} & =\frac{1}{8}\,i\,g^{4}N\,\int\frac{d^{3}p}{\left(2\pi\right)^{3}}\,d^{2}\theta_{1}d^{2}\theta_{2}d^{2}\theta_{3}\frac{d^{D}kd^{D}q}{\left(2\pi\right)^{2D}}\,\Gamma^{\alpha}\left(p,\theta_{1}\right)\Gamma_{\gamma}\left(-p,\theta_{2}\right)\nonumber \\
 & \left\{ +D_{2}^{\gamma}\left(k+p\right)\frac{D^{2}}{\left(k+p\right)^{2}}\delta_{12}D_{3}^{\rho}\left(k\right)\frac{D^{2}}{k^{2}}\delta_{23}\frac{D^{2}}{\left(k+q\right)^{2}}\delta_{31}\frac{\left(-b\,q_{\rho\alpha}+a\,C_{\alpha\rho}D^{2}\right)}{q^{2}}\delta_{13}\right.\nonumber \\
 & -D_{2}^{\gamma}\left(k+p\right)\frac{D^{2}}{\left(k+p\right)^{2}}\delta_{12}\frac{D^{2}}{k^{2}}\delta_{23}D_{3}^{\rho}\left(-\left(k+q\right)\right)\frac{D^{2}}{\left(k+q\right)^{2}}\delta_{31}\frac{\left(-b\,q_{\rho\alpha}+a\,C_{\alpha\rho}D^{2}\right)}{q^{2}}\delta_{13}\nonumber \\
 & -\frac{D^{2}}{\left(k+p\right)^{2}}\delta_{12}D_{2}^{\gamma}\left(-k\right)D_{3}^{\rho}\left(k\right)\frac{D^{2}}{k^{2}}\delta_{23}\frac{D^{2}}{\left(k+q\right)^{2}}\delta_{31}\frac{\left(-b\,q_{\rho\alpha}+a\,C_{\alpha\rho}D^{2}\right)}{q^{2}}\delta_{13}\nonumber \\
 & \left.+\frac{D^{2}}{\left(k+p\right)^{2}}\delta_{12}D_{2}^{\gamma}\left(-k\right)\frac{D^{2}}{k^{2}}\delta_{23}D_{3}^{\rho}\left(-\left(k+q\right)\right)\frac{D^{2}}{\left(k+q\right)^{2}}\delta_{31}\frac{\left(-b\,q_{\rho\alpha}+a\,C_{\alpha\rho}D^{2}\right)}{q^{2}}\delta_{13}\right\} \,,\label{eq:D1a}
\end{align}
where we used a similar procedure as in the one loop case to find
the correct sign and the symmetric factor. Now, we can use the \textsc{SusyMath}
package to perform the integration by parts, then Eq.\,(\ref{eq:D1a})
is reduced to, 
\begin{align}
\mathcal{S}_{\Gamma\Gamma}^{\left(a\right)} & =\frac{1}{8}i\,g^{4}N\,\int\frac{d^{3}p}{\left(2\pi\right)^{3}}\,d^{2}\theta\,\int\frac{d^{D}kd^{D}q}{\left(2\pi\right)^{2D}}\Gamma^{\alpha}\left(p,\theta\right)\left\{ \frac{\Delta_{\alpha\beta}^{\Gamma\Gamma}+\Delta_{\alpha\beta}^{\Gamma D^{2}\Gamma}\,D^{2}}{\left(k+p\right)^{2}\left(k+q\right)^{2}q^{2}k^{2}}\right\} \Gamma^{\beta}\left(-p,\theta\right)\,,
\end{align}
where
\begin{align}
\Delta_{\alpha\beta}^{\Gamma\Gamma} & =-2\,a\,\left(k\cdot p\right)\,k_{\alpha\beta}-2\,a\,\left(k\cdot q\right)\,k_{\alpha\beta}-2\,a\,\left(k\cdot p\right)\,q_{\alpha\beta}-\,b\,p_{\beta\gamma}\,q_{\alpha\delta}\,k^{\gamma\delta}-a\,C_{\beta\gamma}\,C^{\delta\epsilon}\,k_{\alpha\epsilon}\,q_{\delta\zeta}\,k^{\gamma\zeta}\nonumber \\
 & +\,a\,k_{\alpha\gamma}\,k_{\beta\delta}\,p^{\gamma\delta}+a\,k_{\beta\gamma}\,q_{\alpha\delta}\,p^{\gamma\delta}-4\,a\,k_{\alpha\beta}\,k^{2}-\,a\,p_{\alpha\beta}\,k^{2}-\left(a+2\,b\right)\,q_{\alpha\beta}\,k^{2},\\
\Delta_{\alpha\beta}^{\Gamma D^{2}\Gamma} & =-\left(a+b\right)\,C_{\beta\gamma}\,q_{\alpha\delta}\,k^{\gamma\delta}-2\,a\,C_{\beta\alpha}\,k^{2}.
\end{align}
Finally, by doing the two loops integration with help of the integrals
in Table\,\ref{tab:Integrals-of-diagram-S-a}, we obtain
\begin{align}
\mathcal{S}_{\Gamma\Gamma}^{\left(a\right)}=\mathcal{S}_{\Gamma\Gamma}^{\left(b\right)} & =\frac{\left(3\,a-b\right)}{8\left(192\pi^{2}\epsilon\right)}\,N\,i\,g^{4}\int\frac{d^{3}p}{\left(2\pi\right)^{3}}\,d^{2}\theta\,\Gamma^{\alpha}\left(p,\theta\right)\left\{ -p_{\alpha\beta}+3\,C_{\beta\alpha}\,D^{2}\right\} \Gamma^{\beta}\left(-p,\theta\right)\,.\label{eq:S-GG-a}
\end{align}

The diagram $\mathcal{S}_{\Gamma\Gamma}^{\left(c\right)}$ in Figure\,\ref{fig:SGG}
is
\begin{align}
\mathcal{S}_{\Gamma\Gamma}^{\left(c\right)} & =\frac{1}{8}\,i\,g^{4}\,N\,\int\frac{d^{3}p}{\left(2\pi\right)^{3}}\,d^{2}\theta\int\frac{d^{D}kd^{D}q}{\left(2\pi\right)^{2D}}\Gamma^{\alpha}\left(p,\theta\right)\left\{ \frac{\Delta_{\alpha\beta}^{\Gamma\Gamma}+\Delta_{\alpha\beta}^{\Gamma D^{2}\Gamma}\,D^{2}}{\left(k+p\right)^{2}\left(k+q\right)^{2}q^{2}k^{2}}\right\} \Gamma^{\beta}\left(-p,\theta\right)\,,
\end{align}
where
\begin{align}
\Delta_{\alpha\beta}^{\Gamma D^{2}\Gamma} & =\left(a+b\right)\,C_{\alpha\gamma}\,q_{\beta\delta}\,k^{\gamma\delta}-2\,a\,C_{\beta\alpha}k^{2}\,,
\end{align}
and
\begin{align}
\Delta_{\alpha\beta}^{\Gamma\Gamma} & =-2\,a\,\left(k\cdot p\right)\,k_{\alpha\beta}-\left(\frac{1}{2}\,a+b\right)\,p_{\alpha\gamma}\,q_{\beta\delta}\,k^{\gamma\delta}+\frac{1}{2}\,a\,C_{\alpha\gamma}\,C^{\delta\epsilon}\,k_{\beta\delta}\,p_{\epsilon\zeta}\,k^{\gamma\zeta}+\frac{1}{2}\,a\,k_{\alpha\gamma}\,k_{\beta\delta}\,p^{\gamma\delta}\nonumber \\
 & +\frac{1}{2}\,a\,C_{\alpha\gamma}\,C^{\delta\epsilon}\,k_{\delta\zeta}\,q_{\beta\epsilon}\,p^{\gamma\zeta}-4\,a\,k_{\alpha\beta}\,k^{2}-a\,p_{\alpha\beta}\,k^{2}-2\left(a+b\right)\,q_{\alpha\beta}\,k^{2}\,,
\end{align}
then, using Table\,\ref{tab:Integrals-of-diagram-S-b}, we find
\begin{align}
\mathcal{S}_{\Gamma\Gamma}^{\left(c\right)}=\mathcal{S}_{\Gamma\Gamma}^{\left(d\right)} & =\frac{1}{8}\left(\frac{\left(3\,a-b\right)}{192\pi^{2}\epsilon}\right)N\,i\,g^{4}\int\frac{d^{3}p}{\left(2\pi\right)^{3}}\,d^{2}\theta\Gamma^{\alpha}\left(p,\theta\right)\left\{ -p_{\alpha\beta}+3\,C_{\beta\alpha}\,D^{2}\right\} \Gamma^{\beta}\left(-p,\theta\right)\,.\label{eq:S-GG-c}
\end{align}

The diagram $\mathcal{S}_{\Gamma\Gamma}^{\left(e\right)}$ in the
Figure\,\ref{fig:SGG} is
\begin{align}
\mathcal{S}_{\Gamma\Gamma}^{\left(e\right)} & =-\frac{1}{32}\,i\,g^{4}N\,\int\frac{d^{3}p}{\left(2\pi\right)^{3}}\,d^{2}\theta\int\frac{d^{D}kd^{D}q}{\left(2\pi\right)^{2D}}\Gamma^{\alpha}\left(p,\theta\right)\frac{\left\{ \Delta_{\alpha\beta}^{\Gamma\Gamma}+\Delta_{\alpha\beta}^{\Gamma D^{2}\Gamma}\,D^{2}\right\} }{\left(k+p\right)^{2}\left(k+q\right)^{2}q^{2}\left(k^{2}\right)^{2}}\Gamma^{\beta}\left(-p,\theta\right)\,,
\end{align}
where
\begin{align*}
\Delta_{\alpha\beta}^{\Gamma\Gamma}= & \left(-20\,a+\,4\,b\right)\,\left(k\cdot p\right)\,\left(k\cdot q\right)\,k_{\alpha\beta}+\left(2\,a+2\,b\right)\,C_{\alpha\gamma}\,C_{\beta\delta}\,\left(k\cdot q\right)\,p_{\epsilon\zeta}\,k^{\gamma\zeta}\,k^{\delta\epsilon}\\
 & +\left(-2\,a-2\,b\right)\,C_{\alpha\gamma}\,C_{\beta\delta}\,\left(k\cdot p\right)\,q_{\epsilon\zeta}\,k^{\gamma\epsilon}\,k^{\delta\zeta}+\,\frac{1}{2}\,a\,C_{\alpha\gamma}\,C_{\beta\delta}\,p_{\epsilon\zeta}\,q_{\eta\theta}\,k^{\gamma\theta}\,k^{\delta\zeta}\,k^{\epsilon\eta}\\
 & +\left(\,8\,a-4\,b\right)\,\left(k\cdot q\right)\,k_{\alpha\gamma}\,k_{\beta\delta}\,p^{\gamma\delta}-\left(\frac{1}{2}\,a+\,b\right)\,C_{\alpha\gamma}\,C^{\delta\epsilon}\,k_{\beta\zeta}\,k_{\epsilon\eta}\,q_{\delta\theta}\,k^{\gamma\theta}\,p^{\zeta\eta}\\
 & -2\,a\,\left(k\cdot p\right)\,k_{\alpha\gamma}\,k_{\beta\delta}\,q^{\delta\gamma}+\,a\,k_{\alpha\gamma}\,k_{\beta\delta}\,k_{\epsilon\zeta}\,p^{\delta\epsilon}\,q^{\zeta\gamma}-20\,a\,\left(k\cdot p\right)\,k_{\alpha\beta}\,k^{2}\\
 & +\left(-24\,a+\,6\,b\right)\,\left(k\cdot q\right)\,k_{\alpha\beta}\,k^{2}-\left(8\,a-2\,b\right)\left(k\cdot q\right)\,p_{\alpha\beta}\,k^{2}-2\,a\,\left(k\cdot p\right)\,q_{\alpha\beta}\,k^{2}\\
 & +\frac{7}{2}\,a\,C_{\alpha\gamma}\,C_{\beta\delta}\,p_{\epsilon\zeta}\,k^{\gamma\zeta}\,k^{\delta\epsilon}\,k^{2}-\left(\frac{5}{2}\,a+3\,b\right)\,C_{\alpha\gamma}\,C_{\beta\delta}\,q_{\epsilon\zeta}\,k^{\gamma\epsilon}\,k^{\delta\zeta}\,k^{2}\\
 & +\frac{13}{2}\,a\,k_{\alpha\gamma}\,k_{\beta\delta}\,p^{\gamma\delta}\,k^{2}+\,\frac{1}{2}\,a\,k_{\beta\gamma}\,q_{\alpha\delta}\,p^{\gamma\delta}\,k^{2}-\left(\frac{1}{2}\,a+\,b\right)\,C_{\alpha\gamma}\,C_{\beta\delta}\,q_{\epsilon\zeta}\,k^{\delta\epsilon}\,p^{\gamma\zeta}\,k^{2}
\end{align*}
\begin{align}
 & -\frac{1}{2}\,a\,k_{\beta\gamma}\,p_{\alpha\delta}\,q^{\gamma\delta}\,k^{2}-\left(a+b\right)\,C_{\alpha\gamma}\,C_{\beta\delta}\,k_{\epsilon\zeta}\,p^{\delta\zeta}\,q^{\gamma\epsilon}\,k^{2}+\,\frac{1}{2}\,a\,C_{\alpha\gamma}\,C_{\beta\delta}\,p_{\epsilon\zeta}\,k^{\delta\epsilon}\,q^{\gamma\zeta}\,k^{2}\nonumber \\
 & -\frac{3}{2}\,a\,k_{\alpha\gamma}\,k_{\beta\delta}\,q^{\delta\gamma}\,k^{2}-\left(a+b\right)\,C_{\alpha\gamma}\,C_{\beta\delta}\,k_{\epsilon\zeta}\,p^{\gamma\zeta}\,q^{\delta\epsilon}\,k^{2}-32\,a\,k_{\alpha\beta}\,k^{4}-14\,a\,p_{\alpha\beta}\,k^{4}\nonumber \\
 & -\left(4\,a+3\,b\right)q_{\alpha\beta}\,k^{4}\,,
\end{align}
\begin{align}
\Delta_{\alpha\beta}^{\Gamma D^{2}\Gamma}= & -\left(6\,a-2\,b\right)C_{\beta\alpha}\,\left(k\cdot q\right)\,k^{2}+\left(\frac{3}{2}\,a-b\right)\,C_{\beta\gamma}\,q_{\alpha\delta}\,k^{\gamma\delta}\,k^{2}+\frac{3}{2}\,a\,C_{\alpha\gamma}\,q_{\beta\delta}\,k^{\gamma\delta}\,k^{2}\nonumber \\
 & +\left(\frac{1}{2}\,a-b\right)\,C_{\beta\gamma}\,k_{\alpha\delta}\,q^{\gamma\delta}\,k^{2}+\frac{5}{2}\,a\,C_{\alpha\gamma}\,k_{\beta\delta}\,q^{\gamma\delta}\,k^{2}-8\,a\,C_{\beta\alpha}\,k^{4}\,,
\end{align}
using Table\,\ref{tab:Integrals-of-diagram-S-c}, we find
\begin{align}
\mathcal{S}_{\Gamma\Gamma}^{\left(e\right)}=\mathcal{S}_{\Gamma\Gamma}^{\left(f\right)} & =-\frac{1}{4}\left(\frac{a\,N}{192\pi^{2}\epsilon}\right)\,i\,g^{4}\,\int\frac{d^{3}p}{\left(2\pi\right)^{3}}\,d^{2}\theta\Gamma^{\alpha}\left(p,\theta\right)\left\{ p_{\alpha\beta}+3\,C_{\beta\alpha}\,D^{2}\right\} \Gamma^{\beta}\left(-p,\theta\right)\,.\label{eq:S-GG-e}
\end{align}

The diagram $\mathcal{S}_{\Gamma\Gamma}^{\left(g\right)}$ in the
Figure\,\ref{fig:SGG} is
\begin{align}
\mathcal{S}_{\Gamma\Gamma}^{\left(g\right)} & =\frac{1}{32}\,i\,g^{4}N\,\int\frac{d^{3}p}{\left(2\pi\right)^{3}}\,d^{2}\theta\int\frac{d^{D}kd^{D}q}{\left(2\pi\right)^{2D}}\Gamma^{\alpha}\left(p,\theta\right)\frac{\left\{ \Delta_{\alpha\beta\left(1\right)}^{\Gamma\Gamma}+\Delta_{\alpha\beta\left(2\right)}^{\Gamma\Gamma}+\Delta_{\alpha\beta}^{\Gamma D^{2}\Gamma}\,D^{2}\right\} }{\left(k+p\right)^{2}\left(q+p\right)^{2}\left(k-q\right)^{2}q^{2}k^{2}}\Gamma^{\beta}\left(-p,\theta\right)\,,
\end{align}
where
\begin{align}
\Delta_{\alpha\beta\left(1\right)}^{\Gamma\Gamma} & =-\frac{1}{2}\,a\,C_{\alpha\gamma}\,C^{\delta\epsilon}\,p_{\epsilon\zeta}\,q_{\beta\eta}\,q_{\delta\theta}\,k^{\gamma\theta}\,k^{\zeta\eta}+3\,b\,C_{\alpha\gamma}\,C^{\delta\epsilon}\,k_{\epsilon\zeta}\,q_{\beta\eta}\,q_{\delta\theta}\,k^{\gamma\theta}\,p^{\zeta\eta}\nonumber \\
 & +\frac{1}{2}\,a\,C_{\beta\gamma}\,C^{\delta\epsilon}\,k_{\alpha\zeta}\,p_{\epsilon\eta}\,q_{\delta\theta}\,k^{\eta\theta}\,q^{\gamma\zeta}-a\,C_{\beta\gamma}\,C^{\delta\epsilon}\,k_{\alpha\zeta}\,k_{\epsilon\eta}\,q_{\delta\theta}\,p^{\eta\theta}\,q^{\gamma\zeta}\nonumber \\
 & +\frac{1}{2}\,a\,C_{\beta\gamma}\,C^{\delta\epsilon}\,k_{\alpha\zeta}\,k_{\delta\eta}\,q_{\epsilon\theta}\,p^{\eta\theta}\,q^{\gamma\zeta}+\left(a+3\,b\,\right)C_{\beta\gamma}\,C^{\delta\epsilon}\,k_{\alpha\zeta}\,k_{\delta\eta}\,q_{\epsilon\theta}\,p^{\zeta\theta}\,q^{\gamma\eta}\nonumber \\
 & -a\,k_{\alpha\gamma}\,k_{\delta\epsilon}\,q_{\beta\zeta}\,p^{\epsilon\zeta}\,q^{\delta\gamma}-2\,a\,C_{\alpha\gamma}\,C_{\beta\delta}\,\left(k\cdot q\right)\,p_{\epsilon\zeta}\,k^{\gamma\zeta}\,q^{\delta\epsilon}+\left(4\,a-4\,b\right)\,\left(p\cdot q\right)\,k_{\alpha\beta}\,k^{2}\nonumber \\
 & -\left(\frac{1}{2}\,a+b\right)\,C_{\alpha\gamma}\,C_{\beta\delta}\,p_{\epsilon\zeta}\,q_{\eta\theta}\,k^{\gamma\eta}\,k^{\zeta\theta}\,q^{\delta\epsilon}+\frac{1}{2}\,a\,C_{\alpha\gamma}\,C_{\beta\delta}\,k_{\epsilon\zeta}\,q_{\eta\theta}\,k^{\gamma\theta}\,p^{\zeta\eta}\,q^{\delta\epsilon}\nonumber \\
 & +\frac{1}{2}\,a\,C_{\alpha\gamma}\,C_{\beta\delta}\,k_{\epsilon\zeta}\,q_{\eta\theta}\,k^{\gamma\eta}\,p^{\epsilon\theta}\,q^{\delta\zeta}-b\,C_{\alpha\gamma}\,C_{\beta\delta}\,k_{\epsilon\zeta}\,p_{\eta\theta}\,k^{\gamma\theta}\,q^{\delta\zeta}\,q^{\epsilon\eta}-2\,a\,\left(k\cdot q\right)\,q_{\alpha\beta}\,k^{2}\,,\nonumber \\
 & -\frac{1}{2}\,a\,C_{\beta\gamma}\,C^{\delta\epsilon}\,k_{\alpha\zeta}\,k_{\delta\eta}\,p_{\epsilon\theta}\,q^{\gamma\theta}\,q^{\zeta\eta}-\frac{1}{2}\,a\,C_{\beta\gamma}\,C^{\delta\epsilon}\,k_{\alpha\zeta}\,k_{\delta\eta}\,p_{\epsilon\theta}\,q^{\gamma\eta}\,q^{\zeta\theta}\nonumber \\
 & +\frac{1}{2}\,a\,p_{\alpha\gamma}\,q_{\beta\delta}\,k^{\gamma\delta}\,k^{2}-\left(\frac{3}{2}\,a-b\right)\,k_{\alpha\gamma}\,q_{\beta\delta}\,p^{\gamma\delta}\,k^{2}-\frac{1}{2}\,a\,q_{\alpha\gamma}\,q_{\beta\delta}\,p^{\delta\gamma}\,k^{2}-4\,a\,\left(k\cdot p\right)\,q_{\alpha\beta}\,k^{2}
\end{align}
\begin{align}
\Delta_{\alpha\beta\left(2\right)}^{\Gamma\Gamma}= & -\left(2\,a-2\,b\right)\,C_{\alpha\gamma}\,C_{\beta\delta}\,q_{\epsilon\zeta}\,k^{\gamma\epsilon}\,p^{\delta\zeta}\,k^{2}+\frac{3}{2}\,a\,C_{\alpha\gamma}\,C_{\beta\delta}\,p_{\epsilon\zeta}\,k^{\gamma\zeta}\,q^{\delta\epsilon}\,k^{2}+4\,b\,\left(p\cdot q\right)\,q_{\alpha\beta}\,k^{2}\nonumber \\
 & +a\,C_{\alpha\gamma}\,C_{\beta\delta}\,k_{\epsilon\zeta}\,p^{\gamma\zeta}\,q^{\delta\epsilon}\,k^{2}+a\,C_{\alpha\gamma}\,C_{\beta\delta}\,k_{\epsilon\zeta}\,q^{\gamma\zeta}\,q^{\delta\epsilon}\,k^{2}+2\,a\,C_{\alpha\gamma}\,C_{\beta\delta}\,p_{\epsilon\zeta}\,q^{\gamma\zeta}\,q^{\delta\epsilon}\,k^{2}\nonumber \\
 & +b\,C_{\alpha\gamma}\,C_{\beta\delta}\,p_{\epsilon\zeta}\,k^{\gamma\epsilon}\,q^{\delta\zeta}\,k^{2}+\frac{1}{2}\,a\,C_{\alpha\gamma}\,C_{\beta\delta}\,k_{\epsilon\zeta}\,p^{\gamma\epsilon}\,q^{\delta\zeta}\,k^{2}+\left(6\,a+4\,b\right)\,\left(k\cdot p\right)\,k_{\alpha\beta}\,q^{2}\nonumber \\
 & +\left(\frac{1}{2}\,a+2\,b\right)C_{\alpha\gamma}\,C_{\beta\delta}\,p_{\epsilon\zeta}\,q^{\gamma\epsilon}\,q^{\delta\zeta}\,k^{2}+\left(4\,a-4\,b\right)\,\left(k\cdot p\right)\,q_{\alpha\beta}\,q^{2}+\frac{1}{2}\,a\,p_{\alpha\gamma}\,q_{\beta\delta}\,k^{\gamma\delta}\,q^{2}\nonumber \\
 & +\left(a+2\,b\right)\,C_{\alpha\gamma}\,C_{\beta\delta}\,p_{\epsilon\zeta}\,k^{\gamma\zeta}\,k^{\delta\epsilon}\,q^{2}-\frac{1}{2}\,a\,C_{\alpha\gamma}\,C_{\beta\delta}\,p_{\epsilon\zeta}\,k^{\gamma\epsilon}\,k^{\delta\zeta}\,q^{2}+\frac{1}{2}\,a\,k_{\alpha\gamma}\,k_{\beta\delta}\,p^{\gamma\delta}\,q^{2}\nonumber \\
 & -\left(\frac{7}{4}\,a-2\,b\right)\,k_{\alpha\gamma}\,q_{\beta\delta}\,p^{\gamma\delta}\,q^{2}-a\,k_{\alpha\gamma}\,k_{\beta\delta}\,p^{\delta\gamma}\,q^{2}+\left(\frac{3}{4}\,a+b\right)\,C_{\alpha\gamma}\,C_{\beta\delta}\,p_{\epsilon\zeta}\,k^{\gamma\zeta}\,q^{\delta\epsilon}\,q^{2}\nonumber \\
 & -\left(\frac{3}{2}\,a-b\right)\,C_{\alpha\gamma}\,C_{\beta\delta}\,k_{\epsilon\zeta}\,p^{\gamma\zeta}\,q^{\delta\epsilon}\,q^{2}-\left(a+b\right)\,C_{\alpha\gamma}\,C_{\beta\delta}\,p_{\epsilon\zeta}\,k^{\gamma\epsilon}\,q^{\delta\zeta}\,q^{2}\nonumber \\
 & -\left(a-b\right)\,C_{\alpha\gamma}\,C_{\beta\delta}\,k_{\epsilon\zeta}\,p^{\gamma\epsilon}\,q^{\delta\zeta}\,q^{2}+a\,k_{\alpha\beta}\,k^{2}\,q^{2}-\left(13\,a+4\,b\right)\,p_{\alpha\beta}\,k^{2}\,q^{2}\,,
\end{align}
and
\begin{align}
\Delta_{\alpha\beta}^{\Gamma D^{2}\Gamma} & =-\left(a+2\,b\right)\,C_{\alpha\gamma}\,q_{\beta\delta}\,q_{\epsilon\zeta}\,k^{\gamma\zeta}\,k^{\delta\epsilon}+4\,a\,C_{\beta\gamma}\,\left(k\cdot q\right)\,k_{\alpha\delta}\,q^{\gamma\delta}+\frac{1}{2}\,a\,C_{\beta\gamma}\,k_{\alpha\delta}\,k_{\epsilon\zeta}\,q^{\gamma\zeta}\,q^{\delta\epsilon}\nonumber \\
 & -\left(\frac{5}{2}\,a+4\,b\right)\,C_{\alpha\gamma}\,C_{\beta\delta}\,C^{\epsilon\zeta}\,k_{\zeta\eta}\,q_{\epsilon\theta}\,k^{\gamma\theta}\,q^{\delta\eta}+2\left(a+b\right)C_{\beta\gamma}\,k_{\alpha\delta}\,k_{\epsilon\zeta}\,q^{\gamma\zeta}\,q^{\epsilon\delta}\nonumber \\
 & +2\,b\,C_{\alpha\gamma}\,q_{\beta\delta}\,k^{\gamma\delta}\,k^{2}+2\,b\,C_{\beta\gamma}\,k_{\alpha\delta}\,q^{\gamma\delta}\,k^{2}+\left(\frac{1}{2}\,a-b\right)\,C_{\alpha\gamma}\,q_{\beta\delta}\,k^{\gamma\delta}\,q^{2}\nonumber \\
 & +\left(\frac{1}{2}\,a-\,b\right)\,C_{\beta\gamma}\,k_{\alpha\delta}\,q^{\gamma\delta}\,q^{2}-\left(2\,a+8\,b\right)C_{\beta\alpha}\,k^{2}\,q^{2}\,,
\end{align}
using Table\,\ref{tab:Integrals-of-diagram-S-d}, we find,
\begin{align}
\mathcal{S}_{\Gamma\Gamma}^{\left(g\right)} & =\frac{1}{2}\left(\frac{N}{192\pi^{2}\epsilon}\right)i\,g^{4}\int\frac{d^{3}p}{\left(2\pi\right)^{3}}\,d^{2}\theta\Gamma^{\alpha}\left(p,\theta\right)\left\{ \left(4\,a+b\right)p_{\alpha\beta}+3\,b\,C_{\beta\alpha}\,D^{2}\right\} \Gamma^{\beta}\left(-p,\theta\right)\,.\label{eq:SS-GG-f}
\end{align}

The diagram $\mathcal{S}_{\Gamma\Gamma}^{\left(h\right)}$ in the
Figure\,\ref{fig:SGG} is
\begin{align}
\mathcal{S}_{\Gamma\Gamma}^{\left(h\right)} & =-\frac{1}{2}\,i\,g^{4}\,N\,\int\frac{d^{3}p}{\left(2\pi\right)^{3}}\,d^{2}\theta\frac{d^{D}kd^{D}q}{\left(2\pi\right)^{2D}}\,\Gamma^{\alpha}\left(p,\theta\right)\frac{\left\{ -b\,q_{\alpha\beta}-a\,C_{\beta\alpha}\,D^{2}\right\} }{\left(k+p\right)^{2}\left(k+q\right)^{2}q^{2}}\Gamma^{\beta}\left(-p,\theta\right)\,,
\end{align}
using Eq.\,(\ref{eq:Int 6}) and Eq.\,(\ref{eq:Int 7}), we find
\begin{align}
\mathcal{S}_{\Gamma\Gamma}^{\left(h\right)} & =-\left(\frac{N}{192\pi^{2}\epsilon}\right)i\,g^{4}\int\frac{d^{3}p}{\left(2\pi\right)^{3}}\,d^{2}\theta\Gamma^{\alpha}\left(p,\theta\right)\left\{ b\,p_{\alpha\beta}+3\,a\,C_{\beta\alpha}\,D^{2}\right\} \Gamma^{\beta}\left(-p,\theta\right)\,.\label{eq:S-GG-g}
\end{align}

Finally, $\mathcal{S}_{\Gamma\Gamma}^{\left(i\right)}$ in the Figure\,\ref{fig:SGG}
is
\begin{align}
\mathcal{S}_{\Gamma\Gamma}^{\left(i\right)} & =0\,,\label{eq:SS-GG-h}
\end{align}

\section{Two-point vertex function to the scalar superfield}

In this section we give details on the two-point vertex functions
associated to the scalar superfield, a set of diagrams represented
in Figure\,\ref{fig:SPHPH}. 
\begin{center}
\begin{figure}
\begin{centering}
\subfloat[]{\centering{}\includegraphics[scale=0.5]{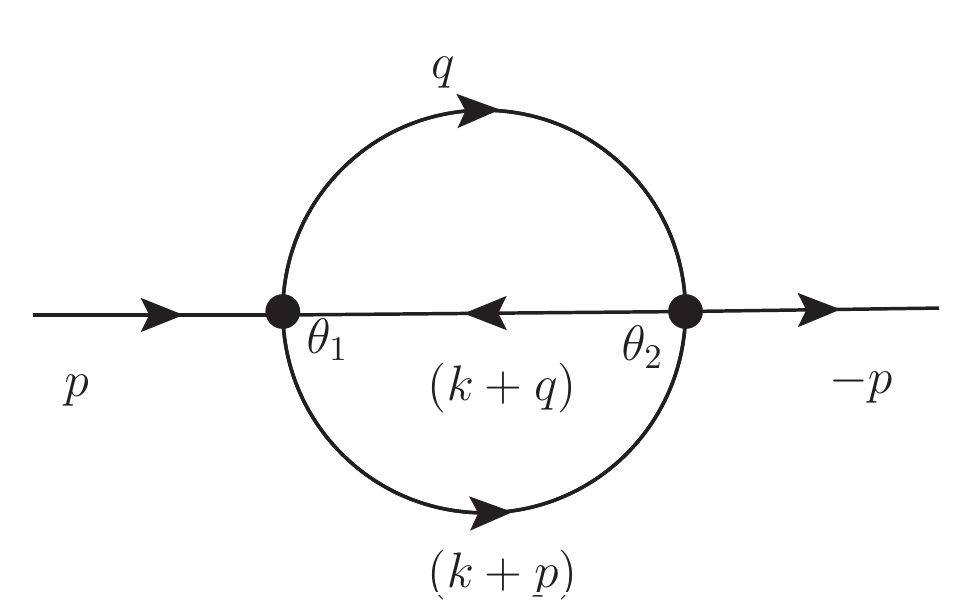}}\subfloat[]{\centering{}\includegraphics[scale=0.45]{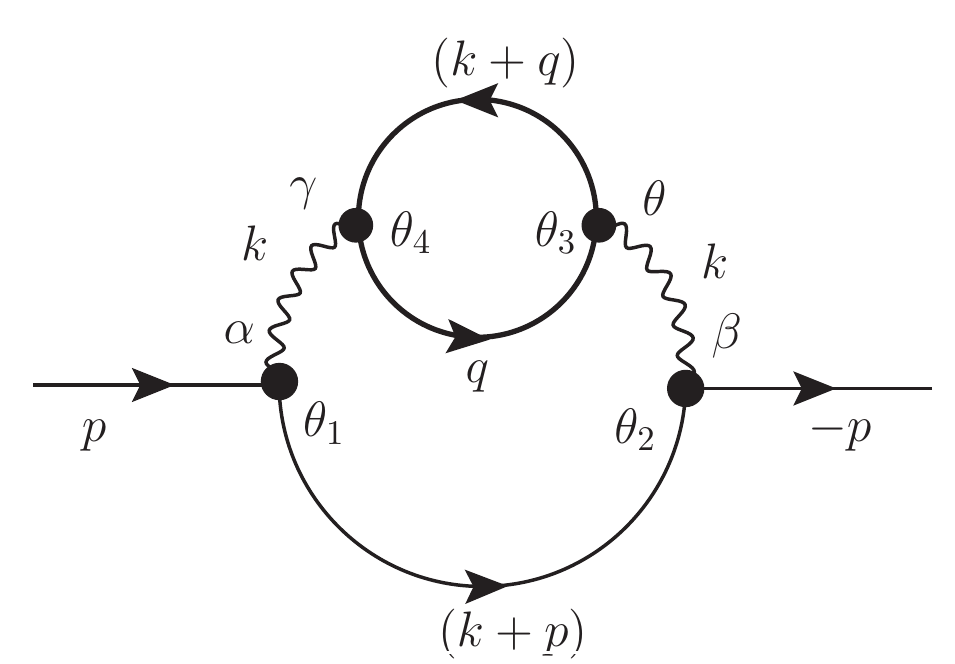}}\subfloat[]{\centering{}\includegraphics[scale=0.45]{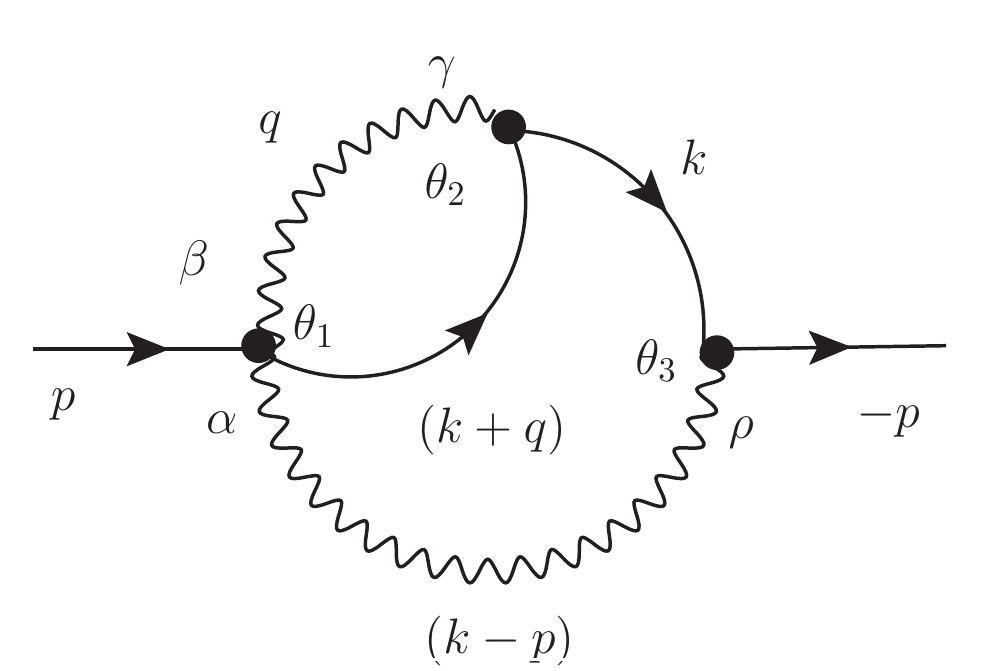}}
\par\end{centering}
\begin{centering}
\subfloat[]{\centering{}\includegraphics[scale=0.45]{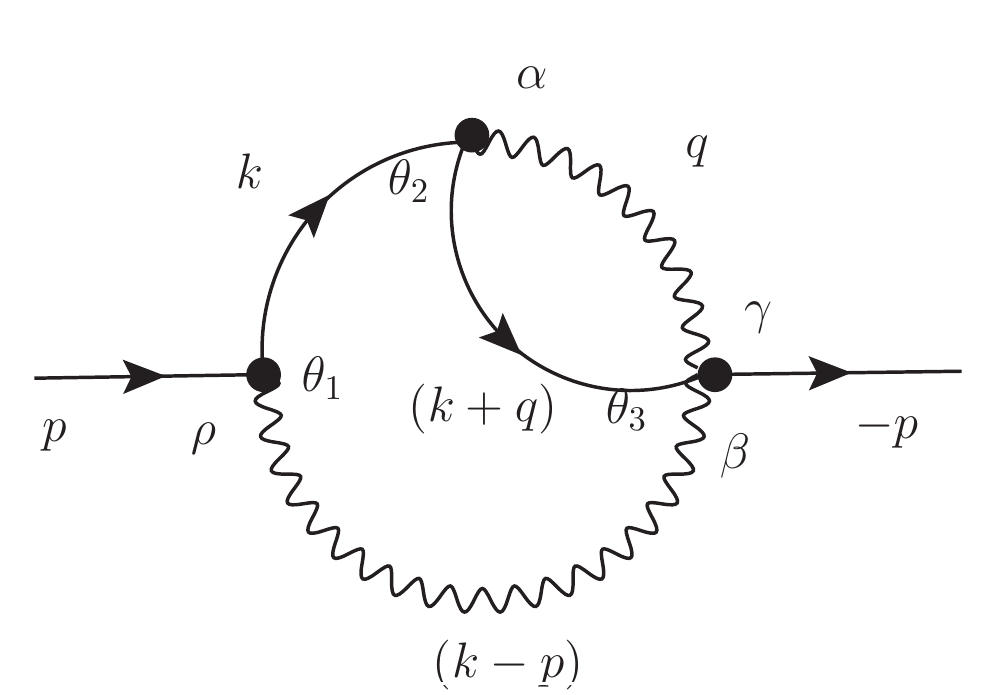}}\subfloat[]{\centering{}\includegraphics[scale=0.45]{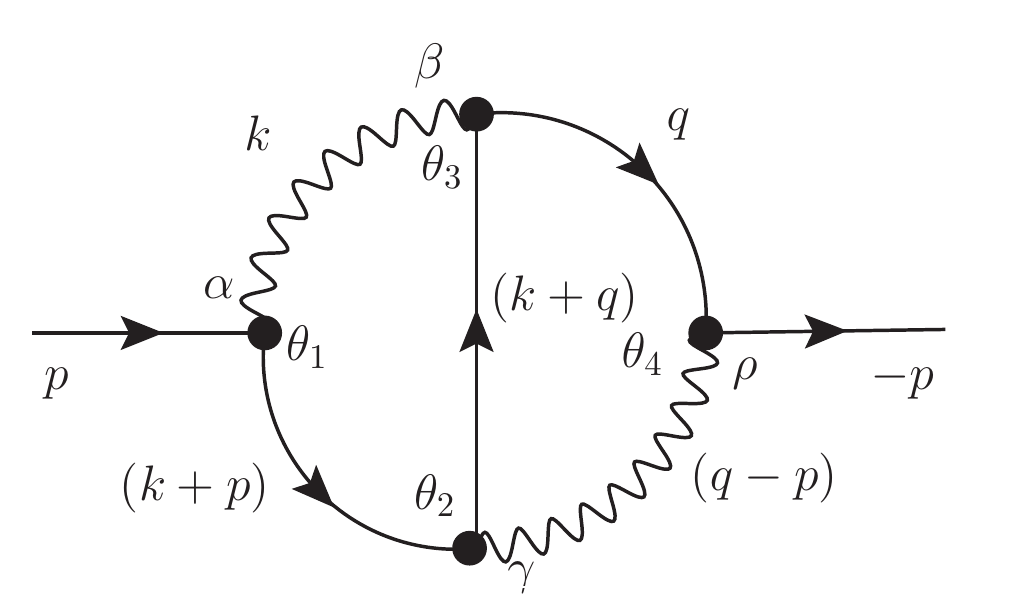}}\subfloat[]{\centering{}\includegraphics[scale=0.45]{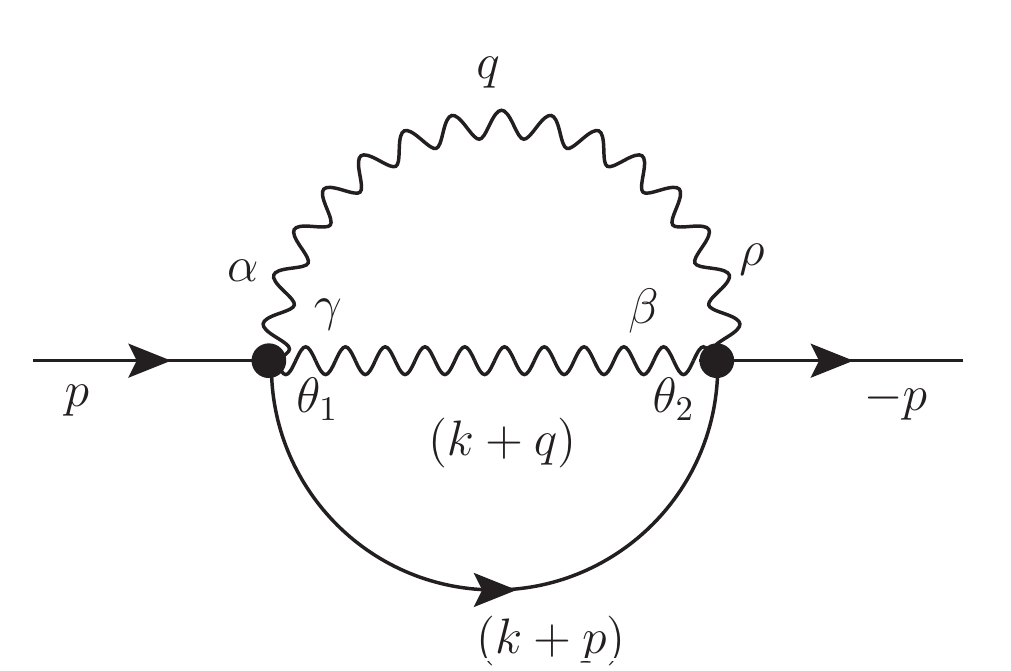}}
\par\end{centering}
\begin{centering}
\subfloat[]{\centering{}\includegraphics[scale=0.45]{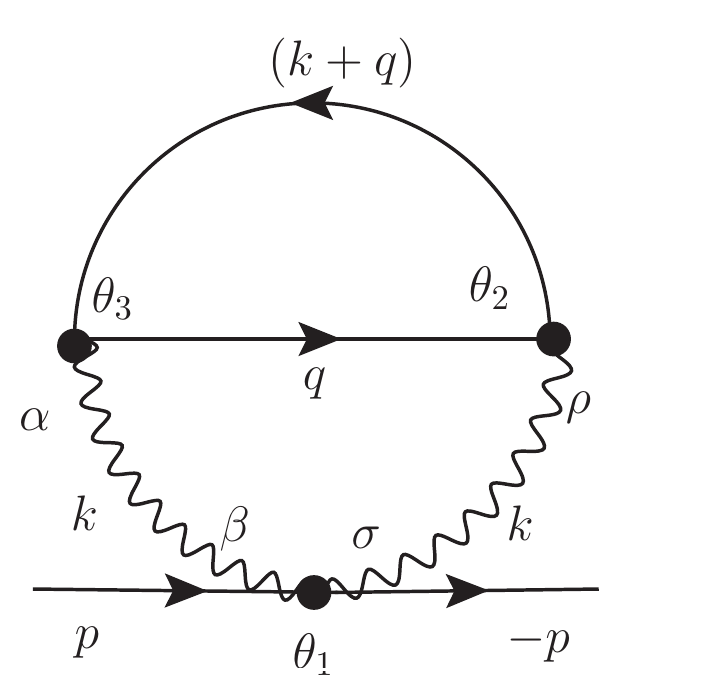}}~~~~~\subfloat[]{\centering{}\includegraphics[scale=0.45]{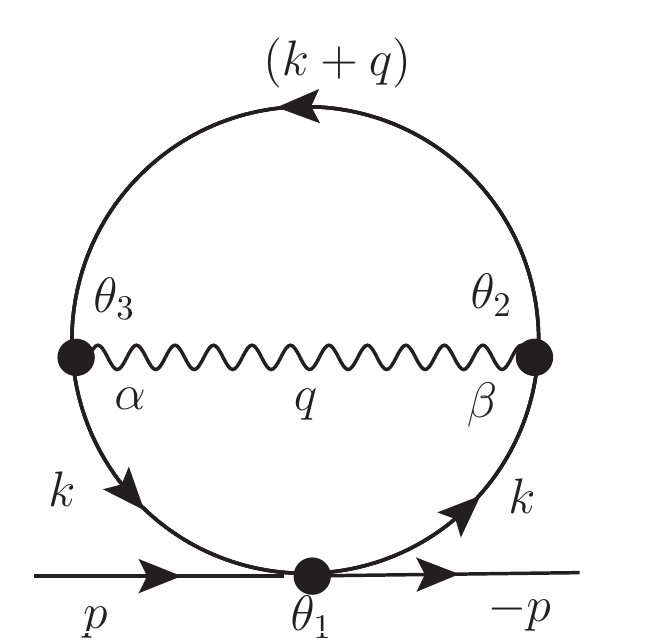}}~~~~\subfloat[]{\centering{}\includegraphics[scale=0.45]{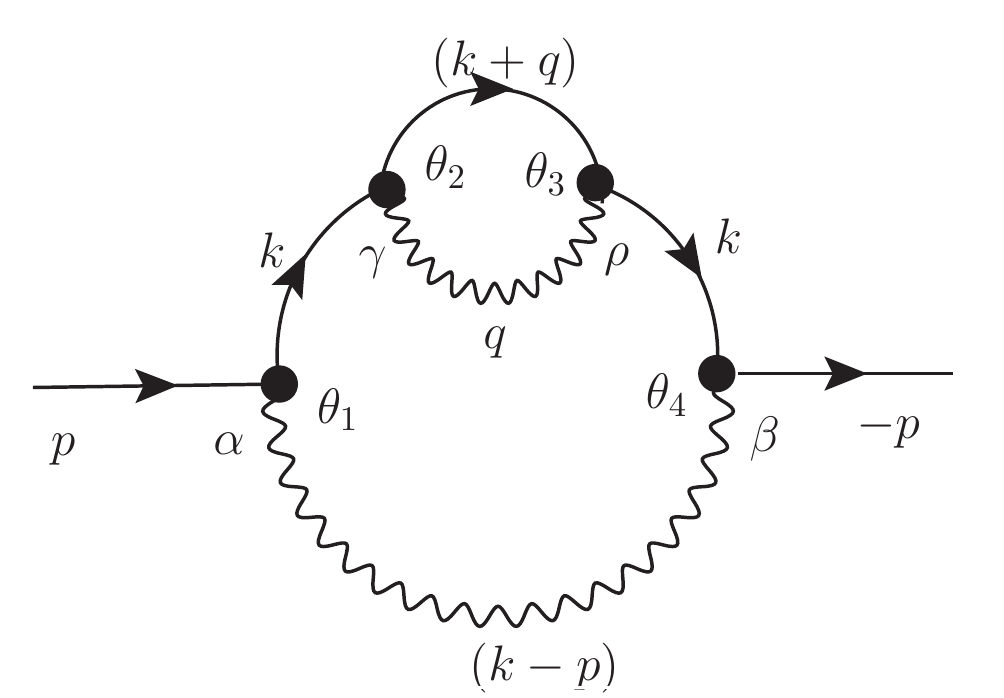}}
\par\end{centering}
\centering{}\subfloat[]{\centering{}\includegraphics[scale=0.45]{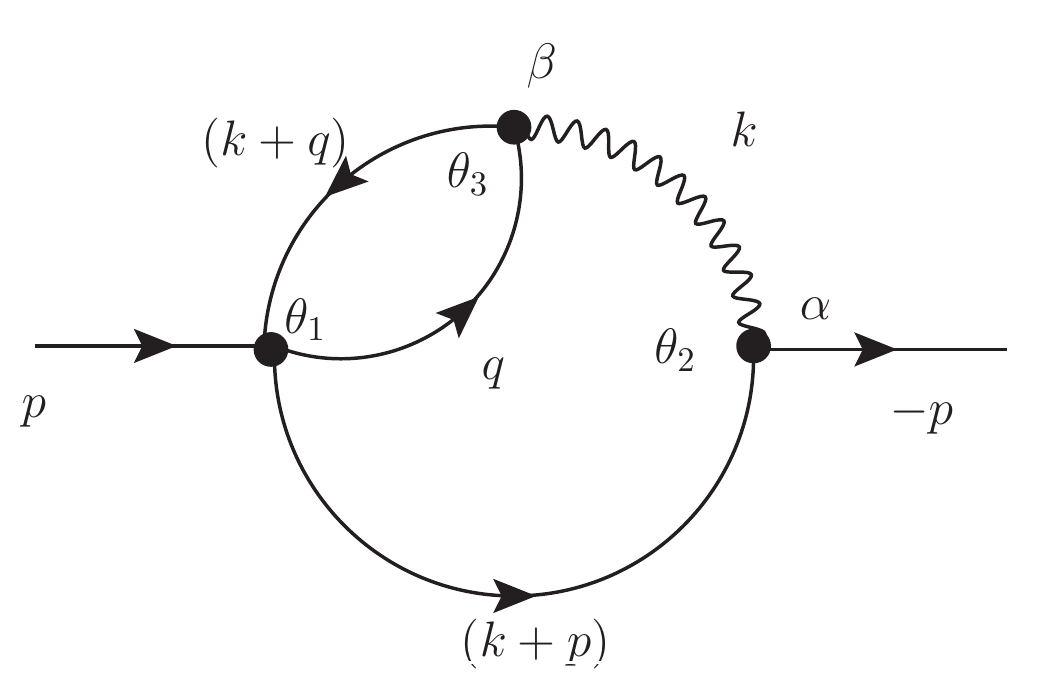}}\subfloat[]{\centering{}\includegraphics[scale=0.45]{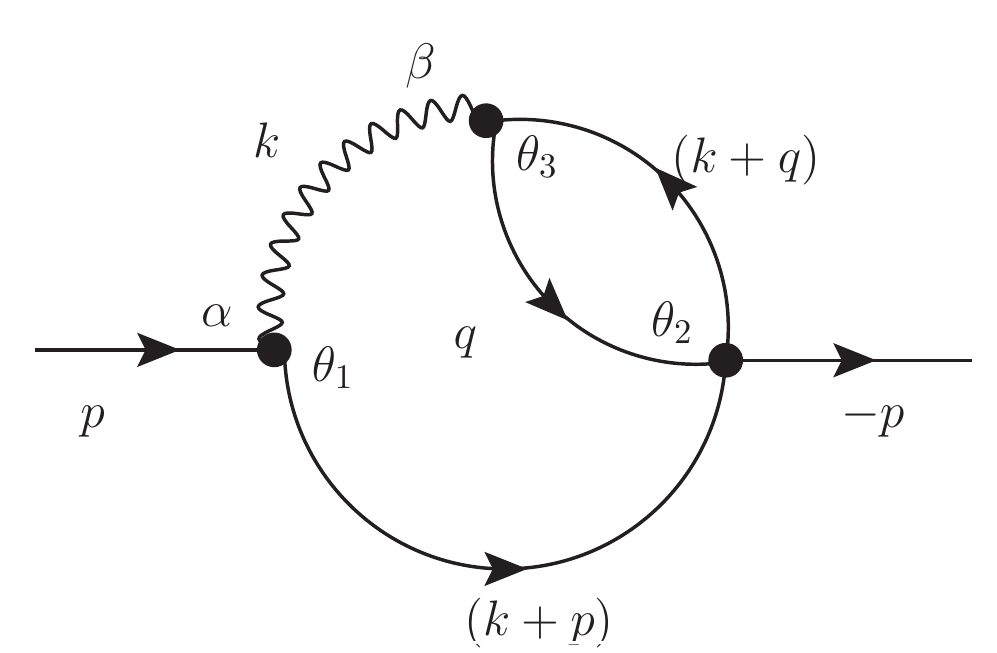}}\caption{\label{fig:SPHPH}$\mathcal{S}_{\overline{\Phi}\Phi}$}
\end{figure}
\par\end{center}

We start with the computation of diagram $\mathcal{S}_{\overline{\Phi}\Phi}^{\left(a\right)}$
in Figure\,\ref{fig:SPHPH},
\begin{align}
\mathcal{S}_{\overline{\Phi}\Phi}^{\left(a\right)}= & i\,\left(N+1\right)\frac{\lambda^{2}}{4}\int\frac{d^{3}p}{\left(2\pi\right)^{3}}d^{2}\theta\frac{d^{D}kd^{D}q}{\left(2\pi\right)^{2D}}\frac{\overline{\Phi}_{i}\left(p,\theta\right)D^{2}\Phi_{i}\left(-p,\theta\right)}{\left(k+p\right)^{2}\left(k+q\right)^{2}q^{2}}\,,
\end{align}
using Eq.\,(\ref{eq:Int 7}), we obtain
\begin{align}
\mathcal{S}_{\overline{\Phi}\Phi}^{\left(a\right)}= & -\frac{i\lambda^{2}\left(N+1\right)}{4\left(32\pi^{2}\epsilon\right)}\int\frac{d^{3}p}{\left(2\pi\right)^{3}}d^{2}\theta\overline{\Phi}_{i}\left(p,\theta\right)D^{2}\Phi_{i}\left(-p,\theta\right)\,.\label{eq:S-PHPH-a}
\end{align}

The diagram $\mathcal{S}_{\overline{\Phi}\Phi}^{\left(b\right)}$
in the Figure\,\ref{fig:SPHPH} is
\begin{align}
\mathcal{S}_{\overline{\Phi}\Phi}^{\left(b\right)} & =-\frac{1}{64}\,i\,g^{4}N\int\frac{d^{3}p}{\left(2\pi\right)^{3}}\,d^{2}\theta\,\overline{\Phi}_{i}\left(p,\theta\right)D^{2}\Phi_{i}\left(-p,\theta\right)\times\nonumber \\
 & \int\frac{d^{D}kd^{D}q}{\left(2\pi\right)^{2D}}\left\{ \frac{-8\left(a^{2}+6\,a\,b+b^{2}\right)\left[k\cdot q+q^{2}\right]k^{2}+4\left(a-b\right)^{2}\left(k^{2}\right)^{2}}{\left(k+p\right)^{2}\left(k+q\right)^{2}\left(k^{2}\right)^{2}q^{2}}\right\} \,,
\end{align}
using Eq.\,\,(\ref{eq:Int 7}) and Eq.\,(\ref{eq:Int 9}), by power
counting one of these three integrals is finite, then 
\begin{align}
\mathcal{S}_{\overline{\Phi}\Phi}^{\left(b\right)} & =\frac{1}{8}\left(\frac{\left(a+b\right)^{2}}{32\pi^{2}\epsilon}\right)\,i\,g^{4}N\int\frac{d^{3}p}{\left(2\pi\right)^{3}}\,d^{2}\theta\,\overline{\Phi}_{i}\left(p,\theta\right)D^{2}\Phi_{i}\left(-p,\theta\right)\,.\label{eq:S-PHPH-b}
\end{align}

Following the same procedure explained before, we consider the next
diagrams $\mathcal{S}_{\overline{\Phi}\Phi}^{\left(c\right)}$ and
$\mathcal{S}_{\overline{\Phi}\Phi}^{\left(d\right)}$ in the Figure\,\ref{fig:SPHPH},
\begin{align}
\mathcal{S}_{\overline{\Phi}\Phi}^{\left(c\right)}=\mathcal{S}_{\overline{\Phi}\Phi}^{\left(d\right)} & =\frac{1}{16}\,i\,g^{4}\int\frac{d^{3}p}{\left(2\pi\right)^{3}}\,d^{2}\theta\int\frac{d^{D}kd^{D}q}{\left(2\pi\right)^{2D}}\,\frac{4\left(a^{2}+a\,b\right)k\cdot q+8\,a^{2}k^{2}}{\left(k-p\right)^{2}\left(k+q\right)^{2}k^{2}q^{2}}\overline{\Phi}_{i}\left(p,\theta\right)D^{2}\Phi_{i}\left(-p,\theta\right)\,,
\end{align}
using Eq.\,(\ref{eq:Int 7}) and Eq.\,(\ref{eq:Int 9}), we find
\begin{align}
\mathcal{S}_{\overline{\Phi}\Phi}^{\left(c\right)} & =\mathcal{S}_{\overline{\Phi}\Phi}^{\left(d\right)}=\frac{1}{8}\left(\frac{a\,b-3\,a^{2}}{32\pi^{2}\epsilon}\right)\,i\,g^{4}\int\frac{d^{3}p}{\left(2\pi\right)^{3}}\,d^{2}\theta\overline{\Phi}_{i}\left(p,\theta\right)D^{2}\Phi_{i}\left(-p,\theta\right)\,.\label{eq:S-PHPH-c and -d}
\end{align}

The diagram $\mathcal{S}_{\overline{\Phi}\Phi}^{\left(e\right)}$
in the Figure\,\ref{fig:SPHPH} is
\begin{align}
\mathcal{S}_{\overline{\Phi}\Phi}^{\left(e\right)} & =-\frac{1}{64}\,i\,g^{4}\,\int\frac{d^{3}p}{\left(2\pi\right)^{3}}\,d^{2}\theta\int\frac{d^{D}kd^{D}q}{\left(2\pi\right)^{2D}}\,\frac{\Delta\left(k,q\right)}{\left(k+p\right)^{2}\left(q-p\right)^{2}\left(k+q\right)^{2}k^{2}q^{2}}\nonumber \\
 & \times\overline{\Phi}_{i}\left(p,\theta\right)D^{2}\Phi_{i}\left(-p,\theta\right)\,,
\end{align}
where
\begin{align}
\Delta\left(k,q\right) & =-4\left(b^{2}+2\,a\,b+4\,a^{2}\right)\left(k\cdot q\right)^{2}-3\left(2\,a\,b+b^{2}\right)\,q_{\alpha\beta}\,q_{\gamma\delta}\,k^{\alpha\delta}\,k^{\beta\gamma}\nonumber \\
 & +\left(20\,a^{2}+12\,a\,b+14\,b^{2}\right)k^{2}\,q^{2}\,,
\end{align}
using Eq.\,(\ref{eq:Int 7}) and Eq.\,(\ref{eq: Int 13}), we find
\begin{align}
\mathcal{S}_{\overline{\Phi}\Phi}^{\left(e\right)} & =\frac{1}{16}\left(\frac{3\,a^{2}+2\,a\,b+3\,b^{2}}{32\pi^{2}\epsilon}\right)\,i\,g^{4}\,\int\frac{d^{3}p}{\left(2\pi\right)^{3}}\,d^{2}\theta\overline{\Phi}_{i}\left(p,\theta\right)D^{2}\Phi_{i}\left(-p,\theta\right)\,.\label{eq:S-PHPH-e}
\end{align}

The diagram $\mathcal{S}_{\overline{\Phi}\Phi}^{\left(f\right)}$
in the Figure\,\ref{fig:SPHPH} is
\begin{align}
\mathcal{S}_{\overline{\Phi}\Phi}^{\left(f\right)} & =\frac{1}{8}\,i\,g^{4}\int\frac{d^{3}p}{\left(2\pi\right)^{3}}\,d^{2}\theta\int\frac{d^{D}kd^{D}q}{\left(2\pi\right)^{2D}}\,\frac{-2\,a^{2}}{q^{2}\left(k+p\right)^{2}\left(k+q\right)^{2}}\overline{\Phi}_{i}\left(p,\theta\right)D^{2}\Phi_{i}\left(-p,\theta\right)\,,
\end{align}
using Eq.\,(\ref{eq:Int 7}), we find
\begin{align}
\mathcal{S}_{\overline{\Phi}\Phi}^{\left(f\right)} & =\frac{1}{4}\left(\frac{a^{2}}{32\pi^{2}\epsilon}\right)\,i\,g^{4}\,\int\frac{d^{3}p}{\left(2\pi\right)^{3}}\,d^{2}\theta\,\overline{\Phi}_{i}\left(p,\theta\right)D^{2}\Phi_{i}\left(-p,\theta\right)\,.\label{eq:S-PHPH-f}
\end{align}

The diagrams $\mathcal{S}_{\overline{\Phi}\Phi}^{\left(g\right)}$
and $\mathcal{S}_{\overline{\Phi}\Phi}^{\left(h\right)}$ in the Figure\,\ref{fig:SPHPH}
are
\begin{align}
\mathcal{S}_{\overline{\Phi}\Phi}^{\left(g\right)}=\mathcal{S}_{\overline{\Phi}\Phi}^{\left(h\right)} & =0\,,\label{eq:S-PHPH-g and -h}
\end{align}
after $D$'s manipulations.

The diagram $\mathcal{S}_{\overline{\Phi}\Phi}^{\left(i\right)}$
in the Figure\,\ref{fig:SPHPH} is 
\begin{align}
\mathcal{S}_{\overline{\Phi}\Phi}^{\left(i\right)} & =\frac{1}{64}\,i\,g^{4}\int\frac{d^{3}p}{\left(2\pi\right)^{3}}\,d^{2}\theta\int\frac{d^{D}kd^{D}q}{\left(2\pi\right)^{2D}}\,\frac{-64\,a^{2}\left(k\cdot q\right)\,k^{2}-64\,a^{2}\,k^{4}-16\,\left(a^{2}-a\,b\right)k^{2}\,q^{2}}{\left(k-p\right)^{2}\left(k+q\right)^{2}q^{2}\left(k^{2}\right)^{2}}\nonumber \\
 & \times\overline{\Phi}_{i}\left(p,\theta\right)D^{2}\Phi_{i}\left(-p,\theta\right)\,,
\end{align}
using Eqs.\,(\ref{eq:Int 7}), (\ref{eq:Int 9}) and by power counting,
we find
\begin{align}
\mathcal{S}_{\overline{\Phi}\Phi}^{\left(i\right)} & =\frac{1}{2}\left(\frac{a^{2}}{32\pi^{2}\epsilon}\right)\,i\,g^{4}\int\frac{d^{3}p}{\left(2\pi\right)^{3}}\,d^{2}\theta\overline{\Phi}_{i}\left(p,\theta\right)D^{2}\Phi_{i}\left(-p,\theta\right)\,.\label{eq:S-PHPH-i}
\end{align}

Finally, the diagrams $\mathcal{S}_{\overline{\Phi}\Phi}^{\left(j\right)}$
and $\mathcal{S}_{\overline{\Phi}\Phi}^{\left(k\right)}$ in the Figure\,\ref{fig:SPHPH}
are
\begin{align}
\mathcal{S}_{\overline{\Phi}\Phi}^{\left(j\right)}=-\mathcal{S}_{\overline{\Phi}\Phi}^{\left(k\right)} & =-\frac{1}{16}i\,\lambda\,g^{2}\left(1+N\right)\int\frac{d^{3}p}{\left(2\pi\right)^{3}}\,d^{2}\theta\int\frac{d^{D}kd^{D}q}{\left(2\pi\right)^{2D}}\,\frac{-4\left(a-b\right)\,k\cdot q-2\left(a-b\right)\,k^{2}}{\left(k+p\right)^{2}\left(k+q\right)^{2}k^{2}q^{2}}\nonumber \\
 & \times\overline{\Phi}_{i}\left(p,\theta\right)D^{2}\Phi_{i}\left(-p,\theta\right)\,,
\end{align}
using Eqs.\,(\ref{eq:Int 7}) and\,(\ref{eq:Int 9}), we find
\begin{align}
\mathcal{S}_{\overline{\Phi}\Phi}^{\left(j\right)}=\mathcal{S}_{\overline{\Phi}\Phi}^{\left(k\right)} & =0\,.\label{eq:S-PHPH-j and -k}
\end{align}

\section{\label{sec:Four-points-vertex}Four-point vertex function to the
scalar superfield}

In this section we show all the diagrams that contribute to the four-point
vertex function. 
\begin{center}
\begin{figure}
\begin{centering}
\subfloat[]{\begin{centering}
\includegraphics[scale=0.5]{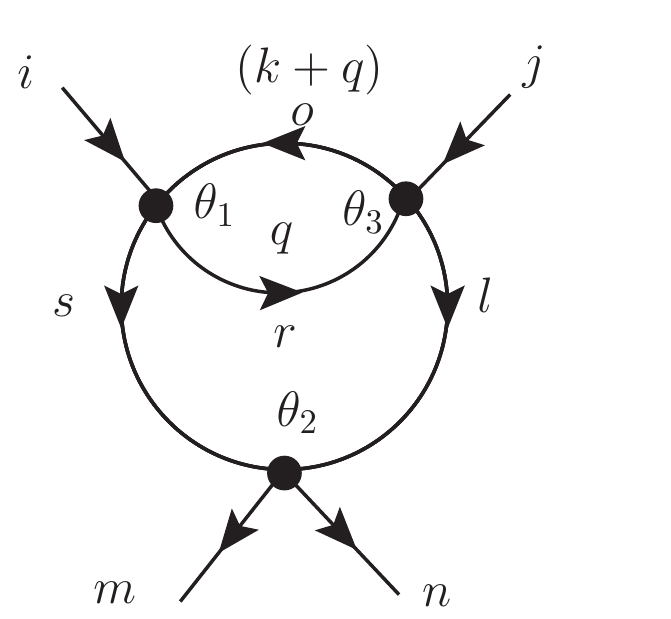}
\par\end{centering}
}\subfloat[]{\centering{}\includegraphics[scale=0.5]{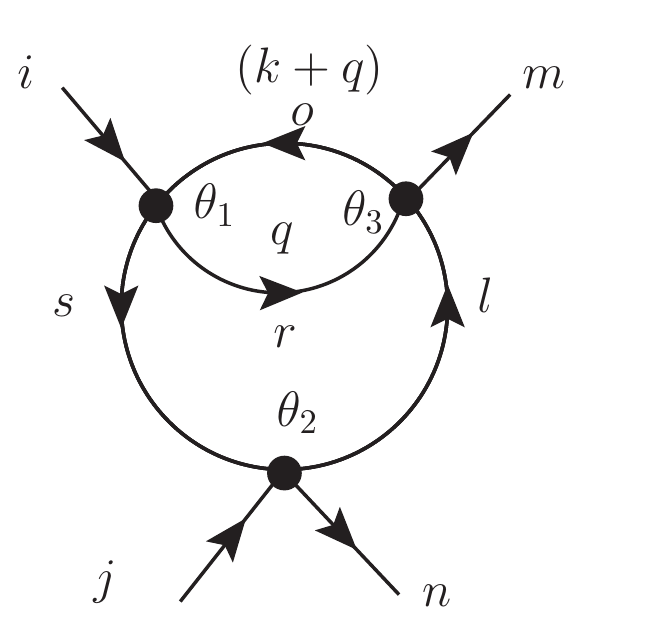}}\subfloat[]{\centering{}\includegraphics[scale=0.5]{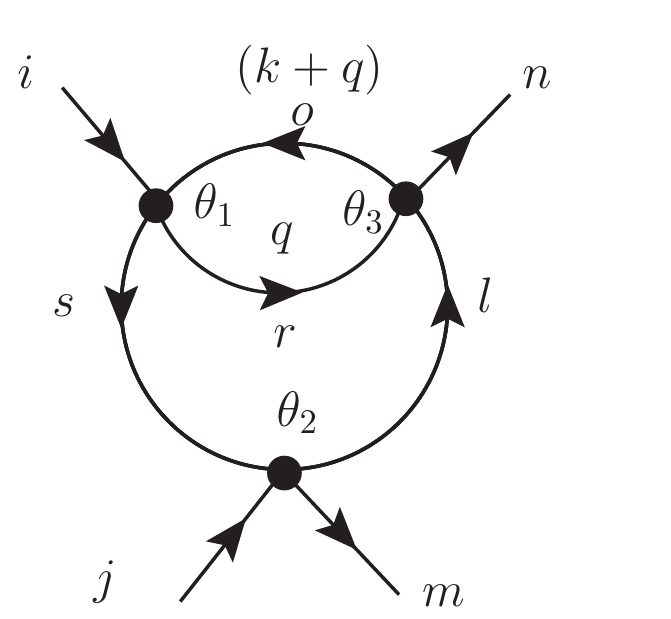}}\subfloat[]{\centering{}\includegraphics[scale=0.5]{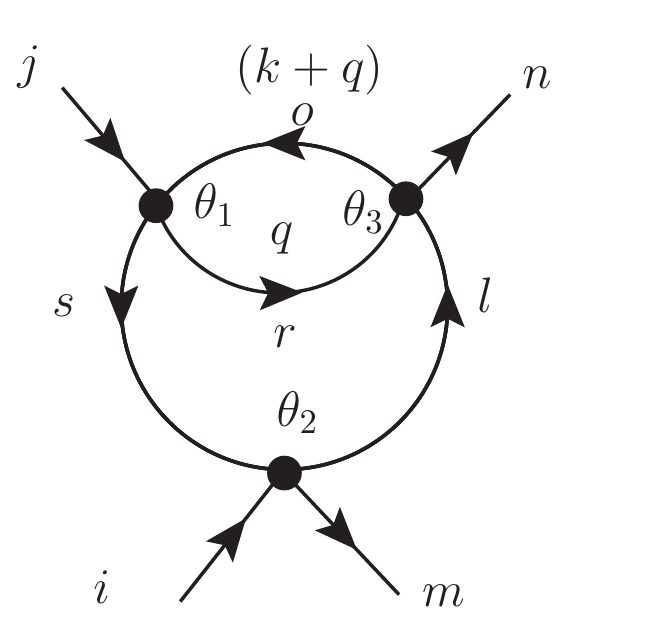}}
\par\end{centering}
\begin{centering}
\subfloat[]{\centering{}\includegraphics[scale=0.5]{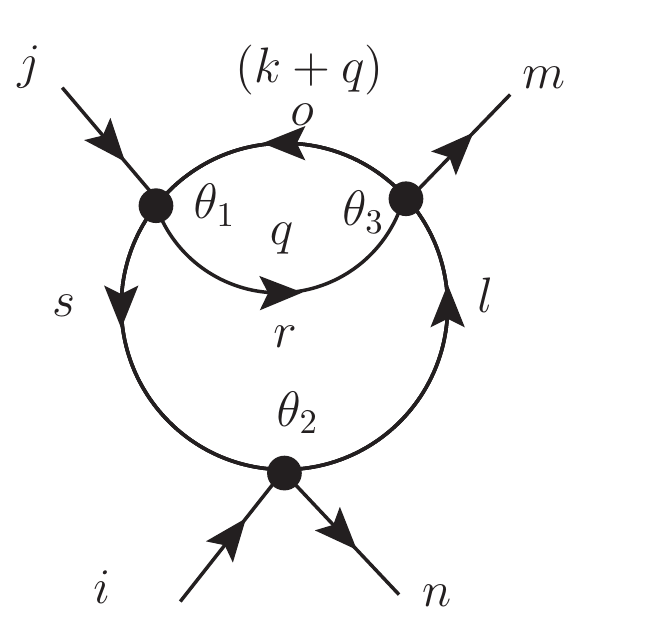}}\subfloat[]{\centering{}\includegraphics[scale=0.5]{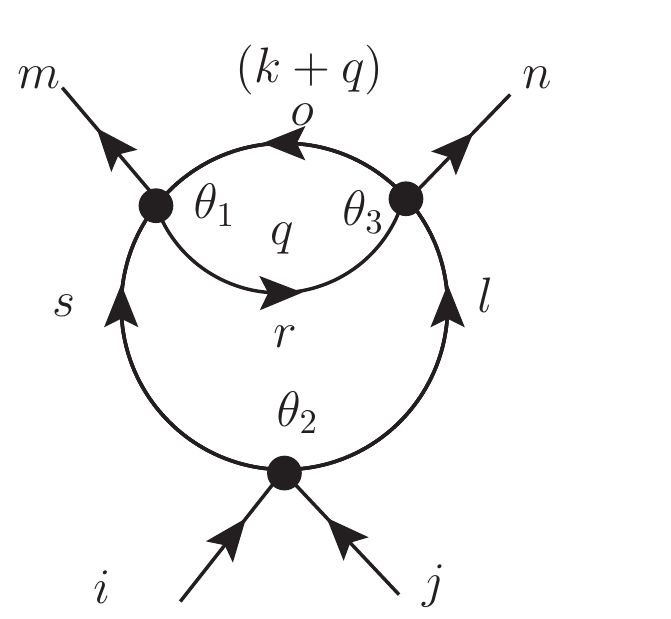}}\subfloat[]{\centering{}\includegraphics[scale=0.5]{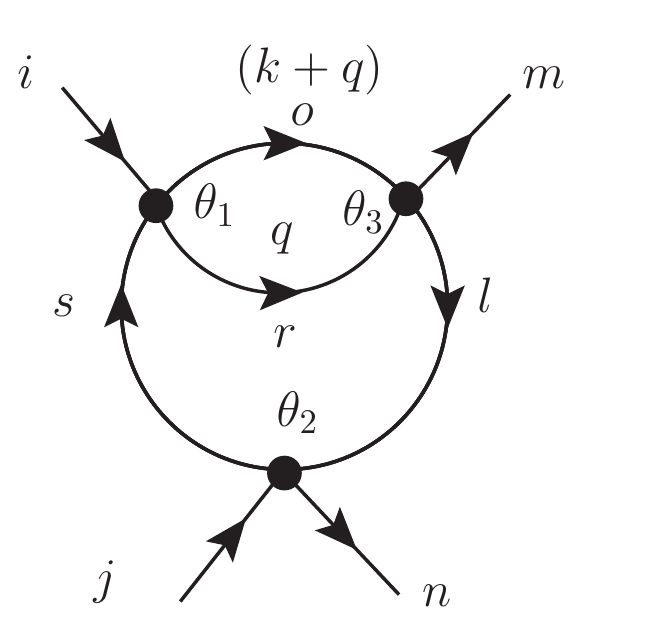}}\subfloat[]{\centering{}\includegraphics[scale=0.5]{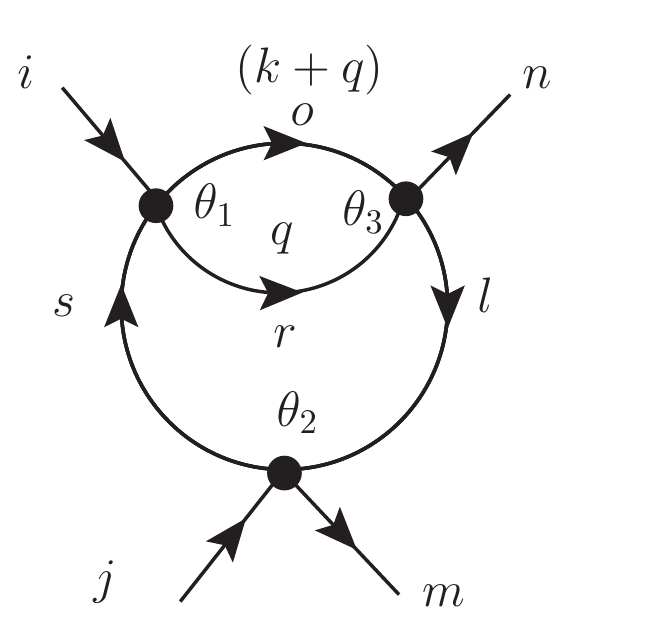}}
\par\end{centering}
\centering{}\subfloat[]{\centering{}\includegraphics[scale=0.5]{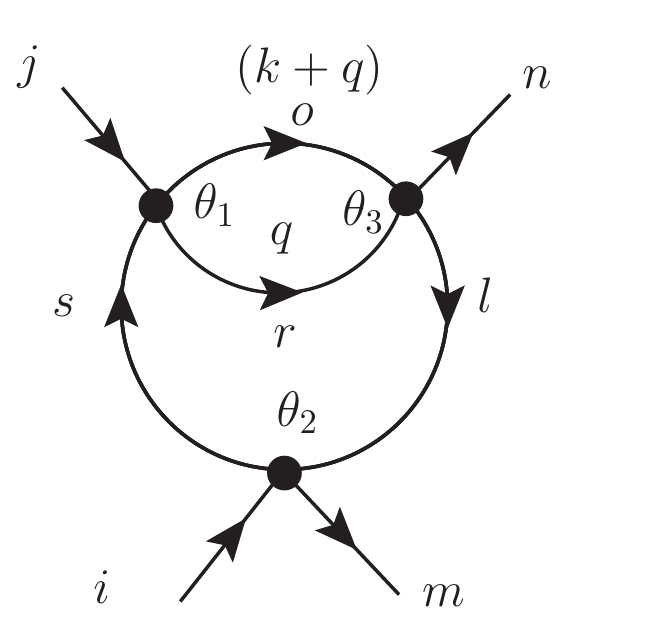}}\subfloat[]{\centering{}\includegraphics[scale=0.5]{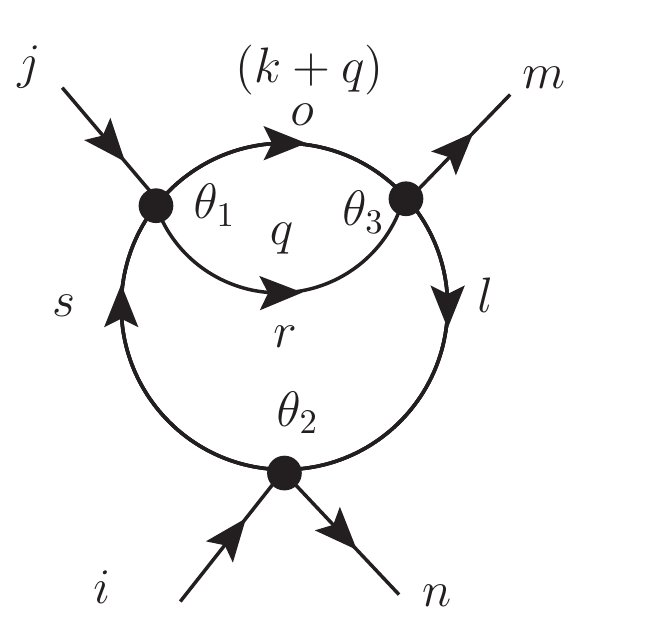}}\caption{\label{fig:D1-order-lambda-3}$\mathcal{S}_{\left(\overline{\Phi}\Phi\right)^{2}}^{\left(D1\right)}$}
\end{figure}
\par\end{center}

\begin{center}
\begin{table}
\centering{}%
\begin{tabular}{lcc}
 &  & \tabularnewline
\hline 
\hline 
$D1-a$ &  & $\left(3+N\right)\left(\delta_{in}\delta_{jm}+\delta_{im}\delta_{jn}\right)$\tabularnewline
$D1-b$ &  & $\left(2N+3\right)\delta_{im}\delta_{jn}+\left(N+2\right)\delta_{in}\delta_{jm}$\tabularnewline
$D1-c$ &  & $\left(N+2\right)\delta_{im}\delta_{jn}+\left(2N+3\right)\delta_{in}\delta_{jm}$\tabularnewline
$D1-d$ &  & $\left(2N+3\right)\,\delta_{jn}\delta_{im}+\left(N+2\right)\delta_{jm}\delta_{in}$\tabularnewline
$D1-e$ &  & $\left(N+2\right)\,\delta_{jn}\delta_{im}+\left(2\,N+3\right)\delta_{jm}\delta_{in}$\tabularnewline
$D1-f$ &  & $\left(3+N\right)\left(\delta_{in}\delta_{jm}+\delta_{im}\delta_{jn}\right)$\tabularnewline
$D1-g$ &  & $\delta_{in}\delta_{jm}+\left(2+N\right)\delta_{im}\delta_{jn}$\tabularnewline
$D1-h$ &  & $\left(2+N\right)\delta_{in}\delta_{jm}+\delta_{im}\delta_{jn}$\tabularnewline
$D1-i$ &  & $\delta_{in}\delta_{jm}+\left(N+2\right)\delta_{jn}\delta_{im}$\tabularnewline
$D1-j$ &  & $\left(N+2\right)\delta_{jm}\delta_{in}+\delta_{jn}\delta_{im}$\tabularnewline
\hline 
\hline 
 &  & \tabularnewline
\end{tabular}\caption{\label{tab:S-PPPP-1}Values of the diagrams in Figure\,\ref{fig:D1-order-lambda-3}
with common factor\protect \\
 $-\frac{1}{32\pi^{2}\epsilon}\,i\,\frac{\lambda^{3}}{8}\int_{\theta}\overline{\Phi}_{i}\Phi_{m}\Phi_{n}\overline{\Phi}_{j}$. }
\end{table}
\par\end{center}

We start with the diagrams that contribute with the order $\mathcal{O}\left(\lambda^{3}\right)$,
in this order there is only one topology that is equivalent to $10$
diagrams, the first of then is the diagram $\mathcal{S}_{\left(\overline{\Phi}\Phi\right)^{2}}^{\left(D1-a\right)}$
in the Figure\,\ref{fig:D1-order-lambda-3},
\begin{align}
\mathcal{S}_{\left(\overline{\Phi}\Phi\right)^{2}}^{\left(D1-a\right)} & =-\left(3+N\right)\,i\,\frac{\lambda^{3}}{8}\left(\delta_{in}\delta_{jm}+\delta_{im}\delta_{jn}\right)\int\frac{d^{D}kd^{D}q}{\left(2\pi\right)^{2D}}\frac{-k^{2}}{\left(k^{2}\right)^{2}q^{2}\left(k+q\right)^{2}}\,\int_{\theta}\overline{\Phi}_{i}\Phi_{m}\Phi_{n}\overline{\Phi}_{j}
\end{align}
where $\int_{\theta}\overline{\Phi}_{i}\Phi_{m}\Phi_{n}\overline{\Phi}_{j}\equiv\int d^{2}\theta\overline{\Phi}_{i}\left(0,\theta\right)\Phi_{m}\left(0,\theta\right)\Phi_{n}\left(0,\theta\right)\overline{\Phi}_{j}\left(0,\theta\right)$
and using Eq.\,(\ref{eq:Int 7}) , we find
\begin{align}
\mathcal{S}_{\left(\overline{\Phi}\Phi\right)^{2}}^{\left(D1-a\right)} & =-\frac{1}{32\pi^{2}\epsilon}\,i\,\frac{\lambda^{3}}{8}\left(3+N\right)\left(\delta_{in}\delta_{jm}+\delta_{im}\delta_{jn}\right)\int_{\theta}\overline{\Phi}_{i}\Phi_{m}\Phi_{n}\overline{\Phi}_{j}.
\end{align}

Then, we can repeat the same procedure for the other diagrams, with
the difference between diagrams given only by the manipulations of
$\delta$'s that can change for each diagram. We provide these values
in the Table\,\ref{tab:S-PPPP-1}, for each diagram in Figure\,\ref{fig:D1-order-lambda-3}.
Therefore, the total contribution of diagram $\mathcal{S}_{\left(\overline{\Phi}\Phi\right)^{2}}^{\left(D1\right)}$
turns out to be
\begin{align}
\mathcal{S}_{\left(\overline{\Phi}\Phi\right)^{2}}^{\left(D1\right)} & =-i\,\frac{1}{4\left(32\pi^{2}\epsilon\right)}\left(5N+11\right)\,\lambda^{3}\left(\delta_{in}\delta_{jm}+\delta_{im}\delta_{jn}\right)\int_{\theta}\overline{\Phi}_{i}\Phi_{m}\Phi_{n}\overline{\Phi}_{j}\,.\label{eq:D1-lambda-3}
\end{align}

Following with the process, now we consider all diagrams that contribute
to the order $\mathcal{O}\left(\lambda^{2}g^{2}\right)$. In this
order there are $5$ topologies that are equivalent to $44$ diagrams.
\begin{center}
\begin{figure}
\centering{}\subfloat[]{\begin{centering}
\includegraphics[scale=0.5]{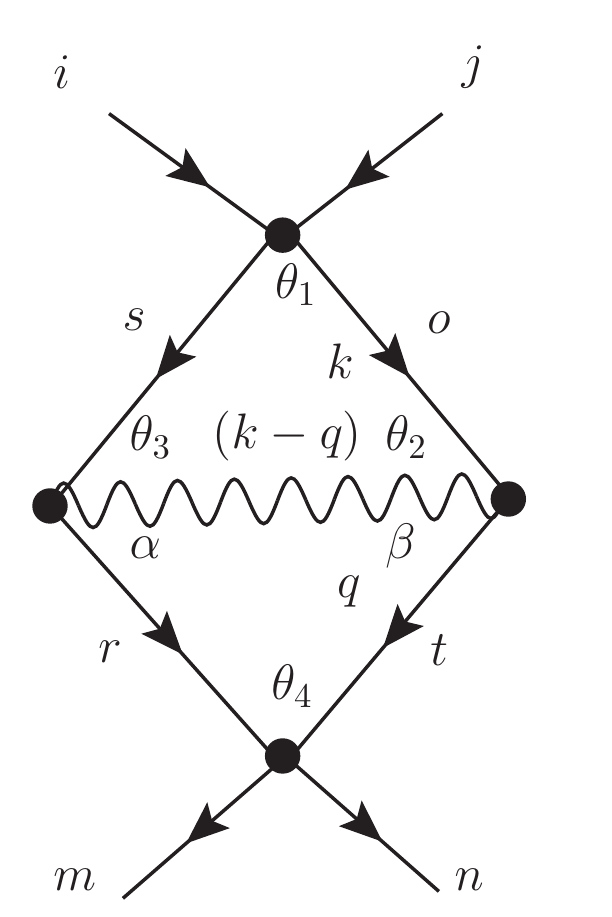}
\par\end{centering}
}\subfloat[]{\centering{}\includegraphics[scale=0.5]{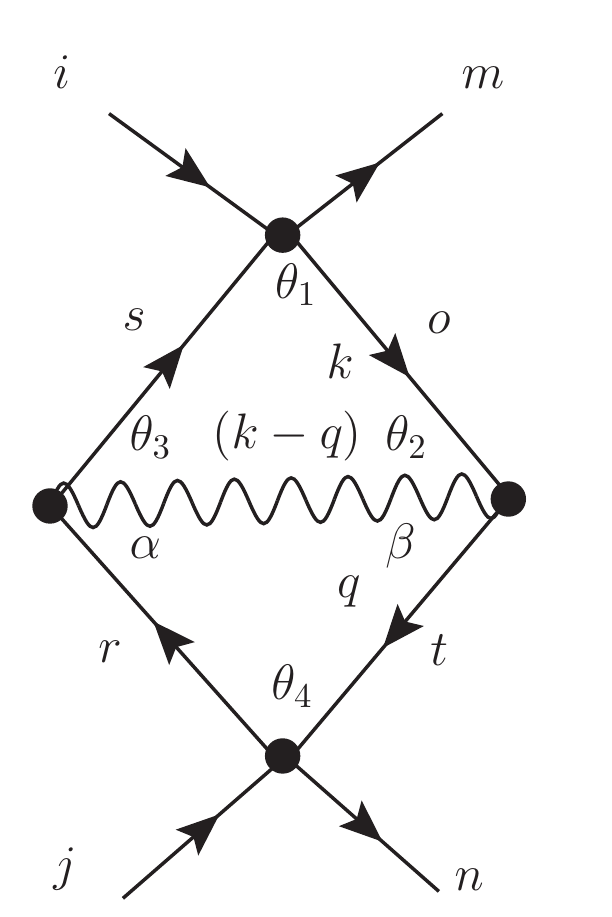}}\subfloat[]{\centering{}\includegraphics[scale=0.5]{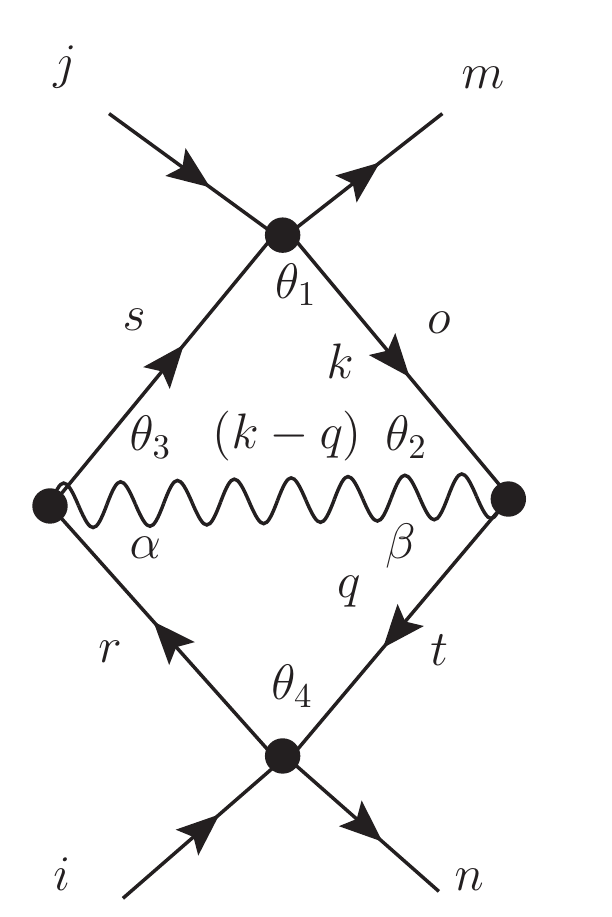}}\caption{\label{fig:D2-order-g-2-lambda-2}$\mathcal{S}_{\left(\overline{\Phi}\Phi\right)^{2}}^{\left(D2\right)}$}
\end{figure}
\par\end{center}

\begin{center}
\begin{table}
\centering{}%
\begin{tabular}{lcccccc}
 &  &  &  &  &  & \tabularnewline
\hline 
\hline 
$D2-a$ &  & $-\left(\delta_{in}\delta_{jm}+\delta_{im}\delta_{jn}\right)$ &  & $D2-b$ &  & $\left(N+2\right)\delta_{im}\delta_{jn}+\delta_{in}\delta_{mj}$\tabularnewline
$D2-c$ &  & $\left(N+2\right)\delta_{jm}\delta_{in}+\delta_{jn}\delta_{mi}$ &  &  &  & \tabularnewline
\hline 
\hline 
 &  &  &  &  &  & \tabularnewline
\end{tabular}\caption{\label{tab:S-PPPP-2}Values of the diagrams in Figure\,\ref{fig:D2-order-g-2-lambda-2}
with common factor\protect \\
 $-\frac{a}{8\left(32\pi^{2}\epsilon\right)}\,i\,\lambda^{2}\,g^{2}\int_{\theta}\overline{\Phi}_{i}\Phi_{m}\Phi_{n}\overline{\Phi}_{j}$
.}
\end{table}
\par\end{center}

We start with $\mathcal{S}_{\left(\overline{\Phi}\Phi\right)^{2}}^{\left(D2-a\right)}$
in the Figure\,\ref{fig:D2-order-g-2-lambda-2}
\begin{align}
\mathcal{S}_{\left(\overline{\Phi}\Phi\right)^{2}}^{\left(D2-a\right)} & =-\frac{1}{32}\,i\,\lambda^{2}\,g^{2}\left(\delta_{im}\delta_{jn}+\delta_{in}\delta_{jm}\right)\int_{\theta}\overline{\Phi}_{i}\Phi_{m}\Phi_{n}\overline{\Phi}_{j}\nonumber \\
 & \times\int\frac{d^{D}kd^{D}q}{\left(2\pi\right)^{2D}}\left\{ \frac{-8\,a\,k^{2}\,q^{2}}{\left(k^{2}\right)^{2}\left(q^{2}\right)^{2}\left(k-q\right)^{2}}\right\} \,,
\end{align}
using Eq.\,(\ref{eq:Int 7}) we find,
\begin{align}
\mathcal{S}_{\left(\overline{\Phi}\Phi\right)^{2}}^{\left(D2\right)} & =-\frac{a}{4\left(32\pi^{2}\epsilon\right)}\,i\,\lambda^{2}\,g^{2}\left(\delta_{im}\delta_{jn}+\delta_{in}\delta_{jm}\right)\int_{\theta}\overline{\Phi}_{i}\Phi_{m}\Phi_{n}\overline{\Phi}_{j}\,.
\end{align}

In Table\,\ref{tab:S-PPPP-2}, we give the factors associated to
each diagram in Figure\,\ref{fig:D2-order-g-2-lambda-2}. Therefore,
the total contribution of diagram $\mathcal{S}_{\left(\overline{\Phi}\Phi\right)^{2}}^{\left(D2\right)}$
is, 
\begin{align}
\mathcal{S}_{\left(\overline{\Phi}\Phi\right)^{2}}^{\left(D2\right)} & =\frac{a}{4\left(32\pi^{2}\epsilon\right)}\,\left(N+2\right)\,i\,\lambda^{2}\,g^{2}\left(\delta_{im}\delta_{jn}+\delta_{in}\delta_{mj}\right)\int_{\theta}\overline{\Phi}_{i}\Phi_{m}\Phi_{n}\overline{\Phi}_{j}\,.\label{eq:S-D2}
\end{align}

\begin{center}
\begin{figure}
\begin{centering}
\subfloat[]{\begin{centering}
\includegraphics[scale=0.5]{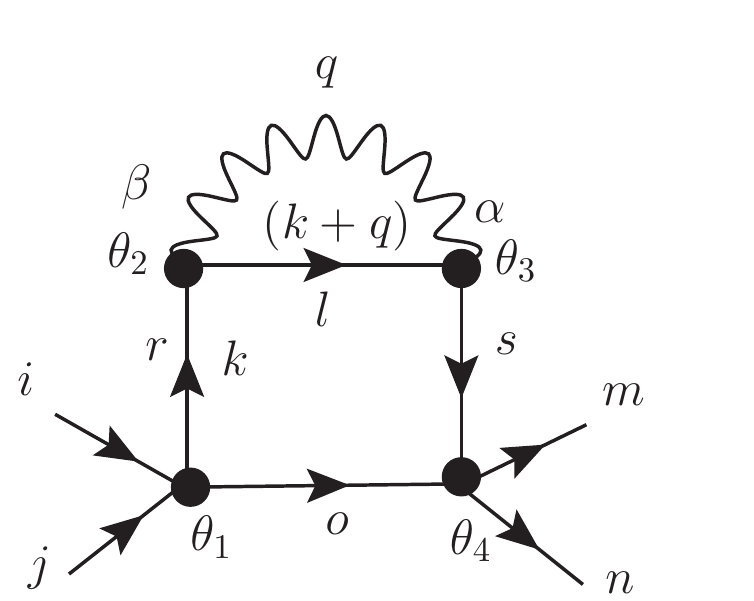}
\par\end{centering}
}\subfloat[]{\centering{}\includegraphics[scale=0.5]{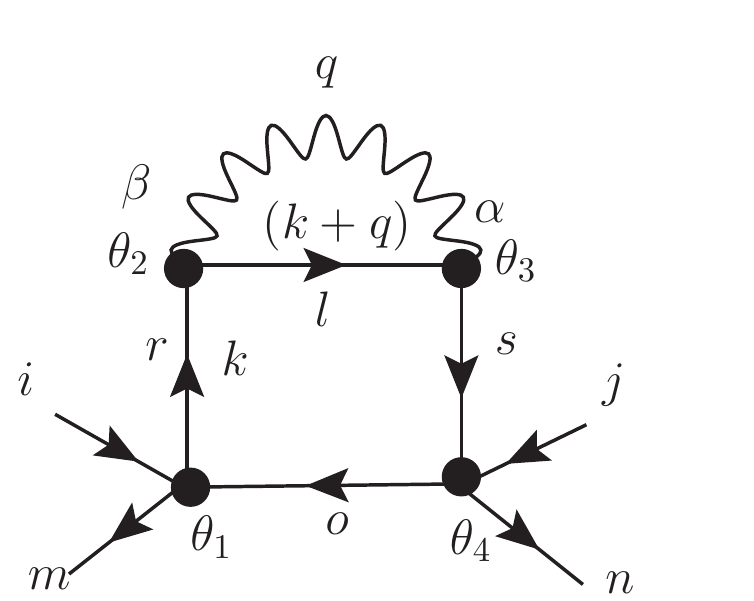}}\subfloat[]{\centering{}\includegraphics[scale=0.5]{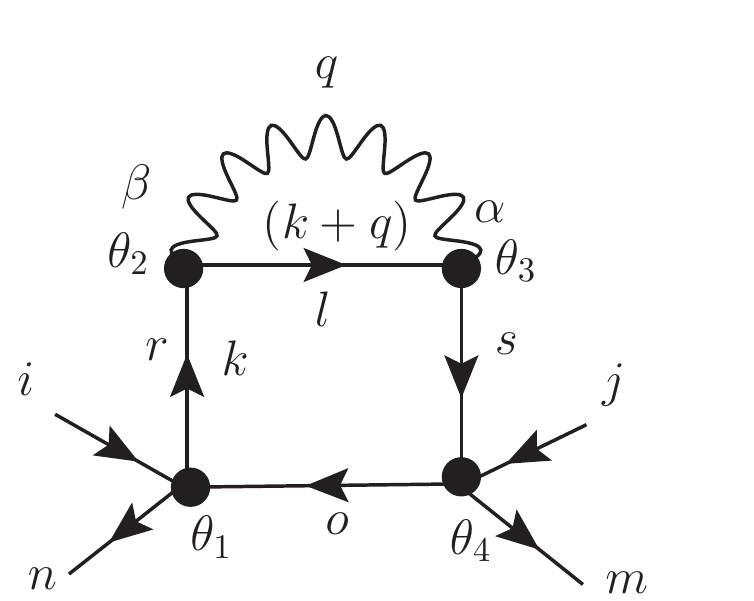}}
\par\end{centering}
\centering{}\subfloat[]{\centering{}\includegraphics[scale=0.5]{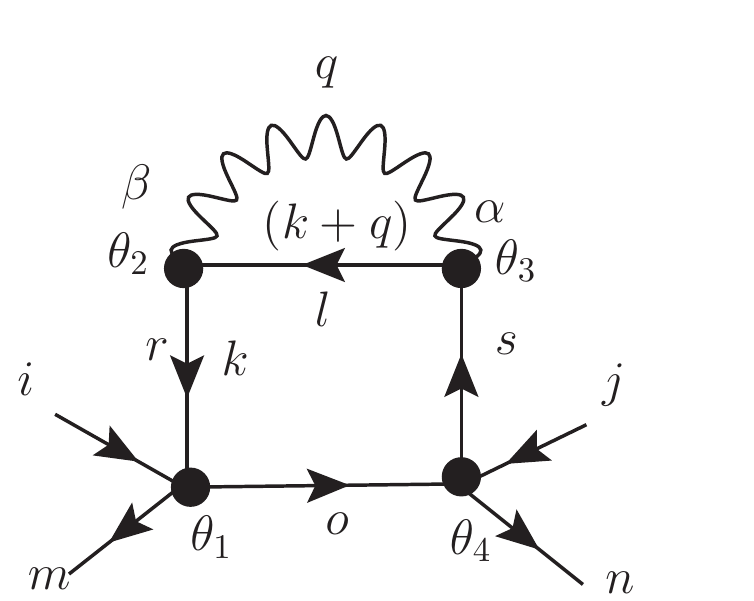}}\subfloat[]{\centering{}\includegraphics[scale=0.5]{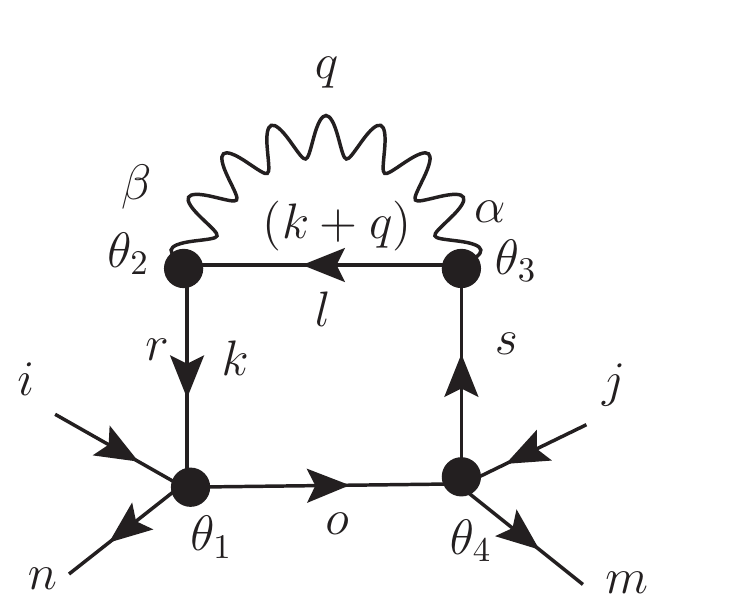}}\caption{\label{fig:D3-order-lambda-2-g-2}$\mathcal{S}_{\left(\overline{\Phi}\Phi\right)^{2}}^{\left(D3\right)}$}
\end{figure}
\par\end{center}

\begin{center}
\begin{table}
\centering{}%
\begin{tabular}{lcccccc}
 &  &  &  &  &  & \tabularnewline
\hline 
\hline 
$D3-a$ &  & $2\left(\delta_{in}\delta_{jm}+\delta_{im}\delta_{jn}\right)$ &  & $D3-b$ &  & $\left(N+2\right)\delta_{mi}\delta_{jn}+\delta_{mj}\delta_{in}$\tabularnewline
$D3-c$ &  & $\left(N+2\right)\delta_{ni}\delta_{jm}+\delta_{nj}\delta_{im}$ &  & $D3-d$ &  & $\left(N+2\right)\delta_{mi}\delta_{jn}+\delta_{mj}\delta_{in}$\tabularnewline
$D3-e$ &  & $\left(N+2\right)\delta_{ni}\delta_{jm}+\delta_{nj}\delta_{im}$ &  &  &  & \tabularnewline
\hline 
\hline 
 &  &  &  &  &  & \tabularnewline
\end{tabular}\caption{\label{tab:S-PPPP-3}Values of the diagrams in Figure\,\ref{fig:D3-order-lambda-2-g-2}
with common factor\protect \\
 $-\frac{a}{8\left(32\pi^{2}\epsilon\right)}\,i\,\lambda^{2}\,g^{2}\int_{\theta}\overline{\Phi}_{i}\Phi_{m}\Phi_{n}\overline{\Phi}_{j}$
.}
\end{table}
\par\end{center}

$\mathcal{S}_{\left(\overline{\Phi}\Phi\right)^{2}}^{\left(D3-a\right)}$
in the Figure\,\ref{fig:D3-order-lambda-2-g-2} is
\begin{align}
\mathcal{S}_{\left(\overline{\Phi}\Phi\right)^{2}}^{\left(D3-a\right)} & =-\frac{1}{16}\,i\,\lambda^{2}\,g^{2}\left(\delta_{in}\delta_{jm}+\delta_{im}\delta_{jn}\right)\int_{\theta}\overline{\Phi}_{i}\Phi_{m}\Phi_{n}\overline{\Phi}_{j}\nonumber \\
 & \times\int\frac{d^{D}kd^{D}q}{\left(2\pi\right)^{2D}}\left\{ \frac{-8\,a\,\left(k\cdot q\right)\,k^{2}-8\,a\,\left(k^{2}\right)^{2}+2\left(b-a\right)k^{2}\,q^{2}}{\left(k^{2}\right)^{3}q^{2}\left(k+q\right)^{2}}\right\} \,,
\end{align}
using Eqs.\,(\ref{eq:Int 7}), (\ref{eq:Int 9}) , we find
\begin{align}
\mathcal{S}_{\left(\overline{\Phi}\Phi\right)^{2}}^{\left(D3-a\right)} & =-\frac{a}{4\left(32\pi^{2}\epsilon\right)}\,i\,\lambda^{2}\,g^{2}\left(\delta_{in}\delta_{jm}+\delta_{im}\delta_{jn}\right)\int_{\theta}\overline{\Phi}_{i}\Phi_{m}\Phi_{n}\overline{\Phi}_{j}\,,
\end{align}
adding the diagrams $\mathcal{S}_{\left(\overline{\Phi}\Phi\right)^{2}}^{\left(D3-a\right)}$
to $\mathcal{S}_{\left(\overline{\Phi}\Phi\right)^{2}}^{\left(D3-e\right)}$
using the values in Table\,\ref{tab:S-PPPP-3} we find
\begin{align}
\mathcal{S}_{\left(\overline{\Phi}\Phi\right)^{2}}^{\left(D3\right)} & =-\frac{a}{4\left(32\pi^{2}\epsilon\right)}\,\left(N+4\right)\,i\,\lambda^{2}\,g^{2}\left(\delta_{mi}\delta_{jn}+\delta_{mj}\delta_{in}\right)\int_{\theta}\overline{\Phi}_{i}\Phi_{m}\Phi_{n}\overline{\Phi}_{j}\,.\label{eq:S-D3}
\end{align}

\begin{center}
\begin{figure}
\begin{centering}
\subfloat[]{\begin{centering}
\includegraphics[scale=0.5]{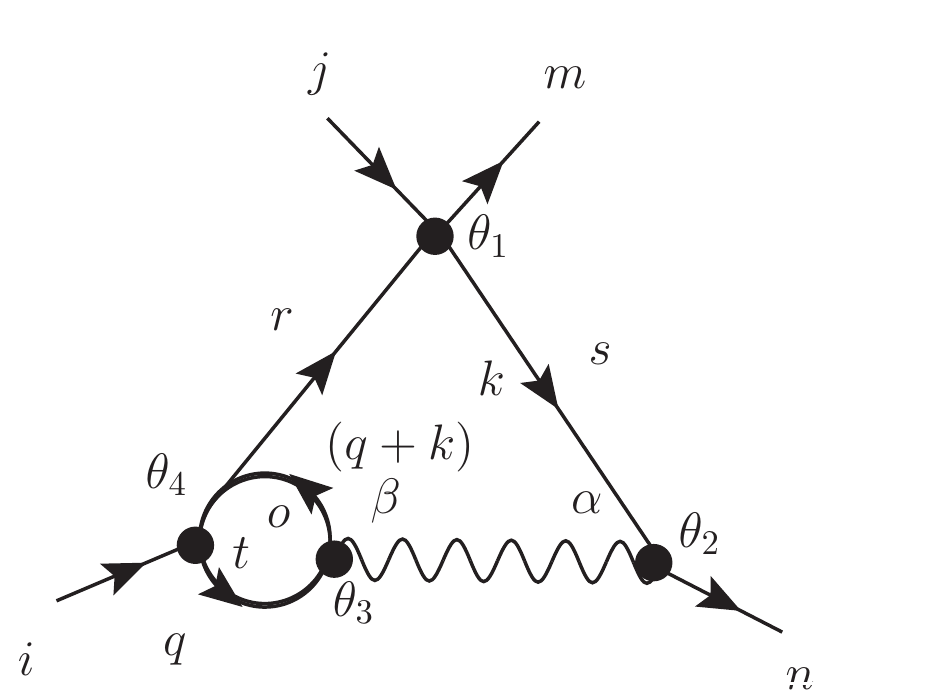}
\par\end{centering}
}\subfloat[]{\centering{}\includegraphics[scale=0.5]{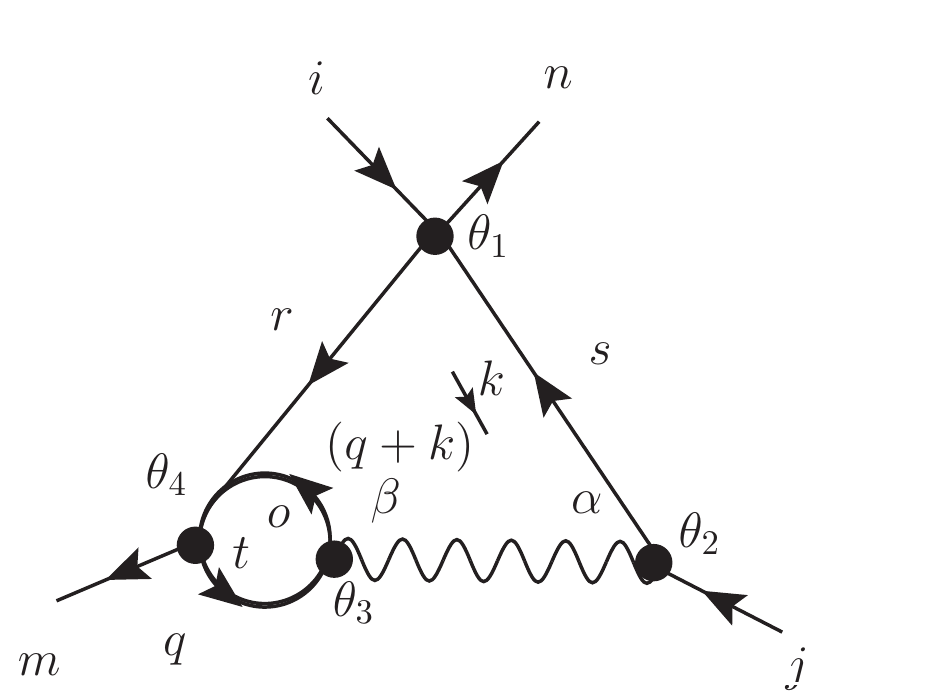}}\subfloat[]{\centering{}\includegraphics[scale=0.5]{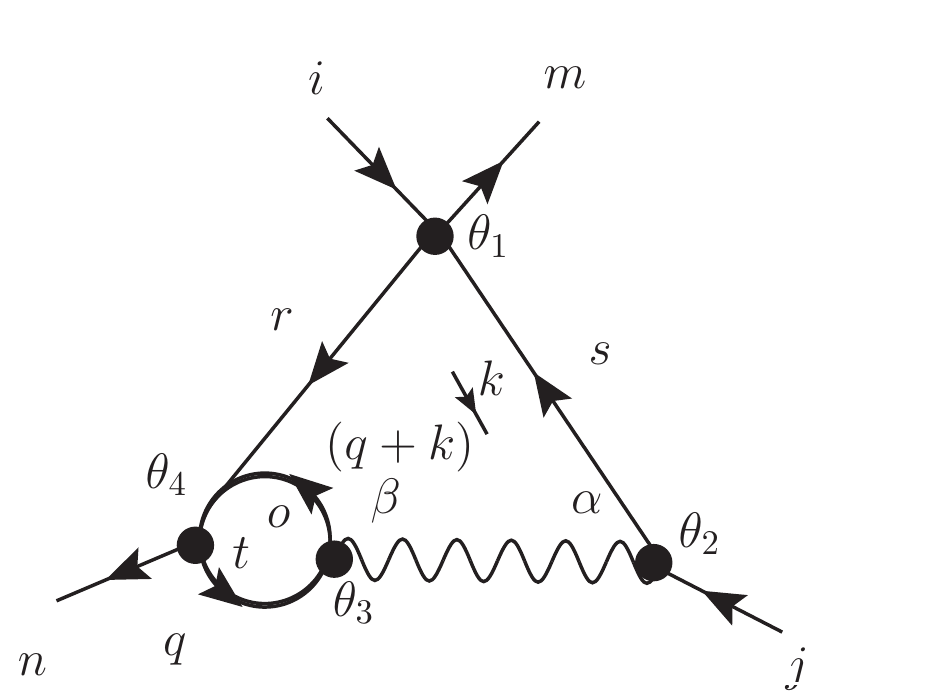}}
\par\end{centering}
\begin{centering}
\subfloat[]{\centering{}\includegraphics[scale=0.5]{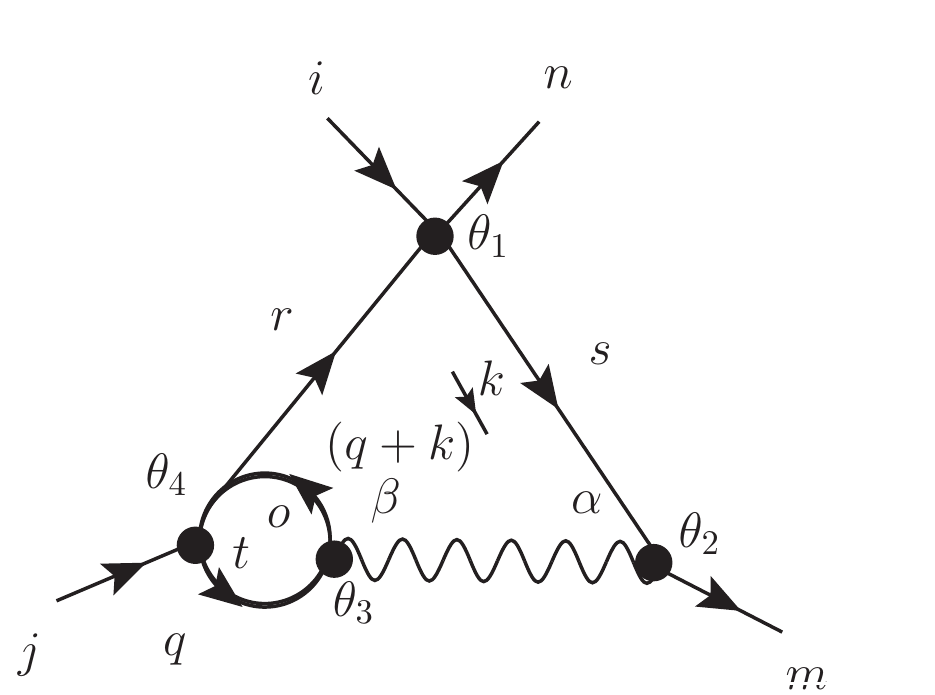}}\subfloat[]{\centering{}\includegraphics[scale=0.5]{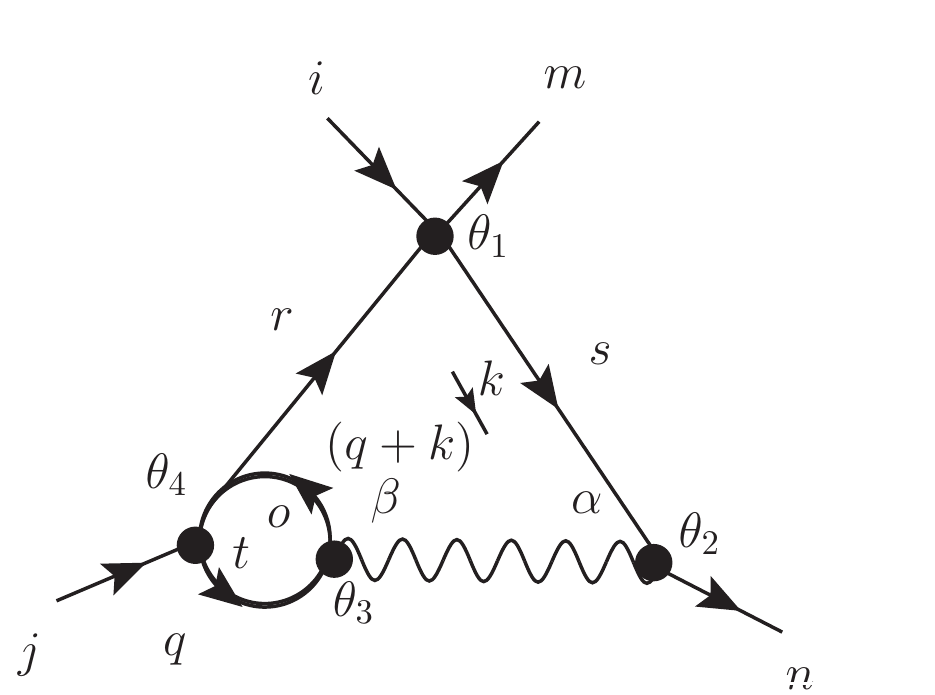}}\subfloat[]{\centering{}\includegraphics[scale=0.5]{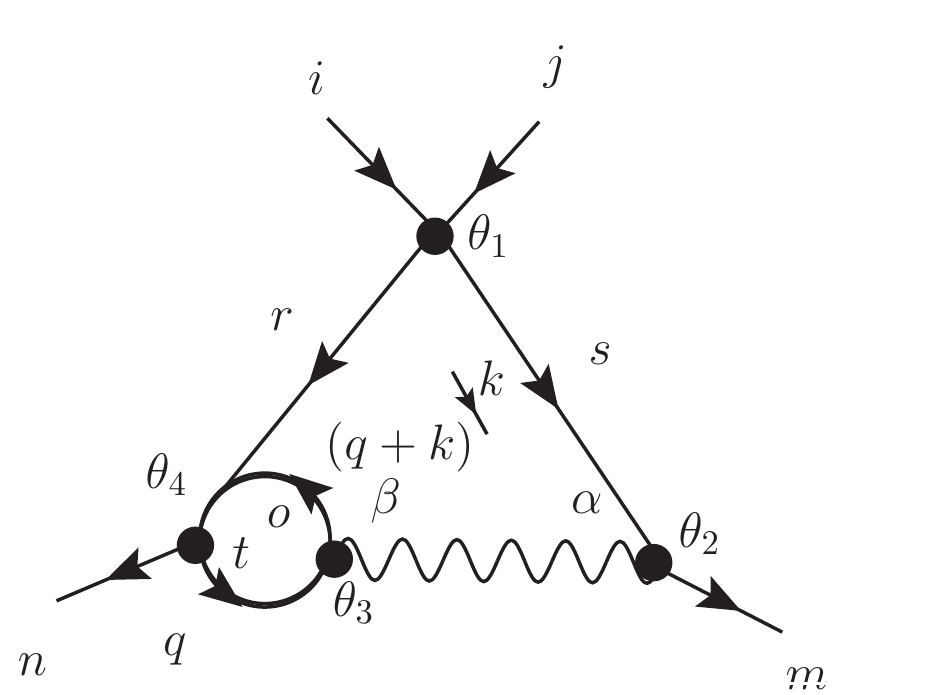}}
\par\end{centering}
\begin{centering}
\subfloat[]{\centering{}\includegraphics[scale=0.5]{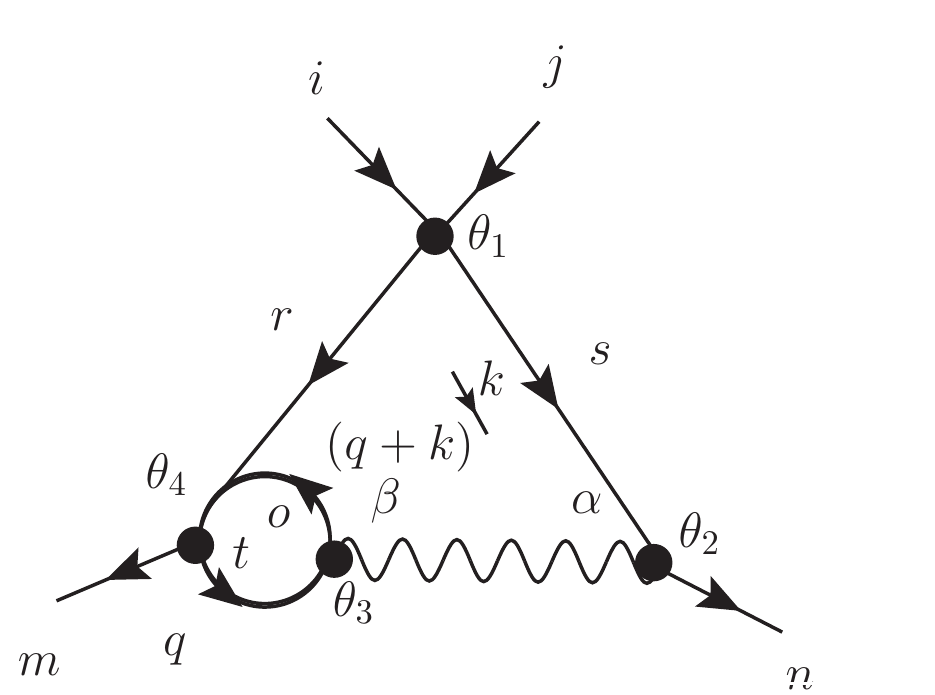}}\subfloat[]{\centering{}\includegraphics[scale=0.5]{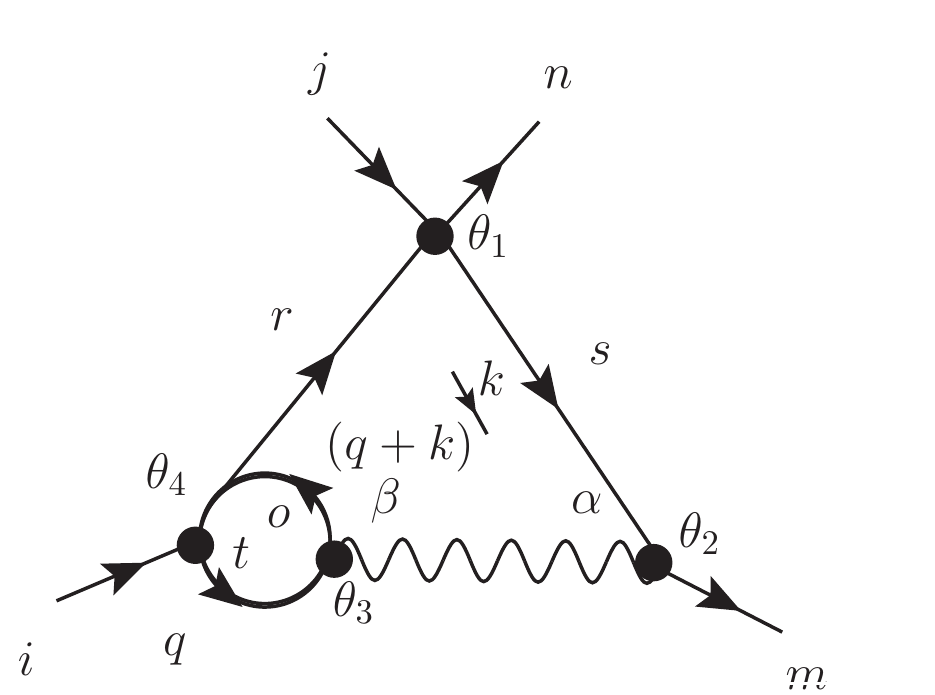}}\subfloat[]{\centering{}\includegraphics[scale=0.5]{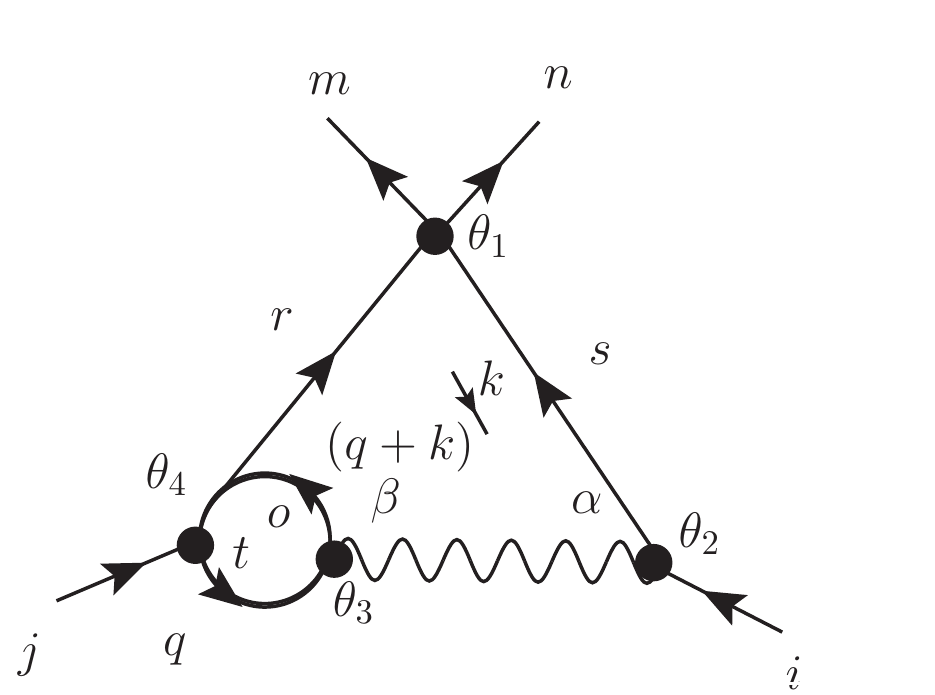}}
\par\end{centering}
\centering{}\subfloat[]{\centering{}\includegraphics[scale=0.5]{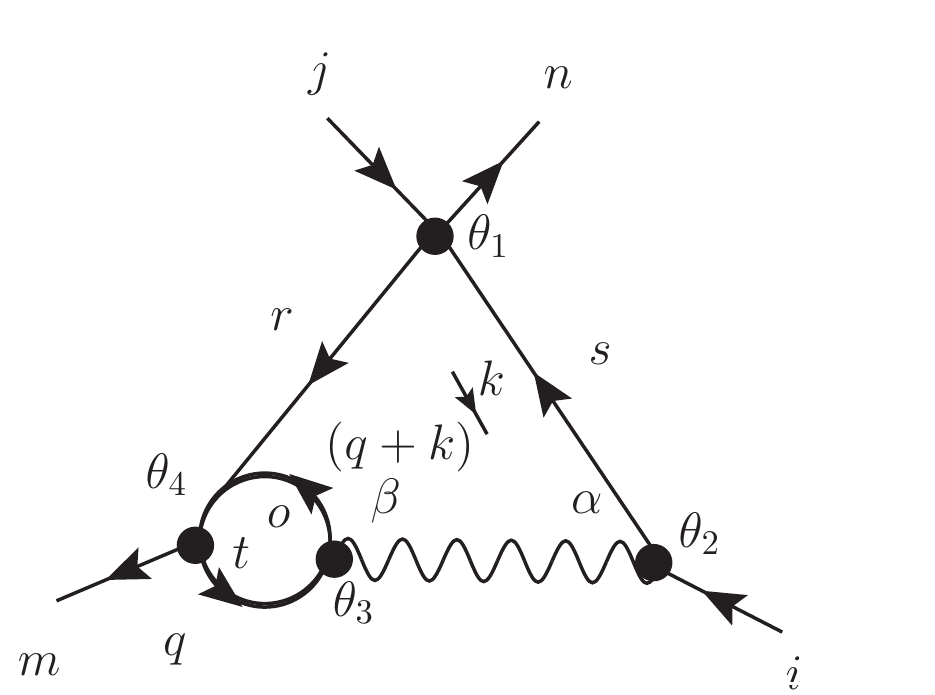}}\subfloat[]{\centering{}\includegraphics[scale=0.5]{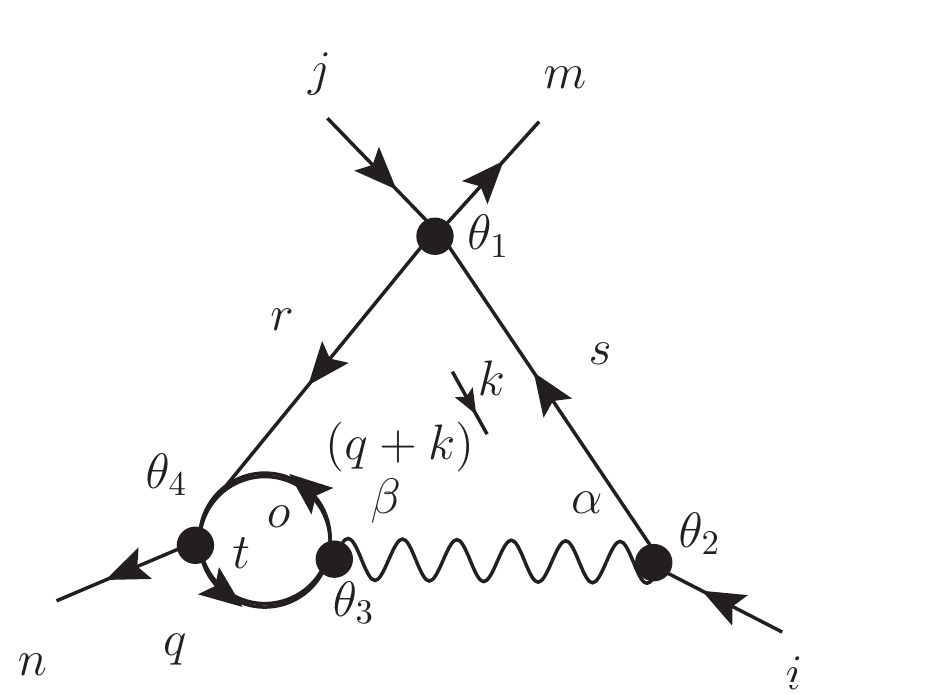}}\subfloat[]{\centering{}\includegraphics[scale=0.5]{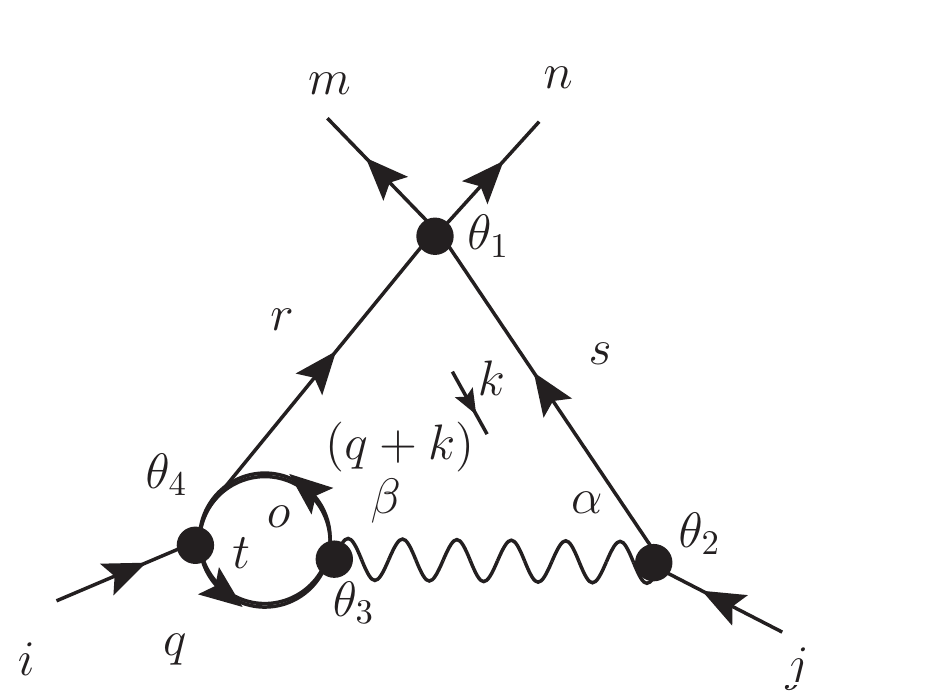}}\caption{$\mathcal{S}_{\left(\overline{\Phi}\Phi\right)^{2}}^{\left(D4\right)}$}
\end{figure}
\par\end{center}

$\mathcal{S}_{\left(\overline{\Phi}\Phi\right)^{2}}^{\left(D4-a\right)}$
in the Figure\,\ref{fig:D3-order-lambda-2-g-2} is
\begin{align}
\mathcal{S}_{\left(\overline{\Phi}\Phi\right)^{2}}^{\left(D4-a\right)} & =\frac{1}{32}\,i\,\lambda^{2}g^{2}\,\left(N+1\right)\left(\delta_{jm}\delta_{in}+\delta_{im}\delta_{jn}\right)\int_{\theta}\overline{\Phi}_{i}\Phi_{m}\Phi_{n}\overline{\Phi}_{j}\nonumber \\
 & \times\int\frac{d^{D}kd^{D}q}{\left(2\pi\right)^{2D}}\left\{ \frac{4\left(a-b\right)k\cdot q\,k^{2}+2\left(a-b\right)\,\left(k^{2}\right)^{2}}{\left(k^{2}\right)^{3}q^{2}\left(k+q\right)^{2}}\right\} \,,
\end{align}
using Eqs.\,(\ref{eq:Int 7}) and\,(\ref{eq:Int 9}), we find 
\begin{align}
\mathcal{S}_{\left(\overline{\Phi}\Phi\right)^{2}}^{\left(D4-a\right)} & =\mathcal{S}_{\left(\overline{\Phi}\Phi\right)^{2}}^{\left(D4\right)}=0\,.\label{eq:S-D4}
\end{align}

\begin{center}
\begin{figure}
\begin{centering}
\subfloat[]{\begin{centering}
\includegraphics[scale=0.5]{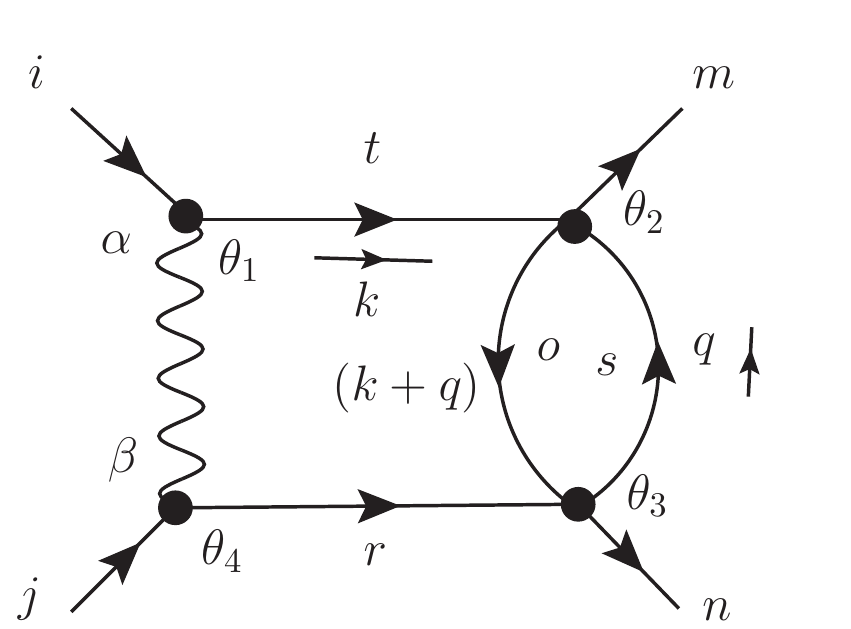}
\par\end{centering}
}\subfloat[]{\centering{}\includegraphics[scale=0.5]{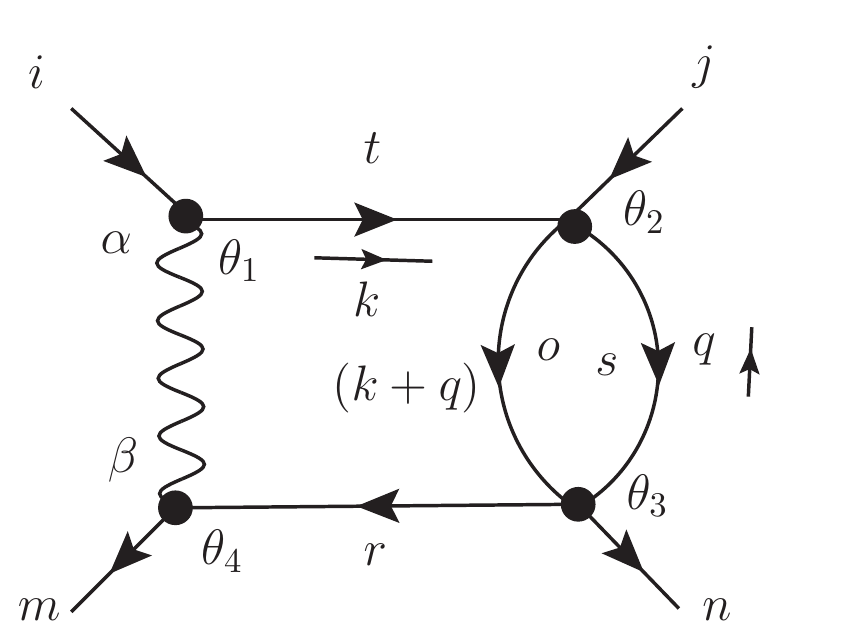}}\subfloat[]{\centering{}\includegraphics[scale=0.5]{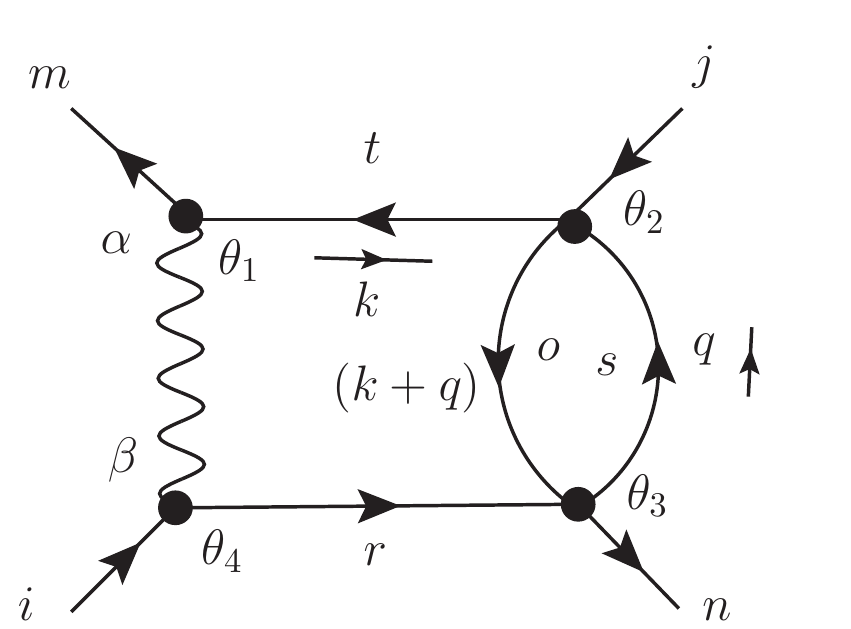}}
\par\end{centering}
\begin{centering}
\subfloat[]{\centering{}\includegraphics[scale=0.5]{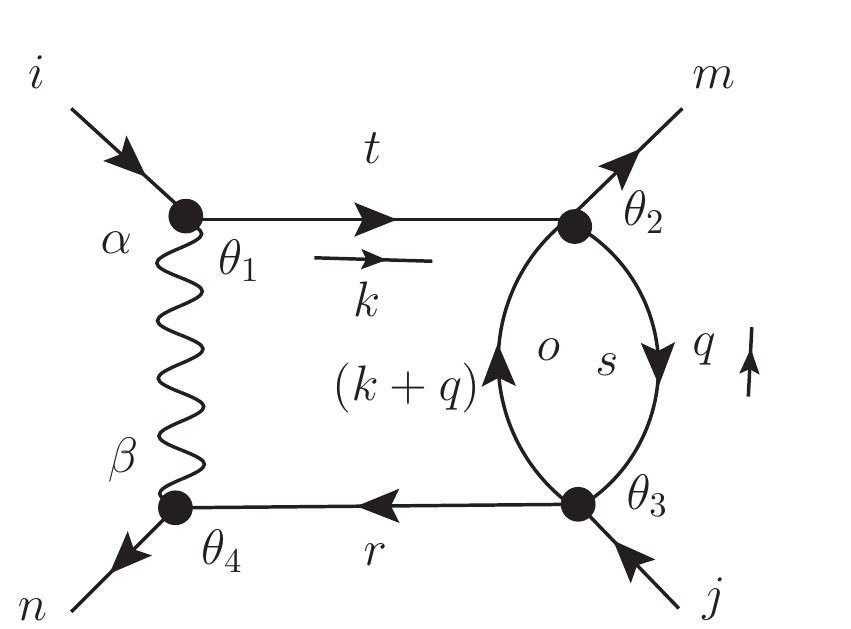}}\subfloat[]{\centering{}\includegraphics[scale=0.5]{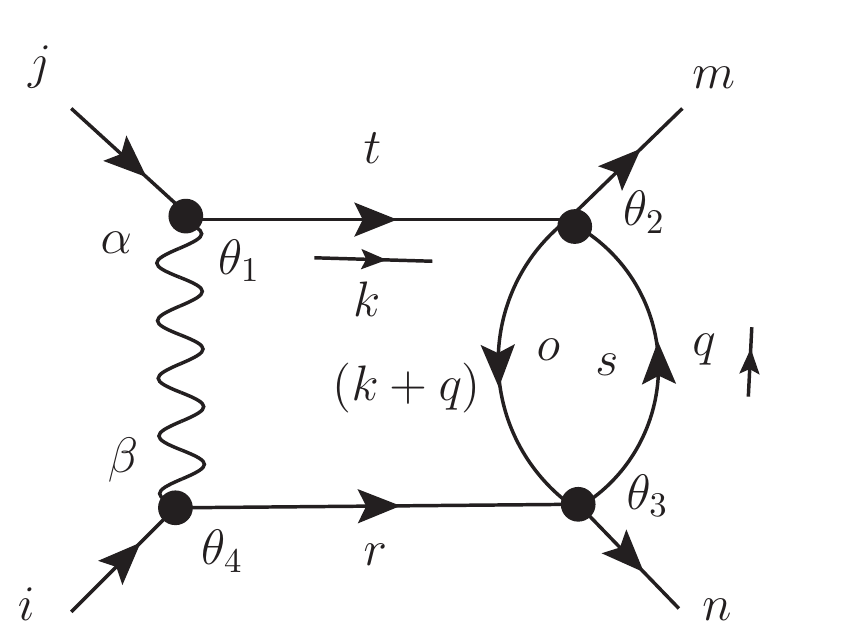}}\subfloat[]{\centering{}\includegraphics[scale=0.5]{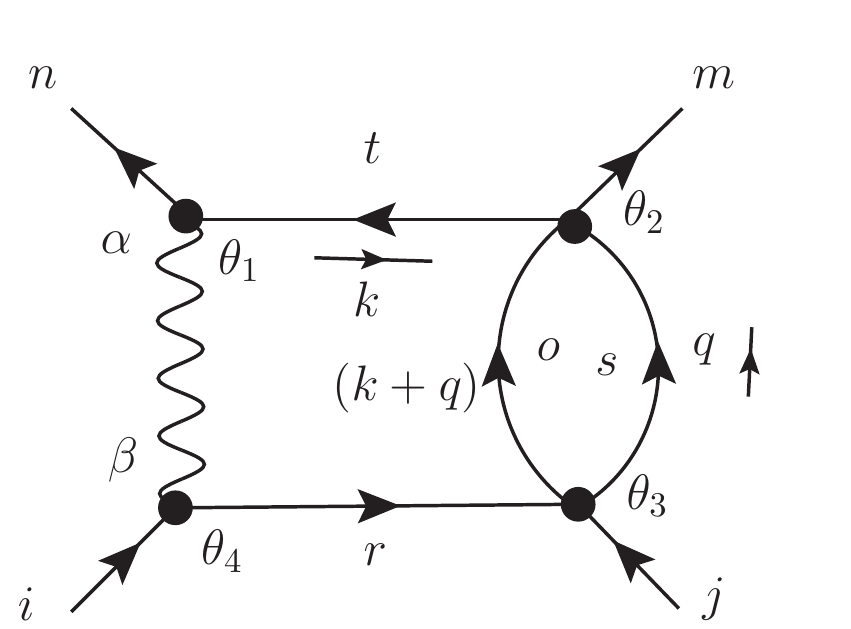}}
\par\end{centering}
\begin{centering}
\subfloat[]{\centering{}\includegraphics[scale=0.5]{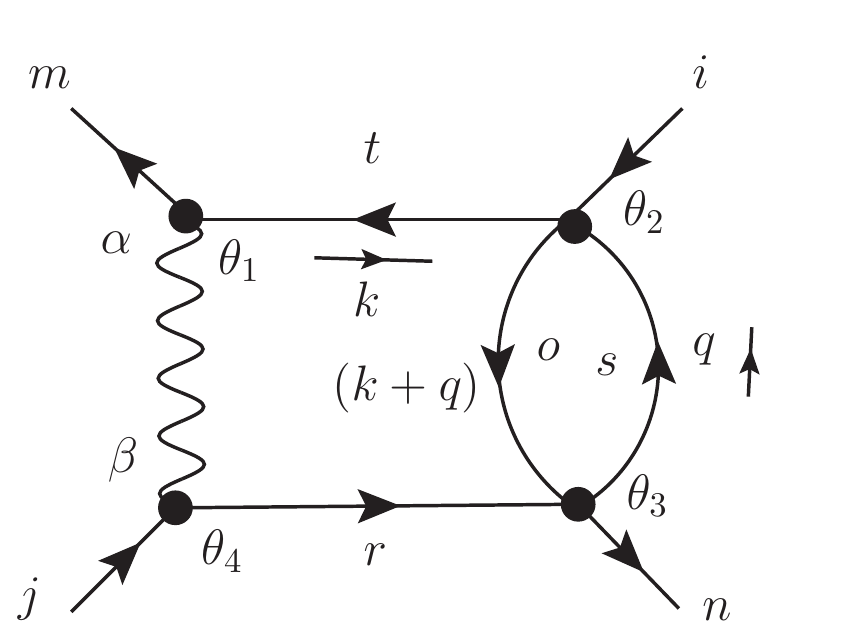}}\subfloat[]{\centering{}\includegraphics[scale=0.5]{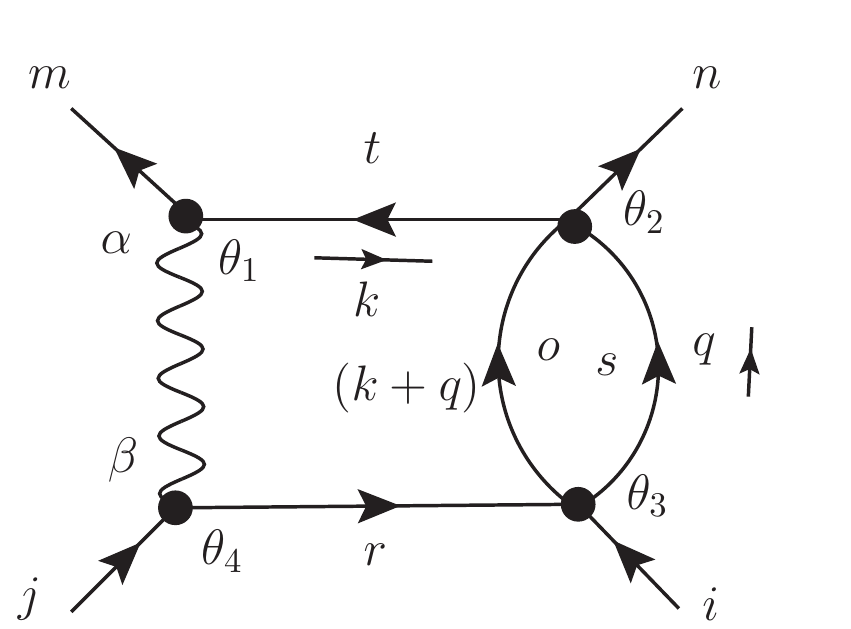}}\subfloat[]{\centering{}\includegraphics[scale=0.5]{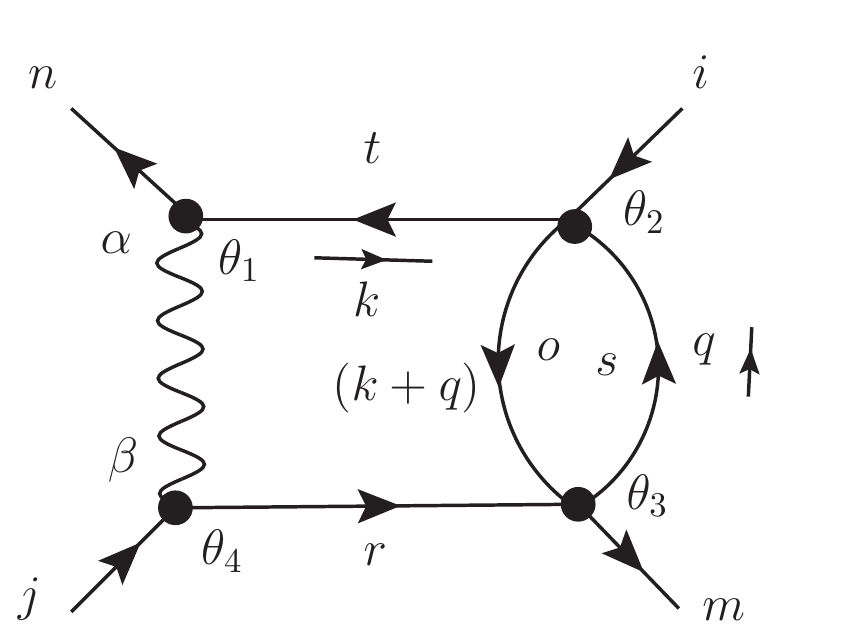}}
\par\end{centering}
\centering{}\subfloat[]{\centering{}\includegraphics[scale=0.5]{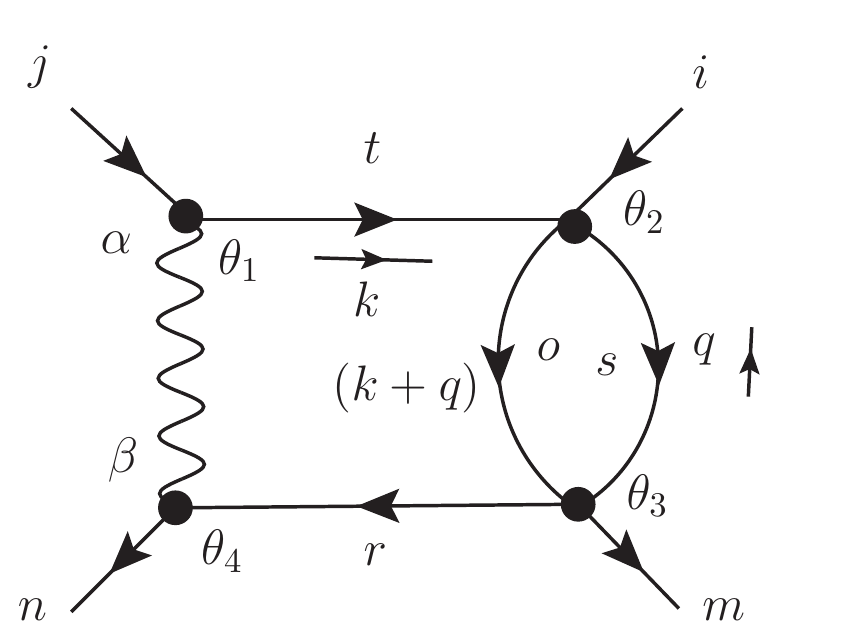}}\subfloat[]{\centering{}\includegraphics[scale=0.5]{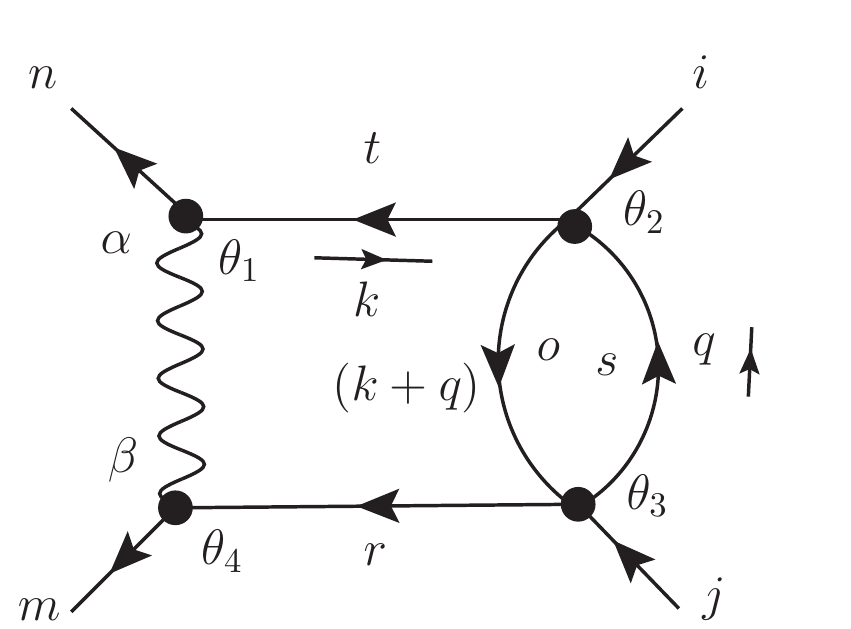}}\subfloat[]{\centering{}\includegraphics[scale=0.5]{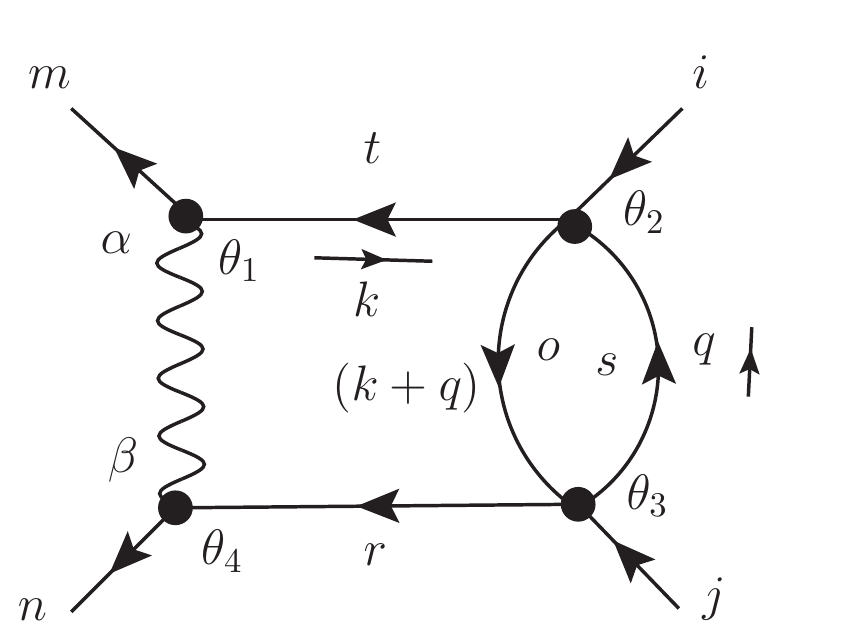}}\caption{\label{fig:D5-order-lambda-2-g-2}$\mathcal{S}_{\left(\overline{\Phi}\Phi\right)^{2}}^{\left(D5\right)}$}
\end{figure}
\par\end{center}

\begin{center}
\begin{table}
\centering{}%
\begin{tabular}{lcccccc}
 &  &  &  &  &  & \tabularnewline
\hline 
\hline 
$D5-a$ &  & $\delta_{in}\delta_{mj}+\left(N+2\right)\delta_{im}\delta_{nj}$ &  & $D5-b$ &  & $-\left(\delta_{in}\delta_{jm}+\delta_{im}\delta_{jn}\right)$\tabularnewline
$D5-c$ &  & $-\left(\delta_{mi}\delta_{jn}+\left(N+2\right)\delta_{mj}\delta_{ni}\right)$ &  & $D5-d$ &  & $-\left(\left(N+2\right)\delta_{im}\delta_{nj}+\delta_{in}\delta_{mj}\right)$\tabularnewline
$D5-e$ &  & $\left(N+2\right)\delta_{jm}\delta_{in}+\delta_{jn}\delta_{mi}$ &  & $D5-f$ &  & $-\left(\delta_{ni}\delta_{jm}+\delta_{nj}\delta_{mi}\right)$\tabularnewline
$D5-g$ &  & $-\left(\delta_{mj}\delta_{in}+\left(N+2\right)\,\delta_{mi}\delta_{jn}\right)$ &  & $D5-h$ &  & $-\left(\delta_{mj}\delta_{in}+\delta_{mi}\delta_{jn}\right)$\tabularnewline
$D5-i$ &  & $-\left(\left(N+2\right)\delta_{ni}\delta_{jm}+\delta_{im}\delta_{jn}\right)$ &  & $D5-j$ &  & $-\left(\delta_{in}\delta_{jm}+\delta_{nj}\delta_{im}\right)$\tabularnewline
$D5-k$ &  & $\left(N+2\right)\delta_{ni}\delta_{mj}+\delta_{im}\delta_{jn}$ &  & $D5-l$ &  & $\left(N+2\right)\delta_{mi}\delta_{nj}+\delta_{in}\delta_{jm}$\tabularnewline
\hline 
\hline 
 &  &  &  &  &  & \tabularnewline
\end{tabular}\caption{\label{tab:S-PPPP-5}Values of the diagrams in Figure\,\ref{fig:D5-order-lambda-2-g-2}
with common factor\protect \\
 $\frac{\left(a-b\right)}{16\left(32\pi^{2}\epsilon\right)}\,i\,\lambda^{2}\,g^{2}\int_{\theta}\overline{\Phi}_{i}\Phi_{m}\Phi_{n}\overline{\Phi}_{j}$
.}
\end{table}
\par\end{center}

$\mathcal{S}_{\left(\overline{\Phi}\Phi\right)^{2}}^{\left(D5-a\right)}$
in the Figure\,\ref{fig:D3-order-lambda-2-g-2} is
\begin{align}
\mathcal{S}_{\left(\overline{\Phi}\Phi\right)^{2}}^{\left(D5-a\right)} & =-\frac{1}{32}\,i\,\lambda^{2}\,g^{2}\,\left[\delta_{in}\delta_{mj}+\left(N+2\right)\delta_{im}\delta_{nj}\right]\int_{\theta}\overline{\Phi}_{i}\Phi_{m}\Phi_{n}\overline{\Phi}_{j}\nonumber \\
 & \times\int\frac{d^{D}kd^{D}q}{\left(2\pi\right)^{2D}}\left\{ \frac{2\left(a-b\right)\left(k^{2}\right)^{2}}{\left(k^{2}\right)^{3}q^{2}\left(q+k\right)^{2}}\right\} \,,
\end{align}
using Eq.\,(\ref{eq:Int 7}) and adding $\mathcal{S}_{\left(\overline{\Phi}\Phi\right)^{2}}^{\left(D5-a\right)}$
to $\mathcal{S}_{\left(\overline{\Phi}\Phi\right)^{2}}^{\left(D5-l\right)}$
with the values in the Table\,\ref{tab:S-PPPP-5}, we find 
\begin{align}
\mathcal{S}_{\left(\overline{\Phi}\Phi\right)^{2}}^{\left(D5\right)} & =-\frac{\left(a-b\right)}{4\left(32\pi^{2}\epsilon\right)}\,i\,\lambda^{2}\,g^{2}\left(\delta_{in}\delta_{jm}+\delta_{mi}\delta_{nj}\right)\int_{\theta}\overline{\Phi}_{i}\Phi_{m}\Phi_{n}\overline{\Phi}_{j}\,.\label{eq:S-D5}
\end{align}

\begin{center}
\begin{figure}
\begin{centering}
\subfloat[]{\begin{centering}
\includegraphics[scale=0.5]{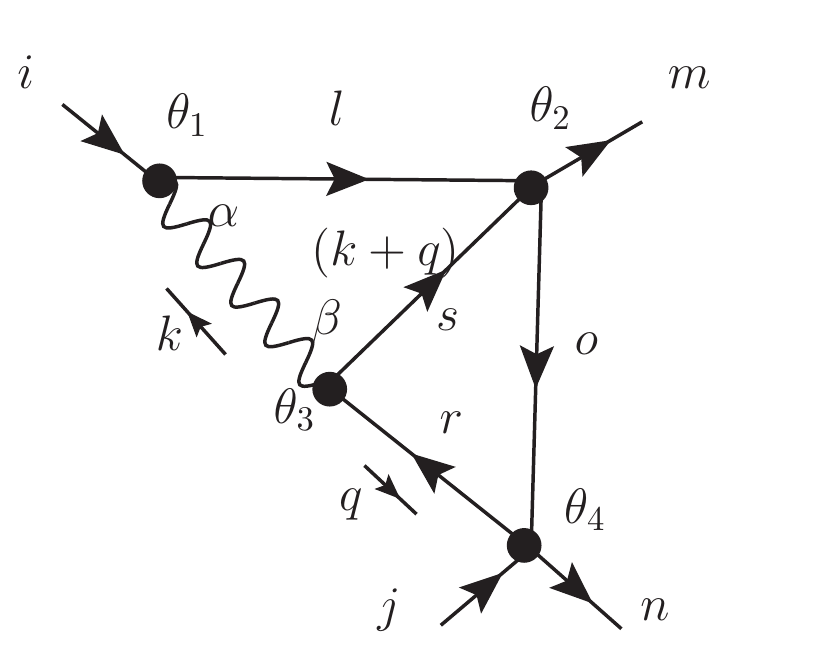}
\par\end{centering}
}\subfloat[]{\centering{}\includegraphics[scale=0.5]{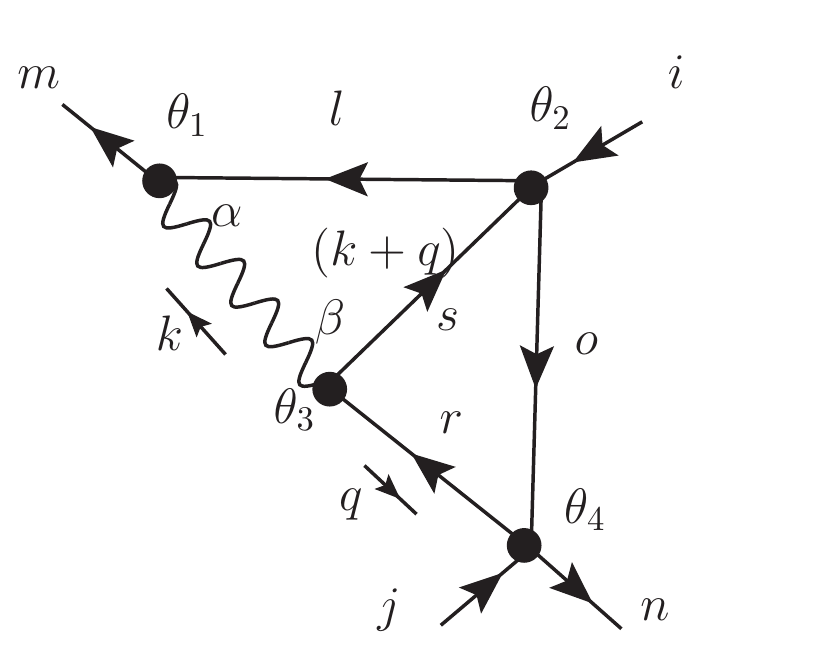}}\subfloat[]{\centering{}\includegraphics[scale=0.5]{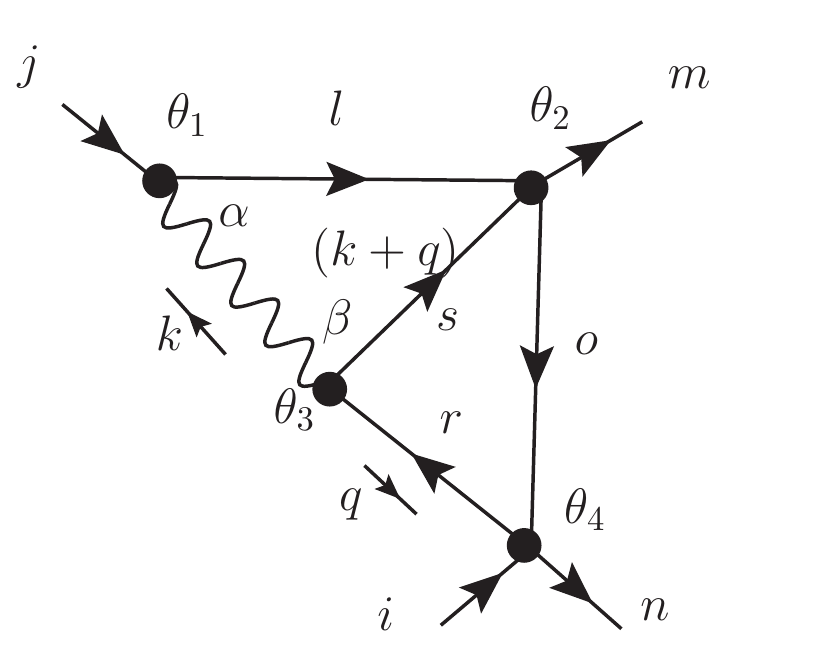}}
\par\end{centering}
\begin{centering}
\subfloat[]{\centering{}\includegraphics[scale=0.5]{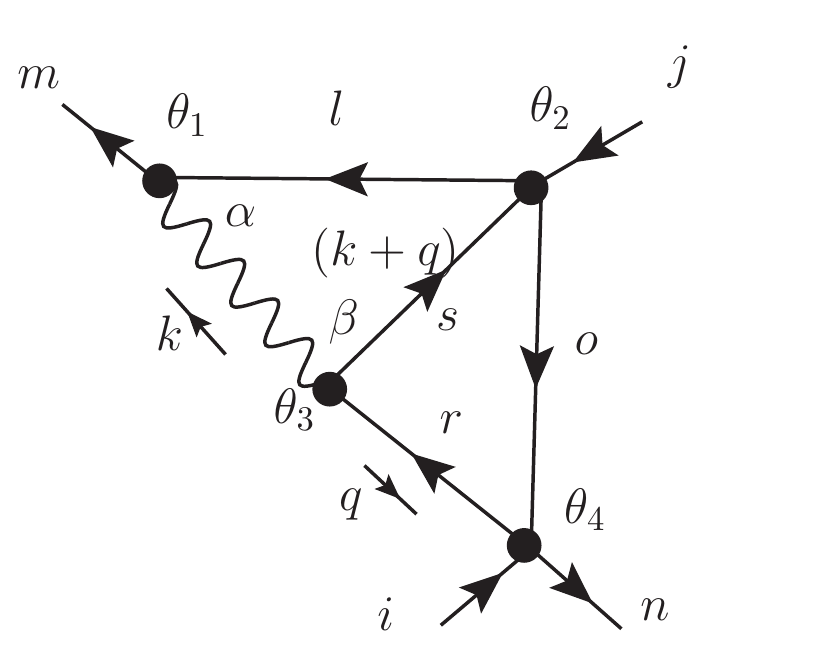}}\subfloat[]{\centering{}\includegraphics[scale=0.5]{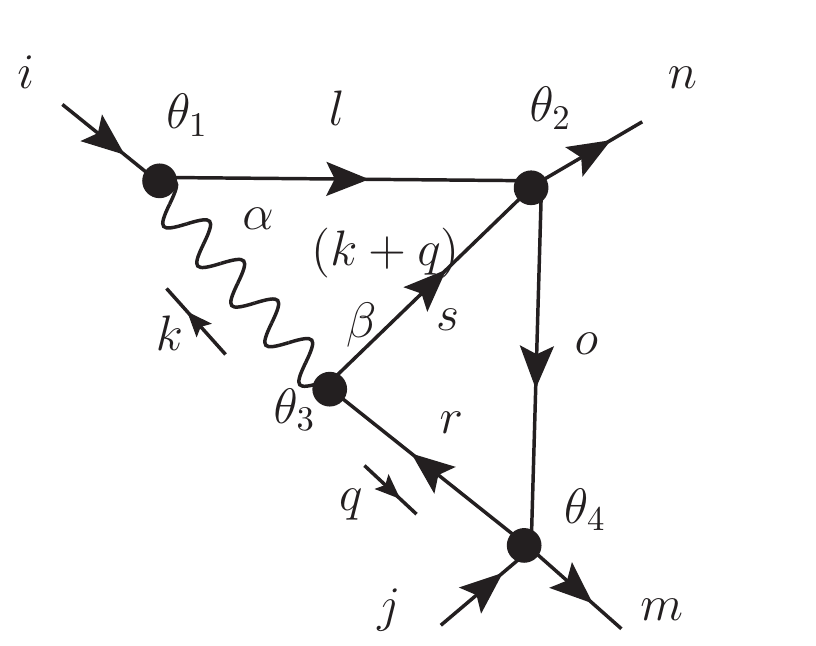}}\subfloat[]{\centering{}\includegraphics[scale=0.5]{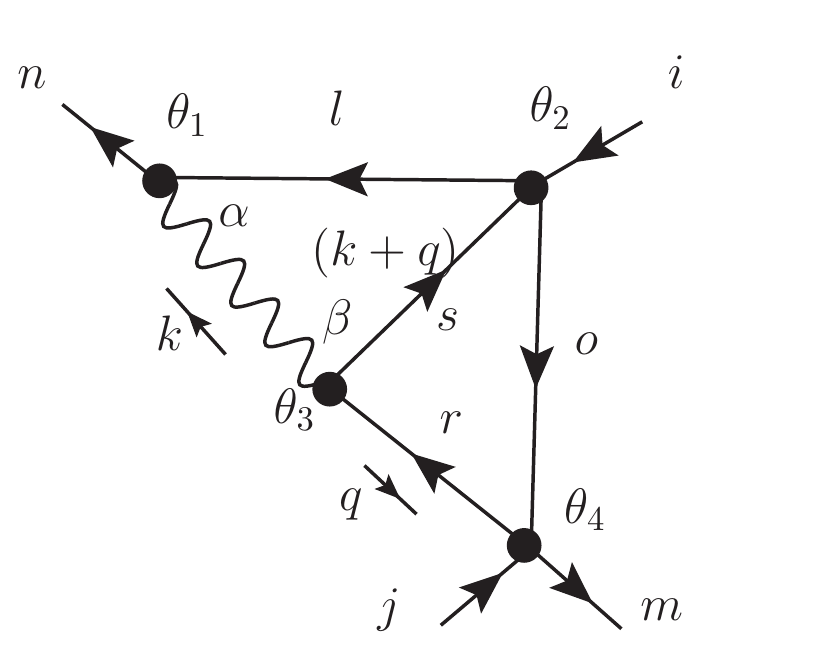}}
\par\end{centering}
\begin{centering}
\subfloat[]{\centering{}\includegraphics[scale=0.5]{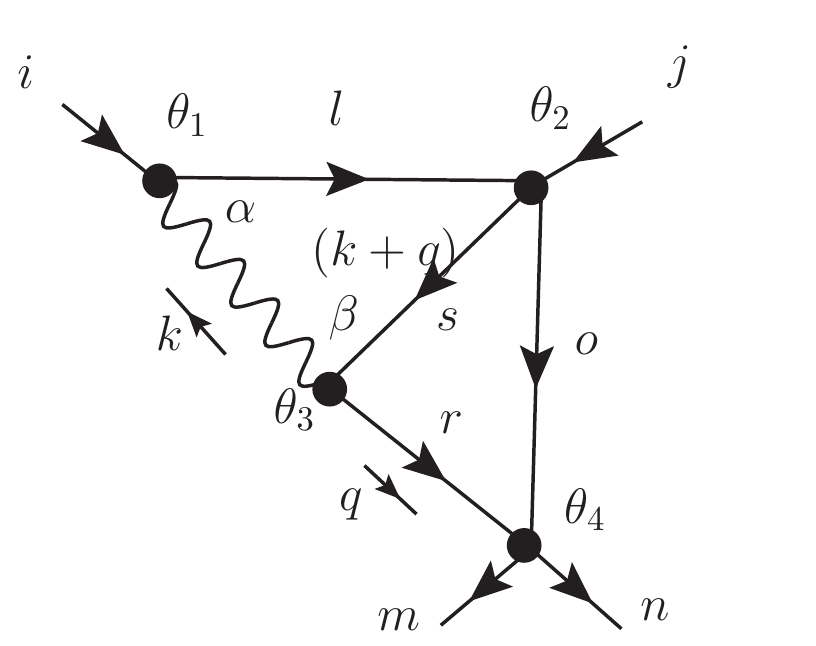}}\subfloat[]{\centering{}\includegraphics[scale=0.5]{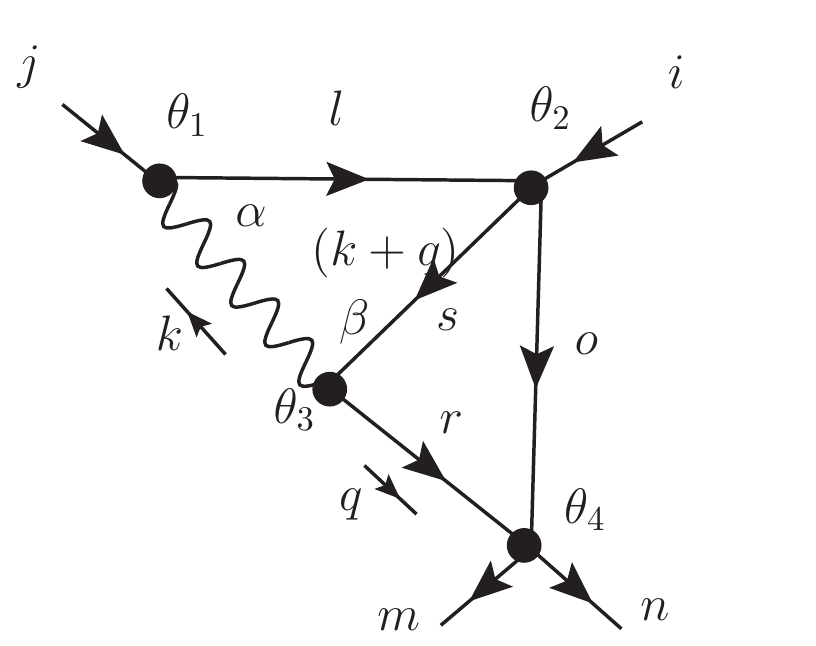}}\subfloat[]{\centering{}\includegraphics[scale=0.5]{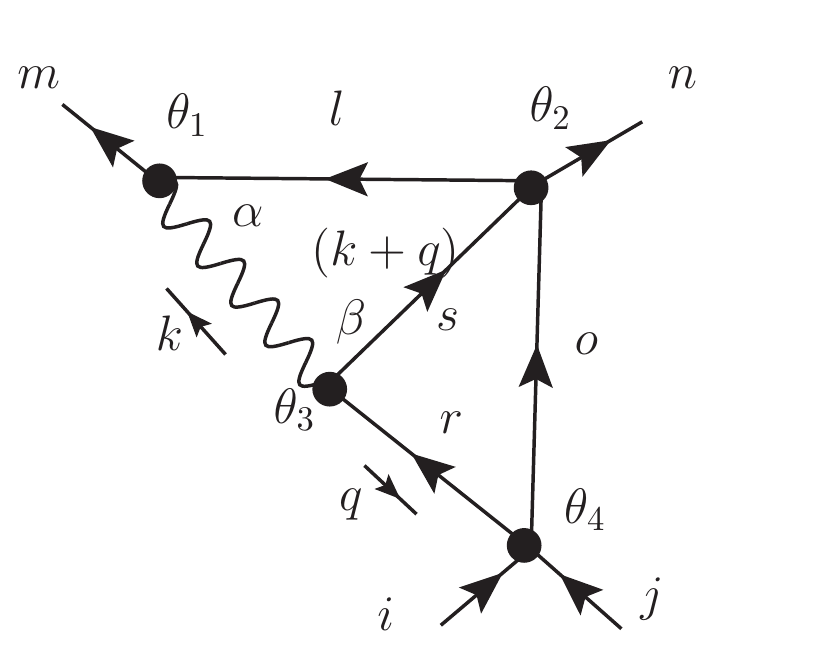}}
\par\end{centering}
\centering{}\subfloat[]{\centering{}\includegraphics[scale=0.5]{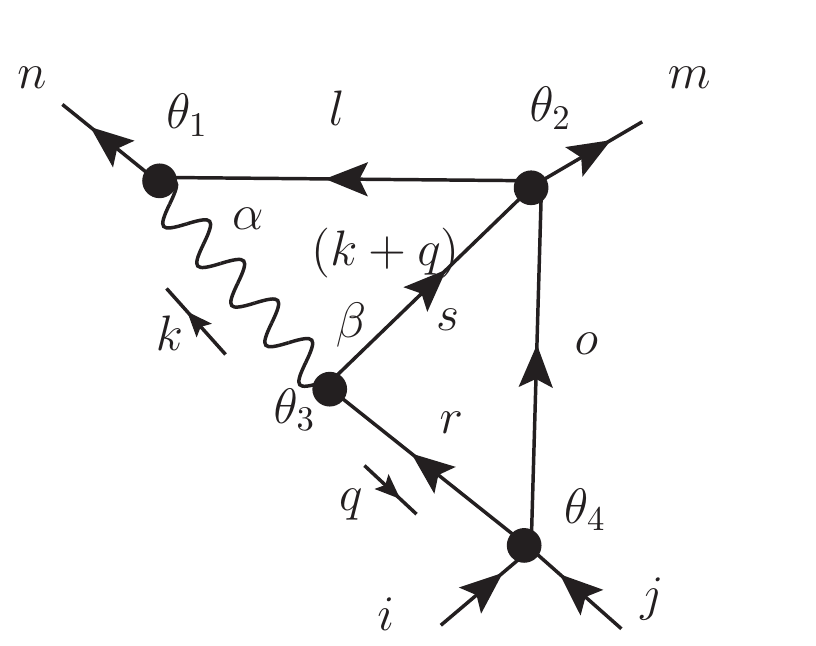}}\subfloat[]{\centering{}\includegraphics[scale=0.5]{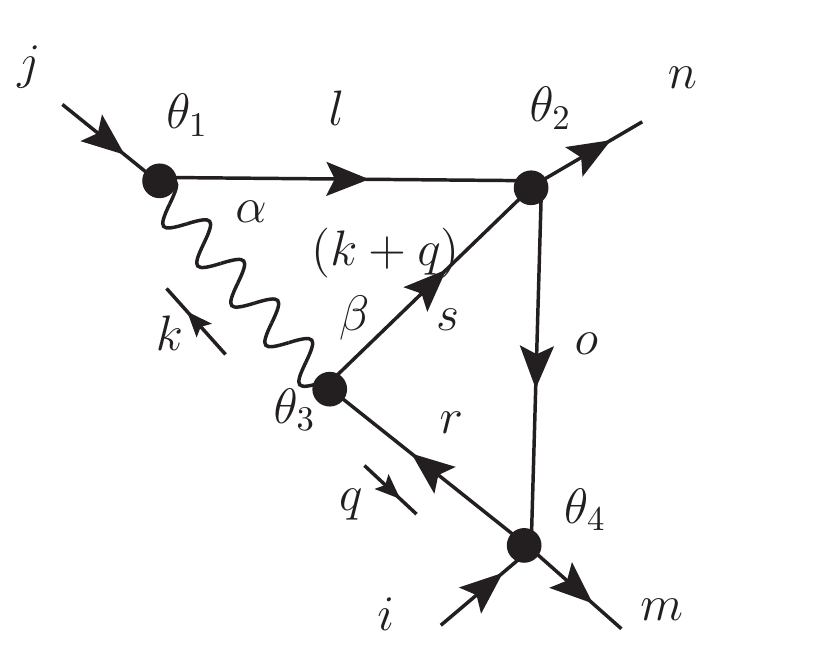}}\subfloat[]{\centering{}\includegraphics[scale=0.5]{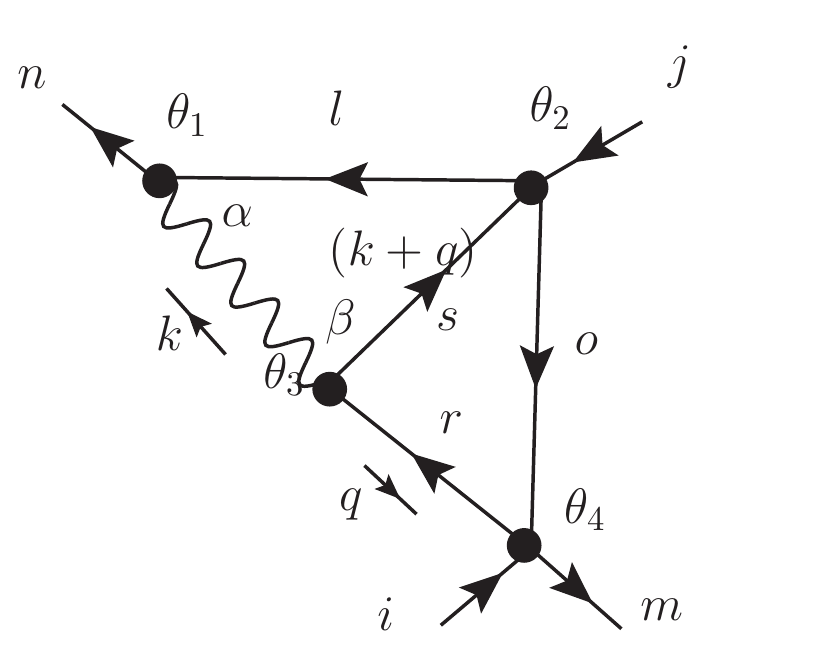}}\caption{\label{fig:D6-order-lambda-2-g-2}$\mathcal{S}_{\left(\overline{\Phi}\Phi\right)^{2}}^{\left(D6\right)}$}
\end{figure}
\par\end{center}

\begin{center}
\begin{table}
\centering{}%
\begin{tabular}{lcccccc}
 &  &  &  &  &  & \tabularnewline
\hline 
\hline 
$D6-a$ &  & $-\left(\left(N+2\right)\delta_{im}\delta_{jn}+\delta_{mj}\delta_{ni}\right)$ &  & $D6-b$ &  & $\left(N+2\right)\delta_{im}\delta_{jn}+\delta_{mj}\delta_{ni}$\tabularnewline
$D6-c$ &  & $-\left(\delta_{jn}\delta_{im}+\left(N+2\right)\delta_{jm}\delta_{in}\right)$ &  & $D6-d$ &  & $\delta_{jn}\delta_{im}+\left(N+2\right)\delta_{jm}\delta_{in}$\tabularnewline
$D6-e$ &  & $-\left(\delta_{jn}\delta_{im}+\left(N+2\right)\delta_{jm}\delta_{in}\right)$ &  & $D6-f$ &  & $\delta_{jn}\delta_{im}+\left(N+2\right)\delta_{jm}\delta_{in}$\tabularnewline
$D6-g$ &  & $2\left(\delta_{jm}\delta_{in}+\delta_{im}\delta_{jn}\right)$ &  & $D6-h$ &  & $2\left(\delta_{jm}\delta_{in}+\delta_{im}\delta_{jn}\right)$\tabularnewline
$D6-i$ &  & $2\left(\delta_{jm}\delta_{in}+\delta_{im}\delta_{jn}\right)$ &  & $D6-j$ &  & $2\left(\delta_{jm}\delta_{in}+\delta_{im}\delta_{jn}\right)$\tabularnewline
$D6-k$ &  & $-\left(\left(N+2\right)\delta_{jn}\delta_{im}+\delta_{in}\delta_{jm}\right)$ &  & $D6-l$ &  & $\left(N+2\right)\delta_{jn}\delta_{im}+\delta_{in}\delta_{jm}$\tabularnewline
\hline 
\hline 
 &  &  &  &  &  & \tabularnewline
\end{tabular}\caption{\label{tab:S-PPPP-6}Values of the diagrams in Figure\,\ref{fig:D6-order-lambda-2-g-2}
with common factor\protect \\
 $\frac{\left(a-b\right)}{16\left(32\pi^{2}\epsilon\right)}\,i\,\lambda^{2}\,g^{2}\int_{\theta}\overline{\Phi}_{i}\Phi_{m}\Phi_{n}\overline{\Phi}_{j}$
.}
\end{table}
\par\end{center}

Finally, $\mathcal{S}_{\left(\overline{\Phi}\Phi\right)^{2}}^{\left(D6-a\right)}$
in the Figure\,\ref{fig:D6-order-lambda-2-g-2} is
\begin{align}
\mathcal{S}_{\left(\overline{\Phi}\Phi\right)^{2}}^{\left(D6-a\right)} & =\frac{1}{32}\,i\,\lambda^{2}\,g^{2}\left[\left(N+2\right)\delta_{im}\delta_{jn}+\delta_{mj}\delta_{ni}\right]\int_{\theta}\overline{\Phi}_{i}\Phi_{m}\Phi_{n}\overline{\Phi}_{j}\nonumber \\
 & \int\frac{d^{D}kd^{D}q}{\left(2\pi\right)^{2D}}\,\left\{ \frac{2\left(a-b\right)k^{2}q^{2}}{\left(k^{2}\right)^{2}\left(k+q\right)^{2}\left(q^{2}\right)^{2}}\right\} \,,
\end{align}
using Eq.\,(\ref{eq:Int 7}) and adding the diagrams $\mathcal{S}_{\left(\overline{\Phi}\Phi\right)^{2}}^{\left(D6-a\right)}$
to $\mathcal{S}_{\left(\overline{\Phi}\Phi\right)^{2}}^{\left(D6-l\right)}$
with the values in Table\,\ref{tab:S-PPPP-6}, we find 
\begin{align}
\mathcal{S}_{\left(\overline{\Phi}\Phi\right)^{2}}^{\left(D6\right)} & =\frac{\left(a-b\right)}{2\left(32\pi^{2}\epsilon\right)}\,i\,\lambda^{2}\,g^{2}\left(\delta_{jn}\delta_{im}+\delta_{in}\delta_{jm}\right)\int_{\theta}\overline{\Phi}_{i}\Phi_{m}\Phi_{n}\overline{\Phi}_{j}\,.\label{eq:S-D6}
\end{align}

Now, we will consider all the diagrams that contribute with the order
$\mathcal{O}\left(\lambda g^{4}\right)$, in this order there are
$14$ topologies that are equivalent to $139$ diagrams. 
\begin{center}
\begin{figure}
\begin{centering}
\subfloat[]{\begin{centering}
\includegraphics[scale=0.5]{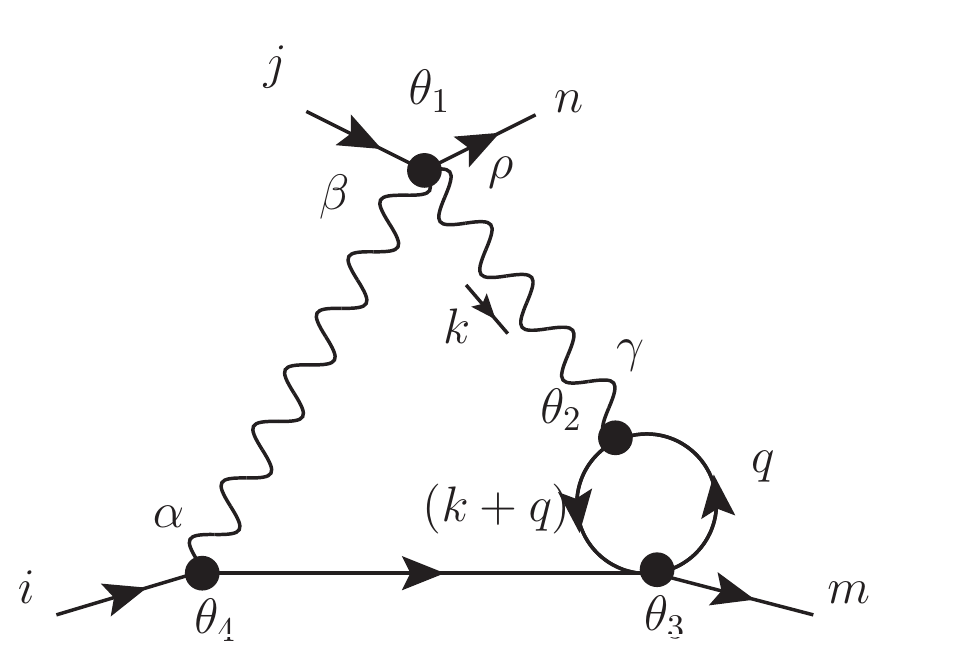}
\par\end{centering}
}\subfloat[]{\centering{}\includegraphics[scale=0.5]{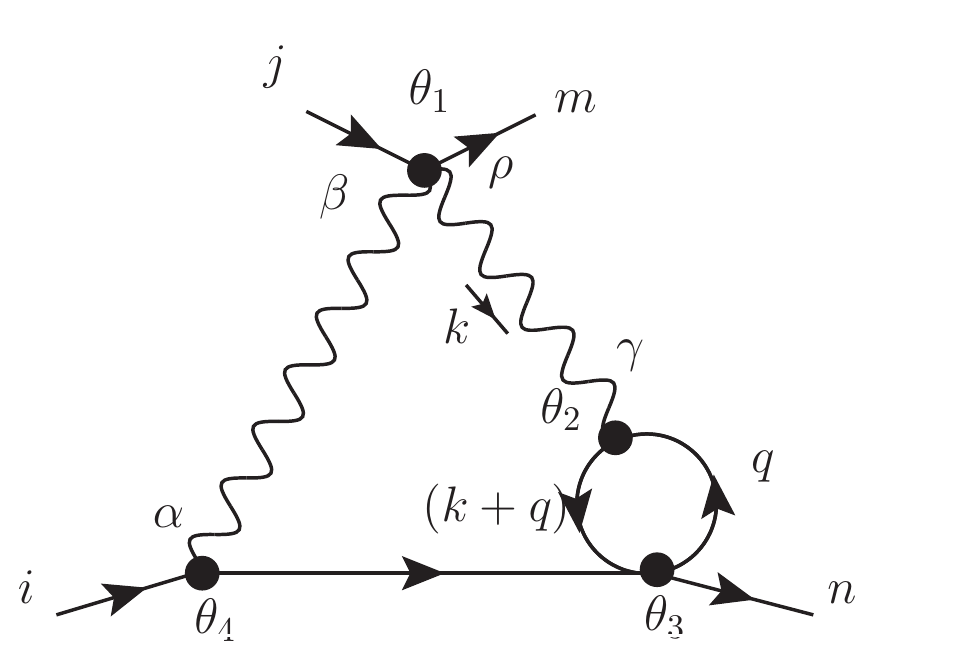}}\subfloat[]{\centering{}\includegraphics[scale=0.5]{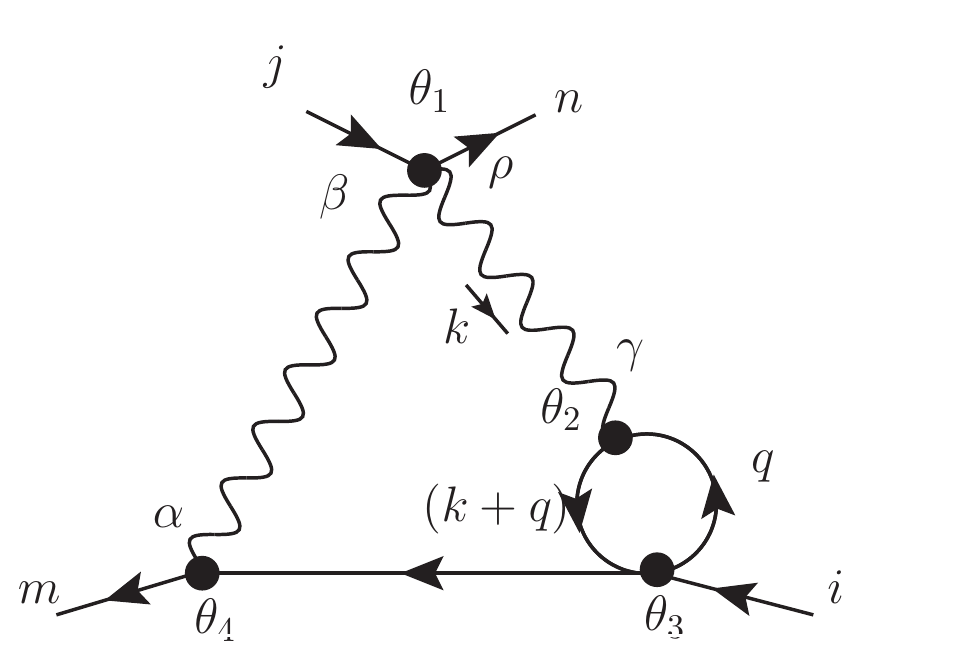}}
\par\end{centering}
\begin{centering}
\subfloat[]{\centering{}\includegraphics[scale=0.5]{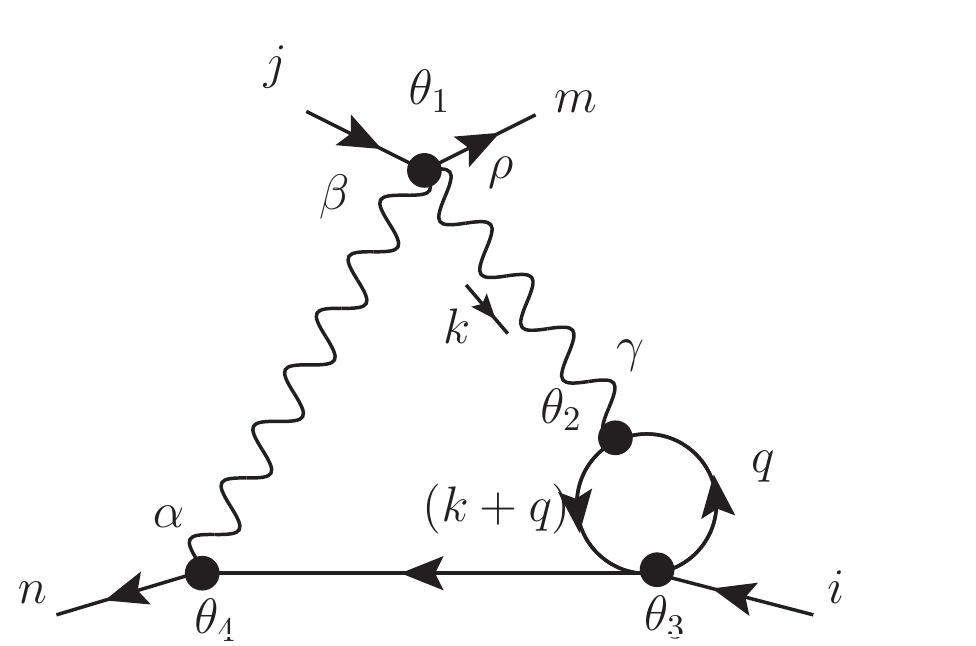}}\subfloat[]{\centering{}\includegraphics[scale=0.5]{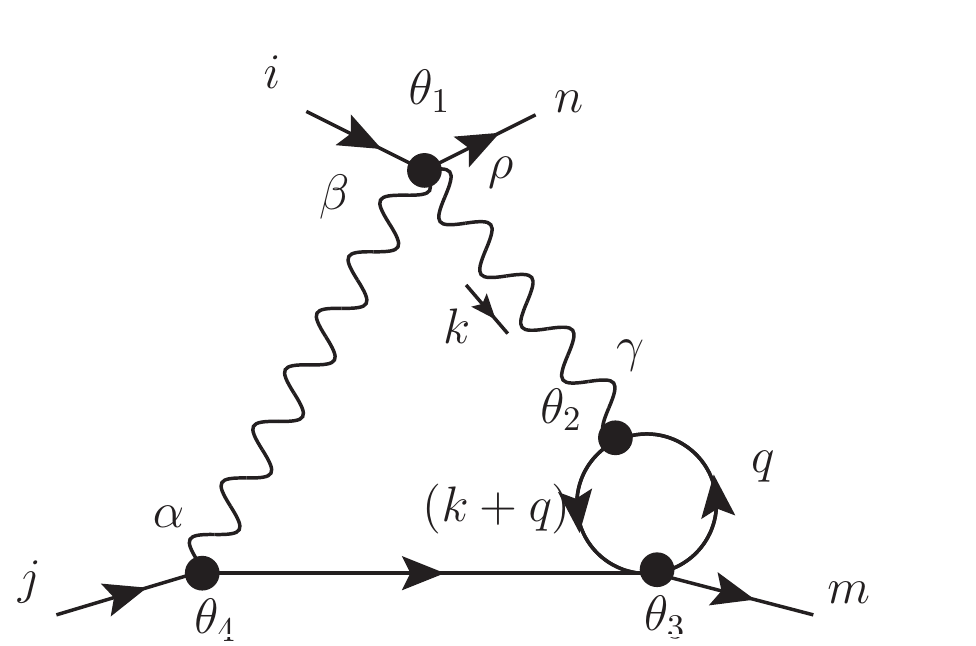}}\subfloat[]{\centering{}\includegraphics[scale=0.5]{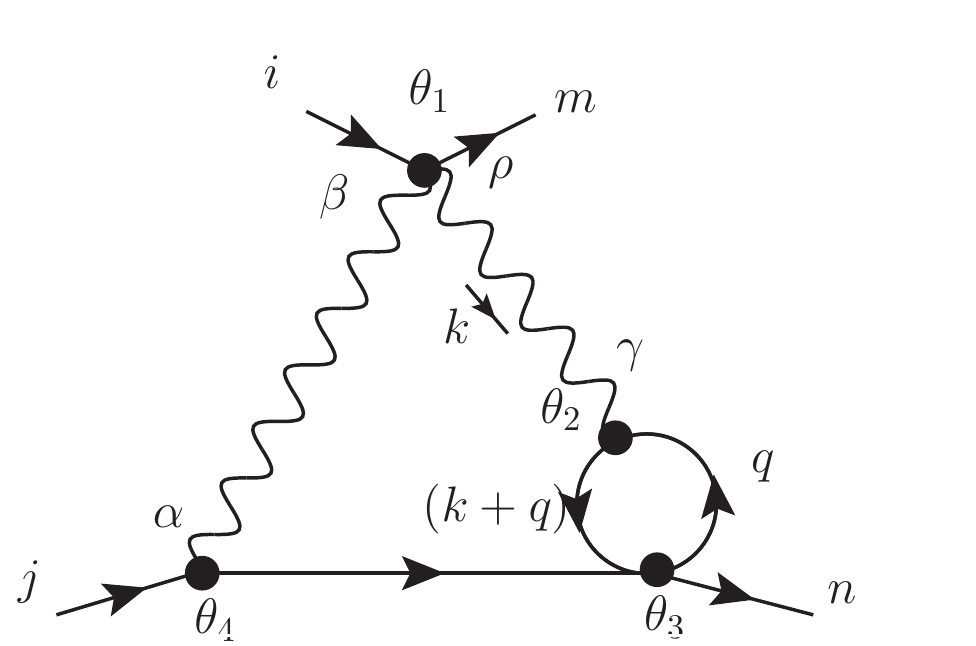}}
\par\end{centering}
\begin{centering}
\subfloat[]{\centering{}\includegraphics[scale=0.5]{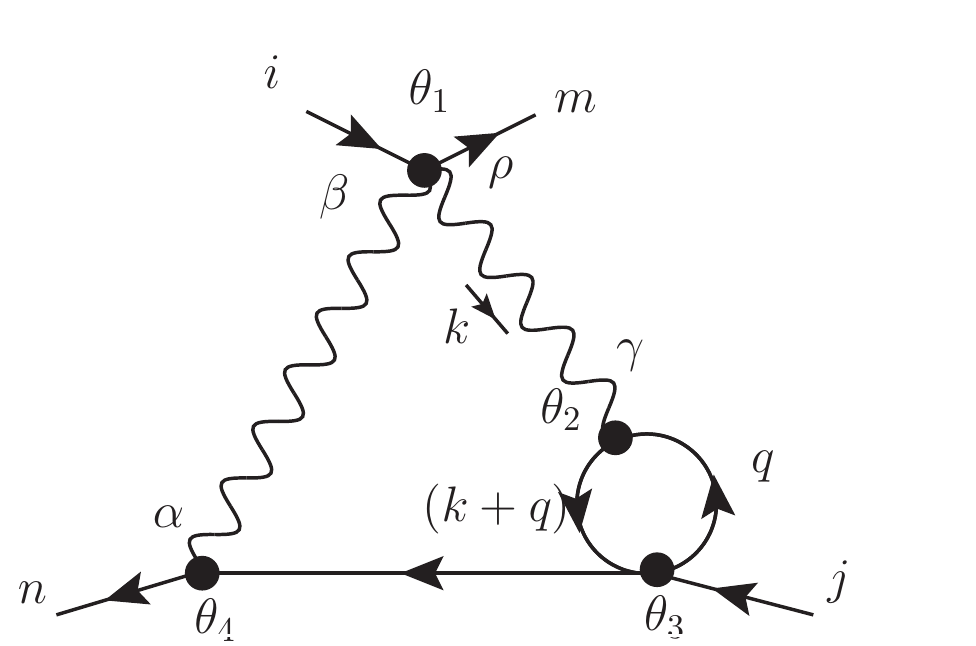}}\subfloat[]{\centering{}\includegraphics[scale=0.5]{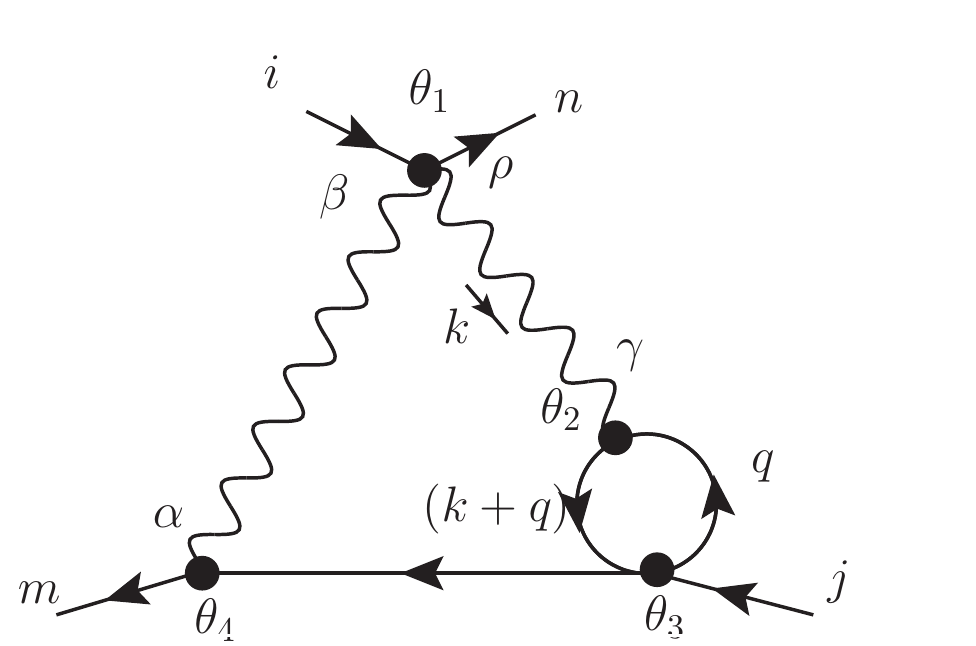}}
\par\end{centering}
\centering{}\caption{\label{fig:D7-order-lambda-g-4}$\mathcal{S}_{\left(\overline{\Phi}\Phi\right)^{2}}^{\left(D7\right)}$}
\end{figure}
\par\end{center}

We start with $\mathcal{S}_{\left(\overline{\Phi}\Phi\right)^{2}}^{\left(D7-a\right)}$
in the Figure\,\ref{fig:D7-order-lambda-g-4},
\begin{align}
\mathcal{S}_{\left(\overline{\Phi}\Phi\right)^{2}}^{\left(D7-a\right)} & =-\frac{1}{32}\,i\,\lambda\,g^{4}\left(N+1\right)\delta_{im}\delta_{jn}\int_{\theta}\overline{\Phi}_{i}\Phi_{m}\Phi_{n}\overline{\Phi}_{j}\nonumber \\
 & \times\int\frac{d^{D}kd^{D}q}{\left(2\pi\right)^{2D}}\left\{ \frac{-4\left(a-b\right)^{2}\left(k\cdot q\right)k^{2}-2\left(a-b\right)^{2}\left(k^{2}\right)^{2}}{\left(k^{2}\right)^{3}q^{2}\left(k+q\right)^{2}}\right\} \,,
\end{align}
using Eqs.\,(\ref{eq:Int 7}) and\,(\ref{eq:Int 9}), we find 
\begin{align}
\mathcal{S}_{\left(\overline{\Phi}\Phi\right)^{2}}^{\left(D7-a\right)} & =\mathcal{S}_{\left(\overline{\Phi}\Phi\right)^{2}}^{\left(D7\right)}=0\,.\label{eq:S-D7}
\end{align}

\begin{center}
\begin{figure}
\begin{centering}
\subfloat[]{\begin{centering}
\includegraphics[scale=0.5]{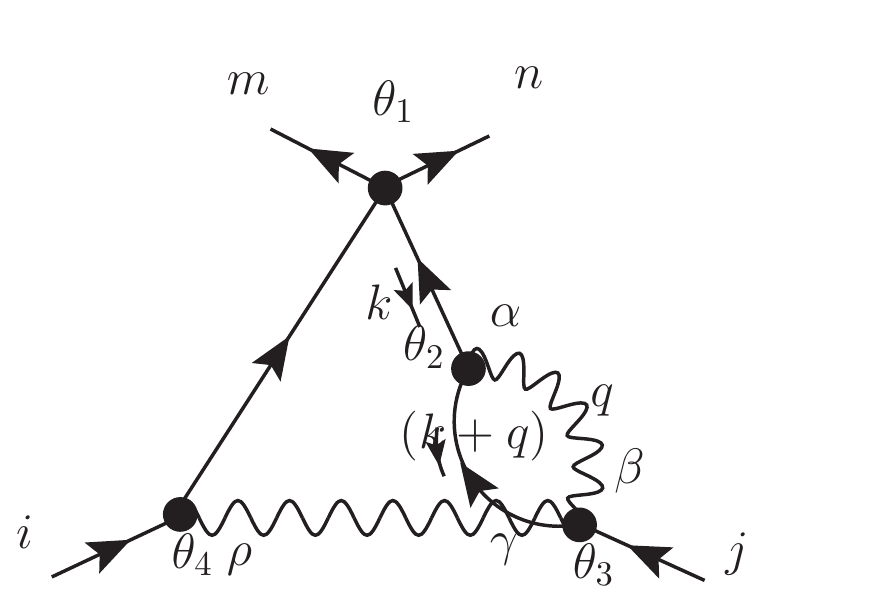}
\par\end{centering}
}\subfloat[]{\centering{}\includegraphics[scale=0.5]{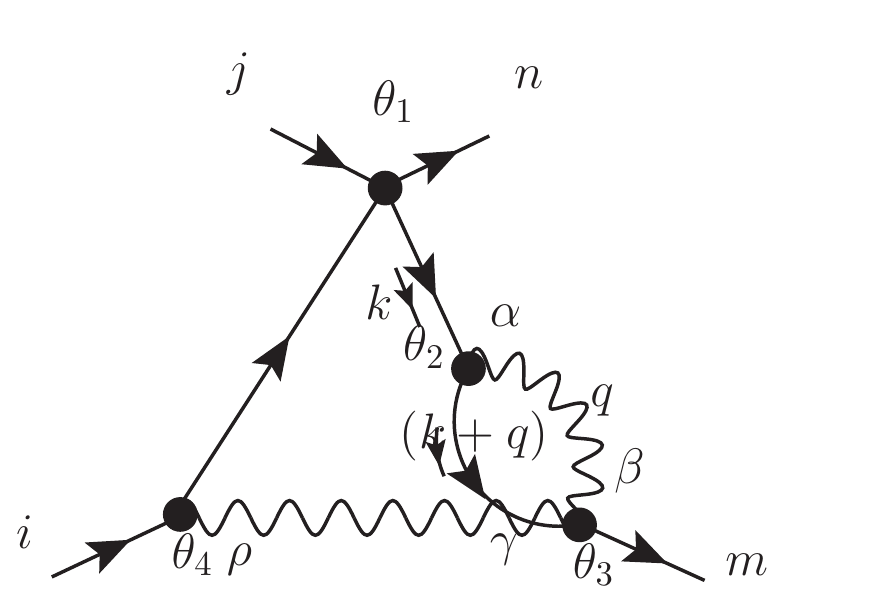}}\subfloat[]{\centering{}\includegraphics[scale=0.5]{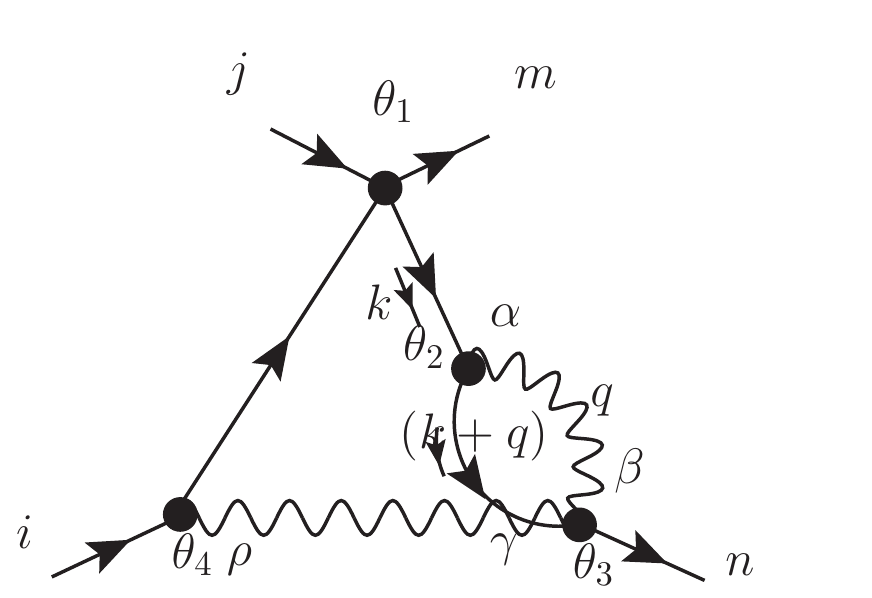}}
\par\end{centering}
\begin{centering}
\subfloat[]{\centering{}\includegraphics[scale=0.5]{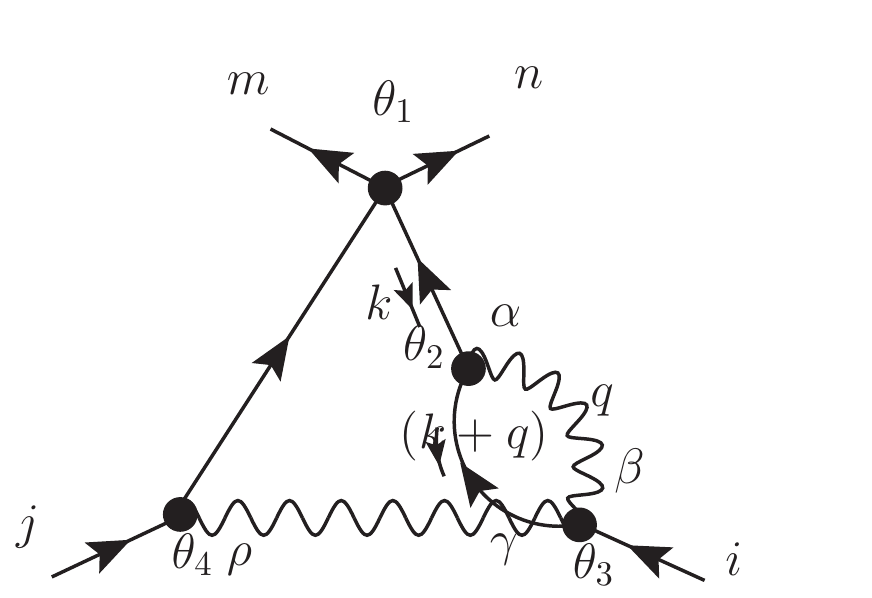}}\subfloat[]{\centering{}\includegraphics[scale=0.5]{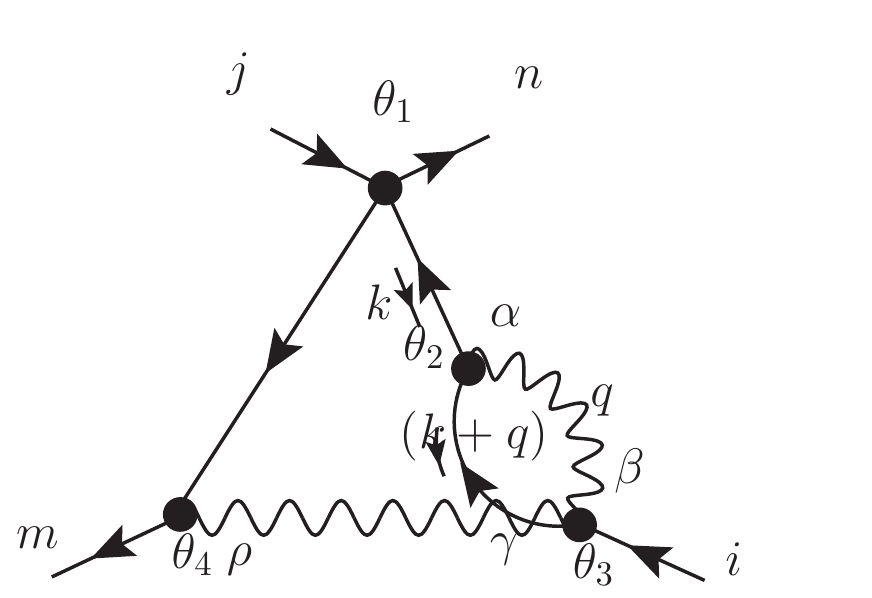}}\subfloat[]{\centering{}\includegraphics[scale=0.5]{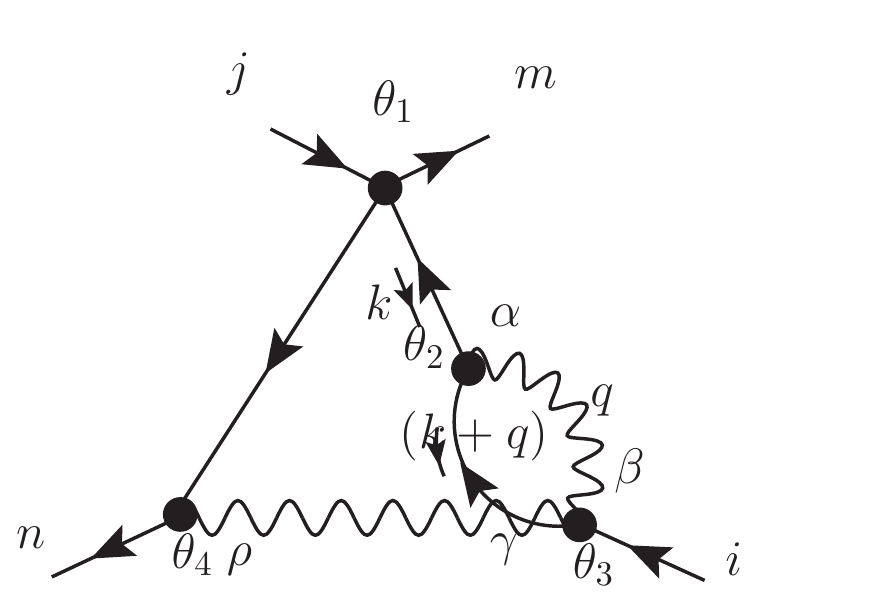}}
\par\end{centering}
\begin{centering}
\subfloat[]{\centering{}\includegraphics[scale=0.5]{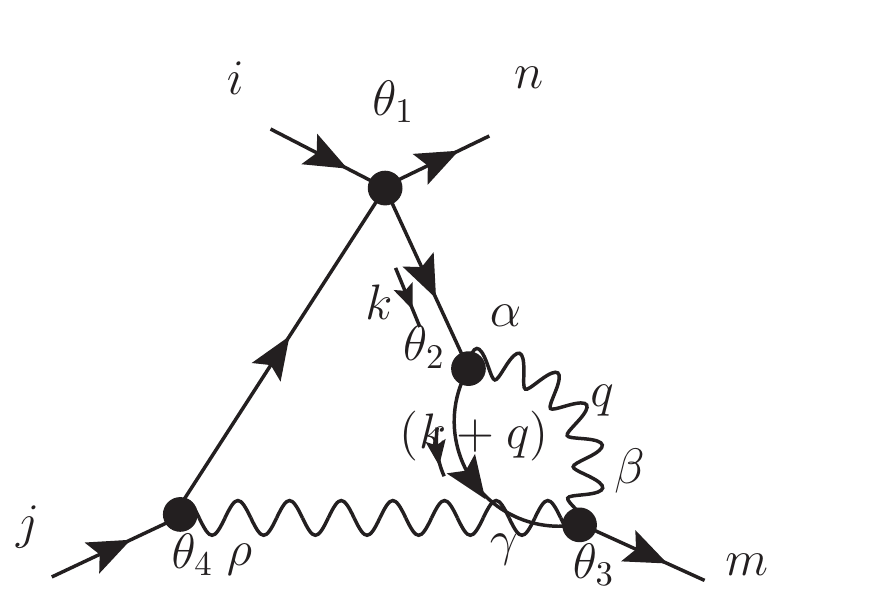}}\subfloat[]{\centering{}\includegraphics[scale=0.5]{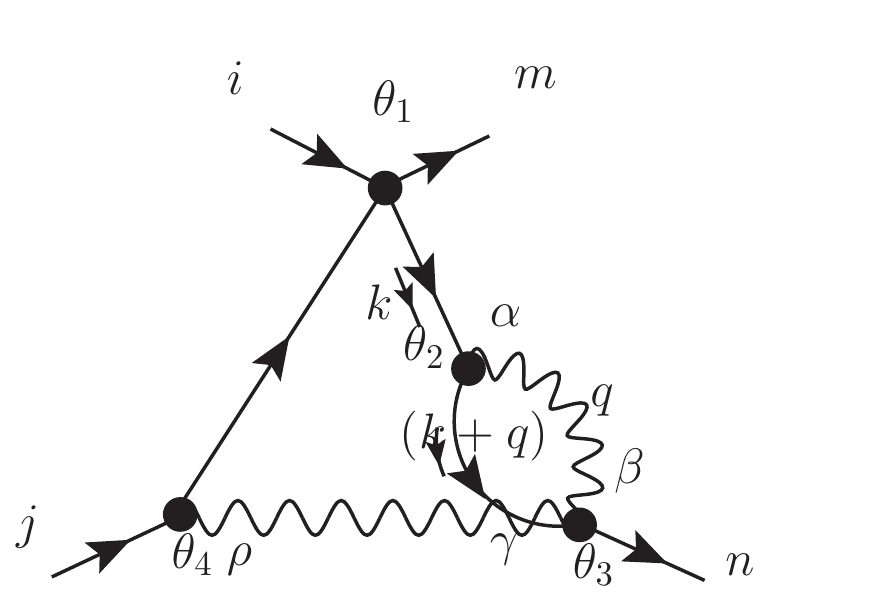}}\subfloat[]{\centering{}\includegraphics[scale=0.5]{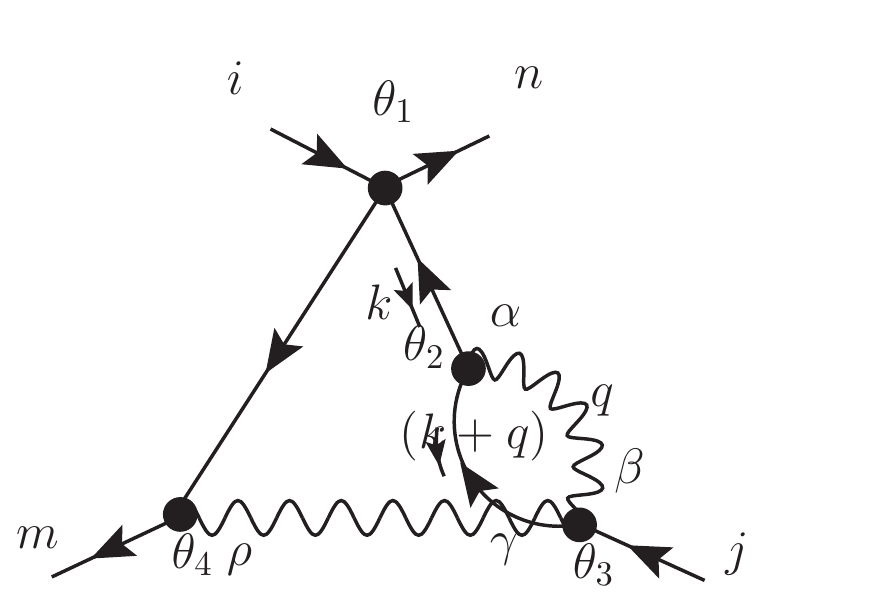}}
\par\end{centering}
\begin{centering}
\subfloat[]{\centering{}\includegraphics[scale=0.5]{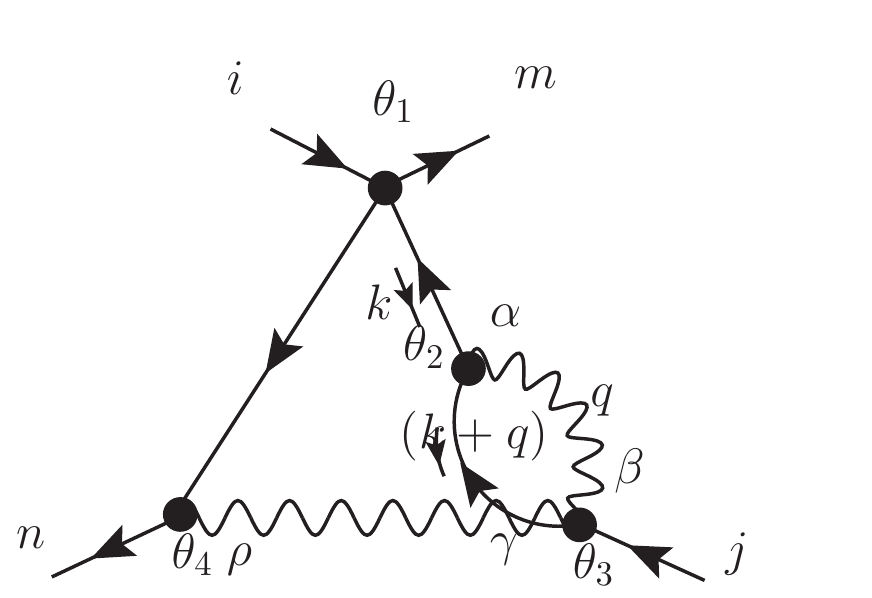}}\subfloat[]{\centering{}\includegraphics[scale=0.5]{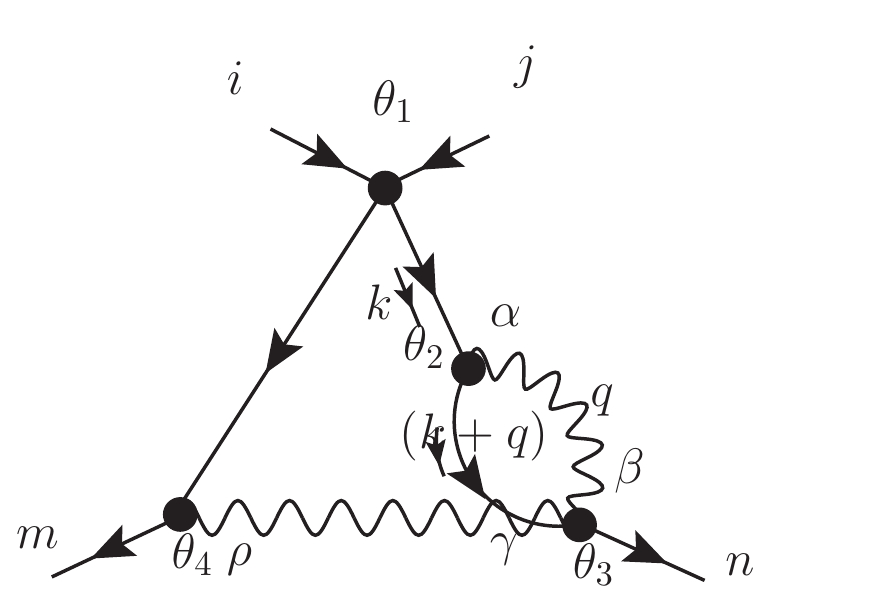}}\subfloat[]{\centering{}\includegraphics[scale=0.5]{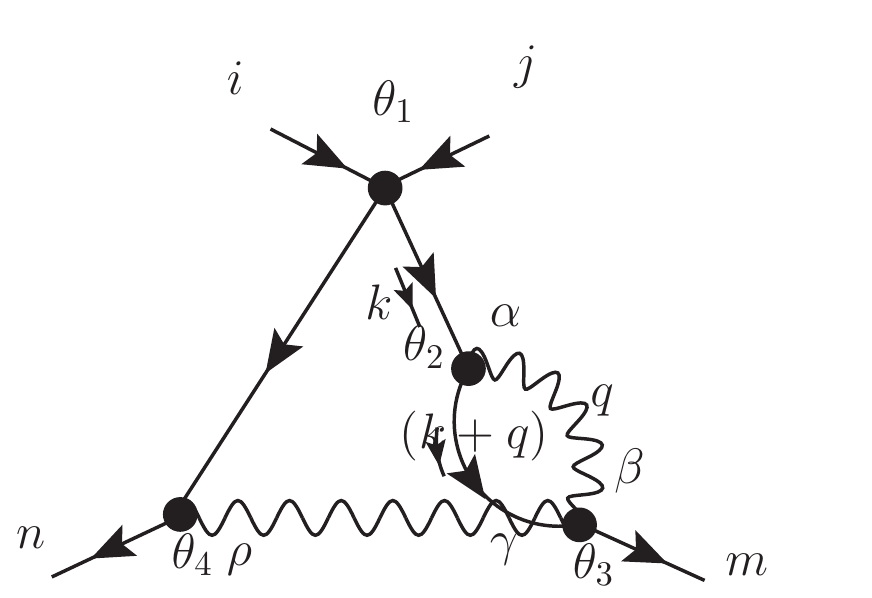}}
\par\end{centering}
\centering{}\caption{\label{fig:D8-order-lambda-g-4}$\mathcal{S}_{\left(\overline{\Phi}\Phi\right)^{2}}^{\left(D8\right)}$}
\end{figure}
\par\end{center}

\begin{center}
\begin{table}
\centering{}%
\begin{tabular}{lcccccc}
 &  &  &  &  &  & \tabularnewline
\hline 
\hline 
$D8-a$ &  & $-\left(\delta_{jn}\delta_{im}+\delta_{mj}\delta_{ni}\right)$ &  & $D8-b$ &  & $\delta_{jn}\delta_{im}+\delta_{mj}\delta_{ni}$\tabularnewline
$D8-c$ &  & $\delta_{jn}\delta_{im}+\delta_{mj}\delta_{ni}$ &  & $D8-d$ &  & $-\left(\delta_{jn}\delta_{im}+\delta_{mj}\delta_{ni}\right)$\tabularnewline
$D8-e$ &  & $\delta_{jn}\delta_{im}+\delta_{mj}\delta_{ni}$ &  & $D8-f$ &  & $\delta_{jn}\delta_{im}+\delta_{mj}\delta_{ni}$\tabularnewline
$D8-g$ &  & $\delta_{jn}\delta_{im}+\delta_{mj}\delta_{ni}$ &  & $D8-h$ &  & $\delta_{jn}\delta_{im}+\delta_{mj}\delta_{ni}$\tabularnewline
$D8-i$ &  & $\delta_{jn}\delta_{im}+\delta_{mj}\delta_{ni}$ &  & $D8-j$ &  & $\delta_{jn}\delta_{im}+\delta_{mj}\delta_{ni}$\tabularnewline
$D8-k$ &  & $-\left(\delta_{jn}\delta_{im}+\delta_{mj}\delta_{ni}\right)$ &  & $D8-l$ &  & $-\left(\delta_{jn}\delta_{im}+\delta_{mj}\delta_{ni}\right)$\tabularnewline
\hline 
\hline 
 &  &  &  &  &  & \tabularnewline
\end{tabular}\caption{\label{tab:S-PPPP-8}Values of the diagrams in Figure\,\ref{fig:D8-order-lambda-g-4}
with common factor\protect \\
 $\frac{1}{32}\left(\frac{b^{2}-4\,a\,b+3\,a^{2}}{32\pi^{2}\epsilon}\right)\,i\,\lambda\,g^{4}\int_{\theta}\overline{\Phi}_{i}\Phi_{m}\Phi_{n}\overline{\Phi}_{j}$
.}
\end{table}
\par\end{center}

$\mathcal{S}_{\left(\overline{\Phi}\Phi\right)^{2}}^{\left(D8-a\right)}$
in the Figure\,\ref{fig:D8-order-lambda-g-4} is
\begin{align}
\mathcal{S}_{\left(\overline{\Phi}\Phi\right)^{2}}^{\left(D8-a\right)} & =-\frac{1}{32}\,i\,\lambda\,g^{4}\left(\delta_{jn}\delta_{im}+\delta_{mj}\delta_{ni}\right)\int_{\theta}\overline{\Phi}_{i}\Phi_{m}\Phi_{n}\overline{\Phi}_{j}\nonumber \\
 & \times\int\frac{d^{D}kd^{D}q}{\left(2\pi\right)^{2D}}\left\{ \frac{-2\left(a^{2}-b^{2}\right)\left(k\cdot q\right)k^{2}-4\left(a^{2}-a\,b\right)\left(k^{2}\right)^{2}}{\left(k^{2}\right)^{3}\left(k+q\right)^{2}q^{2}}\right\} \,,
\end{align}
using Eqs.\,(\ref{eq:Int 7}), (\ref{eq:Int 9}) and adding $\mathcal{S}_{\left(\overline{\Phi}\Phi\right)^{2}}^{\left(D8-a\right)}$
to $\mathcal{S}_{\left(\overline{\Phi}\Phi\right)^{2}}^{\left(D8-l\right)}$
with the values in the Table\,\ref{tab:S-PPPP-8}, we find
\begin{align}
\mathcal{S}_{\left(\overline{\Phi}\Phi\right)^{2}}^{\left(D8\right)} & =\frac{1}{8}\left(\frac{b^{2}-4\,a\,b+3\,a^{2}}{32\pi^{2}\epsilon}\right)\,i\,\lambda\,g^{4}\left(\delta_{jn}\delta_{im}+\delta_{mj}\delta_{ni}\right)\int_{\theta}\overline{\Phi}_{i}\Phi_{m}\Phi_{n}\overline{\Phi}_{j}\,.\label{eq:S-D8}
\end{align}

\begin{center}
\begin{figure}
\begin{centering}
\subfloat[]{\begin{centering}
\includegraphics[scale=0.5]{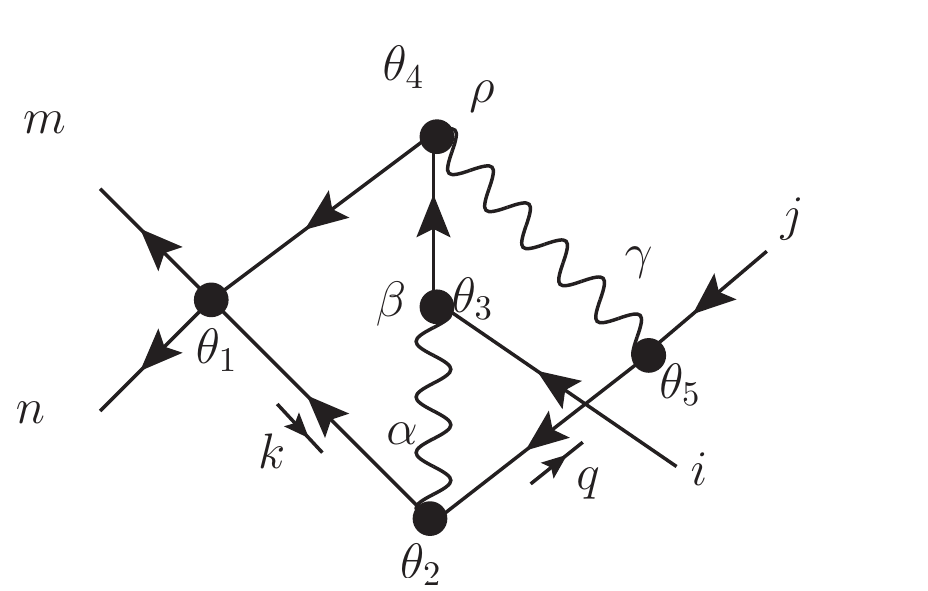}
\par\end{centering}
}\subfloat[]{\centering{}\includegraphics[scale=0.5]{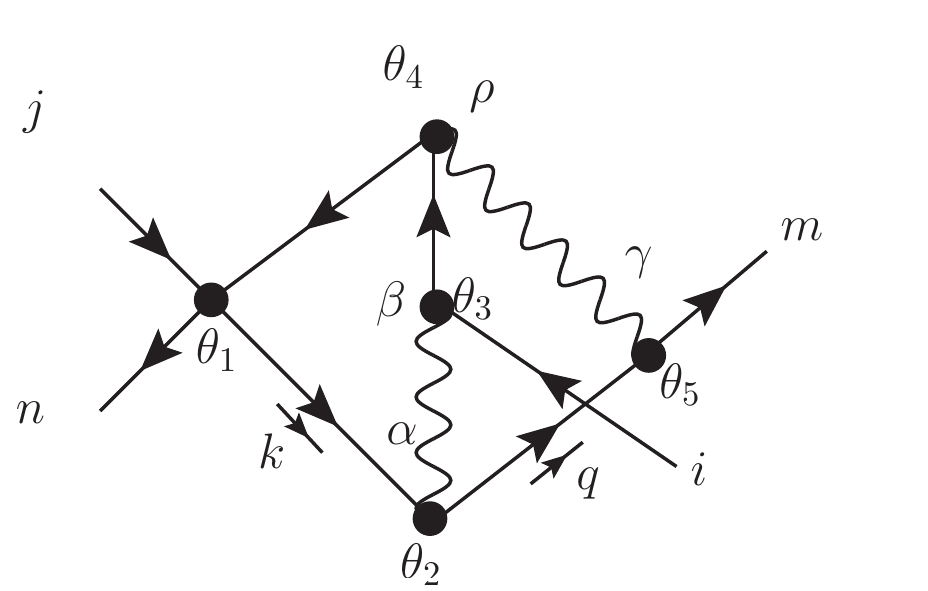}}\subfloat[]{\centering{}\includegraphics[scale=0.5]{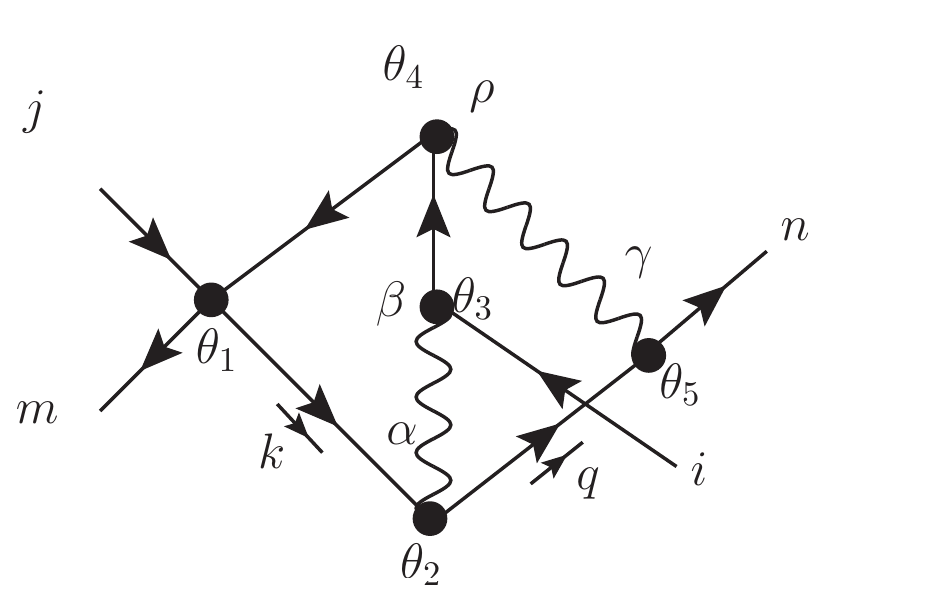}}
\par\end{centering}
\begin{centering}
\subfloat[]{\centering{}\includegraphics[scale=0.5]{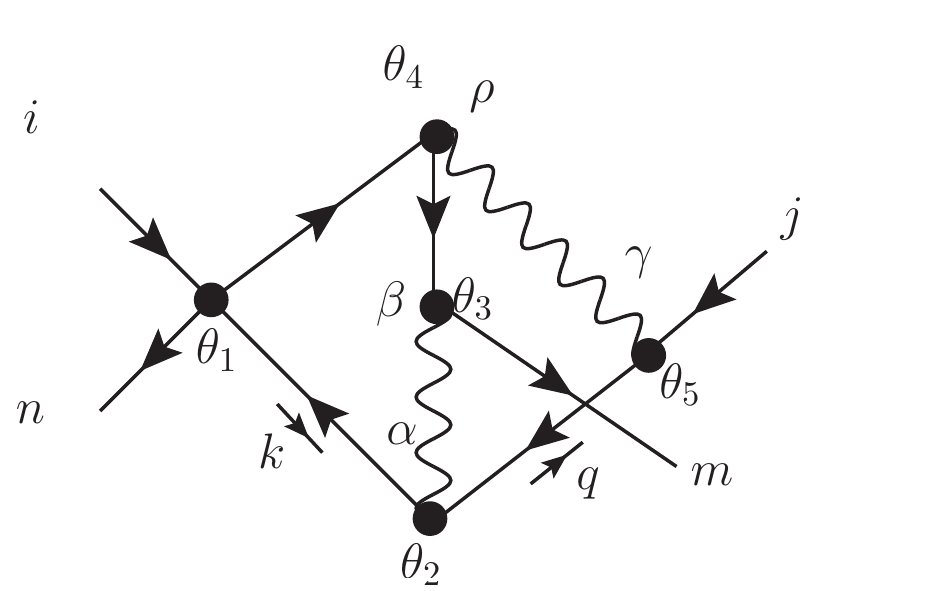}}\subfloat[]{\centering{}\includegraphics[scale=0.5]{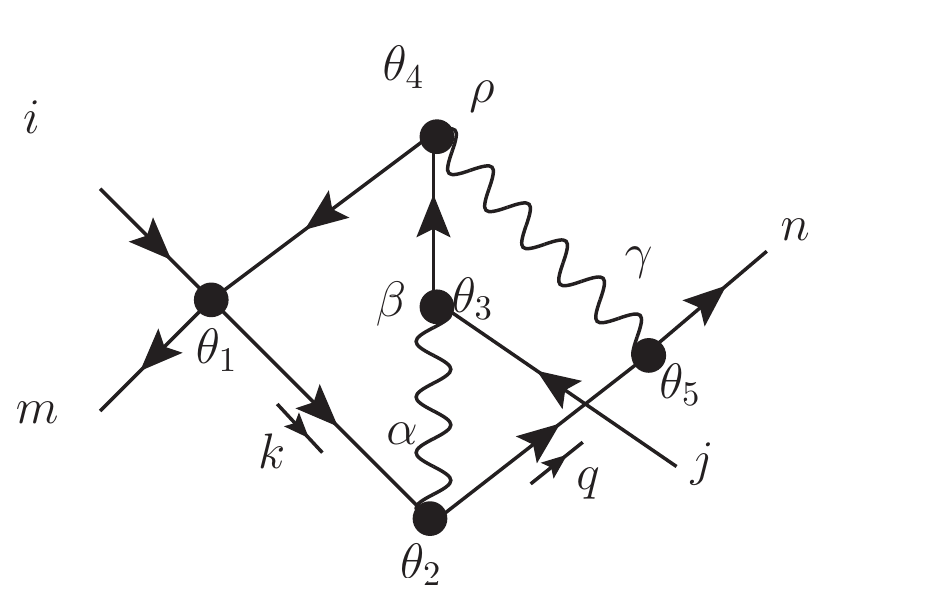}}\subfloat[]{\centering{}\includegraphics[scale=0.5]{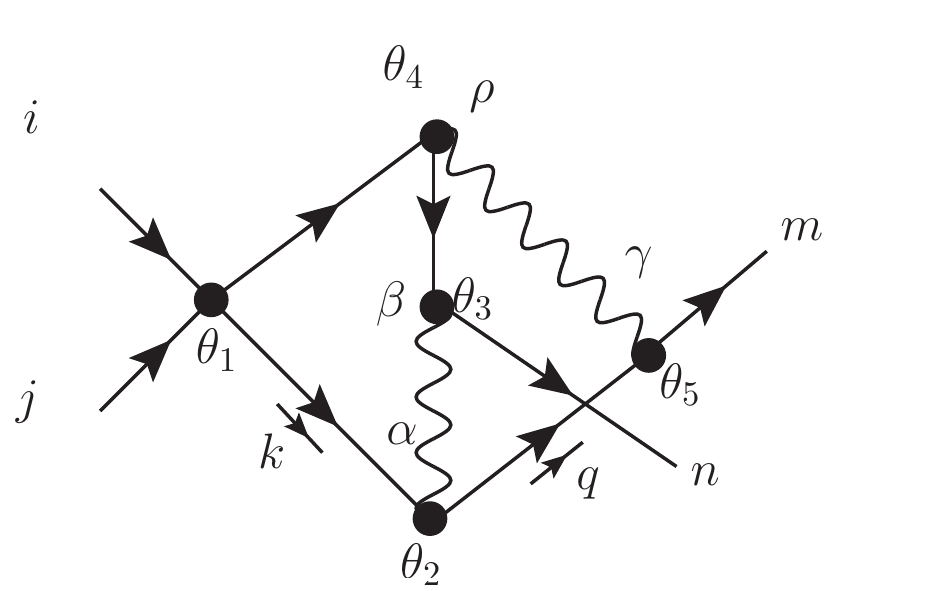}}
\par\end{centering}
\centering{}\caption{\label{fig:D9-order-lambda-g-4}$\mathcal{S}_{\left(\overline{\Phi}\Phi\right)^{2}}^{\left(D9\right)}$}
\end{figure}
\par\end{center}

\begin{center}
\begin{table}
\centering{}%
\begin{tabular}{lcccccc}
 &  &  &  &  &  & \tabularnewline
\hline 
\hline 
$D9-a$ &  & $\delta_{jn}\delta_{im}+\delta_{mj}\delta_{ni}$ &  & $D9-b$ &  & $\delta_{jn}\delta_{im}+\delta_{mj}\delta_{ni}$\tabularnewline
$D9-c$ &  & $\delta_{jn}\delta_{im}+\delta_{mj}\delta_{ni}$ &  & $D9-d$ &  & $\delta_{jn}\delta_{im}+\delta_{mj}\delta_{ni}$\tabularnewline
$D9-e$ &  & $\delta_{jn}\delta_{im}+\delta_{mj}\delta_{ni}$ &  & $D9-f$ &  & $\delta_{jn}\delta_{im}+\delta_{mj}\delta_{ni}$\tabularnewline
\hline 
\hline 
 &  &  &  &  &  & \tabularnewline
\end{tabular}\caption{\label{tab:S-PPPP-9}Values of the diagrams in Figure\,\ref{fig:D9-order-lambda-g-4}
with common factor\protect \\
 $\frac{1}{32}\left(\frac{\left(a-b\right)^{2}}{32\pi^{2}\epsilon}\right)\,i\,\lambda\,g^{4}\int_{\theta}\overline{\Phi}_{i}\Phi_{m}\Phi_{n}\overline{\Phi}_{j}$
.}
\end{table}
\par\end{center}

$\mathcal{S}_{\left(\overline{\Phi}\Phi\right)^{2}}^{\left(D9-a\right)}$
in the Figure\,\ref{fig:D9-order-lambda-g-4} is
\begin{align}
\mathcal{S}_{\left(\overline{\Phi}\Phi\right)^{2}}^{\left(D9-a\right)} & =\frac{1}{128}\,i\,\lambda\,g^{4}\left(\delta_{jn}\delta_{im}+\delta_{mj}\delta_{ni}\right)\int_{\theta}\overline{\Phi}_{i}\Phi_{m}\Phi_{n}\overline{\Phi}_{j}\nonumber \\
 & \times\int\frac{d^{D}kd^{D}q}{\left(2\pi\right)^{2D}}\,\left\{ \frac{-4\left(a-b\right)^{2}k^{2}q^{2}\left(k-q\right)^{2}}{\left(k-q\right)^{4}\left(k^{2}\right)^{2}\left(q^{2}\right)^{2}}\right\} \,,
\end{align}
using Eq.\,(\ref{eq:Int 7}) and adding $\mathcal{S}_{\left(\overline{\Phi}\Phi\right)^{2}}^{\left(D9-a\right)}$
to $\mathcal{S}_{\left(\overline{\Phi}\Phi\right)^{2}}^{\left(D9-f\right)}$
with the values in Table\,\ref{tab:S-PPPP-9}, we find 
\begin{align}
\mathcal{S}_{\left(\overline{\Phi}\Phi\right)^{2}}^{\left(D9\right)} & =\frac{3}{16}\left(\frac{\left(a-b\right)^{2}}{32\pi^{2}\epsilon}\right)\,i\,\lambda\,g^{4}\left(\delta_{jn}\delta_{im}+\delta_{mj}\delta_{ni}\right)\int_{\theta}\overline{\Phi}_{i}\Phi_{m}\Phi_{n}\overline{\Phi}_{j}\,.\label{eq:S-D9}
\end{align}

\begin{center}
\begin{figure}
\begin{centering}
\subfloat[]{\begin{centering}
\includegraphics[scale=0.5]{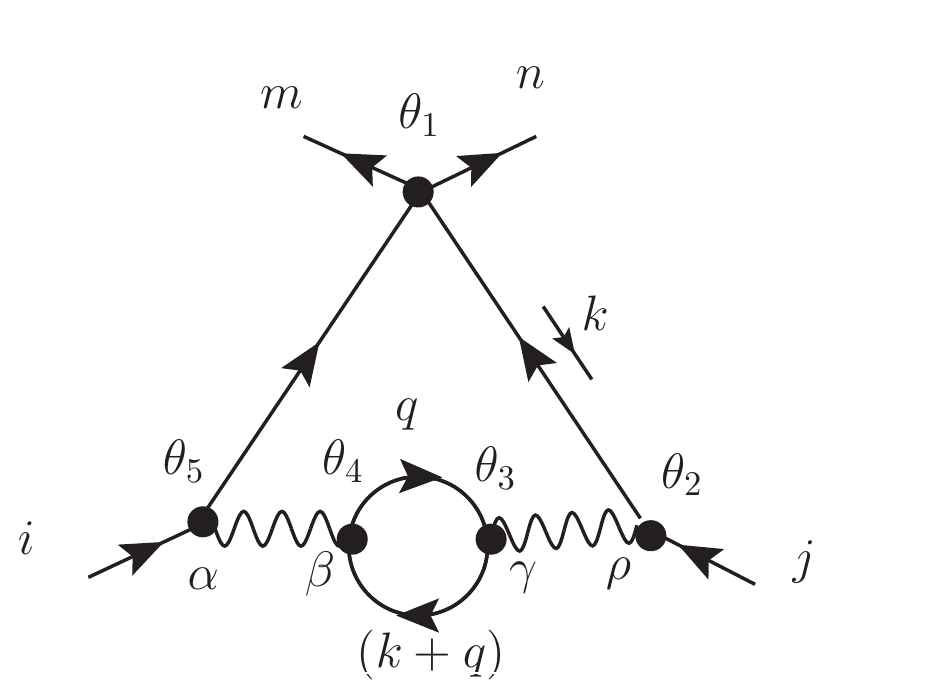}
\par\end{centering}
}\subfloat[]{\centering{}\includegraphics[scale=0.5]{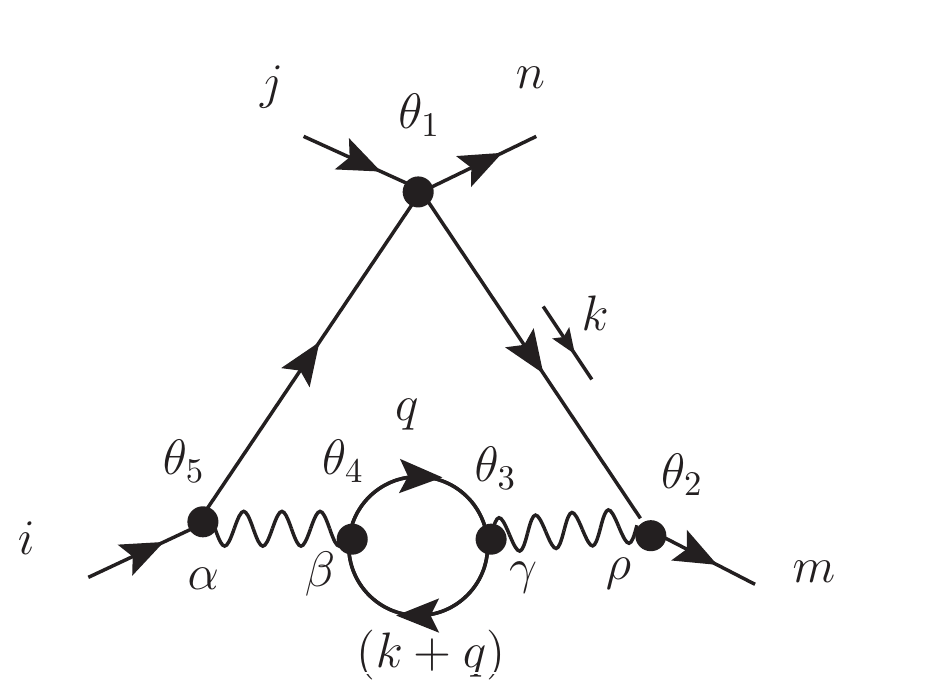}}\subfloat[]{\centering{}\includegraphics[scale=0.5]{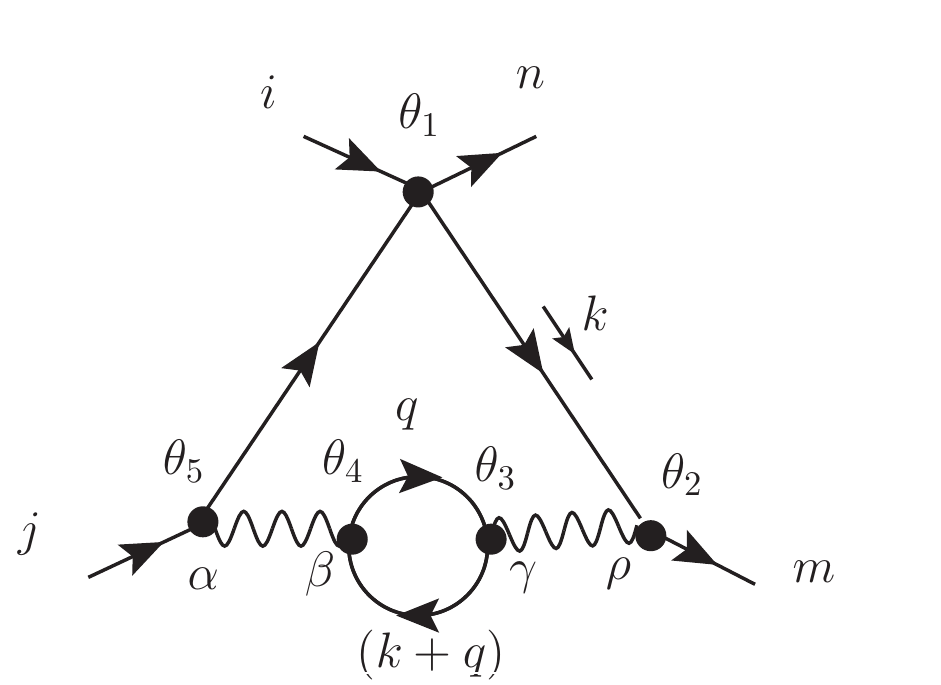}}
\par\end{centering}
\begin{centering}
\subfloat[]{\centering{}\includegraphics[scale=0.5]{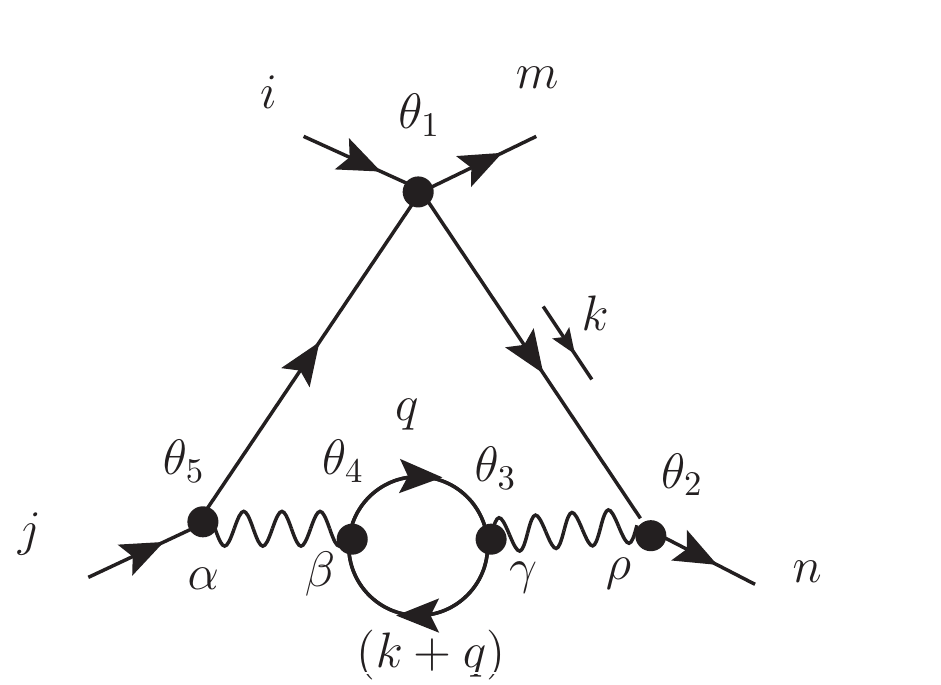}}\subfloat[]{\centering{}\includegraphics[scale=0.5]{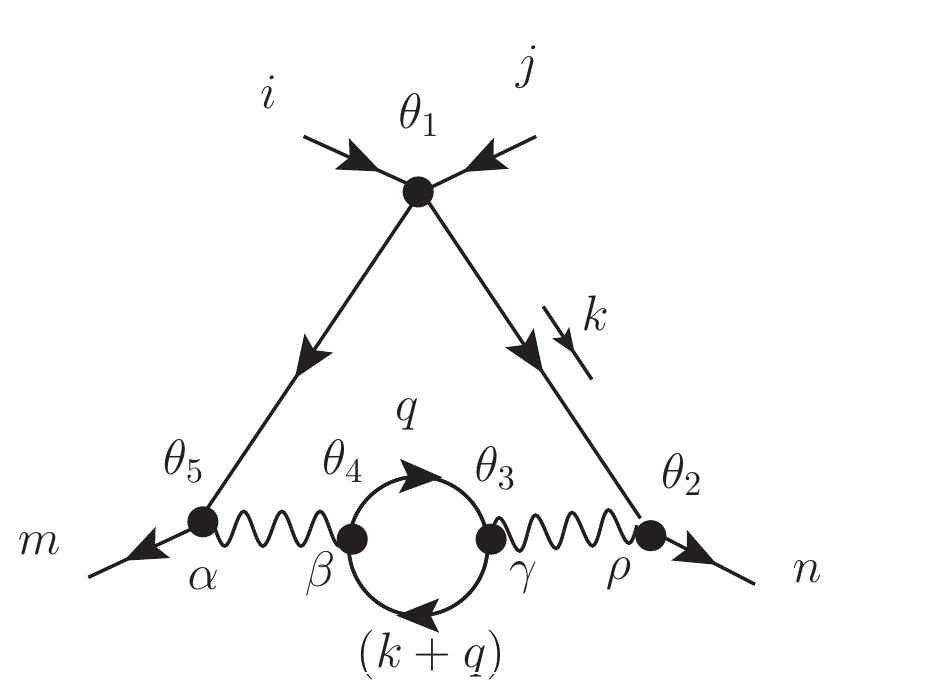}}
\par\end{centering}
\centering{}\caption{\label{fig:D10-orde-lambda-g-4}$\mathcal{S}_{\left(\overline{\Phi}\Phi\right)^{2}}^{\left(D10\right)}$}
\end{figure}
\par\end{center}

$\mathcal{S}_{\left(\overline{\Phi}\Phi\right)^{2}}^{\left(D10-a\right)}$
in the Figure\,\ref{fig:D10-orde-lambda-g-4} is
\begin{align}
\mathcal{S}_{\left(\overline{\Phi}\Phi\right)^{2}}^{\left(D10-a\right)} & =\frac{1}{128}\,i\,\lambda\,g^{4}\,N\left(\delta_{jn}\delta_{im}+\delta_{mj}\delta_{ni}\right)\int_{\theta}\overline{\Phi}_{i}\Phi_{m}\Phi_{n}\overline{\Phi}_{j}\nonumber \\
 & \times\int\frac{d^{D}kd^{D}q}{\left(2\pi\right)^{2D}}\left\{ \frac{-4\left(a-b\right)^{2}\left(k^{2}\right)^{2}\left(\left(k+q\right)^{2}+q^{2}\right)}{\left(k^{2}\right)^{4}\left(k+q\right)^{2}q^{2}}\right\} \,,
\end{align}
using Eqs.\,(\ref{eq:Int 7}), (\ref{eq:Int 9}), we find 
\begin{align}
\mathcal{S}_{\left(\overline{\Phi}\Phi\right)^{2}}^{\left(D10-a\right)} & =\mathcal{S}_{\left(\overline{\Phi}\Phi\right)^{2}}^{\left(D10\right)}=0\,.\label{eq:S-D10}
\end{align}

\begin{center}
\begin{figure}
\begin{centering}
\subfloat[]{\begin{centering}
\includegraphics[scale=0.5]{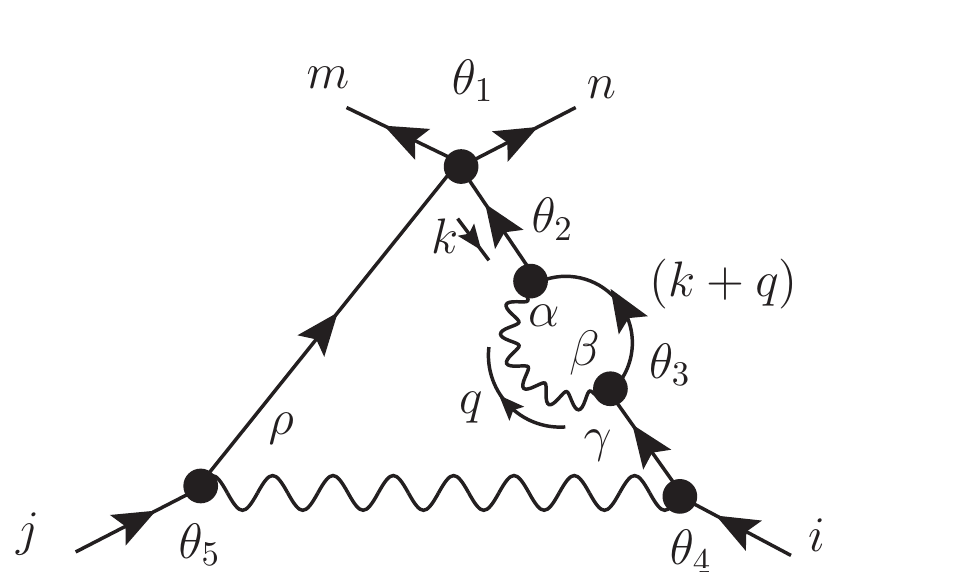}
\par\end{centering}
}\subfloat[]{\centering{}\includegraphics[scale=0.5]{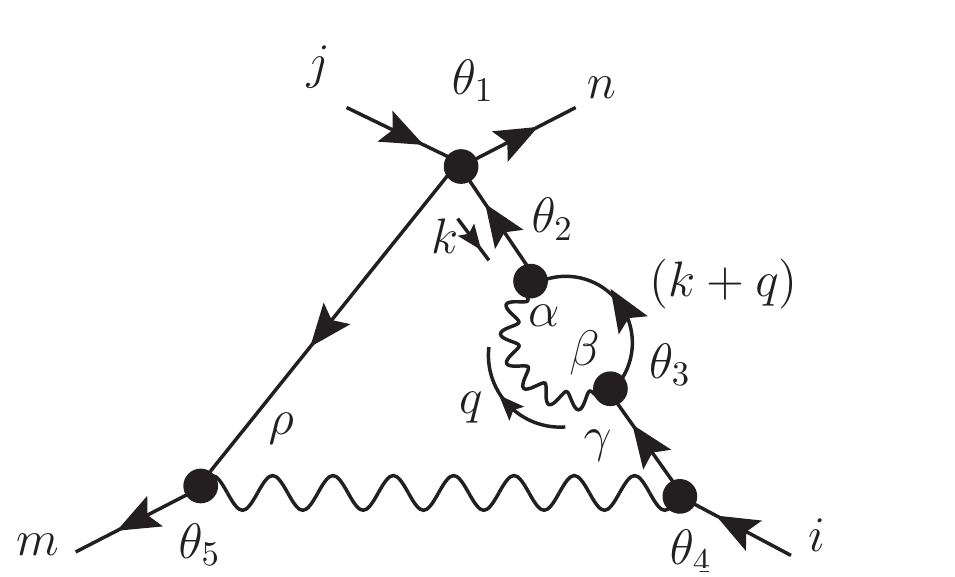}}\subfloat[]{\centering{}\includegraphics[scale=0.5]{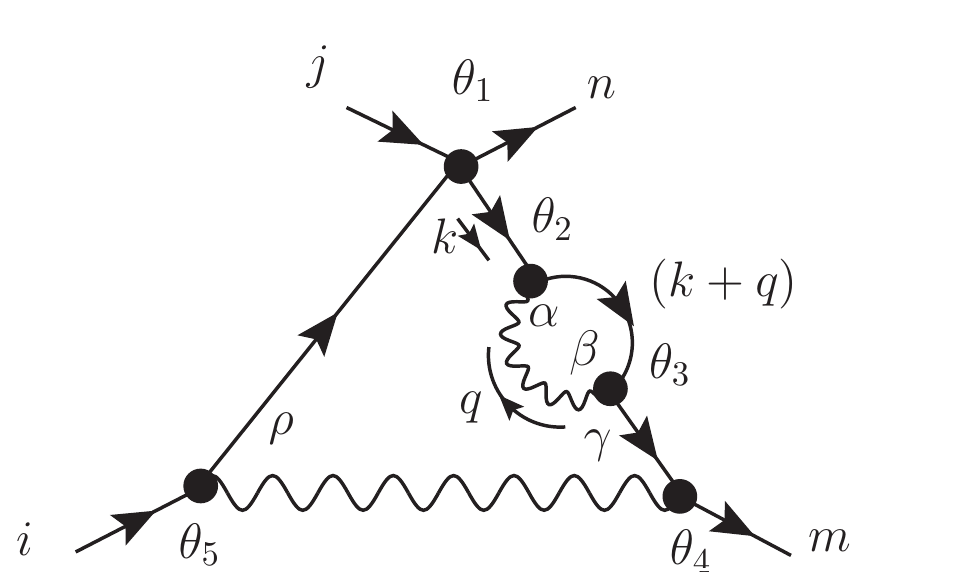}}
\par\end{centering}
\begin{centering}
\subfloat[]{\centering{}\includegraphics[scale=0.5]{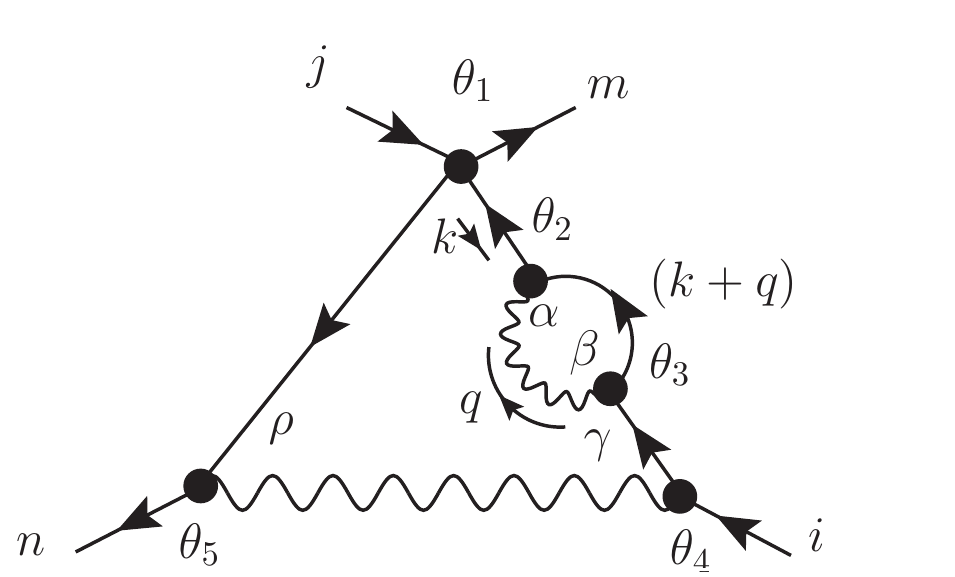}}\subfloat[]{\centering{}\includegraphics[scale=0.5]{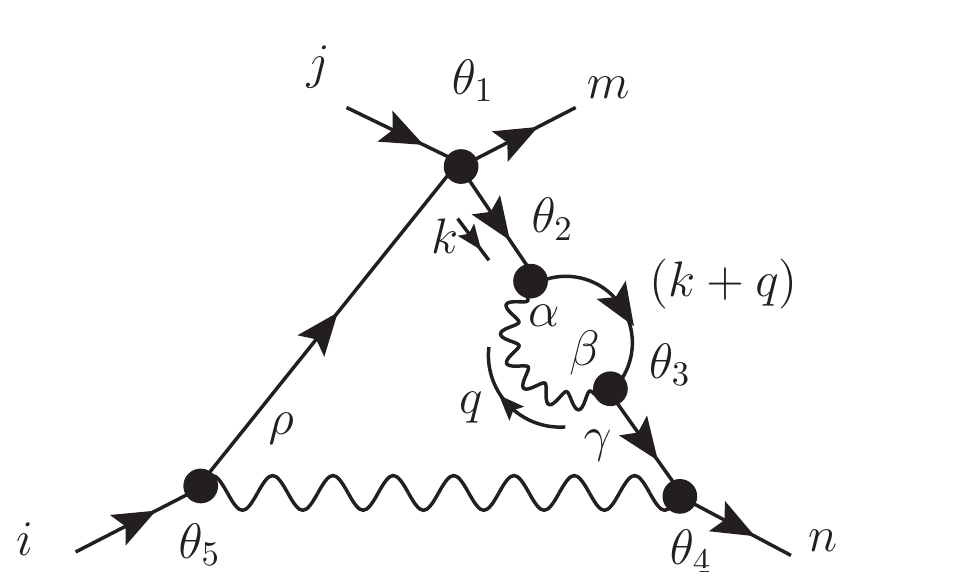}}\subfloat[]{\centering{}\includegraphics[scale=0.5]{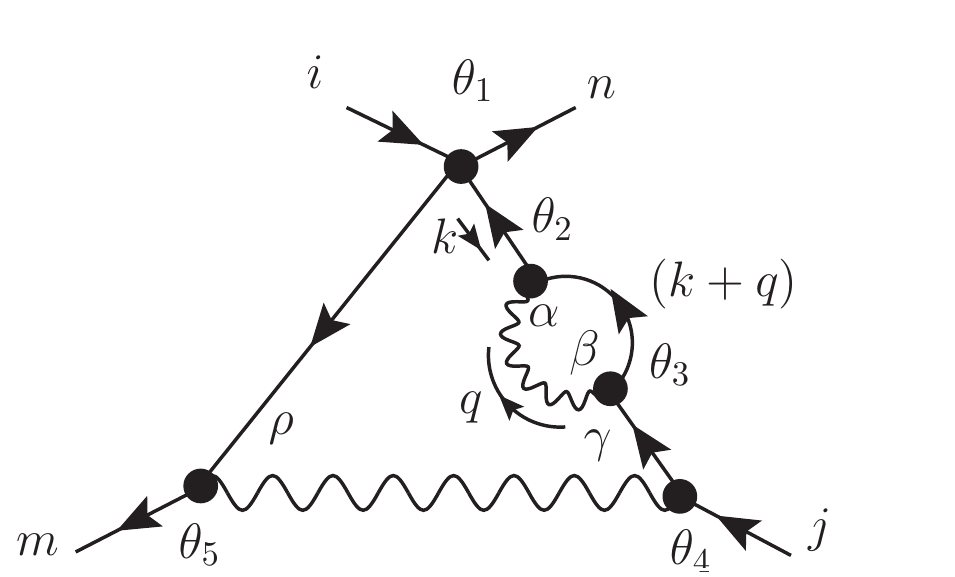}}
\par\end{centering}
\begin{centering}
\subfloat[]{\centering{}\includegraphics[scale=0.5]{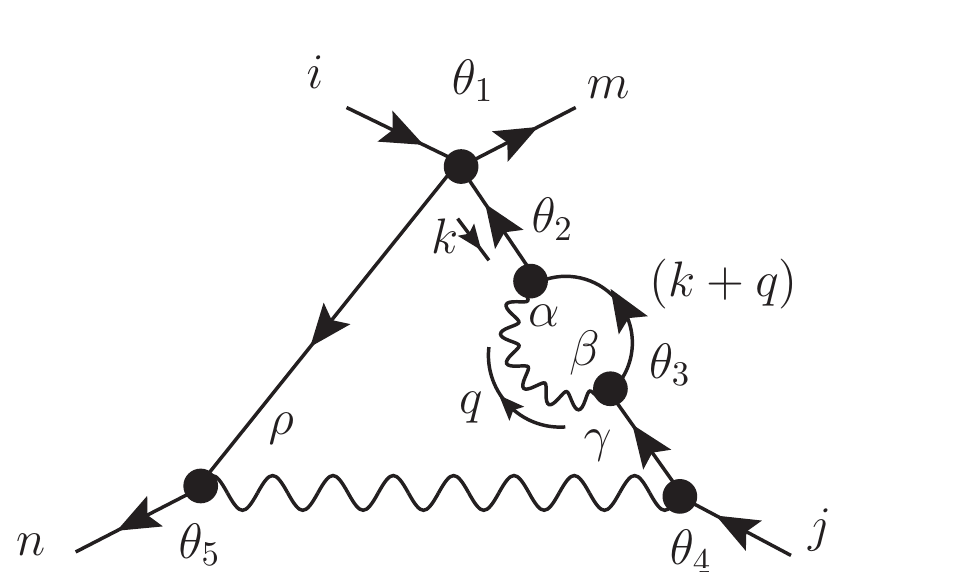}}\subfloat[]{\centering{}\includegraphics[scale=0.5]{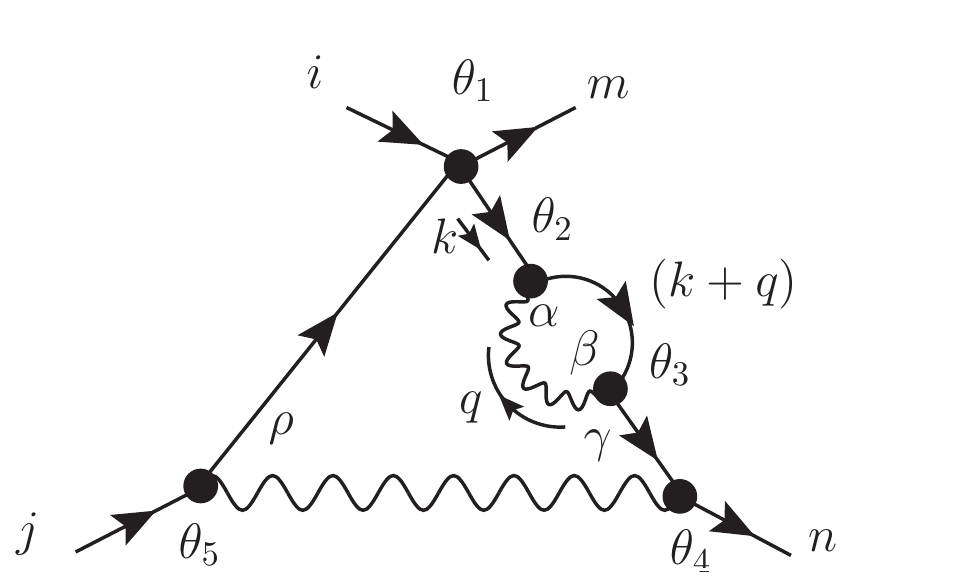}}\subfloat[]{\centering{}\includegraphics[scale=0.5]{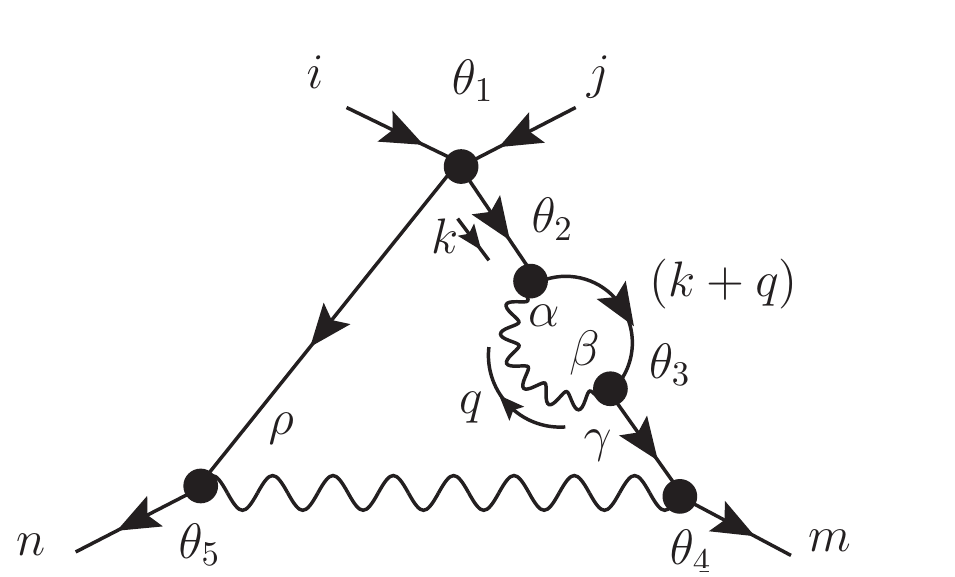}}
\par\end{centering}
\begin{centering}
\subfloat[]{\centering{}\includegraphics[scale=0.5]{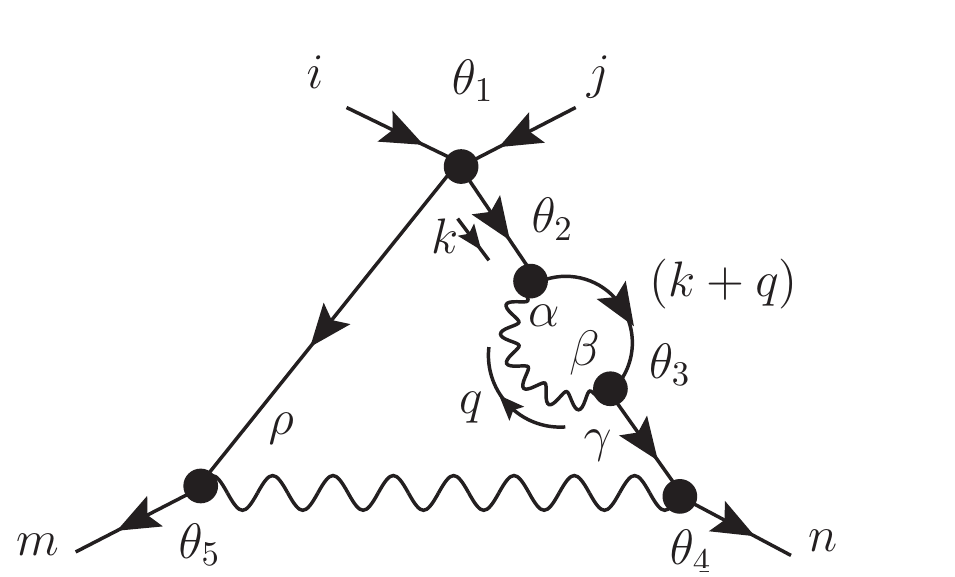}}\subfloat[]{\centering{}\includegraphics[scale=0.5]{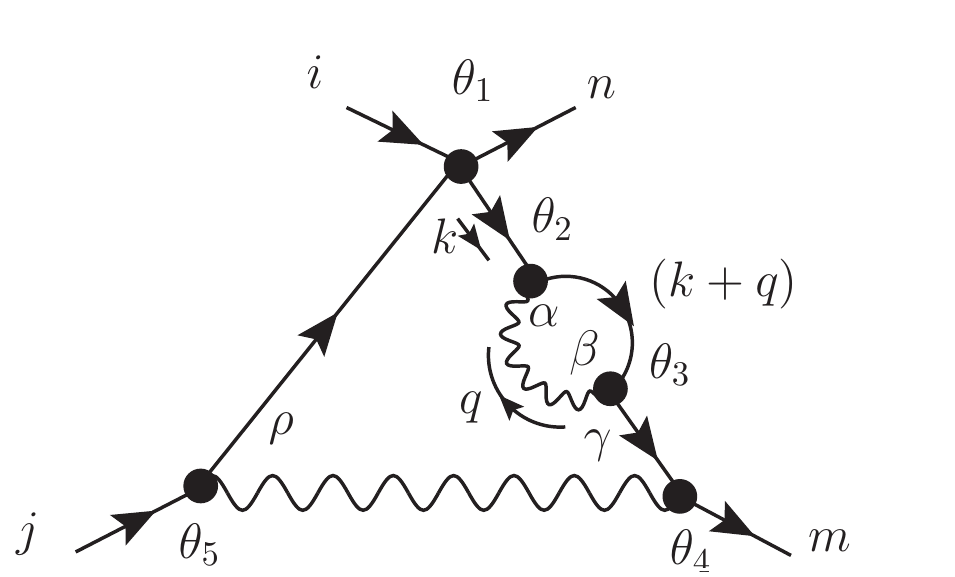}}\subfloat[]{\centering{}\includegraphics[scale=0.5]{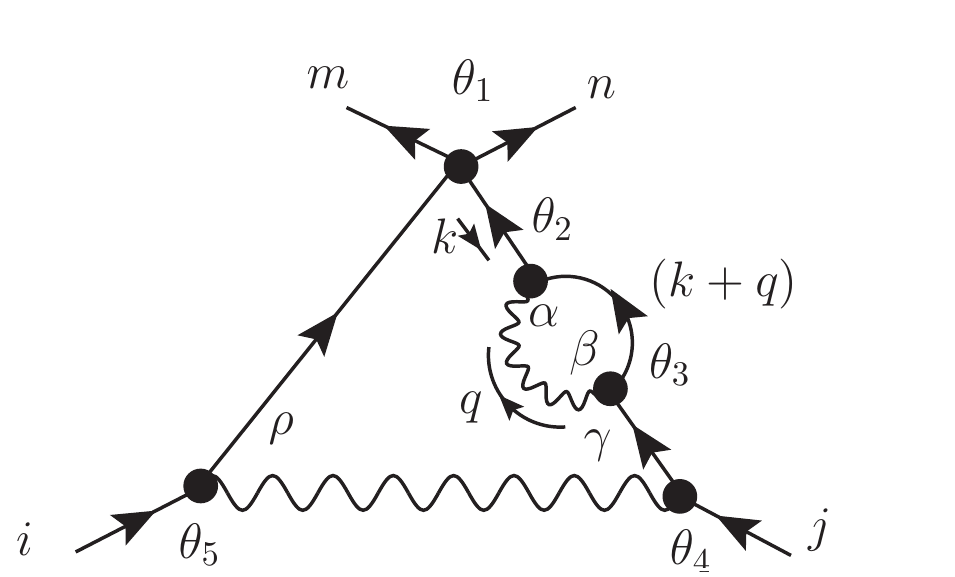}}
\par\end{centering}
\centering{}\caption{\label{fig:D11-order-lambda-g-4}$\mathcal{S}_{\left(\overline{\Phi}\Phi\right)^{2}}^{\left(D11\right)}$}
\end{figure}
\par\end{center}

\begin{center}
\begin{table}
\centering{}%
\begin{tabular}{lcccccc}
 &  &  &  &  &  & \tabularnewline
\hline 
\hline 
$D11-a$ &  & $-\left(\delta_{jn}\delta_{im}+\delta_{mj}\delta_{ni}\right)$ &  & $D11-b$ &  & $\delta_{jn}\delta_{im}+\delta_{mj}\delta_{ni}$\tabularnewline
$D11-c$ &  & $\delta_{jn}\delta_{im}+\delta_{mj}\delta_{ni}$ &  & $D11-d$ &  & $\delta_{jn}\delta_{im}+\delta_{mj}\delta_{ni}$\tabularnewline
$D11-e$ &  & $\delta_{jn}\delta_{im}+\delta_{mj}\delta_{ni}$ &  & $D11-f$ &  & $\delta_{jn}\delta_{im}+\delta_{mj}\delta_{ni}$\tabularnewline
$D11-g$ &  & $\delta_{jn}\delta_{im}+\delta_{mj}\delta_{ni}$ &  & $D11-h$ &  & $\delta_{jn}\delta_{im}+\delta_{mj}\delta_{ni}$\tabularnewline
$D11-i$ &  & $-\left(\delta_{jn}\delta_{im}+\delta_{mj}\delta_{ni}\right)$ &  & $D11-j$ &  & $-\left(\delta_{jn}\delta_{im}+\delta_{mj}\delta_{ni}\right)$\tabularnewline
$D11-k$ &  & $\delta_{jn}\delta_{im}+\delta_{mj}\delta_{ni}$ &  & $D11-l$ &  & $-\left(\delta_{jn}\delta_{im}+\delta_{mj}\delta_{ni}\right)$\tabularnewline
\hline 
\hline 
 &  &  &  &  &  & \tabularnewline
\end{tabular}\caption{\label{tab:S-PPPP-11}Values of the diagrams in Figure\,\ref{fig:D11-order-lambda-g-4}
with common factor\protect \\
 $\frac{8}{128}\left(\frac{a\,b-a^{2}}{32\pi^{2}\epsilon}\right)\,i\,\lambda\,g^{4}\int_{\theta}\overline{\Phi}_{i}\Phi_{m}\Phi_{n}\overline{\Phi}_{j}$
.}
\end{table}
\par\end{center}

$\mathcal{S}_{\left(\overline{\Phi}\Phi\right)^{2}}^{\left(D11-a\right)}$
in the Figure\,\ref{fig:D11-order-lambda-g-4} is
\begin{align}
\mathcal{S}_{\left(\overline{\Phi}\Phi\right)^{2}}^{\left(D11-a\right)} & =-\frac{1}{128}\,i\,\lambda\,g^{4}\left(\delta_{in}\delta_{jm}+\delta_{mi}\delta_{nj}\right)\int_{\theta}\overline{\Phi}_{i}\Phi_{m}\Phi_{n}\overline{\Phi}_{j}\nonumber \\
 & \times\int\frac{d^{D}kd^{D}q}{\left(2\pi\right)^{2D}}\left\{ \frac{16\left(a^{2}-a\,b\right)\left(k^{2}\right)^{2}\left(\left(k\cdot q\right)+k^{2}\right)+4\left(a-b\right)^{2}\left(k^{2}\right)^{2}q^{2}}{\left(k^{2}\right)^{4}q^{2}\left(k+q\right)^{2}}\right\} \,,
\end{align}
using Eqs.\,(\ref{eq:Int 7}), (\ref{eq:Int 9}) and the momentum
power counting, then adding $\mathcal{S}_{\left(\overline{\Phi}\Phi\right)^{2}}^{\left(D11-a\right)}$
to $\mathcal{S}_{\left(\overline{\Phi}\Phi\right)^{2}}^{\left(D11-l\right)}$
with the values in the Table\,\ref{tab:S-PPPP-11}, we find 
\begin{align}
\mathcal{S}_{\left(\overline{\Phi}\Phi\right)^{2}}^{\left(D11\right)} & =\frac{1}{4}\left(\frac{a\,b-a^{2}}{32\pi^{2}\epsilon}\right)i\,\lambda\,g^{4}\left(\delta_{in}\delta_{jm}+\delta_{mi}\delta_{nj}\right)\int_{\theta}\overline{\Phi}_{i}\Phi_{m}\Phi_{n}\overline{\Phi}_{j}\,.\label{eq:S-D11}
\end{align}

\begin{center}
\begin{figure}
\begin{centering}
\subfloat[]{\begin{centering}
\includegraphics[scale=0.5]{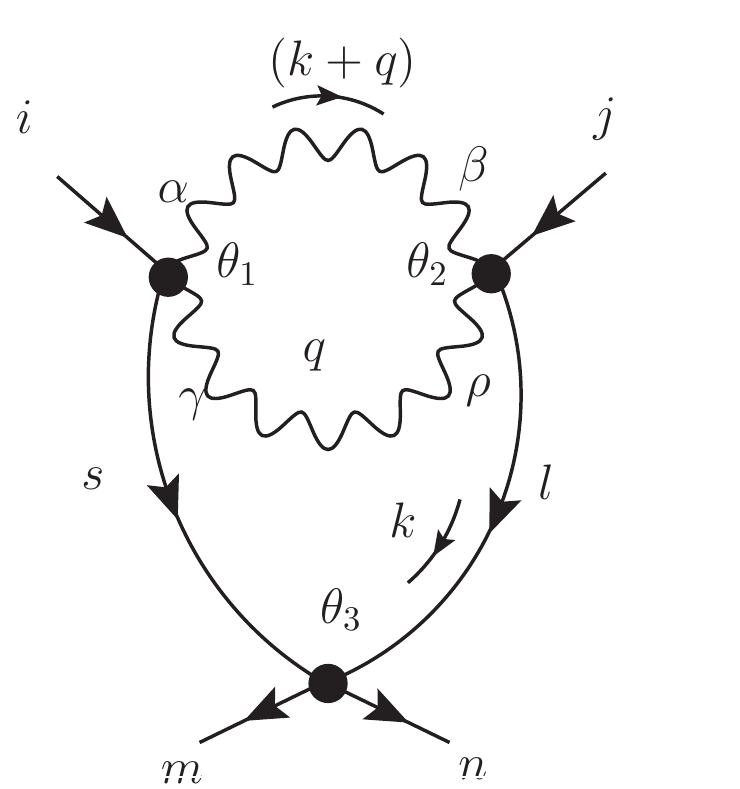}
\par\end{centering}
}\subfloat[]{\centering{}\includegraphics[scale=0.5]{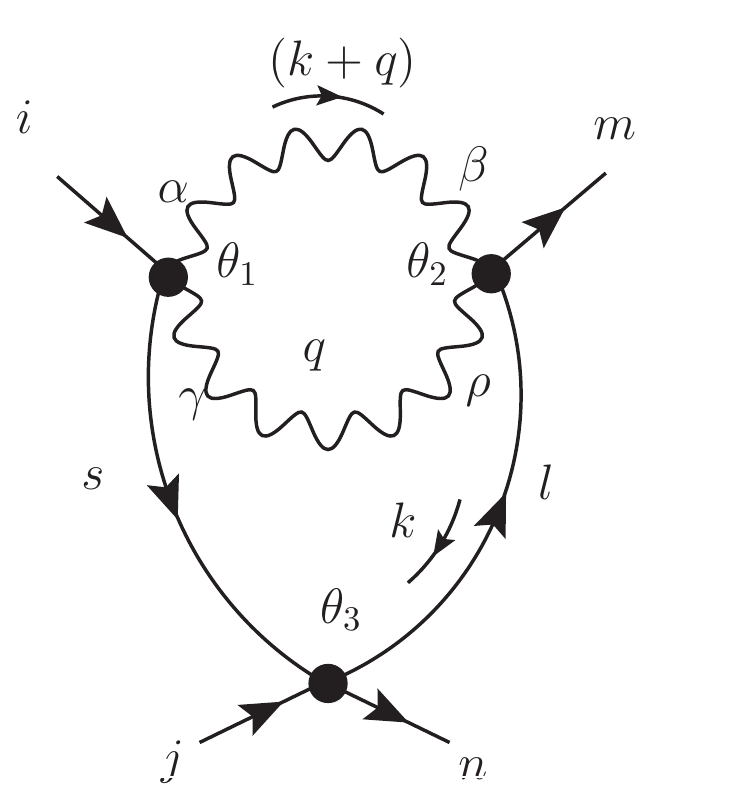}}\subfloat[]{\centering{}\includegraphics[scale=0.5]{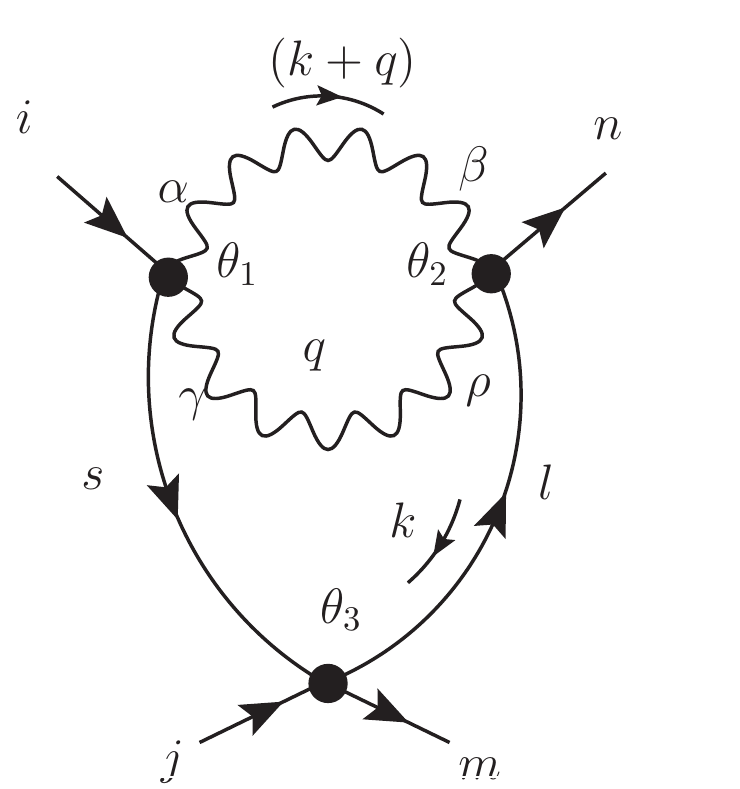}}
\par\end{centering}
\begin{centering}
\subfloat[]{\centering{}\includegraphics[scale=0.5]{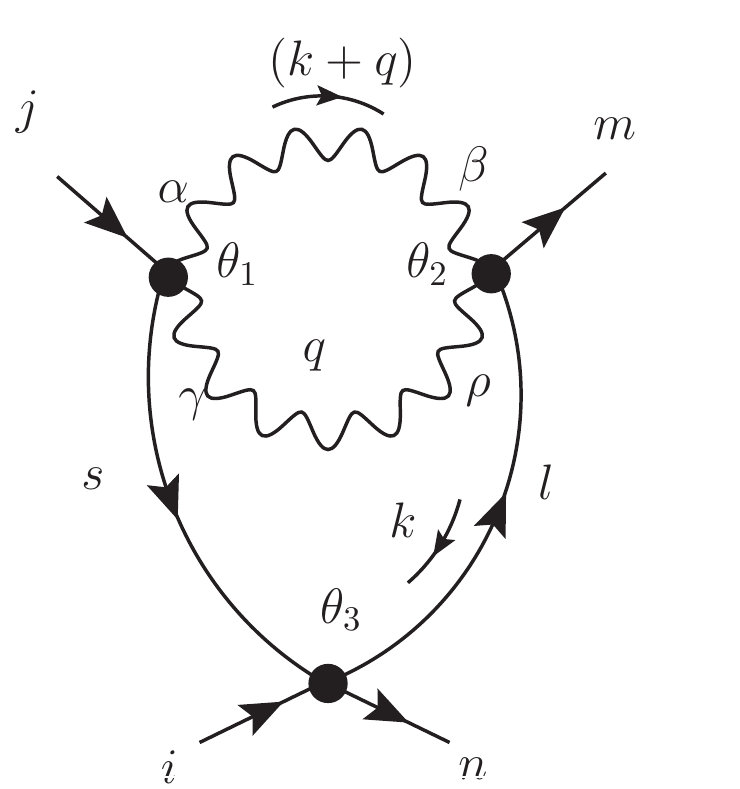}}\subfloat[]{\centering{}\includegraphics[scale=0.5]{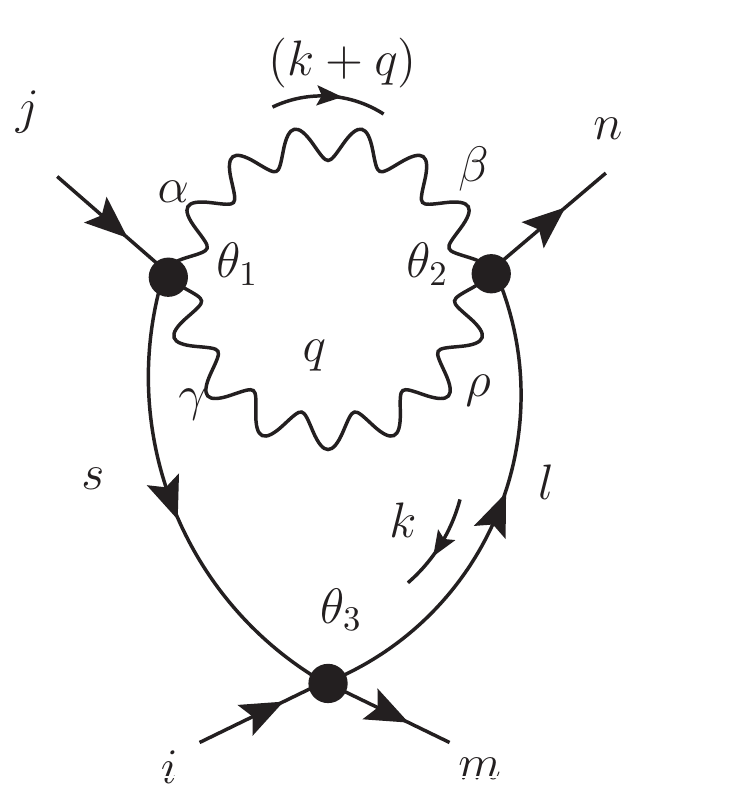}}\subfloat[]{\centering{}\includegraphics[scale=0.5]{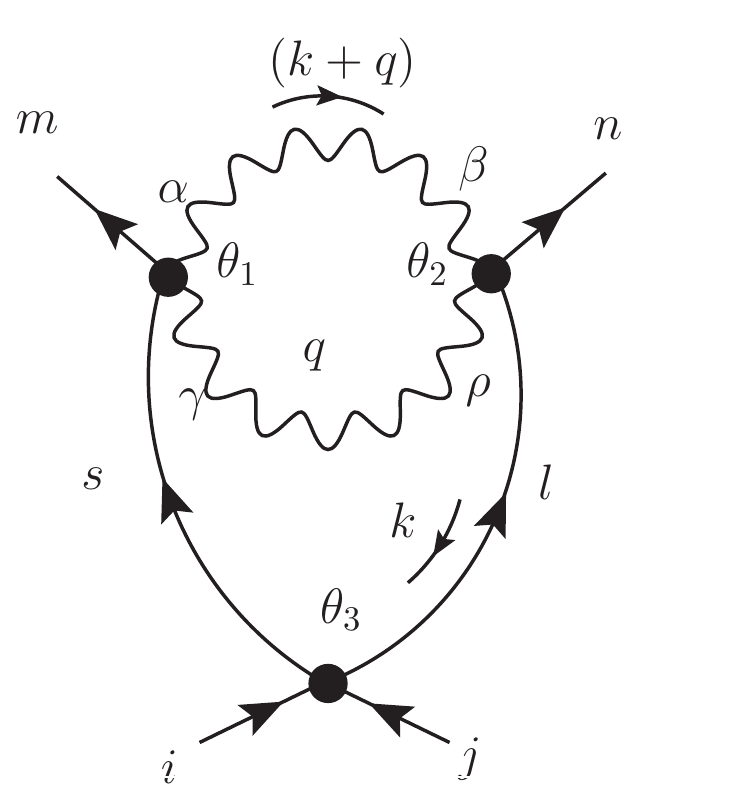}}
\par\end{centering}
\centering{}\caption{\label{fig:D12-order-lambda-g-4}$\mathcal{S}_{\left(\overline{\Phi}\Phi\right)^{2}}^{\left(D12\right)}$}
\end{figure}
\par\end{center}

$\mathcal{S}_{\left(\overline{\Phi}\Phi\right)^{2}}^{\left(D12-a\right)}$
in the Figure\,\ref{fig:D12-order-lambda-g-4} is
\begin{align}
\mathcal{S}_{\left(\overline{\Phi}\Phi\right)^{2}}^{\left(D12-a\right)} & =-\frac{1}{16}\,i\,\lambda\,g^{4}\left(\delta_{jn}\delta_{im}+\delta_{mj}\delta_{ni}\right)\int_{\theta}\overline{\Phi}_{i}\Phi_{m}\Phi_{n}\overline{\Phi}_{j}\int\frac{d^{D}kd^{D}q}{\left(2\pi\right)^{2D}}\left\{ \frac{2\,a^{2}\,k^{2}}{\left(k^{2}\right)^{2}q^{2}\left(k+q\right)^{2}}\right\} \,,
\end{align}
using Eq.\,(\ref{eq:Int 7}), then adding $\mathcal{S}_{\left(\overline{\Phi}\Phi\right)^{2}}^{\left(D12-a\right)}$
to $\mathcal{S}_{\left(\overline{\Phi}\Phi\right)^{2}}^{\left(D12-f\right)}$,
we have
\begin{align}
\mathcal{S}_{\left(\overline{\Phi}\Phi\right)^{2}}^{\left(D12\right)} & =\frac{3}{4}\left(\frac{a^{2}}{32\pi^{2}\epsilon}\right)\,i\,\lambda\,g^{4}\left(\delta_{jn}\delta_{im}+\delta_{mj}\delta_{ni}\right)\int_{\theta}\overline{\Phi}_{i}\Phi_{m}\Phi_{n}\overline{\Phi}_{j}\,.\label{eq:S-D12}
\end{align}

\begin{center}
\begin{figure}
\begin{centering}
\subfloat[]{\centering{}\includegraphics[scale=0.5]{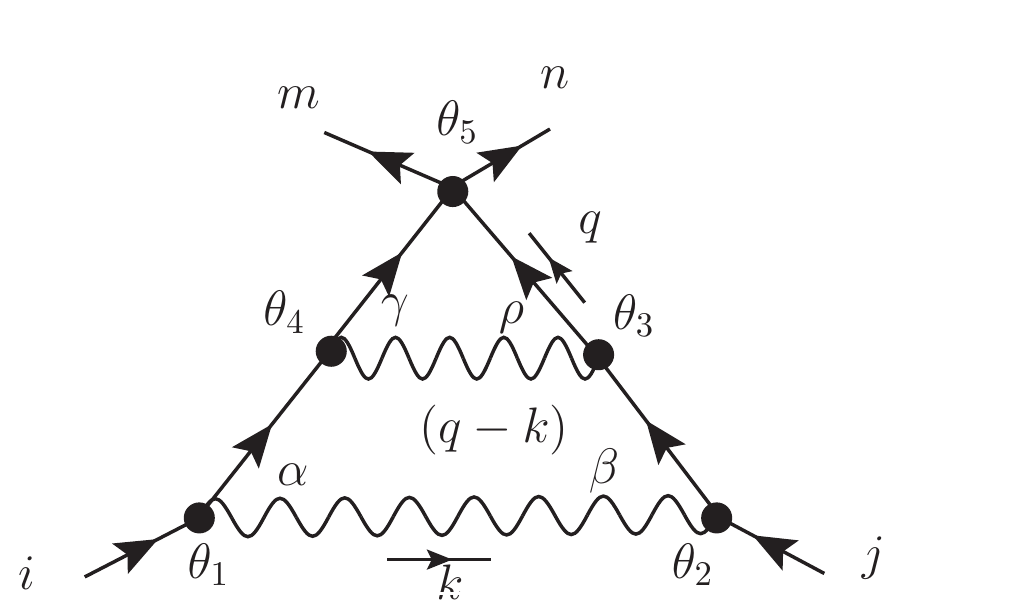}}\subfloat[]{\centering{}\includegraphics[scale=0.5]{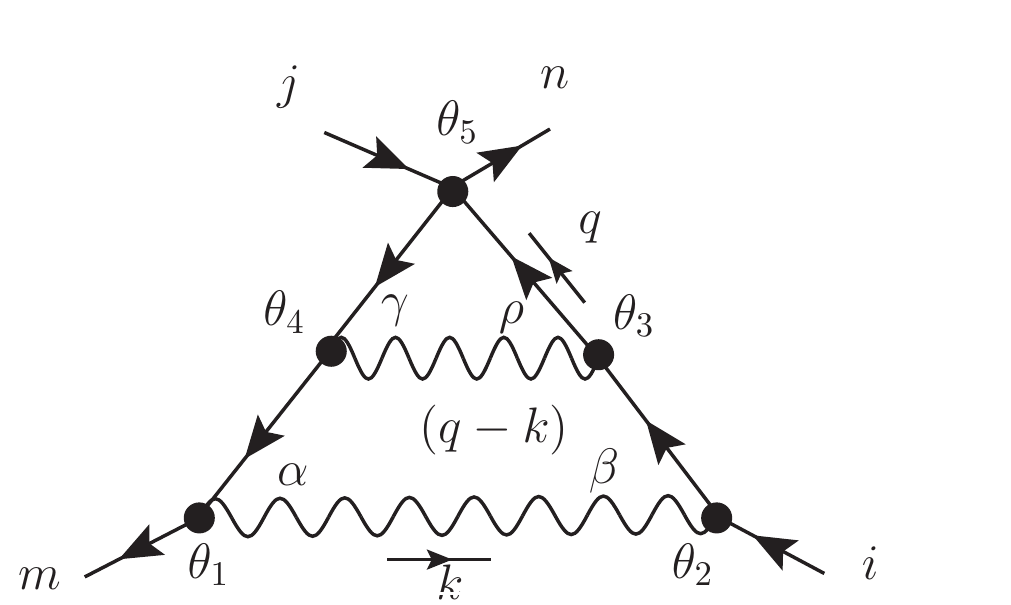}}\subfloat[]{\centering{}\includegraphics[scale=0.5]{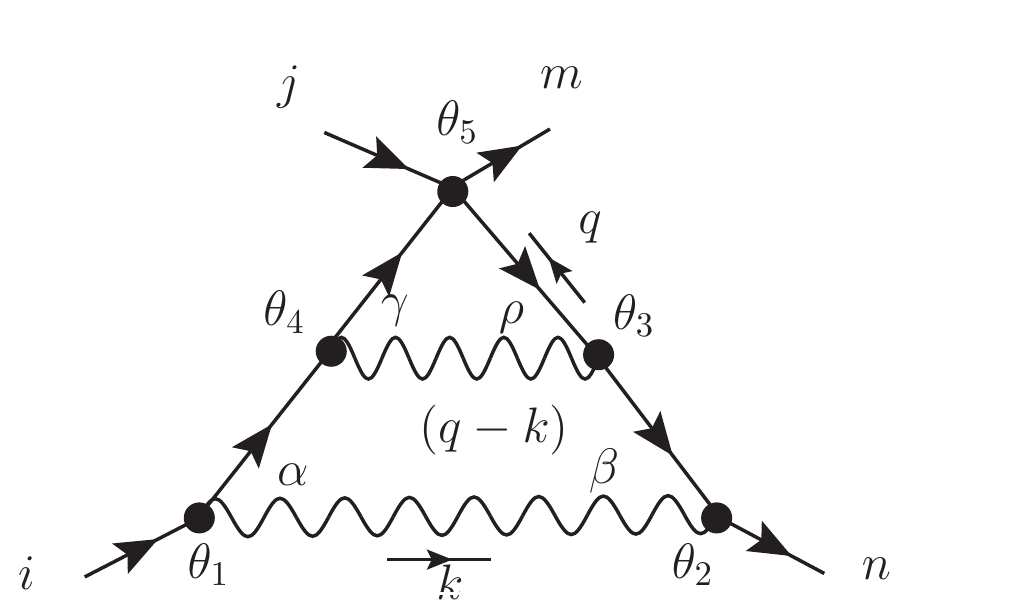}}
\par\end{centering}
\begin{centering}
\subfloat[]{\centering{}\includegraphics[scale=0.5]{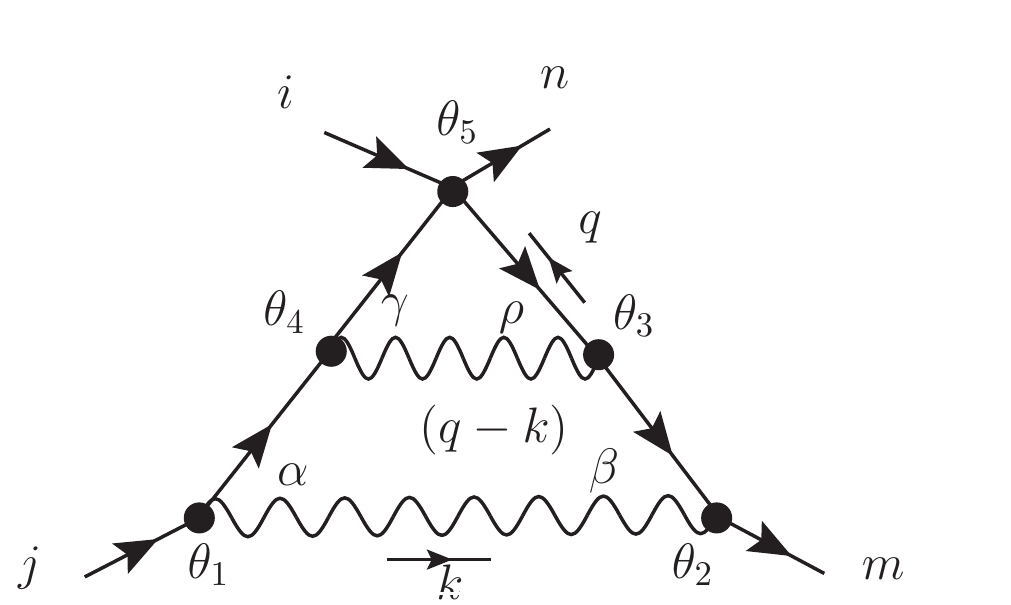}}\subfloat[]{\centering{}\includegraphics[scale=0.5]{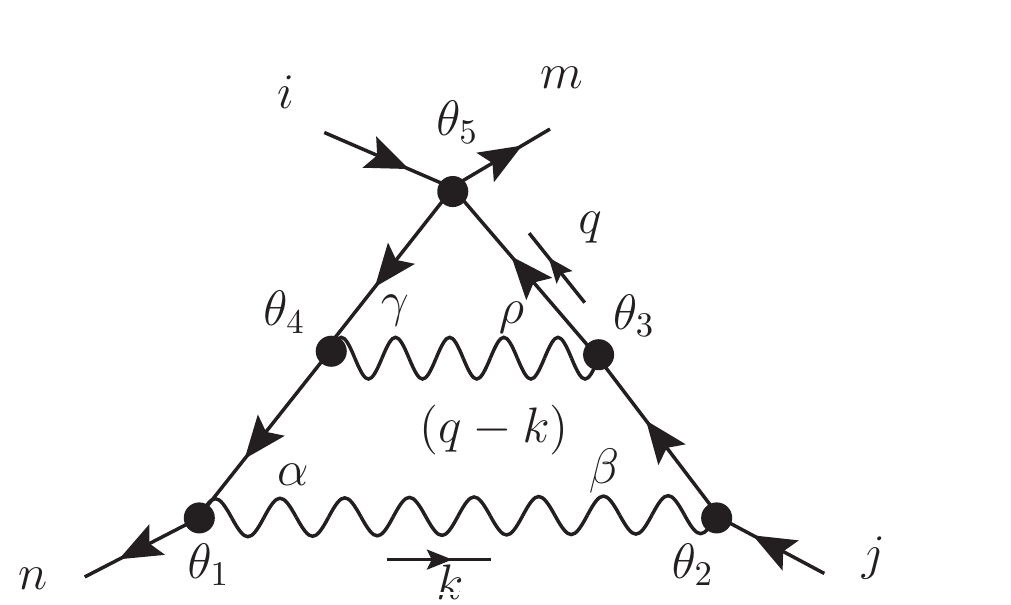}}\subfloat[]{\centering{}\includegraphics[scale=0.5]{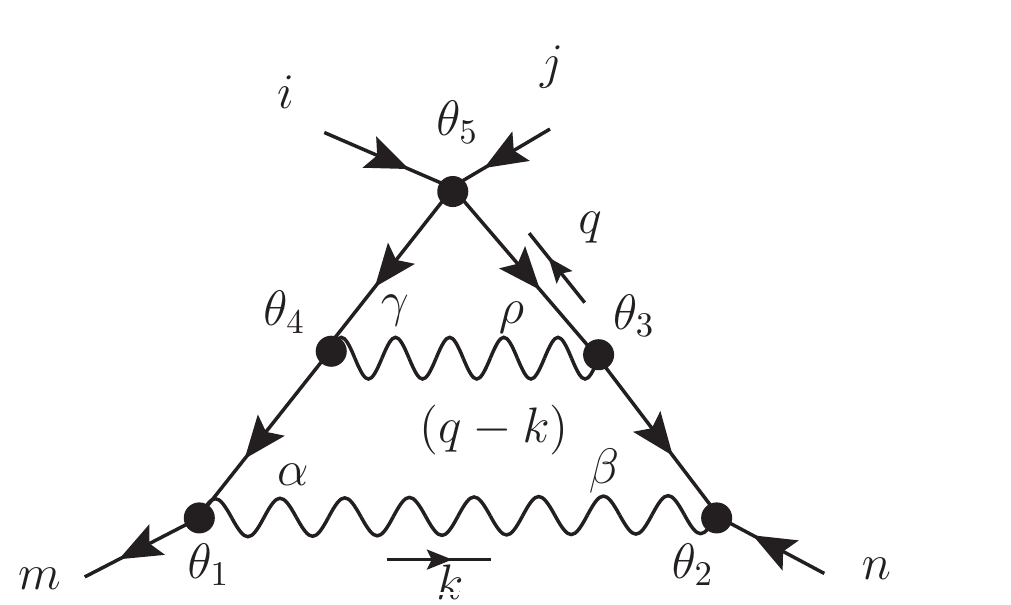}}
\par\end{centering}
\centering{}\caption{\label{fig:D13-order-lambda-g-4}$\mathcal{S}_{\left(\overline{\Phi}\Phi\right)^{2}}^{\left(D13\right)}$}
\end{figure}
\par\end{center}

$\mathcal{S}_{\left(\overline{\Phi}\Phi\right)^{2}}^{\left(D13-a\right)}$
in the Figure\,\ref{fig:D13-order-lambda-g-4} is
\begin{align}
\mathcal{S}_{\left(\overline{\Phi}\Phi\right)^{2}}^{\left(D13-a\right)} & =\frac{1}{128}\,i\,\lambda\,g^{4}\left(\delta_{jn}\delta_{im}+\delta_{mj}\delta_{ni}\right)\int_{\theta}\overline{\Phi}_{i}\Phi_{m}\Phi_{n}\overline{\Phi}_{j}\nonumber \\
 & \times\int\frac{d^{D}kd^{D}q}{\left(2\pi\right)^{2D}}\left\{ \frac{16\left(a\,b-a^{2}\right)\,\left(k^{2}\right)^{2}q^{2}}{\left(k^{2}\right)^{3}\left(q^{2}\right)^{2}\left(q-k\right)^{2}}\right\} \,,
\end{align}
using Eq.\,(\ref{eq:Int 7}), then adding $\mathcal{S}_{\left(\overline{\Phi}\Phi\right)^{2}}^{\left(D13-a\right)}$
to $\mathcal{S}_{\left(\overline{\Phi}\Phi\right)^{2}}^{\left(D13-f\right)}$,
we have 
\begin{align}
\mathcal{S}_{\left(\overline{\Phi}\Phi\right)^{2}}^{\left(D13\right)} & =\frac{3}{4}\,a\left(\frac{a-b}{32\pi^{2}\epsilon}\right)i\,\lambda\,g^{4}\left(\delta_{jn}\delta_{im}+\delta_{mj}\delta_{ni}\right)\int_{\theta}\overline{\Phi}_{i}\Phi_{m}\Phi_{n}\overline{\Phi}_{j}\,.\label{eq:S-D13}
\end{align}

\begin{center}
\begin{figure}
\begin{centering}
\subfloat[]{\centering{}\includegraphics[scale=0.5]{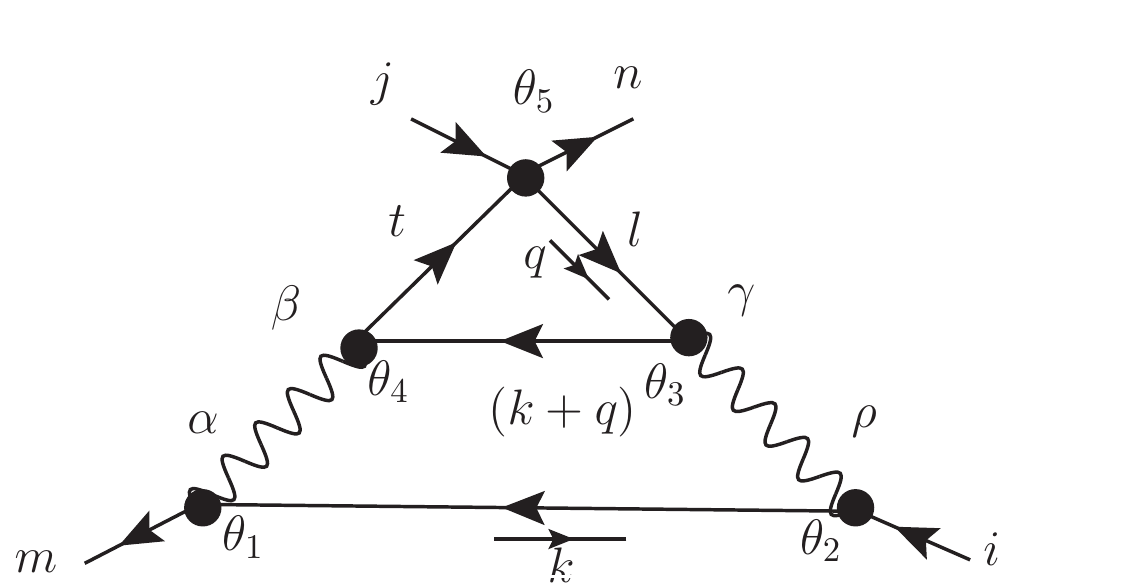}}\subfloat[]{\centering{}\includegraphics[scale=0.5]{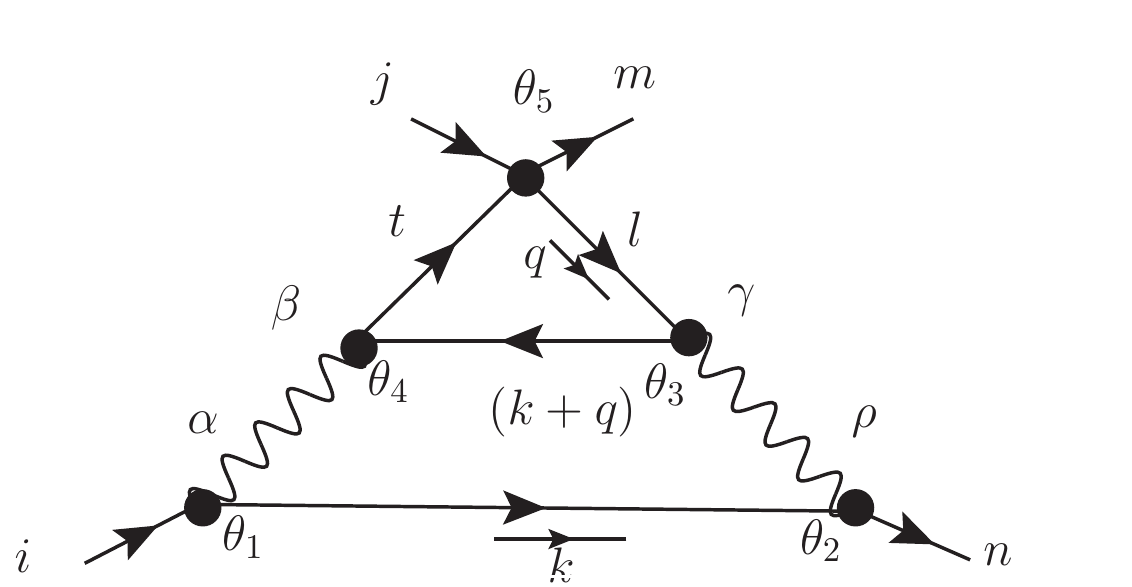}}\subfloat[]{\centering{}\includegraphics[scale=0.5]{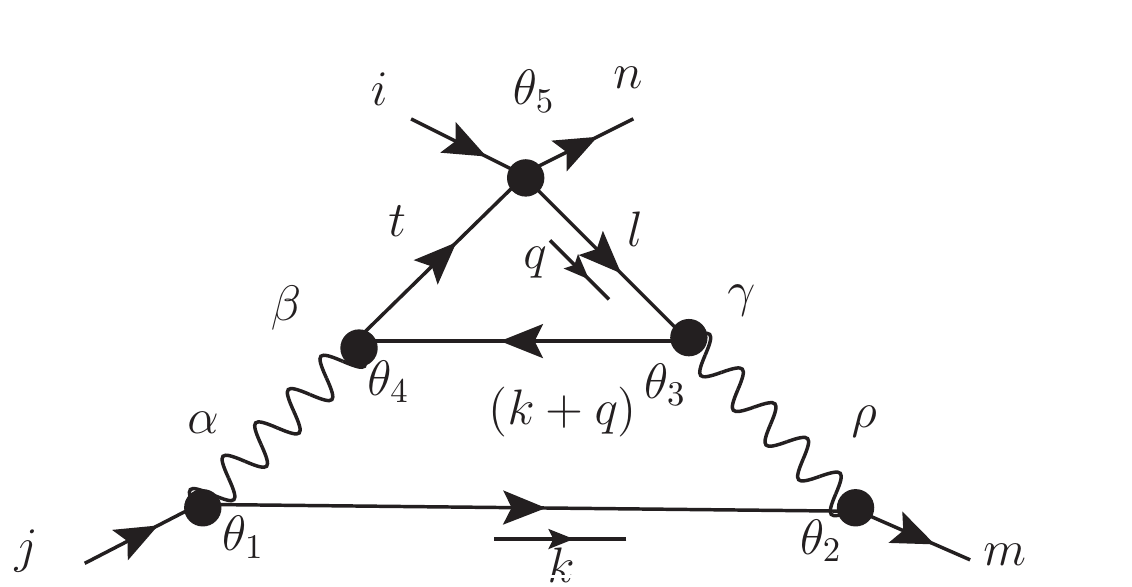}}
\par\end{centering}
\begin{centering}
\subfloat[]{\centering{}\includegraphics[scale=0.5]{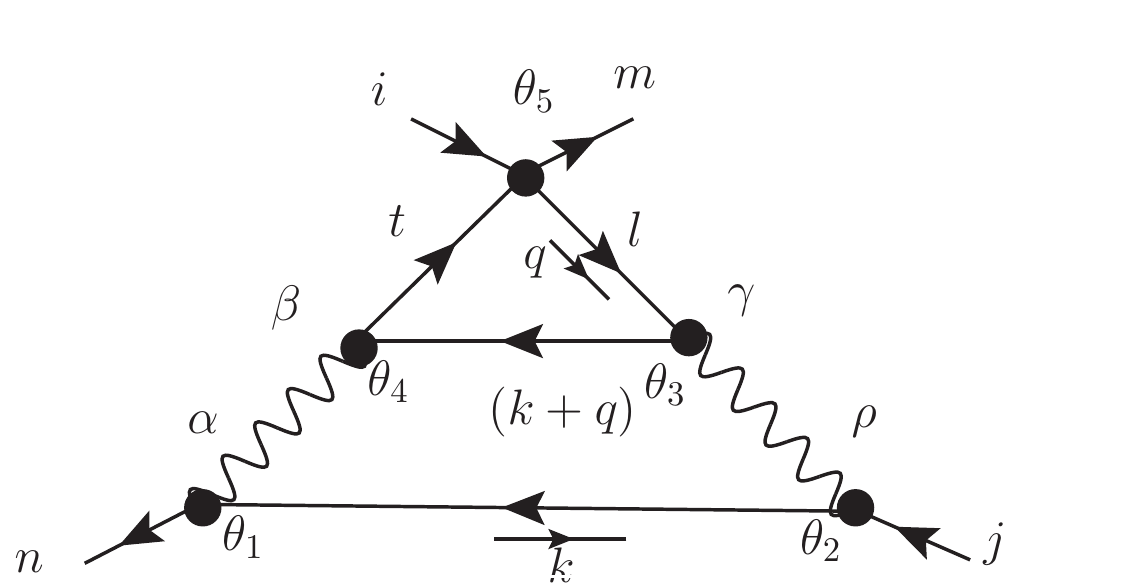}}
\par\end{centering}
\centering{}\caption{\label{fig:D14-order-lambda-g-4}$\mathcal{S}_{\left(\overline{\Phi}\Phi\right)^{2}}^{\left(D14\right)}$}
\end{figure}
\par\end{center}

$\mathcal{S}_{\left(\overline{\Phi}\Phi\right)^{2}}^{\left(D14-a\right)}$
in the Figure\,\ref{fig:D14-order-lambda-g-4} is
\begin{align}
\mathcal{S}_{\left(\overline{\Phi}\Phi\right)^{2}}^{\left(D14-a\right)} & =\frac{1}{128}\,i\,\lambda\,g^{4}\left(N+1\right)\delta_{jn}\delta_{mi}\int_{\theta}\overline{\Phi}_{i}\Phi_{m}\Phi_{n}\overline{\Phi}_{j}\int\frac{d^{D}kd^{D}q}{\left(2\pi\right)^{2D}}\left\{ \frac{-4\left(a-b\right)^{2}k^{2}q^{2}\left(2\left(k\cdot q\right)+k^{2}\right)}{\left(k^{2}\right)^{3}\left(k+q\right)^{2}\left(q^{2}\right)^{2}}\right\} \,,
\end{align}
using Eqs.\,(\ref{eq:Int 7}), (\ref{eq:Int 9}) and momentum power
counting, we find, 
\begin{align}
\mathcal{S}_{\left(\overline{\Phi}\Phi\right)^{2}}^{\left(D14-a\right)} & =\mathcal{S}_{\left(\overline{\Phi}\Phi\right)^{2}}^{\left(D14\right)}=0\,.\label{eq:S-D14}
\end{align}

\begin{center}
\begin{figure}
\begin{centering}
\subfloat[]{\centering{}\includegraphics[scale=0.45]{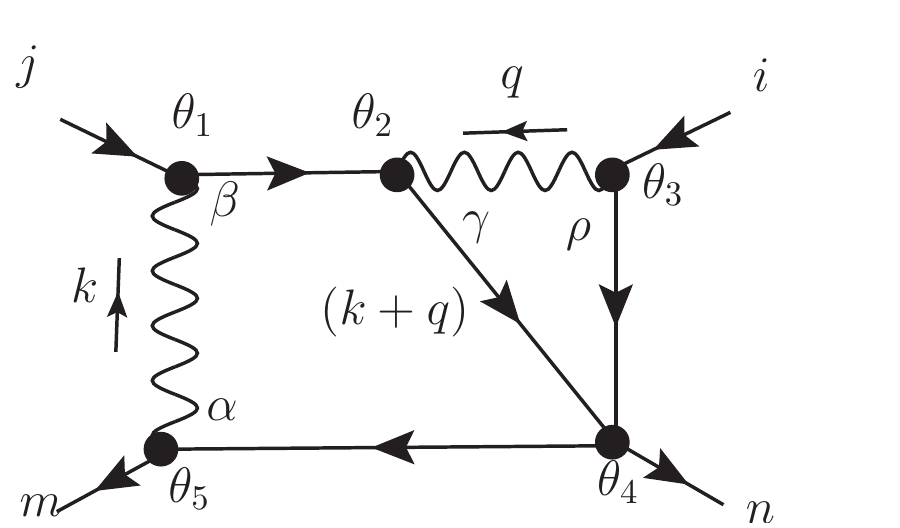}}\subfloat[]{\centering{}\includegraphics[scale=0.45]{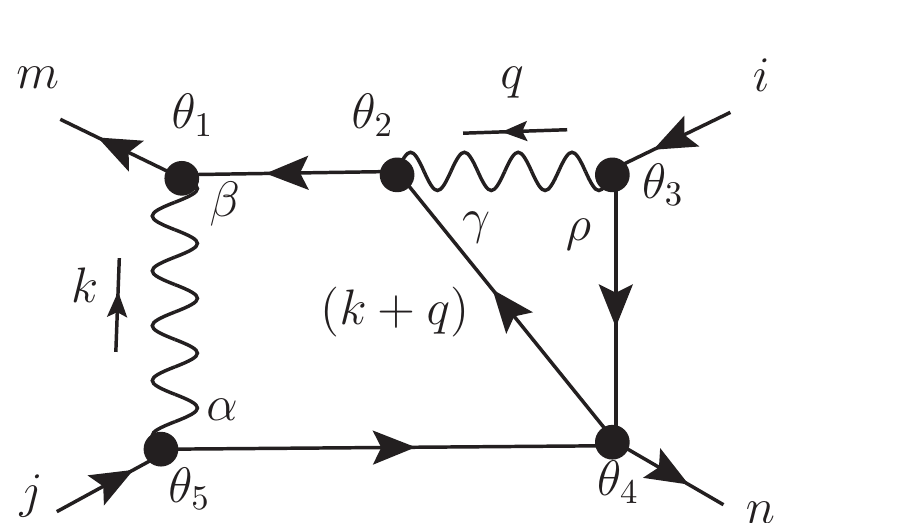}}\subfloat[]{\centering{}\includegraphics[scale=0.45]{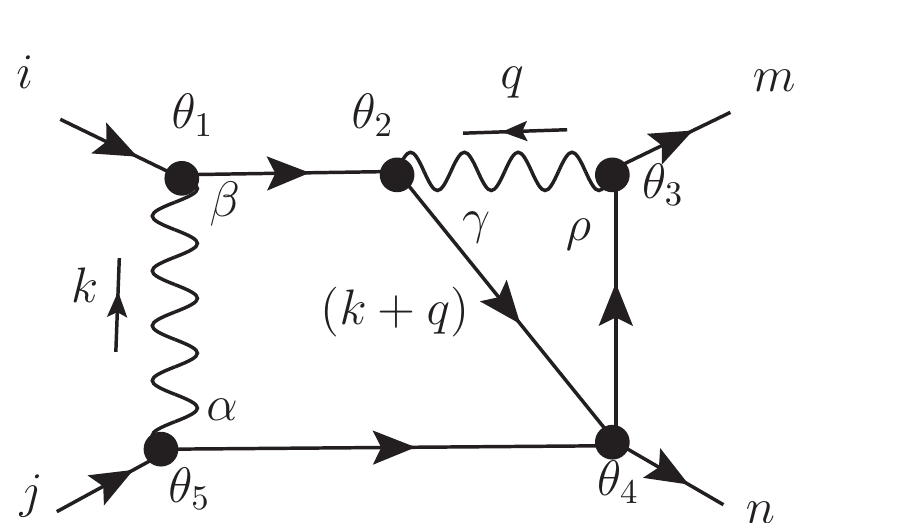}}\subfloat[]{\centering{}\includegraphics[scale=0.45]{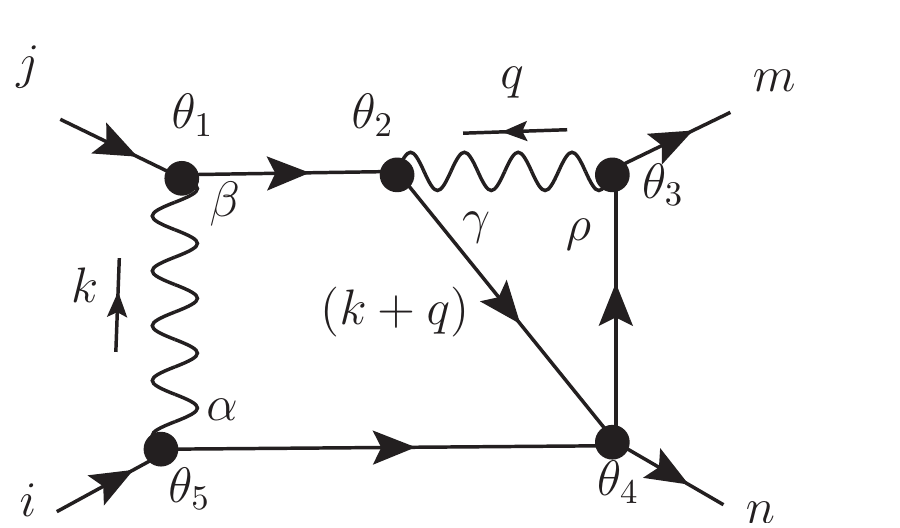}}
\par\end{centering}
\begin{centering}
\subfloat[]{\centering{}\includegraphics[scale=0.45]{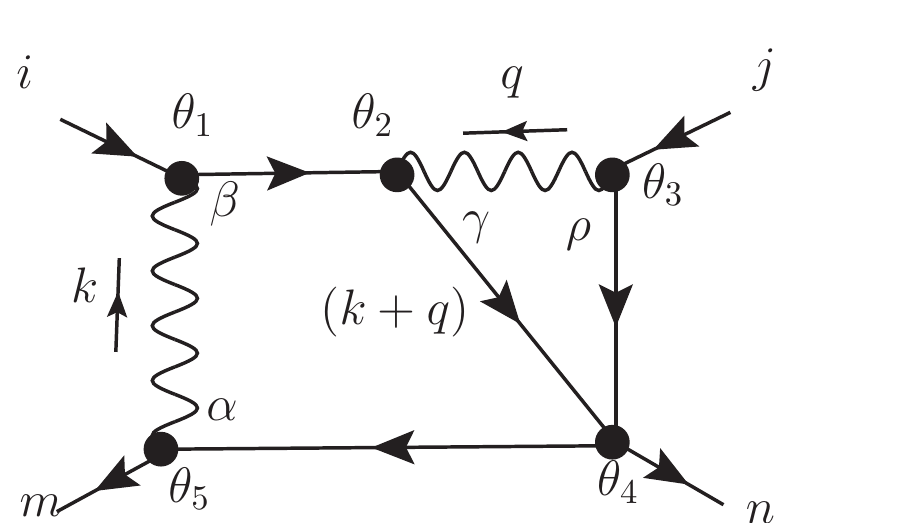}}\subfloat[]{\centering{}\includegraphics[scale=0.45]{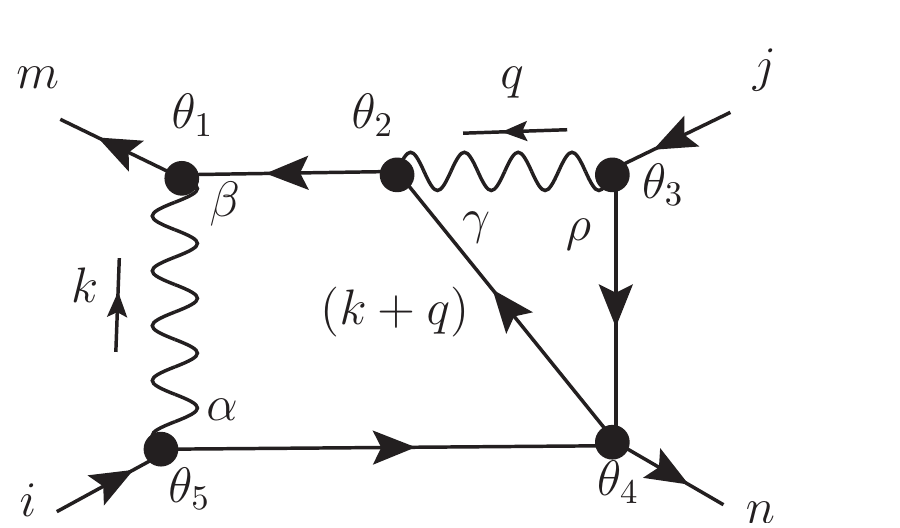}}\subfloat[]{\centering{}\includegraphics[scale=0.45]{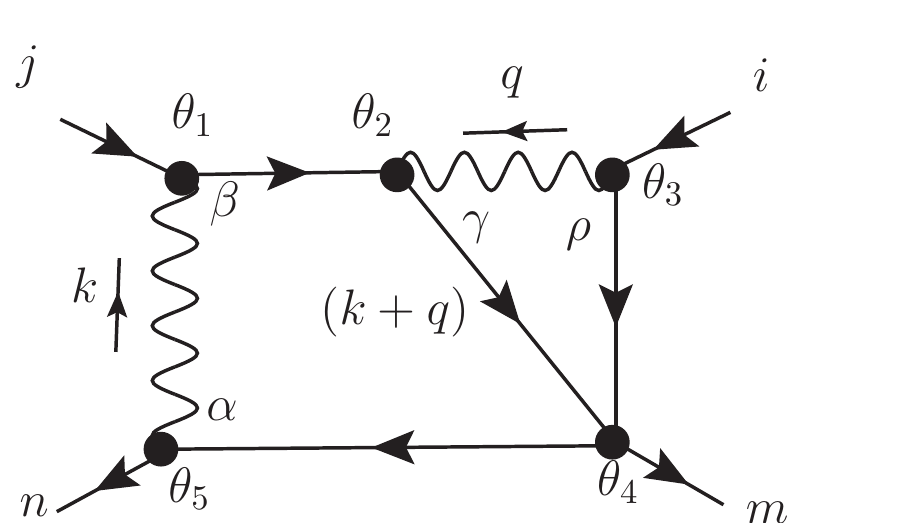}}\subfloat[]{\centering{}\includegraphics[scale=0.45]{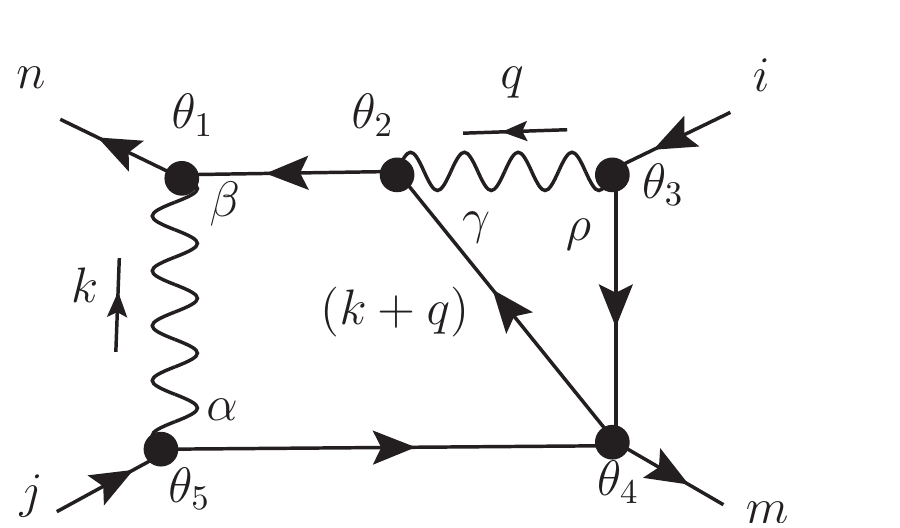}}
\par\end{centering}
\begin{centering}
\subfloat[]{\centering{}\includegraphics[scale=0.45]{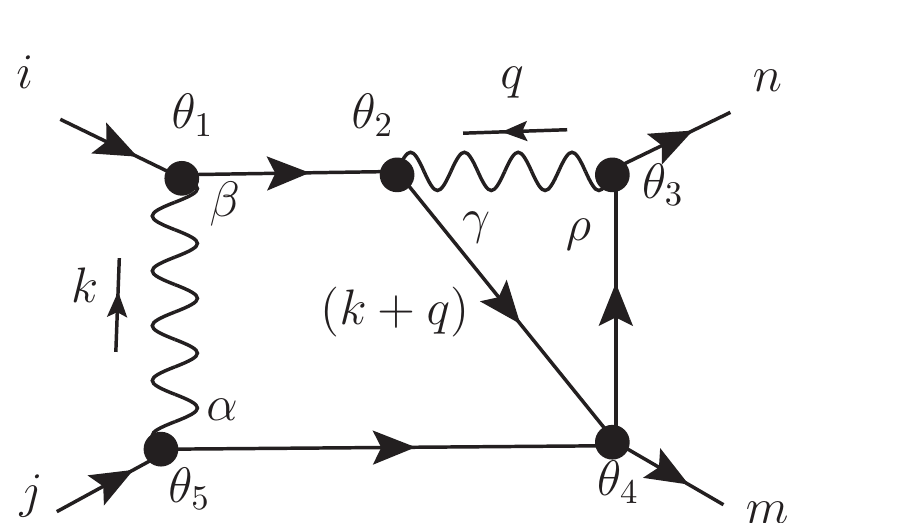}}\subfloat[]{\centering{}\includegraphics[scale=0.45]{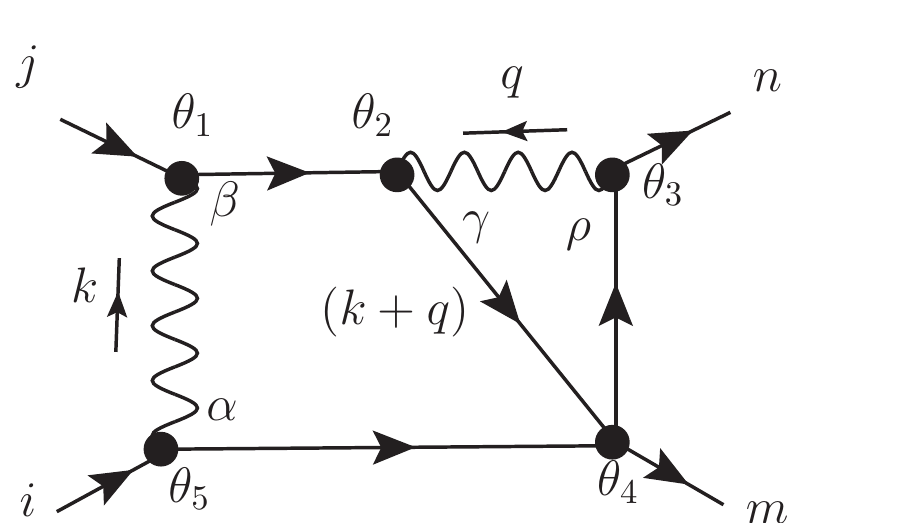}}\subfloat[]{\centering{}\includegraphics[scale=0.45]{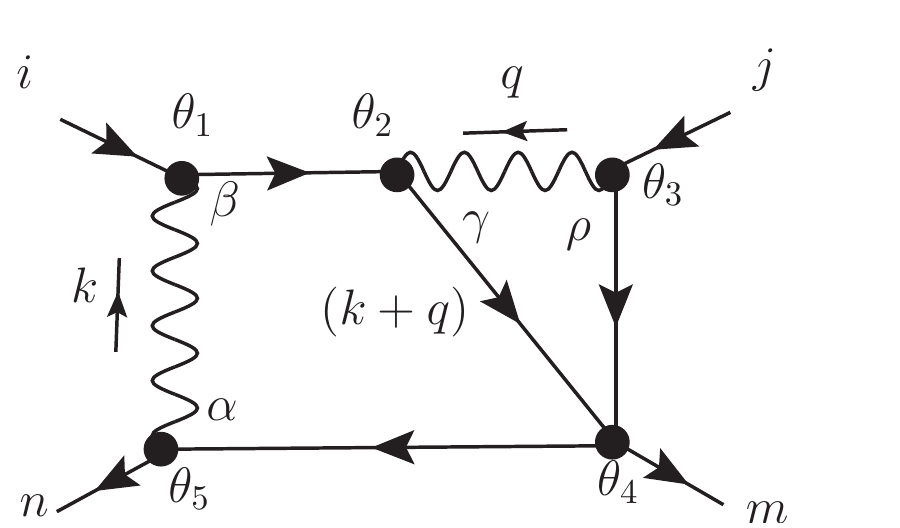}}\subfloat[]{\centering{}\includegraphics[scale=0.45]{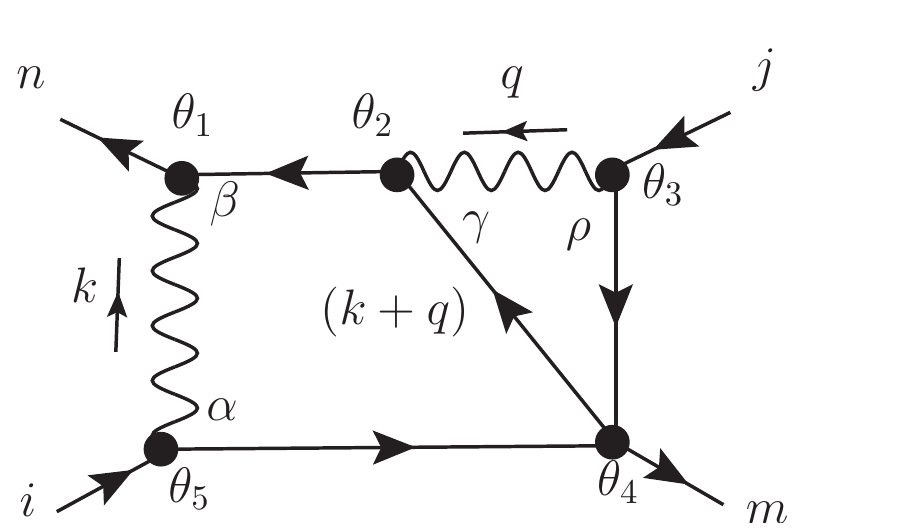}}
\par\end{centering}
\begin{centering}
\subfloat[]{\centering{}\includegraphics[scale=0.45]{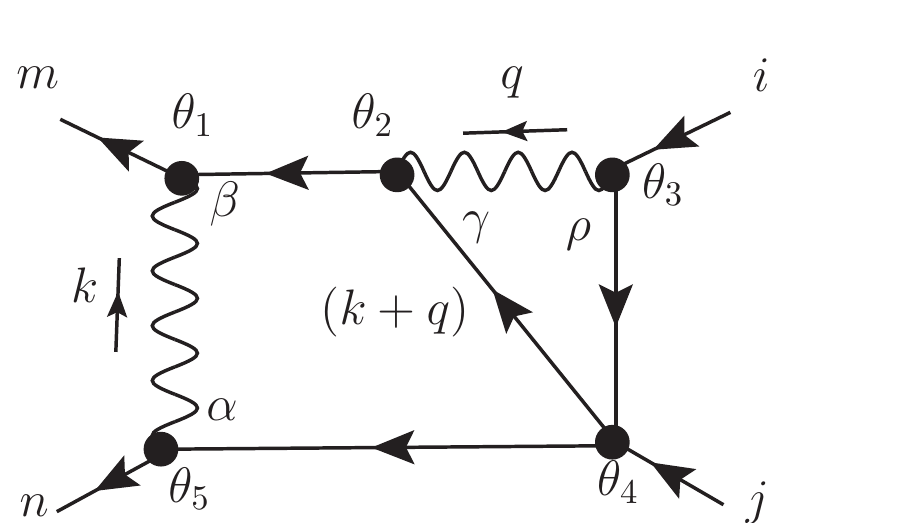}}\subfloat[]{\centering{}\includegraphics[scale=0.45]{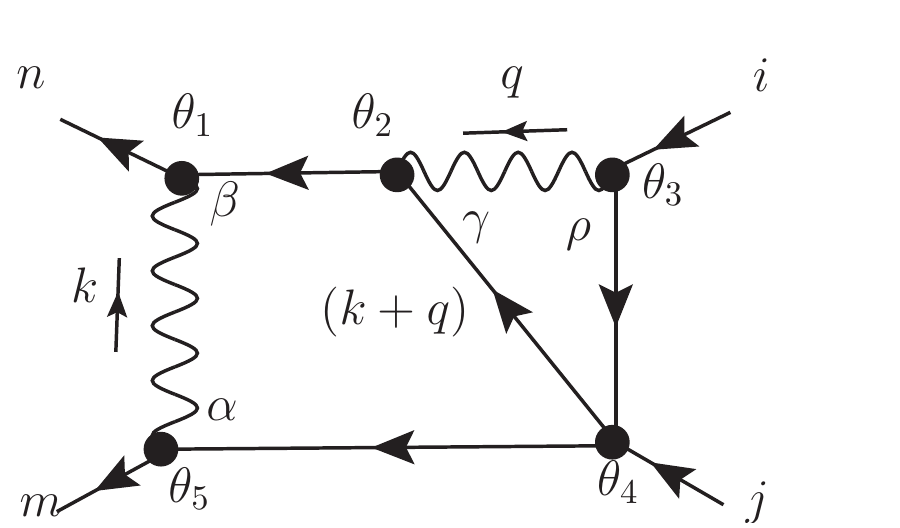}}\subfloat[]{\centering{}\includegraphics[scale=0.45]{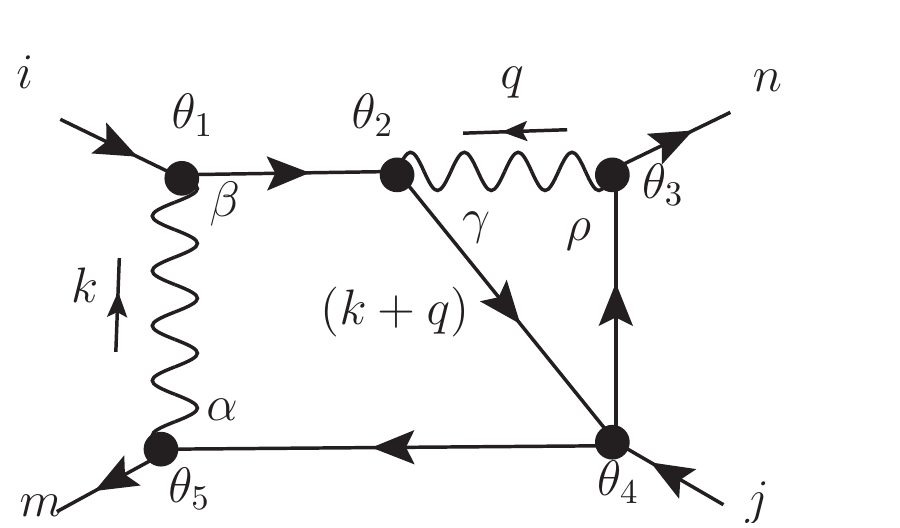}}\subfloat[]{\centering{}\includegraphics[scale=0.45]{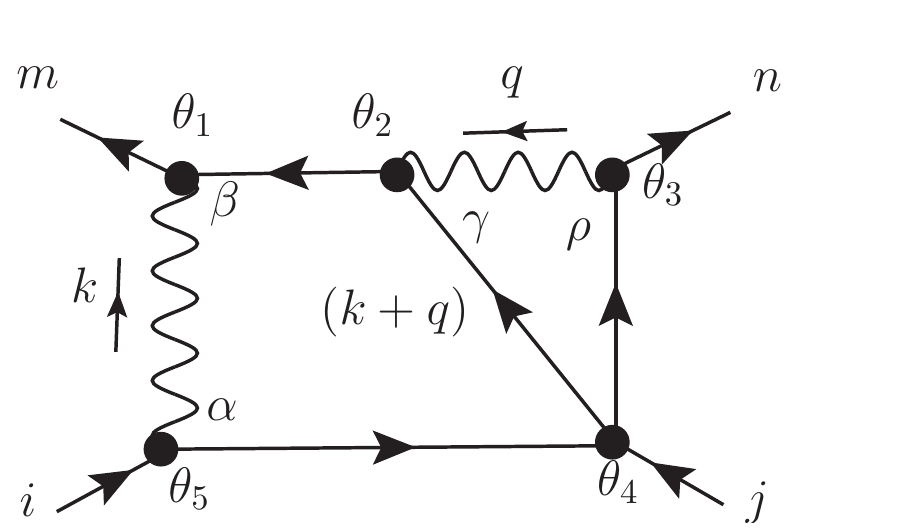}}
\par\end{centering}
\begin{centering}
\subfloat[]{\centering{}\includegraphics[scale=0.45]{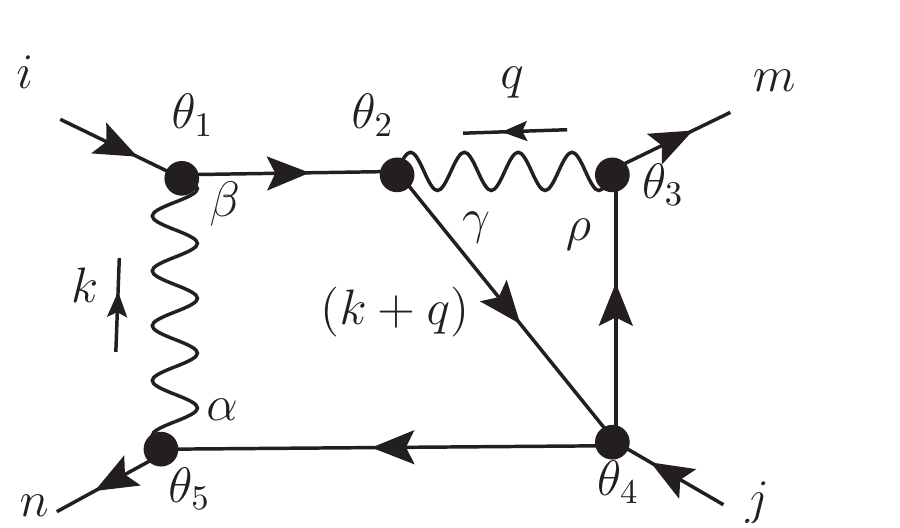}}\subfloat[]{\centering{}\includegraphics[scale=0.45]{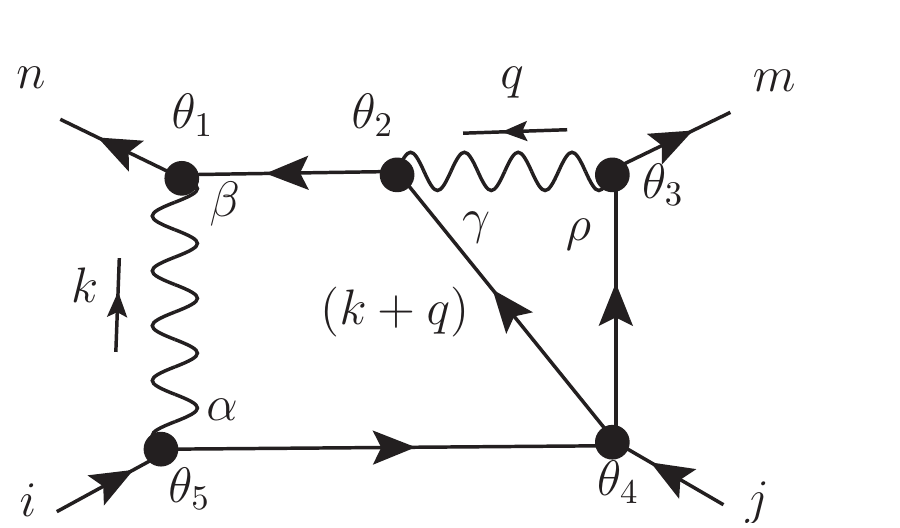}}\subfloat[]{\centering{}\includegraphics[scale=0.45]{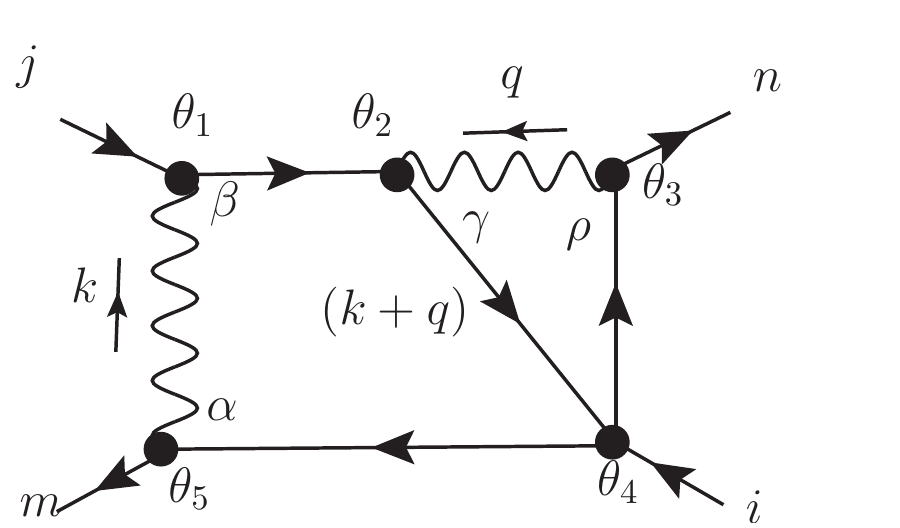}}\subfloat[]{\centering{}\includegraphics[scale=0.45]{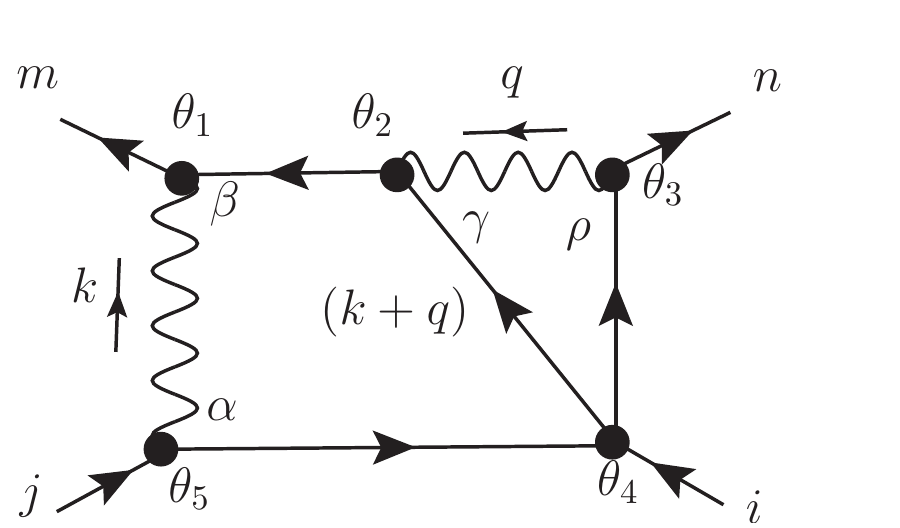}}
\par\end{centering}
\begin{centering}
\subfloat[]{\centering{}\includegraphics[scale=0.45]{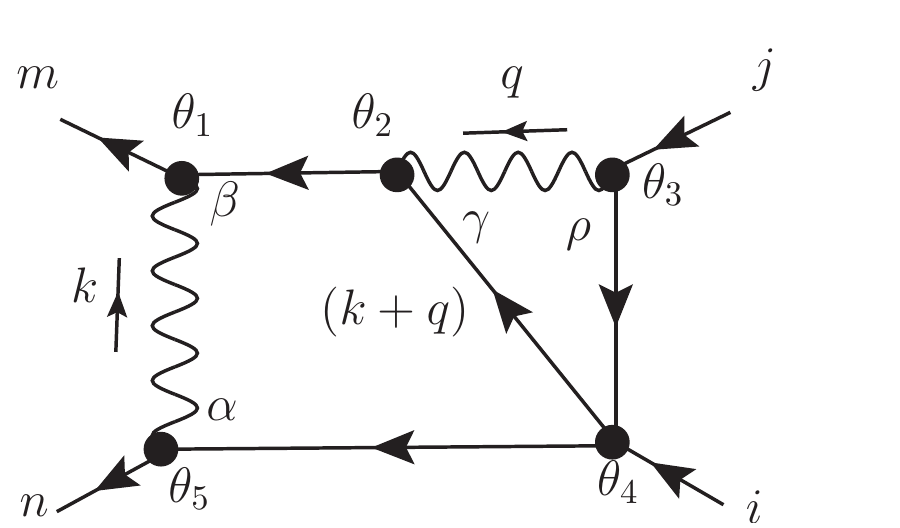}}\subfloat[]{\centering{}\includegraphics[scale=0.45]{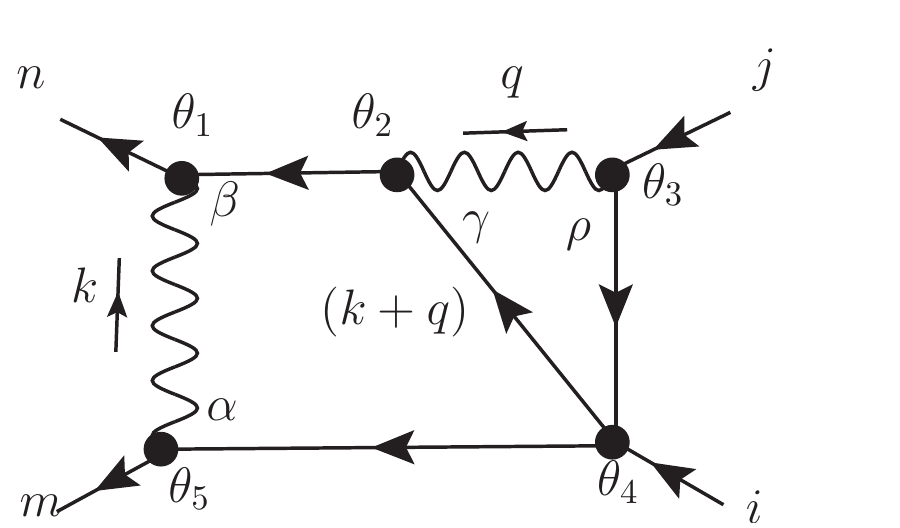}}\subfloat[]{\centering{}\includegraphics[scale=0.45]{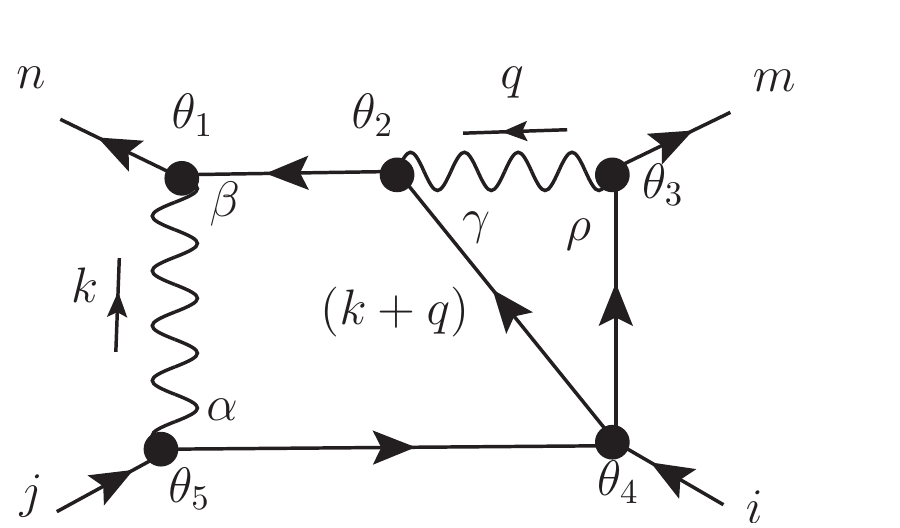}}\subfloat[]{\centering{}\includegraphics[scale=0.45]{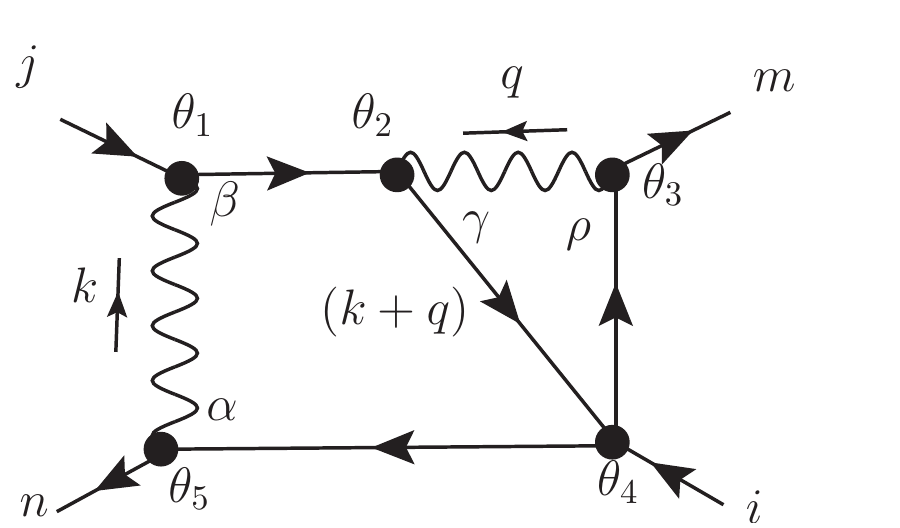}}
\par\end{centering}
\centering{}\caption{\label{fig:D15-order-lambda-g-4}$\mathcal{S}_{\left(\overline{\Phi}\Phi\right)^{2}}^{\left(D15\right)}$}
\end{figure}
\par\end{center}

\begin{center}
\begin{table}
\centering{}%
\begin{tabular}{lcccccc}
 &  &  &  &  &  & \tabularnewline
\hline 
\hline 
$D15-a$ &  & $-\left(\delta_{jn}\delta_{im}+\delta_{mj}\delta_{ni}\right)$ &  & $D15-b$ &  & $\delta_{jn}\delta_{im}+\delta_{mj}\delta_{ni}$\tabularnewline
$D15-c$ &  & $-\left(\delta_{jn}\delta_{im}+\delta_{mj}\delta_{ni}\right)$ &  & $D15-d$ &  & $-\left(\delta_{jn}\delta_{im}+\delta_{mj}\delta_{ni}\right)$\tabularnewline
$D15-e$ &  & $-\left(\delta_{jn}\delta_{im}+\delta_{mj}\delta_{ni}\right)$ &  & $D15-f$ &  & $\delta_{jn}\delta_{im}+\delta_{mj}\delta_{ni}$\tabularnewline
$D15-g$ &  & $-\left(\delta_{jn}\delta_{im}+\delta_{mj}\delta_{ni}\right)$ &  & $D15-h$ &  & $\delta_{jn}\delta_{im}+\delta_{mj}\delta_{ni}$\tabularnewline
$D15-i$ &  & $-\left(\delta_{jn}\delta_{im}+\delta_{mj}\delta_{ni}\right)$ &  & $D15-j$ &  & $-\left(\delta_{jn}\delta_{im}+\delta_{mj}\delta_{ni}\right)$\tabularnewline
$D15-k$ &  & $-\left(\delta_{jn}\delta_{im}+\delta_{mj}\delta_{ni}\right)$ &  & $D15-l$ &  & $\delta_{jn}\delta_{im}+\delta_{mj}\delta_{ni}$\tabularnewline
$D15-m$ &  & $-\left(\delta_{jn}\delta_{im}+\delta_{mj}\delta_{ni}\right)$ &  & $D15-n$ &  & $-\left(\delta_{jn}\delta_{im}+\delta_{mj}\delta_{ni}\right)$\tabularnewline
$D15-o$ &  & $\delta_{jn}\delta_{im}+\delta_{mj}\delta_{ni}$ &  & $D15-p$ &  & $-\left(\delta_{jn}\delta_{im}+\delta_{mj}\delta_{ni}\right)$\tabularnewline
$D15-q$ &  & $\delta_{jn}\delta_{im}+\delta_{mj}\delta_{ni}$ &  & $D15-r$ &  & $-\left(\delta_{jn}\delta_{im}+\delta_{mj}\delta_{ni}\right)$\tabularnewline
$D15-s$ &  & $\delta_{jn}\delta_{im}+\delta_{mj}\delta_{ni}$ &  & $D15-t$ &  & $-\left(\delta_{jn}\delta_{im}+\delta_{mj}\delta_{ni}\right)$\tabularnewline
$D15-u$ &  & $-\left(\delta_{jn}\delta_{im}+\delta_{mj}\delta_{ni}\right)$ &  & $D15-v$ &  & $-\left(\delta_{jn}\delta_{im}+\delta_{mj}\delta_{ni}\right)$\tabularnewline
$D15-w$ &  & $-\left(\delta_{jn}\delta_{im}+\delta_{mj}\delta_{ni}\right)$ &  & $D15-x$ &  & $-\left(\delta_{jn}\delta_{im}+\delta_{mj}\delta_{ni}\right)$\tabularnewline
\hline 
\hline 
 &  &  &  &  &  & \tabularnewline
\end{tabular}\caption{\label{tab:S-PPPP-15}Values of the diagrams in Figure\,\ref{fig:D15-order-lambda-g-4}
with common factor\protect \\
 $\frac{1}{32}\left(\frac{\left(a-b\right)^{2}}{32\pi^{2}\epsilon}\right)\,i\,\lambda\,g^{4}\int_{\theta}\overline{\Phi}_{i}\Phi_{m}\Phi_{n}\overline{\Phi}_{j}$
.}
\end{table}
\par\end{center}

$\mathcal{S}_{\left(\overline{\Phi}\Phi\right)^{2}}^{\left(D15-a\right)}$
in the Figure\,\ref{fig:D15-order-lambda-g-4} is
\begin{align}
\mathcal{S}_{\left(\overline{\Phi}\Phi\right)^{2}}^{\left(D15-a\right)} & =\frac{1}{128}i\,\lambda\,g^{4}\left(\delta_{jn}\delta_{im}+\delta_{jm}\delta_{in}\right)\int_{\theta}\overline{\Phi}_{i}\Phi_{m}\Phi_{n}\overline{\Phi}_{j}\int\frac{d^{D}kd^{D}q}{\left(2\pi\right)^{2D}}\left\{ \frac{4\left(a-b\right)^{2}\left(k^{2}\right)^{2}q^{2}}{\left(k^{2}\right)^{3}\left(k+q\right)^{2}\left(q^{2}\right)^{2}}\right\} \,,
\end{align}
using Eq.\,(\ref{eq:Int 7}) and adding $\mathcal{S}_{\left(\overline{\Phi}\Phi\right)^{2}}^{\left(D15-a\right)}$
to $\mathcal{S}_{\left(\overline{\Phi}\Phi\right)^{2}}^{\left(D15-x\right)}$
with the values in the Table\,\ref{tab:S-PPPP-15}, we find 
\begin{align}
\mathcal{S}_{\left(\overline{\Phi}\Phi\right)^{2}}^{\left(D15\right)} & =-\frac{1}{4}\left(\frac{\left(a-b\right)^{2}}{32\pi^{2}\epsilon}\right)i\,\lambda\,g^{4}\left(\delta_{jn}\delta_{im}+\delta_{jm}\delta_{in}\right)\int_{\theta}\overline{\Phi}_{i}\Phi_{m}\Phi_{n}\overline{\Phi}_{j}\,.\label{eq:S-D15}
\end{align}

\begin{center}
\begin{figure}
\begin{centering}
\subfloat[]{\begin{centering}
\includegraphics[scale=0.5]{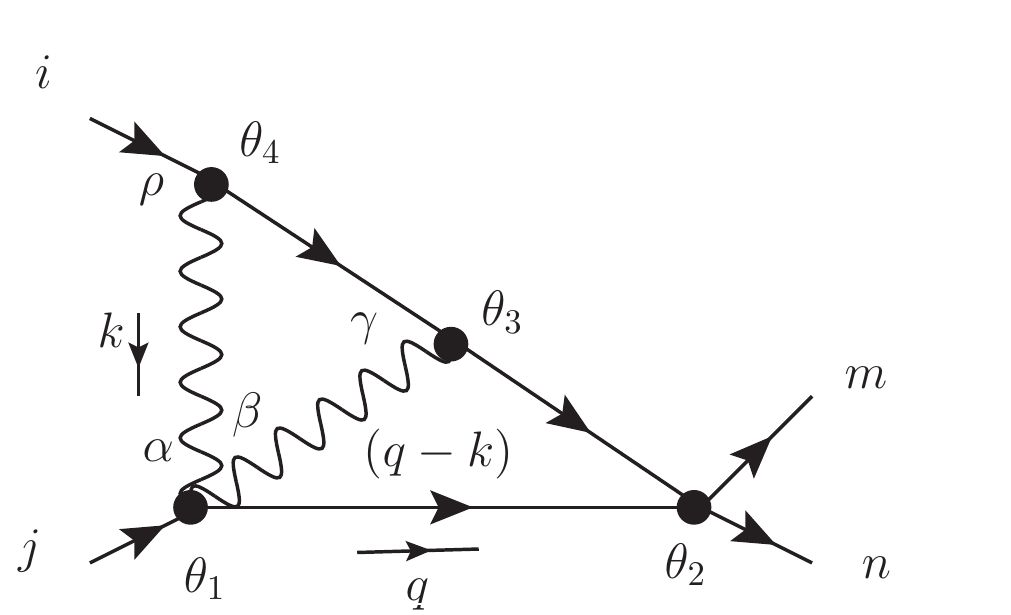}
\par\end{centering}
}\subfloat[]{\centering{}\includegraphics[scale=0.5]{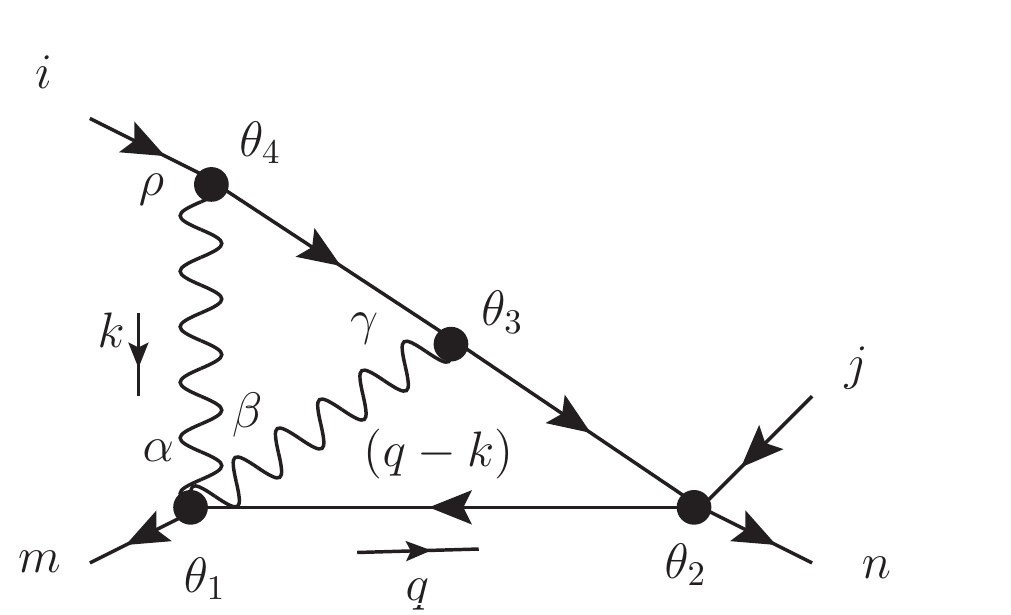}}\subfloat[]{\centering{}\includegraphics[scale=0.5]{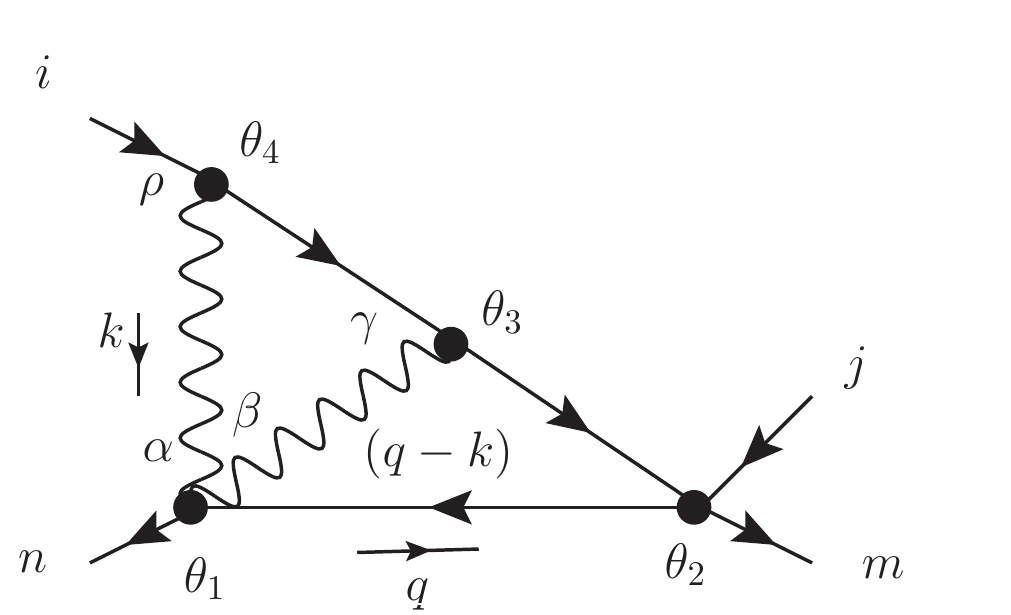}}
\par\end{centering}
\begin{centering}
\subfloat[]{\centering{}\includegraphics[scale=0.5]{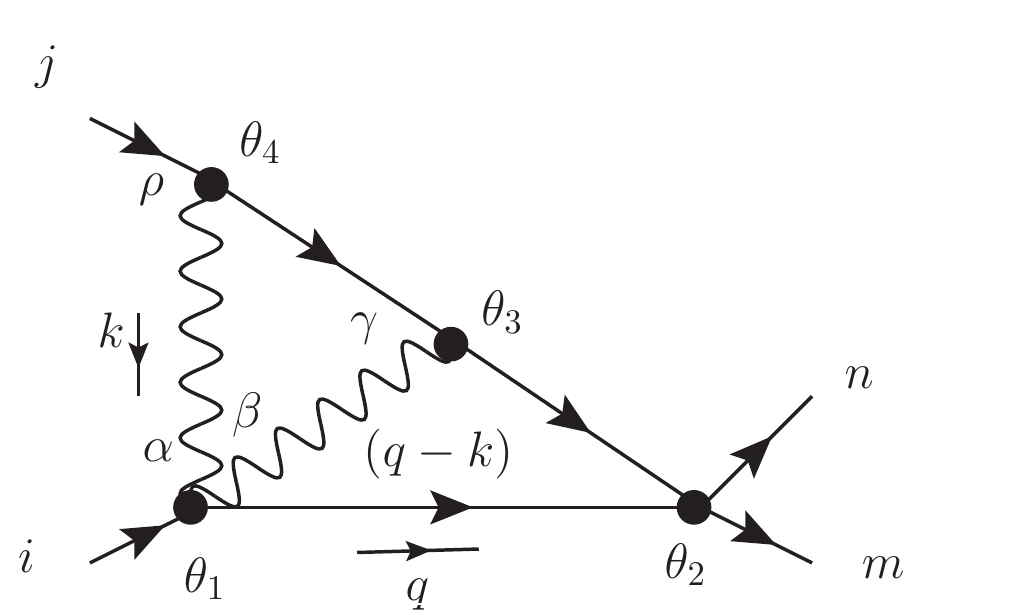}}\subfloat[]{\centering{}\includegraphics[scale=0.5]{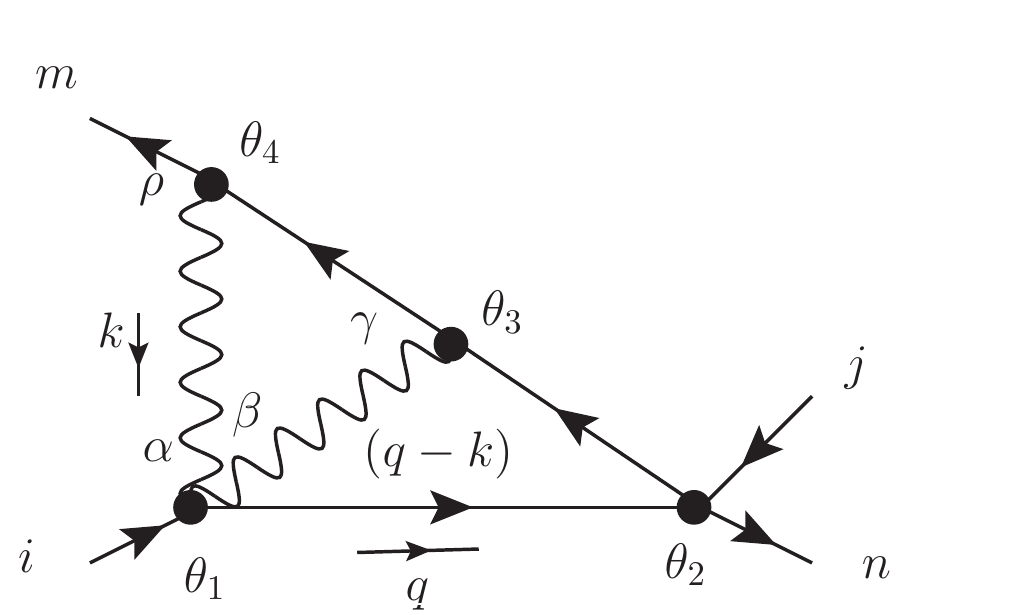}}\subfloat[]{\centering{}\includegraphics[scale=0.5]{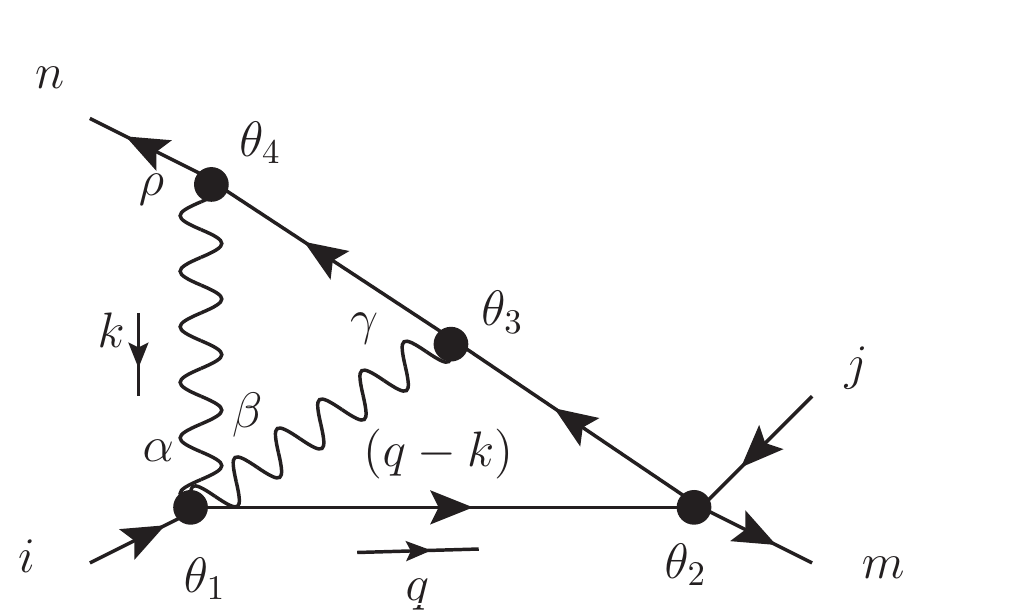}}
\par\end{centering}
\begin{centering}
\subfloat[]{\centering{}\includegraphics[scale=0.5]{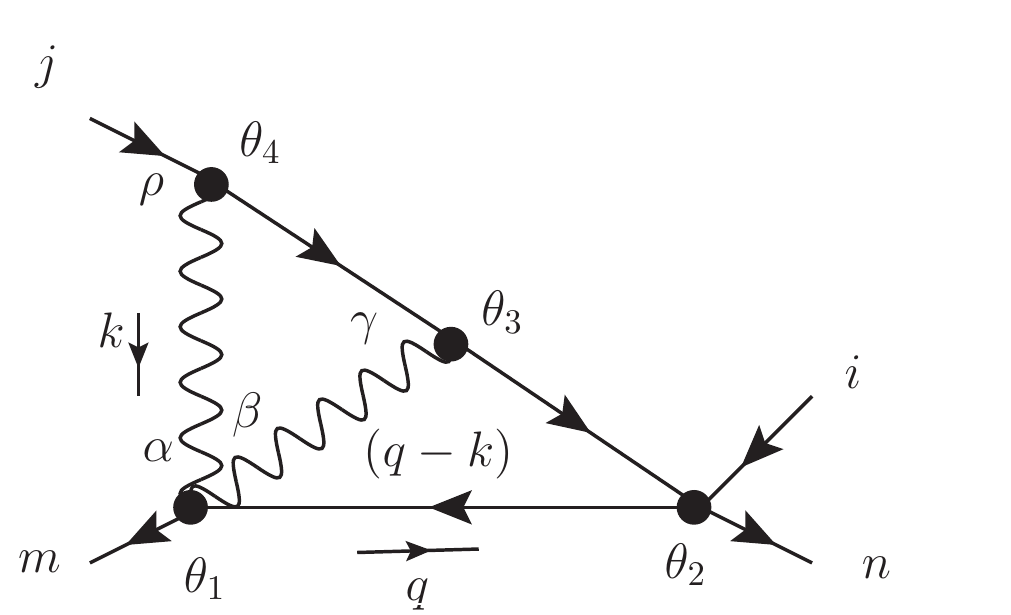}}\subfloat[]{\centering{}\includegraphics[scale=0.5]{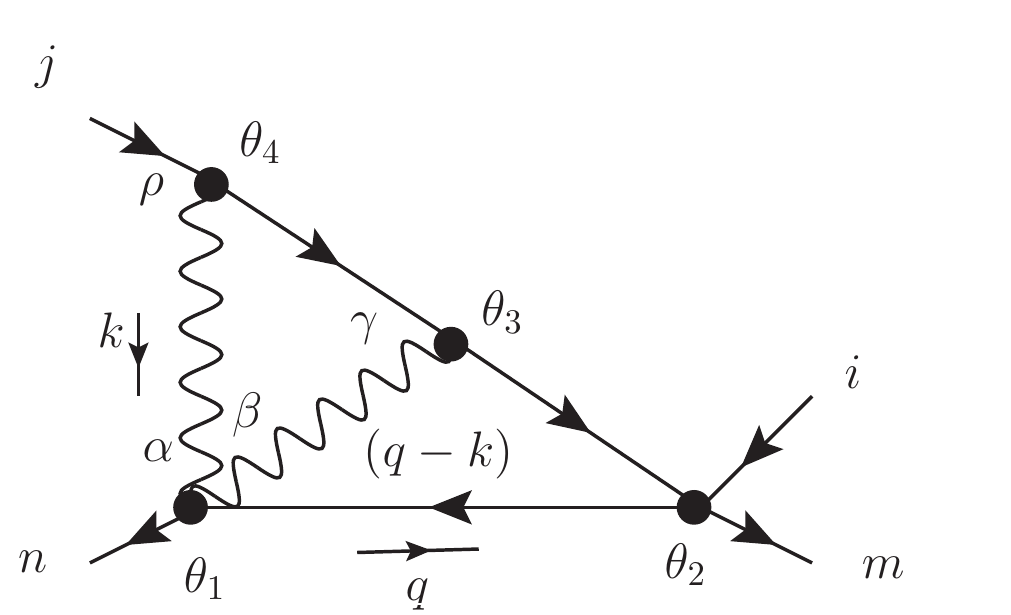}}\subfloat[]{\centering{}\includegraphics[scale=0.5]{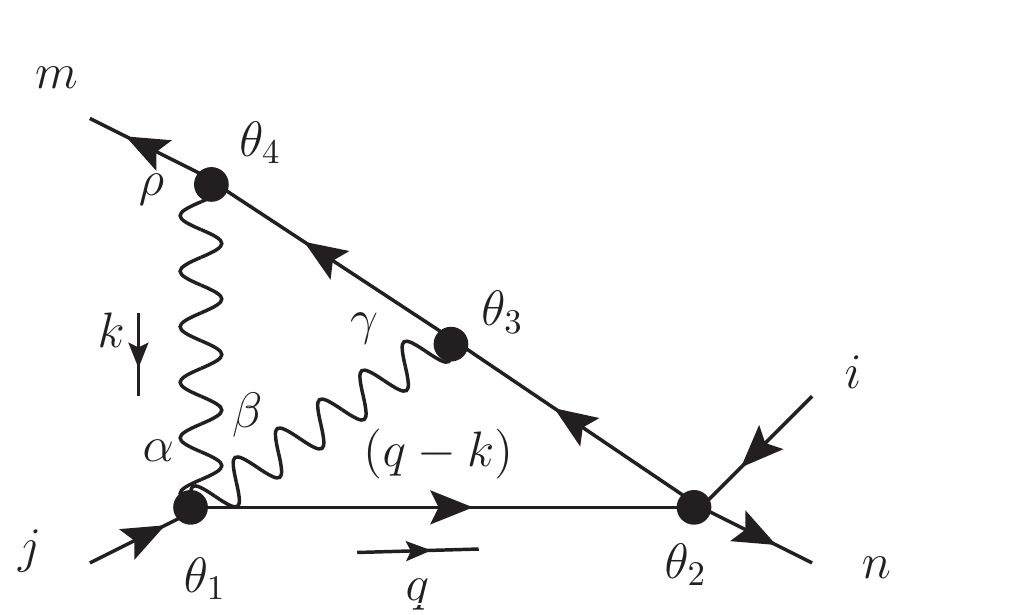}}
\par\end{centering}
\begin{centering}
\subfloat[]{\centering{}\includegraphics[scale=0.5]{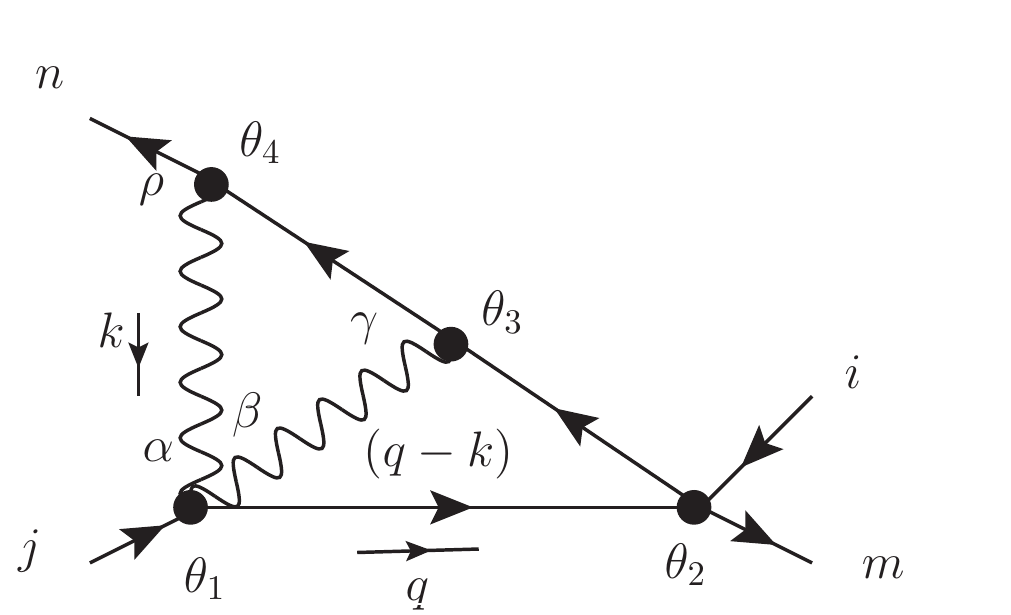}}\subfloat[]{\centering{}\includegraphics[scale=0.5]{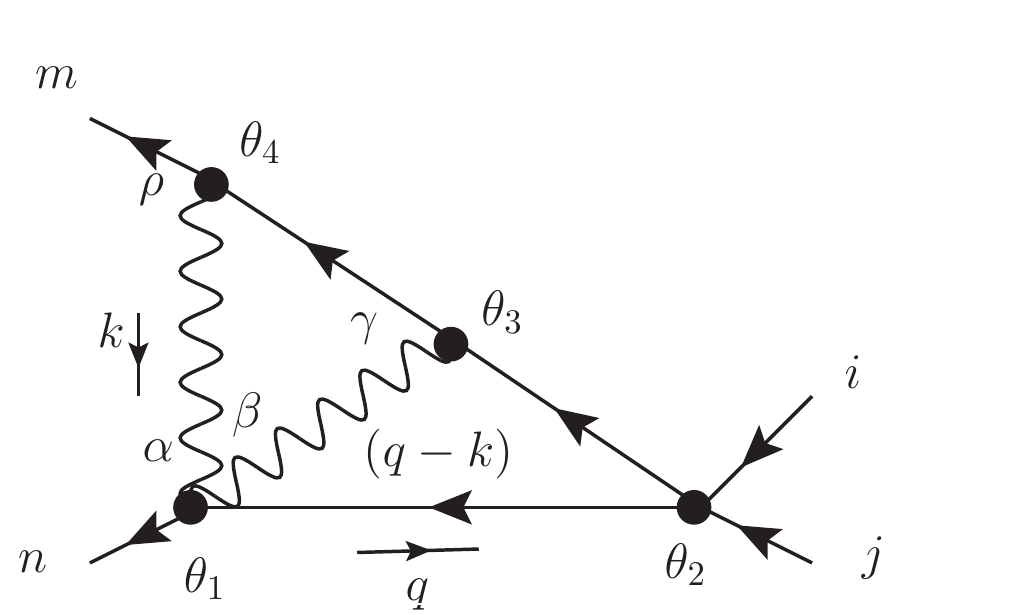}}\subfloat[]{\centering{}\includegraphics[scale=0.5]{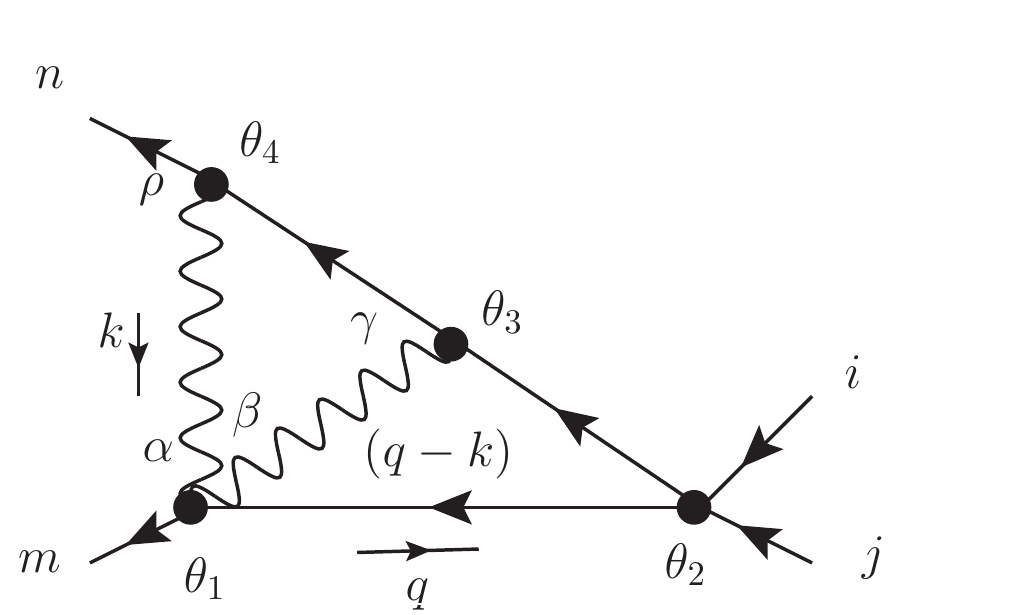}}
\par\end{centering}
\centering{}\caption{\label{fig:D16-order-lambda-g-4}$\mathcal{S}_{\left(\overline{\Phi}\Phi\right)^{2}}^{\left(D16\right)}$}
\end{figure}
\par\end{center}

$\mathcal{S}_{\left(\overline{\Phi}\Phi\right)^{2}}^{\left(D16-a\right)}$
in the Figure\,\ref{fig:D16-order-lambda-g-4} is
\begin{align}
\mathcal{S}_{\left(\overline{\Phi}\Phi\right)^{2}}^{\left(D16-a\right)} & =-\frac{1}{32}\,i\,\lambda\,g^{4}\left(\delta_{jn}\delta_{im}+\delta_{mj}\delta_{ni}\right)\int_{\theta}\overline{\Phi}_{i}\Phi_{m}\Phi_{n}\overline{\Phi}_{j}\int\frac{d^{D}kd^{D}q}{\left(2\pi\right)^{2D}}\left\{ \frac{\left(4\,a\,b-4\,a^{2}\right)\,k^{2}\,q^{2}}{\left(k^{2}\right)^{2}\left(q-k\right)^{2}\left(q^{2}\right)^{2}}\right\} \,,
\end{align}
using Eq.\,(\ref{eq:Int 7}) and adding $\mathcal{S}_{\left(\overline{\Phi}\Phi\right)^{2}}^{\left(D15-a\right)}$
to $\mathcal{S}_{\left(\overline{\Phi}\Phi\right)^{2}}^{\left(D15-l\right)}$
we find, 
\begin{align}
\mathcal{S}_{\left(\overline{\Phi}\Phi\right)^{2}}^{\left(D16\right)} & =\frac{3}{2}\,a\,\left(\frac{b-a}{32\pi^{2}\epsilon}\right)\,i\,\lambda\,g^{4}\left(\delta_{jn}\delta_{im}+\delta_{mj}\delta_{ni}\right)\int_{\theta}\overline{\Phi}_{i}\Phi_{m}\Phi_{n}\overline{\Phi}_{j}\,.\label{eq:S-D16}
\end{align}

\begin{center}
\begin{figure}
\begin{centering}
\subfloat[]{\begin{centering}
\includegraphics[scale=0.5]{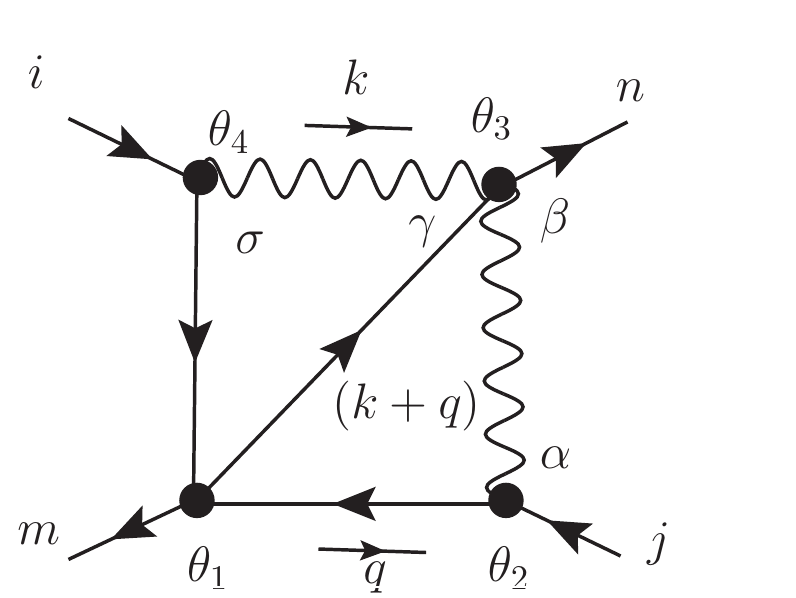}
\par\end{centering}
}\subfloat[]{\centering{}\includegraphics[scale=0.5]{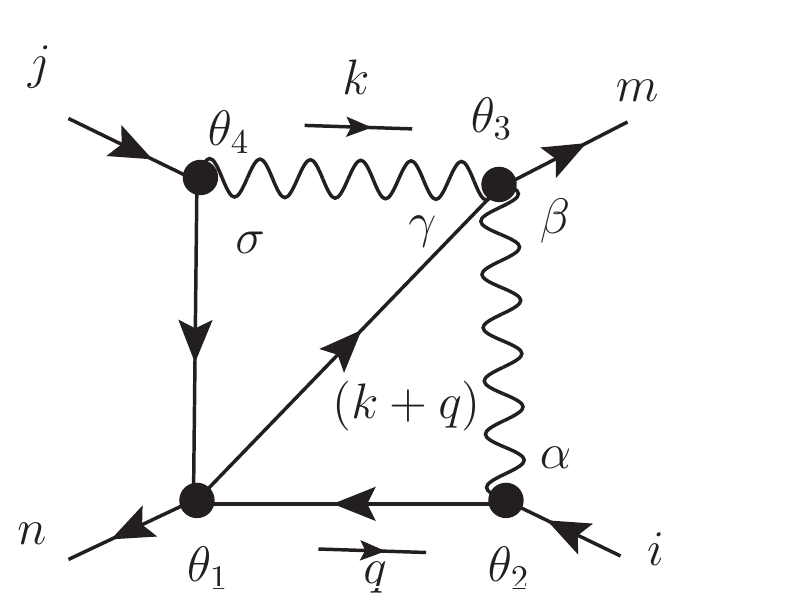}}\subfloat[]{\centering{}\includegraphics[scale=0.5]{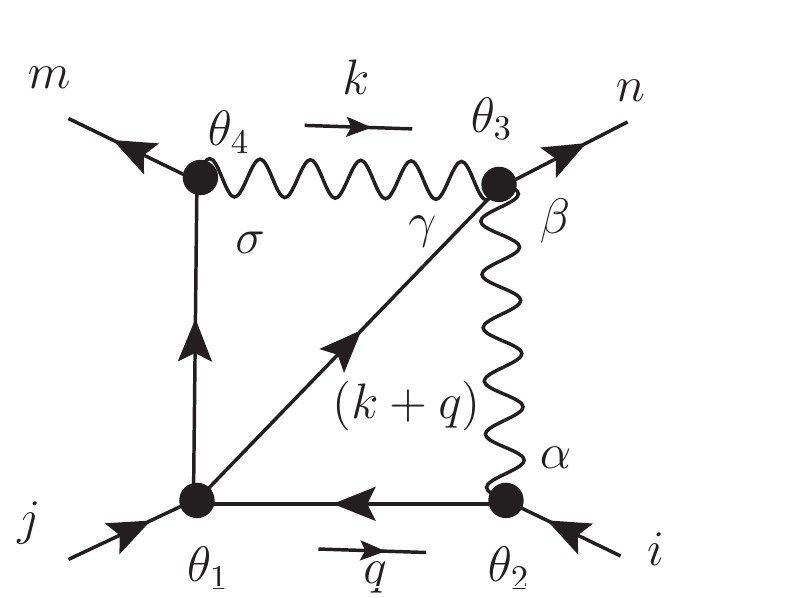}}\subfloat[]{\centering{}\includegraphics[scale=0.5]{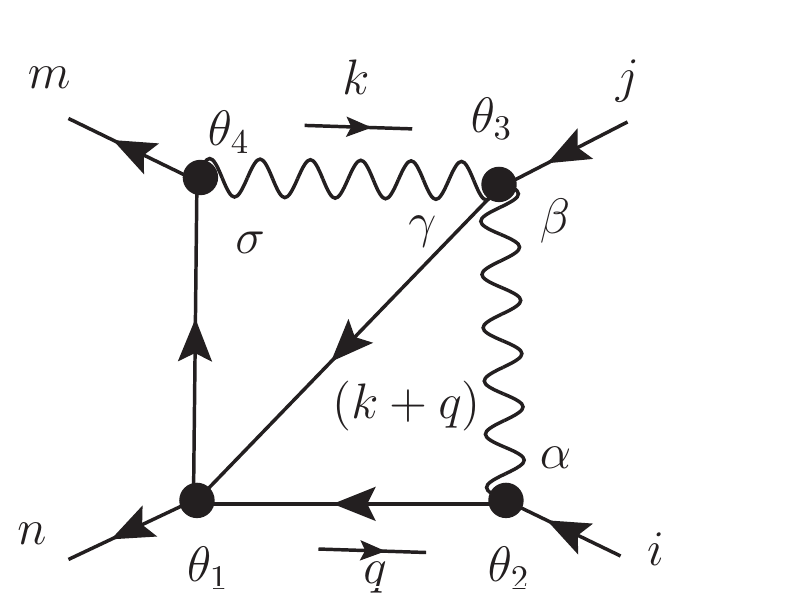}}
\par\end{centering}
\begin{centering}
\subfloat[]{\centering{}\includegraphics[scale=0.5]{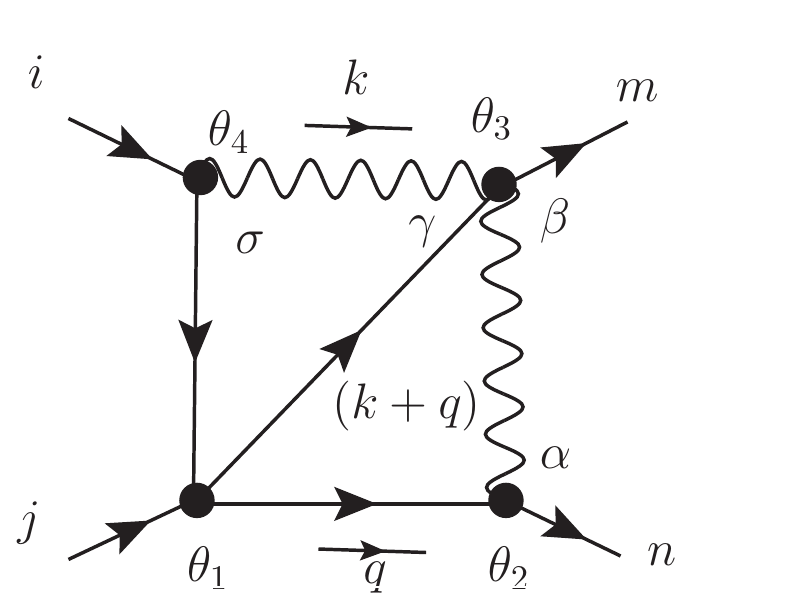}}\subfloat[]{\centering{}\includegraphics[scale=0.5]{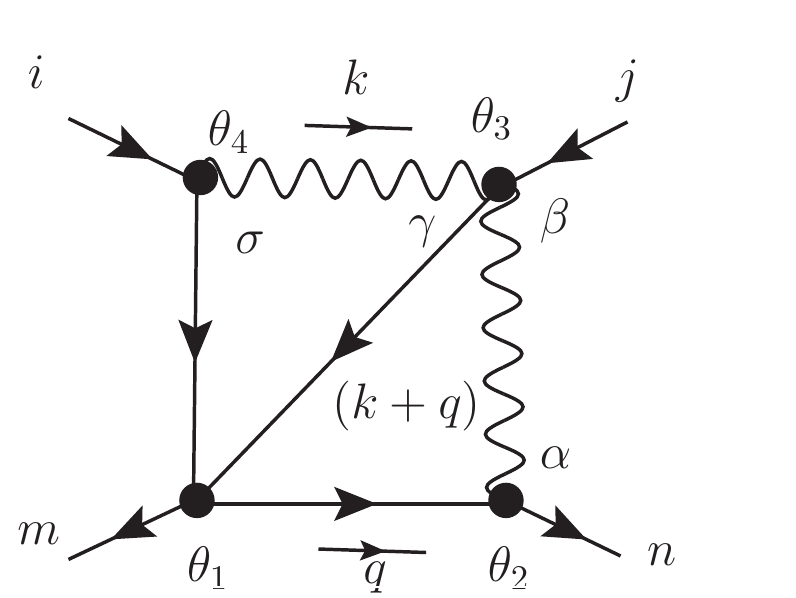}}\subfloat[]{\centering{}\includegraphics[scale=0.5]{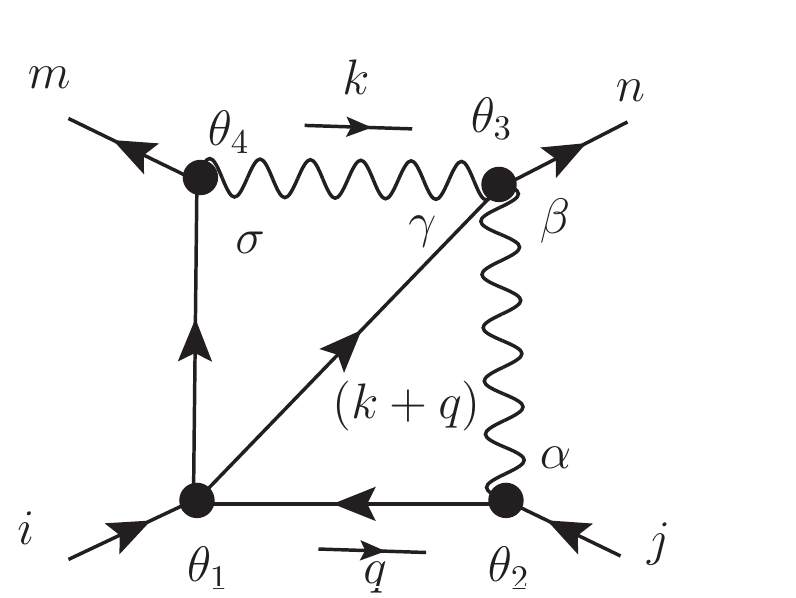}}\subfloat[]{\centering{}\includegraphics[scale=0.5]{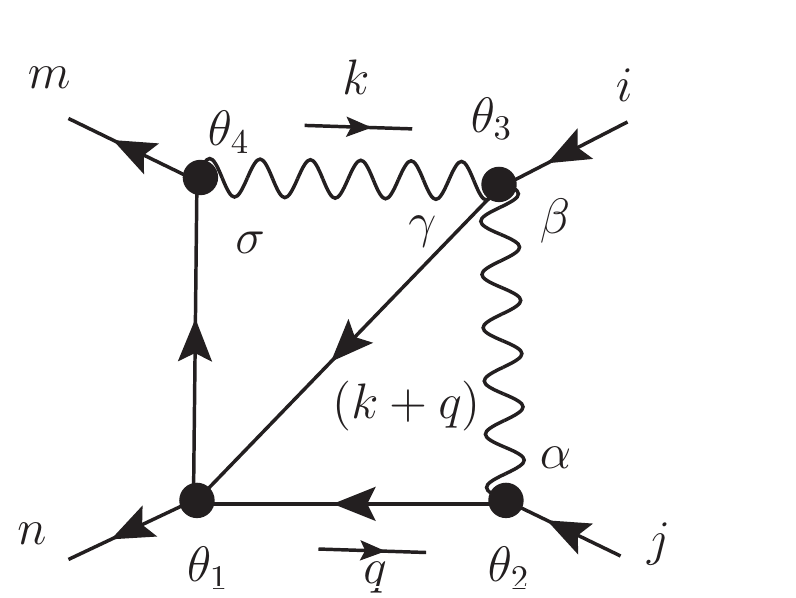}}
\par\end{centering}
\begin{centering}
\subfloat[]{\centering{}\includegraphics[scale=0.5]{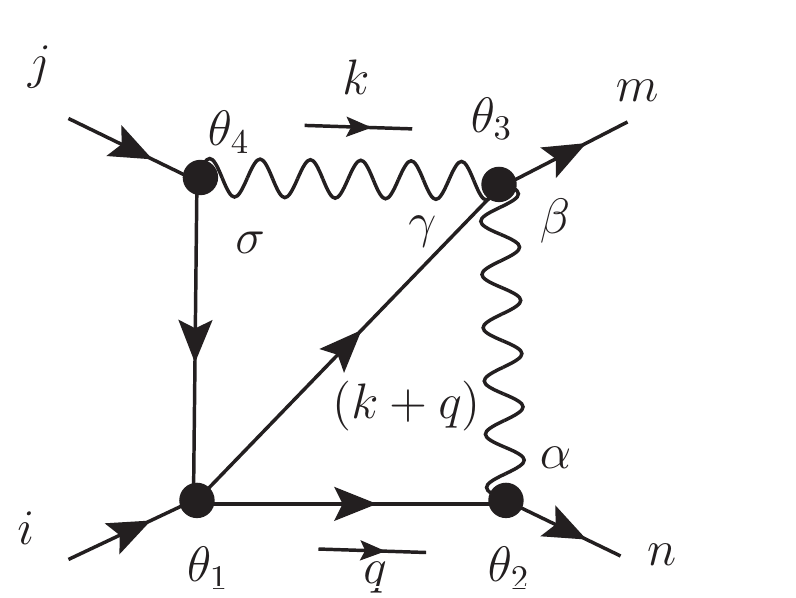}}\subfloat[]{\centering{}\includegraphics[scale=0.5]{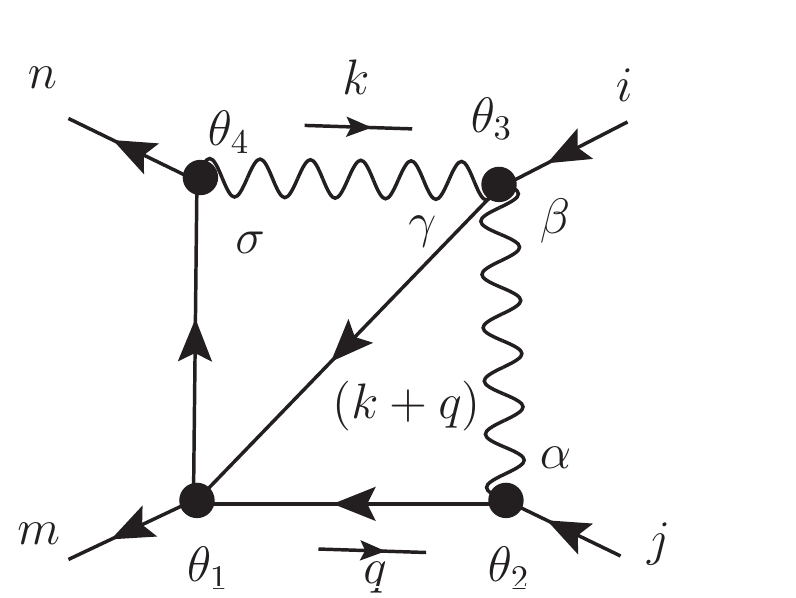}}\subfloat[]{\centering{}\includegraphics[scale=0.5]{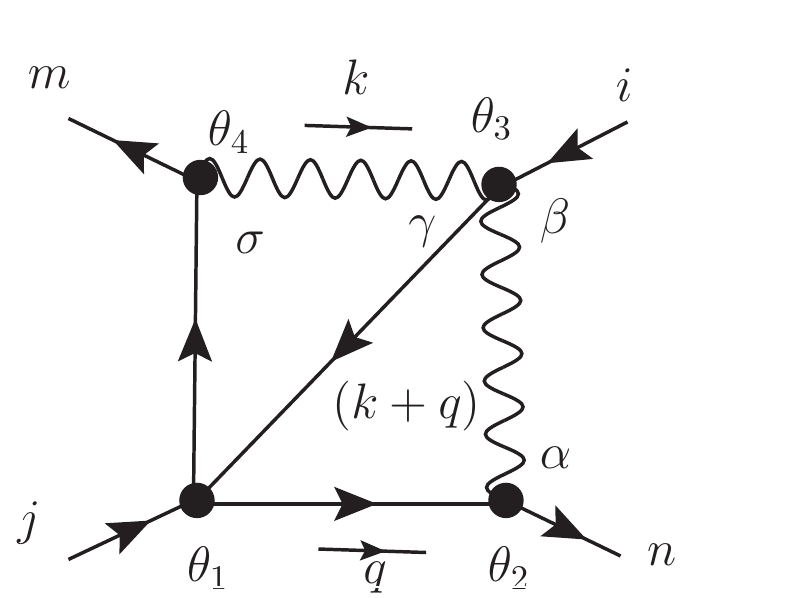}}\subfloat[]{\centering{}\includegraphics[scale=0.5]{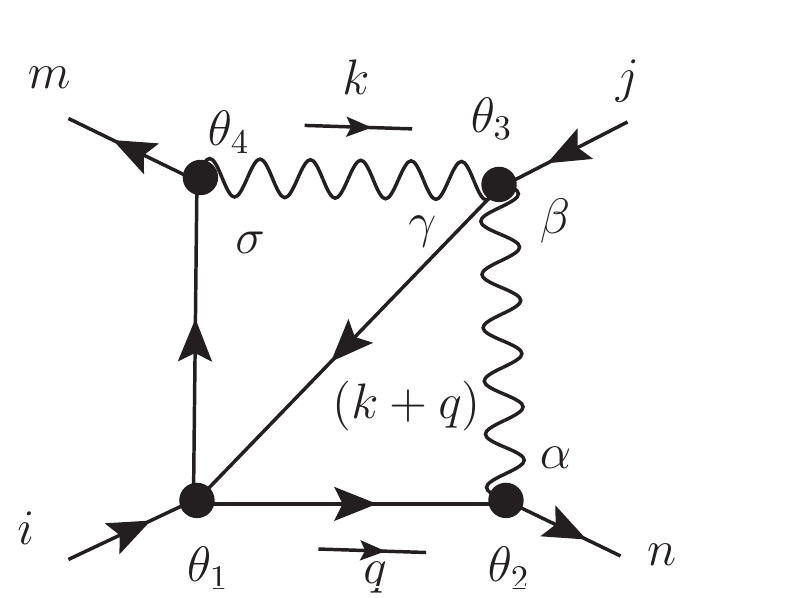}}
\par\end{centering}
\centering{}\caption{\label{fig:D17-order-lambda-g-4}$\mathcal{S}_{\left(\overline{\Phi}\Phi\right)^{2}}^{\left(D17\right)}$}
\end{figure}
\par\end{center}

\begin{center}
\begin{table}
\centering{}%
\begin{tabular}{lcccccc}
 &  &  &  &  &  & \tabularnewline
\hline 
\hline 
$D17-a$ &  & $-\left(\delta_{jn}\delta_{im}+\delta_{mj}\delta_{ni}\right)$ &  & $D17-b$ &  & $-\left(\delta_{jn}\delta_{im}+\delta_{mj}\delta_{ni}\right)$\tabularnewline
$D17-c$ &  & $\delta_{jn}\delta_{im}+\delta_{mj}\delta_{ni}$ &  & $D17-d$ &  & $\delta_{jn}\delta_{im}+\delta_{mj}\delta_{ni}$\tabularnewline
$D17-e$ &  & $\delta_{jn}\delta_{im}+\delta_{mj}\delta_{ni}$ &  & $D17-f$ &  & $\delta_{jn}\delta_{im}+\delta_{mj}\delta_{ni}$\tabularnewline
$D17-g$ &  & $\delta_{jn}\delta_{im}+\delta_{mj}\delta_{ni}$ &  & $D17-h$ &  & $\delta_{jn}\delta_{im}+\delta_{mj}\delta_{ni}$\tabularnewline
$D17-i$ &  & $\delta_{jn}\delta_{im}+\delta_{mj}\delta_{ni}$ &  & $D17-j$ &  & $\delta_{jn}\delta_{im}+\delta_{mj}\delta_{ni}$\tabularnewline
$D17-k$ &  & $-\left(\delta_{jn}\delta_{im}+\delta_{mj}\delta_{ni}\right)$ &  & $D17-l$ &  & $-\left(\delta_{jn}\delta_{im}+\delta_{mj}\delta_{ni}\right)$\tabularnewline
\hline 
\hline 
 &  &  &  &  &  & \tabularnewline
\end{tabular}\caption{\label{tab:S-PPPP-17}Values of the diagrams in Figure\,\ref{fig:D17-order-lambda-g-4}
with common factor\protect \\
 $\frac{1}{16}\left(\frac{\left(a-b\right)^{2}}{32\pi^{2}\epsilon}\right)\,i\,\lambda\,g^{4}\int_{\theta}\overline{\Phi}_{i}\Phi_{m}\Phi_{n}\overline{\Phi}_{j}$
.}
\end{table}
\par\end{center}

$\mathcal{S}_{\left(\overline{\Phi}\Phi\right)^{2}}^{\left(D17-a\right)}$
in the Figure\,\ref{fig:D17-order-lambda-g-4} is
\begin{align}
\mathcal{S}_{\left(\overline{\Phi}\Phi\right)^{2}}^{\left(D17-a\right)} & =-\frac{1}{32}\,i\,\lambda\,g^{4}\left(\delta_{jn}\delta_{im}+\delta_{mj}\delta_{ni}\right)\int_{\theta}\overline{\Phi}_{i}\Phi_{m}\Phi_{n}\overline{\Phi}_{j}\int\frac{d^{D}kd^{D}q}{\left(2\pi\right)^{2D}}\left\{ \frac{-2\left(a-b\right)^{2}k^{2}q^{2}}{\left(k^{2}\right)^{2}\left(k+q\right)^{2}\left(q^{2}\right)^{2}}\right\} \,,
\end{align}
using Eq.\,(\ref{eq:Int 7}) and adding $\mathcal{S}_{\left(\overline{\Phi}\Phi\right)^{2}}^{\left(D17-a\right)}$
to $\mathcal{S}_{\left(\overline{\Phi}\Phi\right)^{2}}^{\left(D17-l\right)}$
with the values in the Table\,\ref{tab:S-PPPP-17}, we find 
\begin{align}
\mathcal{S}_{\left(\overline{\Phi}\Phi\right)^{2}}^{\left(D17\right)} & =\frac{1}{4}\left(\frac{\left(a-b\right)^{2}}{32\pi^{2}\epsilon}\right)\,i\,\lambda\,g^{4}\left(\delta_{jn}\delta_{im}+\delta_{mj}\delta_{ni}\right)\int_{\theta}\overline{\Phi}_{i}\Phi_{m}\Phi_{n}\overline{\Phi}_{j}\,.\label{eq:S-D17}
\end{align}

\begin{center}
\begin{figure}
\begin{centering}
\subfloat[]{\begin{centering}
\includegraphics[scale=0.5]{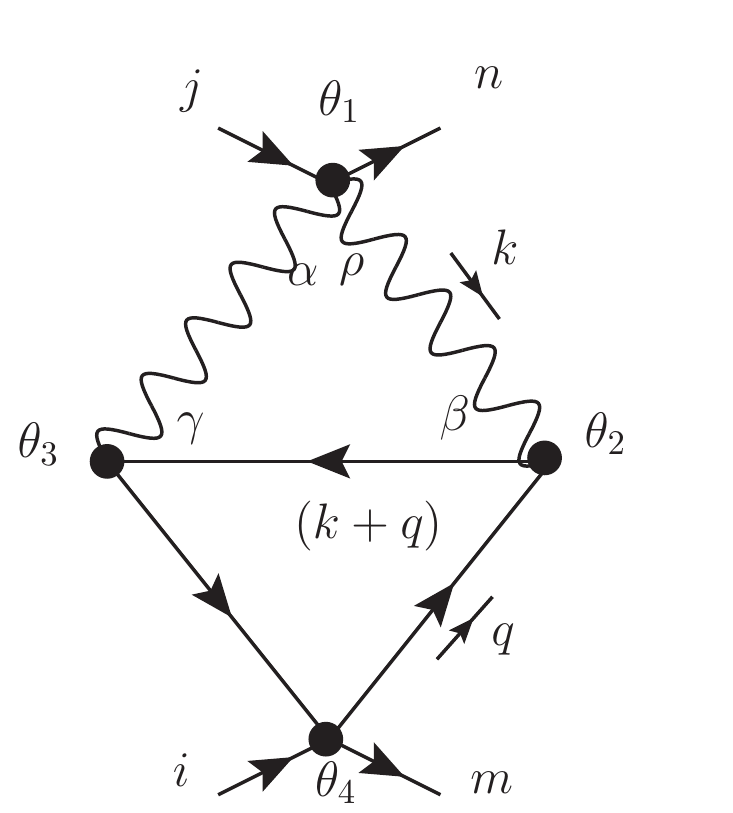}
\par\end{centering}
}\subfloat[]{\centering{}\includegraphics[scale=0.5]{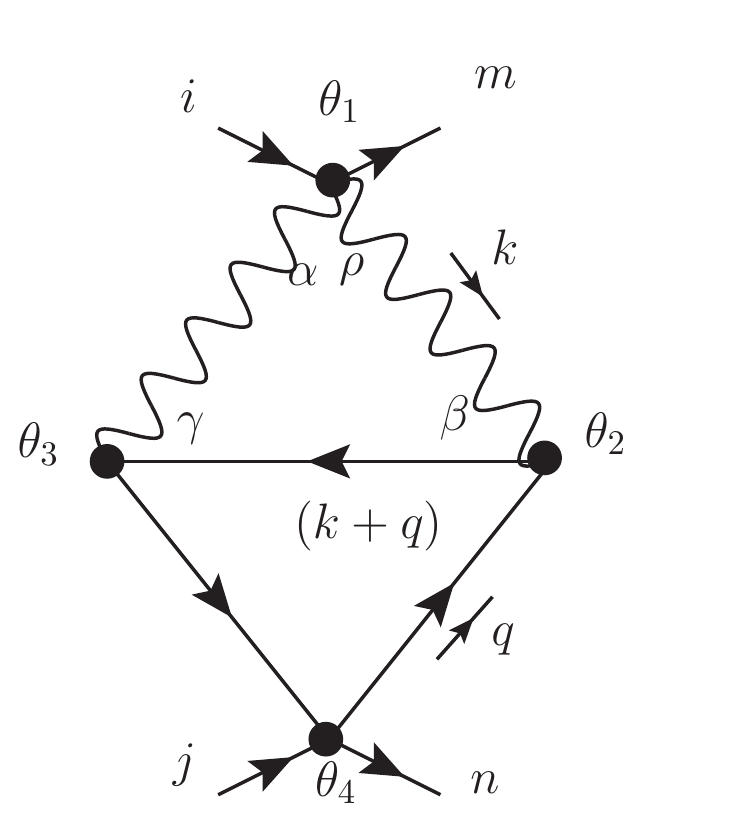}}\subfloat[]{\centering{}\includegraphics[scale=0.5]{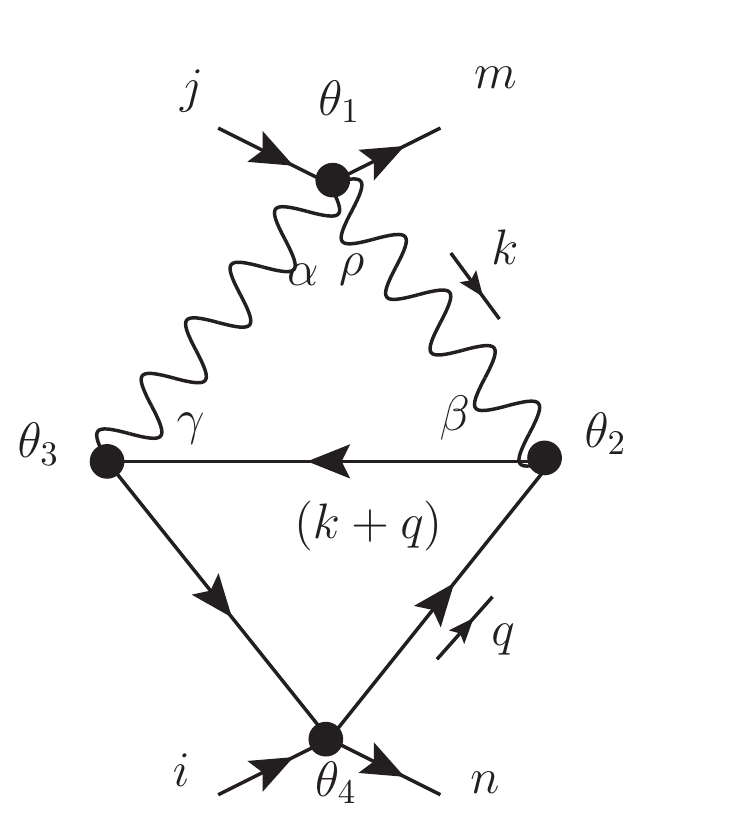}}\subfloat[]{\centering{}\includegraphics[scale=0.5]{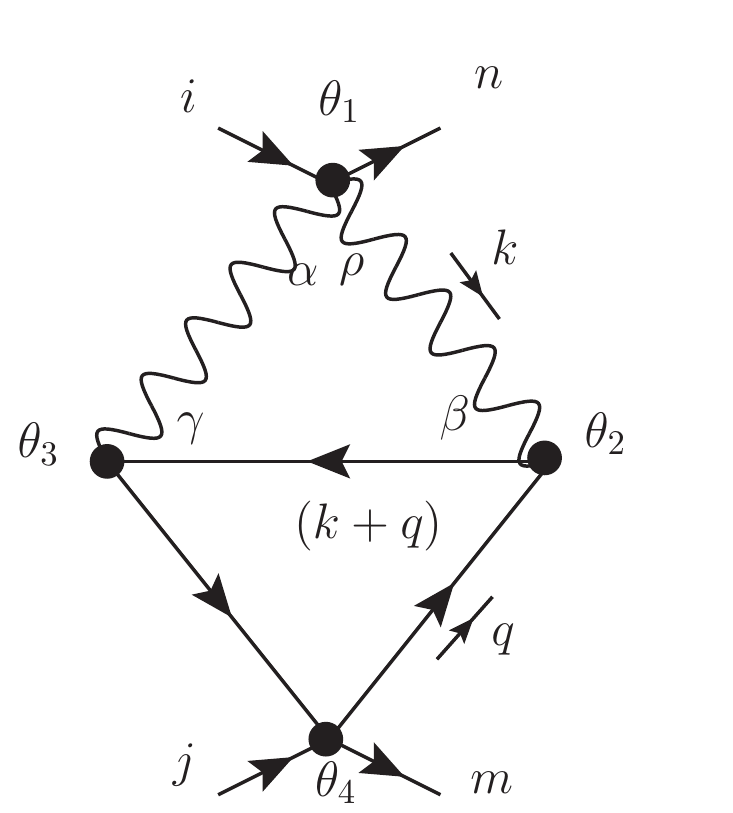}}
\par\end{centering}
\centering{}\caption{\label{fig:D18-order-lambda-g-4}$\mathcal{S}_{\left(\overline{\Phi}\Phi\right)^{2}}^{\left(D18\right)}$}
\end{figure}
\par\end{center}

\begin{center}
\begin{table}
\centering{}%
\begin{tabular}{lcccccccccc}
 &  &  &  &  &  &  &  &  &  & \tabularnewline
\hline 
\hline 
$D18-a$ &  & $\delta_{jn}\delta_{im}$ &  & $D18-b$ &  & $\delta_{jn}\delta_{im}$ &  & $D18-c$ &  & $\delta_{mj}\delta_{ni}$\tabularnewline
$D18-d$ &  & $\delta_{mj}\delta_{ni}$ &  &  &  &  &  &  &  & \tabularnewline
\hline 
\hline 
 &  &  &  &  &  &  &  &  &  & \tabularnewline
\end{tabular}\caption{\label{tab:S-PPPP-18}Values of the diagrams in Figure\,\ref{fig:D18-order-lambda-g-4}
with common factor\protect \\
 $\frac{1}{16}\left(\frac{\left(a+b\right)^{2}}{32\pi^{2}\epsilon}\right)\left(1+N\right)\,i\,\lambda\,g^{4}\int_{\theta}\overline{\Phi}_{i}\Phi_{m}\Phi_{n}\overline{\Phi}_{j}$
.}
\end{table}
\par\end{center}

$\mathcal{S}_{\left(\overline{\Phi}\Phi\right)^{2}}^{\left(D18-a\right)}$
in the Figure\,\ref{fig:D18-order-lambda-g-4} is
\begin{align}
\mathcal{S}_{\left(\overline{\Phi}\Phi\right)^{2}}^{\left(D18-a\right)} & =\frac{1}{32}\,i\,\lambda\,g^{4}\left(1+N\right)\delta_{nj}\delta_{mi}\int_{\theta}\overline{\Phi}_{i}\Phi_{m}\Phi_{n}\overline{\Phi}_{j}\nonumber \\
 & \times\int\frac{d^{D}kd^{D}q}{\left(2\pi\right)^{2D}}\left\{ \frac{-16\,a\,b\left(k\cdot q\right)q^{2}-2\left(a^{2}+6\,a\,b+b^{2}\right)k^{2}q^{2}}{\left(k^{2}\right)^{2}\left(k+q\right)^{2}\left(q^{2}\right)^{2}}\right\} \,,
\end{align}
using Eqs.\,(\ref{eq:Int 7}) and\,(\ref{eq:Int 9}), then adding
$\mathcal{S}_{\left(\overline{\Phi}\Phi\right)^{2}}^{\left(D18-a\right)}$
to $\mathcal{S}_{\left(\overline{\Phi}\Phi\right)^{2}}^{\left(D18-d\right)}$
with the values in Table\,\ref{tab:S-PPPP-18}, we find 
\begin{align}
\mathcal{S}_{\left(\overline{\Phi}\Phi\right)^{2}}^{\left(D18\right)} & =\frac{1}{8}\left(\frac{\left(a+b\right)^{2}}{32\pi^{2}\epsilon}\right)\left(1+N\right)\,i\,\lambda\,g^{4}\left(\delta_{im}\delta_{jn}+\delta_{jm}\delta_{in}\right)\int_{\theta}\overline{\Phi}_{i}\Phi_{m}\Phi_{n}\overline{\Phi}_{j}\,.\label{eq:S-D18}
\end{align}

\begin{center}
\begin{figure}
\begin{centering}
\subfloat[]{\begin{centering}
\includegraphics[scale=0.5]{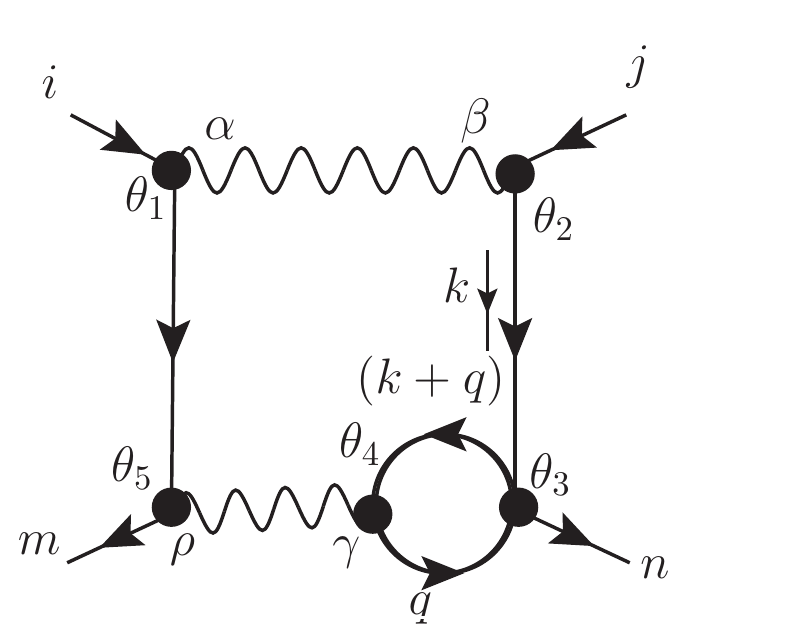}
\par\end{centering}
}\subfloat[]{\centering{}\includegraphics[scale=0.5]{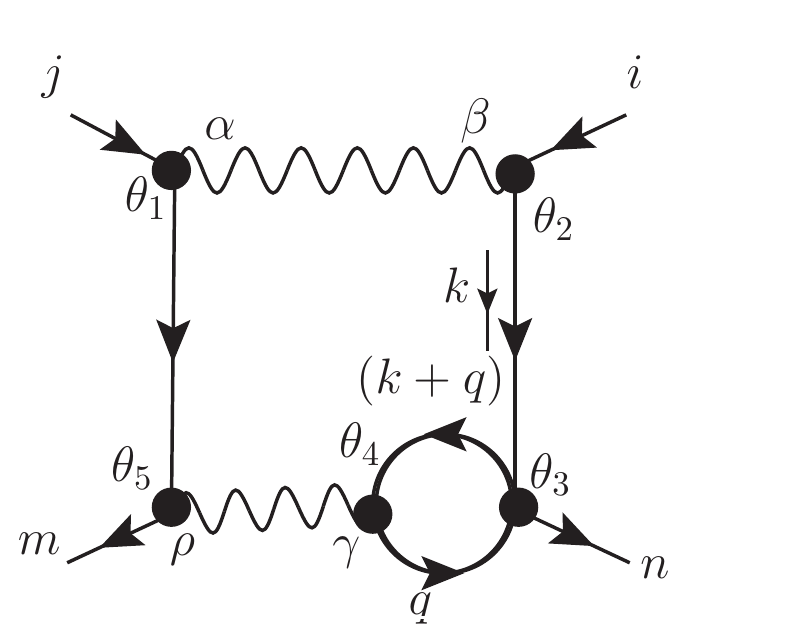}}\subfloat[]{\centering{}\includegraphics[scale=0.5]{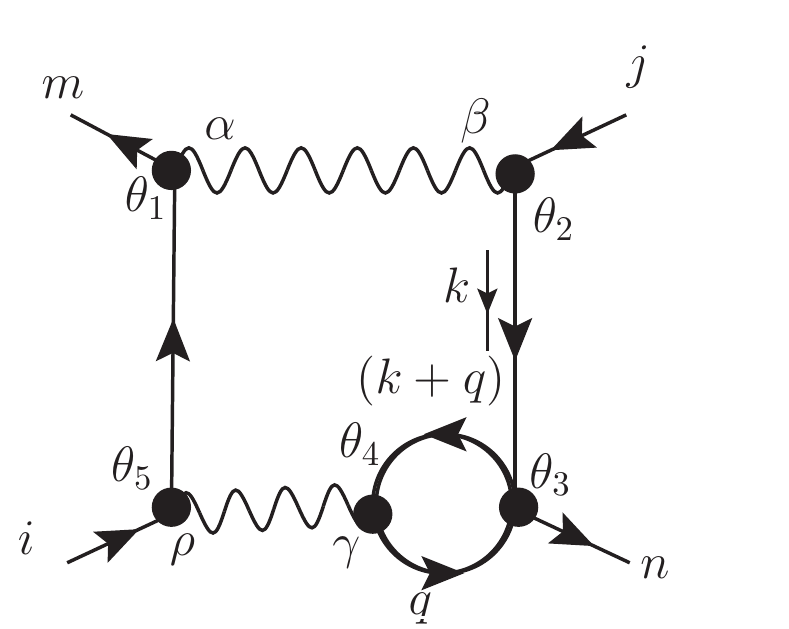}}\subfloat[]{\centering{}\includegraphics[scale=0.5]{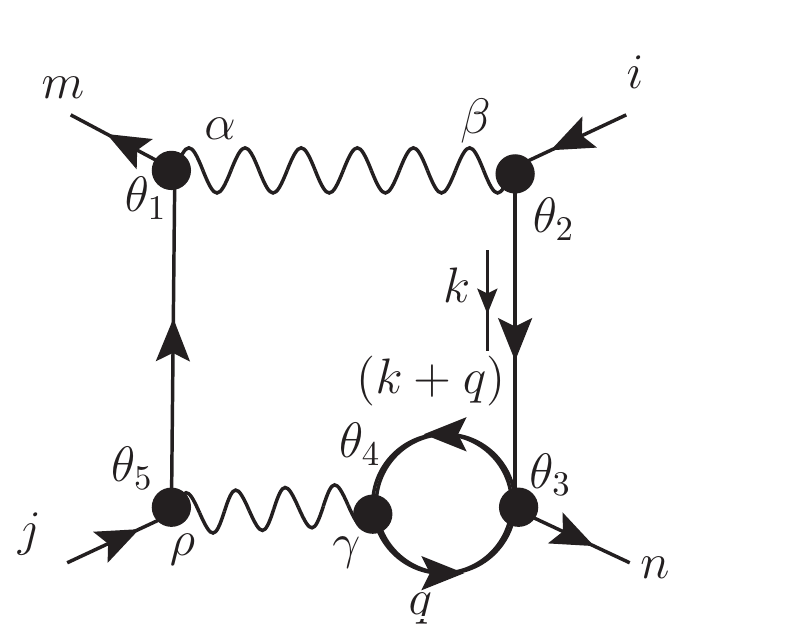}}
\par\end{centering}
\begin{centering}
\subfloat[]{\centering{}\includegraphics[scale=0.5]{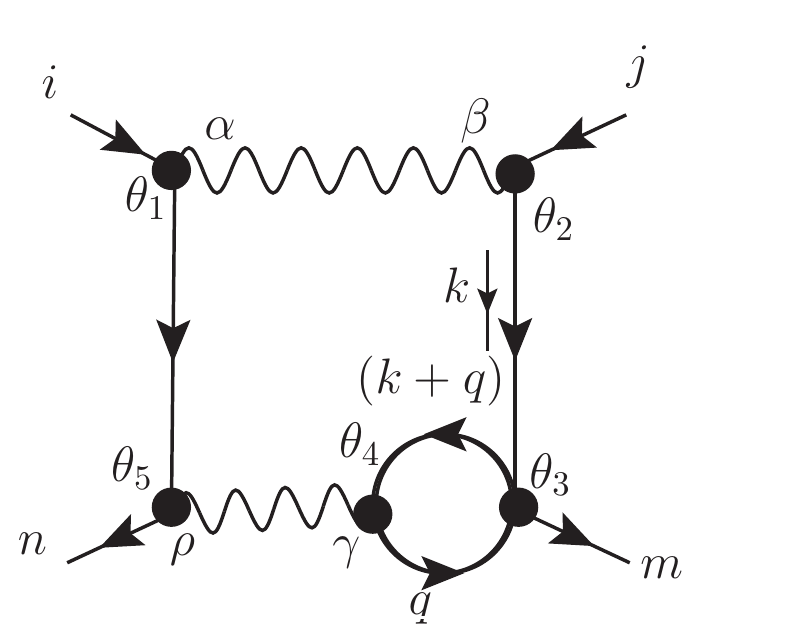}}\subfloat[]{\centering{}\includegraphics[scale=0.5]{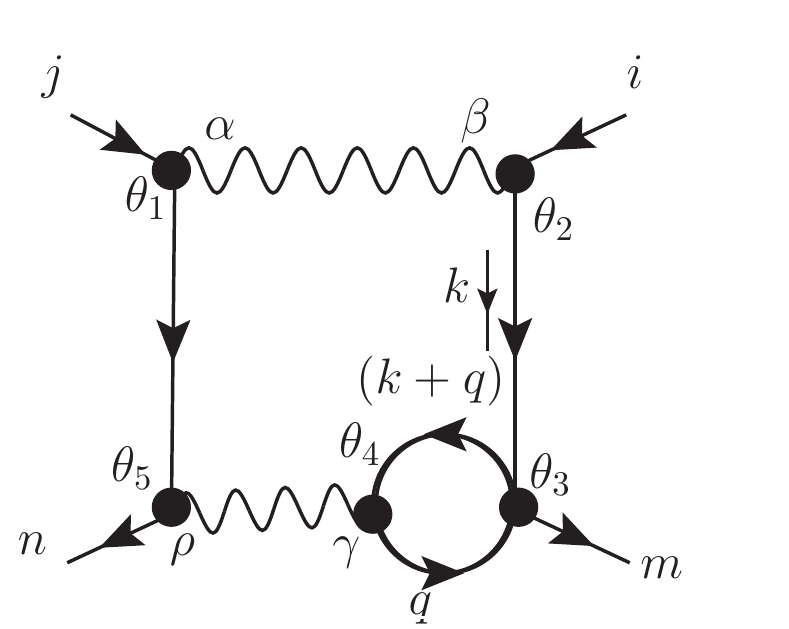}}\subfloat[]{\centering{}\includegraphics[scale=0.5]{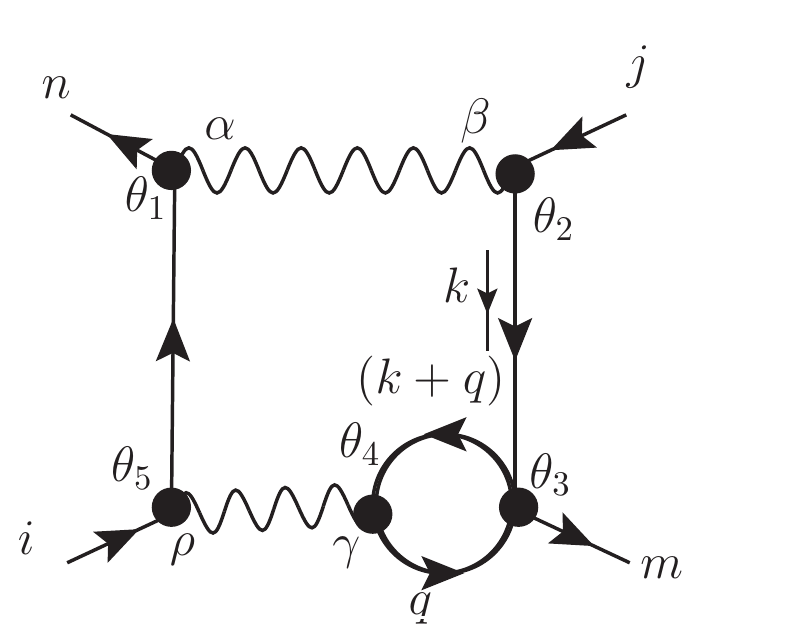}}\subfloat[]{\centering{}\includegraphics[scale=0.5]{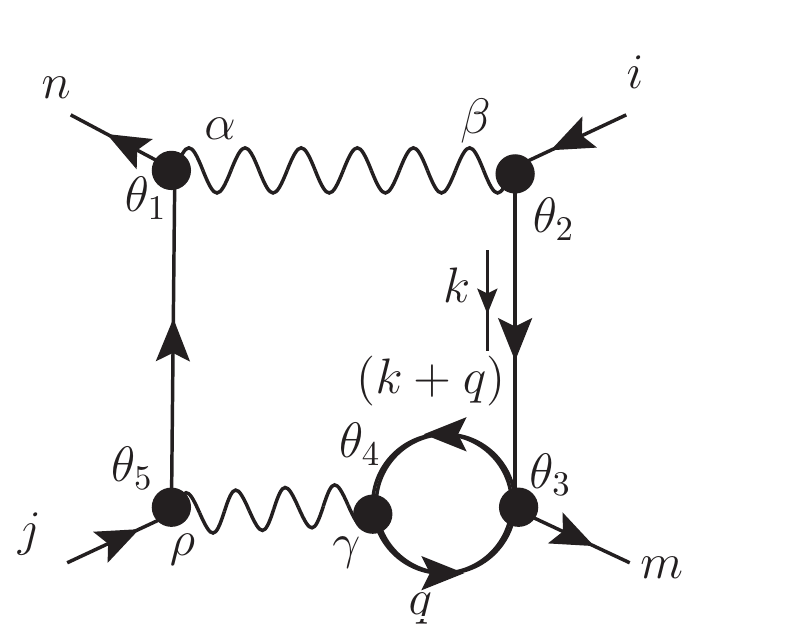}}
\par\end{centering}
\begin{centering}
\subfloat[]{\centering{}\includegraphics[scale=0.5]{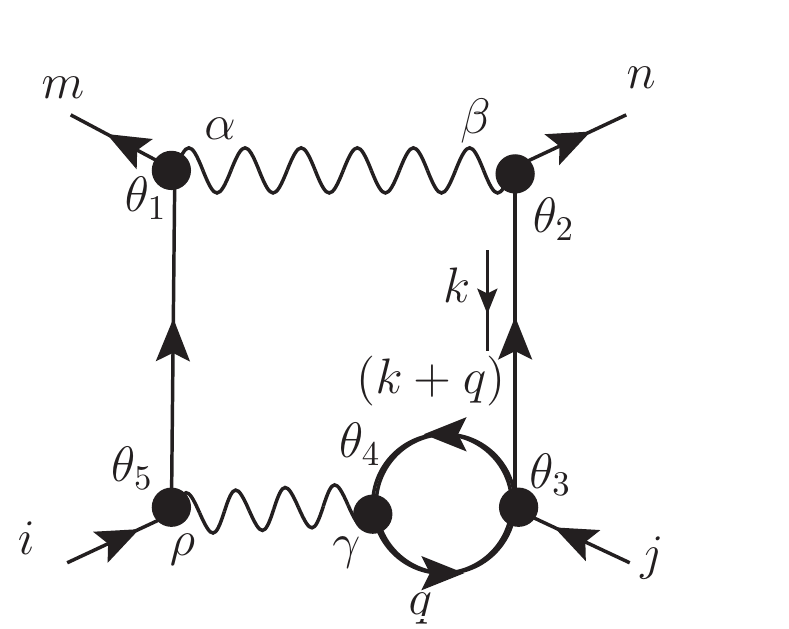}}\subfloat[]{\centering{}\includegraphics[scale=0.5]{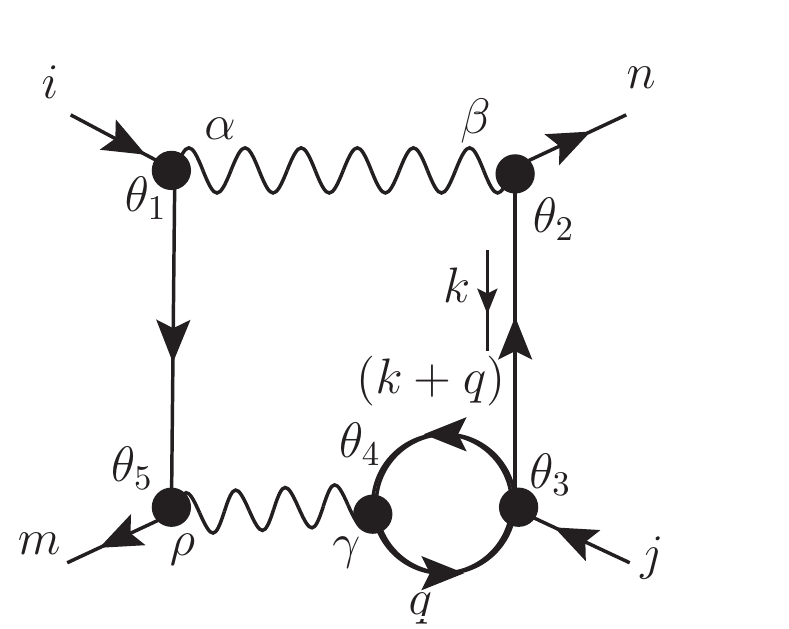}}\subfloat[]{\centering{}\includegraphics[scale=0.5]{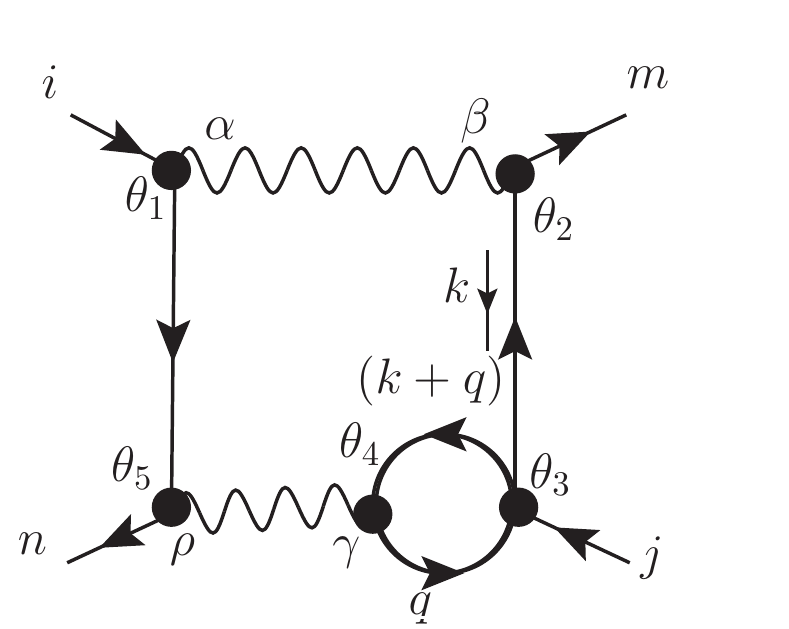}}\subfloat[]{\centering{}\includegraphics[scale=0.5]{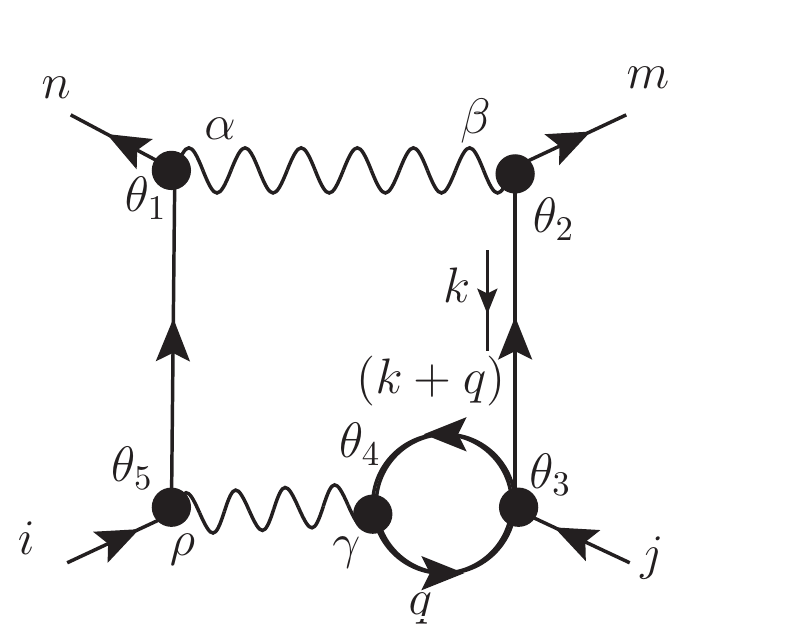}}
\par\end{centering}
\begin{centering}
\subfloat[]{\centering{}\includegraphics[scale=0.5]{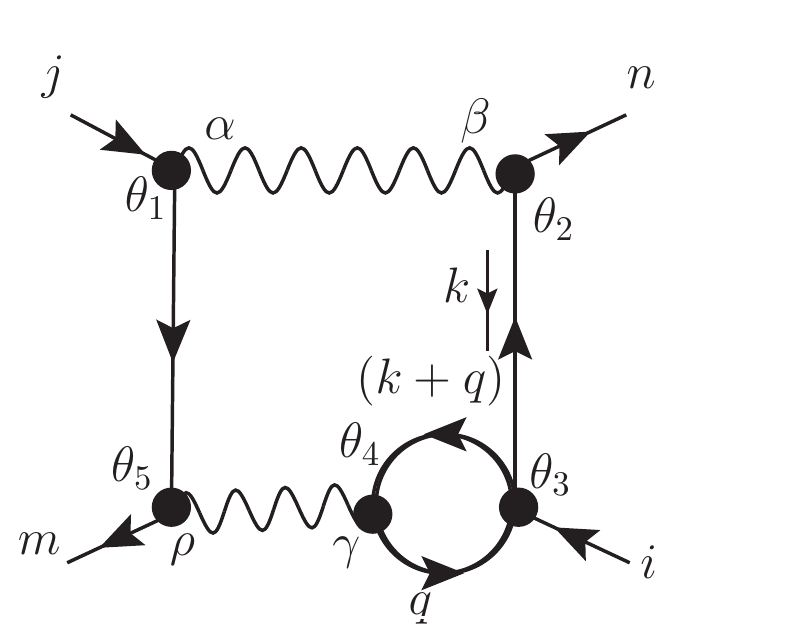}}\subfloat[]{\centering{}\includegraphics[scale=0.5]{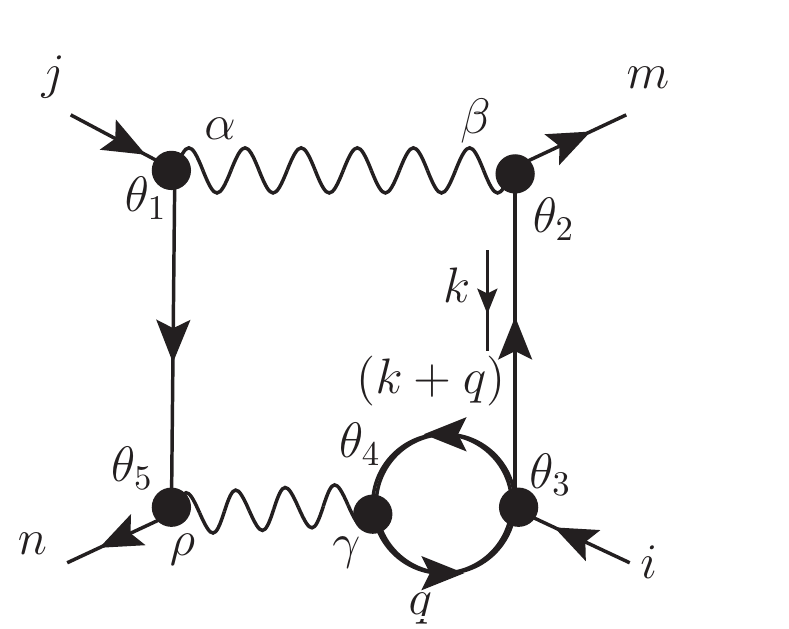}}\subfloat[]{\centering{}\includegraphics[scale=0.5]{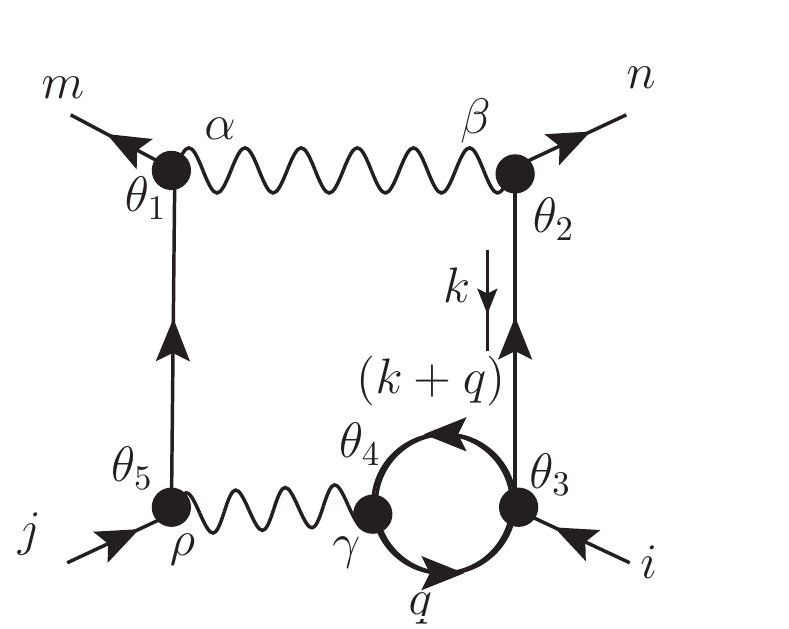}}\subfloat[]{\centering{}\includegraphics[scale=0.5]{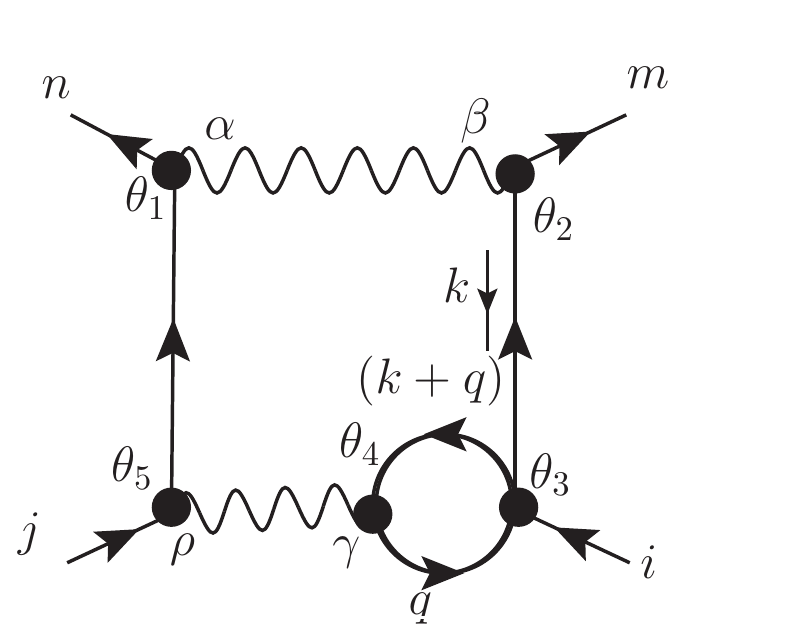}}
\par\end{centering}
\centering{}\caption{\label{fig:D19-order-lambda-g-4}$\mathcal{S}_{\left(\overline{\Phi}\Phi\right)^{2}}^{\left(D19\right)}$}
\end{figure}
\par\end{center}

$\mathcal{S}_{\left(\overline{\Phi}\Phi\right)^{2}}^{\left(D19-a\right)}$
in the Figure\,\ref{fig:D19-order-lambda-g-4} is
\begin{align}
\mathcal{S}_{\left(\overline{\Phi}\Phi\right)^{2}}^{\left(D19-a\right)} & =-\frac{1}{128}\,i\,\lambda\,g^{4}\left(1+N\right)\delta_{jn}\delta_{mi}\int_{\theta}\overline{\Phi}_{i}\Phi_{m}\Phi_{n}\overline{\Phi}_{j}\int\frac{d^{D}kd^{D}q}{\left(2\pi\right)^{2D}}\left\{ \frac{-4\left(a-b\right)^{2}\left(k^{2}\right)^{2}\left(2\left(k\cdot q\right)+k^{2}\right)}{\left(k^{2}\right)^{4}\left(k+q\right)^{2}q^{2}}\right\} ,\,
\end{align}
using Eqs.\,(\ref{eq:Int 7}), (\ref{eq:Int 9}), we find 
\begin{align}
\mathcal{S}_{\left(\overline{\Phi}\Phi\right)^{2}}^{\left(D19-a\right)} & =\mathcal{S}_{\left(\overline{\Phi}\Phi\right)^{2}}^{\left(D19\right)}=0\,.\label{eq:S-D19}
\end{align}

\begin{center}
\begin{figure}
\begin{centering}
\subfloat[]{\begin{centering}
\includegraphics[scale=0.5]{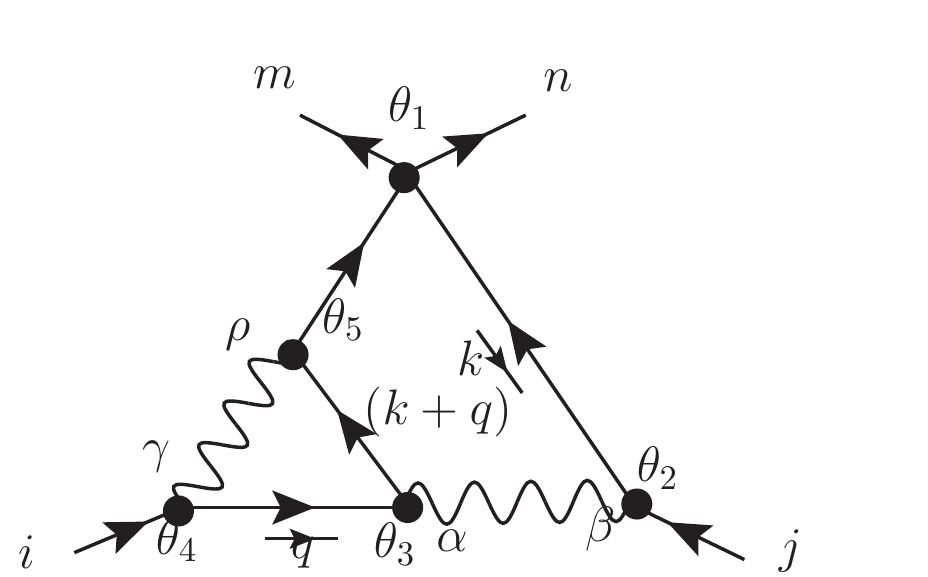}
\par\end{centering}
}\subfloat[]{\centering{}\includegraphics[scale=0.5]{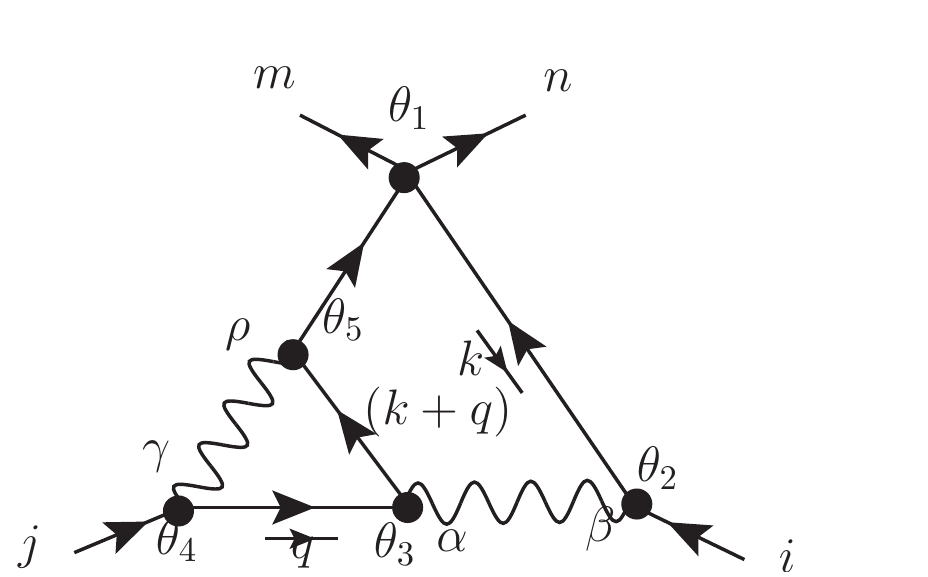}}\subfloat[]{\centering{}\includegraphics[scale=0.5]{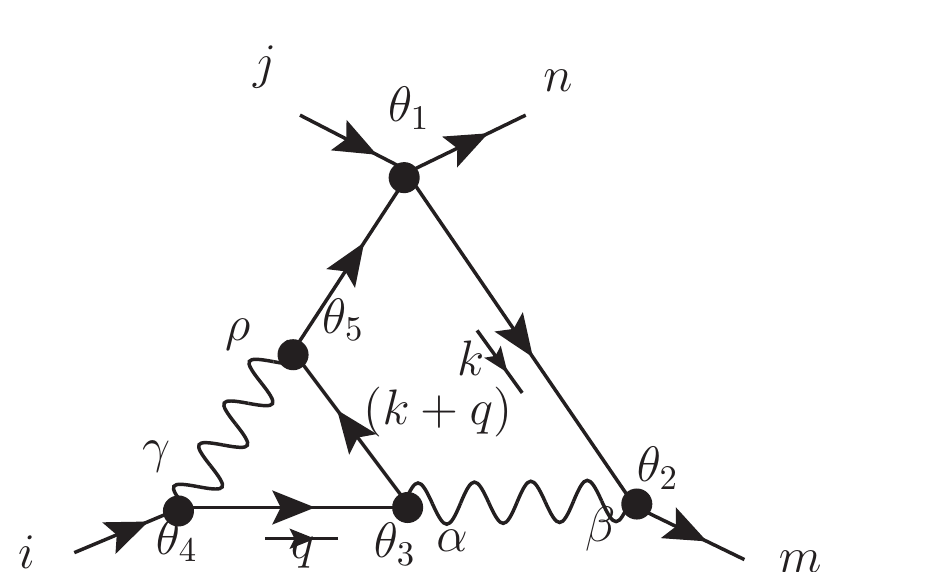}}
\par\end{centering}
\begin{centering}
\subfloat[]{\centering{}\includegraphics[scale=0.5]{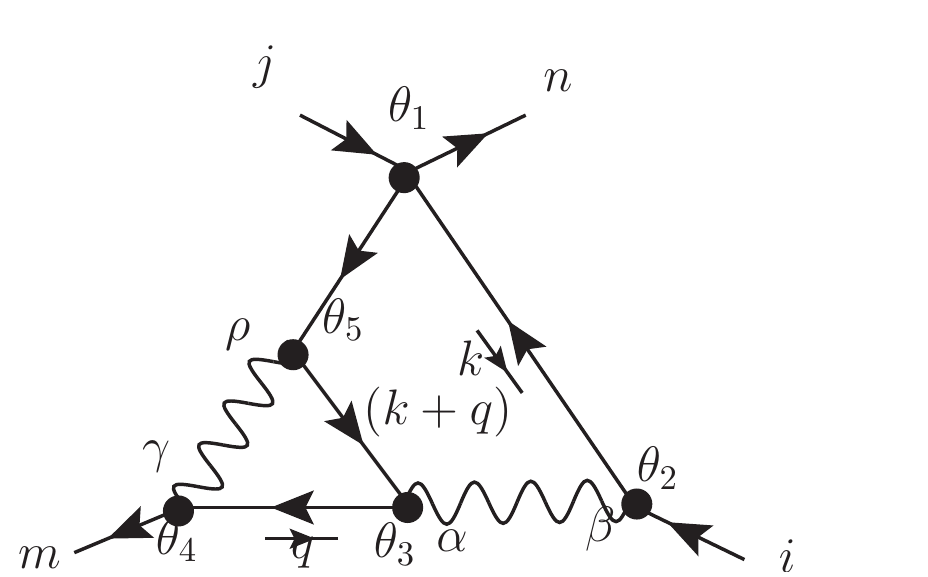}}\subfloat[]{\centering{}\includegraphics[scale=0.5]{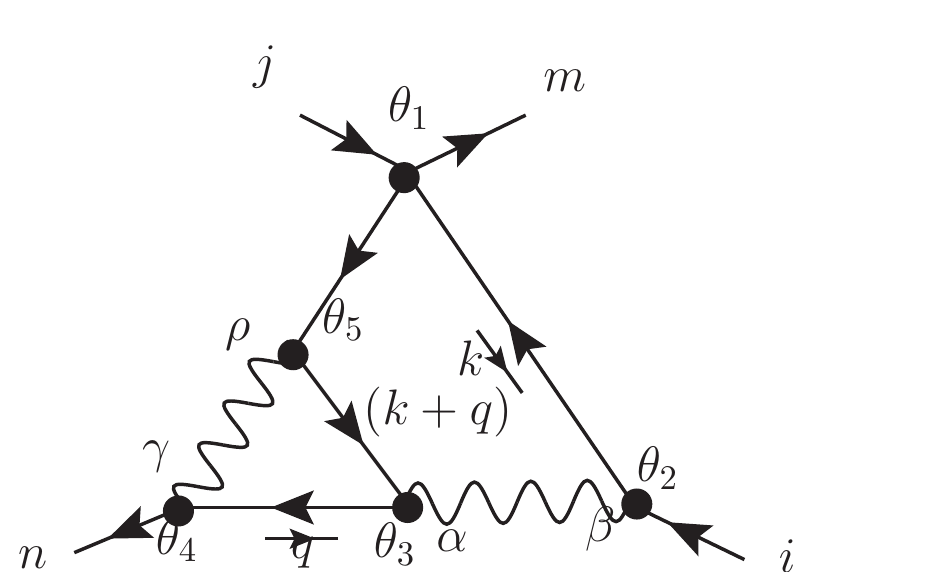}}\subfloat[]{\centering{}\includegraphics[scale=0.5]{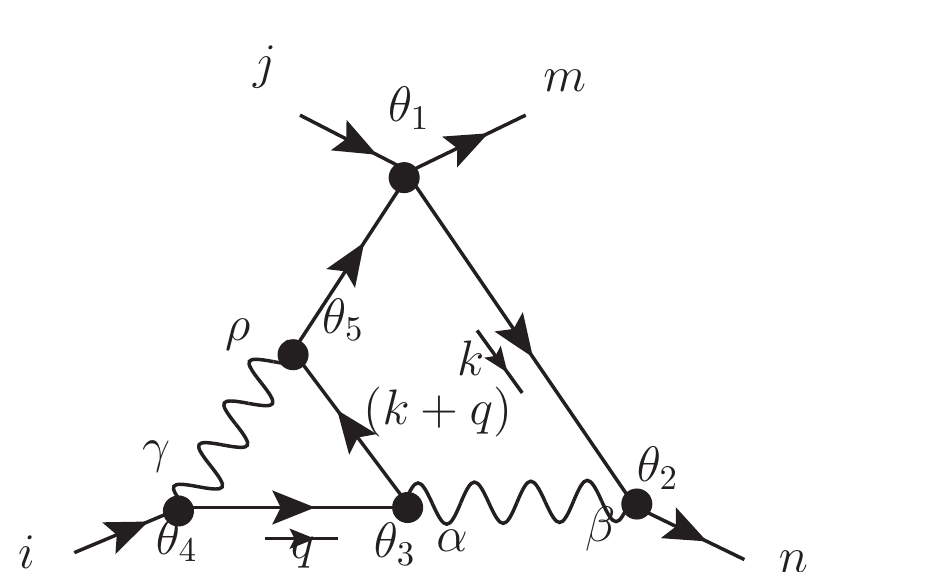}}
\par\end{centering}
\begin{centering}
\subfloat[]{\centering{}\includegraphics[scale=0.5]{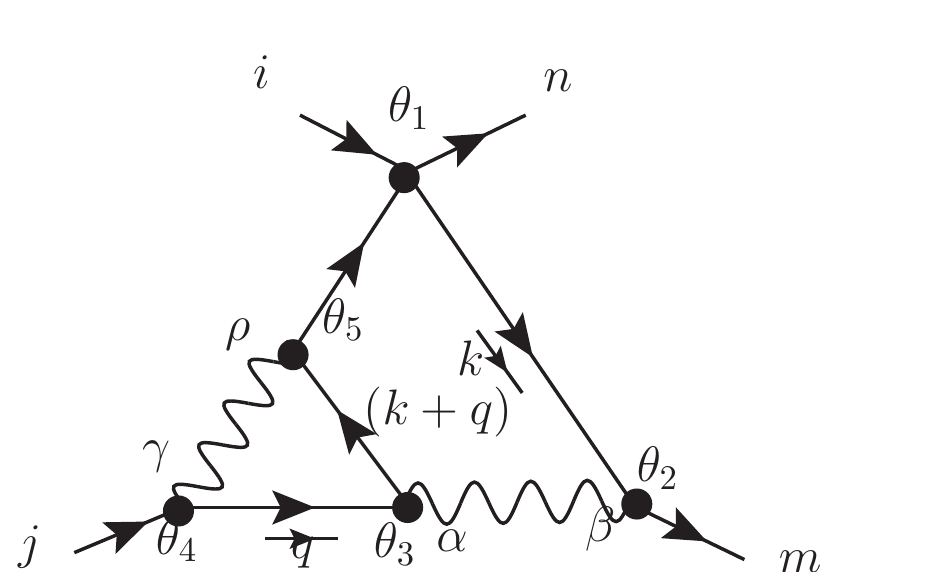}}\subfloat[]{\centering{}\includegraphics[scale=0.5]{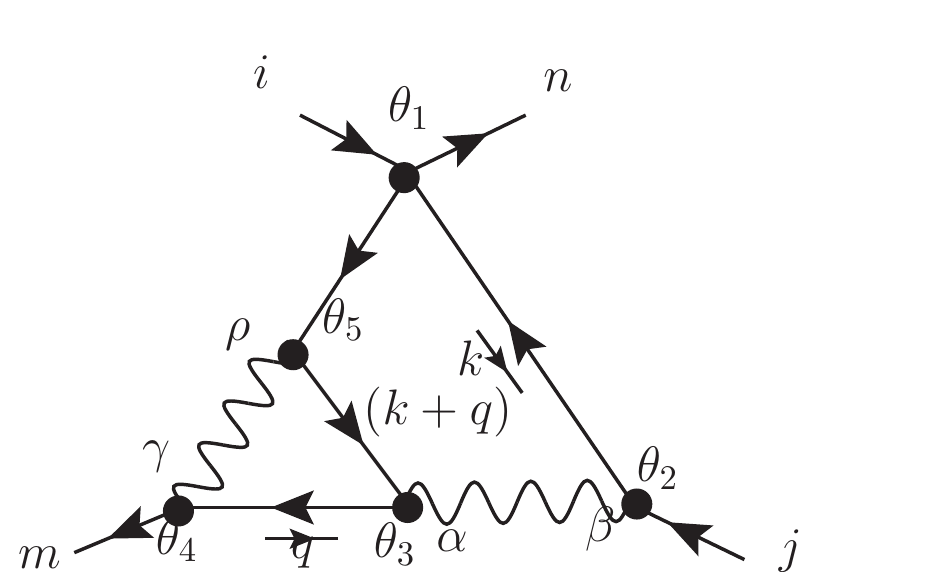}}\subfloat[]{\centering{}\includegraphics[scale=0.5]{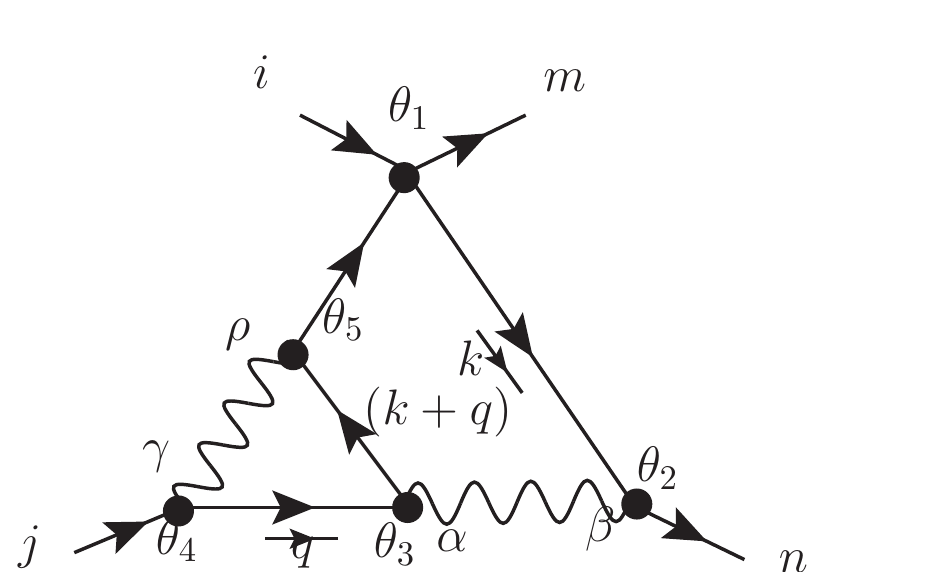}}
\par\end{centering}
\begin{centering}
\subfloat[]{\centering{}\includegraphics[scale=0.5]{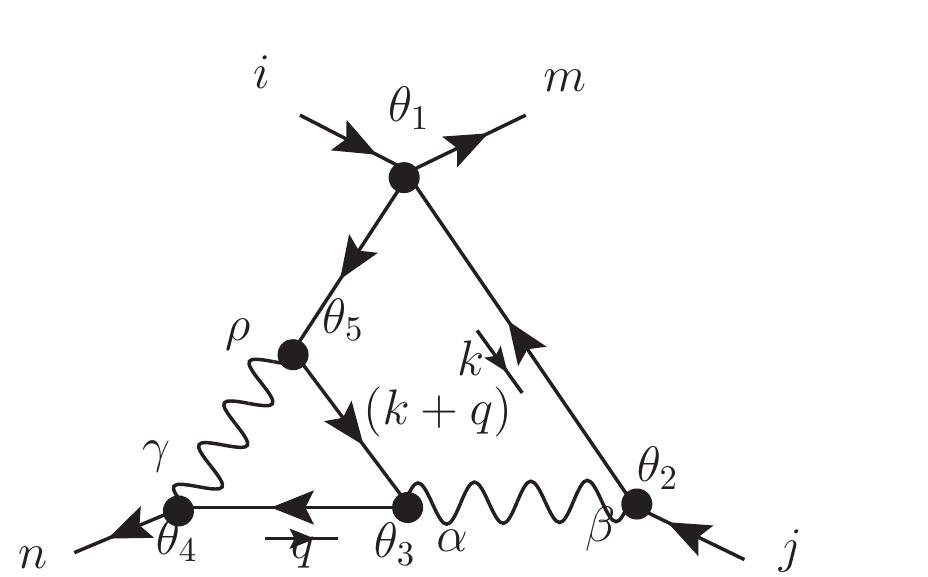}}\subfloat[]{\centering{}\includegraphics[scale=0.5]{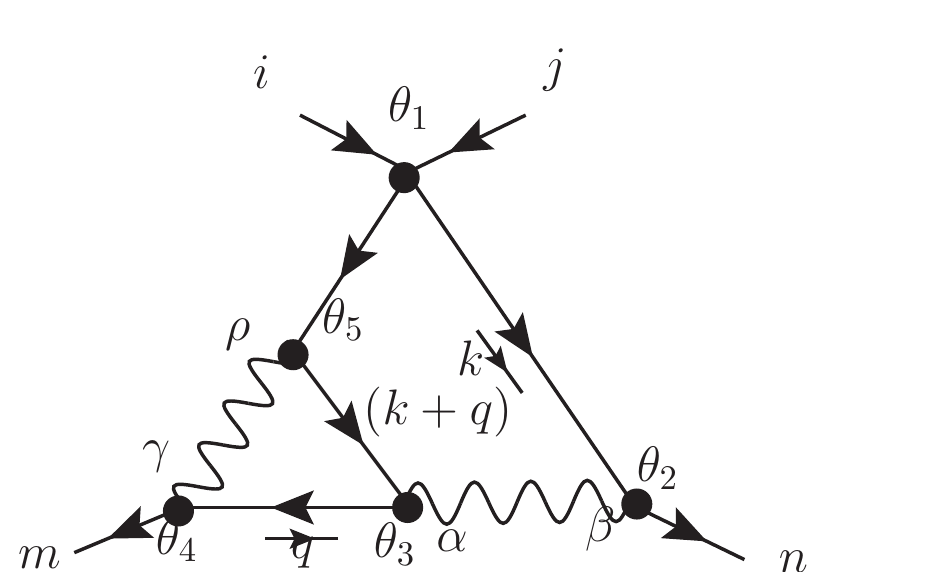}}\subfloat[]{\centering{}\includegraphics[scale=0.5]{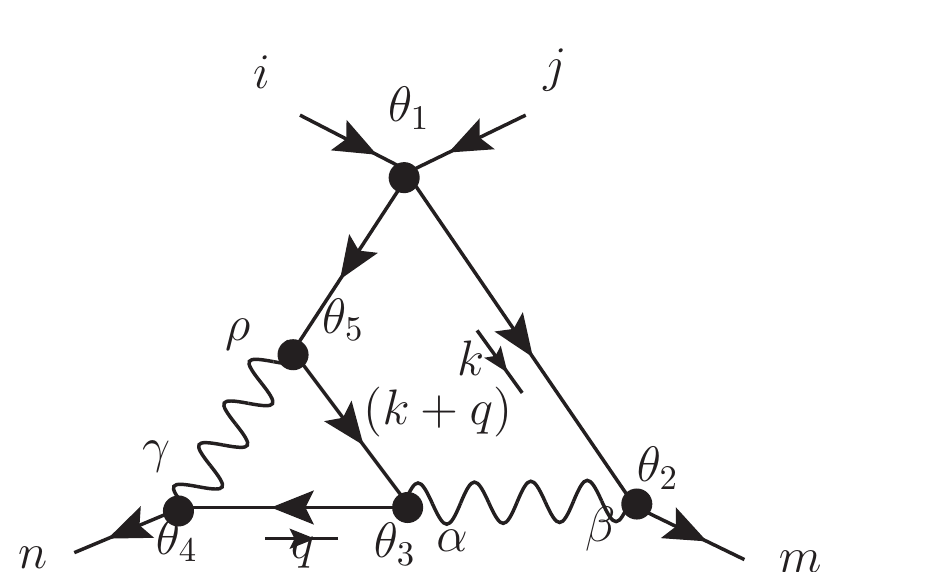}}
\par\end{centering}
\centering{}\caption{\label{fig:D20-order-lambda-g-4}$\mathcal{S}_{\left(\overline{\Phi}\Phi\right)^{2}}^{\left(D20\right)}$ }
\end{figure}
\par\end{center}

\begin{center}
\begin{table}
\centering{}%
\begin{tabular}{lcccccc}
 &  &  &  &  &  & \tabularnewline
\hline 
\hline 
$D20-a$ &  & $\delta_{jn}\delta_{im}+\delta_{mj}\delta_{ni}$ &  & $D20-b$ &  & $\delta_{jn}\delta_{im}+\delta_{mj}\delta_{ni}$\tabularnewline
$D20-c$ &  & $-\left(\delta_{jn}\delta_{im}+\delta_{mj}\delta_{ni}\right)$ &  & $D20-d$ &  & $-\left(\delta_{jn}\delta_{im}+\delta_{mj}\delta_{ni}\right)$\tabularnewline
$D20-e$ &  & $-\left(\delta_{jn}\delta_{im}+\delta_{mj}\delta_{ni}\right)$ &  & $D20-f$ &  & $-\left(\delta_{jn}\delta_{im}+\delta_{mj}\delta_{ni}\right)$\tabularnewline
$D20-g$ &  & $-\left(\delta_{jn}\delta_{im}+\delta_{mj}\delta_{ni}\right)$ &  & $D20-h$ &  & $-\left(\delta_{jn}\delta_{im}+\delta_{mj}\delta_{ni}\right)$\tabularnewline
$D20-i$ &  & $-\left(\delta_{jn}\delta_{im}+\delta_{mj}\delta_{ni}\right)$ &  & $D20-j$ &  & $-\left(\delta_{jn}\delta_{im}+\delta_{mj}\delta_{ni}\right)$\tabularnewline
$D20-k$ &  & $\delta_{jn}\delta_{im}+\delta_{mj}\delta_{ni}$ &  & $D20-l$ &  & $\delta_{jn}\delta_{im}+\delta_{mj}\delta_{ni}$\tabularnewline
\hline 
\hline 
 &  &  &  &  &  & \tabularnewline
\end{tabular}\caption{\label{tab:S-PPPP-20}Values of the diagrams in Figure\,\ref{fig:D20-order-lambda-g-4}
with common factor\protect \\
 $\frac{1}{32}\left(\frac{\left(a-b\right)^{2}}{32\pi^{2}\epsilon}\right)\,i\,\lambda\,g^{4}\int_{\theta}\overline{\Phi}_{i}\Phi_{m}\Phi_{n}\overline{\Phi}_{j}$
.}
\end{table}
\par\end{center}

$\mathcal{S}_{\left(\overline{\Phi}\Phi\right)^{2}}^{\left(D20-a\right)}$
in the Figure\,\ref{fig:D20-order-lambda-g-4} is
\begin{align}
\mathcal{S}_{\left(\overline{\Phi}\Phi\right)^{2}}^{\left(D20-a\right)} & =-\frac{1}{128}\,i\,\lambda\,g^{4}\left(\delta_{jn}\delta_{im}+\delta_{jm}\delta_{in}\right)\int_{\theta}\overline{\Phi}_{i}\Phi_{m}\Phi_{n}\overline{\Phi}_{j}\int\frac{d^{D}kd^{D}q}{\left(2\pi\right)^{2D}}\left\{ \frac{4\left(a-b\right)^{2}\left(k^{2}\right)^{2}q^{2}}{\left(k^{2}\right)^{3}\left(k+q\right)^{2}\left(q^{2}\right)^{2}}\right\} \,,
\end{align}
using Eq.\,(\ref{eq:Int 7}), then adding $\mathcal{S}_{\left(\overline{\Phi}\Phi\right)^{2}}^{\left(D20-a\right)}$
to $\mathcal{S}_{\left(\overline{\Phi}\Phi\right)^{2}}^{\left(D20-l\right)}$
with the values in Table\,\ref{tab:S-PPPP-20}, we find
\begin{align}
\mathcal{S}_{\left(\overline{\Phi}\Phi\right)^{2}}^{\left(D20\right)} & =-\frac{1}{8}\left(\frac{\left(a-b\right)^{2}}{32\pi^{2}\epsilon}\right)\,i\,\lambda\,g^{4}\left(\delta_{jn}\delta_{im}+\delta_{jm}\delta_{in}\right)\int_{\theta}\overline{\Phi}_{i}\Phi_{m}\Phi_{n}\overline{\Phi}_{j}\,.\label{eq:S-D20}
\end{align}

Now, we consider all the diagrams that contribute with the order $\mathcal{O}\left(g^{6}\right)$,
in this order there are $21$ topologies that are equivalent to $178$
diagrams. 
\begin{center}
\begin{figure}
\begin{centering}
\subfloat[]{\centering{}\includegraphics[scale=0.5]{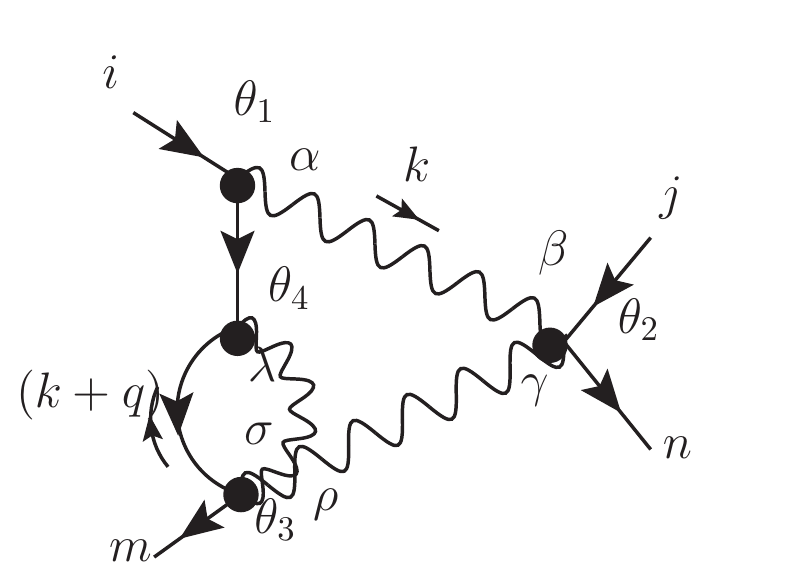}}\subfloat[]{\centering{}\includegraphics[scale=0.5]{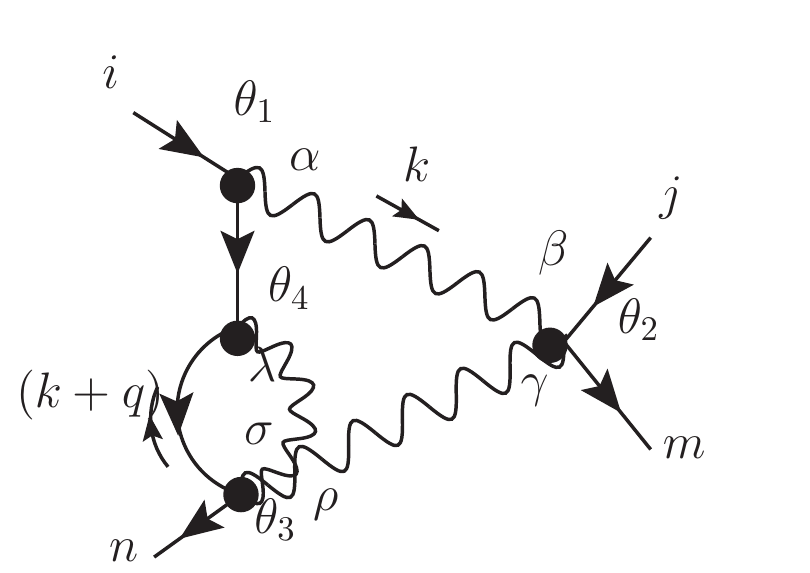}}\subfloat[]{\centering{}\includegraphics[scale=0.5]{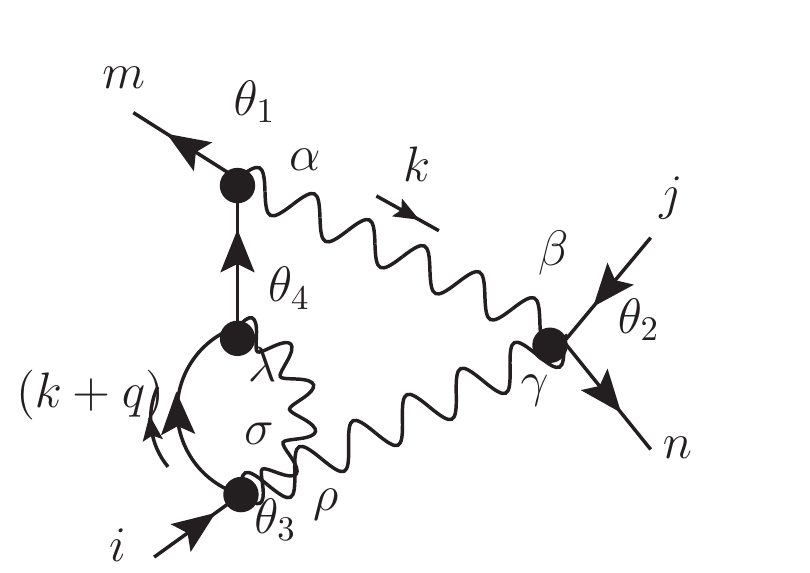}}\subfloat[]{\centering{}\includegraphics[scale=0.5]{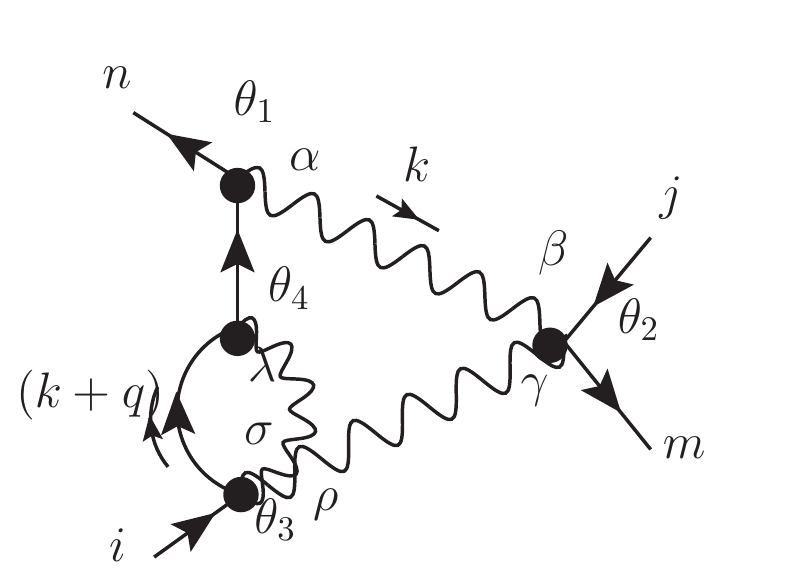}}
\par\end{centering}
\centering{}\subfloat[]{\centering{}\includegraphics[scale=0.5]{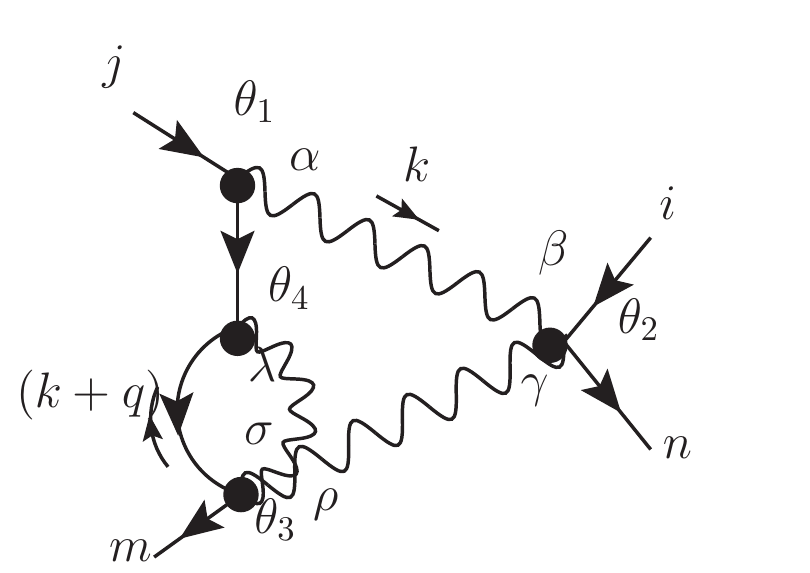}}\subfloat[]{\centering{}\includegraphics[scale=0.5]{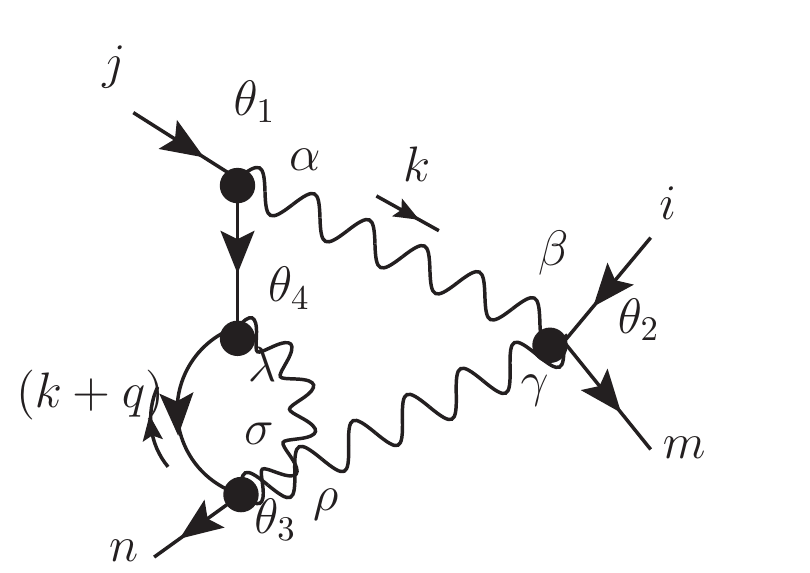}}\subfloat[]{\centering{}\includegraphics[scale=0.5]{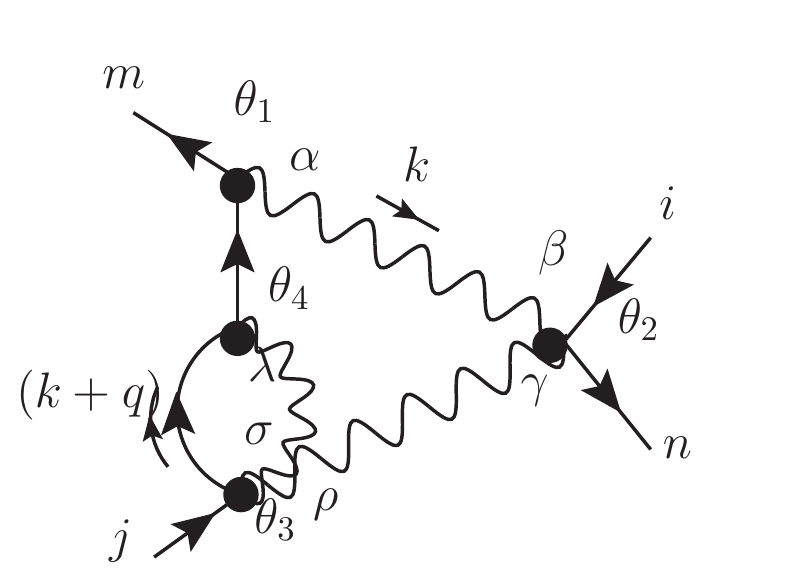}}\subfloat[]{\centering{}\includegraphics[scale=0.5]{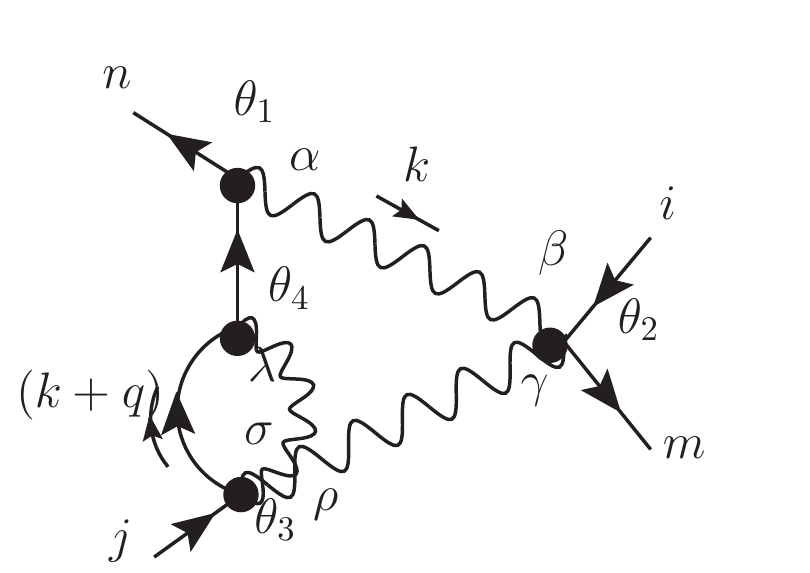}}\caption{\label{fig:D21-order-g-6}$\mathcal{S}_{\left(\overline{\Phi}\Phi\right)^{2}}^{\left(D21\right)}$}
\end{figure}
\par\end{center}

\begin{center}
\begin{table}
\centering{}%
\begin{tabular}{lcccccccccc}
 &  &  &  &  &  &  &  &  &  & \tabularnewline
\hline 
\hline 
$D21-a$ &  & $\delta_{jn}\delta_{im}$ &  & $D21-b$ &  & $\delta_{mj}\delta_{ni}$ &  & $D21-c$ &  & $\delta_{jn}\delta_{im}$\tabularnewline
$D21-d$ &  & $\delta_{mj}\delta_{ni}$ &  & $D21-e$ &  & $\delta_{mj}\delta_{ni}$ &  & $D21-f$ &  & $\delta_{jn}\delta_{im}$\tabularnewline
$D21-g$ &  & $\delta_{mj}\delta_{ni}$ &  & $D21-h$ &  & $\delta_{jn}\delta_{im}$ &  &  &  & \tabularnewline
\hline 
\hline 
 &  &  &  &  &  &  &  &  &  & \tabularnewline
\end{tabular}\caption{\label{tab:S-PPPP-21}Values of the diagrams in Figure\,\ref{fig:D21-order-g-6}
with common factor\protect \\
 $\frac{1}{32}\left(\frac{b^{3}-3\,a^{3}+7\,a^{2}\,b-5\,a\,b^{2}}{32\pi^{2}\epsilon}\right)\,i\,g^{6}\int_{\theta}\overline{\Phi}_{i}\Phi_{m}\Phi_{n}\overline{\Phi}_{j}$
.}
\end{table}
\par\end{center}

We start with $\mathcal{S}_{\left(\overline{\Phi}\Phi\right)^{2}}^{\left(D21-a\right)}$
in the Figure\,\ref{fig:D21-order-g-6},
\begin{align}
\mathcal{S}_{\left(\overline{\Phi}\Phi\right)^{2}}^{\left(D21-a\right)} & =-\frac{1}{32}\,i\,g^{6}\,\delta_{mi}\delta_{jn}\int_{\theta}\overline{\Phi}_{i}\Phi_{m}\Phi_{n}\overline{\Phi}_{j}\int\frac{d^{D}kd^{D}q}{\left(2\pi\right)^{2D}}\times\nonumber \\
 & \left\{ \frac{-2\left(a^{3}-a^{2}\,b-a\,b^{2}+b^{3}\right)\left(k\cdot q\right)k^{2}-4\,a\left(a-b\right)^{2}\left(k^{2}\right)^{2}}{\left(k^{2}\right)^{3}\left(k+q\right)^{2}q^{2}}\right\} \,,
\end{align}
using Eqs.\,(\ref{eq:Int 7}) and\,(\ref{eq:Int 9}), then adding
$\mathcal{S}_{\left(\overline{\Phi}\Phi\right)^{2}}^{\left(D21-a\right)}$
to $\mathcal{S}_{\left(\overline{\Phi}\Phi\right)^{2}}^{\left(D21-h\right)}$
with the values in Table\,\ref{tab:S-PPPP-21}, we find
\begin{align}
\mathcal{S}_{\left(\overline{\Phi}\Phi\right)^{2}}^{\left(D21\right)} & =-\frac{1}{8}\left(\frac{\left(a-b\right)^{2}\left(3\,a-b\right)}{32\pi^{2}\epsilon}\right)\,i\,g^{6}\,\left(\delta_{mj}\delta_{in}+\delta_{mi}\delta_{jn}\right)\int_{\theta}\overline{\Phi}_{i}\Phi_{m}\Phi_{n}\overline{\Phi}_{j}\,.\label{eq:S-D21}
\end{align}

\begin{center}
\begin{figure}
\begin{centering}
\subfloat[]{\centering{}\includegraphics[scale=0.5]{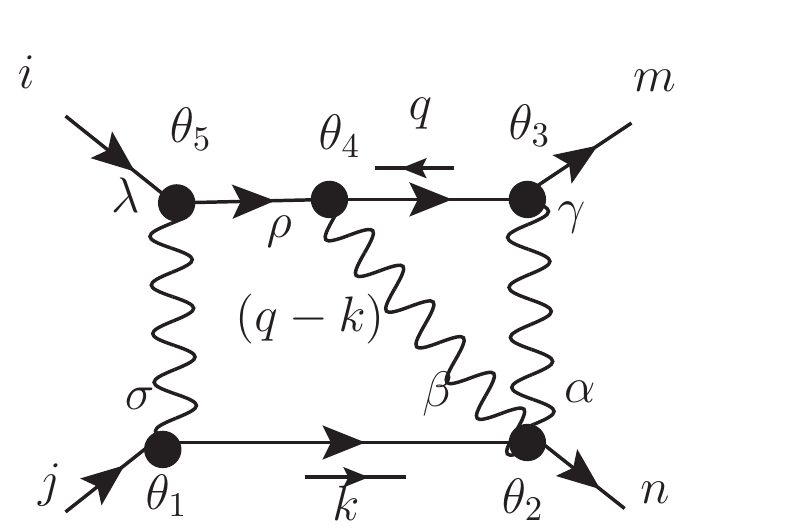}}\subfloat[]{\centering{}\includegraphics[scale=0.5]{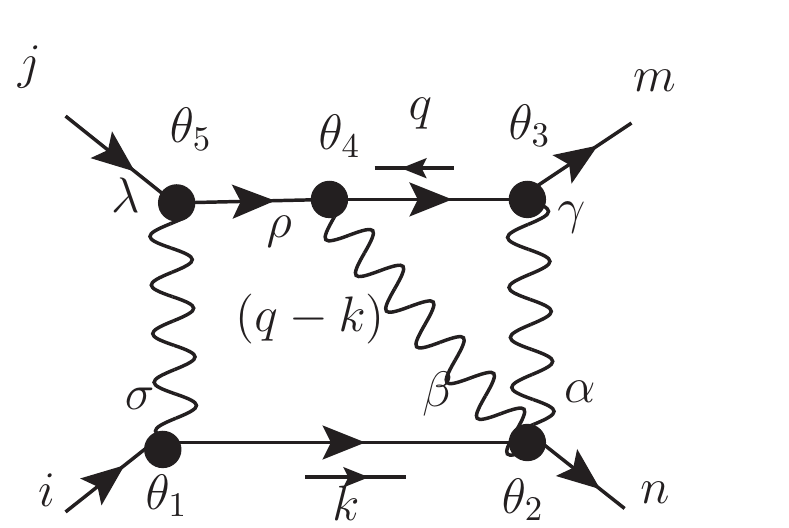}}\subfloat[]{\centering{}\includegraphics[scale=0.5]{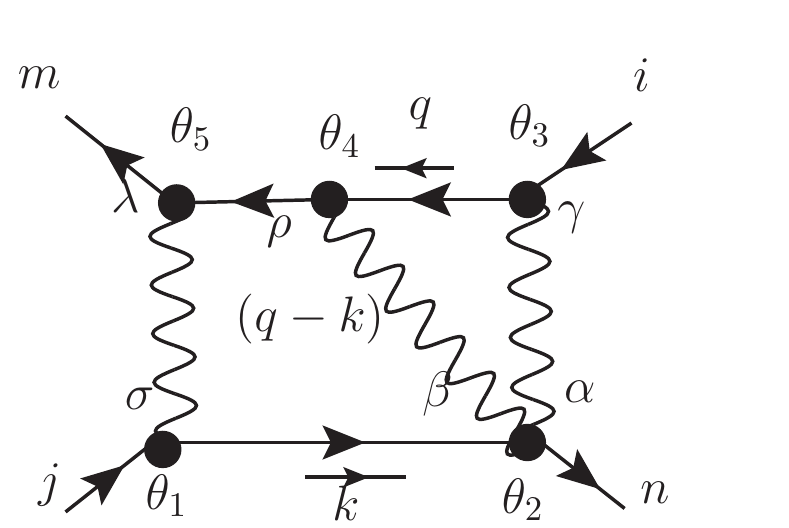}}\subfloat[]{\centering{}\includegraphics[scale=0.5]{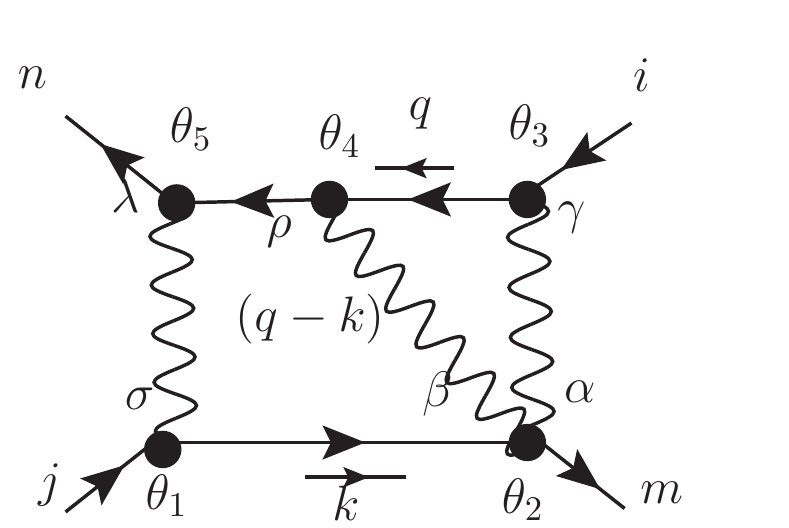}}
\par\end{centering}
\begin{centering}
\subfloat[]{\centering{}\includegraphics[scale=0.5]{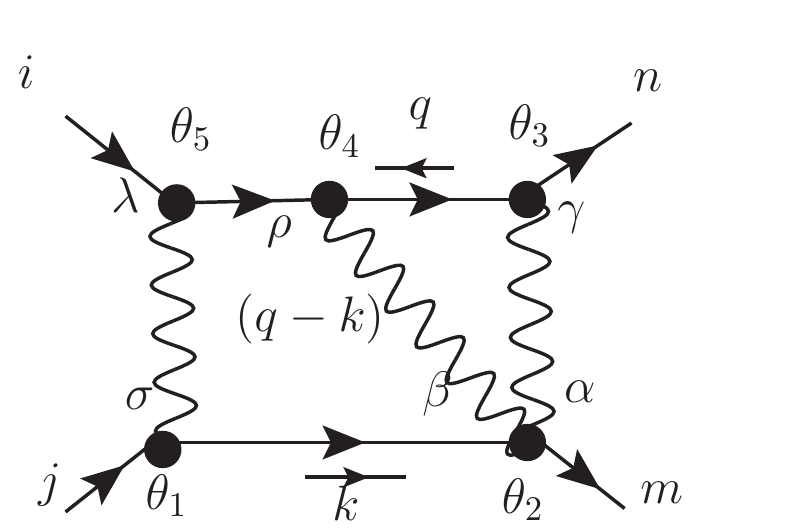}}\subfloat[]{\centering{}\includegraphics[scale=0.5]{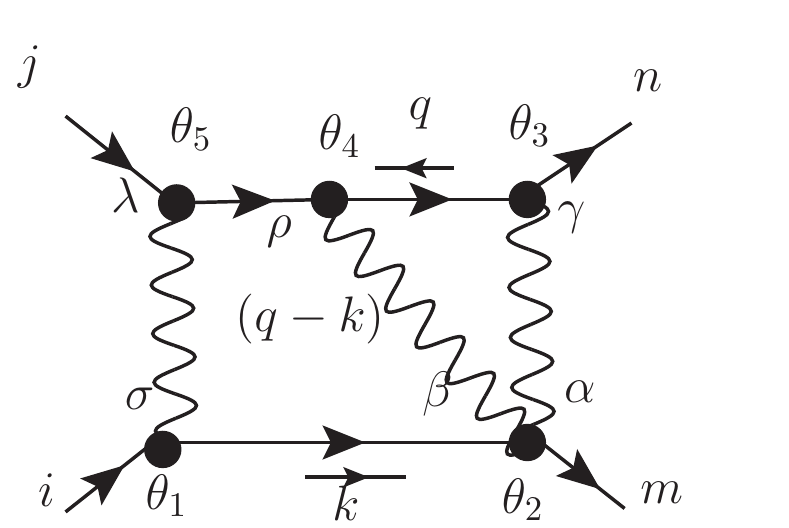}}\subfloat[]{\centering{}\includegraphics[scale=0.5]{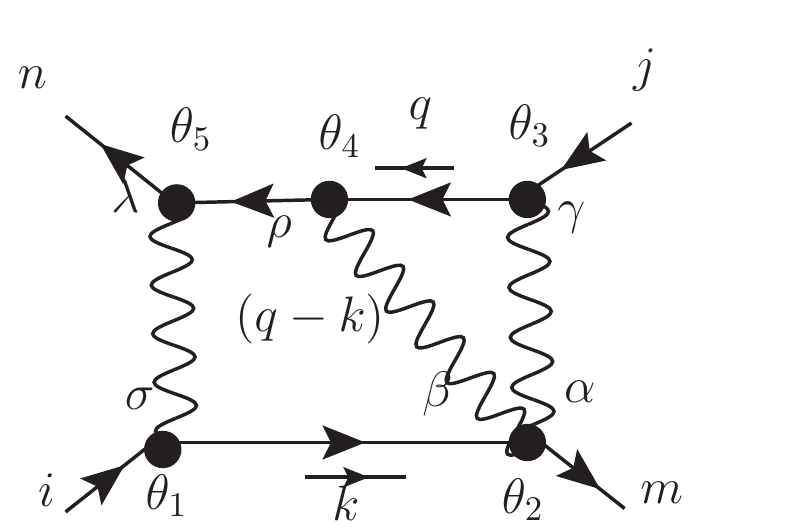}}\subfloat[]{\centering{}\includegraphics[scale=0.5]{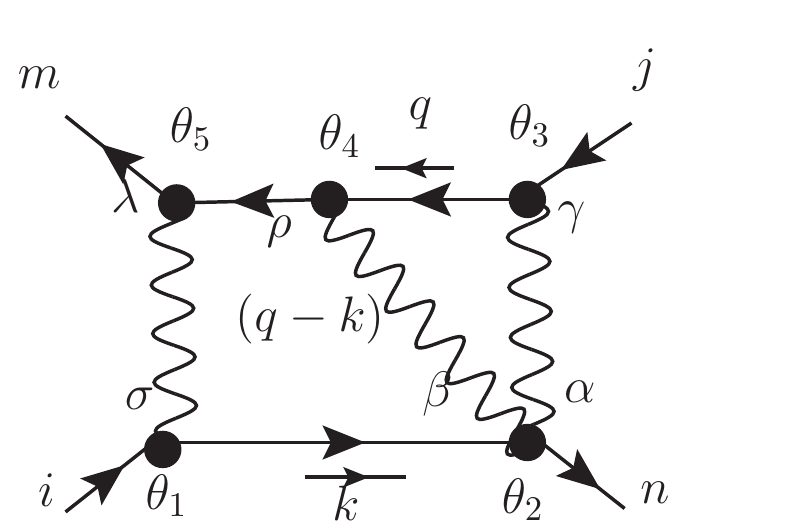}}
\par\end{centering}
\begin{centering}
\subfloat[]{\centering{}\includegraphics[scale=0.5]{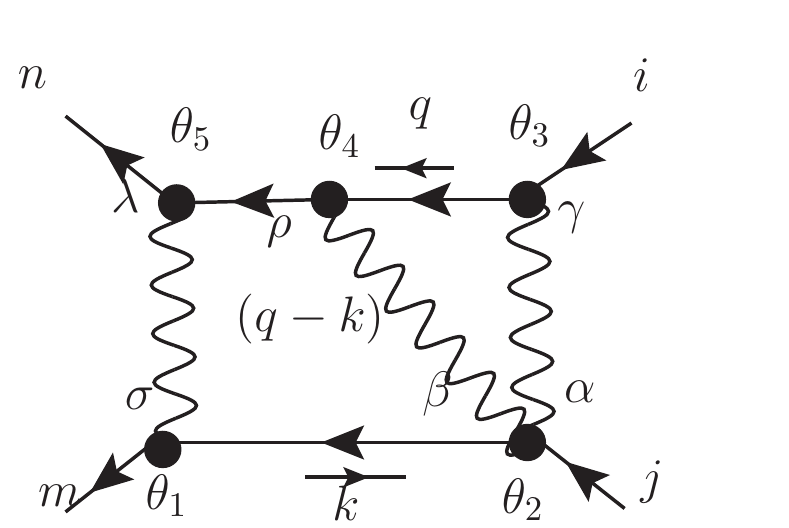}}\subfloat[]{\centering{}\includegraphics[scale=0.5]{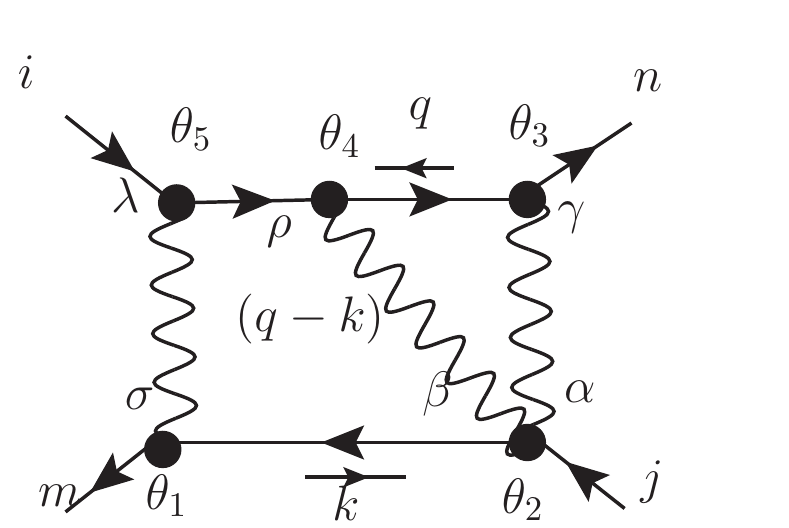}}\subfloat[]{\centering{}\includegraphics[scale=0.5]{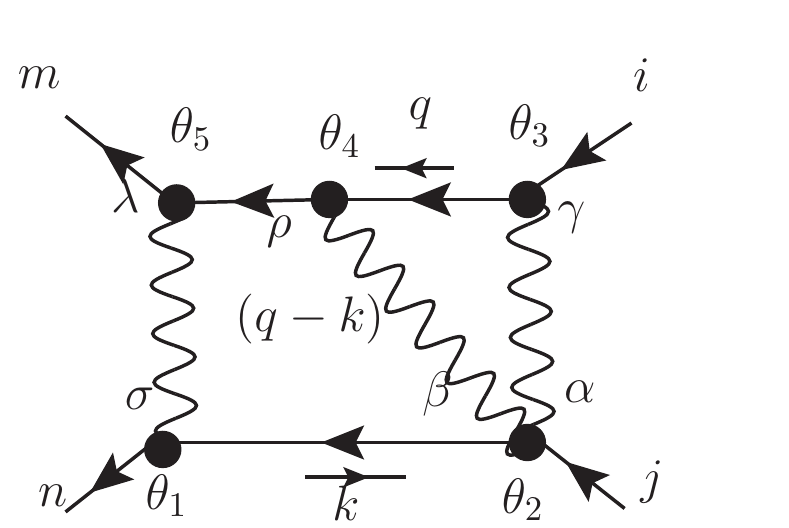}}\subfloat[]{\centering{}\includegraphics[scale=0.5]{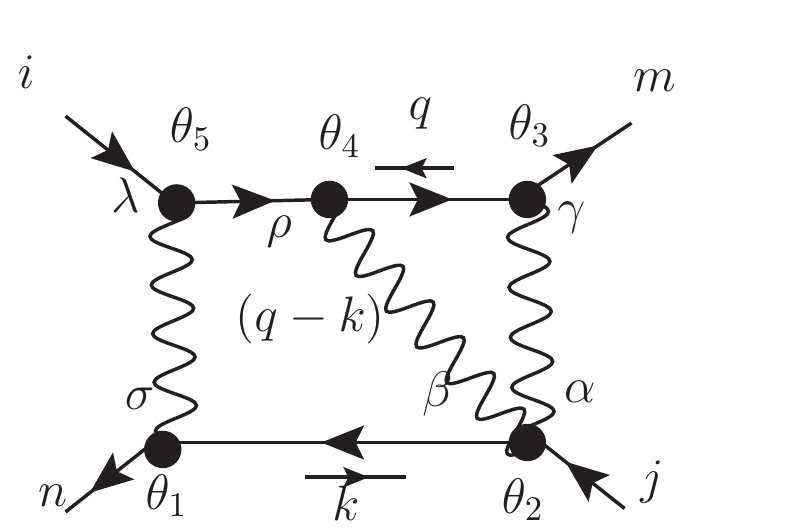}}
\par\end{centering}
\centering{}\subfloat[]{\centering{}\includegraphics[scale=0.5]{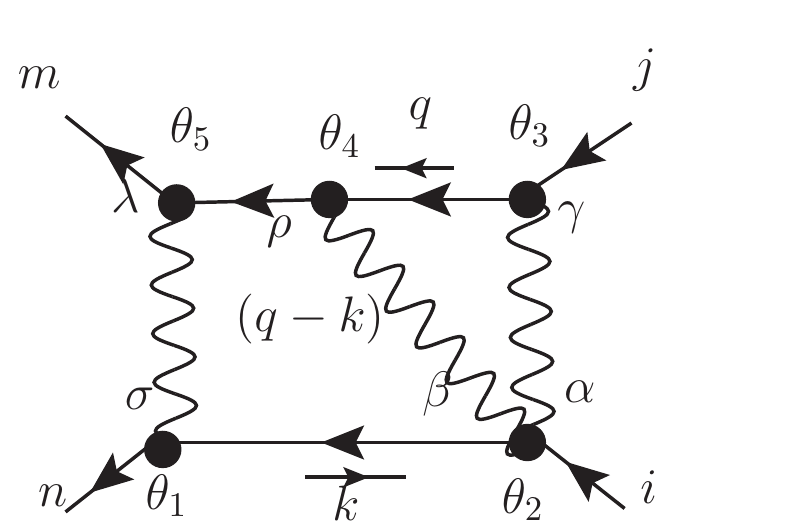}}\subfloat[]{\centering{}\includegraphics[scale=0.5]{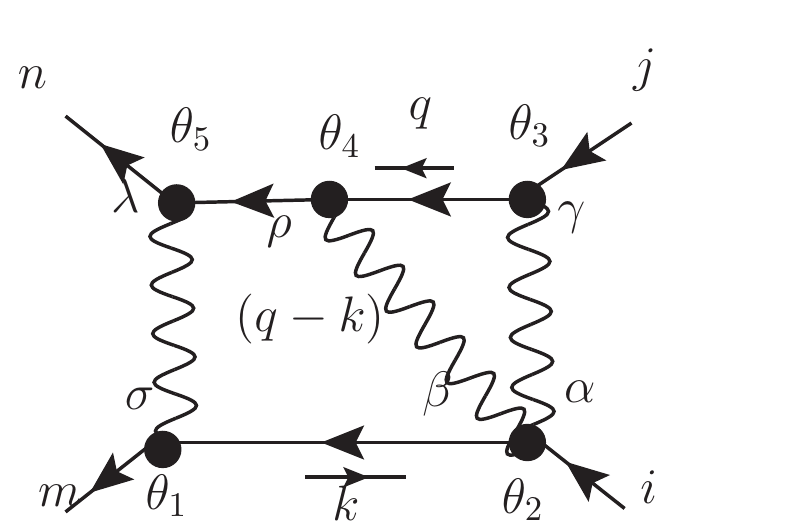}}\subfloat[]{\centering{}\includegraphics[scale=0.5]{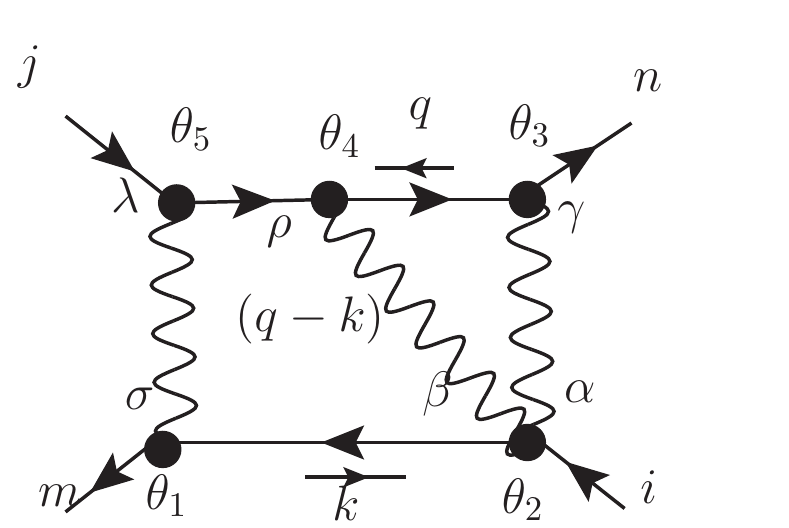}}\subfloat[]{\centering{}\includegraphics[scale=0.5]{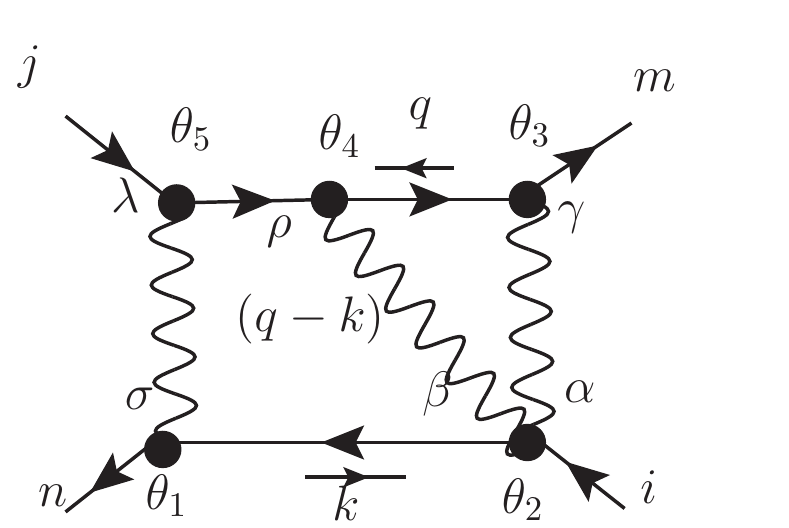}}\caption{\label{fig:D22-order-g-6}$\mathcal{S}_{\left(\overline{\Phi}\Phi\right)^{2}}^{\left(D22\right)}$}
\end{figure}
\par\end{center}

\begin{center}
\begin{table}
\centering{}%
\begin{tabular}{lcccccccccc}
 &  &  &  &  &  &  &  &  &  & \tabularnewline
\hline 
\hline 
$D22-a$ &  & $\delta_{jn}\delta_{im}$ &  & $D22-b$ &  & $\delta_{mj}\delta_{ni}$ &  & $D22-c$ &  & $-\delta_{jn}\delta_{im}$\tabularnewline
$D22-d$ &  & $-\delta_{mj}\delta_{ni}$ &  & $D22-e$ &  & $\delta_{mj}\delta_{ni}$ &  & $D22-f$ &  & $\delta_{jn}\delta_{im}$\tabularnewline
$D22-g$ &  & $-\delta_{jn}\delta_{im}$ &  & $D22-h$ &  & $-\delta_{mj}\delta_{ni}$ &  & $D22-i$ &  & $\delta_{mj}\delta_{ni}$\tabularnewline
$D22-j$ &  & $-\delta_{mj}\delta_{ni}$ &  & $D22-k$ &  & $\delta_{jn}\delta_{im}$ &  & $D22-l$ &  & $-\delta_{jn}\delta_{im}$\tabularnewline
$D22-m$ &  & $\delta_{mj}\delta_{ni}$ &  & $D22-n$ &  & $\delta_{jn}\delta_{im}$ &  & $D22-o$ &  & $-\delta_{jn}\delta_{im}$\tabularnewline
$D22-p$ &  & $-\delta_{mj}\delta_{ni}$ &  &  &  &  &  &  &  & \tabularnewline
\hline 
\hline 
 &  &  &  &  &  &  &  &  &  & \tabularnewline
\end{tabular}\caption{\label{tab:S-PPPP-22}Values of the diagrams in Figure\,\ref{fig:D22-order-g-6}
with common factor\protect \\
 $\frac{1}{16}\left(\frac{a\left(a-b\right)^{2}}{32\pi^{2}\epsilon}\right)\,i\,g^{6}\int_{\theta}\overline{\Phi}_{i}\Phi_{m}\Phi_{n}\overline{\Phi}_{j}$
.}
\end{table}
\par\end{center}

$\mathcal{S}_{\left(\overline{\Phi}\Phi\right)^{2}}^{\left(D22-a\right)}$
in the Figure\,\ref{fig:D22-order-g-6} is
\begin{align}
\mathcal{S}_{\left(\overline{\Phi}\Phi\right)^{2}}^{\left(D22-a\right)} & =\frac{1}{128}\,i\,\delta_{jn}\delta_{mi}\,g^{6}\int_{\theta}\overline{\Phi}_{i}\Phi_{m}\Phi_{n}\overline{\Phi}_{j}\int\frac{d^{D}kd^{D}q}{\left(2\pi\right)^{2D}}\left\{ \frac{-8\,a\left(a-b\right)^{2}\left(k^{2}\right)^{2}q^{2}}{\left(k^{2}\right)^{3}\left(q-k\right)^{2}\left(q^{2}\right)^{2}}\right\} ,\,
\end{align}
using Eq.\,(\ref{eq:Int 7}), then adding $\mathcal{S}_{\left(\overline{\Phi}\Phi\right)^{2}}^{\left(D21-a\right)}$
to $\mathcal{S}_{\left(\overline{\Phi}\Phi\right)^{2}}^{\left(D21-p\right)}$
with the values in Table\,\ref{tab:S-PPPP-22}, we find 
\begin{align}
\mathcal{S}_{\left(\overline{\Phi}\Phi\right)^{2}}^{\left(D22-a\right)} & =\mathcal{S}_{\left(\overline{\Phi}\Phi\right)^{2}}^{\left(D22\right)}=0\,.\label{eq:S-D22}
\end{align}

\begin{center}
\begin{figure}
\centering{}\subfloat[]{\centering{}\includegraphics[scale=0.5]{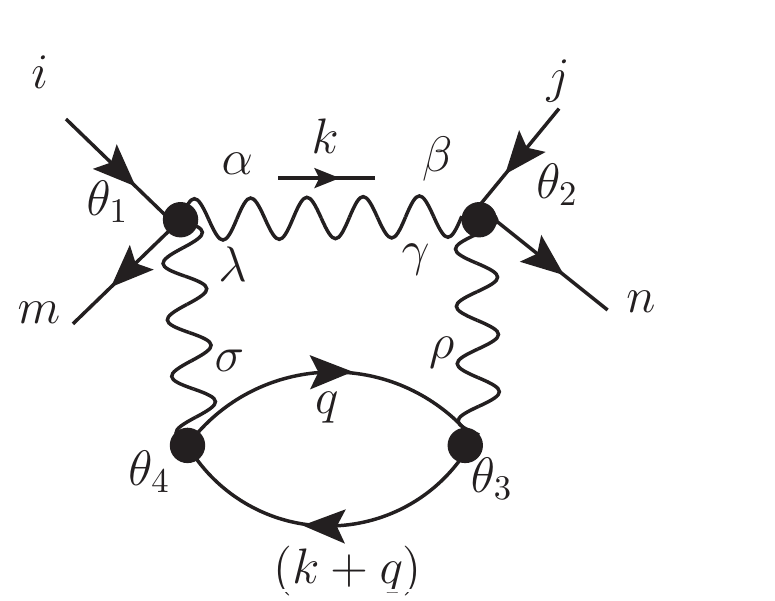}}\subfloat[]{\centering{}\includegraphics[scale=0.5]{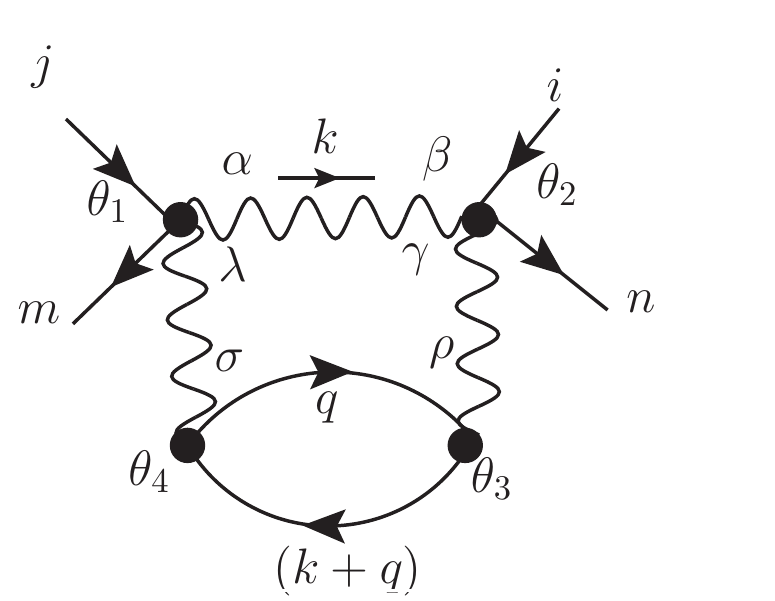}}\caption{\label{fig:D23-order-g-6}$\mathcal{S}_{\left(\overline{\Phi}\Phi\right)^{2}}^{\left(D23\right)}$}
\end{figure}
\par\end{center}

\begin{center}
\begin{table}
\centering{}%
\begin{tabular}{lcccccc}
 &  &  &  &  &  & \tabularnewline
\hline 
\hline 
$D23-a$ &  & $\delta_{jn}\delta_{im}$ &  & $D23-b$ &  & $\delta_{mj}\delta_{ni}$\tabularnewline
\hline 
\hline 
 &  &  &  &  &  & \tabularnewline
\end{tabular}\caption{\label{tab:S-PPPP-23}Values of the diagrams in Figure\,\ref{fig:D23-order-g-6}
with common factor\protect \\
 $\frac{1}{16}\left(\frac{\left(a+b\right)^{3}}{32\pi^{2}\epsilon}\right)N\,i\,g^{6}\int_{\theta}\overline{\Phi}_{i}\Phi_{m}\Phi_{n}\overline{\Phi}_{j}$
.}
\end{table}
\par\end{center}

$\mathcal{S}_{\left(\overline{\Phi}\Phi\right)^{2}}^{\left(D23-a\right)}$
in the Figure\,\ref{fig:D23-order-g-6} is
\begin{align}
\mathcal{S}_{\left(\overline{\Phi}\Phi\right)^{2}}^{\left(D23-a\right)} & =-\frac{1}{32}\,i\,N\,g^{6}\delta_{im}\delta_{jn}\int_{\theta}\overline{\Phi}_{i}\Phi_{m}\Phi_{n}\overline{\Phi}_{j}\nonumber \\
 & \times\int\frac{d^{D}kd^{D}q}{\left(2\pi\right)^{2D}}\left\{ \frac{-8\left(a^{3}+3\,a\,b^{2}\right)k^{2}\left(\left(k\cdot q\right)+q^{2}\right)-2\left(a-b\right)^{3}\left(k^{2}\right)^{2}}{\left(k^{2}\right)^{3}\left(k+q\right)^{2}q^{2}}\right\} \,,
\end{align}
using Eqs.\,(\ref{eq:Int 7}), (\ref{eq:Int 9}), then adding $\mathcal{S}_{\left(\overline{\Phi}\Phi\right)^{2}}^{\left(D23-a\right)}$
to $\mathcal{S}_{\left(\overline{\Phi}\Phi\right)^{2}}^{\left(D23-b\right)}$
with the values in Table\,\ref{tab:S-PPPP-23}, we find 
\begin{align}
\mathcal{S}_{\left(\overline{\Phi}\Phi\right)^{2}}^{\left(D23\right)} & =\frac{1}{16}\left(\frac{\left(a+b\right)^{3}}{32\pi^{2}\epsilon}\right)\,i\,N\,g^{6}\left(\delta_{im}\delta_{jn}+\delta_{jm}\delta_{in}\right)\int_{\theta}\overline{\Phi}_{i}\Phi_{m}\Phi_{n}\overline{\Phi}_{j}\,.\label{eq:S-D23}
\end{align}

\begin{center}
\begin{figure}
\begin{centering}
\subfloat[]{\centering{}\includegraphics[scale=0.5]{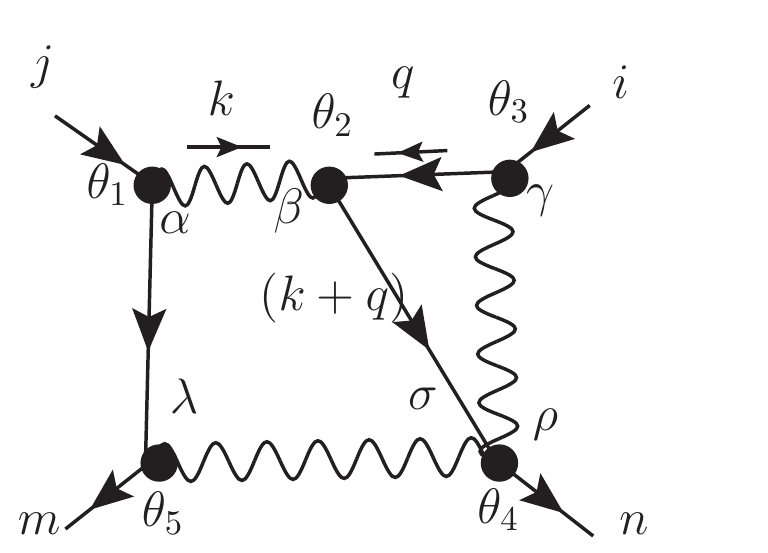}}\subfloat[]{\centering{}\includegraphics[scale=0.5]{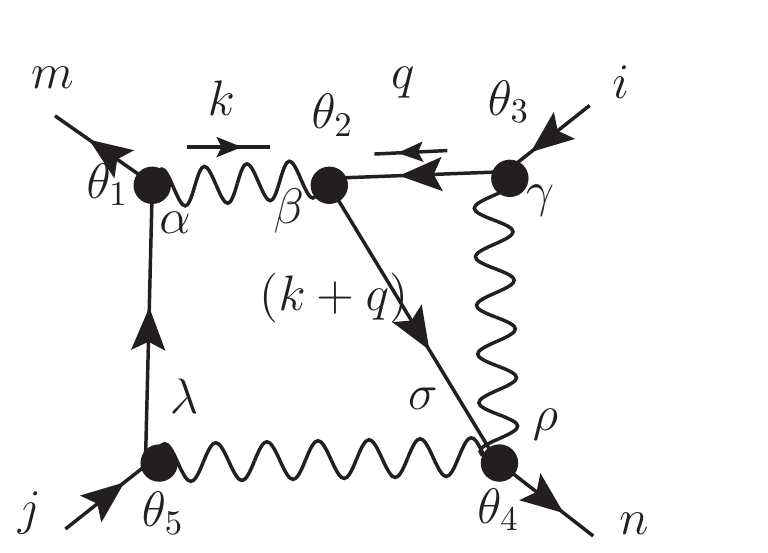}}\subfloat[]{\centering{}\includegraphics[scale=0.5]{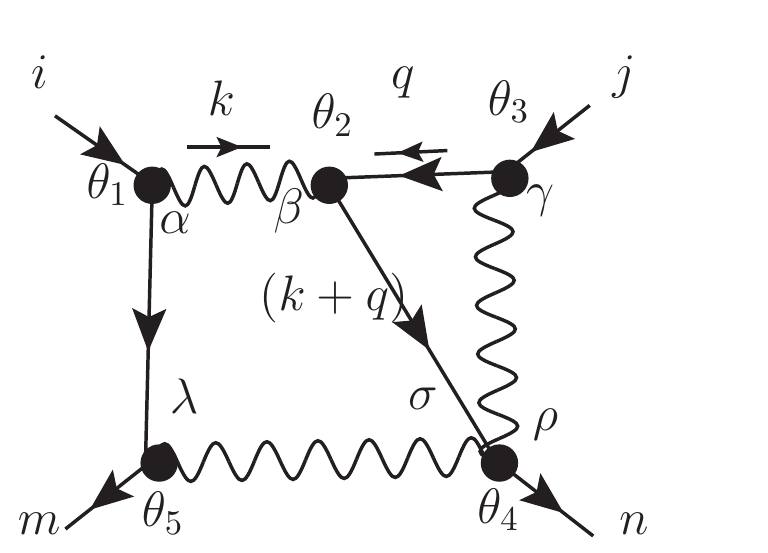}}\subfloat[]{\centering{}\includegraphics[scale=0.5]{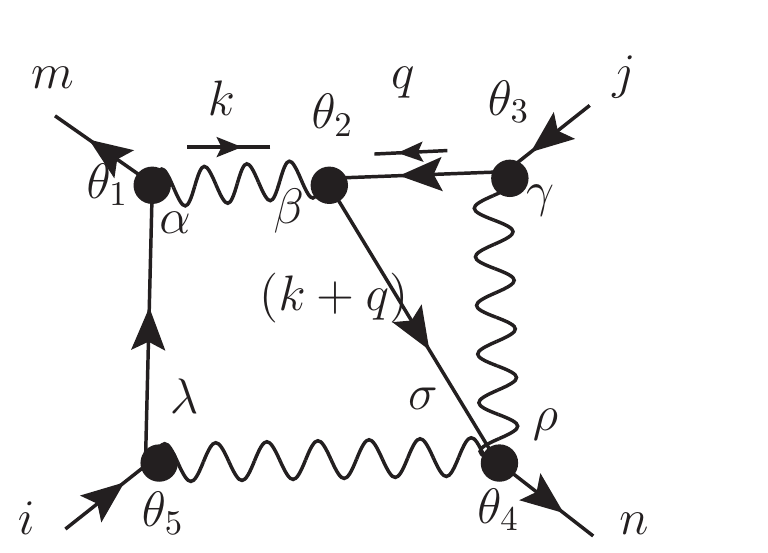}}
\par\end{centering}
\begin{centering}
\subfloat[]{\centering{}\includegraphics[scale=0.5]{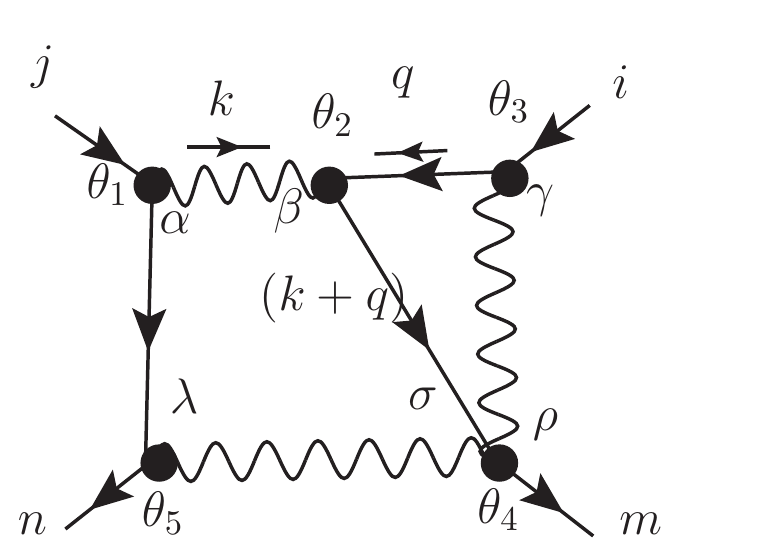}}\subfloat[]{\centering{}\includegraphics[scale=0.5]{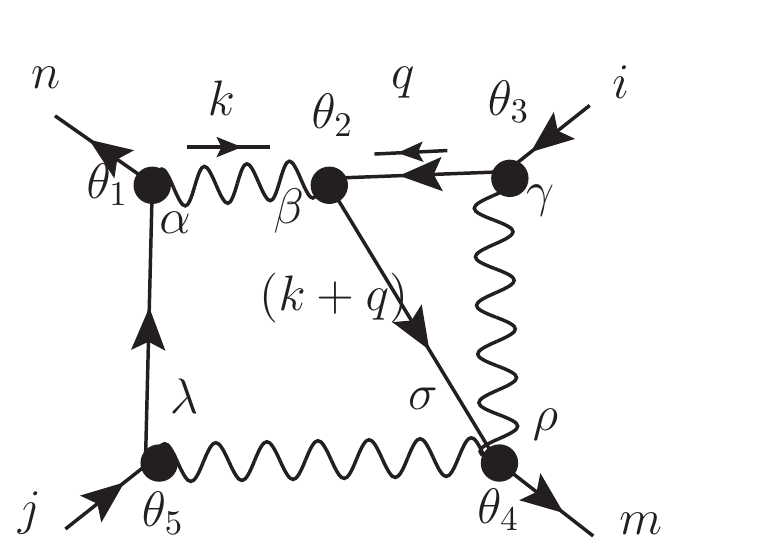}}\subfloat[]{\centering{}\includegraphics[scale=0.5]{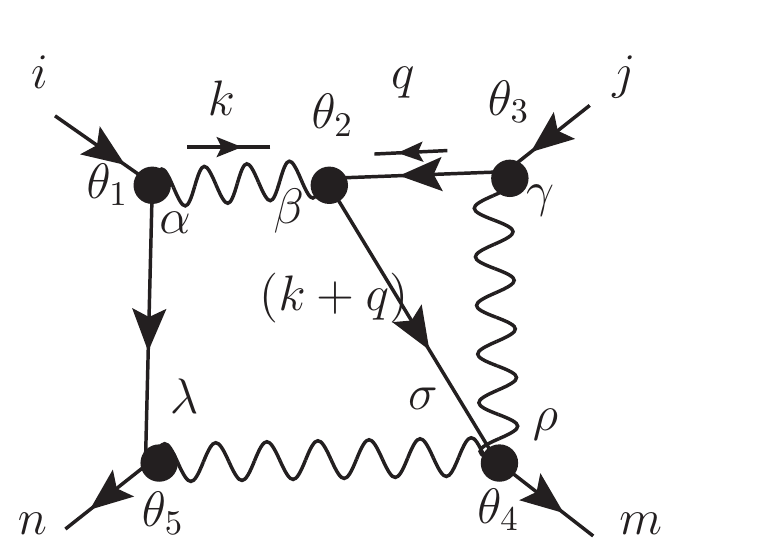}}\subfloat[]{\centering{}\includegraphics[scale=0.5]{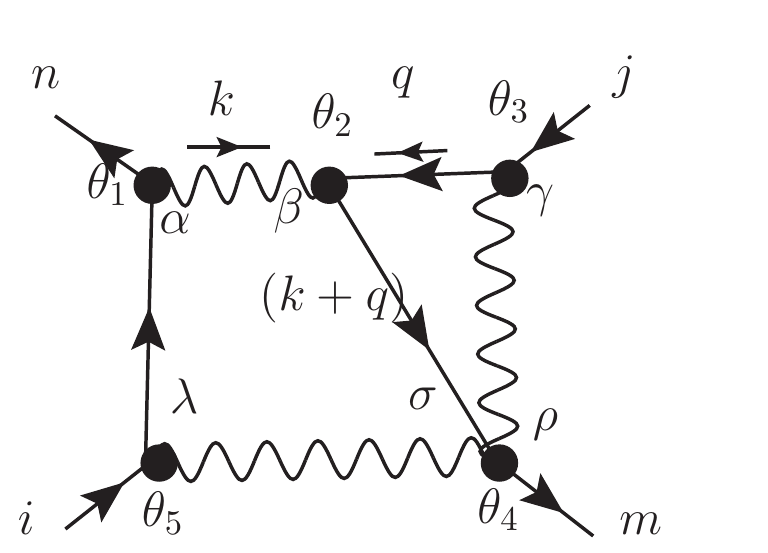}}
\par\end{centering}
\begin{centering}
\subfloat[]{\centering{}\includegraphics[scale=0.5]{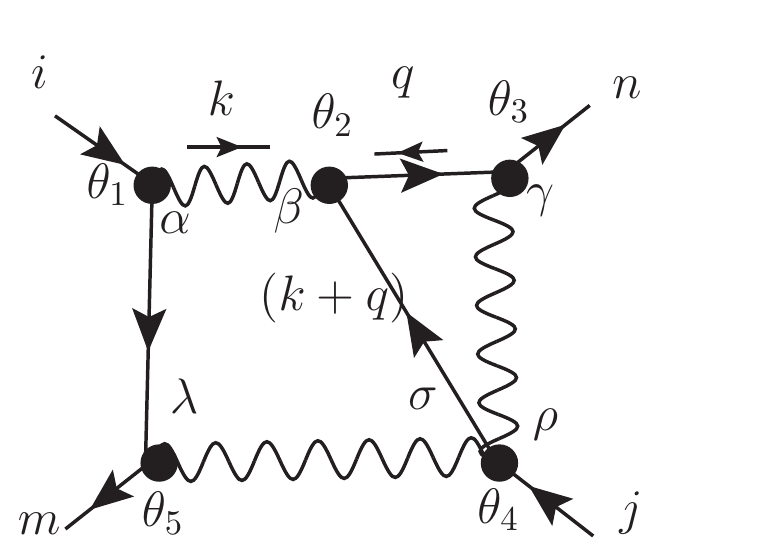}}\subfloat[]{\centering{}\includegraphics[scale=0.5]{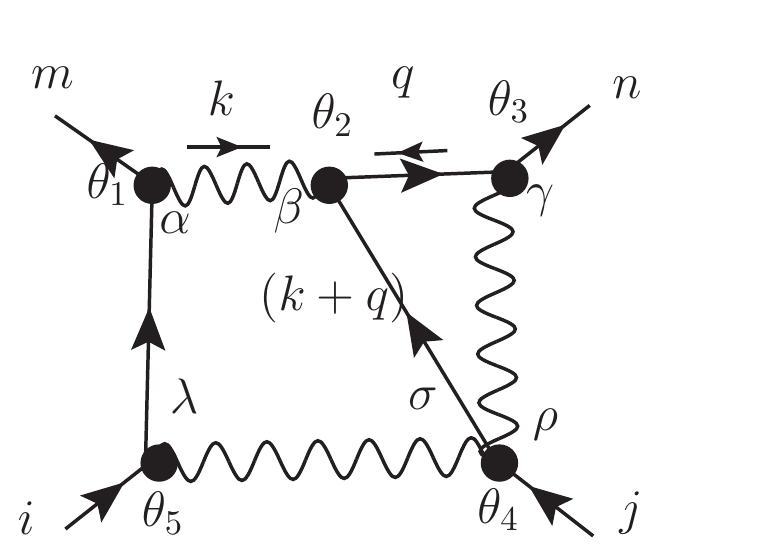}}\subfloat[]{\centering{}\includegraphics[scale=0.5]{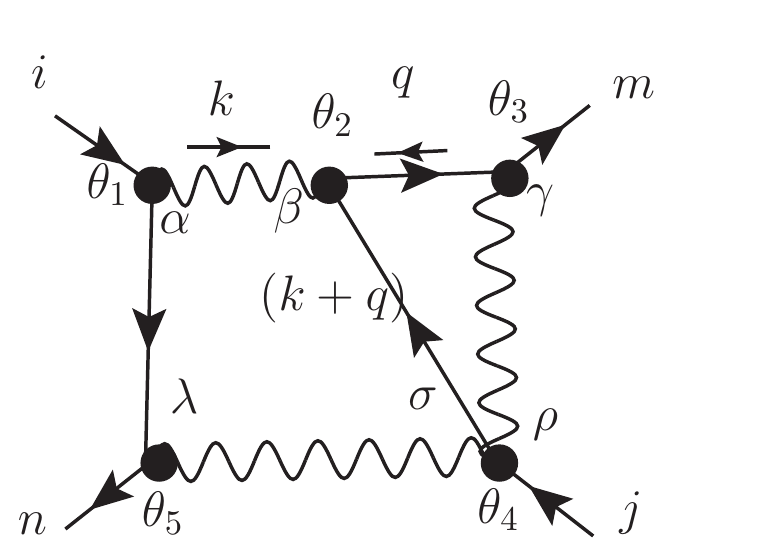}}\subfloat[]{\centering{}\includegraphics[scale=0.5]{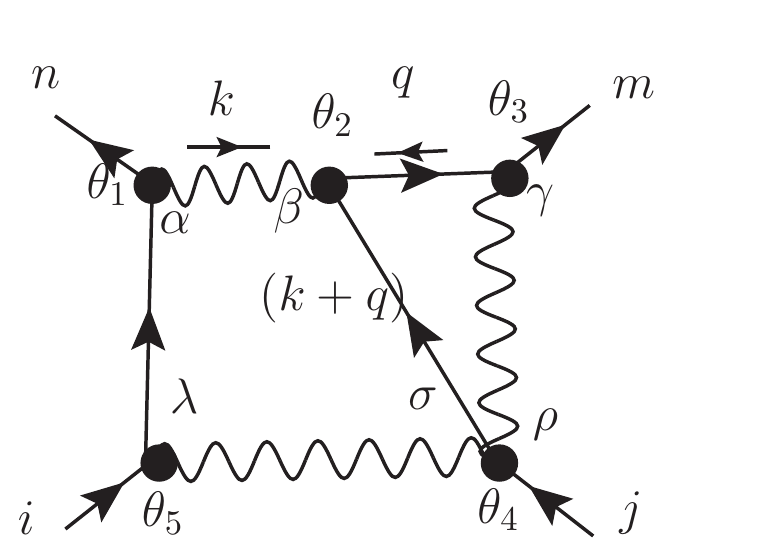}}
\par\end{centering}
\centering{}\subfloat[]{\centering{}\includegraphics[scale=0.5]{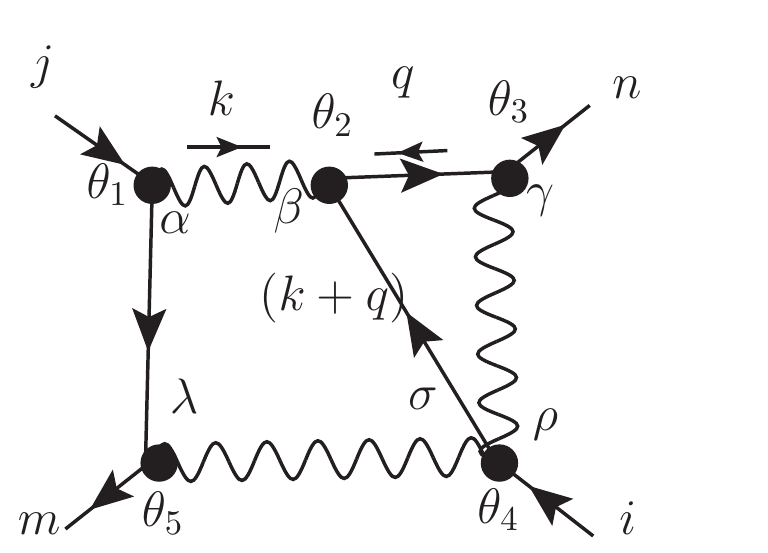}}\subfloat[]{\centering{}\includegraphics[scale=0.5]{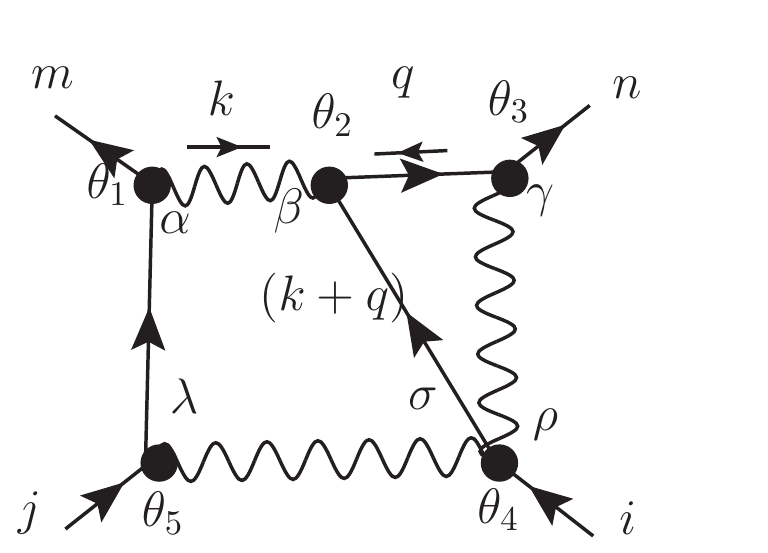}}\subfloat[]{\centering{}\includegraphics[scale=0.5]{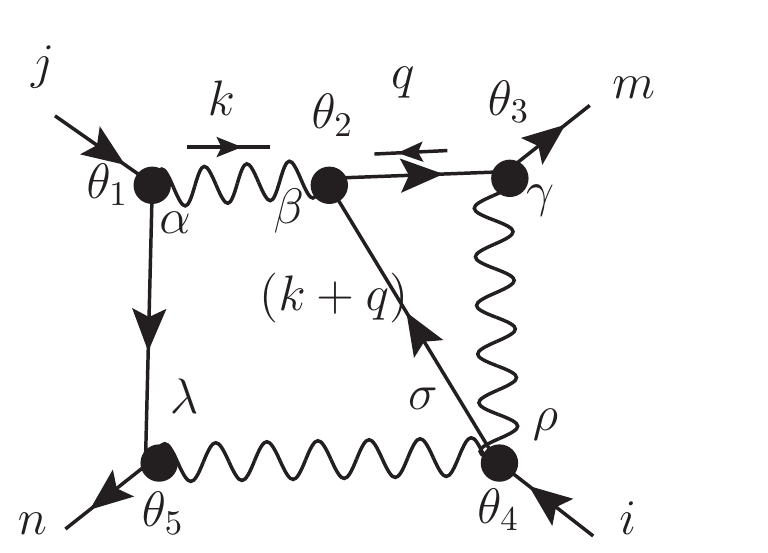}}\subfloat[]{\centering{}\includegraphics[scale=0.5]{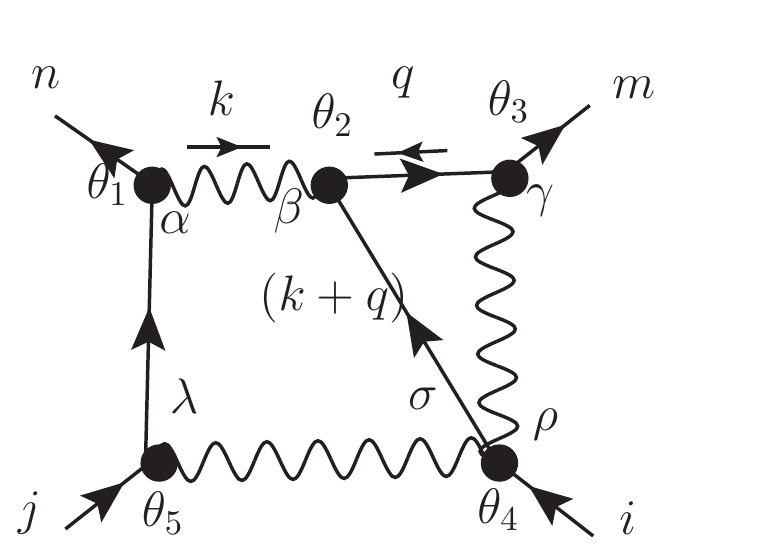}}\caption{\label{fig:D24-order-g-6}$\mathcal{S}_{\left(\overline{\Phi}\Phi\right)^{2}}^{\left(D24\right)}$}
\end{figure}
\par\end{center}

\begin{center}
\begin{table}
\centering{}%
\begin{tabular}{lcccccccccc}
 &  &  &  &  &  &  &  &  &  & \tabularnewline
\hline 
\hline 
$D24-a$ &  & $\delta_{mj}\delta_{ni}$ &  & $D24-b$ &  & $\delta_{mj}\delta_{ni}$ &  & $D24-c$ &  & $\delta_{jn}\delta_{im}$\tabularnewline
$D24-d$ &  & $\delta_{jn}\delta_{im}$ &  & $D24-e$ &  & $\delta_{jn}\delta_{im}$ &  & $D24-f$ &  & $\delta_{jn}\delta_{im}$\tabularnewline
$D24-g$ &  & $\delta_{mj}\delta_{ni}$ &  & $D24-h$ &  & $\delta_{mj}\delta_{ni}$ &  & $D24-i$ &  & $\delta_{jn}\delta_{im}$\tabularnewline
$D24-j$ &  & $\delta_{jn}\delta_{im}$ &  & $D24-k$ &  & $\delta_{mj}\delta_{ni}$ &  & $D24-l$ &  & $\delta_{mj}\delta_{ni}$\tabularnewline
$D24-m$ &  & $\delta_{mj}\delta_{ni}$ &  & $D24-n$ &  & $\delta_{mj}\delta_{ni}$ &  & $D24-o$ &  & $\delta_{jn}\delta_{im}$\tabularnewline
$D24-p$ &  & $\delta_{jn}\delta_{im}$ &  &  &  &  &  &  &  & \tabularnewline
\hline 
\hline 
 &  &  &  &  &  &  &  &  &  & \tabularnewline
\end{tabular}\caption{\label{tab:S-PPPP-24}Values of the diagrams in Figure\,\ref{fig:D24-order-g-6}
with common factor\protect \\
 $\frac{1}{64}\left(\frac{\left(a-b\right)^{3}}{32\pi^{2}\epsilon}\right)\,i\,g^{6}\int_{\theta}\overline{\Phi}_{i}\Phi_{m}\Phi_{n}\overline{\Phi}_{j}$
.}
\end{table}
\par\end{center}

$\mathcal{S}_{\left(\overline{\Phi}\Phi\right)^{2}}^{\left(D24-a\right)}$
in the Figure\,\ref{fig:D24-order-g-6} is
\begin{align}
\mathcal{S}_{\left(\overline{\Phi}\Phi\right)^{2}}^{\left(D24-a\right)} & =\frac{1}{128}\,i\,\delta_{in}\delta_{mj}\,g^{6}\int_{\theta}\overline{\Phi}_{i}\Phi_{m}\Phi_{n}\overline{\Phi}_{j}\int\frac{d^{D}kd^{D}q}{\left(2\pi\right)^{2D}}\left\{ \frac{4\left(a-b\right)^{3}\left(k\cdot q\right)k^{2}\,q^{2}}{\left(k^{2}\right)^{3}\left(k+q\right)^{2}\left(q^{2}\right)^{2}}\right\} \,,
\end{align}
using Eq.\,(\ref{eq:Int 9}), then adding $\mathcal{S}_{\left(\overline{\Phi}\Phi\right)^{2}}^{\left(D24-a\right)}$
to $\mathcal{S}_{\left(\overline{\Phi}\Phi\right)^{2}}^{\left(D24-p\right)}$
with the values in Table\,\ref{tab:S-PPPP-24}, we find 
\begin{align}
\mathcal{S}_{\left(\overline{\Phi}\Phi\right)^{2}}^{\left(D24\right)} & =\frac{1}{8}\left(\frac{\left(a-b\right)^{3}}{32\pi^{2}\epsilon}\right)i\,g^{6}\left(\delta_{in}\delta_{mj}+\delta_{jn}\delta_{mi}\right)\int_{\theta}\overline{\Phi}_{i}\Phi_{m}\Phi_{n}\overline{\Phi}_{j}\,.\label{eq:S-D24}
\end{align}

\begin{center}
\begin{figure}
\begin{centering}
\subfloat[]{\centering{}\includegraphics[scale=0.5]{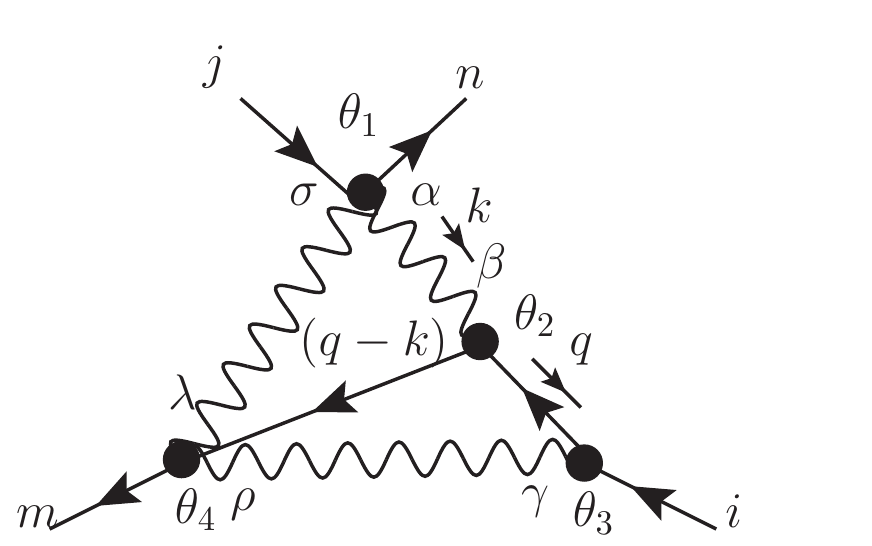}}\subfloat[]{\centering{}\includegraphics[scale=0.5]{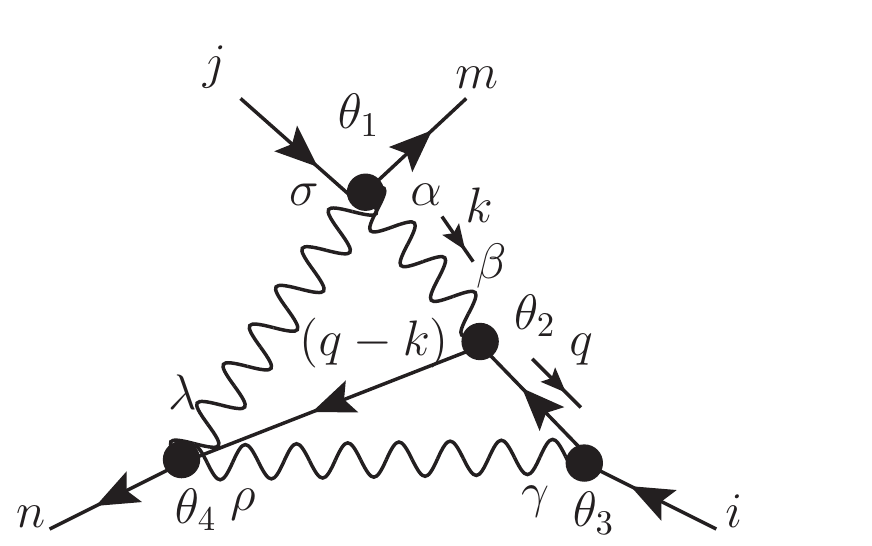}}\subfloat[]{\centering{}\includegraphics[scale=0.5]{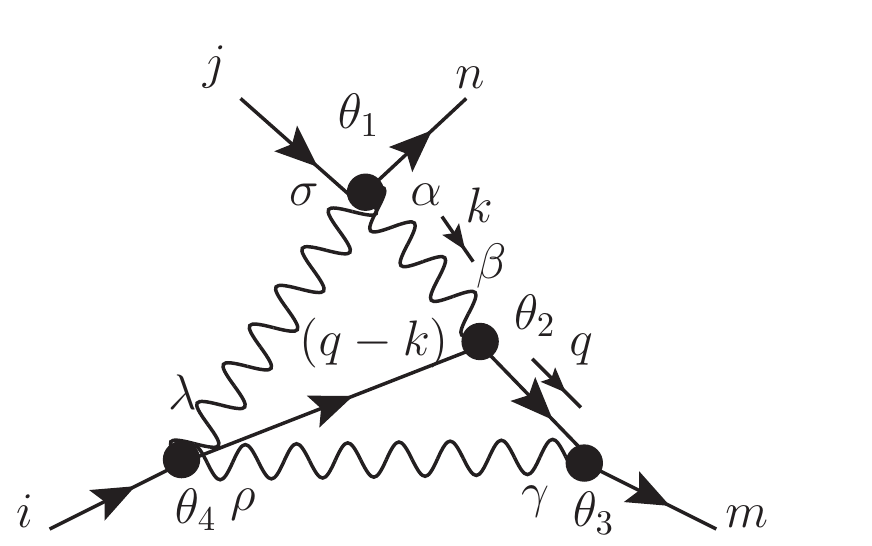}}
\par\end{centering}
\begin{centering}
\subfloat[]{\centering{}\includegraphics[scale=0.5]{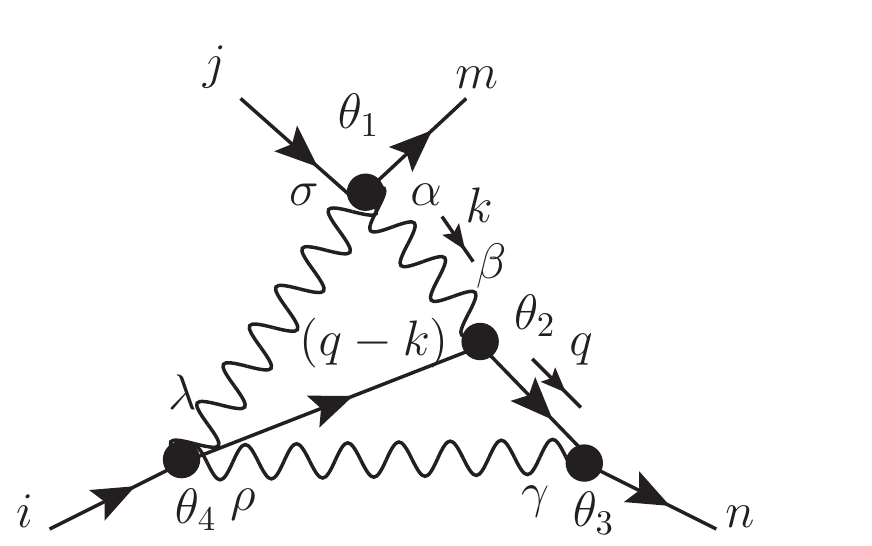}}\subfloat[]{\centering{}\includegraphics[scale=0.5]{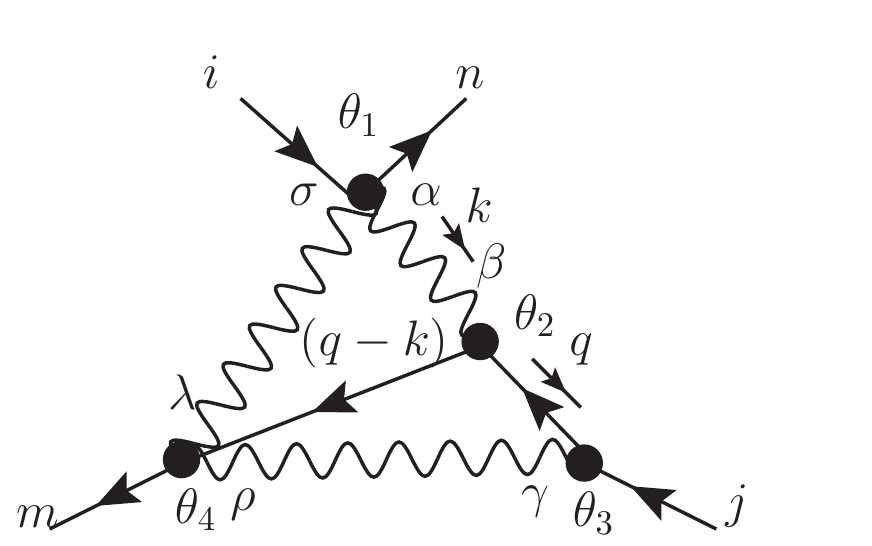}}\subfloat[]{\centering{}\includegraphics[scale=0.5]{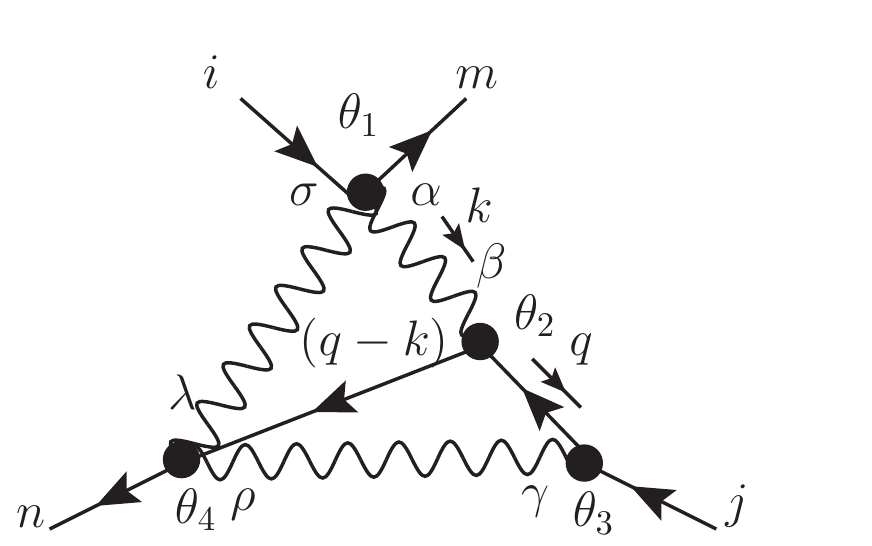}}
\par\end{centering}
\centering{}\subfloat[]{\centering{}\includegraphics[scale=0.5]{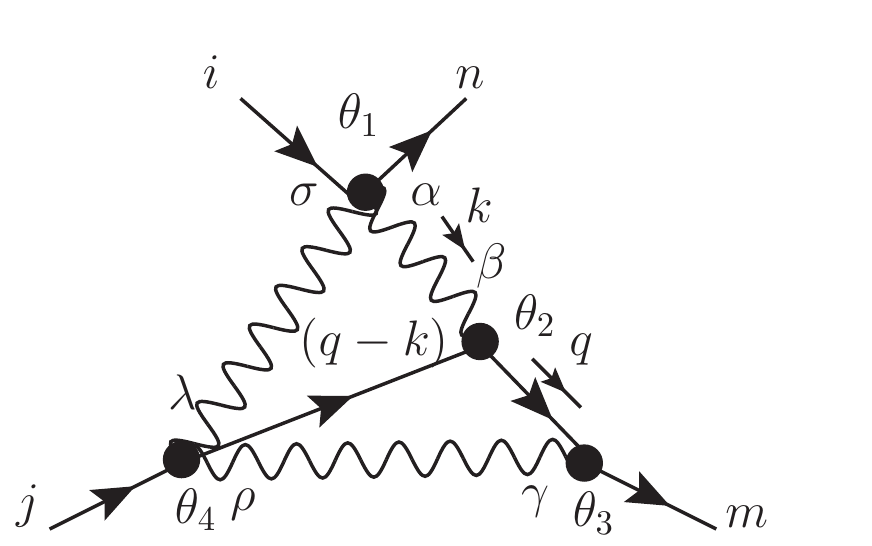}}\subfloat[]{\centering{}\includegraphics[scale=0.5]{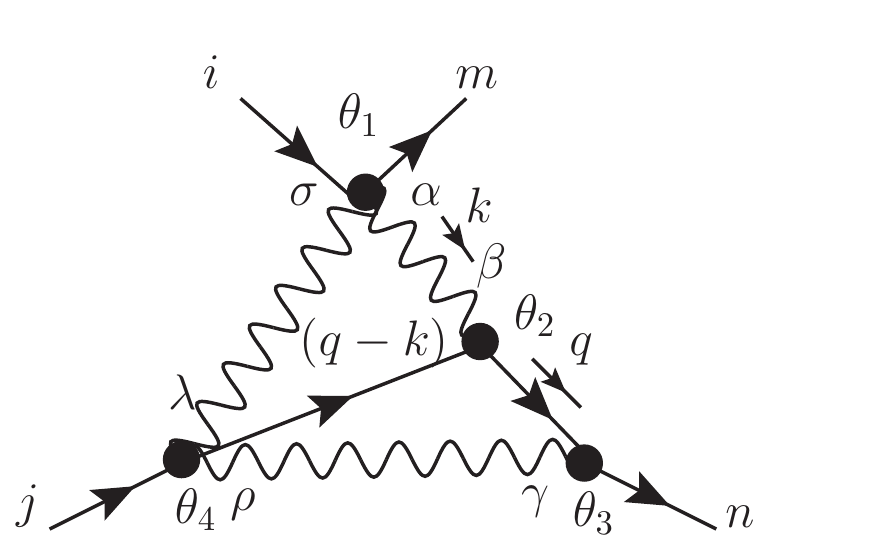}}\caption{\label{fig:D25-order-g-6}$\mathcal{S}_{\left(\overline{\Phi}\Phi\right)^{2}}^{\left(D25\right)}$}
\end{figure}
\par\end{center}

\begin{center}
\begin{table}
\centering{}%
\begin{tabular}{lcccccccccc}
 &  &  &  &  &  &  &  &  &  & \tabularnewline
\hline 
\hline 
$D25-a$ &  & $\delta_{jn}\delta_{im}$ &  & $D25-b$ &  & $\delta_{mj}\delta_{ni}$ &  & $D25-c$ &  & $\delta_{jn}\delta_{im}$\tabularnewline
$D25-d$ &  & $\delta_{mj}\delta_{ni}$ &  & $D25-e$ &  & $\delta_{mj}\delta_{ni}$ &  & $D25-f$ &  & $\delta_{jn}\delta_{im}$\tabularnewline
$D25-g$ &  & $\delta_{mj}\delta_{ni}$ &  & $D25-h$ &  & $\delta_{jn}\delta_{im}$ &  &  &  & \tabularnewline
\hline 
\hline 
 &  &  &  &  &  &  &  &  &  & \tabularnewline
\end{tabular}\caption{\label{tab:S-PPPP-25}Values of the diagrams in Figure\,\ref{fig:D25-order-g-6}
with common factor\protect \\
 $-\frac{1}{16}\left(\frac{\left(a-b\right)\left(a^{2}+b^{2}\right)}{32\pi^{2}\epsilon}\right)\,i\,g^{6}\int_{\theta}\overline{\Phi}_{i}\Phi_{m}\Phi_{n}\overline{\Phi}_{j}$
.}
\end{table}
\par\end{center}

$\mathcal{S}_{\left(\overline{\Phi}\Phi\right)^{2}}^{\left(D25-a\right)}$
in the Figure\,\ref{fig:D25-order-g-6} is
\begin{align}
\mathcal{S}_{\left(\overline{\Phi}\Phi\right)^{2}}^{\left(D25-a\right)} & =\frac{1}{32}\,i\,g^{6}\delta_{im}\delta_{jn}\int_{\theta}\overline{\Phi}_{i}\Phi_{m}\Phi_{n}\overline{\Phi}_{j}\int\frac{d^{D}kd^{D}q}{\left(2\pi\right)^{2D}}\times\nonumber \\
 & \left\{ \frac{8\left(a\,b^{2}-a^{2}\,b\right)\left(k\cdot q\right)q^{2}+2\left(a^{3}+a^{2}\,b-a\,b^{2}-b^{3}\right)k^{2}q^{2}}{\left(k^{2}\right)^{2}\left(q-k\right)^{2}\left(q^{2}\right)^{2}}\right\} \,,
\end{align}
using Eqs.\,(\ref{eq:Int 7}) and\,(\ref{eq:Int 9}), then adding
$\mathcal{S}_{\left(\overline{\Phi}\Phi\right)^{2}}^{\left(D25-a\right)}$
to $\mathcal{S}_{\left(\overline{\Phi}\Phi\right)^{2}}^{\left(D25-h\right)}$
with the values in Table\,\ref{tab:S-PPPP-25}, we find 
\begin{align}
\mathcal{S}_{\left(\overline{\Phi}\Phi\right)^{2}}^{\left(D25\right)} & =-\frac{1}{4}\frac{\left(a-b\right)\left(a^{2}+b^{2}\right)}{32\pi^{2}\epsilon}\,i\,g^{6}\left(\delta_{im}\delta_{jn}+\delta_{jm}\delta_{in}\right)\int_{\theta}\overline{\Phi}_{i}\Phi_{m}\Phi_{n}\overline{\Phi}_{j}\,.\label{eq:S-D25}
\end{align}

\begin{center}
\begin{figure}
\begin{centering}
\subfloat[]{\centering{}\includegraphics[scale=0.5]{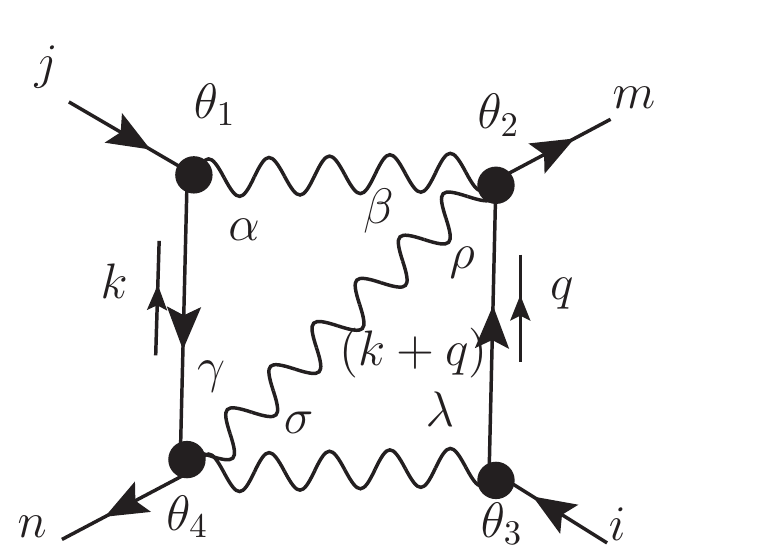}}\subfloat[]{\centering{}\includegraphics[scale=0.5]{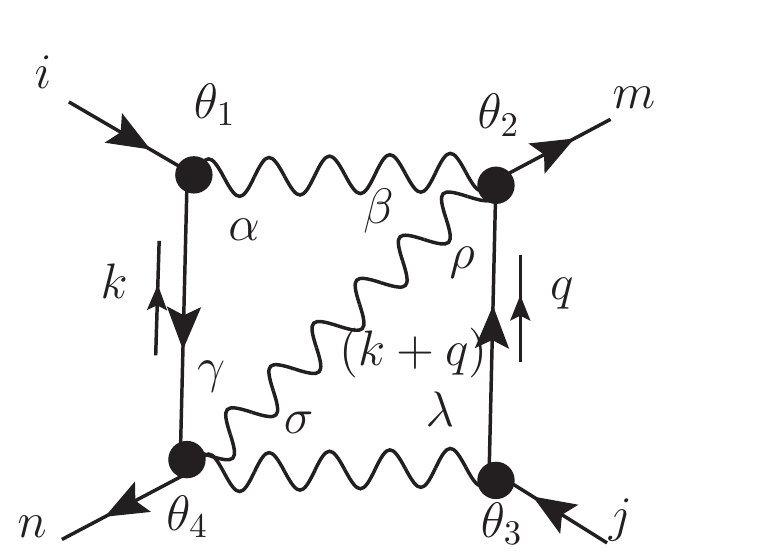}}\subfloat[]{\centering{}\includegraphics[scale=0.5]{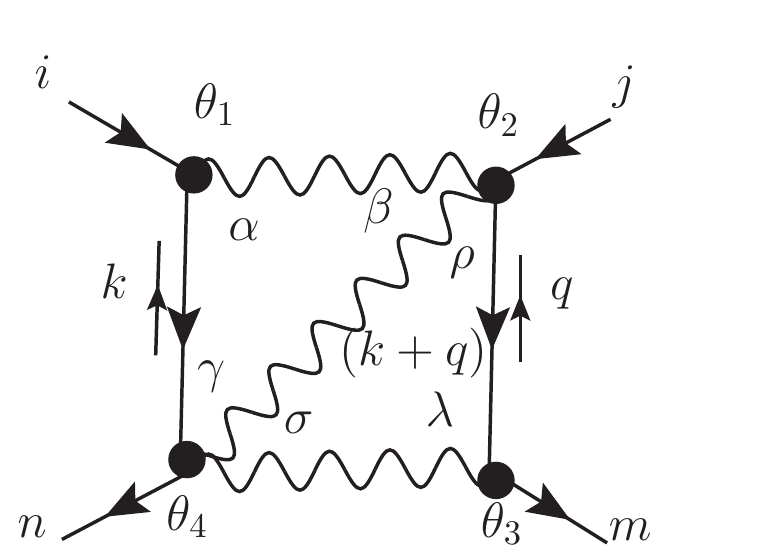}}\subfloat[]{\centering{}\includegraphics[scale=0.5]{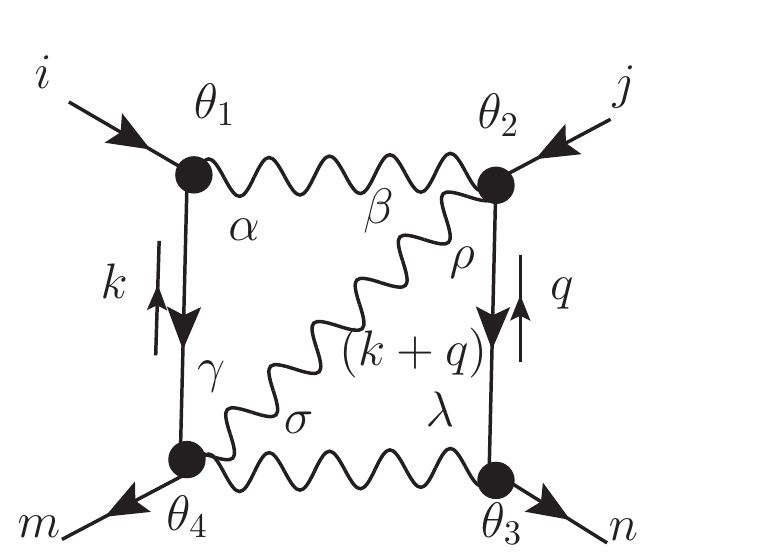}}
\par\end{centering}
\centering{}\subfloat[]{\centering{}\includegraphics[scale=0.5]{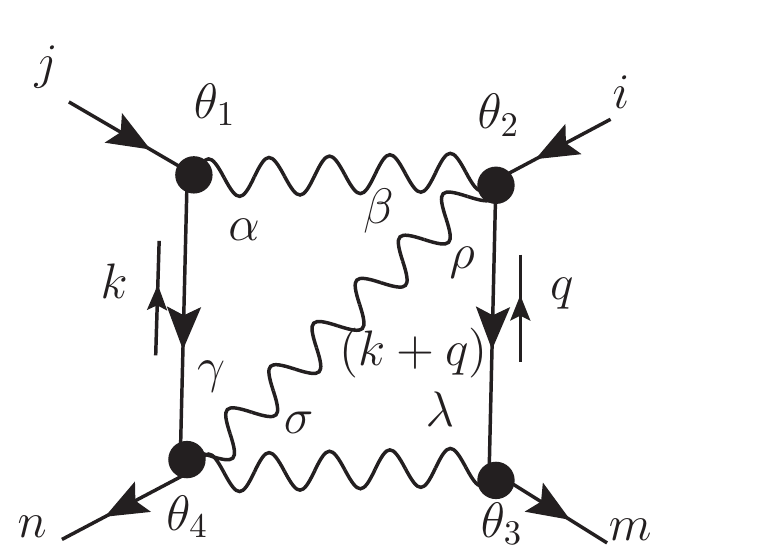}}\subfloat[]{\centering{}\includegraphics[scale=0.5]{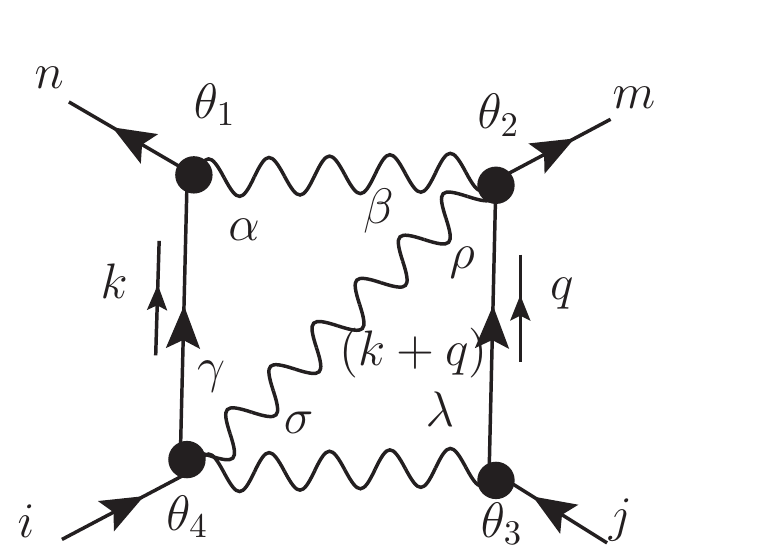}}\subfloat[]{\centering{}\includegraphics[scale=0.5]{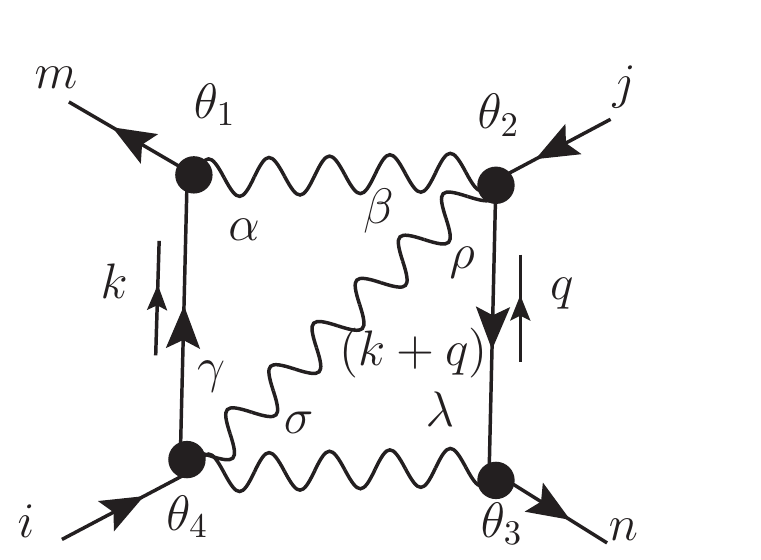}}\subfloat[]{\centering{}\includegraphics[scale=0.5]{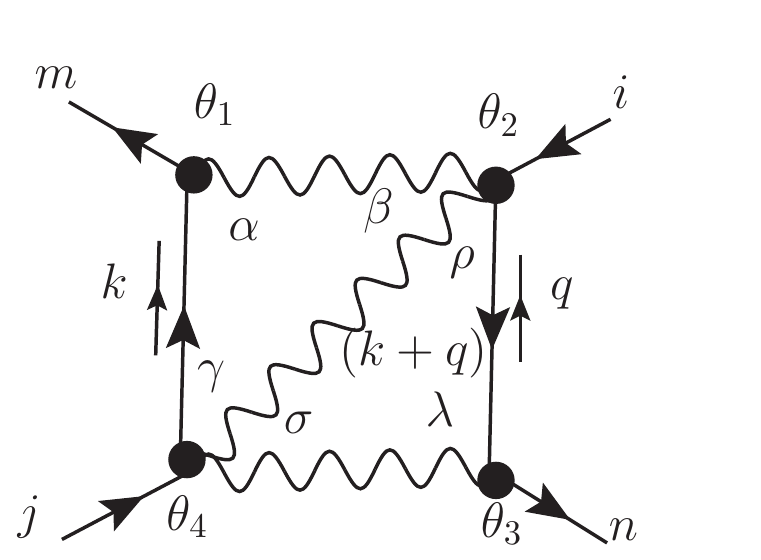}}\caption{\label{fig:D26-order-g-6}$\mathcal{S}_{\left(\overline{\Phi}\Phi\right)^{2}}^{\left(D26\right)}$}
\end{figure}
\par\end{center}

\begin{center}
\begin{table}
\centering{}%
\begin{tabular}{lcccccccccc}
 &  &  &  &  &  &  &  &  &  & \tabularnewline
\hline 
\hline 
$D26-a$ &  & $\delta_{im}\delta_{nj}$ &  & $D26-b$ &  & $\delta_{jm}\delta_{ni}$ &  & $D26-c$ &  & $-\delta_{jm}\delta_{ni}$\tabularnewline
$D26-d$ &  & $-\delta_{im}\delta_{nj}$ &  & $D26-e$ &  & $-\delta_{im}\delta_{nj}$ &  & $D26-f$ &  & $-\delta_{jm}\delta_{ni}$\tabularnewline
$D26-g$ &  & $\delta_{im}\delta_{nj}$ &  & $D26-h$ &  & $\delta_{jm}\delta_{ni}$ &  &  &  & \tabularnewline
\hline 
\hline 
 &  &  &  &  &  &  &  &  &  & \tabularnewline
\end{tabular}\caption{\label{tab:S-PPPP-26}Values of the diagrams in Figure\,\ref{fig:D26-order-g-6}
with common factor\protect \\
 $\frac{1}{16}\left(\frac{a\left(a-b\right)^{2}}{32\pi^{2}\epsilon}\right)\,i\,g^{6}\int_{\theta}\overline{\Phi}_{i}\Phi_{m}\Phi_{n}\overline{\Phi}_{j}$
.}
\end{table}
\par\end{center}

$\mathcal{S}_{\left(\overline{\Phi}\Phi\right)^{2}}^{\left(D26-a\right)}$
in the Figure\,\ref{fig:D26-order-g-6} is
\begin{align}
\mathcal{S}_{\left(\overline{\Phi}\Phi\right)^{2}}^{\left(D26-a\right)} & =\frac{1}{32}\,i\,\delta_{im}\delta_{nj}\,g^{6}\int_{\theta}\overline{\Phi}_{i}\Phi_{m}\Phi_{n}\overline{\Phi}_{j}\,\int\frac{d^{D}kd^{D}q}{\left(2\pi\right)^{2D}}\left\{ \frac{-2\,a\left(a-b\right)^{2}k^{2}q^{2}}{\left(k^{2}\right)^{2}\left(k+q\right)^{2}\left(q^{2}\right)^{2}}\right\} \,,
\end{align}
using Eq.\,(\ref{eq:Int 7}), then adding $\mathcal{S}_{\left(\overline{\Phi}\Phi\right)^{2}}^{\left(D26-a\right)}$
to $\mathcal{S}_{\left(\overline{\Phi}\Phi\right)^{2}}^{\left(D26-h\right)}$
with the values in Table\,\ref{tab:S-PPPP-26}, we find 
\begin{align}
\mathcal{S}_{\left(\overline{\Phi}\Phi\right)^{2}}^{\left(D26-a\right)}=\mathcal{S}_{\left(\overline{\Phi}\Phi\right)^{2}}^{\left(D26\right)} & =0\,.\label{eq:S-D26}
\end{align}

$\mathcal{S}_{\left(\overline{\Phi}\Phi\right)^{2}}^{\left(D27-a\right)}$
in the Figure\,\ref{fig:D27-order-g-6} is
\begin{center}
\begin{figure}
\begin{centering}
\subfloat[]{\centering{}\includegraphics[scale=0.5]{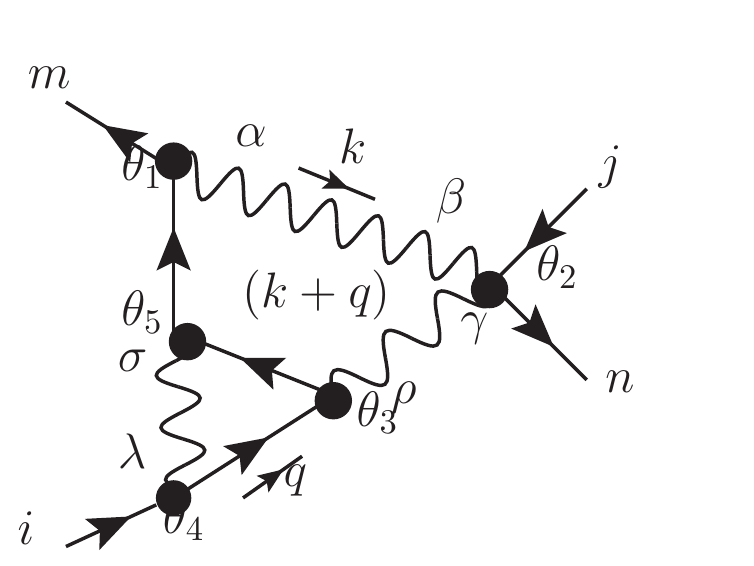}}\subfloat[]{\centering{}\includegraphics[scale=0.5]{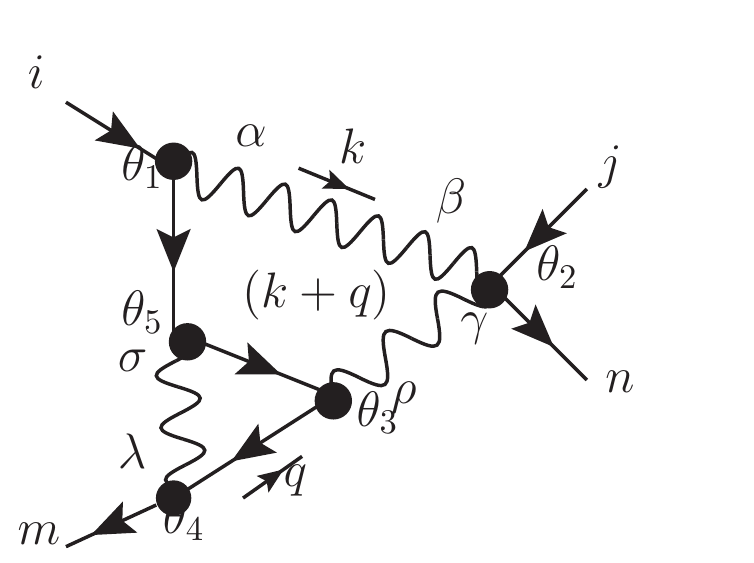}}\subfloat[]{\centering{}\includegraphics[scale=0.5]{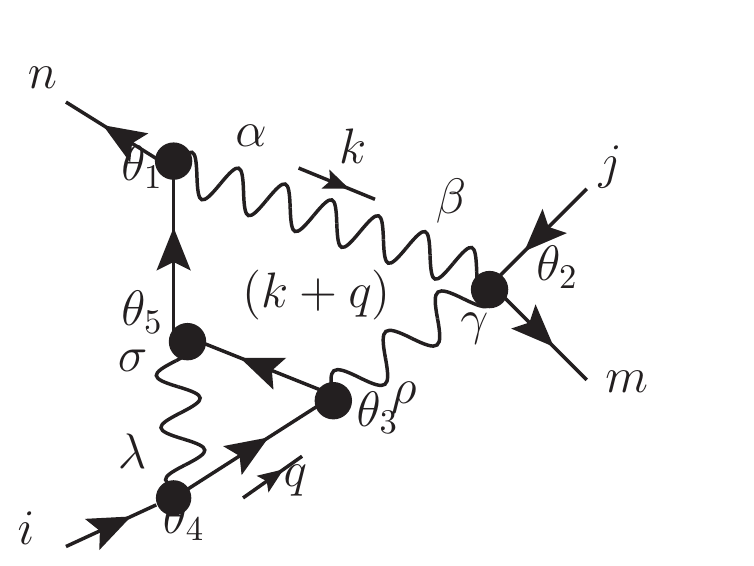}}\subfloat[]{\centering{}\includegraphics[scale=0.5]{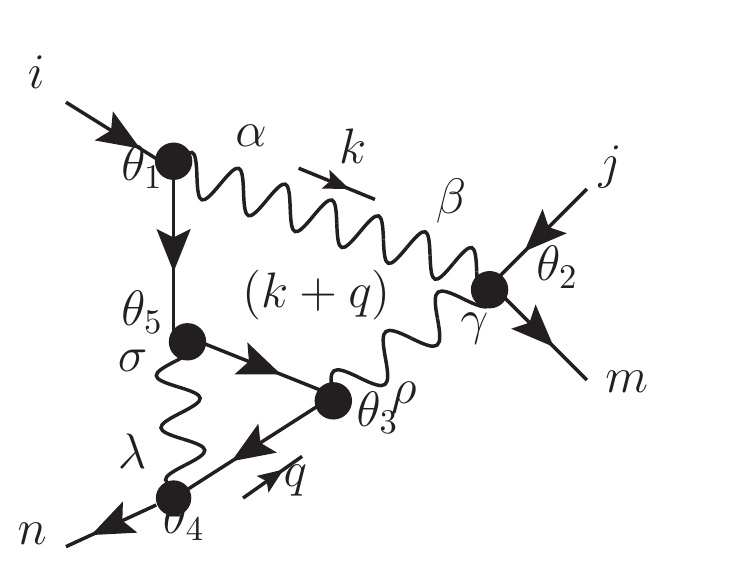}}
\par\end{centering}
\centering{}\subfloat[]{\centering{}\includegraphics[scale=0.5]{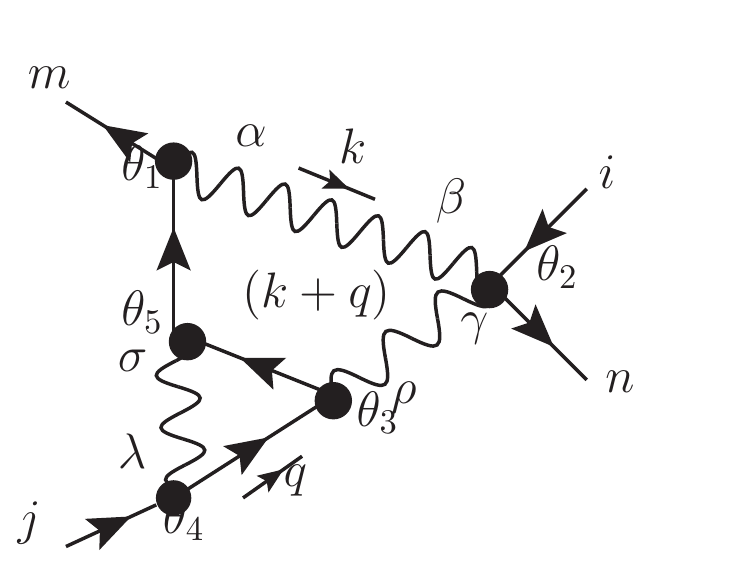}}\subfloat[]{\centering{}\includegraphics[scale=0.5]{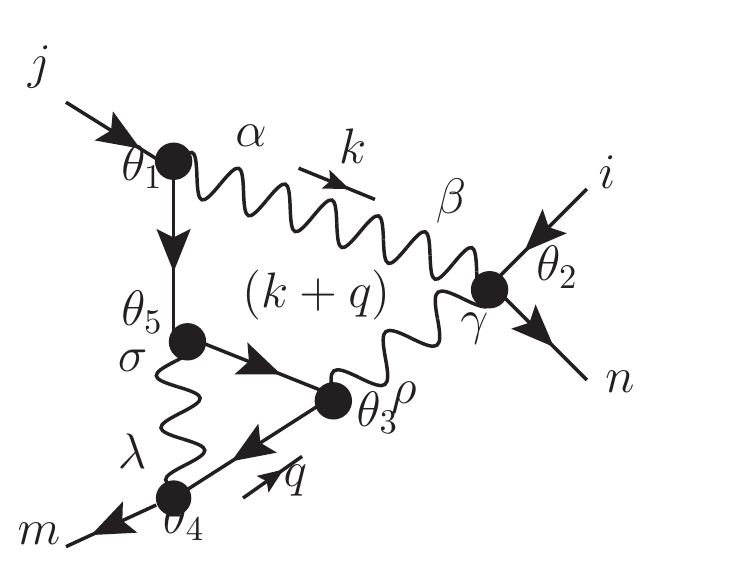}}\subfloat[]{\centering{}\includegraphics[scale=0.5]{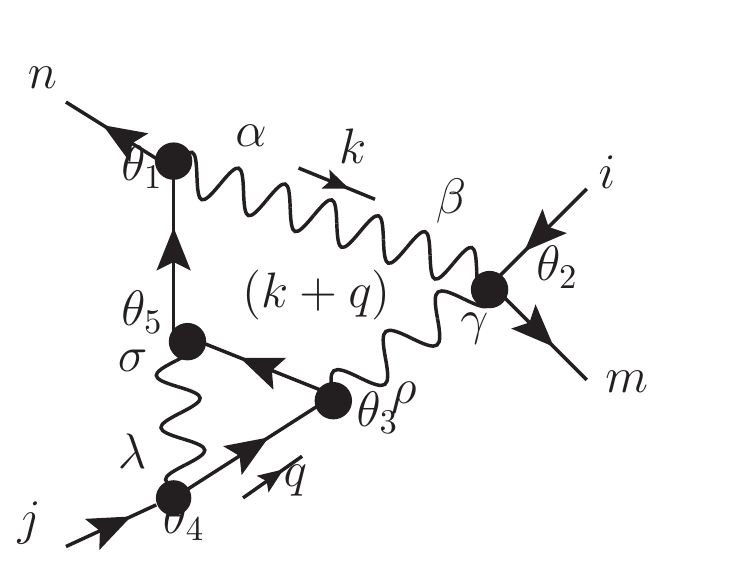}}\subfloat[]{\centering{}\includegraphics[scale=0.5]{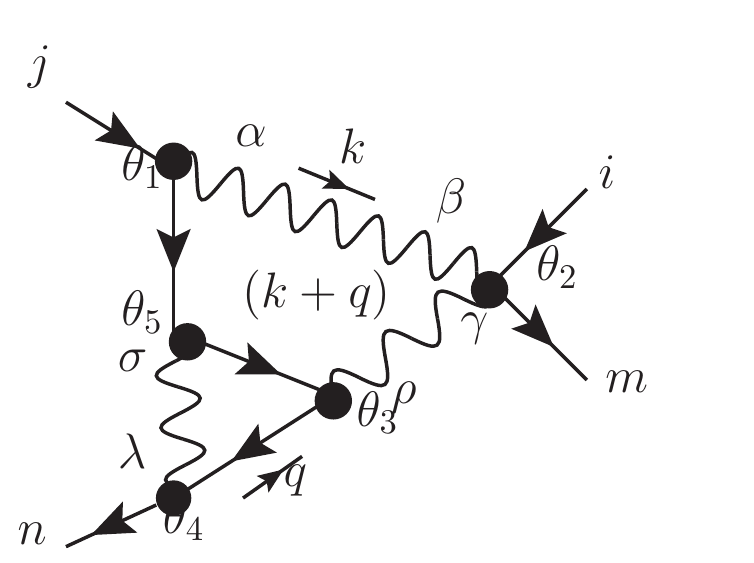}}\caption{\label{fig:D27-order-g-6}$\mathcal{S}_{\left(\overline{\Phi}\Phi\right)^{2}}^{\left(D27\right)}$}
\end{figure}
\par\end{center}

\begin{center}
\begin{table}
\centering{}%
\begin{tabular}{lcccccccccc}
 &  &  &  &  &  &  &  &  &  & \tabularnewline
\hline 
\hline 
$D27-a$ &  & $\delta_{im}\delta_{jn}$ &  & $D27-b$ &  & $\delta_{im}\delta_{jn}$ &  & $D27-c$ &  & $\delta_{jm}\delta_{in}$\tabularnewline
$D27-d$ &  & $\delta_{jm}\delta_{in}$ &  & $D27-e$ &  & $\delta_{jm}\delta_{in}$ &  & $D27-f$ &  & $\delta_{jm}\delta_{in}$\tabularnewline
$D27-g$ &  & $\delta_{im}\delta_{jn}$ &  & $D27-h$ &  & $\delta_{im}\delta_{jn}$ &  &  &  & \tabularnewline
\hline 
\hline 
 &  &  &  &  &  &  &  &  &  & \tabularnewline
\end{tabular}\caption{\label{tab:S-PPPP-27}Values of the diagrams in Figure\,\ref{fig:D27-order-g-6}
with common factor\protect \\
 $\frac{1}{32}\left(\frac{\left(a-b\right)^{3}}{32\pi^{2}\epsilon}\right)\,i\,g^{6}\int_{\theta}\overline{\Phi}_{i}\Phi_{m}\Phi_{n}\overline{\Phi}_{j}$
.}
\end{table}
\par\end{center}

$\mathcal{S}_{\left(\overline{\Phi}\Phi\right)^{2}}^{\left(D26-a\right)}$
in the Figure\,\ref{fig:D26-order-g-6} is
\begin{align}
\mathcal{S}_{\left(\overline{\Phi}\Phi\right)^{2}}^{\left(D27-a\right)} & =\frac{1}{128}\,i\,g^{6}\delta_{im}\delta_{jn}\int_{\theta}\overline{\Phi}_{i}\Phi_{m}\Phi_{n}\overline{\Phi}_{j}\,\int\frac{d^{D}kd^{D}q}{\left(2\pi\right)^{2D}}\left\{ \frac{-4\left(a-b\right)^{3}\left(k^{2}\right)^{2}q^{2}}{\left(k^{2}\right)^{3}\left(k+q\right)^{2}\left(q^{2}\right)^{2}}\right\} \,,
\end{align}
using Eq.\,(\ref{eq:Int 7}), then adding $\mathcal{S}_{\left(\overline{\Phi}\Phi\right)^{2}}^{\left(D27-a\right)}$
to $\mathcal{S}_{\left(\overline{\Phi}\Phi\right)^{2}}^{\left(D27-h\right)}$
with the values in Table\,\ref{tab:S-PPPP-27}, we find 
\begin{align}
\mathcal{S}_{\left(\overline{\Phi}\Phi\right)^{2}}^{\left(D27\right)} & =\frac{1}{8}\left(\frac{\left(a-b\right)^{3}}{32\pi^{2}\epsilon}\right)\,i\,g^{6}\left(\delta_{im}\delta_{jn}+\delta_{jm}\delta_{in}\right)\int_{\theta}\overline{\Phi}_{i}\Phi_{m}\Phi_{n}\overline{\Phi}_{j}\,.\label{eq:S-D27}
\end{align}

\begin{center}
\begin{figure}
\begin{centering}
\subfloat[]{\centering{}\includegraphics[scale=0.5]{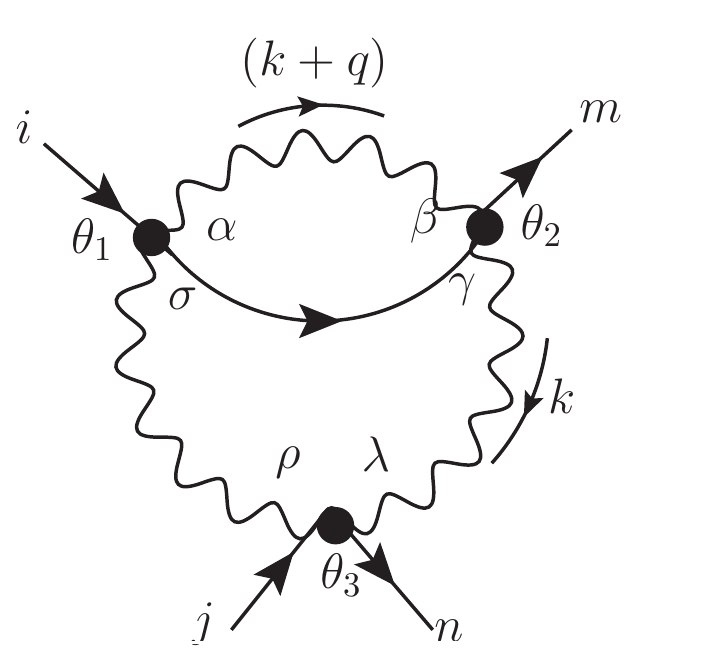}}\subfloat[]{\centering{}\includegraphics[scale=0.5]{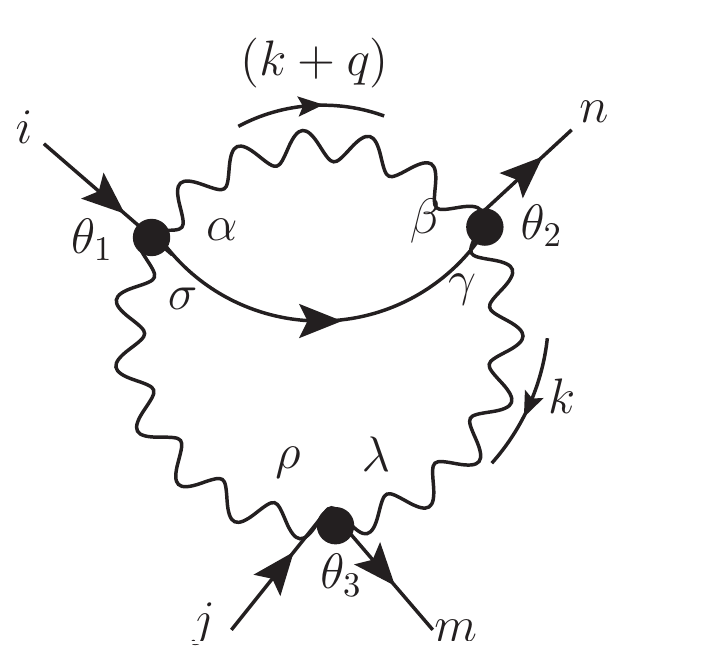}}\subfloat[]{\centering{}\includegraphics[scale=0.5]{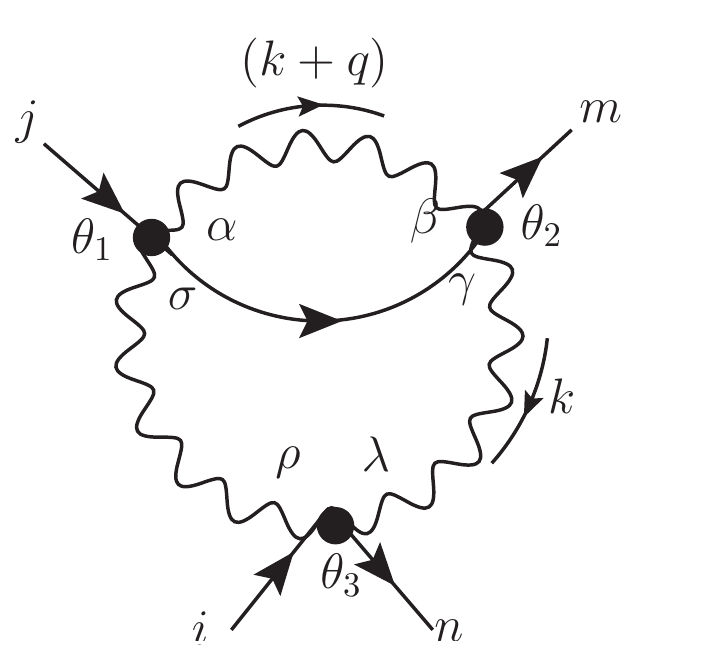}}\subfloat[]{\centering{}\includegraphics[scale=0.5]{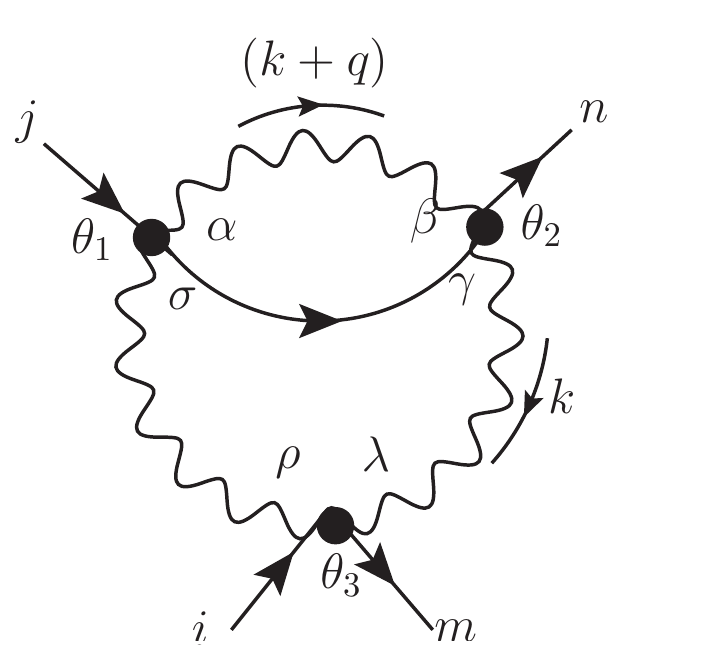}}
\par\end{centering}
\centering{}\caption{\label{fig:D28-order-g-6}$\mathcal{S}_{\left(\overline{\Phi}\Phi\right)^{2}}^{\left(D28\right)}$}
\end{figure}
\par\end{center}

\begin{center}
\begin{table}
\centering{}%
\begin{tabular}{lcccccccccc}
 &  &  &  &  &  &  &  &  &  & \tabularnewline
\hline 
\hline 
$D28-a$ &  & $\delta_{im}\delta_{jn}$ &  & $D28-b$ &  & $\delta_{jm}\delta_{in}$ &  & $D28-c$ &  & $\delta_{jm}\delta_{in}$\tabularnewline
$D28-d$ &  & $\delta_{im}\delta_{jn}$ &  &  &  &  &  &  &  & \tabularnewline
\hline 
\hline 
 &  &  &  &  &  &  &  &  &  & \tabularnewline
\end{tabular}\caption{\label{tab:S-PPPP-28}Values of the diagrams in Figure\,\ref{fig:D28-order-g-6}
with common factor\protect \\
 $\frac{1}{4}\left(\frac{a^{3}+2a\,b^{2}}{32\pi^{2}\epsilon}\right)\,i\,g^{6}\int_{\theta}\overline{\Phi}_{i}\Phi_{m}\Phi_{n}\overline{\Phi}_{j}$
.}
\end{table}
\par\end{center}

$\mathcal{S}_{\left(\overline{\Phi}\Phi\right)^{2}}^{\left(D28-a\right)}$
in the Figure\,\ref{fig:D28-order-g-6} is
\begin{align}
\mathcal{S}_{\left(\overline{\Phi}\Phi\right)^{2}}^{\left(D28-a\right)} & =-\frac{1}{8}\,i\,g^{6}\,\delta_{im}\delta_{jn}\int_{\theta}\overline{\Phi}_{i}\Phi_{m}\Phi_{n}\overline{\Phi}_{j}\int\frac{d^{D}kd^{D}q}{\left(2\pi\right)^{2D}}\left\{ \frac{4\,a\,b^{2}\left(k\cdot q\right)+\left(2\,a^{3}+6\,a\,b^{2}\right)\,k^{2}}{\left(k^{2}\right)^{2}\left(k+q\right)^{2}q^{2}}\right\} \,,
\end{align}
using Eqs.\,(\ref{eq:Int 7}) and\,(\ref{eq:Int 9}), then adding
$\mathcal{S}_{\left(\overline{\Phi}\Phi\right)^{2}}^{\left(D28-a\right)}$
to $\mathcal{S}_{\left(\overline{\Phi}\Phi\right)^{2}}^{\left(D28-d\right)}$
with the values in Table\,\ref{tab:S-PPPP-28}, we find 
\begin{align}
\mathcal{S}_{\left(\overline{\Phi}\Phi\right)^{2}}^{\left(D28\right)} & =\frac{1}{2}\,a\,\left(\frac{a^{2}+2\,b^{2}}{32\pi^{2}\epsilon}\right)\,i\,g^{6}\,\left(\delta_{im}\delta_{jn}+\delta_{jm}\delta_{in}\right)\int_{\theta}\overline{\Phi}_{i}\Phi_{m}\Phi_{n}\overline{\Phi}_{j}\,.\label{eq:S-D28}
\end{align}

\begin{center}
\begin{figure}
\begin{centering}
\subfloat[]{\centering{}\includegraphics[scale=0.5]{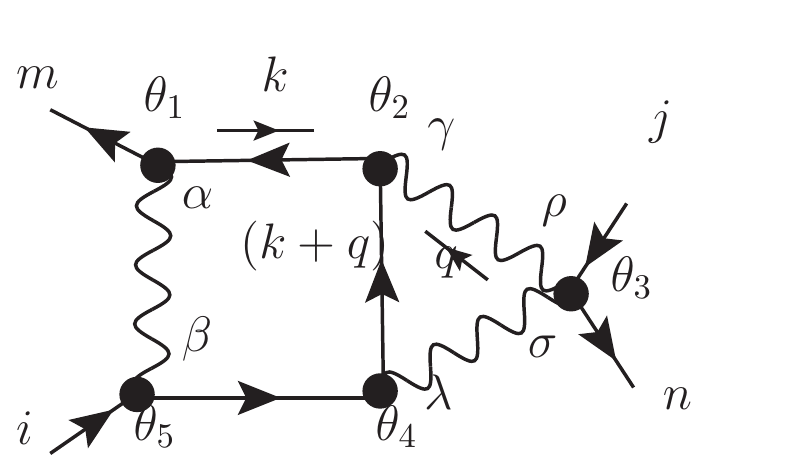}}\subfloat[]{\centering{}\includegraphics[scale=0.5]{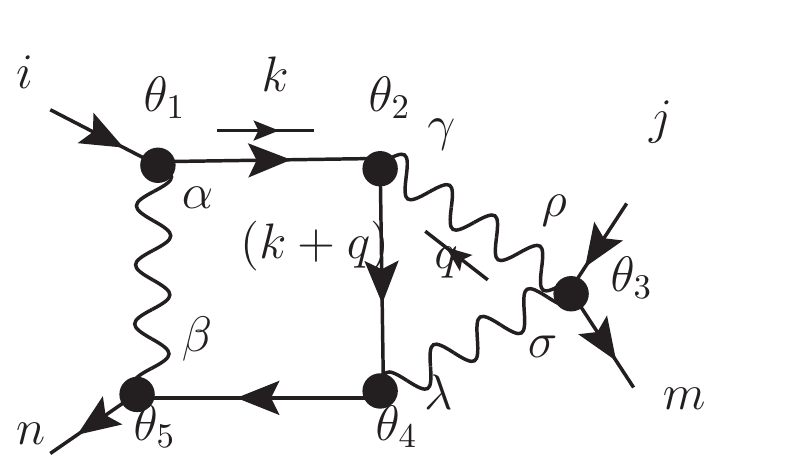}}\subfloat[]{\centering{}\includegraphics[scale=0.5]{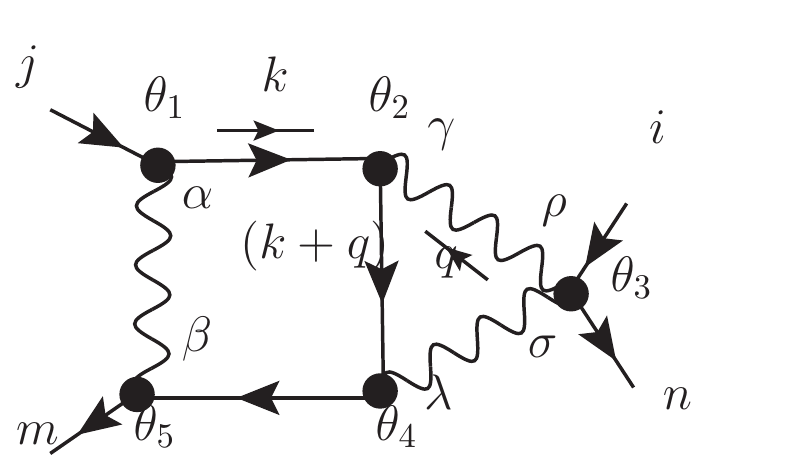}}\subfloat[]{\centering{}\includegraphics[scale=0.5]{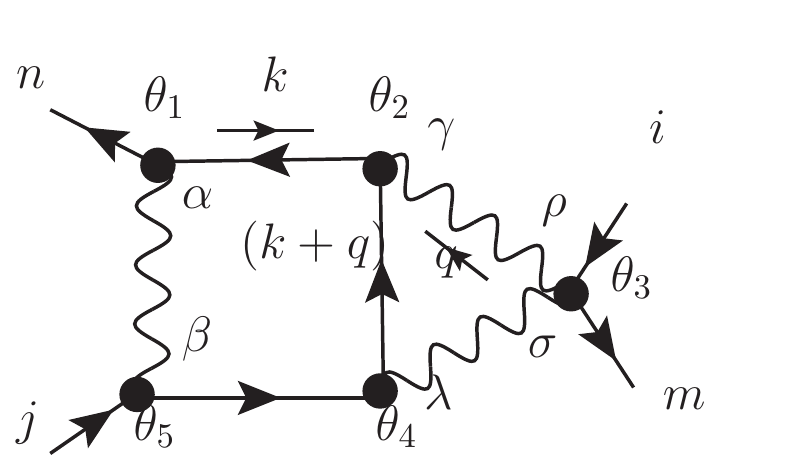}}
\par\end{centering}
\centering{}\caption{\label{fig:D29-order-g-6}$\mathcal{S}_{\left(\overline{\Phi}\Phi\right)^{2}}^{\left(D29\right)}$}
\end{figure}
\par\end{center}

\begin{center}
\begin{table}
\centering{}%
\begin{tabular}{lcccccccccc}
 &  &  &  &  &  &  &  &  &  & \tabularnewline
\hline 
\hline 
$D29-a$ &  & $\delta_{mi}\delta_{jn}$ &  & $D29-b$ &  & $\delta_{mj}\delta_{in}$ &  & $D29-c$ &  & $\delta_{mj}\delta_{in}$\tabularnewline
$D29-d$ &  & $\delta_{mi}\delta_{jn}$ &  &  &  &  &  &  &  & \tabularnewline
\hline 
\hline 
 &  &  &  &  &  &  &  &  &  & \tabularnewline
\end{tabular}\caption{\label{tab:S-PPPP-29}Values of the diagrams in Figure\,\ref{fig:D29-order-g-6}
with common factor\protect \\
 $\frac{1}{32}\left(\frac{\left(a-b\right)\left(a+b\right)^{2}}{32\pi^{2}\epsilon}\right)\,i\,g^{6}\int_{\theta}\overline{\Phi}_{i}\Phi_{m}\Phi_{n}\overline{\Phi}_{j}$
.}
\end{table}
\par\end{center}

$\mathcal{S}_{\left(\overline{\Phi}\Phi\right)^{2}}^{\left(D29-a\right)}$
in the Figure\,\ref{fig:D29-order-g-6} is
\begin{align}
\mathcal{S}_{\left(\overline{\Phi}\Phi\right)^{2}}^{\left(D29-a\right)} & =\frac{1}{128}\,i\,g^{6}\delta_{mi}\delta_{jn}\int_{\theta}\overline{\Phi}_{i}\Phi_{m}\Phi_{n}\overline{\Phi}_{j}\,\int\frac{d^{D}kd^{D}q}{\left(2\pi\right)^{2D}}\times\nonumber \\
 & \left\{ \frac{-32\left(a^{2}b-a\,b^{2}\right)\left(k\cdot q\right)\left(k^{2}\right)^{2}-4\left(a^{3}+5\,a^{2}b-5\,a\,b^{2}-b^{3}\right)\left(k^{2}\right)^{2}q^{2}}{\left(k^{2}\right)^{3}\left(k+q\right)^{2}\left(q^{2}\right)^{2}}\right\} \,,
\end{align}
using Eqs.\,(\ref{eq:Int 7}) and\,(\ref{eq:Int 9}), then adding
$\mathcal{S}_{\left(\overline{\Phi}\Phi\right)^{2}}^{\left(D29-a\right)}$
to $\mathcal{S}_{\left(\overline{\Phi}\Phi\right)^{2}}^{\left(D29-d\right)}$
with the values in Table\,\ref{tab:S-PPPP-29}, we find
\begin{align}
\mathcal{S}_{\left(\overline{\Phi}\Phi\right)^{2}}^{\left(D29\right)} & =\frac{1}{16}\left(\frac{\left(a-b\right)\left(a+b\right)^{2}}{32\pi^{2}\epsilon}\right)i\,g^{6}\left(\delta_{mi}\delta_{jn}+\delta_{mj}\delta_{in}\right)\int_{\theta}\overline{\Phi}_{i}\Phi_{m}\Phi_{n}\overline{\Phi}_{j}\,.\label{eq:S-D29}
\end{align}

\begin{center}
\begin{figure}
\begin{centering}
\subfloat[]{\centering{}\includegraphics[scale=0.5]{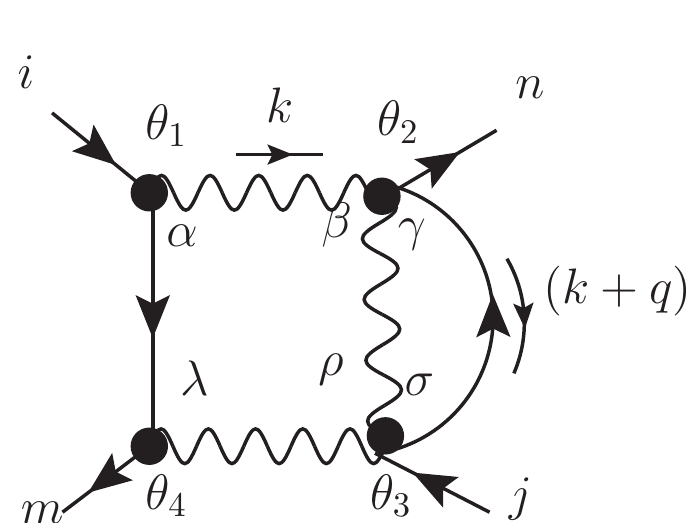}}\subfloat[]{\centering{}\includegraphics[scale=0.5]{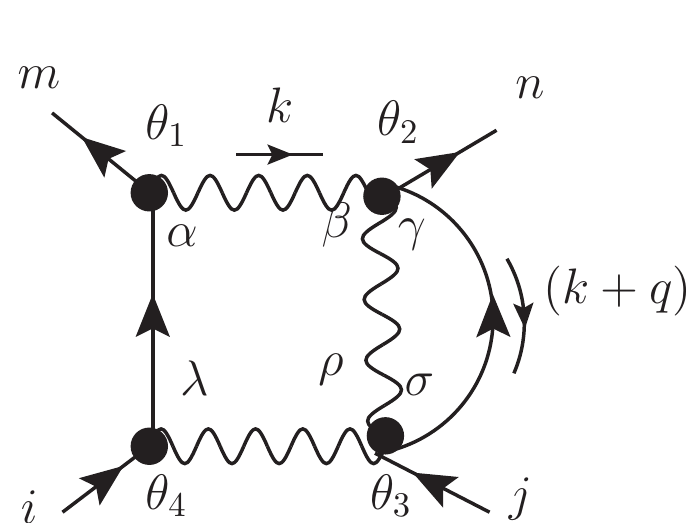}}\subfloat[]{\centering{}\includegraphics[scale=0.5]{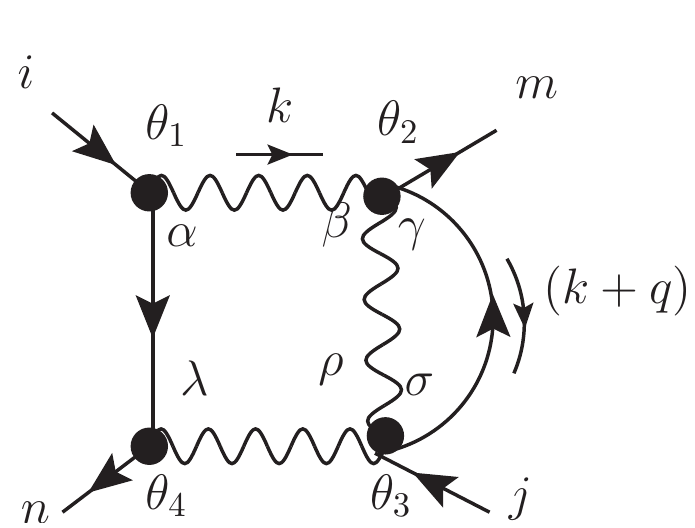}}\subfloat[]{\centering{}\includegraphics[scale=0.5]{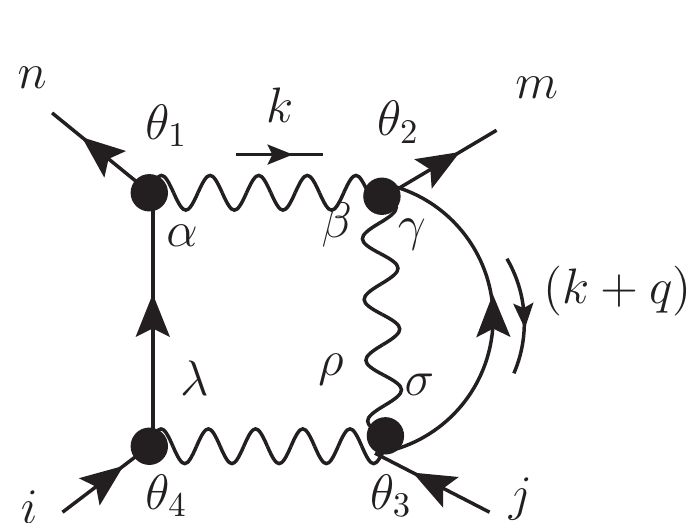}}
\par\end{centering}
\begin{centering}
\subfloat[]{\centering{}\includegraphics[scale=0.5]{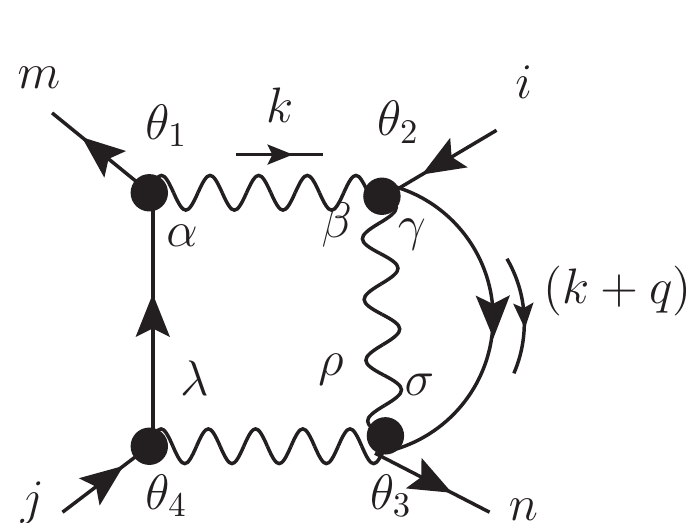}}\subfloat[]{\centering{}\includegraphics[scale=0.5]{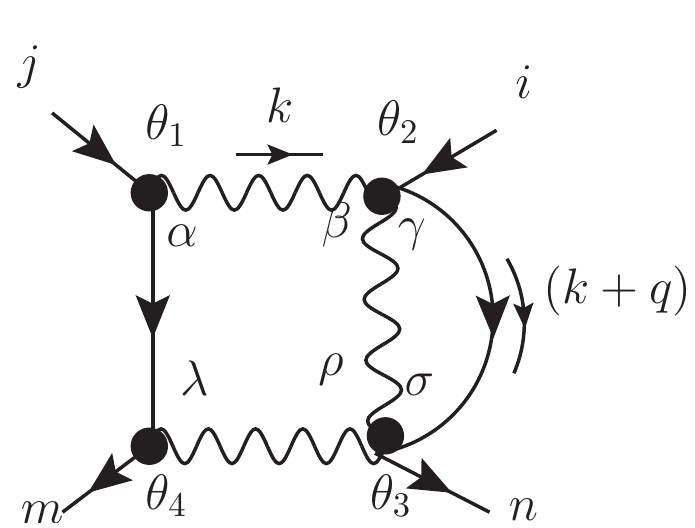}}\subfloat[]{\centering{}\includegraphics[scale=0.5]{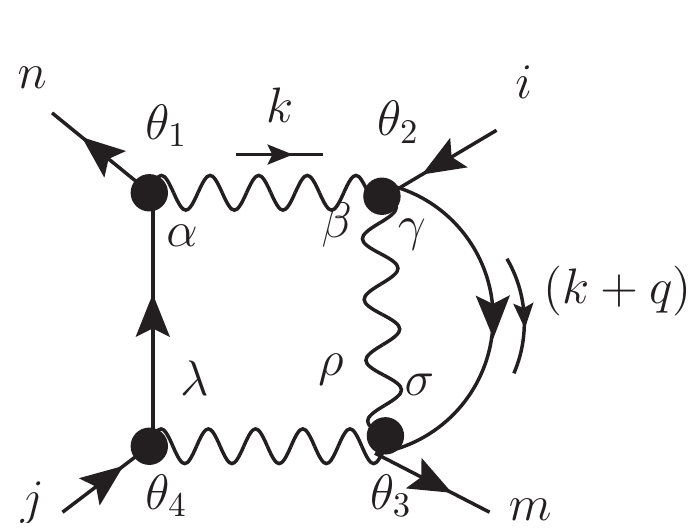}}\subfloat[]{\centering{}\includegraphics[scale=0.5]{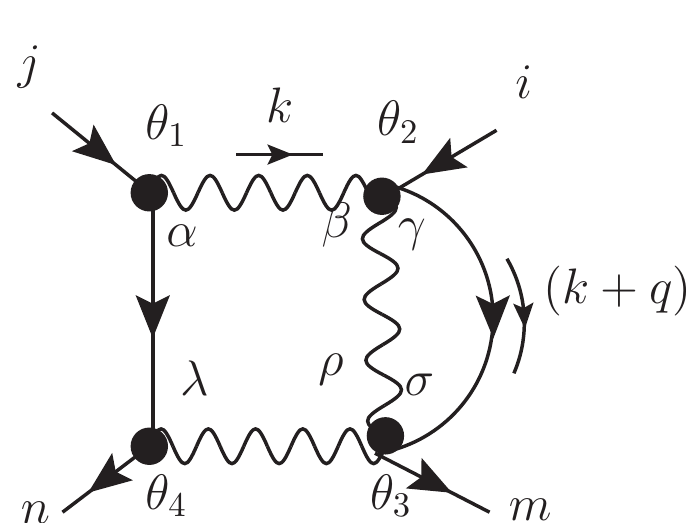}}
\par\end{centering}
\centering{}\caption{\label{fig:D30-order-g-6}$\mathcal{S}_{\left(\overline{\Phi}\Phi\right)^{2}}^{\left(D30\right)}$}
\end{figure}
\par\end{center}

\begin{center}
\begin{table}
\centering{}%
\begin{tabular}{lcccccccccc}
 &  &  &  &  &  &  &  &  &  & \tabularnewline
\hline 
\hline 
$D30-a$ &  & $\delta_{nj}\delta_{mi}$ &  & $D30-b$ &  & $\delta_{nj}\delta_{mi}$ &  & $D30-c$ &  & $\delta_{ni}\delta_{mj}$\tabularnewline
$D30-d$ &  & $\delta_{ni}\delta_{mj}$ &  & $D30-e$ &  & $\delta_{ni}\delta_{mj}$ &  & $D30-f$ &  & $\delta_{ni}\delta_{mj}$\tabularnewline
$D30-g$ &  & $\delta_{nj}\delta_{mi}$ &  & $D30-h$ &  & $\delta_{nj}\delta_{mi}$ &  &  &  & \tabularnewline
\hline 
\hline 
 &  &  &  &  &  &  &  &  &  & \tabularnewline
\end{tabular}\caption{\label{tab:S-PPPP-30}Values of the diagrams in Figure\,\ref{fig:D30-order-g-6}
with common factor\protect \\
 $\frac{1}{32}\left(\frac{b^{3}+5\,a^{2}\,b-4\,a\,b^{2}-2\,a^{3}}{32\pi^{2}\epsilon}\right)\,i\,g^{6}\int_{\theta}\overline{\Phi}_{i}\Phi_{m}\Phi_{n}\overline{\Phi}_{j}$
.}
\end{table}
\par\end{center}

$\mathcal{S}_{\left(\overline{\Phi}\Phi\right)^{2}}^{\left(D30-a\right)}$
in the Figure\,\ref{fig:D30-order-g-6} is
\begin{align}
\mathcal{S}_{\left(\overline{\Phi}\Phi\right)^{2}}^{\left(D30-a\right)} & =-\frac{1}{32}\,i\,\delta_{nj}\delta_{mi}\,g^{6}\int_{\theta}\overline{\Phi}_{i}\Phi_{m}\Phi_{n}\overline{\Phi}_{j}\int\frac{d^{D}kd^{D}q}{\left(2\pi\right)^{2D}}\left\{ \frac{-2\,b\left(a-b\right)^{2}\left(k\cdot q\right)k^{2}-2\,a\left(a-b\right)^{2}\left(k^{2}\right)^{2}}{\left(k^{2}\right)^{3}\left(k+q\right)^{2}q^{2}}\right\} \,,
\end{align}
using Eqs.\,(\ref{eq:Int 7}) and\,(\ref{eq:Int 9}), then adding
$\mathcal{S}_{\left(\overline{\Phi}\Phi\right)^{2}}^{\left(D30-a\right)}$
to $\mathcal{S}_{\left(\overline{\Phi}\Phi\right)^{2}}^{\left(D30-h\right)}$
with the values in Table\,\ref{tab:S-PPPP-30}, we find 
\begin{align}
\mathcal{S}_{\left(\overline{\Phi}\Phi\right)^{2}}^{\left(D30\right)} & =-\frac{1}{8}\left(\frac{\left(a-b\right)^{2}\left(2\,a-b\right)}{32\pi^{2}\epsilon}\right)\,i\,g^{6}\left(\delta_{nj}\delta_{mi}+\delta_{ni}\delta_{mj}\right)\int_{\theta}\overline{\Phi}_{i}\Phi_{m}\Phi_{n}\overline{\Phi}_{j}\,.\label{eq:S-D30}
\end{align}

\begin{center}
\begin{figure}
\begin{centering}
\subfloat[]{\centering{}\includegraphics[scale=0.5]{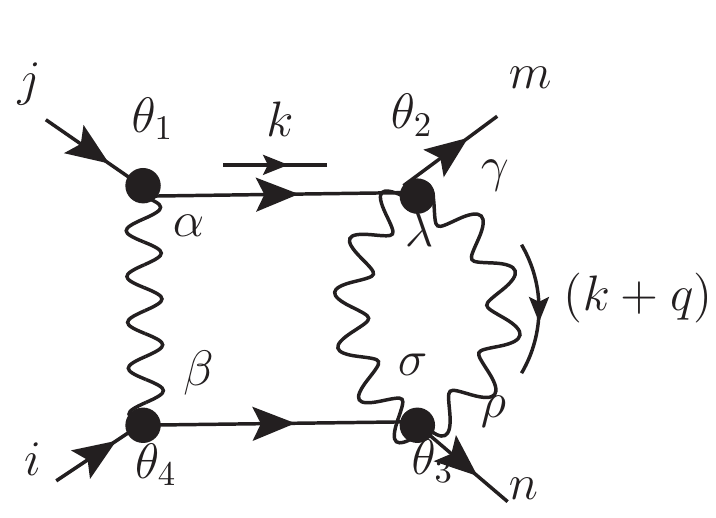}}\subfloat[]{\centering{}\includegraphics[scale=0.5]{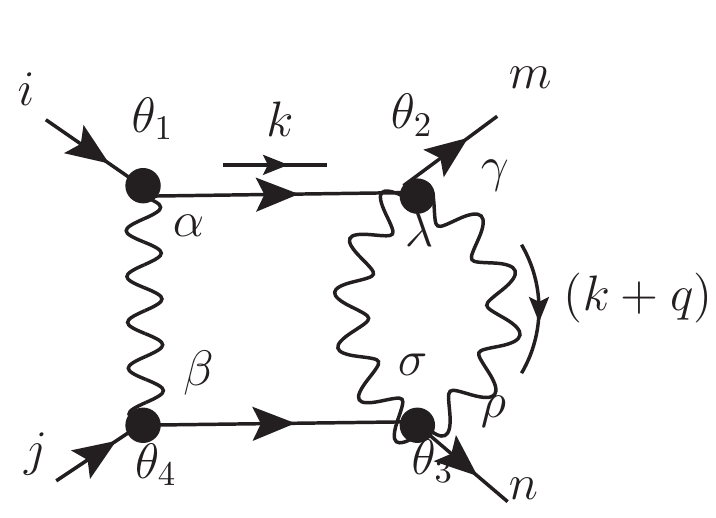}}\subfloat[]{\centering{}\includegraphics[scale=0.5]{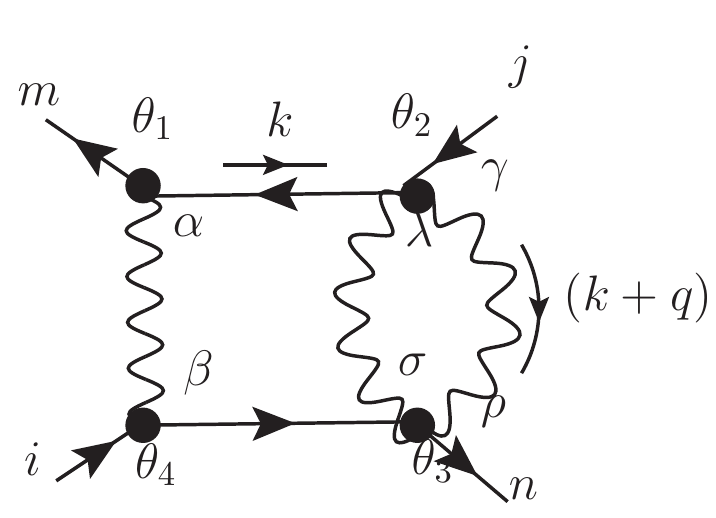}}\subfloat[]{\centering{}\includegraphics[scale=0.5]{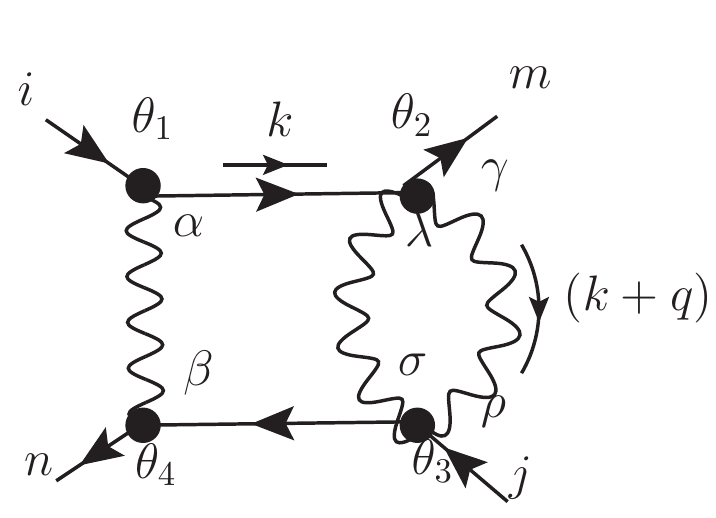}}
\par\end{centering}
\begin{centering}
\subfloat[]{\centering{}\includegraphics[scale=0.5]{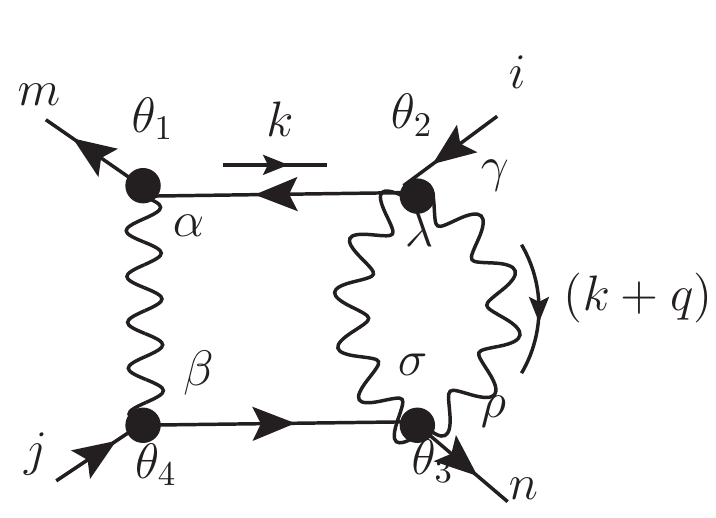}}\subfloat[]{\centering{}\includegraphics[scale=0.5]{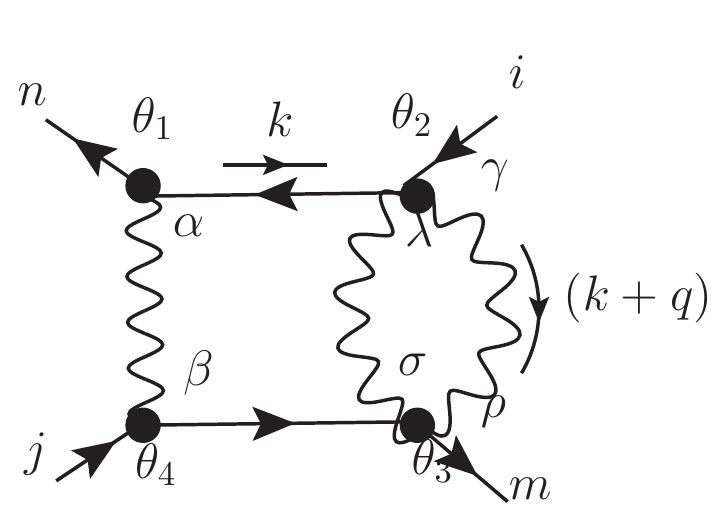}}\subfloat[]{\centering{}\includegraphics[scale=0.5]{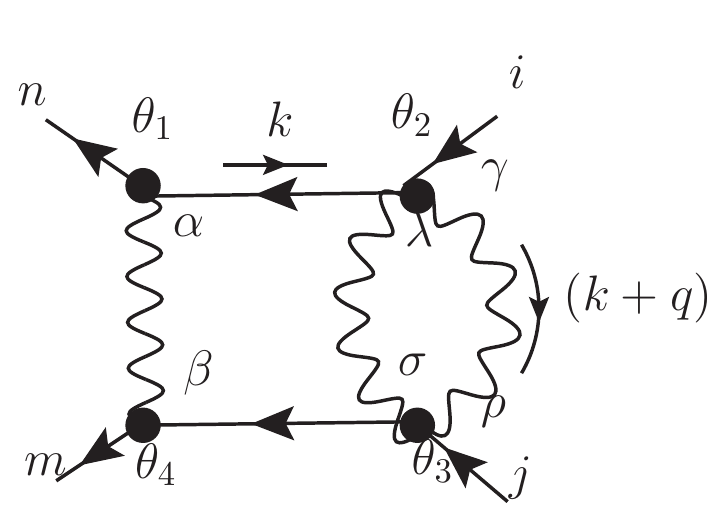}}\subfloat[]{\centering{}\includegraphics[scale=0.5]{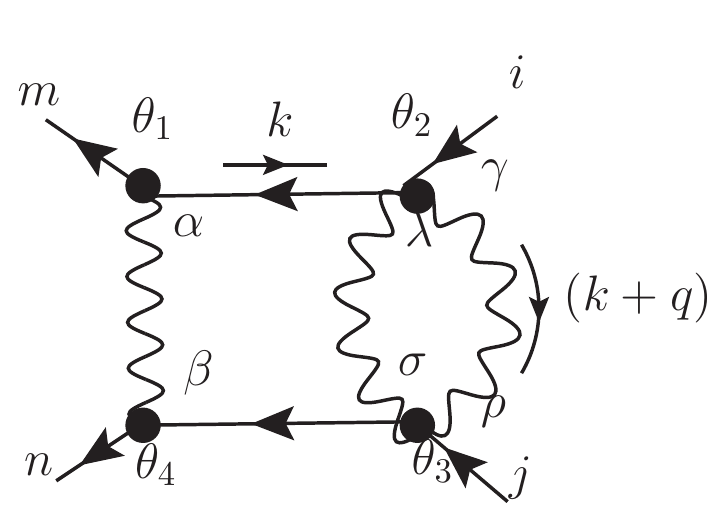}}
\par\end{centering}
\centering{}\caption{\label{fig:D31-order-g-6}$\mathcal{S}_{\left(\overline{\Phi}\Phi\right)^{2}}^{\left(D31\right)}$}
\end{figure}
\par\end{center}

\begin{center}
\begin{table}
\centering{}%
\begin{tabular}{lcccccccccc}
 &  &  &  &  &  &  &  &  &  & \tabularnewline
\hline 
\hline 
$D31-a$ &  & $\delta_{ni}\delta_{jm}$ &  & $D31-b$ &  & $\delta_{nj}\delta_{im}$ &  & $D31-c$ &  & $-\delta_{ni}\delta_{jm}$\tabularnewline
$D31-d$ &  & $-\delta_{nj}\delta_{im}$ &  & $D31-e$ &  & $-\delta_{nj}\delta_{im}$ &  & $D31-f$ &  & $-\delta_{ni}\delta_{jm}$\tabularnewline
$D31-g$ &  & $\delta_{ni}\delta_{jm}$ &  & $D31-h$ &  & $\delta_{nj}\delta_{im}$ &  &  &  & \tabularnewline
\hline 
\hline 
 &  &  &  &  &  &  &  &  &  & \tabularnewline
\end{tabular}\caption{\label{tab:S-PPPP-31}Values of the diagrams in Figure\,\ref{fig:D31-order-g-6}
with common factor\protect \\
 $\frac{1}{16}\left(\frac{a^{2}\,b-a^{3}}{32\pi^{2}\epsilon}\right)\,i\,g^{6}\int_{\theta}\overline{\Phi}_{i}\Phi_{m}\Phi_{n}\overline{\Phi}_{j}$
.}
\end{table}
\par\end{center}

$\mathcal{S}_{\left(\overline{\Phi}\Phi\right)^{2}}^{\left(D31-a\right)}$
in the Figure\,\ref{fig:D31-order-g-6} is
\begin{align}
\mathcal{S}_{\left(\overline{\Phi}\Phi\right)^{2}}^{\left(D31-a\right)} & =\frac{1}{64}\,i\,\delta_{ni}\delta_{jm}\,g^{6}\int_{\theta}\overline{\Phi}_{i}\Phi_{m}\Phi_{n}\overline{\Phi}_{j}\,\int\frac{d^{D}kd^{D}q}{\left(2\pi\right)^{2D}}\,\left\{ \frac{-4\left(a^{3}-a^{2}\,b\right)\left(k^{2}\right)^{2}}{\left(k^{2}\right)^{3}\left(k+q\right)^{2}q^{2}}\right\} \,,
\end{align}
using Eq.\,(\ref{eq:Int 7}), then adding $\mathcal{S}_{\left(\overline{\Phi}\Phi\right)^{2}}^{\left(D31-a\right)}$
to $\mathcal{S}_{\left(\overline{\Phi}\Phi\right)^{2}}^{\left(D31-h\right)}$
with the values in Table\,\ref{tab:S-PPPP-31}, we find 
\begin{align}
\mathcal{S}_{\left(\overline{\Phi}\Phi\right)^{2}}^{\left(D31\right)} & =0\,.\label{eq:S-D31}
\end{align}

\begin{center}
\begin{figure}
\begin{centering}
\subfloat[]{\centering{}\includegraphics[scale=0.5]{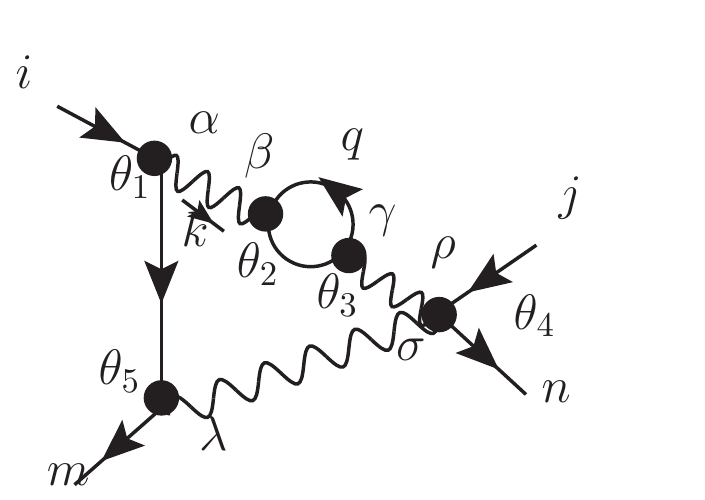}}\subfloat[]{\centering{}\includegraphics[scale=0.5]{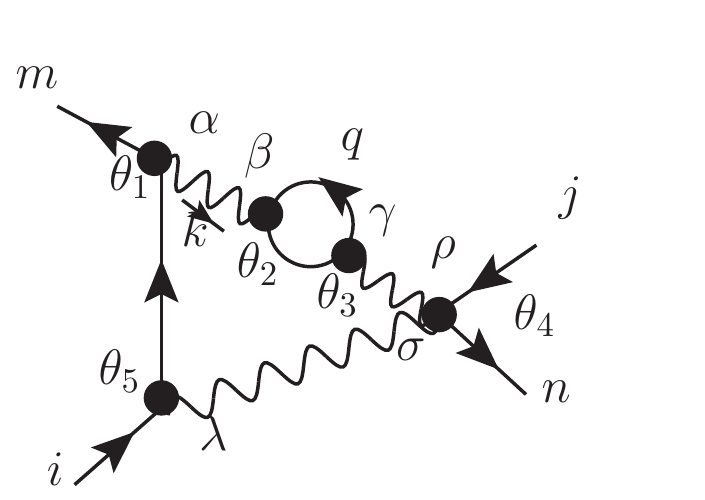}}\subfloat[]{\centering{}\includegraphics[scale=0.5]{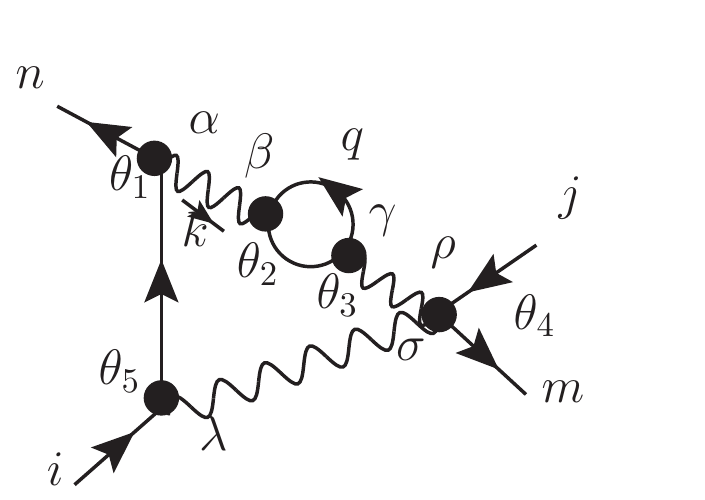}}\subfloat[]{\centering{}\includegraphics[scale=0.5]{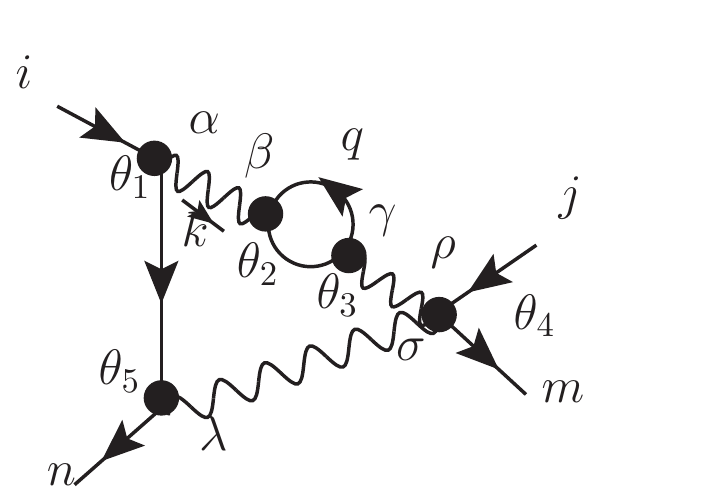}}
\par\end{centering}
\begin{centering}
\subfloat[]{\centering{}\includegraphics[scale=0.5]{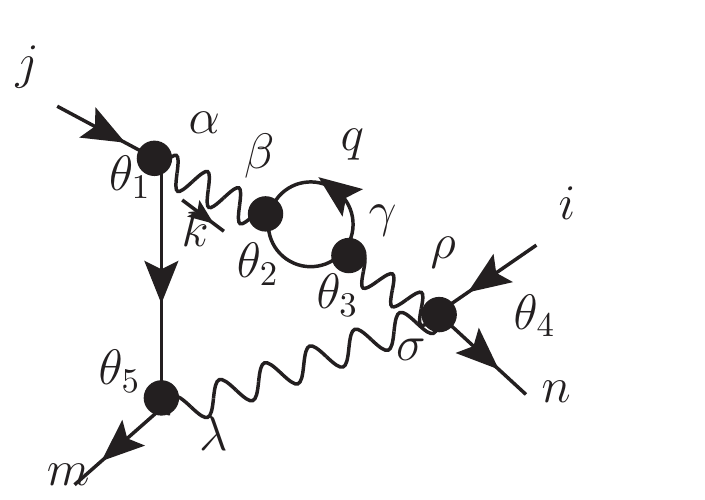}}\subfloat[]{\centering{}\includegraphics[scale=0.5]{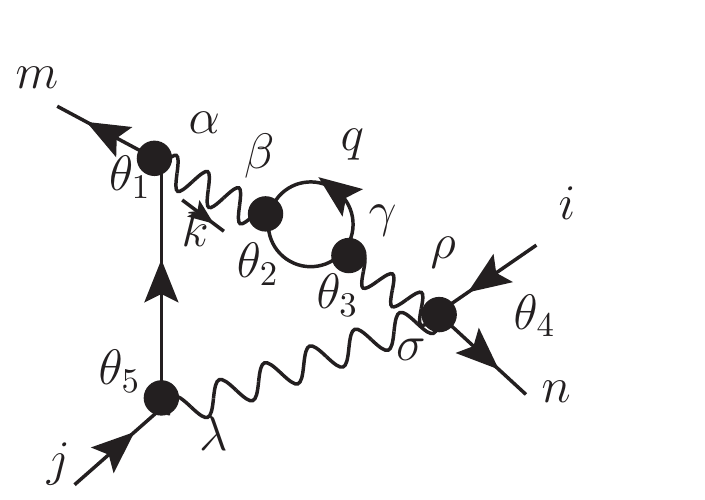}}\subfloat[]{\centering{}\includegraphics[scale=0.5]{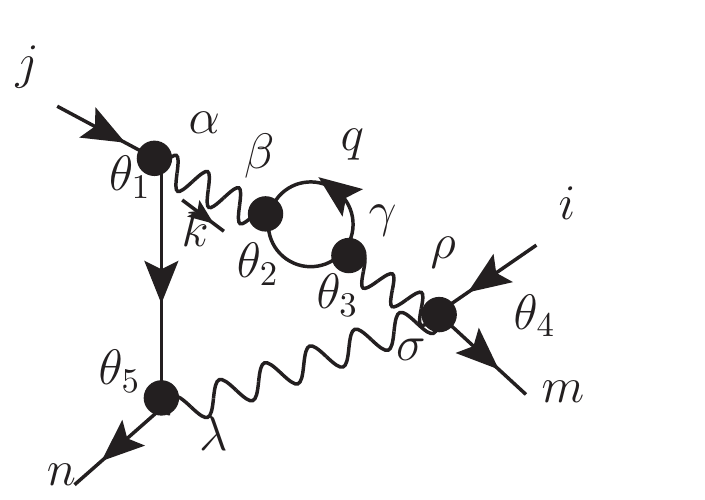}}\subfloat[]{\centering{}\includegraphics[scale=0.5]{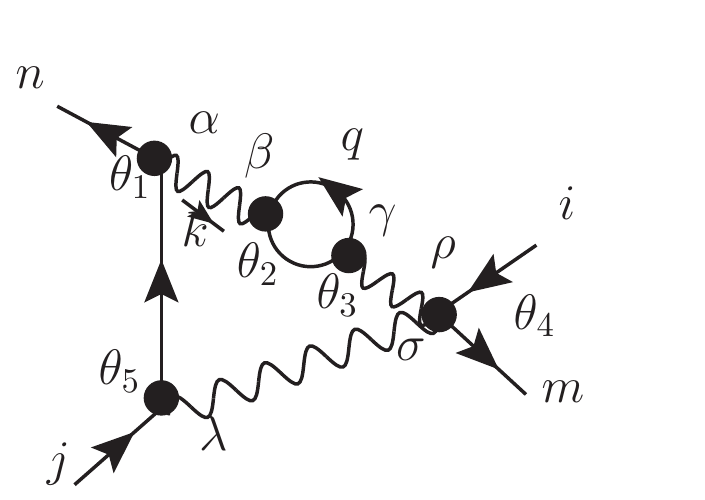}}
\par\end{centering}
\centering{}\caption{\label{fig:D32-order-g-6}$\mathcal{S}_{\left(\overline{\Phi}\Phi\right)^{2}}^{\left(D32\right)}$}
\end{figure}
\par\end{center}

$\mathcal{S}_{\left(\overline{\Phi}\Phi\right)^{2}}^{\left(D32-a\right)}$
in the Figure\,\ref{fig:D32-order-g-6} is
\begin{align}
\mathcal{S}_{\left(\overline{\Phi}\Phi\right)^{2}}^{\left(D32-a\right)} & =-\frac{1}{128}i\,N\,g^{6}\delta_{mi}\delta_{jn}\int_{\theta}\overline{\Phi}_{i}\Phi_{m}\Phi_{n}\overline{\Phi}_{j}\int\frac{d^{D}kd^{D}q}{\left(2\pi\right)^{2D}}\left\{ \frac{4\left(a-b\right)^{3}\left(k^{2}\right)^{2}\left(2\left(k\cdot q\right)+k^{2}+2q^{2}\right)}{\left(k^{2}\right)^{4}\left(k+q\right)^{2}q^{2}}\right\} \,,
\end{align}
using Eqs.\,(\ref{eq:Int 7}), (\ref{eq:Int 9}), then adding $\mathcal{S}_{\left(\overline{\Phi}\Phi\right)^{2}}^{\left(D32-a\right)}$
to $\mathcal{S}_{\left(\overline{\Phi}\Phi\right)^{2}}^{\left(D32-h\right)}$,
we find 
\begin{align}
\mathcal{S}_{\left(\overline{\Phi}\Phi\right)^{2}}^{\left(D32-a\right)}=\mathcal{S}_{\left(\overline{\Phi}\Phi\right)^{2}}^{\left(D32\right)} & =0\,.\label{eq:S-D32}
\end{align}

\begin{center}
\begin{figure}
\begin{centering}
\subfloat[]{\centering{}\includegraphics[scale=0.5]{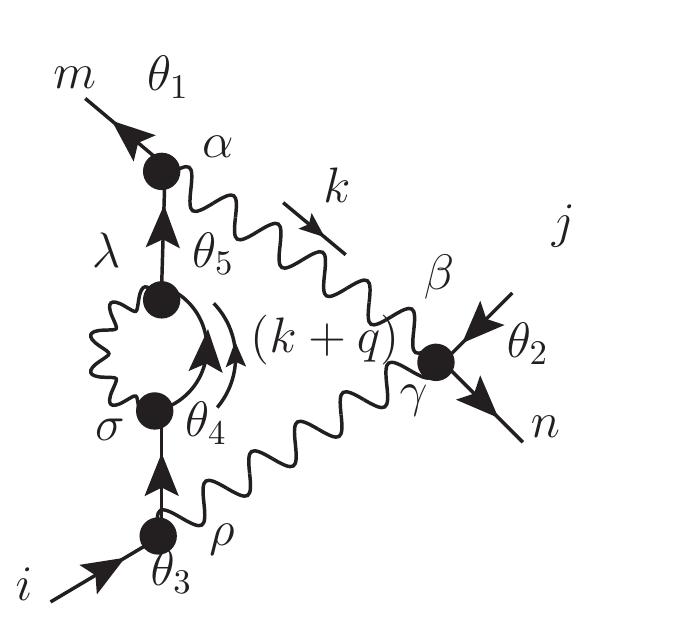}}\subfloat[]{\centering{}\includegraphics[scale=0.5]{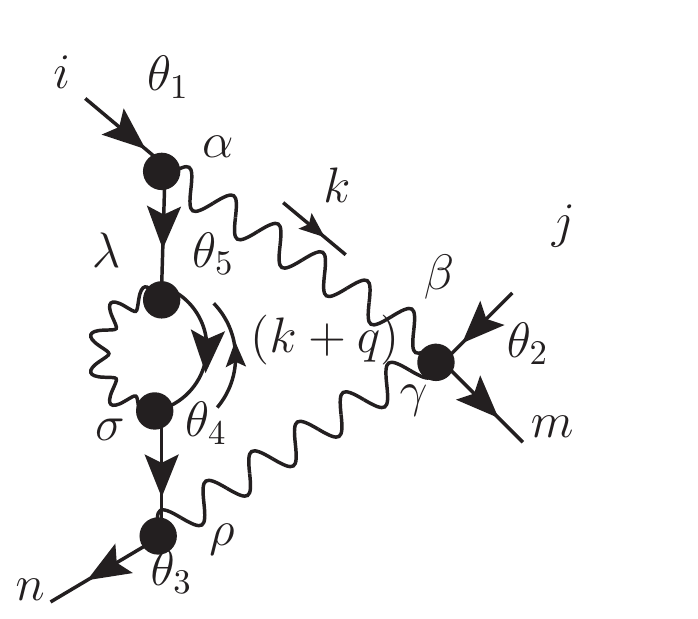}}\subfloat[]{\centering{}\includegraphics[scale=0.5]{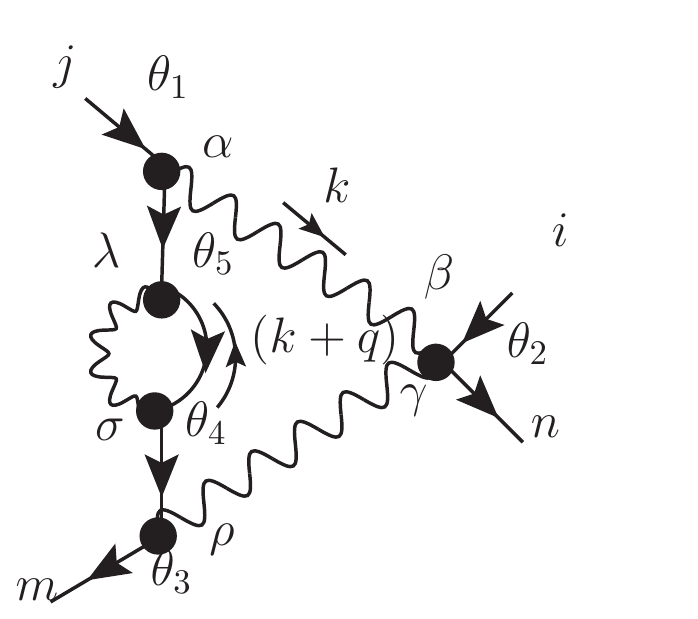}}\subfloat[]{\centering{}\includegraphics[scale=0.5]{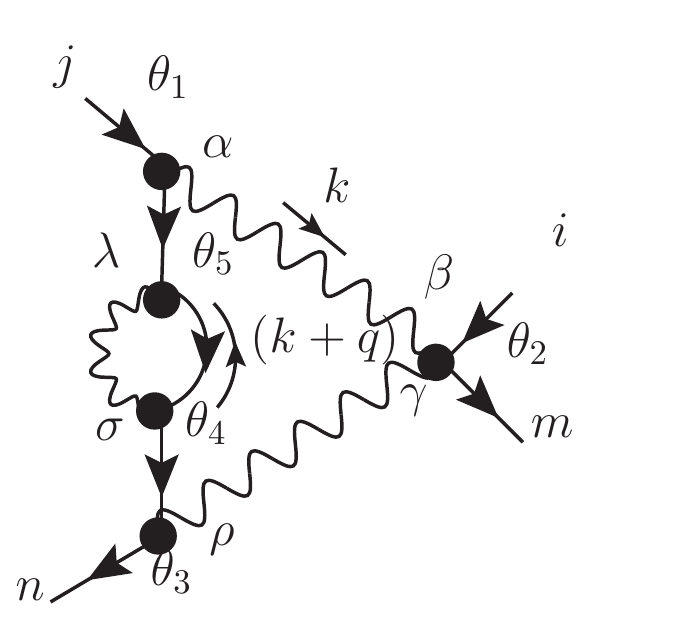}}
\par\end{centering}
\centering{}\caption{\label{fig:D33-order-g-6}$\mathcal{S}_{\left(\overline{\Phi}\Phi\right)^{2}}^{\left(D33\right)}$}
\end{figure}
\par\end{center}

\begin{center}
\begin{table}
\centering{}%
\begin{tabular}{lcccccccccc}
 &  &  &  &  &  &  &  &  &  & \tabularnewline
\hline 
\hline 
$D33-a$ &  & $\delta_{im}\delta_{jn}$ &  & $D33-b$ &  & $\delta_{jm}\delta_{in}$ &  & $D33-c$ &  & $\delta_{jm}\delta_{in}$\tabularnewline
$D33-d$ &  & $\delta_{im}\delta_{jn}$ &  &  &  &  &  &  &  & \tabularnewline
\hline 
\hline 
 &  &  &  &  &  &  &  &  &  & \tabularnewline
\end{tabular}\caption{\label{tab:S-PPPP-33}Values of the diagrams in Figure\,\ref{fig:D33-order-g-6}
with common factor\protect \\
 $\frac{1}{16}\left(\frac{a\left(a-b\right)^{2}}{32\pi^{2}\epsilon}\right)\,i\,g^{6}\int_{\theta}\overline{\Phi}_{i}\Phi_{m}\Phi_{n}\overline{\Phi}_{j}$
.}
\end{table}
\par\end{center}

$\mathcal{S}_{\left(\overline{\Phi}\Phi\right)^{2}}^{\left(D33-a\right)}$
in the Figure\,\ref{fig:D33-order-g-6} is
\begin{align}
\mathcal{S}_{\left(\overline{\Phi}\Phi\right)^{2}}^{\left(D33-a\right)} & =\frac{1}{128}\,i\,g^{6}\delta_{im}\delta_{jn}\int_{\theta}\overline{\Phi}_{i}\Phi_{m}\Phi_{n}\overline{\Phi}_{j}\nonumber \\
 & \times\int\frac{d^{D}kd^{D}q}{\left(2\pi\right)^{2D}}\left\{ \frac{-4\left(a-b\right)^{3}\left(k^{2}\right)^{2}q^{2}-16a\,\left(a-b\right)^{2}\left(k^{2}\right)^{2}\left(k^{2}+\left(k\cdot q\right)\right)}{\left(k^{2}\right)^{4}\left(k+q\right)^{2}q^{2}}\right\} \,,
\end{align}
using Eqs.\,(\ref{eq:Int 7}), (\ref{eq:Int 9}), then adding $\mathcal{S}_{\left(\overline{\Phi}\Phi\right)^{2}}^{\left(D33-a\right)}$
to $\mathcal{S}_{\left(\overline{\Phi}\Phi\right)^{2}}^{\left(D33-d\right)}$
with the values in the Table\,\ref{tab:S-PPPP-33}, we find
\begin{align}
\mathcal{S}_{\left(\overline{\Phi}\Phi\right)^{2}}^{\left(D33\right)} & =\frac{1}{8}\left(\frac{a\left(a-b\right)^{2}}{32\pi^{2}\epsilon}\right)\,i\,g^{6}\left(\delta_{im}\delta_{jn}+\delta_{jm}\delta_{in}\right)\int_{\theta}\overline{\Phi}_{i}\Phi_{m}\Phi_{n}\overline{\Phi}_{j}\,.\label{eq:S-D33}
\end{align}

\begin{center}
\begin{figure}
\begin{centering}
\subfloat[]{\centering{}\includegraphics[scale=0.5]{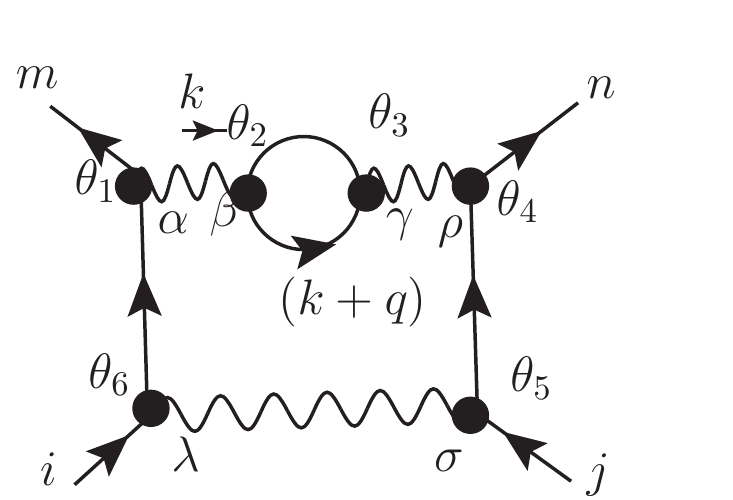}}\subfloat[]{\centering{}\includegraphics[scale=0.5]{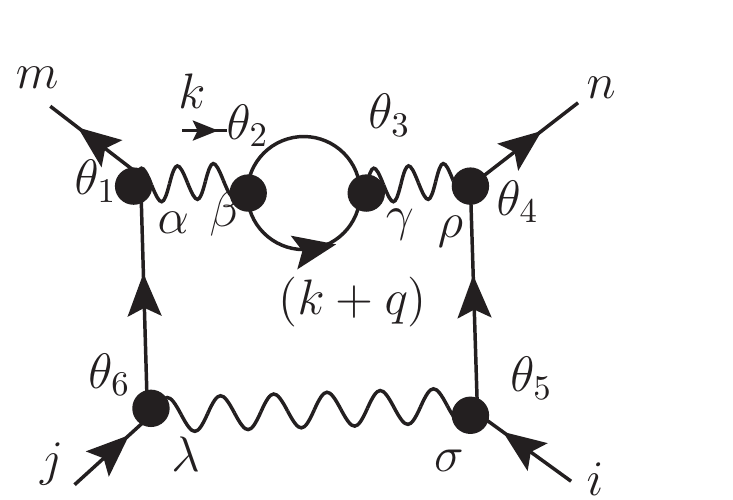}}\subfloat[]{\centering{}\includegraphics[scale=0.5]{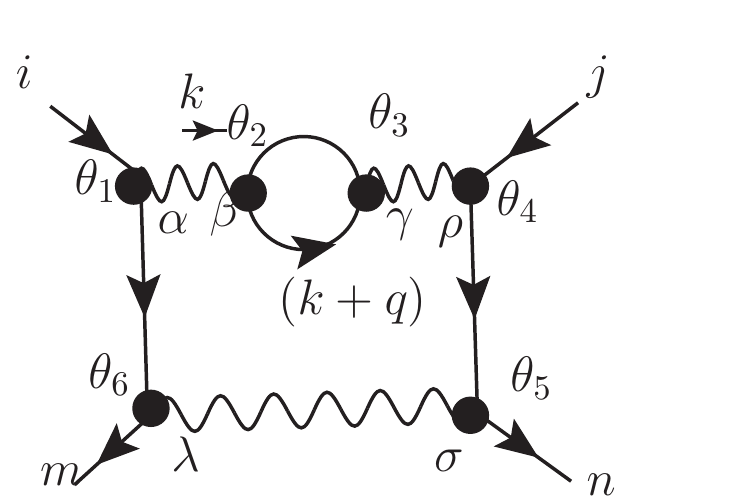}}\subfloat[]{\centering{}\includegraphics[scale=0.5]{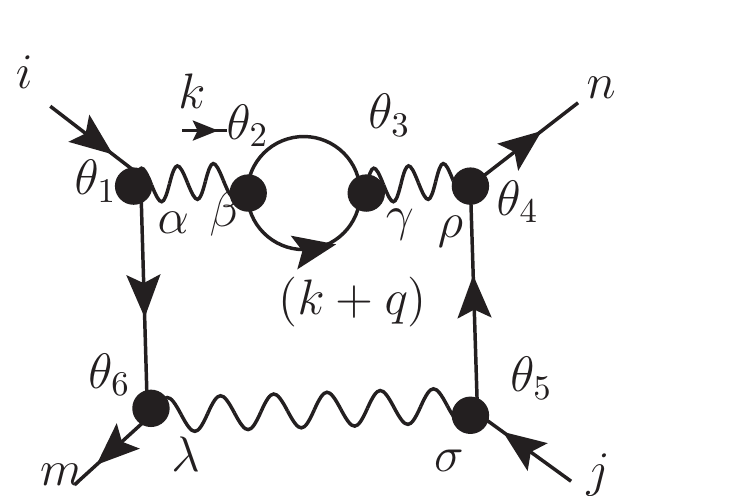}}
\par\end{centering}
\begin{centering}
\subfloat[]{\centering{}\includegraphics[scale=0.5]{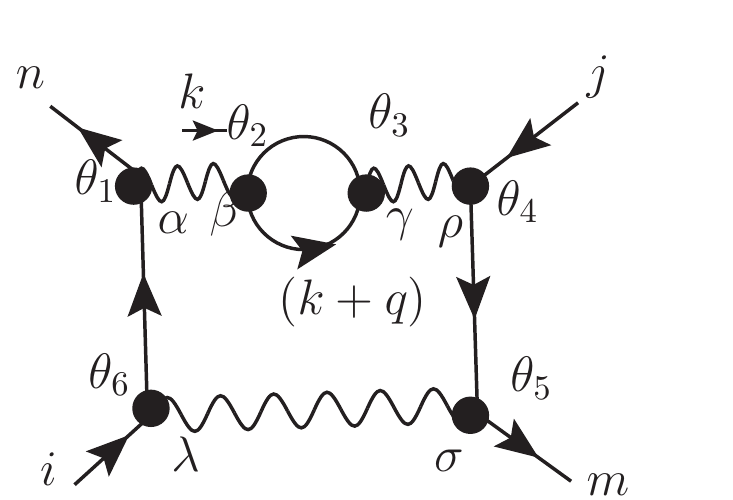}}\subfloat[]{\centering{}\includegraphics[scale=0.5]{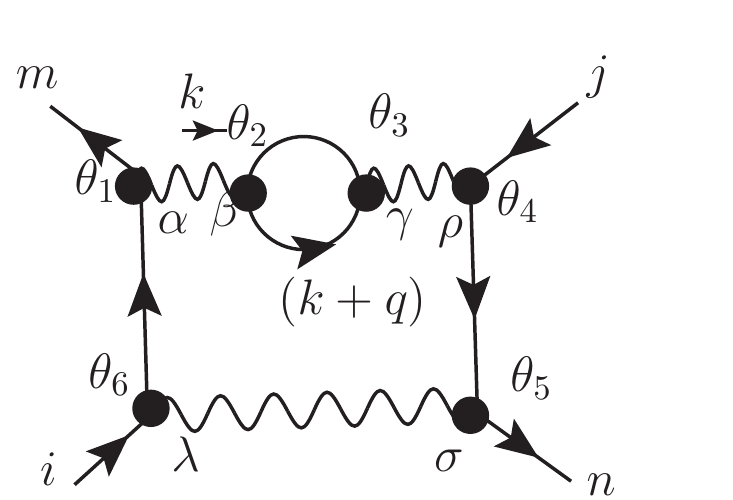}}\subfloat[]{\centering{}\includegraphics[scale=0.5]{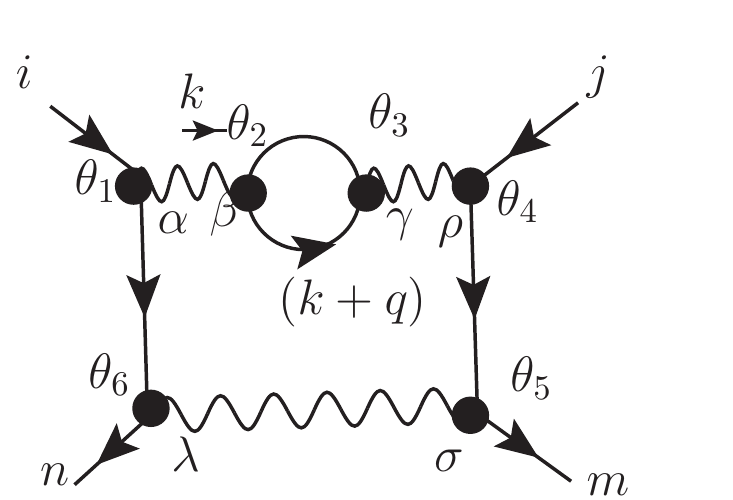}}\subfloat[]{\centering{}\includegraphics[scale=0.5]{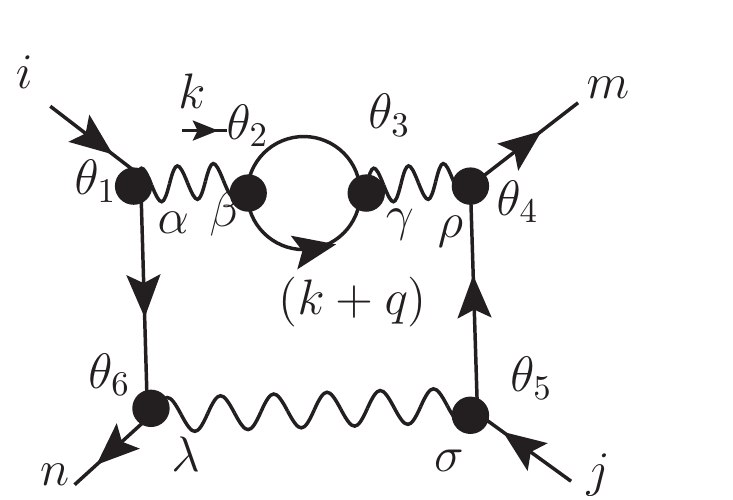}}
\par\end{centering}
\centering{}\caption{\label{fig:D34-order-g-6}$\mathcal{S}_{\left(\overline{\Phi}\Phi\right)^{2}}^{\left(D34\right)}$}
\end{figure}
\par\end{center}

$\mathcal{S}_{\left(\overline{\Phi}\Phi\right)^{2}}^{\left(D34-a\right)}$
in the Figure\,\ref{fig:D34-order-g-6} is
\begin{align}
\mathcal{S}_{\left(\overline{\Phi}\Phi\right)^{2}}^{\left(D34-a\right)} & =\frac{1}{64\left(8\right)}\,i\,g^{6}N\,\delta_{im}\,\delta_{nj}\int_{\theta}\overline{\Phi}_{i}\Phi_{m}\Phi_{n}\overline{\Phi}_{j}\int\frac{d^{D}kd^{D}q}{\left(2\pi\right)^{2D}}\left\{ \frac{8\left(a-b\right)^{3}\left(k^{2}\right)^{3}\left(2\left(k\cdot q\right)+k^{2}+2\,q^{2}\right)}{\left(k^{2}\right)^{5}\left(k+q\right)^{2}q^{2}}\right\} \,,
\end{align}
using Eqs.\,(\ref{eq:Int 7}), (\ref{eq:Int 9}), then adding $\mathcal{S}_{\left(\overline{\Phi}\Phi\right)^{2}}^{\left(D34-a\right)}$
to $\mathcal{S}_{\left(\overline{\Phi}\Phi\right)^{2}}^{\left(D34-h\right)}$,
we find 
\begin{align}
\mathcal{S}_{\left(\overline{\Phi}\Phi\right)^{2}}^{\left(D34-a\right)}=\mathcal{S}_{\left(\overline{\Phi}\Phi\right)^{2}}^{\left(D34\right)} & =0\,.\label{eq:S-D34}
\end{align}

\begin{center}
\begin{figure}
\begin{centering}
\subfloat[]{\centering{}\includegraphics[scale=0.5]{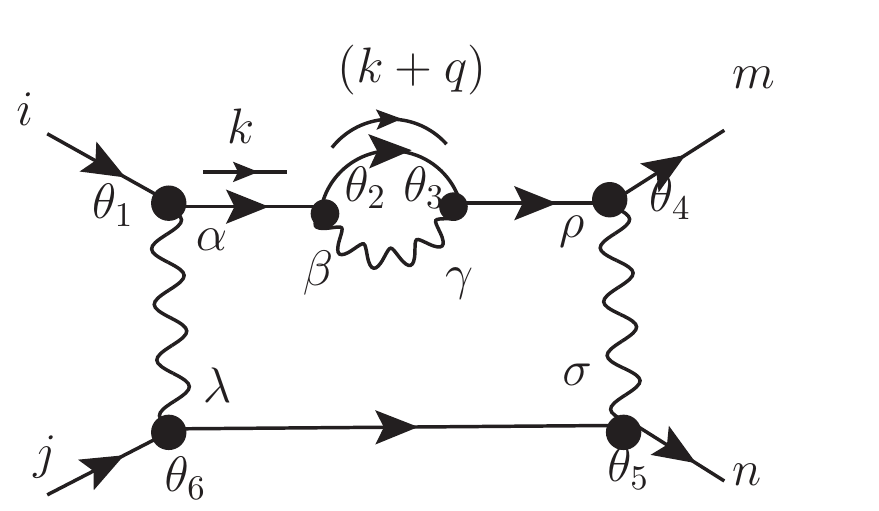}}\subfloat[]{\centering{}\includegraphics[scale=0.5]{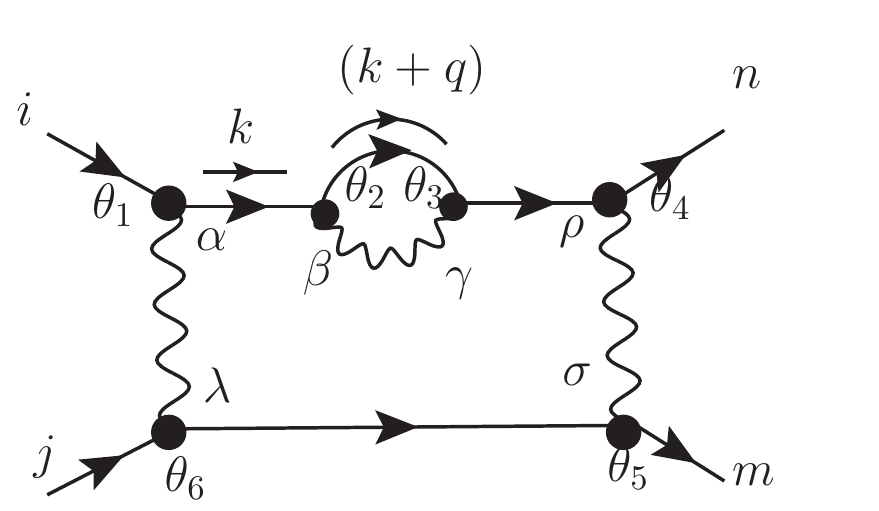}}\subfloat[]{\centering{}\includegraphics[scale=0.5]{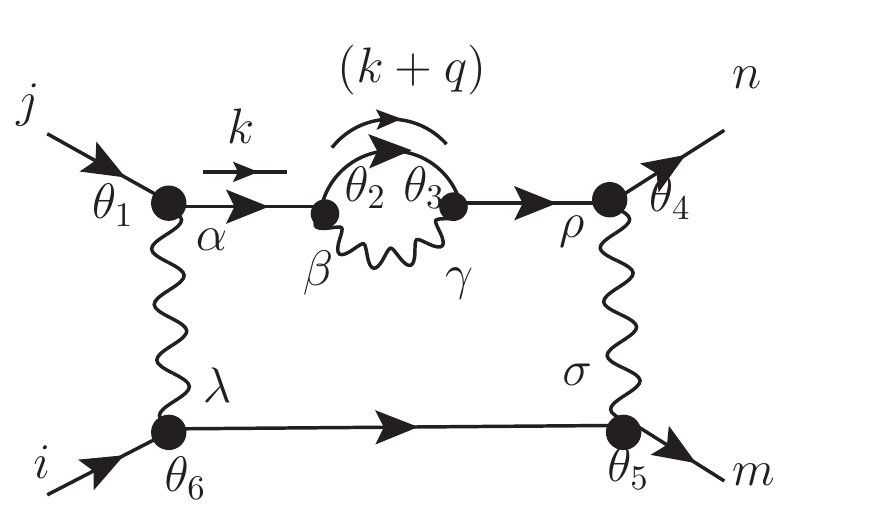}}
\par\end{centering}
\begin{centering}
\subfloat[]{\centering{}\includegraphics[scale=0.5]{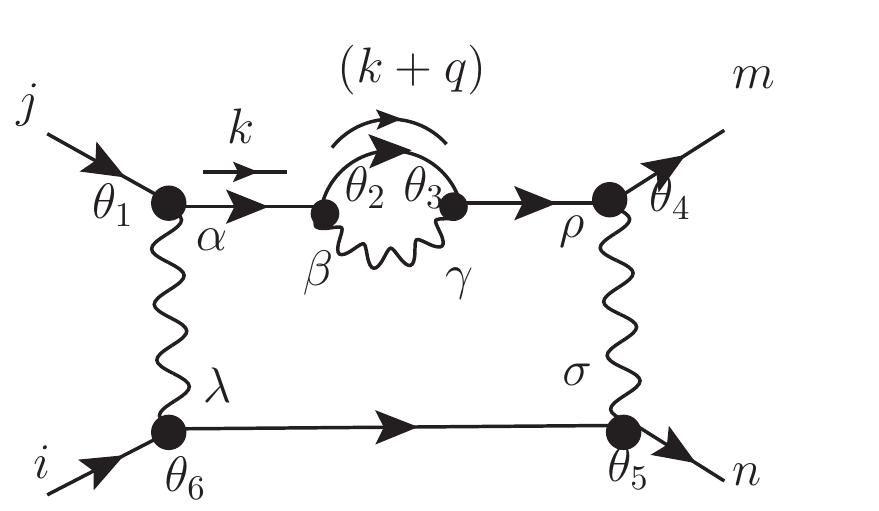}}\subfloat[]{\centering{}\includegraphics[scale=0.5]{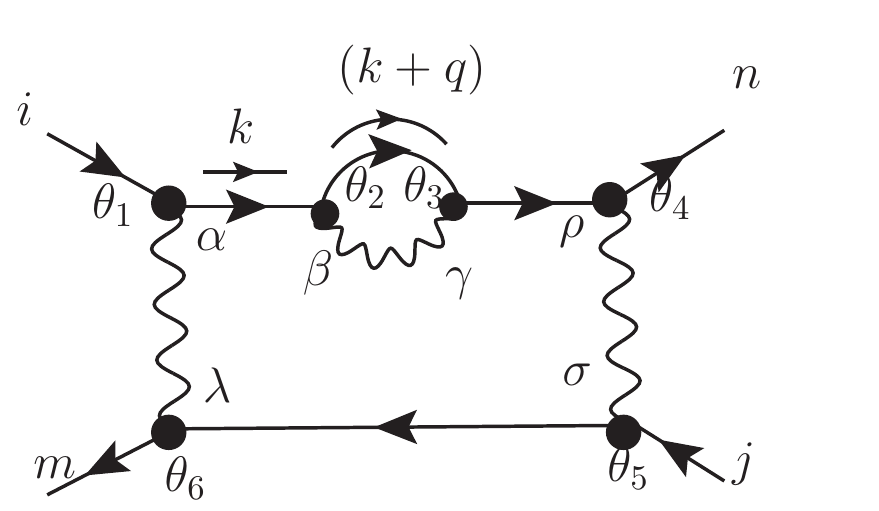}}\subfloat[]{\centering{}\includegraphics[scale=0.5]{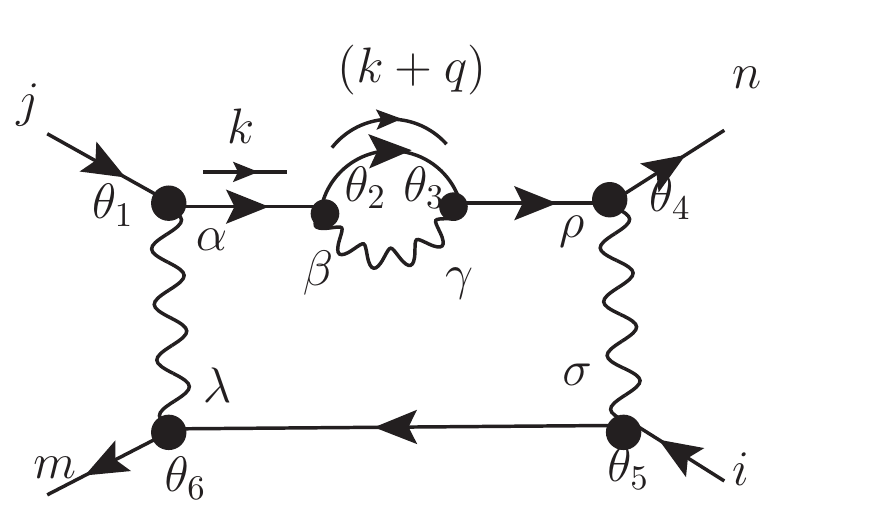}}
\par\end{centering}
\begin{centering}
\subfloat[]{\centering{}\includegraphics[scale=0.5]{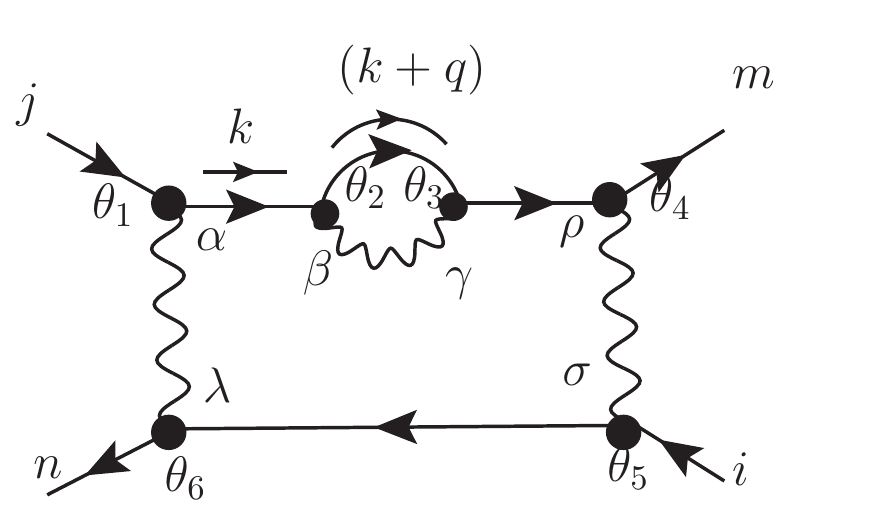}}\subfloat[]{\centering{}\includegraphics[scale=0.5]{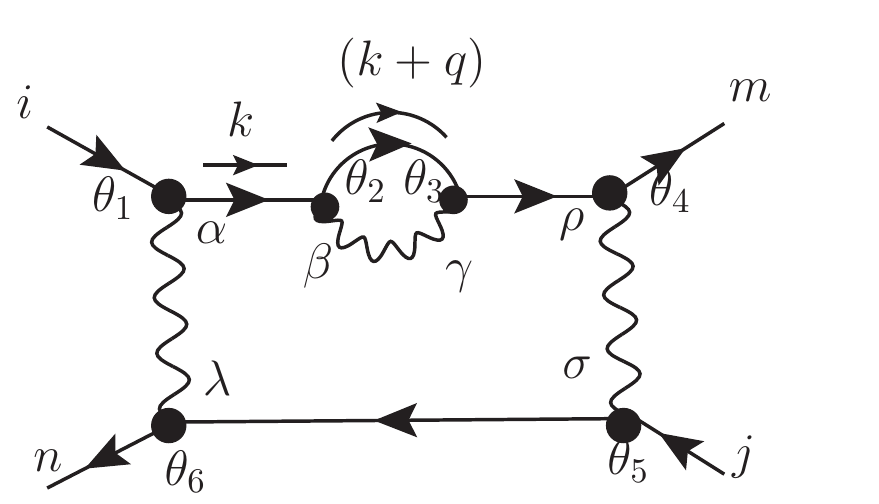}}
\par\end{centering}
\centering{}\caption{\label{fig:D35-order-g-6}$\mathcal{S}_{\left(\overline{\Phi}\Phi\right)^{2}}^{\left(D35\right)}$}
\end{figure}
\par\end{center}

\begin{center}
\begin{table}
\centering{}%
\begin{tabular}{lcccccccccc}
 &  &  &  &  &  &  &  &  &  & \tabularnewline
\hline 
\hline 
$D35-a$ &  & $\delta_{im}\,\delta_{nj}$ &  & $D35-b$ &  & $\delta_{jm}\,\delta_{ni}$ &  & $D35-c$ &  & $\delta_{im}\,\delta_{nj}$\tabularnewline
$D35-d$ &  & $\delta_{jm}\,\delta_{ni}$ &  & $D35-e$ &  & $\delta_{jm}\,\delta_{ni}$ &  & $D35-f$ &  & $\delta_{im}\,\delta_{nj}$\tabularnewline
$D35-g$ &  & $\delta_{jm}\,\delta_{ni}$ &  & $D35-h$ &  & $\delta_{im}\,\delta_{nj}$ &  &  &  & \tabularnewline
\hline 
\hline 
 &  &  &  &  &  &  &  &  &  & \tabularnewline
\end{tabular}\caption{\label{tab:S-PPPP-35}Values of the diagrams in Figure\,\ref{fig:D35-order-g-6}
with common factor\protect \\
 $-\frac{1}{32}\left(\frac{a\left(a-b\right)^{2}}{32\pi^{2}\epsilon}\right)\,i\,g^{6}\int_{\theta}\overline{\Phi}_{i}\Phi_{m}\Phi_{n}\overline{\Phi}_{j}$
.}
\end{table}
\par\end{center}

$\mathcal{S}_{\left(\overline{\Phi}\Phi\right)^{2}}^{\left(D35-a\right)}$
in the Figure\,\ref{fig:D35-order-g-6} is
\begin{align}
\mathcal{S}_{\left(\overline{\Phi}\Phi\right)^{2}}^{\left(D35-a\right)} & =\frac{1}{64\left(8\right)}\,i\,\delta_{im}\,\delta_{nj}\,g^{6}\int_{\theta}\overline{\Phi}_{i}\Phi_{m}\Phi_{n}\overline{\Phi}_{j}\nonumber \\
 & \times\int\frac{d^{D}kd^{D}q}{\left(2\pi\right)^{2D}}\left\{ \frac{32\,a\left(a-b\right)^{2}\left(k^{2}\right)^{3}\left(\left(k\cdot q\right)+k^{2}\right)+8\left(a-b\right)^{3}\left(k^{2}\right)^{3}q^{2}}{\left(k^{2}\right)^{5}\left(k+q\right)^{2}q^{2}}\right\} \,,
\end{align}
using Eqs.\,(\ref{eq:Int 7}), (\ref{eq:Int 9}), then adding $\mathcal{S}_{\left(\overline{\Phi}\Phi\right)^{2}}^{\left(D35-a\right)}$
to $\mathcal{S}_{\left(\overline{\Phi}\Phi\right)^{2}}^{\left(D35-h\right)}$
with the values in Table\,\ref{tab:S-PPPP-35}, we find
\begin{align}
\mathcal{S}_{\left(\overline{\Phi}\Phi\right)^{2}}^{\left(D35\right)} & =-\frac{1}{8}\left(\frac{a\left(a-b\right)^{2}}{32\pi^{2}\epsilon}\right)\,i\,g^{6}\left(\delta_{im}\,\delta_{nj}+\delta_{jm}\,\delta_{ni}\right)\int_{\theta}\overline{\Phi}_{i}\Phi_{m}\Phi_{n}\overline{\Phi}_{j}\,.\label{eq:S-D35}
\end{align}

\begin{center}
\begin{figure}
\begin{centering}
\subfloat[]{\centering{}\includegraphics[scale=0.5]{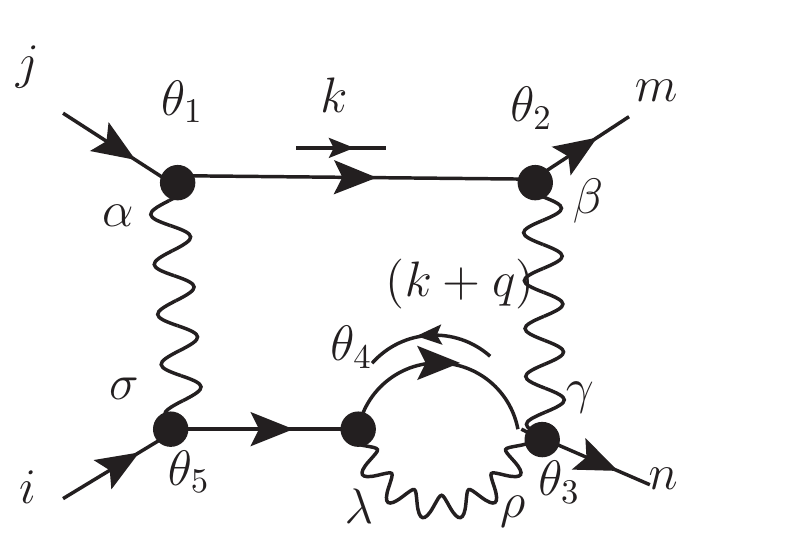}}\subfloat[]{\centering{}\includegraphics[scale=0.5]{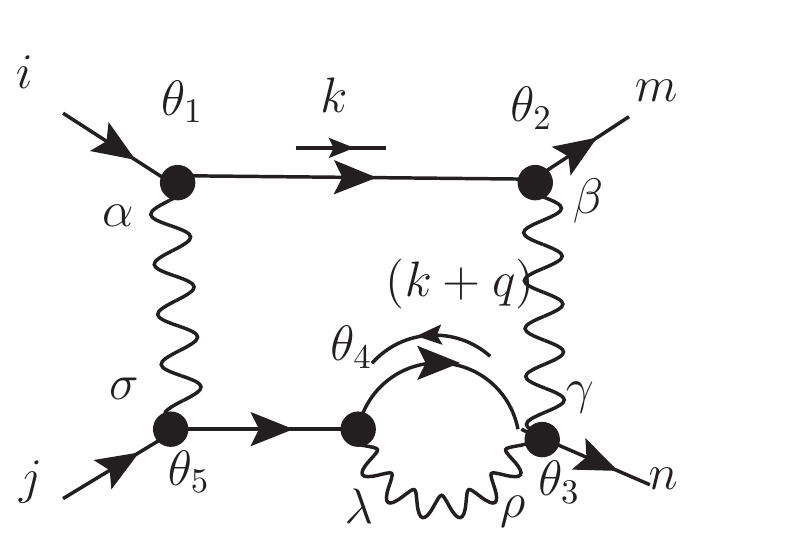}}\subfloat[]{\centering{}\includegraphics[scale=0.5]{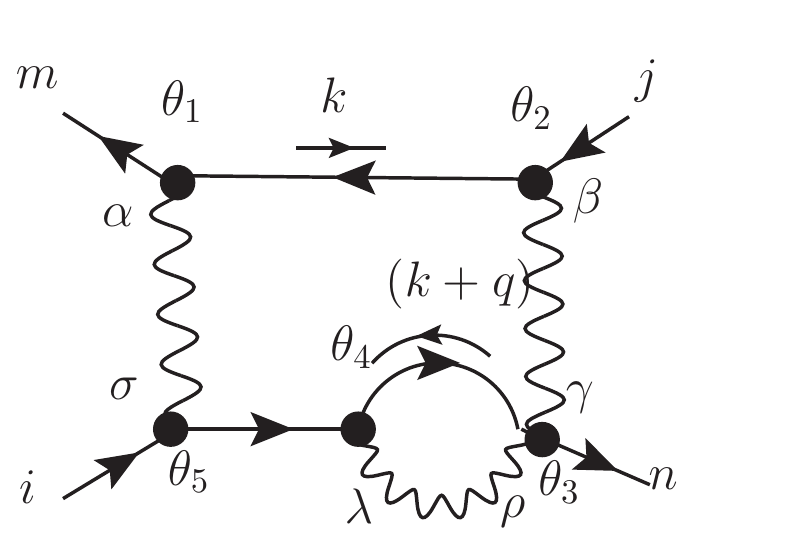}}\subfloat[]{\centering{}\includegraphics[scale=0.5]{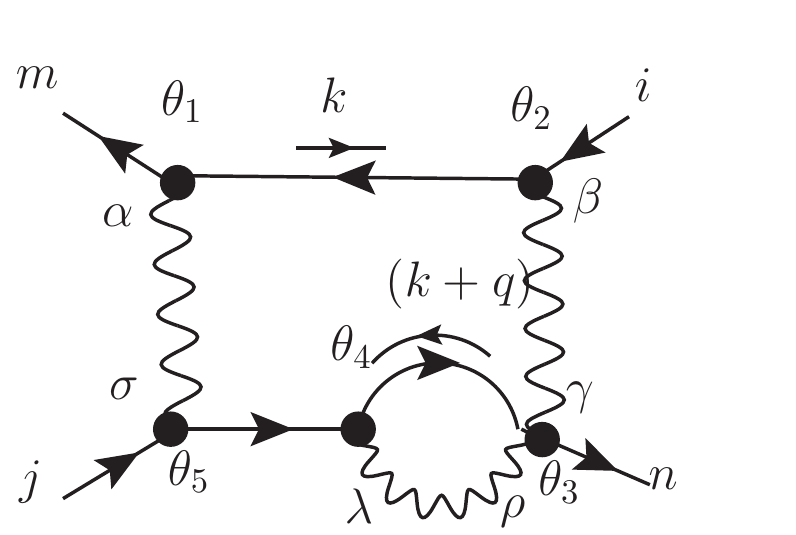}}
\par\end{centering}
\begin{centering}
\subfloat[]{\centering{}\includegraphics[scale=0.5]{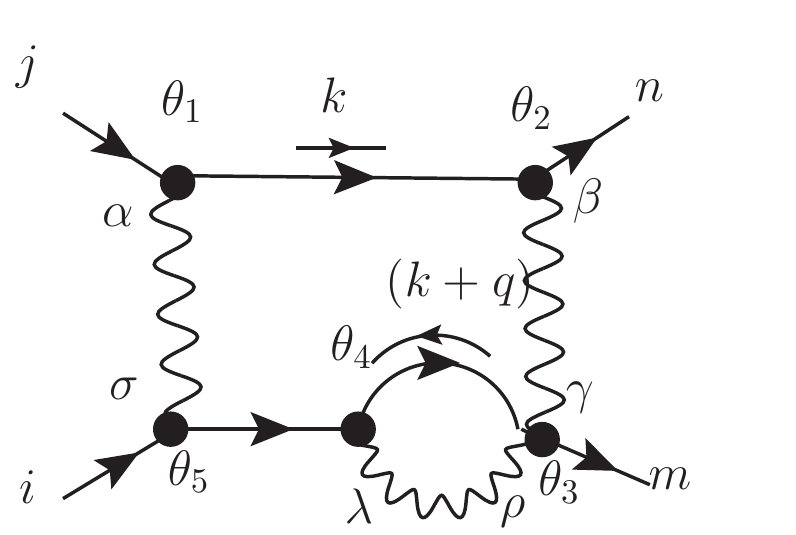}}\subfloat[]{\centering{}\includegraphics[scale=0.5]{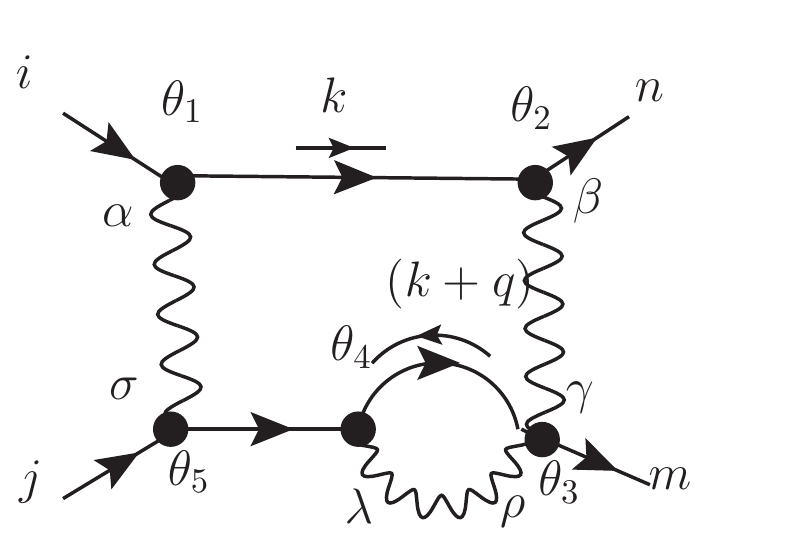}}\subfloat[]{\centering{}\includegraphics[scale=0.5]{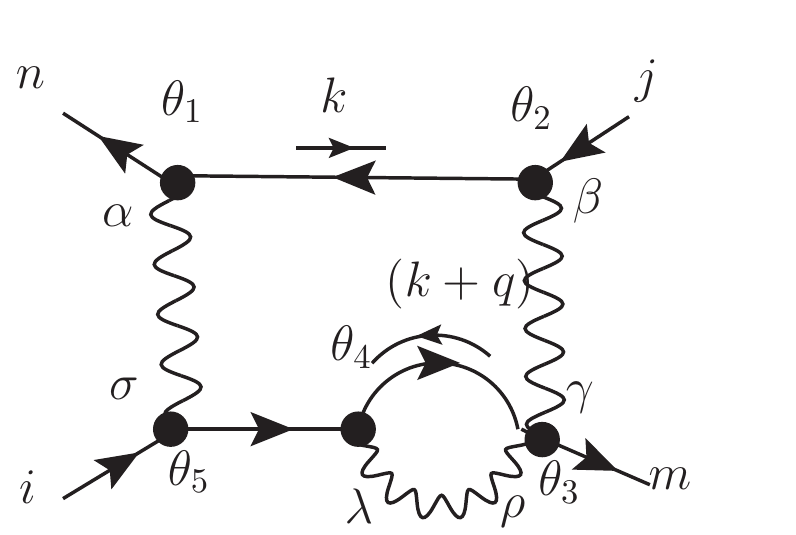}}\subfloat[]{\centering{}\includegraphics[scale=0.5]{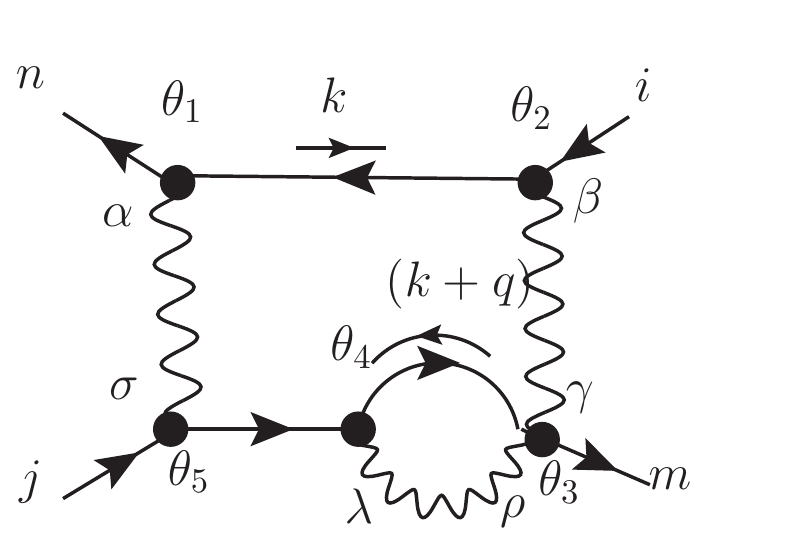}}
\par\end{centering}
\begin{centering}
\subfloat[]{\centering{}\includegraphics[scale=0.5]{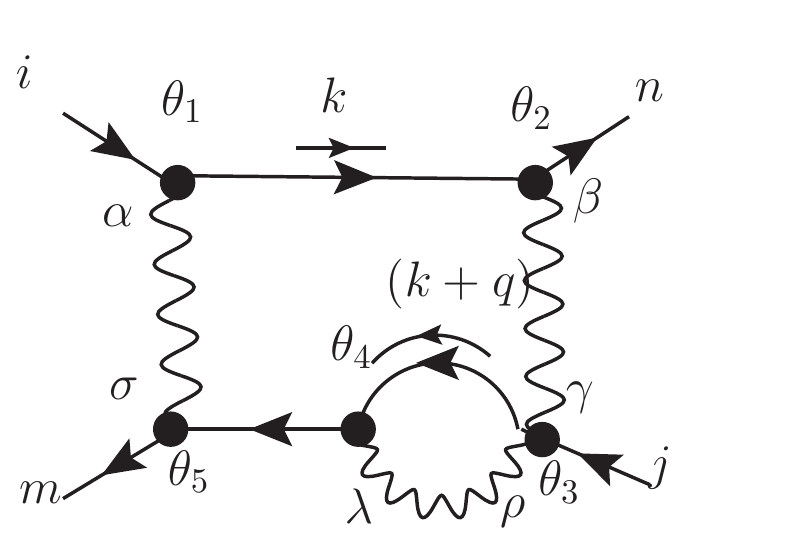}}\subfloat[]{\centering{}\includegraphics[scale=0.5]{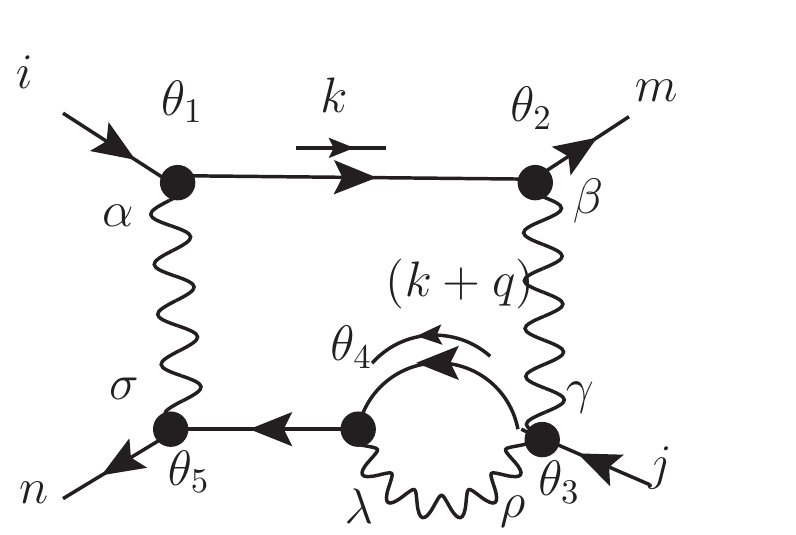}}\subfloat[]{\centering{}\includegraphics[scale=0.5]{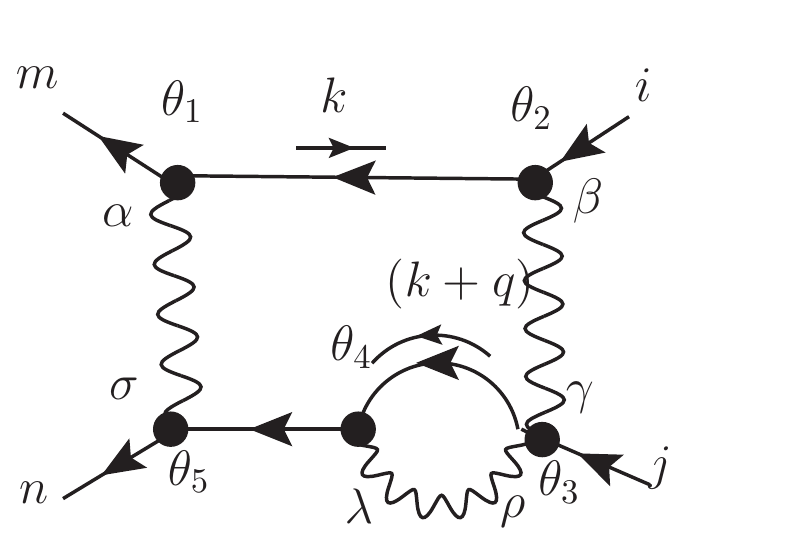}}\subfloat[]{\centering{}\includegraphics[scale=0.5]{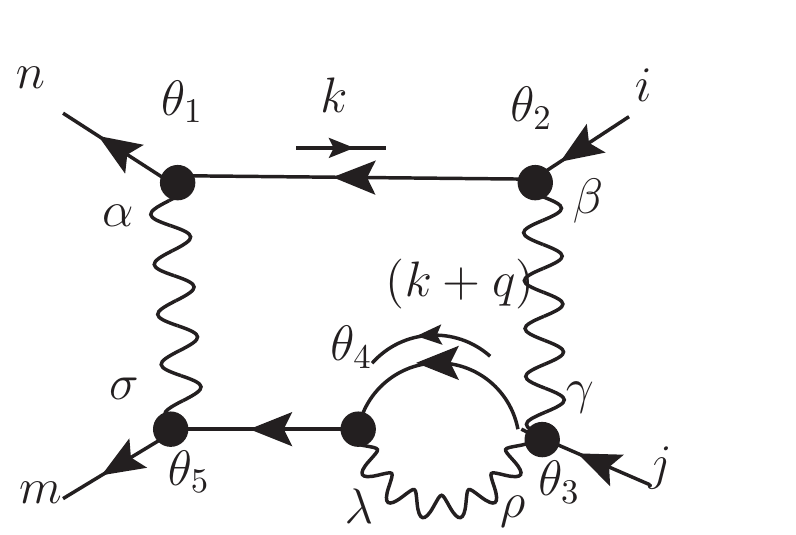}}
\par\end{centering}
\begin{centering}
\subfloat[]{\centering{}\includegraphics[scale=0.5]{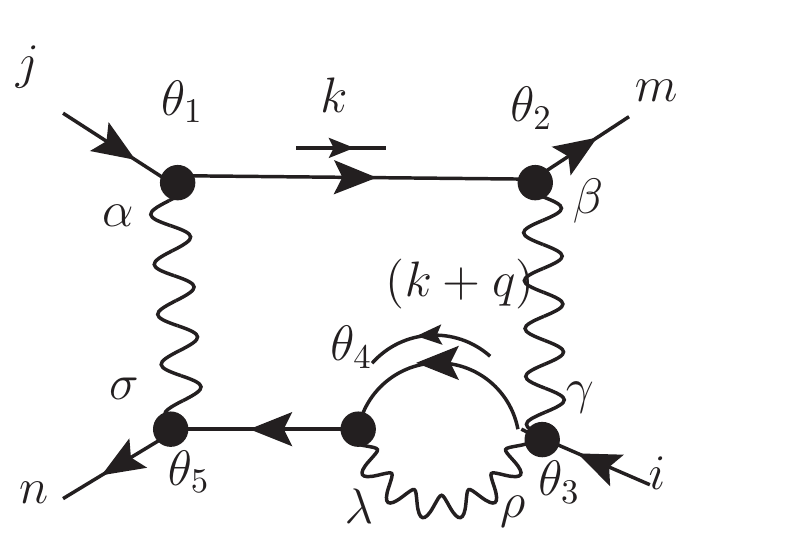}}\subfloat[]{\centering{}\includegraphics[scale=0.5]{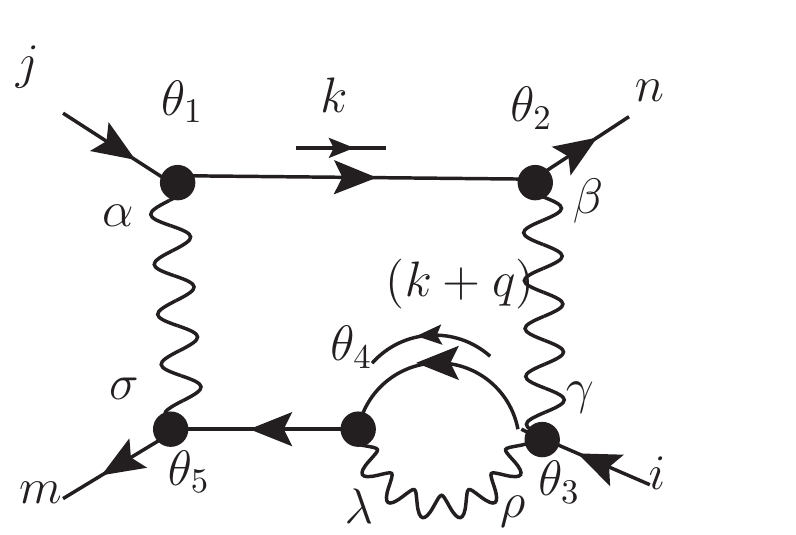}}\subfloat[]{\centering{}\includegraphics[scale=0.5]{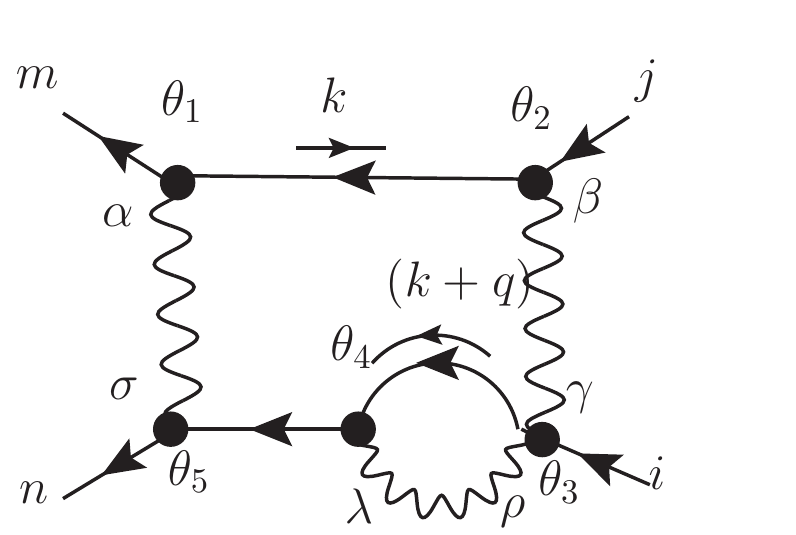}}\subfloat[]{\centering{}\includegraphics[scale=0.5]{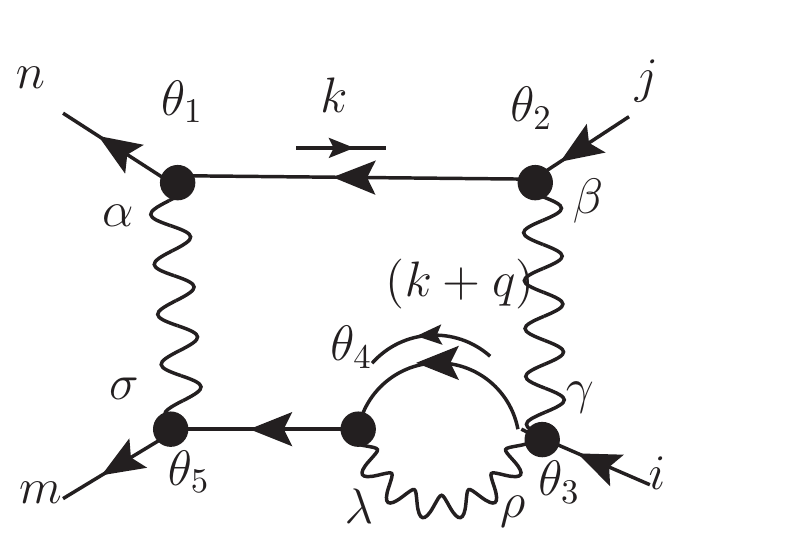}}
\par\end{centering}
\centering{}\caption{\label{fig:D36-order-g-6}$\mathcal{S}_{\left(\overline{\Phi}\Phi\right)^{2}}^{\left(D36\right)}$}
\end{figure}
\par\end{center}

\begin{center}
\begin{table}
\centering{}%
\begin{tabular}{lcccccccccc}
 &  &  &  &  &  &  &  &  &  & \tabularnewline
\hline 
\hline 
$D36-a$ &  & $\delta_{mj}\delta_{ni}$ &  & $D36-b$ &  & $\delta_{mi}\delta_{nj}$ &  & $D36-c$ &  & $\delta_{mj}\delta_{ni}$\tabularnewline
$D36-d$ &  & $\delta_{mi}\delta_{nj}$ &  & $D36-e$ &  & $\delta_{mi}\delta_{nj}$ &  & $D36-f$ &  & $\delta_{mj}\delta_{ni}$\tabularnewline
$D36-g$ &  & $\delta_{mi}\delta_{nj}$ &  & $D36-h$ &  & $\delta_{mj}\delta_{ni}$ &  & $D36-i$ &  & $\delta_{mj}\delta_{ni}$\tabularnewline
$D36-j$ &  & $\delta_{mi}\delta_{nj}$ &  & $D36-k$ &  & $\delta_{mi}\delta_{nj}$ &  & $D36-l$ &  & $\delta_{mj}\delta_{ni}$\tabularnewline
$D36-m$ &  & $\delta_{mj}\delta_{ni}$ &  & $D36-n$ &  & $\delta_{mi}\delta_{nj}$ &  & $D36-o$ &  & $\delta_{mj}\delta_{ni}$\tabularnewline
$D36-p$ &  & $\delta_{mi}\delta_{nj}$ &  &  &  &  &  &  &  & \tabularnewline
\hline 
\hline 
 &  &  &  &  &  &  &  &  &  & \tabularnewline
\end{tabular}\caption{\label{tab:S-PPPP-36}Values of the diagrams in Figure\,\ref{fig:D36-order-g-6}
with common factor\protect \\
 $\frac{1}{64}\left(\frac{3\,a^{3}-7\,a^{2}\,b+5\,a\,b^{2}-b^{3}}{32\pi^{2}\epsilon}\right)\,i\,g^{6}\int_{\theta}\overline{\Phi}_{i}\Phi_{m}\Phi_{n}\overline{\Phi}_{j}$
.}
\end{table}
\par\end{center}

$\mathcal{S}_{\left(\overline{\Phi}\Phi\right)^{2}}^{\left(D36-a\right)}$
in the Figure\,\ref{fig:D36-order-g-6} is 
\begin{align}
\mathcal{S}_{\left(\overline{\Phi}\Phi\right)^{2}}^{\left(D36-a\right)} & =-\frac{1}{128}i\,g^{6}\delta_{mj}\delta_{ni}\int_{\theta}\overline{\Phi}_{i}\Phi_{m}\Phi_{n}\overline{\Phi}_{j}\times\nonumber \\
 & \,\int\frac{d^{D}kd^{D}q}{\left(2\pi\right)^{2D}}\left\{ \frac{8a\,\left(a-b\right)^{2}\left(k^{2}\right)^{3}+4\left(a^{3}-a^{2}\,b-a\,b^{2}+b^{3}\right)\left(k\cdot q\right)\left(k^{2}\right)^{2}}{\left(k^{2}\right)^{4}\left(q+k\right)^{2}q^{2}}\right\} \,,
\end{align}
using Eqs.\,(\ref{eq:Int 7}) and\,(\ref{eq:Int 9}), then adding
$\mathcal{S}_{\left(\overline{\Phi}\Phi\right)^{2}}^{\left(D36-a\right)}$
to $\mathcal{S}_{\left(\overline{\Phi}\Phi\right)^{2}}^{\left(D36-p\right)}$
with the values in Table\,\ref{tab:S-PPPP-36}, we find
\begin{align}
\mathcal{S}_{\left(\overline{\Phi}\Phi\right)^{2}}^{\left(D36\right)} & =\frac{1}{8}\left(\frac{\left(a-b\right)^{2}\left(3\,a-b\right)}{32\pi^{2}\epsilon}\right)i\,g^{6}\left(\delta_{mi}\delta_{nj}+\delta_{mj}\delta_{ni}\right)\int_{\theta}\overline{\Phi}_{i}\Phi_{m}\Phi_{n}\overline{\Phi}_{j}\,.\label{eq:S-D36}
\end{align}

\begin{center}
\begin{figure}
\begin{centering}
\subfloat[]{\centering{}\includegraphics[scale=0.5]{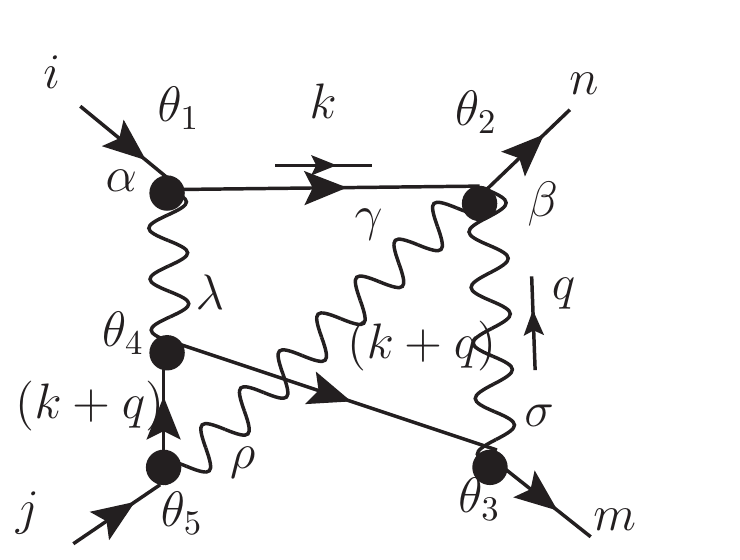}}\subfloat[]{\centering{}\includegraphics[scale=0.5]{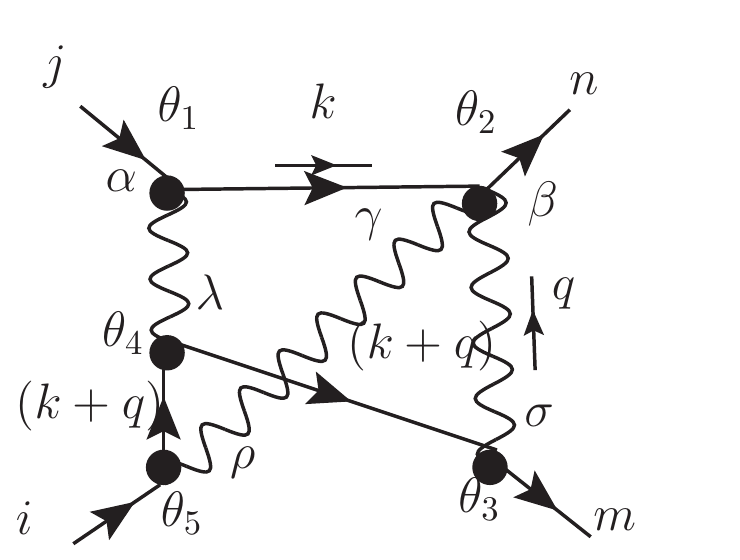}}\subfloat[]{\centering{}\includegraphics[scale=0.5]{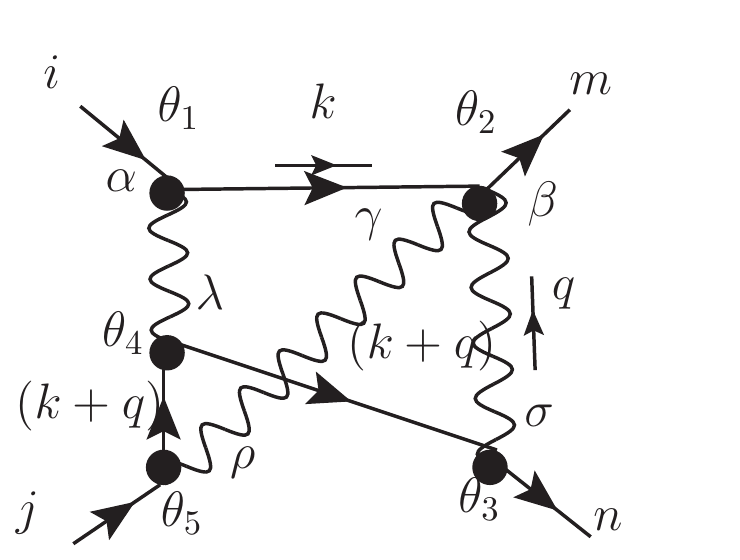}}\subfloat[]{\centering{}\includegraphics[scale=0.5]{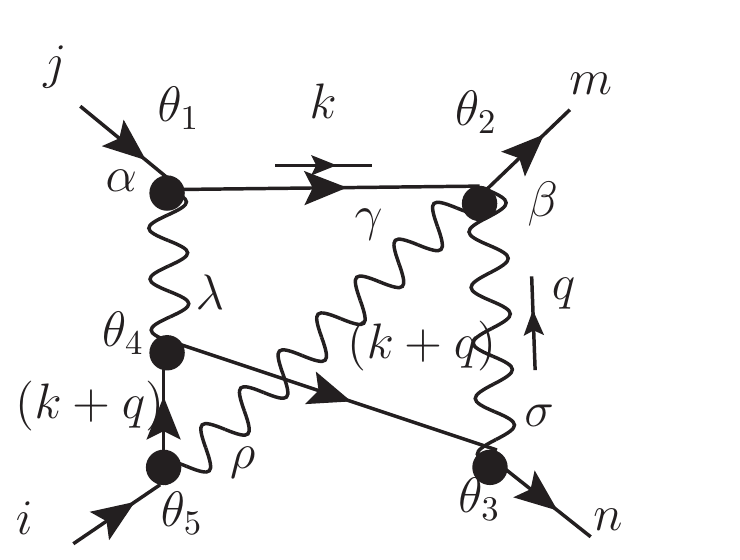}}
\par\end{centering}
\begin{centering}
\subfloat[]{\centering{}\includegraphics[scale=0.5]{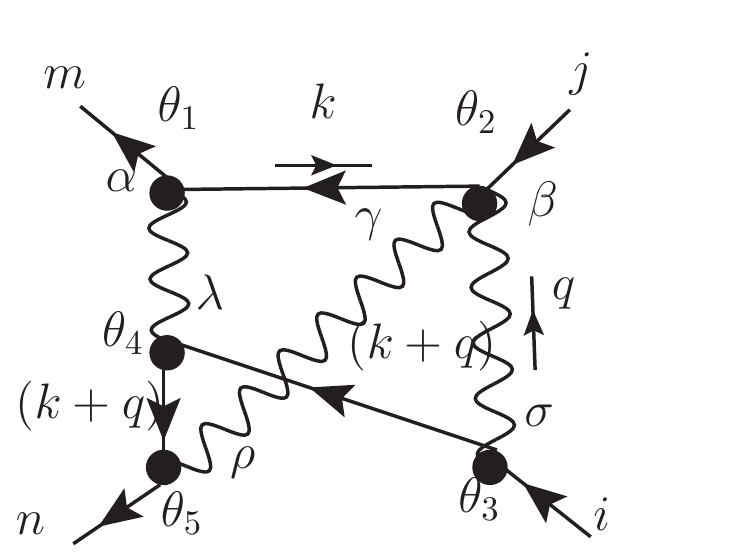}}\subfloat[]{\centering{}\includegraphics[scale=0.5]{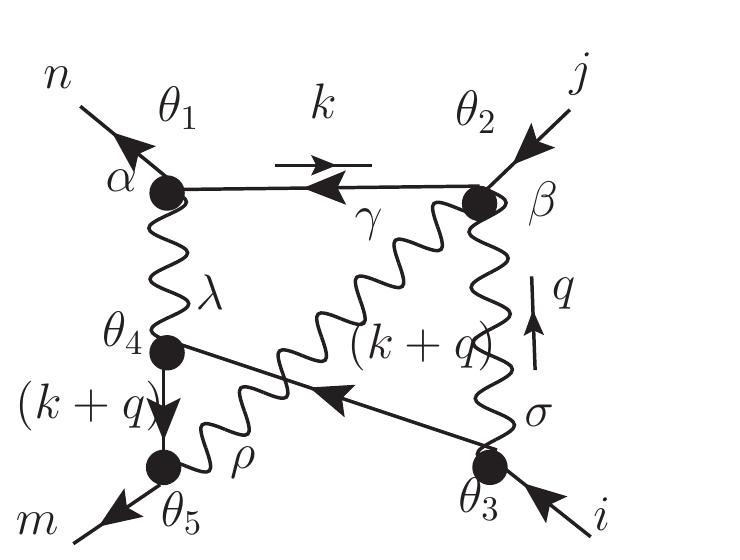}}\subfloat[]{\centering{}\includegraphics[scale=0.5]{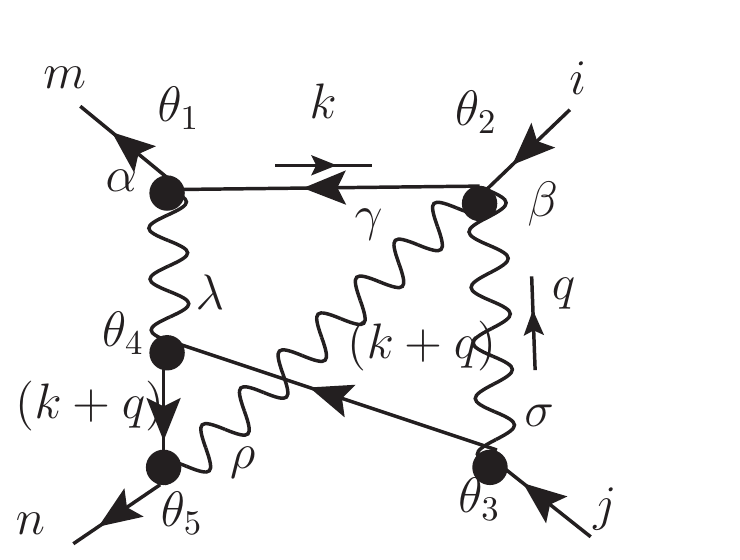}}\subfloat[]{\centering{}\includegraphics[scale=0.5]{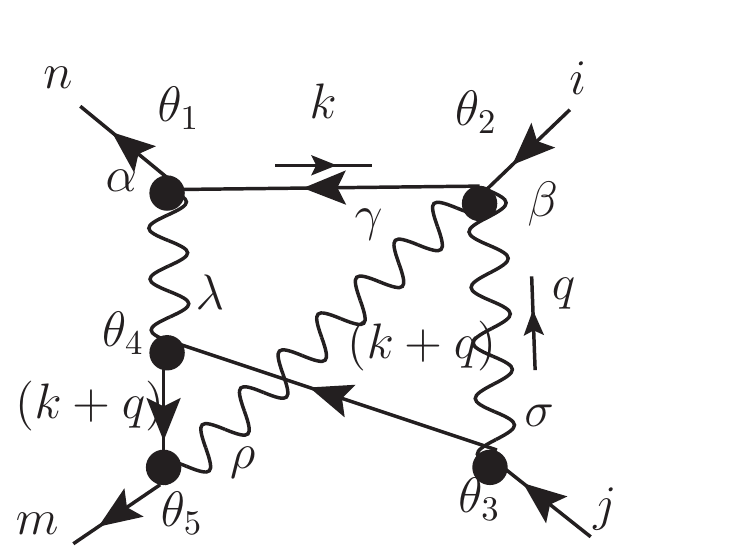}}
\par\end{centering}
\centering{}\caption{\label{fig:D37-order-g-6}$\mathcal{S}_{\left(\overline{\Phi}\Phi\right)^{2}}^{\left(D37\right)}$}
\end{figure}
\par\end{center}

$\mathcal{S}_{\left(\overline{\Phi}\Phi\right)^{2}}^{\left(D37-a\right)}$
in the Figure\,\ref{fig:D37-order-g-6} is
\begin{align}
\mathcal{S}_{\left(\overline{\Phi}\Phi\right)^{2}}^{\left(D37-a\right)} & =\mathcal{S}_{\left(\overline{\Phi}\Phi\right)^{2}}^{\left(D37\right)}=0\,,\label{eq:S-D37}
\end{align}
because of $D$'s manipulations.
\begin{center}
\begin{figure}
\begin{centering}
\subfloat[]{\centering{}\includegraphics[scale=0.5]{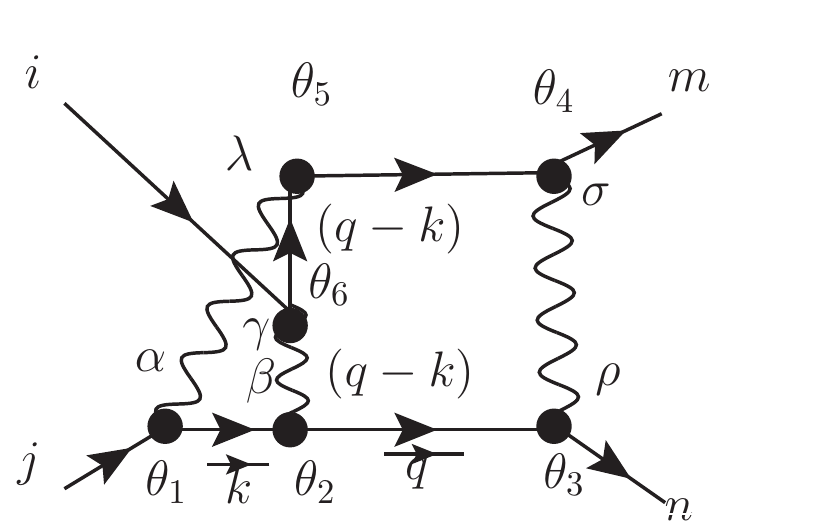}}\subfloat[]{\centering{}\includegraphics[scale=0.5]{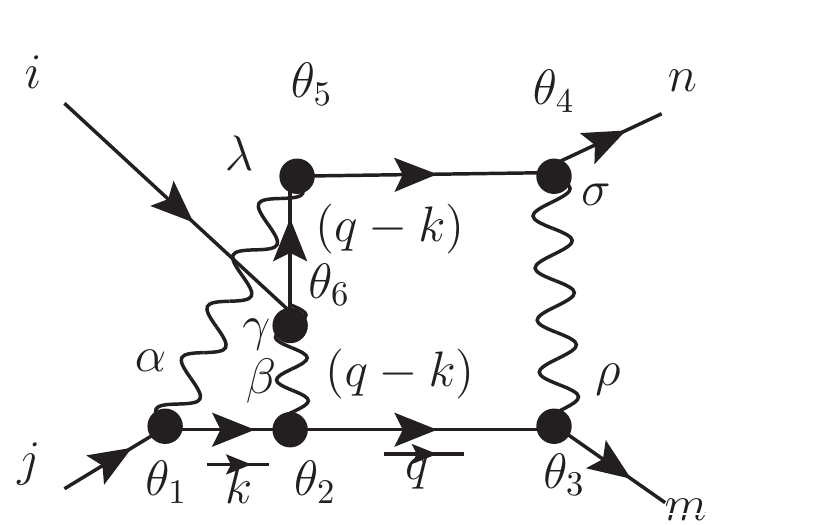}}\subfloat[]{\centering{}\includegraphics[scale=0.5]{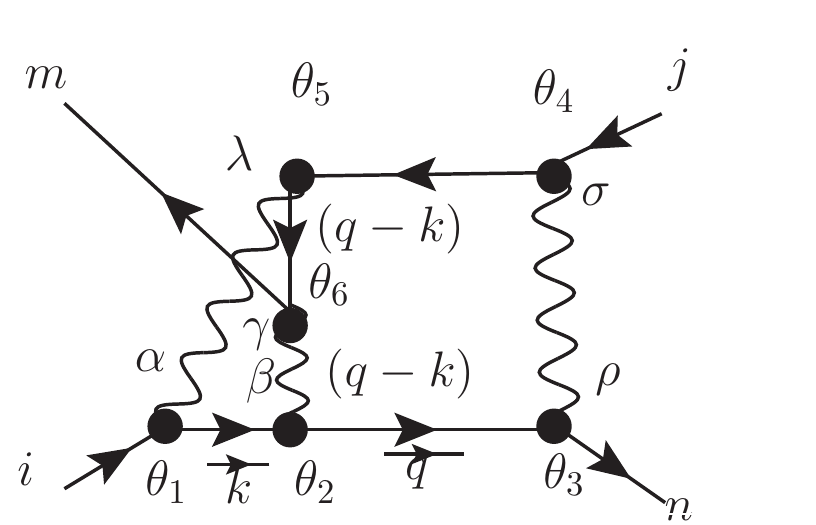}}\subfloat[]{\centering{}\includegraphics[scale=0.5]{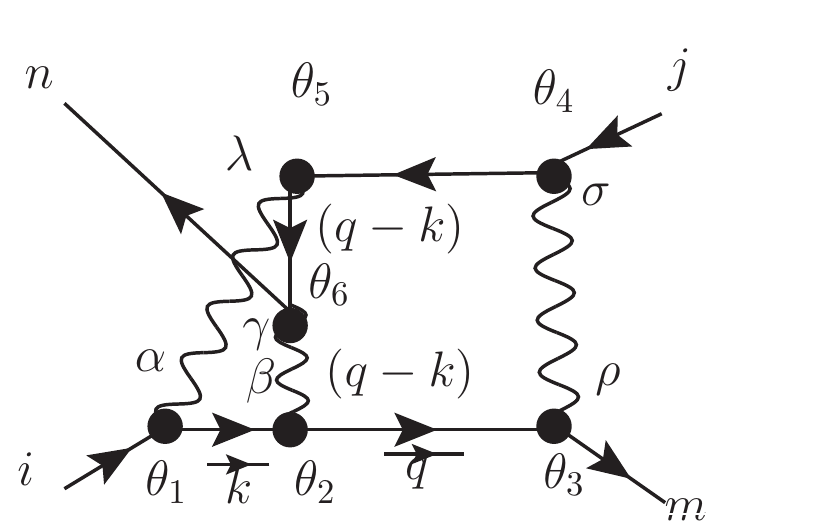}}
\par\end{centering}
\begin{centering}
\subfloat[]{\centering{}\includegraphics[scale=0.5]{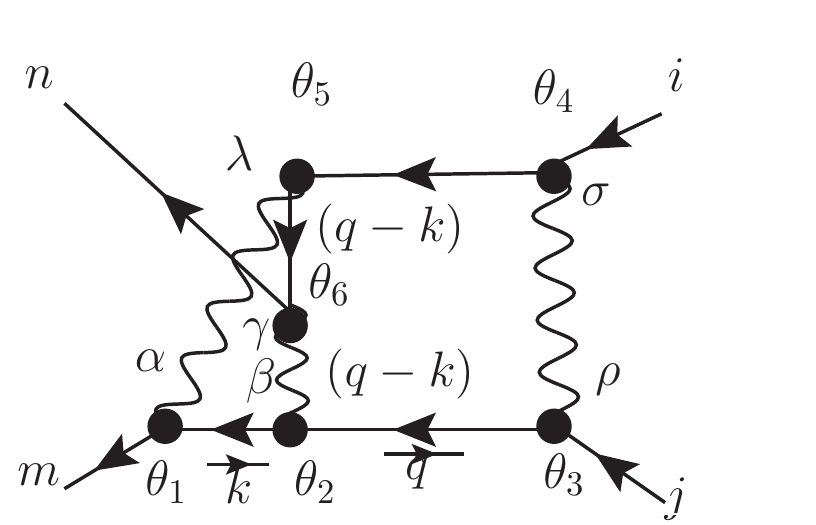}}\subfloat[]{\centering{}\includegraphics[scale=0.5]{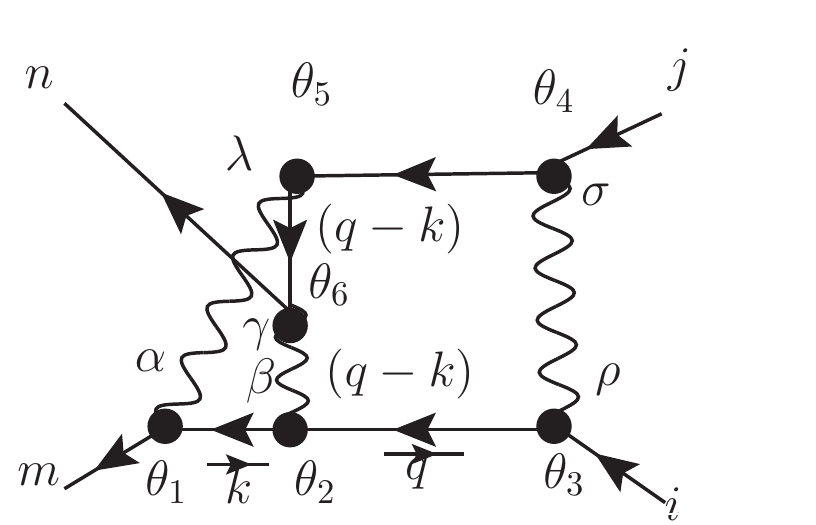}}\subfloat[]{\centering{}\includegraphics[scale=0.5]{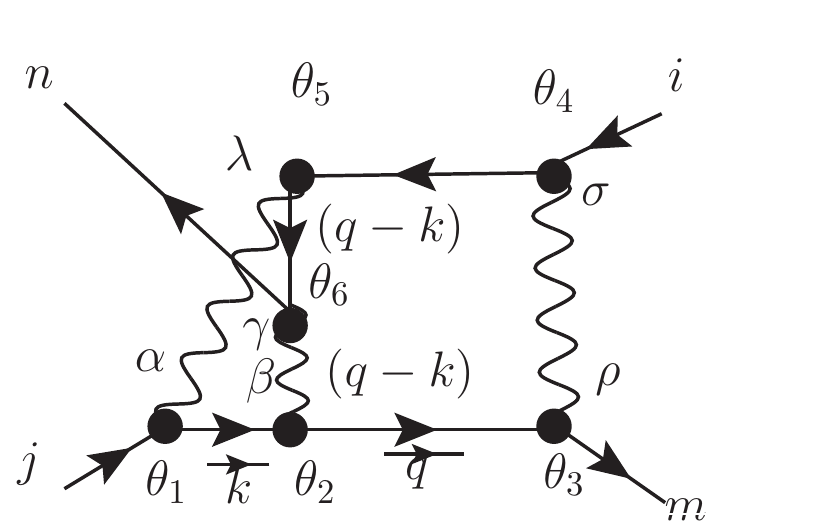}}\subfloat[]{\centering{}\includegraphics[scale=0.5]{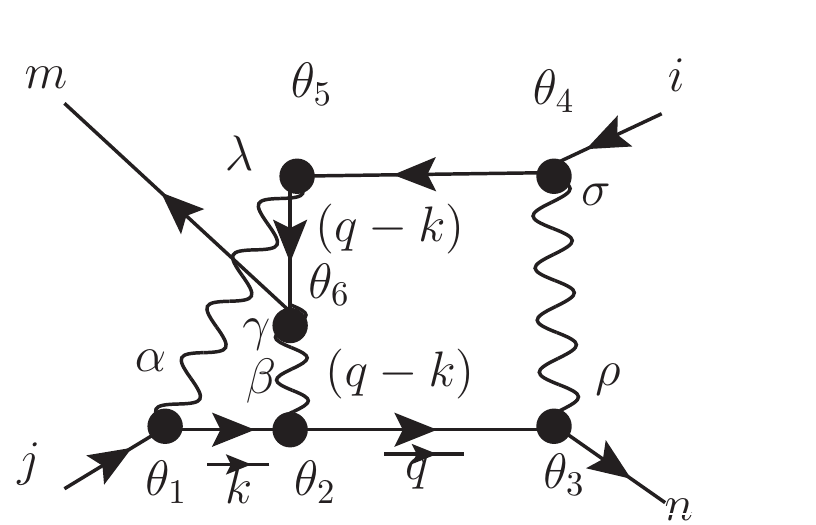}}
\par\end{centering}
\centering{}\caption{\label{fig:D38-order-g-6}$\mathcal{S}_{\left(\overline{\Phi}\Phi\right)^{2}}^{\left(D38\right)}$}
\end{figure}
\par\end{center}

\begin{center}
\begin{table}
\centering{}%
\begin{tabular}{lcccccccccc}
 &  &  &  &  &  &  &  &  &  & \tabularnewline
\hline 
\hline 
$D38-a$ &  & $-\delta_{jn}\delta_{im}$ &  & $D38-b$ &  & $-\delta_{jm}\delta_{in}$ &  & $D38-c$ &  & $\delta_{jm}\delta_{in}$\tabularnewline
$D38-d$ &  & $\delta_{im}\delta_{jn}$ &  & $D38-e$ &  & $-\delta_{jm}\delta_{in}$ &  & $D38-f$ &  & $-\delta_{im}\delta_{jn}$\tabularnewline
$D38-g$ &  & $\delta_{jm}\delta_{in}$ &  & $D38-h$ &  & $\delta_{im}\delta_{jn}$ &  &  &  & \tabularnewline
\hline 
\hline 
 &  &  &  &  &  &  &  &  &  & \tabularnewline
\end{tabular}\caption{\label{tab:S-PPPP-38}Values of the diagrams in Figure\,\ref{fig:D38-order-g-6}
with common factor\protect \\
 $\frac{1}{64}\left(\frac{\left(a-b\right)^{3}}{32\pi^{2}\epsilon}\right)\,i\,g^{6}\int_{\theta}\overline{\Phi}_{i}\Phi_{m}\Phi_{n}\overline{\Phi}_{j}$
.}
\end{table}
\par\end{center}

$\mathcal{S}_{\left(\overline{\Phi}\Phi\right)^{2}}^{\left(D38-a\right)}$
in the Figure\,\ref{fig:D38-order-g-6} is
\begin{align}
\mathcal{S}_{\left(\overline{\Phi}\Phi\right)^{2}}^{\left(D38-a\right)} & =\frac{1}{64\left(8\right)}\,i\,\delta_{jn}\delta_{im}\,g^{6}\int_{\theta}\overline{\Phi}_{i}\Phi_{m}\Phi_{n}\overline{\Phi}_{j}\int\frac{d^{D}kd^{D}q}{\left(2\pi\right)^{2D}}\left\{ -4\,a^{2}\,b\left(k\cdot q\right)^{2}k^{2}q^{2}\right.\nonumber \\
 & +4\,a^{2}\,b\left(k\cdot q\right)^{2}\left(q^{2}\right)^{2}+a^{2}\,b\,k_{\alpha\beta}\,k_{\gamma\delta}\,q^{\alpha\delta}\,q^{\beta\gamma}k^{2}q^{2}-a^{2}\,b\,k_{\alpha\beta}\,k_{\gamma\delta}\,q^{\alpha\delta}\,q^{\beta\gamma}\left(q^{2}\right)^{2}\nonumber \\
 & \left.+8\,\left(a-b\right)^{3}k^{2}\left(q^{2}\right)^{2}\left(k-q\right)^{2}+2\,a^{2}\,b\,\left(k^{2}\right)^{2}\left(q^{2}\right)^{2}-2\,a^{2}\,b\,k^{2}\left(q^{2}\right)^{3}\right\} \nonumber \\
 & \times\frac{1}{\left(k^{2}\right)^{2}\left(q^{2}\right)^{3}\left[\left(q-k\right)^{2}\right]^{2}}\,,
\end{align}
using Eqs.\,(\ref{eq:Int 7}) and\,(\ref{eq: Int 17}), then adding
$\mathcal{S}_{\left(\overline{\Phi}\Phi\right)^{2}}^{\left(D38-a\right)}$
to $\mathcal{S}_{\left(\overline{\Phi}\Phi\right)^{2}}^{\left(D38-h\right)}$
with the values in Table\,\ref{tab:S-PPPP-38}, we find
\begin{align}
\mathcal{S}_{\left(\overline{\Phi}\Phi\right)^{2}}^{\left(D38\right)} & =0\,.\label{eq:S-D38}
\end{align}

\begin{center}
\begin{figure}
\begin{centering}
\subfloat[]{\centering{}\includegraphics[scale=0.5]{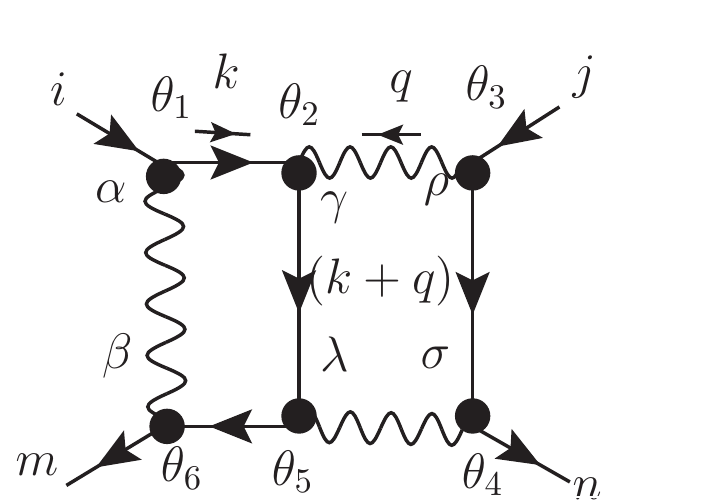}}\subfloat[]{\centering{}\includegraphics[scale=0.5]{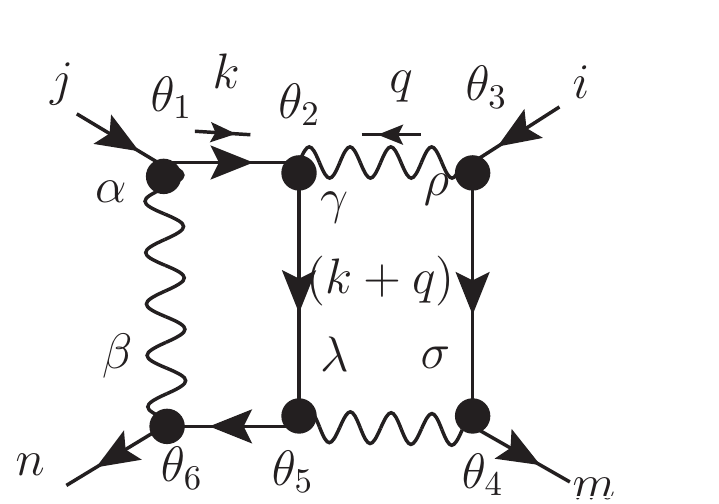}}\subfloat[]{\centering{}\includegraphics[scale=0.5]{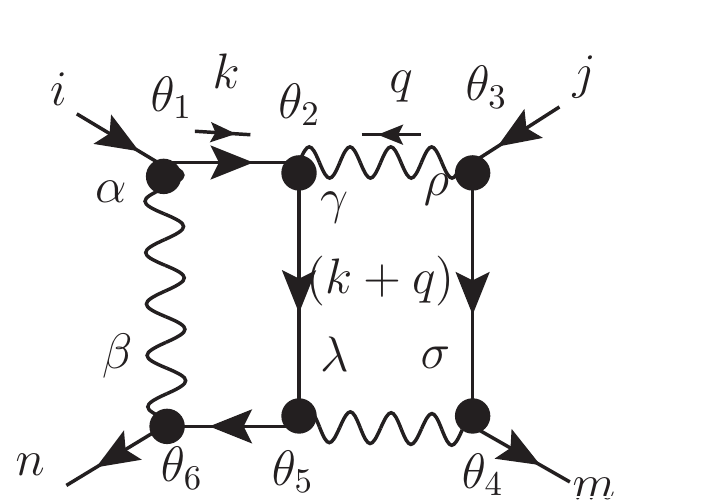}}\subfloat[]{\centering{}\includegraphics[scale=0.5]{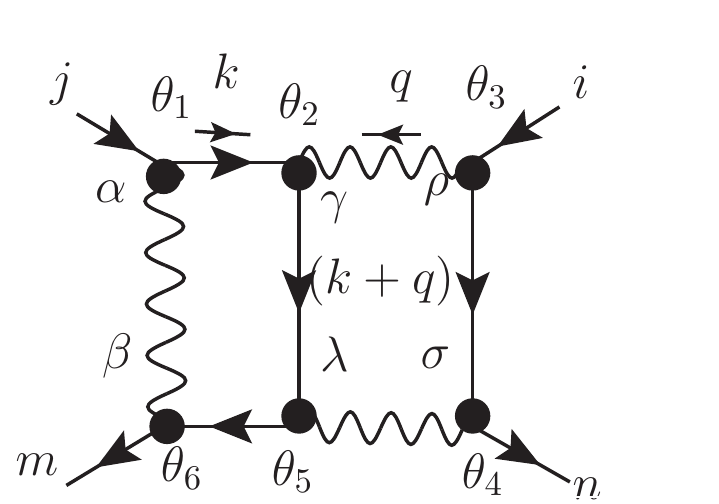}}
\par\end{centering}
\begin{centering}
\subfloat[]{\centering{}\includegraphics[scale=0.5]{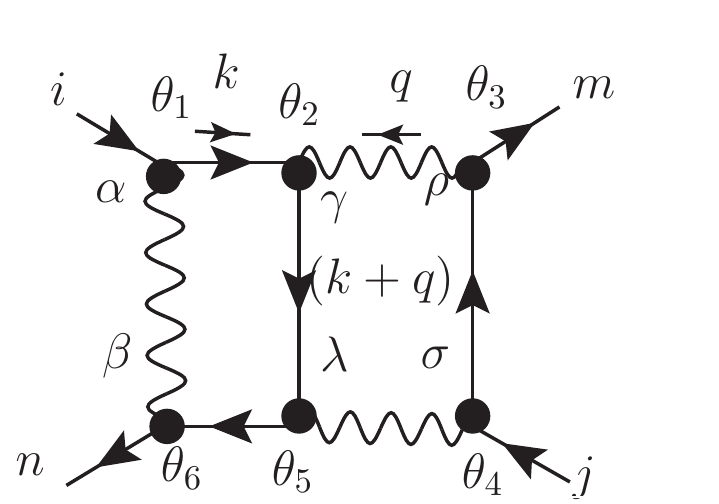}}\subfloat[]{\centering{}\includegraphics[scale=0.5]{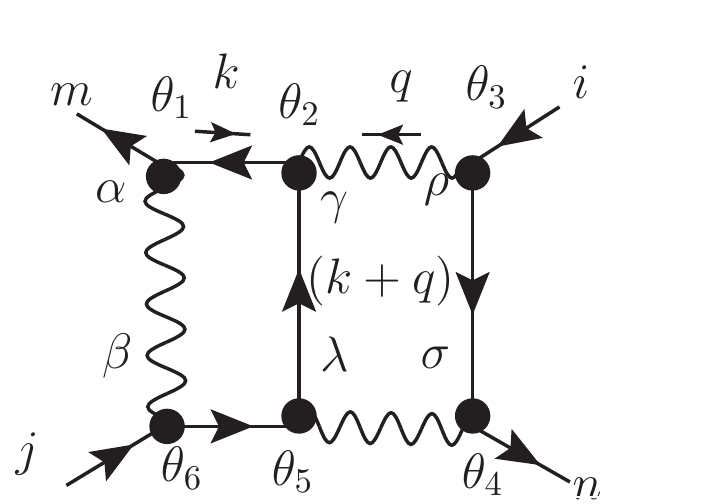}}\subfloat[]{\centering{}\includegraphics[scale=0.5]{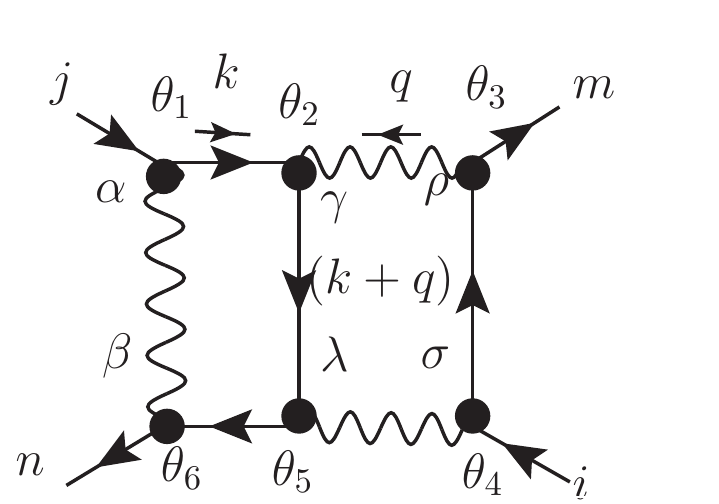}}\subfloat[]{\centering{}\includegraphics[scale=0.5]{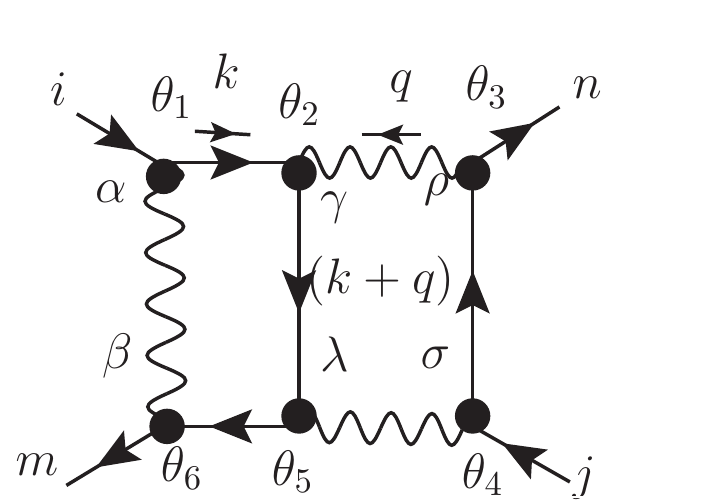}}
\par\end{centering}
\centering{}\caption{\label{fig:D39-order-g-6}$\mathcal{S}_{\left(\overline{\Phi}\Phi\right)^{2}}^{\left(D39\right)}$}
\end{figure}
\par\end{center}

$\mathcal{S}_{\left(\overline{\Phi}\Phi\right)^{2}}^{\left(D39-a\right)}$
in the Figure\,\ref{fig:D39-order-g-6} is
\begin{align}
\mathcal{S}_{\left(\overline{\Phi}\Phi\right)^{2}}^{\left(D39-a\right)} & =-\frac{1}{64\left(8\right)}\,i\,\delta_{im}\,\delta_{jn}\,g^{6}\int_{\theta}\overline{\Phi}_{i}\Phi_{m}\Phi_{n}\overline{\Phi}_{j}\,\int\frac{d^{D}kd^{D}q}{\left(2\pi\right)^{2D}}\left\{ 16\left(a-b\right)^{3}\left(k\cdot q\right)\left(k^{2}\right)^{2}q^{2}\right.\nonumber \\
 & +\left(8\,a^{3}-22\,a^{2}\,b+24\,a\,b^{2}-8\,b^{3}\right)\left(k^{2}\right)^{2}\left(q^{2}\right)^{2}-4\,a^{2}\,b\left(k\cdot q\right)^{2}k^{2}q^{2}\nonumber \\
 & \left.+a^{2}\,b\,q_{\alpha\beta}q_{\gamma\delta}k^{\alpha\gamma}k^{\beta\delta}k^{2}q^{2}\right\} \frac{1}{\left(k^{2}\right)^{3}\left(k+q\right)^{2}\left(q^{2}\right)^{3}}\,,
\end{align}
using Eqs.\,(\ref{eq:Int 7}), (\ref{eq:Int 9}) and\,(\ref{eq: Int 13}),
then adding $\mathcal{S}_{\left(\overline{\Phi}\Phi\right)^{2}}^{\left(D39-a\right)}$
to $\mathcal{S}_{\left(\overline{\Phi}\Phi\right)^{2}}^{\left(D39-h\right)}$,
we find
\begin{align}
\mathcal{S}_{\left(\overline{\Phi}\Phi\right)^{2}}^{\left(D39-a\right)} & =\mathcal{S}_{\left(\overline{\Phi}\Phi\right)^{2}}^{\left(D39\right)}=0\,.\label{eq:S-D39}
\end{align}

\begin{center}
\begin{figure}
\begin{centering}
\subfloat[]{\centering{}\includegraphics[scale=0.5]{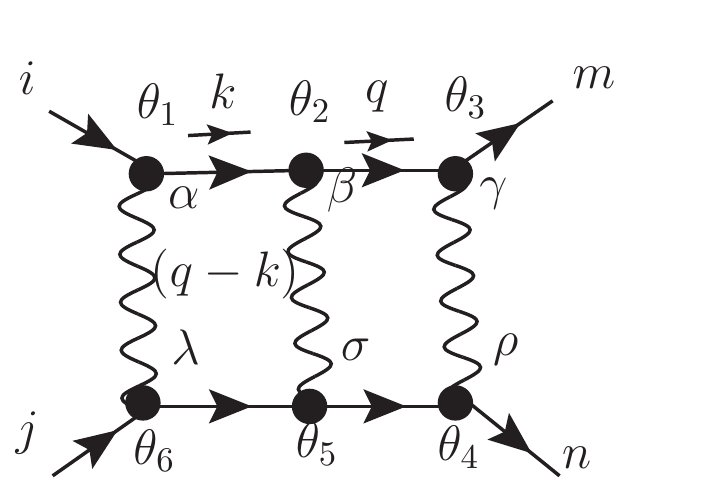}}\subfloat[]{\centering{}\includegraphics[scale=0.5]{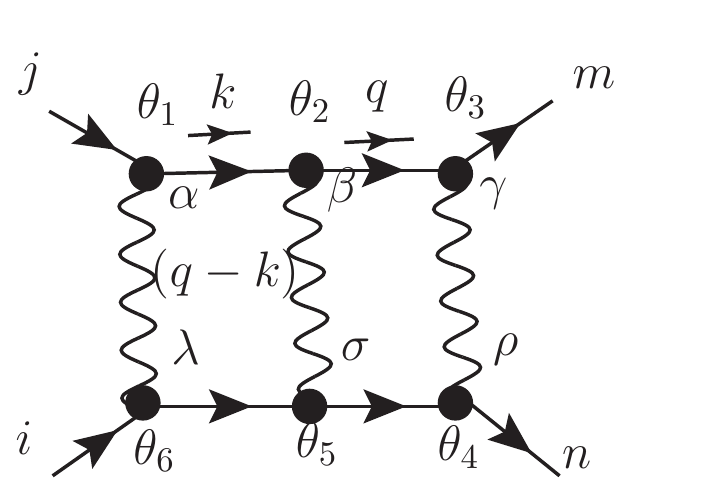}}\subfloat[]{\centering{}\includegraphics[scale=0.5]{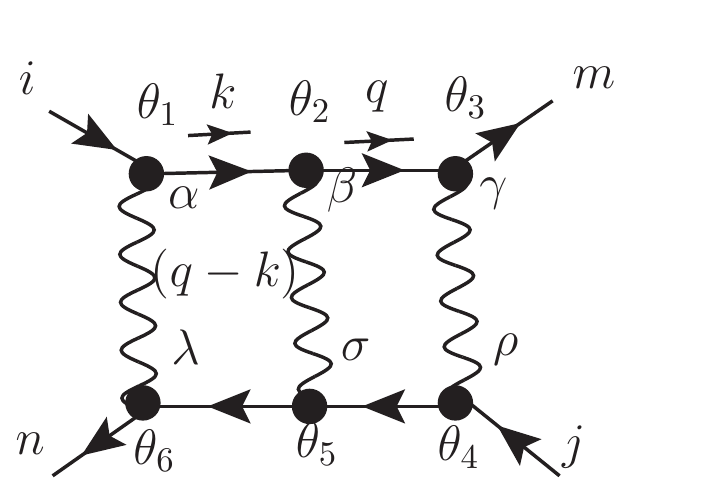}}\subfloat[]{\centering{}\includegraphics[scale=0.5]{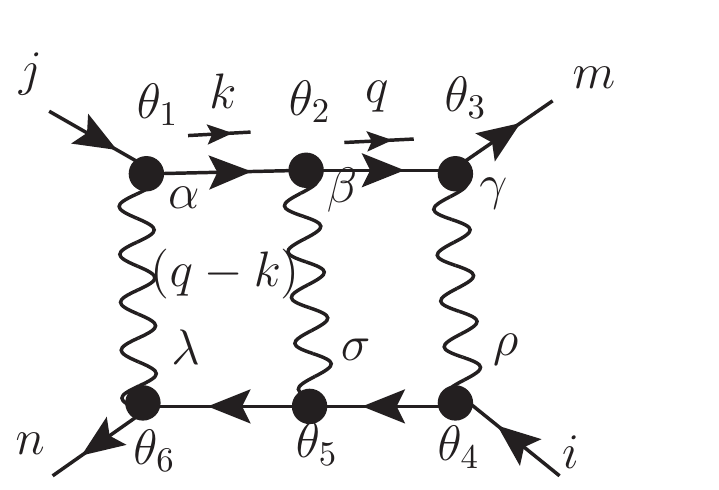}}
\par\end{centering}
\centering{}\caption{\label{fig:D40-order-g-6}$\mathcal{S}_{\left(\overline{\Phi}\Phi\right)^{2}}^{\left(D40\right)}$}
\end{figure}
\par\end{center}

\begin{center}
\begin{table}
\centering{}%
\begin{tabular}{lcccccccccc}
 &  &  &  &  &  &  &  &  &  & \tabularnewline
\hline 
\hline 
$D40-a$ &  & $-\delta_{im}\delta_{jn}$ &  & $D40-b$ &  & $-\delta_{jm}\delta_{in}$ &  & $D40-c$ &  & $\delta_{im}\delta_{jn}$\tabularnewline
$D40-d$ &  & $\delta_{jm}\delta_{in}$ &  &  &  &  &  &  &  & \tabularnewline
\hline 
\hline 
 &  &  &  &  &  &  &  &  &  & \tabularnewline
\end{tabular}\caption{\label{tab:S-PPPP-40}Values of the diagrams in Figure\,\ref{fig:D40-order-g-6}
with common factor\protect \\
 $\frac{1}{16}\left(\frac{a\left(a-b\right)^{2}}{32\pi^{2}\epsilon}\right)\,i\,g^{6}\int_{\theta}\overline{\Phi}_{i}\Phi_{m}\Phi_{n}\overline{\Phi}_{j}$
.}
\end{table}
\par\end{center}

$\mathcal{S}_{\left(\overline{\Phi}\Phi\right)^{2}}^{\left(D40-a\right)}$
in the Figure\,\ref{fig:D40-order-g-6} is
\begin{align}
\mathcal{S}_{\left(\overline{\Phi}\Phi\right)^{2}}^{\left(D40-a\right)} & =-\frac{1}{64\left(8\right)}\,i\,\delta_{im}\delta_{jn}\,g^{6}\int_{\theta}\overline{\Phi}_{i}\Phi_{m}\Phi_{n}\overline{\Phi}_{j}\int\frac{d^{D}kd^{D}q}{\left(2\pi\right)^{2D}}\left\{ \frac{-32\,a\left(a-b\right)^{2}\left(k^{2}\right)^{2}\left(q^{2}\right)^{2}}{\left(k^{2}\right)^{3}\left(q-k\right)^{2}\left(q^{2}\right)^{3}}\right\} \,,
\end{align}
using Eq.\,(\ref{eq:Int 7}), then adding $\mathcal{S}_{\left(\overline{\Phi}\Phi\right)^{2}}^{\left(D40-a\right)}$
to $\mathcal{S}_{\left(\overline{\Phi}\Phi\right)^{2}}^{\left(D40-d\right)}$
with the values in Table\,\ref{tab:S-PPPP-40}, we find
\begin{align}
\mathcal{S}_{\left(\overline{\Phi}\Phi\right)^{2}}^{\left(D40\right)} & =0\,.\label{eq:S-D40}
\end{align}

\begin{center}
\begin{figure}
\begin{centering}
\subfloat[]{\centering{}\includegraphics[scale=0.5]{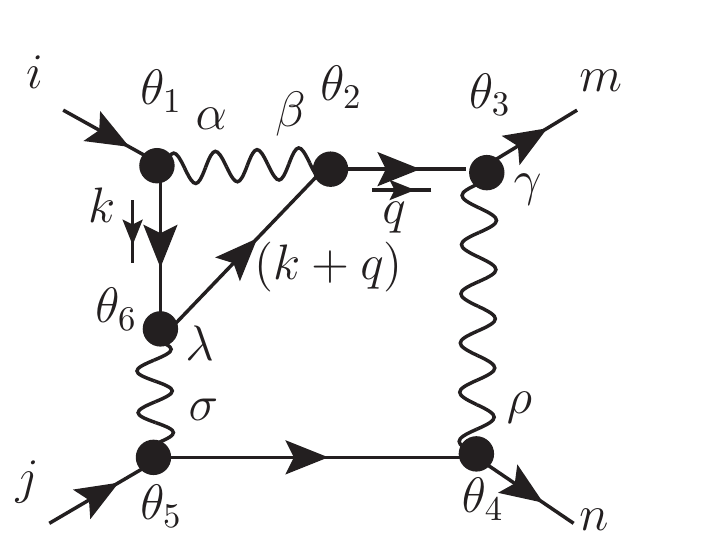}}\subfloat[]{\centering{}\includegraphics[scale=0.5]{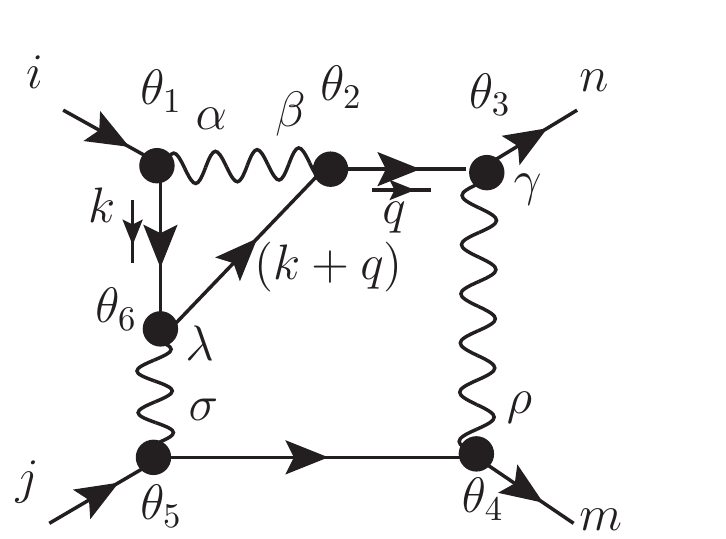}}\subfloat[]{\centering{}\includegraphics[scale=0.5]{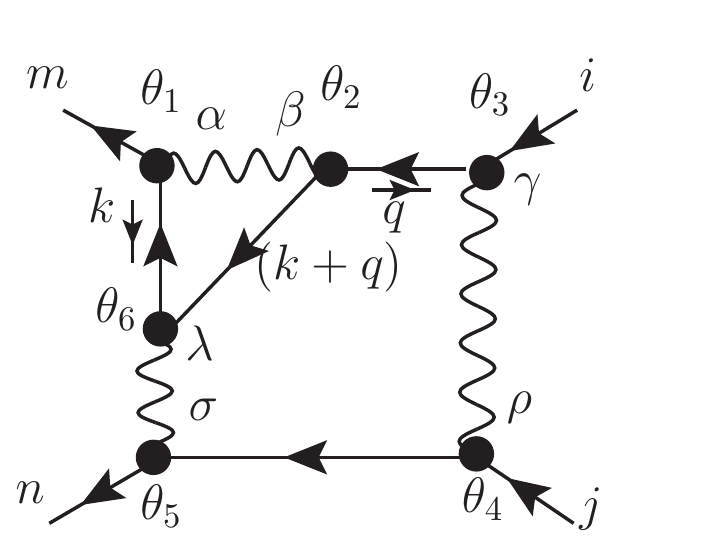}}\subfloat[]{\centering{}\includegraphics[scale=0.5]{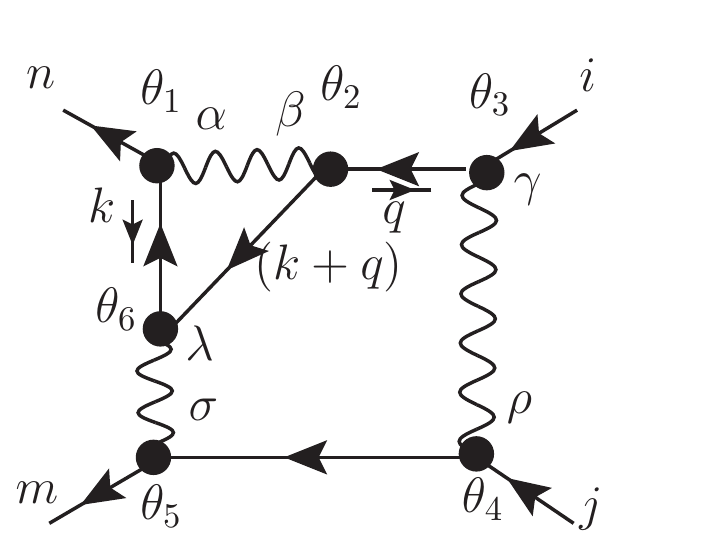}}
\par\end{centering}
\begin{centering}
\subfloat[]{\centering{}\includegraphics[scale=0.5]{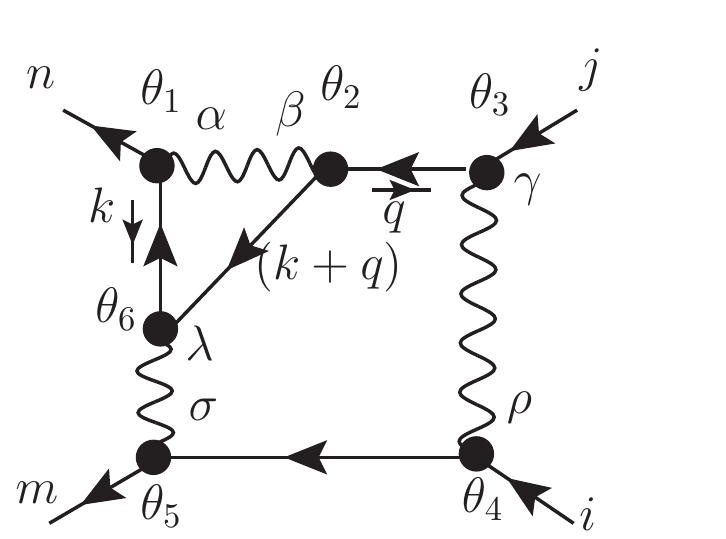}}\subfloat[]{\centering{}\includegraphics[scale=0.5]{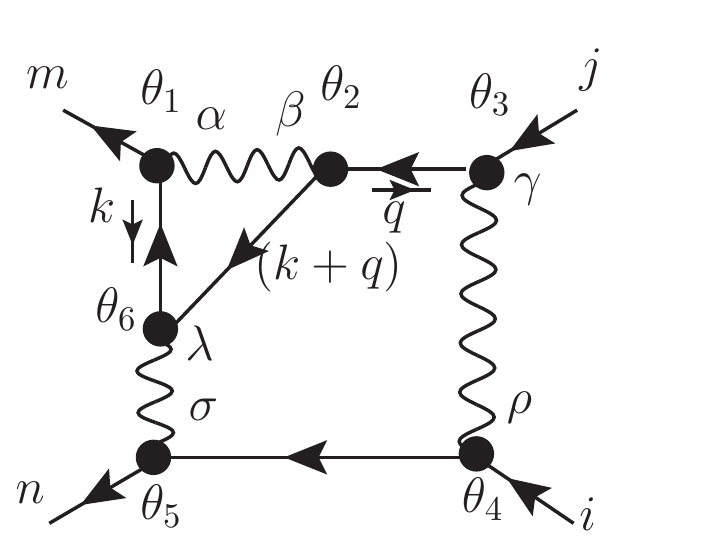}}\subfloat[]{\centering{}\includegraphics[scale=0.5]{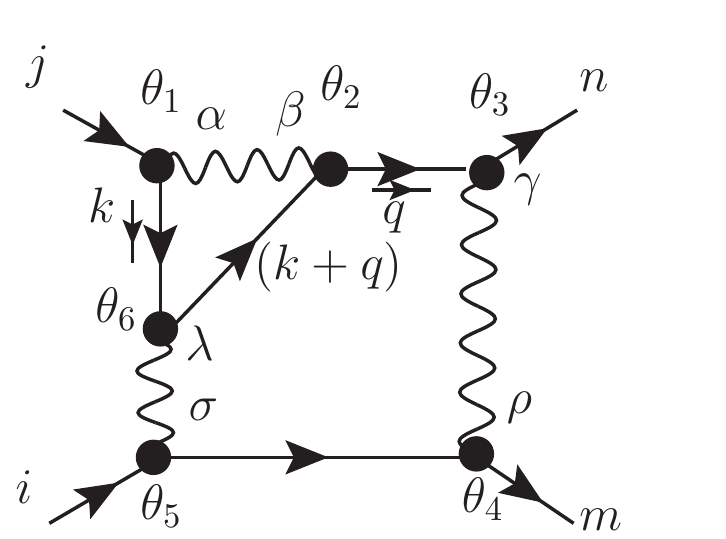}}\subfloat[]{\centering{}\includegraphics[scale=0.5]{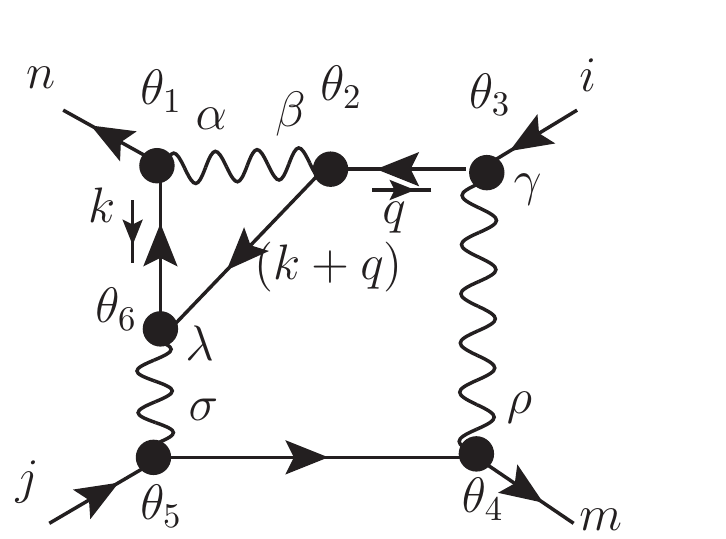}}
\par\end{centering}
\begin{centering}
\subfloat[]{\centering{}\includegraphics[scale=0.5]{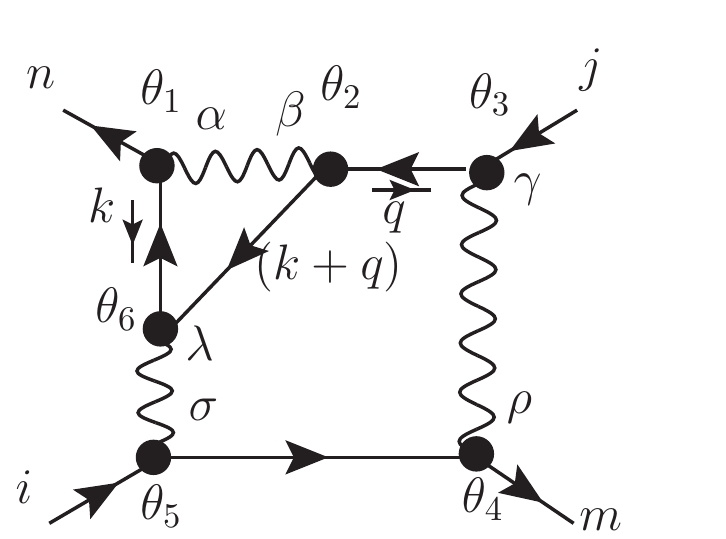}}\subfloat[]{\centering{}\includegraphics[scale=0.5]{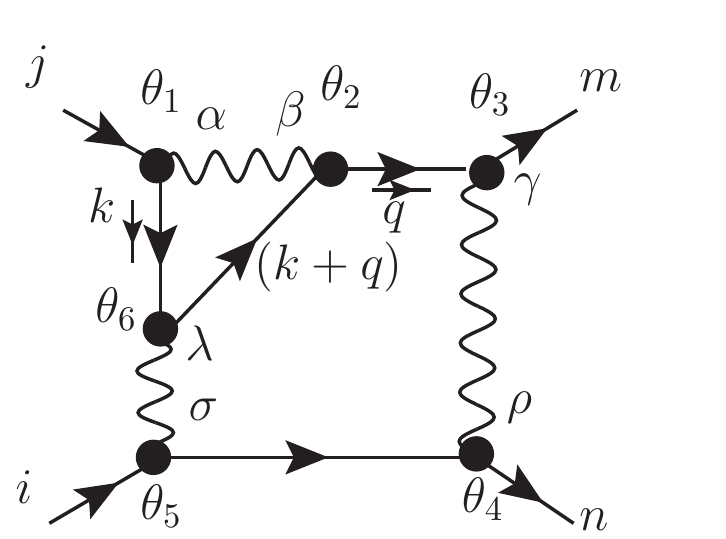}}\subfloat[]{\centering{}\includegraphics[scale=0.5]{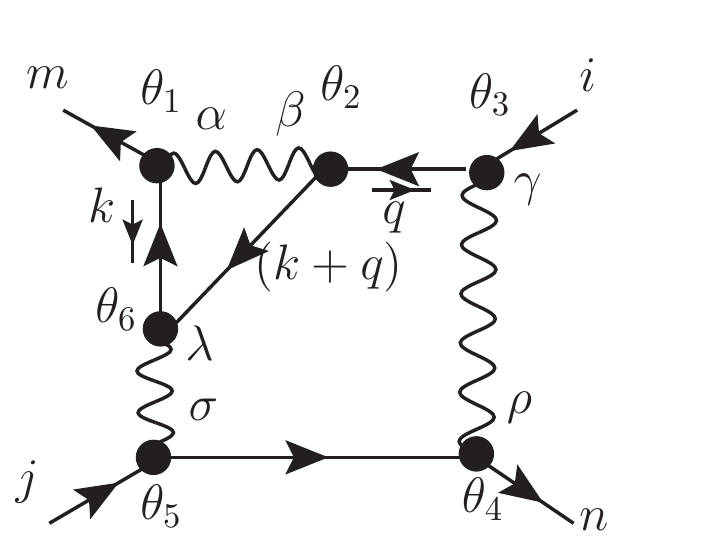}}\subfloat[]{\centering{}\includegraphics[scale=0.5]{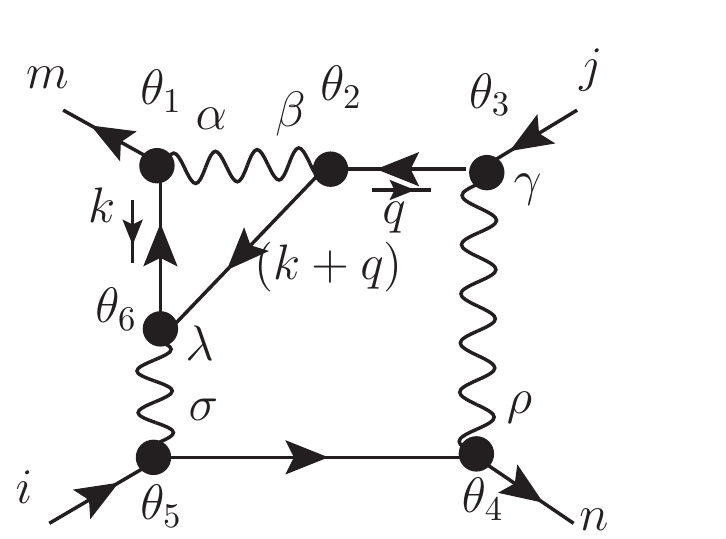}}
\par\end{centering}
\begin{centering}
\subfloat[]{\centering{}\includegraphics[scale=0.5]{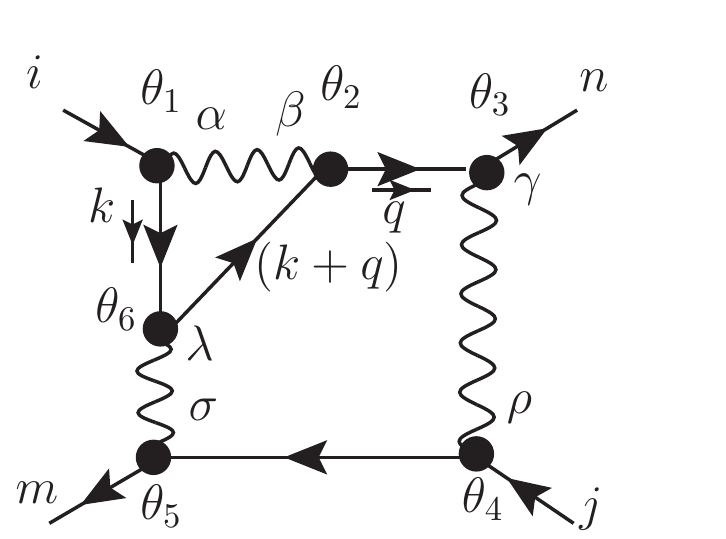}}\subfloat[]{\centering{}\includegraphics[scale=0.5]{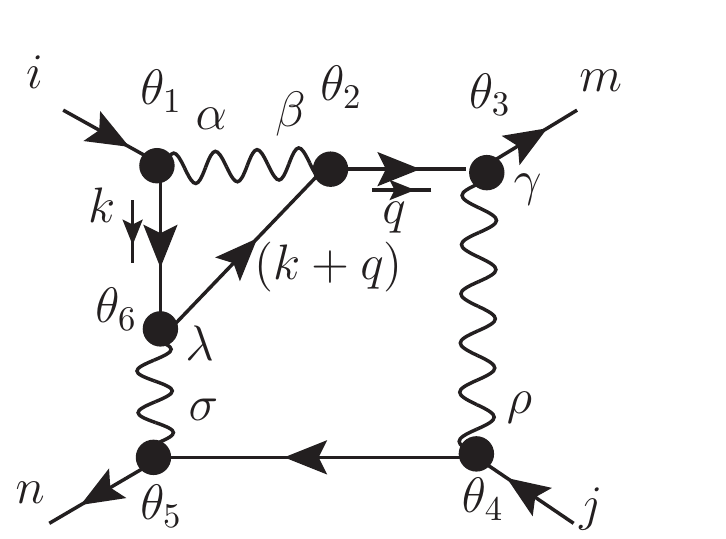}}\subfloat[]{\centering{}\includegraphics[scale=0.5]{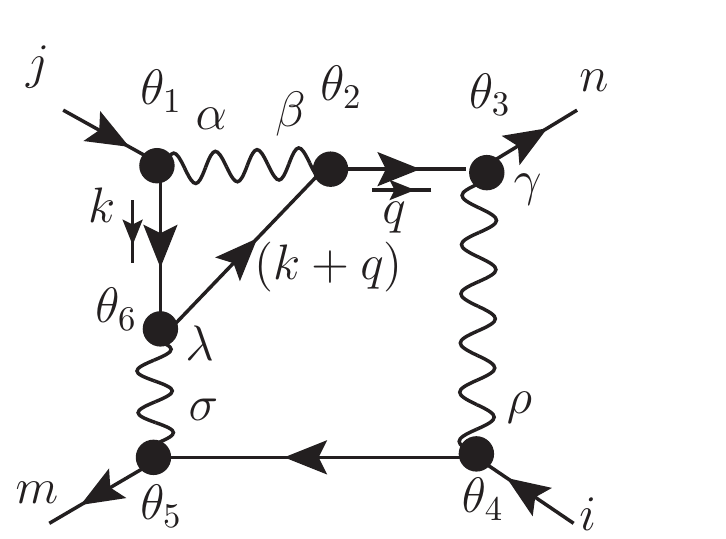}}\subfloat[]{\centering{}\includegraphics[scale=0.5]{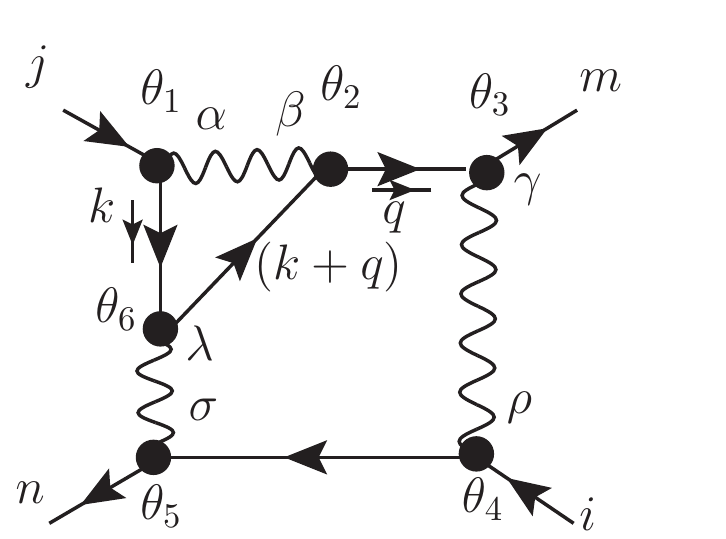}}
\par\end{centering}
\centering{}\caption{\label{fig:D41-order-g-6}$\mathcal{S}_{\left(\overline{\Phi}\Phi\right)^{2}}^{\left(D41\right)}$}
\end{figure}
\par\end{center}

\begin{center}
\begin{table}
\centering{}%
\begin{tabular}{lcccccccccc}
 &  &  &  &  &  &  &  &  &  & \tabularnewline
\hline 
\hline 
$D41-a$ &  & $\delta_{im}\,\delta_{nj}$ &  & $D41-b$ &  & $\delta_{jm}\,\delta_{ni}$ &  & $D41-c$ &  & $\delta_{im}\,\delta_{nj}$\tabularnewline
$D41-d$ &  & $\delta_{jm}\,\delta_{ni}$ &  & $D41-e$ &  & $\delta_{im}\,\delta_{nj}$ &  & $D41-f$ &  & $\delta_{jm}\,\delta_{ni}$\tabularnewline
$D41-g$ &  & $\delta_{im}\,\delta_{nj}$ &  & $D41-h$ &  & $\delta_{jm}\,\delta_{ni}$ &  & $D41-i$ &  & $\delta_{im}\,\delta_{nj}$\tabularnewline
$D41-j$ &  & $\delta_{jm}\,\delta_{ni}$ &  & $D41-k$ &  & $\delta_{im}\,\delta_{nj}$ &  & $D41-l$ &  & $\delta_{jm}\,\delta_{ni}$\tabularnewline
$D41-m$ &  & $\delta_{jm}\,\delta_{ni}$ &  & $D41-n$ &  & $\delta_{im}\,\delta_{nj}$ &  & $D41-o$ &  & $\delta_{im}\,\delta_{nj}$\tabularnewline
$D41-p$ &  & $\delta_{jm}\,\delta_{ni}$ &  &  &  &  &  &  &  & \tabularnewline
\hline 
\hline 
 &  &  &  &  &  &  &  &  &  & \tabularnewline
\end{tabular}\caption{\label{tab:S-PPPP-41}Values of the diagrams in Figure\,\ref{fig:D41-order-g-6}
with common factor\protect \\
 $\frac{1}{64}\left(\frac{\left(b-a\right)^{3}}{32\pi^{2}\epsilon}\right)\,i\,g^{6}\int_{\theta}\overline{\Phi}_{i}\Phi_{m}\Phi_{n}\overline{\Phi}_{j}$
.}
\end{table}
\par\end{center}

$\mathcal{S}_{\left(\overline{\Phi}\Phi\right)^{2}}^{\left(D41-a\right)}$
in the Figure\,\ref{fig:D41-order-g-6} is
\begin{align}
\mathcal{S}_{\left(\overline{\Phi}\Phi\right)^{2}}^{\left(D41-a\right)} & =-\frac{1}{64\left(8\right)}\,i\,\delta_{im}\,\delta_{nj}\,g^{6}\int_{\theta}\overline{\Phi}_{i}\Phi_{m}\Phi_{n}\overline{\Phi}_{j}\,\int\frac{d^{D}kd^{D}q}{\left(2\pi\right)^{2D}}\left\{ 4\left(2\,a^{3}-a^{2}\,b\right)\left(k\cdot q\right)^{2}\left(q^{2}\right)^{2}\right.\nonumber \\
 & +\left(8\,b^{3}-12\,a^{3}+26\,a^{2}\,b-24\,a\,b^{2}\right)k^{2}\left(q^{2}\right)^{3}-a^{3}\,q_{\alpha\beta}q_{\gamma\delta}k^{\alpha\delta}k^{\beta\gamma}\nonumber \\
 & \left.+\left(a^{2}\,b-a^{3}\right)\,k_{\alpha\beta}k_{\gamma\delta}q^{\alpha\delta}q^{\beta\gamma}\right\} \frac{1}{\left(k^{2}\right)^{2}\left(k+q\right)^{2}\left(q^{2}\right)^{4}}\,,
\end{align}
using Eqs.\,(\ref{eq:Int 7}) and\,(\ref{eq: Int 13}), then adding
$\mathcal{S}_{\left(\overline{\Phi}\Phi\right)^{2}}^{\left(D41-a\right)}$
to $\mathcal{S}_{\left(\overline{\Phi}\Phi\right)^{2}}^{\left(D41-p\right)}$
with the values in Table\,\ref{tab:S-PPPP-41}, we find 
\begin{align}
\mathcal{S}_{\left(\overline{\Phi}\Phi\right)^{2}}^{\left(D41\right)} & =\frac{1}{8}\left(\frac{\left(b-a\right)^{3}}{32\pi^{2}\epsilon}\right)\,i\,g^{6}\left(\delta_{im}\,\delta_{nj}+\delta_{jm}\,\delta_{ni}\right)\int_{\theta}\overline{\Phi}_{i}\Phi_{m}\Phi_{n}\overline{\Phi}_{j}\,.\label{eq:S-D41}
\end{align}

\begin{center}
\myclearpage
\par\end{center}

\chapter{\label{chap:Useful-integrals}Useful integrals}

In this appendix we compute a set of general integrals that will be
needed in the computation of the Feynman superdiagrams considered
in this thesis.

\section{One loop integrals}

We will start by considering some massive integrals, which will be
the base for the derivation of the massless integrals that we will
need, as shown in the next subsection. 

First we quote the one loop integral in Minkowski space-time\,\cite{Tan:1997ew},
\begin{align}
\mathcal{I}_{1}\left(A,p,m\right) & =\int\frac{d^{D}l}{\left(2\pi\right)^{D}}\frac{1}{\left[l^{2}+m^{2}+x\left(2p\cdot l+p^{2}\right)\right]^{A}}\nonumber \\
 & =i\,\mu^{\epsilon}\frac{\Gamma\left(A-\frac{D}{2}\right)}{\left(4\pi\right)^{D/2}\Gamma\left(A\right)}\left[x\left(1-x\right)p^{2}+m^{2}\right]^{\frac{D}{2}-A},\label{eq:Int 1}
\end{align}
where $d^{D}l\equiv\mu^{\epsilon}d^{3-\epsilon}l$ and $D=3-\epsilon$.
Using this result we can obtain the set of integrals that will be
needed in this thesis. For convenience, we will look for expressions
where the momenta with uncontracted indices always appearing as bispinors.
Let us start with the definitions,
\begin{eqnarray}
p^{2}\equiv\frac{1}{2}\,p^{\alpha\beta}p_{\alpha\beta}\,, &  & p\cdot l\equiv\frac{1}{2}\,p^{\alpha\beta}l_{\alpha\beta}\,.
\end{eqnarray}
Using the properties in Eqs.\,(\ref{eq:S2}) and\,(\ref{eq:S3}),
we have 
\begin{align}
\frac{\partial}{\partial p^{\alpha\beta}}\,p^{\theta\sigma}=\partial_{\alpha\beta}\,p^{\theta\sigma} & =\frac{1}{2}\left(\delta_{\alpha}^{\theta}\delta_{\beta}^{\sigma}+\delta_{\beta}^{\theta}\delta_{\alpha}^{\sigma}\right)\,.\label{eq:dev 1}
\end{align}
Then, by derivation with respect to $p$ in Eq.\,(\ref{eq:Int 1}),
we obtain
\begin{align}
\frac{\partial}{\partial p^{\alpha\beta}}\left(\int\frac{d^{D}l}{\left(2\pi\right)^{D}}\left\{ \frac{1}{\left[l^{2}+m^{2}+x\left(2p\cdot l+p^{2}\right)\right]^{A}}\right\} \right. & =\left.i\,\frac{\Gamma\left(A-\frac{D}{2}\right)}{\left(4\pi\right)^{D/2}\Gamma\left(A\right)}\left[x\left(1-x\right)p^{2}+m^{2}\right]^{\frac{D}{2}-A}\right)\label{eq:dev int 1}
\end{align}
where
\begin{align}
\frac{\partial}{\partial p^{\alpha\beta}}\left\{ \frac{1}{\left[l^{2}+m^{2}+x\left(2p\cdot l+p^{2}\right)\right]^{A}}\right\}  & =-A\,x\left[l^{2}+m^{2}+x\left(2p\cdot l+p^{2}\right)\right]^{-\left(A+1\right)}\left(l_{\theta\sigma}+p_{\theta\sigma}\right),\label{eq:dev esquerda}
\end{align}
and
\begin{align}
\frac{\partial}{\partial p^{\alpha\beta}}\left\{ \left[x\left(1-x\right)p^{2}+m^{2}\right]^{\frac{D}{2}-A}\right\}  & =x\left(1-x\right)p_{\alpha\beta}\left(\frac{D}{2}-A\right)\left[x\left(1-x\right)p^{2}+m^{2}\right]^{\frac{D}{2}-A-1}.\label{eq:dev direita}
\end{align}
Substituting Eqs.\,(\ref{eq:dev esquerda}) and\,(\ref{eq:dev direita})
in Eq.\,(\ref{eq:dev int 1}) and using Eq.\,(\ref{eq:Int 1}),
we obtain,
\begin{align}
\mathcal{I}_{2}\left(A,p,m\right) & =\int\frac{d^{D}l}{\left(2\pi\right)^{D}}\frac{l_{\alpha\beta}}{\left[l^{2}+m^{2}+x\left(2p\cdot l+p^{2}\right)\right]^{A}}\nonumber \\
 & =i\,\mu^{\epsilon}\frac{\Gamma\left(A-\frac{D}{2}\right)}{\left(4\pi\right)^{D/2}\Gamma\left(A\right)}\left\{ \frac{\left(-x\right)\,p_{\alpha\beta}}{\left[x\left(1-x\right)p^{2}+m^{2}\right]^{A-D/2}}\right\} =-x\,p_{\alpha\beta}\,\mathcal{I}_{1}\left(A,p,m\right),\label{eq:Int 2}
\end{align}
where it was used that 
\begin{equation}
A\,\Gamma\left(A\right)=\Gamma\left(A+1\right)\,.\label{eq:Prop-Gamma}
\end{equation}
Using a similar procedure, it is possible to obtain the following
set of integrals:
\begin{align}
\mathcal{I}_{3}\left(A,p,m\right)= & \int\frac{d^{D}l}{\left(2\pi\right)^{D}}\frac{l_{\mu\nu}\,l_{\alpha\beta}}{\left(l^{2}+m^{2}+x\left(2\,l\cdot p+p^{2}\right)\right)^{A}}=x^{2}\,p_{\mu\nu}\,p_{\alpha\beta}\mathcal{I}_{1}\left(A,p,m\right)\nonumber \\
 & +\frac{1}{2\left(A-1\right)}\left(C_{\mu\alpha}C_{\nu\beta}+C_{\nu\alpha}C_{\mu\beta}\right)\mathcal{I}_{1}\left(A-1,p,m\right)\label{eq:Int 3}
\end{align}
\begin{align}
\mathcal{I}_{4}\left(A,p,m\right) & =\int\frac{d^{D}l}{\left(2\pi\right)^{D}}\frac{l_{\mu\nu}\,l_{\rho\sigma}\,l_{\alpha\beta}}{\left(l^{2}+m^{2}+x\left(2\,l\cdot p+p^{2}\right)\right)^{A}}=-x^{3}\,p_{\mu\nu}\,p_{\rho\sigma}\,p_{\alpha\beta}\,\mathcal{I}_{1}\left(A,p,m\right)\nonumber \\
 & -\frac{x}{2\left(A-1\right)}\Biggl\{\left(C_{\mu\alpha}C_{\nu\beta}+C_{\nu\alpha}C_{\mu\beta}\right)p_{\rho\sigma}+\left(C_{\rho\mu}C_{\sigma\nu}+C_{\sigma\mu}C_{\rho\nu}\right)p_{\alpha\beta}\nonumber \\
 & +\left(C_{\rho\alpha}C_{\sigma\beta}+C_{\sigma\alpha}C_{\rho\beta}\right)p_{\mu\nu}\Biggr\}\mathcal{I}_{1}\left(A-1,p,m\right),\label{eq:Int 4}
\end{align}
\begin{align}
\mathcal{I}_{5}\left(A,p,m\right) & =\int\frac{d^{D}l}{\left(2\pi\right)^{D}}\frac{l_{\mu\nu}\,l_{\rho\sigma}\,l_{\alpha\beta}\,l_{\theta\lambda}}{\left(l^{2}+m^{2}+x\left(2\,l\cdot p+p^{2}\right)\right)^{A}}=x^{4}\,p_{\mu\nu}\,p_{\rho\sigma}\,p_{\alpha\beta}\,p_{\theta\lambda}\mathcal{I}_{1}\left(A,p,m\right)\nonumber \\
 & +\frac{x^{2}}{2\left(A-1\right)}\Biggl\{\left(C_{\mu\alpha}C_{\nu\beta}+C_{\nu\alpha}C_{\mu\beta}\right)p_{\rho\sigma}\,p_{\theta\lambda}+\left(C_{\rho\mu}C_{\sigma\nu}+C_{\sigma\mu}C_{\rho\nu}\right)p_{\alpha\beta}\,p_{\theta\lambda}\nonumber \\
 & +\left(C_{\rho\alpha}C_{\sigma\beta}+C_{\sigma\alpha}C_{\rho\beta}\right)p_{\mu\nu}\,p_{\theta\lambda}+\left(C_{\theta\mu}C_{\lambda\nu}+C_{\lambda\mu}C_{\theta\nu}\right)p_{\rho\sigma}\,p_{\alpha\beta}\nonumber \\
 & +\left(C_{\theta\alpha}C_{\lambda\beta}+C_{\lambda\alpha}C_{\theta\beta}\right)p_{\mu\nu}\,p_{\rho\alpha}+\left(C_{\theta\rho}C_{\lambda\sigma}+C_{\lambda\rho}C_{\theta\sigma}\right)p_{\mu\nu}\,p_{\alpha\beta}\Biggr\}\mathcal{I}_{1}\left(A-1,p,m\right)\nonumber \\
 & +\frac{1}{4\,\left(A-1\right)\left(A-2\right)}\Biggl\{\left(C_{\mu\alpha}C_{\nu\beta}+C_{\nu\alpha}C_{\mu\beta}\right)\left(C_{\theta\rho}C_{\lambda\sigma}+C_{\lambda\rho}C_{\theta\sigma}\right)\nonumber \\
 & +\left(C_{\rho\mu}C_{\sigma\beta}+C_{\sigma\mu}C_{\rho\nu}\right)\left(C_{\theta\alpha}C_{\lambda\beta}+C_{\lambda\alpha}C_{\theta\beta}\right)\nonumber \\
 & +\left(C_{\rho\alpha}C_{\sigma\beta}+C_{\sigma\alpha}C_{\rho\beta}\right)\left(C_{\theta\mu}C_{\lambda\nu}+C_{\lambda\mu}C_{\theta\nu}\right)\Biggr\}\mathcal{I}_{1}\left(A-2,p,m\right),\label{eq:Int 5}
\end{align}

\section{Two loops massless integrals\label{sec:Two-loops-massless integrals}}

In this section we obtain a set of integrals that will be needed in
our calculations of the renormalization group functions. Therefore,
we present only the divergent parts of the integrals, and also the
mass parameter is chosen to be zero. Let us start with 
\begin{align*}
\mathcal{I}_{6} & =\int\frac{d^{D}kd^{D}q}{\left(2\pi\right)^{2D}}\frac{q_{\alpha\beta}}{\left(k+p\right)^{2}\left(q+k\right)^{2}q^{2}}=\int_{0}^{1}dx\int\frac{d^{D}kd^{D}q}{\left(2\pi\right)^{2D}}\frac{q_{\alpha\beta}}{\left(k+p\right)^{2}\left[q^{2}x+\left(1-x\right)\left(q+k\right)^{2}\right]^{2}}\\
 & =\int_{0}^{1}dx\int\frac{d^{D}kd^{D}q}{\left(2\pi\right)^{2D}}\frac{q_{\alpha\beta}}{\left(k+p\right)^{2}\left[q^{2}+\left(1-x\right)\left(2q\cdot k+k^{2}\right)\right]^{2}},
\end{align*}
using Eq.\,(\ref{eq:Int 2}), we can calculate the integration in
$q$, 
\begin{align*}
\mathcal{I}_{6} & =i\,\mu^{\epsilon}\frac{\Gamma\left(2-D/2\right)}{\left(4\pi\right)^{D/2}}\int_{0}^{1}dx\frac{\left(x-1\right)}{\left[x\left(1-x\right)\right]^{2-D/2}}\int\frac{d^{D}k}{\left(2\pi\right)^{D}}\frac{k_{\alpha\beta}}{\left(k+p\right)^{2}\left(k^{2}\right)^{2-D/2}},
\end{align*}
then using the Feynman formula, 
\begin{align}
\frac{1}{A^{\alpha}B^{\beta}...} & =\frac{\Gamma\left(\alpha+\beta+...\right)}{\Gamma\left(\alpha\right)\Gamma\left(\beta\right)...}\int_{0}^{1}dy\,dx\,...\,\frac{\delta\left(1-x-y-...\right)x^{\alpha-1}y^{\beta-1}...}{\left(Ax+By+...\right)^{\alpha+\beta+...}},\label{eq:Feyman Id}
\end{align}
we can rewrite $\mathcal{I}_{6}$ as,
\begin{align*}
\mathcal{I}_{6} & =i\,\mu^{\epsilon}\frac{\Gamma\left(3-D/2\right)}{\left(4\pi\right)^{D/2}}\int_{0}^{1}dxdy\frac{\left(x-1\right)y^{1-D/2}}{\left[x\left(1-x\right)\right]^{2-D/2}}\int\frac{d^{D}k}{\left(2\pi\right)^{D}}\frac{k_{\alpha\beta}}{\left[k^{2}+\left(1-y\right)\left(2k\cdot p+p^{2}\right)\right]^{3-D/2}}.
\end{align*}
Using again Eq.\,(\ref{eq:Int 2}) to evaluate the integration in
$k$, we obtain
\begin{align*}
\mathcal{I}_{6} & =-\mu^{2\epsilon}\frac{\Gamma\left(\epsilon\right)}{\left(4\pi\right)^{3-\epsilon}}\int_{0}^{1}dxdy\frac{\left(x-1\right)\left(y-1\right)\sqrt{y}^{\epsilon-1}}{\sqrt{x\left(1-x\right)}^{1+\epsilon}}\frac{p_{\alpha\beta}}{\left[y\left(1-y\right)p^{2}\right]^{\epsilon}}.
\end{align*}
Finally, expanding in $\epsilon$ and integrating in $x$ and $y$,
the integral $I_{6}$ can be written as 
\begin{align*}
\mathcal{I}_{6} & =-\frac{\Gamma\left(\epsilon\right)}{96\pi^{2}}\,p_{\alpha\beta}\left(1-\epsilon\left\{ \mbox{In}\left[\frac{p^{2}}{4\pi\mu^{2}}\right]-3\right\} \right),
\end{align*}
where $\Gamma\left(\epsilon\right)=\frac{1}{\epsilon}-\gamma+\ln4\pi+1$.
Therefore, 
\begin{align}
\mathcal{I}_{6}=\int\frac{d^{D}kd^{D}q}{\left(2\pi\right)^{2D}}\frac{q_{\alpha\beta}}{\left(k+p\right)^{2}\left(q+k\right)^{2}q^{2}} & =-\frac{p_{\alpha\beta}}{96\pi^{2}\epsilon},\label{eq:Int 6}
\end{align}
omitting all finite terms. Using a similar procedure it is possible
to obtain the following integrals:
\begin{align}
\mathcal{I}_{7} & =\int\frac{d^{D}kd^{D}q}{\left(2\pi\right)^{2D}}\frac{1}{\left(k+p\right)^{2}\left(q+k\right)^{2}q^{2}}=-\frac{1}{32\pi^{2}\epsilon}.\label{eq:Int 7}
\end{align}
\begin{align}
\mathcal{I}_{8} & =\int\frac{d^{D}kd^{D}q}{\left(2\pi\right)^{2D}}\frac{k_{\alpha\beta}}{\left(k+p\right)^{2}\left(q+k\right)^{2}q^{2}}=\frac{p_{\alpha\beta}}{48\pi^{2}\epsilon}.\label{eq:Int 8}
\end{align}
\begin{align}
\mathcal{I}_{9} & =\int\frac{d^{D}kd^{D}q}{\left(2\pi\right)^{2D}}\frac{k_{\alpha\beta}\,q_{\theta\lambda}}{\left(k+p\right)^{2}k^{2}\left(q+k\right)^{2}q^{2}}=\frac{1}{192\pi^{2}\epsilon}\left(C_{\alpha\theta}\,C_{\beta\lambda}+C_{\beta\theta}\,C_{\alpha\lambda}\right)\,.\label{eq:Int 9}
\end{align}
\begin{align}
\mathcal{I}_{10} & =\int\frac{d^{D}kd^{D}q}{\left(2\pi\right)^{2D}}\frac{k_{\mu\nu}\,k_{\theta\lambda}}{\left(k+p\right)^{2}k^{2}\left(k+q\right)^{2}q^{2}}=-\frac{1}{96\pi^{2}\epsilon}\left(C_{\mu\theta}\,C_{\nu\lambda}+C_{\nu\theta}\,C_{\mu\lambda}\right)\,.\label{eq:Int 10}
\end{align}
\begin{align}
\mathcal{I}_{11} & =\int\frac{d^{D}kd^{D}q}{\left(2\pi\right)^{2D}}\frac{q_{\mu\lambda}\,k_{\zeta\theta}\,k_{\kappa\rho}}{\left(k+p\right)^{2}k^{2}\left(k+q\right)^{2}q^{2}}=-\frac{1}{480\pi^{2}\epsilon}\left\{ \left(C_{\mu\zeta}\,C_{\lambda\theta}+C_{\lambda\zeta}\,C_{\mu\theta}\right)p_{\kappa\rho}\right.\nonumber \\
 & \left.+\left(C_{\zeta\kappa}\,C_{\theta\rho}+C_{\theta\kappa}\,C_{\zeta\rho}\right)p_{\mu\lambda}+\left(C_{\mu\kappa}\,C_{\lambda\rho}+C_{\lambda\kappa}\,C_{\mu\rho}\right)p_{\zeta\theta}\right\} \,.\label{eq: Int 11}
\end{align}
\begin{align}
\mathcal{I}_{12} & =\int\frac{d^{D}kd^{D}q}{\left(2\pi\right)^{2D}}\frac{k_{\lambda\theta}\,k_{\mu\nu}\,k_{\kappa\rho}\,q_{\sigma\gamma}}{\left(k+p\right)^{2}k^{2}\left(k+q\right)^{2}k^{2}q^{2}}\nonumber \\
 & =\frac{1}{960\pi^{2}\epsilon}\left\{ \left(C_{\lambda\mu}\,C_{\theta\nu}+C_{\theta\mu}\,C_{\lambda\nu}\right)\left(C_{\kappa\sigma}\,C_{\rho\gamma}+C_{\rho\sigma}\,C_{\kappa\rho}\right)+\left(C_{\lambda\kappa}\,C_{\theta\rho}+C_{\theta\kappa}\,C_{\lambda\rho}\right)\times\right.\nonumber \\
 & \left.\left(C_{\mu\sigma}\,C_{\nu\gamma}+C_{\nu\sigma}\,C_{\mu\gamma}\right)+\left(C_{\lambda\sigma}\,C_{\theta\gamma}+C_{\theta\sigma}\,C_{\lambda\gamma}\right)\left(C_{\mu\kappa}\,C_{\nu\rho}+C_{\nu\kappa}\,C_{\mu\rho}\right)\right\} \,.\label{eq: Int 12}
\end{align}

We consider another integral, which will be more complicated, to wit
\begin{align*}
\mathcal{I}_{13} & =\int\frac{d^{D}kd^{D}q}{\left(2\pi\right)^{2D}}\frac{k_{\delta\theta}\,k_{\sigma\rho}\,q_{\beta\gamma}\,q_{\alpha\lambda}}{\left(k+p\right)^{2}\left(p+q\right)^{2}\left(k-q\right)^{2}k^{2}q^{2}}\\
 & =\Gamma\left(3\right)\int_{0}^{1}dx\,dy\,y\int\frac{d^{D}kd^{D}q}{\left(2\pi\right)^{2D}}\frac{k_{\delta\theta}\,k_{\sigma\rho}\,q_{\beta\gamma}\,q_{\alpha\lambda}}{\left(k+p\right)^{2}k^{2}\left[q^{2}+2\,q\cdot a+a^{2}+c^{2}\right]^{3}}\,,
\end{align*}
with $c^{2}=b^{2}-a^{2},$ $a=\left(1-x\right)y\,p-x\,y\,k$ and $b^{2}=y\left(1-x\right)\,p^{2}+x\,y\,k^{2}$.
Using Eq.\,(\ref{eq:Int 3}), we can write
\begin{align*}
\mathcal{I}_{13} & =\frac{i\,\mu^{\epsilon}}{\left(4\pi\right)^{D/2}}\int_{0}^{1}dx\,dy\left\{ \underset{\hexstar}{\underbrace{\int\frac{d^{D}k}{\left(2\pi\right)^{D}}\frac{k_{\delta\theta}\,k_{\sigma\rho}\,k_{\beta\gamma}\,k_{\alpha\lambda}}{\left(k+p\right)^{2}k^{2}\left[k^{2}+l_{1}\,p^{2}+2\,l_{2}\,k\cdot p\right]^{3-D/2}}}}\right.\\
 & \times\frac{x^{2}\,y^{3}\,\Gamma\left(3-D/2\right)}{\left[\sqrt{x\,y\left(1-x\,y\right)}\right]^{6-D}}+\underset{\XBox}{\underbrace{\int\frac{d^{D}k}{\left(2\pi\right)^{D}}\frac{k_{\delta\theta}\,k_{\sigma\rho}}{\left(k+p\right)^{2}k^{2}\left[k^{2}+l_{1}\,p^{2}+2\,l_{2}\,k\cdot p\right]^{2-D/2}}}}\\
 & \left.\times\frac{1}{2}\left(C_{\beta\alpha}\,C_{\gamma\lambda}+C_{\gamma\alpha}\,C_{\beta\lambda}\right)\frac{y\,\Gamma\left(2-D/2\right)}{\left[\sqrt{x\,y\left(1-x\,y\right)}\right]^{4-D}}\right\} +\mathcal{O}\left(p^{2}\right)\,,
\end{align*}
where 
\begin{eqnarray*}
l_{1}=\frac{\left(1-x\right)\left[1-y\left(1-x\right)\right]}{x\left(1-x\,y\right)}\,, &  & l_{2}=\frac{y\left(1-x\right)}{1-x\,y}\,.
\end{eqnarray*}
From the first integral,
\begin{align*}
\hexstar & =\frac{i\,\mu^{\epsilon}\,\Gamma\left(3-D\right)}{4\left(4\pi\right)^{D/2}\Gamma\left(3-D/2\right)}\left\{ \left(C_{\delta\sigma}\,C_{\theta\rho}+C_{\theta\sigma}\,C_{\delta\rho}\right)\left(C_{\beta\alpha}\,C_{\gamma\lambda}+C_{\gamma\alpha}\,C_{\beta\lambda}\right)\right.\\
 & +\left(C_{\delta\beta}\,C_{\theta\gamma}+C_{\theta\beta}\,C_{\delta\gamma}\right)\left(C_{\sigma\alpha}\,C_{\rho\lambda}+C_{\rho\alpha}\,C_{\sigma\lambda}\right)+\left(C_{\delta\alpha}\,C_{\theta\lambda}+C_{\theta\alpha}\,C_{\delta\lambda}\right)\\
 & \left.\times\left(C_{\sigma\beta}\,C_{\rho\gamma}+C_{\rho\beta}\,C_{\sigma\gamma}\right)\right\} \int_{0}^{1}dz\,dw\,\frac{w\,\left[\sqrt{1-w}\right]^{4-D}}{\left[f^{2}\right]^{3-D}}+\mathcal{O}\left(p\right)\,,
\end{align*}
where $f^{2}=d^{2}-e^{2}$, $d^{2}=\left[w\left(1-z\right)+\left(1-w\right)l_{1}\right]p^{2}$
and $e=\left[w\left(1-z\right)+\left(1-w\right)l_{2}\right]p$. From
the second integral,
\begin{align*}
\XBox & =\frac{i\,\mu^{\epsilon}\,\Gamma\left(3-D\right)}{2\left(4\pi\right)^{D/2}\Gamma\left(2-D/2\right)}\left(C_{\delta\sigma}\,C_{\theta\rho}+C_{\theta\sigma}\,C_{\delta\rho}\right)\left(C_{\beta\alpha}\,C_{\gamma\lambda}+C_{\gamma\alpha}\,C_{\beta\lambda}\right)\\
 & \times\int_{0}^{1}dz\,dw\,\frac{w\,\left[\sqrt{1-w}\right]^{2-D}}{\left[f^{2}\right]^{3-D}}+\mathcal{O}\left(p\right)\,.
\end{align*}
Substituting $\hexstar$ and $\XBox$ in $\mathcal{I}_{13}$, we have
\begin{align}
\mathcal{I}_{13} & =\int\frac{d^{D}kd^{D}q}{\left(2\pi\right)^{2D}}\frac{k_{\delta\theta}\,k_{\sigma\rho}\,q_{\beta\gamma}\,q_{\alpha\lambda}}{\left(k+p\right)^{2}\left(p+q\right)^{2}\left(k-q\right)^{2}k^{2}q^{2}}=-\frac{1}{5\left(384\pi^{2}\epsilon\right)}\left\{ 6\left(C_{\delta\sigma}\,C_{\theta\rho}+C_{\theta\sigma}\,C_{\delta\rho}\right)\right.\nonumber \\
 & \times\left(C_{\beta\alpha}\,C_{\gamma\lambda}+C_{\gamma\alpha}\,C_{\beta\lambda}\right)+\left(C_{\delta\beta}\,C_{\theta\gamma}+C_{\theta\beta}\,C_{\delta\gamma}\right)\left(C_{\sigma\alpha}\,C_{\rho\lambda}+C_{\rho\alpha}\,C_{\sigma\lambda}\right)\nonumber \\
 & \left.+\left(C_{\delta\alpha}\,C_{\theta\lambda}+C_{\theta\alpha}\,C_{\delta\lambda}\right)\left(C_{\sigma\beta}\,C_{\rho\gamma}+C_{\rho\beta}\,C_{\sigma\gamma}\right)\right\} .\label{eq: Int 13}
\end{align}
By similar means, we can find the following set of integrals: 
\begin{align}
\mathcal{I}_{14} & =\int\frac{d^{D}kd^{D}q}{\left(2\pi\right)^{2D}}\frac{q_{\lambda\theta}\,k_{\beta\rho}}{\left(k+p\right)^{2}\left(p+q\right)^{2}\left(k-q\right)^{2}q^{2}}=\int\frac{d^{D}kd^{D}q}{\left(2\pi\right)^{2D}}\frac{q_{\lambda\theta}\,k_{\beta\rho}}{\left(k+p\right)^{2}\left(p+q\right)^{2}\left(k-q\right)^{2}k^{2}}\nonumber \\
 & =-\frac{1}{192\pi^{2}\epsilon}\left(C_{\lambda\beta}\,C_{\theta\rho}+C_{\theta\beta}\,C_{\lambda\rho}\right),\label{eq: Int 14}
\end{align}
\begin{align}
\mathcal{I}_{15} & =\int\frac{d^{D}kd^{D}q}{\left(2\pi\right)^{2D}}\frac{q_{\mu\beta}\,q_{\theta\lambda}}{\left(k+p\right)^{2}\left(p+q\right)^{2}\left(k-q\right)^{2}q^{2}}=\int\frac{d^{D}kd^{D}q}{\left(2\pi\right)^{2D}}\frac{k_{\mu\beta}\,k_{\theta\lambda}}{\left(k+p\right)^{2}\left(p+q\right)^{2}\left(k-q\right)^{2}k^{2}}\nonumber \\
 & =-\frac{1}{96\pi^{2}\epsilon}\left(C_{\mu\theta}\,C_{\beta\lambda}+C_{\beta\theta}\,C_{\mu\lambda}\right),\label{eq: Int 15}
\end{align}
\begin{align}
\mathcal{I}_{16} & =\int\frac{d^{D}kd^{D}q}{\left(2\pi\right)^{2D}}\frac{q_{\mu\nu}\,q_{\lambda\theta}\,k_{\rho\sigma}}{\left(k+p\right)^{2}\left(q+p\right)^{2}\left(k-q\right)^{2}q^{2}}=\int\frac{d^{D}kd^{D}q}{\left(2\pi\right)^{2D}}\frac{k_{\mu\nu}\,k_{\lambda\theta}\,q_{\rho\sigma}}{\left(k+p\right)^{2}\left(q+p\right)^{2}\left(k-q\right)^{2}k^{2}}\nonumber \\
 & =\frac{1}{320\pi^{2}\epsilon}\left\{ \left(C_{\lambda\rho}\,C_{\theta\sigma}+C_{\theta\rho}\,C_{\lambda\sigma}\right)p_{\mu\nu}+\left(C_{\mu\rho}\,C_{\nu\sigma}+C_{\nu\rho}\,C_{\mu\sigma}\right)p_{\lambda\theta}\right.\nonumber \\
 & \left.+\frac{8}{3}\left(C_{\mu\lambda}\,C_{\nu\theta}+C_{\nu\lambda}\,C_{\mu\theta}\right)p_{\rho\sigma}\right\} \,,\label{eq: Int 16}
\end{align}
\begin{align}
\mathcal{I}_{17} & =\int\frac{d^{D}kd^{D}q}{\left(2\pi\right)^{2D}}\frac{k_{\delta\theta}\,k_{\sigma\rho}\,q_{\beta\gamma}\,q_{\alpha\lambda}}{\left(k+p\right)^{2}\left(q^{2}\right)^{2}\left[\left(k-q\right)^{2}\right]^{2}}=\int\frac{d^{D}kd^{D}q}{\left(2\pi\right)^{2D}}\frac{k_{\delta\theta}\,k_{\sigma\rho}\,q_{\beta\gamma}\,q_{\alpha\lambda}}{\left(k^{2}\right)^{2}\left(q+p\right)^{2}\left[\left(k-q\right)^{2}\right]^{2}}\nonumber \\
 & =\frac{1}{4\left(160\pi^{2}\epsilon\right)}\left\{ -\frac{2}{3}\left(C_{\delta\sigma}\,C_{\theta\rho}+C_{\theta\sigma}\,C_{\delta\rho}\right)\left(C_{\beta\alpha}\,C_{\gamma\lambda}+C_{\gamma\alpha}\,C_{\beta\lambda}\right)+\left(C_{\delta\beta}\,C_{\theta\gamma}+C_{\theta\beta}\,C_{\delta\gamma}\right)\times\right.\nonumber \\
 & \left.\left(C_{\sigma\alpha}\,C_{\rho\lambda}+C_{\rho\alpha}\,C_{\sigma\lambda}\right)+\left(C_{\delta\alpha}\,C_{\theta\lambda}+C_{\theta\alpha}\,C_{\delta\lambda}\right)\left(C_{\sigma\beta}\,C_{\rho\gamma}+C_{\rho\beta}\,C_{\sigma\gamma}\right)\right\} \,.\label{eq: Int 17}
\end{align}

\section{Integrals used in the gauge superfield corrections}

In this section we compile a set of Tables, with the divergent results
for the specific integrals associated to each of the diagrams that
contribute to the two-point vertex function of the gauge superfield,
based on the results presented in the last subsection.
\begin{center}
\begin{table}
\centering{}%
\begin{tabular}{ccccccc}
$\mathcal{I}^{\left(a\right)}$ &  &  &  & $\mathcal{I}^{\left(a\right)}$ &  & \tabularnewline
\hline 
\hline 
$\left(k\cdot p\right)\,k_{\alpha\beta}$ &  & $-2\,p_{\alpha\beta}\,I_{d}$ &  & $\left(k\cdot q\right)\,k_{\alpha\beta}$ &  & $-2\,p_{\alpha\beta}\,I_{d}$\tabularnewline
$\left(k\cdot p\right)\,q_{\alpha\beta}$ &  & $p_{\alpha\beta}\,I_{d}$ &  & $p_{\beta\gamma}\,q_{\alpha\delta}\,k^{\gamma\delta}$ &  & $3\,p_{\alpha\beta}\,I_{d}$\tabularnewline
$C_{\beta\gamma}\,C^{\delta\epsilon}\,k_{\alpha\epsilon}\,q_{\delta\zeta}\,k^{\gamma\zeta}$ &  & $2\,p_{\alpha\beta}\,I_{d}$ &  & $k_{\alpha\gamma}\,k_{\beta\delta}\,p^{\gamma\delta}$ &  & $2\,p_{\alpha\beta}\,I_{d}$\tabularnewline
$k_{\beta\gamma}\,q_{\alpha\delta}\,p^{\gamma\delta}$ &  & $-p_{\alpha\beta}\,I_{d}$ &  & $k_{\alpha\beta}\,k^{2}$ &  & $4\,p_{\alpha\beta}\,I_{d}$\tabularnewline
$p_{\alpha\beta}\,k^{2}$ &  & $-6\,p_{\alpha\beta}\,I_{d}$ &  & $q_{\alpha\beta}\,k^{2}$ &  & $-2\,p_{\alpha\beta}\,I_{d}$\tabularnewline
$C_{\beta\gamma}\,q_{\alpha\delta}\,k^{\gamma\delta}$ &  & $3\,C_{\beta\alpha}\,I_{d}$ &  & $C_{\beta\alpha}\,k^{2}$ &  & $-6\,C_{\beta\alpha}\,I_{d}$\tabularnewline
\hline 
\hline 
 &  &  &  &  &  & \tabularnewline
\end{tabular}\caption{\label{tab:Integrals-of-diagram-S-a}Integrals of diagram $\mathcal{S}_{\Gamma\Gamma}^{\left(a\right)}$
in the two-point vertex function of gauge superfield, with $I_{d}=\frac{1}{192\pi^{2}\epsilon}$
and $\mathcal{I}^{\left(a\right)}\equiv\int\frac{d^{D}kd^{D}q}{\left(2\pi\right)^{2D}}\frac{1}{\left(k+p\right)^{2}\left(k+q\right)^{2}q^{2}k^{2}}$.}
\end{table}
\par\end{center}

\begin{center}
\begin{table}
\centering{}%
\begin{tabular}{ccccccc}
$\mathcal{I}^{\left(a\right)}$ &  &  &  & $\mathcal{I}^{\left(a\right)}$ &  & \tabularnewline
\hline 
\hline 
$\left(k\cdot p\right)\,k_{\alpha\beta}$ &  & $-2\,p_{\alpha\beta}\,I_{d}$ &  & $p_{\alpha\gamma}\,q_{\beta\delta}\,k^{\gamma\delta}$ &  & $3\,p_{\alpha\beta}\,I_{d}$\tabularnewline
$C_{\alpha\gamma}\,C^{\delta\epsilon}\,k_{\beta\delta}\,p_{\epsilon\zeta}\,k^{\gamma\zeta}$ &  & $2\,p_{\alpha\beta}\,I_{d}$ &  & $k_{\alpha\gamma}\,k_{\beta\delta}\,p^{\gamma\delta}$ &  & $2\,p_{\alpha\beta}\,I_{d}$\tabularnewline
$C_{\alpha\gamma}\,C^{\delta\epsilon}\,k_{\delta\zeta}\,q_{\beta\epsilon}\,p^{\gamma\zeta}$ &  & $-3\,p_{\alpha\beta}\,I_{d}$ &  & $k_{\alpha\beta}\,k^{2}$ &  & $4\,p_{\alpha\beta}\,I_{d}$\tabularnewline
$p_{\alpha\beta}\,k^{2}$ &  & $-6\,p_{\alpha\beta}\,I_{d}$ &  & $q_{\alpha\beta}\,k^{2}$ &  & $-2\,p_{\alpha\beta}\,I_{d}$\tabularnewline
$C_{\alpha\gamma}\,q_{\beta\delta}\,k^{\gamma\delta}$ &  & $3\,C_{\alpha\beta}\,I_{d}$ &  & $C_{\beta\alpha}\,k^{2}$ &  & $-6\,C_{\beta\alpha}\,I_{d}$\tabularnewline
\hline 
\hline 
 &  &  &  &  &  & \tabularnewline
\end{tabular}\caption{\label{tab:Integrals-of-diagram-S-b}Integrals of diagram $\mathcal{S}_{\Gamma\Gamma}^{\left(b\right)}$
in the two-point vertex function of gauge superfield, with $I_{d}=\frac{1}{192\pi^{2}\epsilon}$
and $\mathcal{I}^{\left(a\right)}\equiv\int\frac{d^{D}kd^{D}q}{\left(2\pi\right)^{2D}}\frac{1}{\left(k+p\right)^{2}\left(k+q\right)^{2}q^{2}k^{2}}$.}
\end{table}
\par\end{center}

\begin{center}
\begin{table}
\centering{}%
\begin{tabular}{ccccccc}
$\mathcal{I}^{\left(b\right)}$ &  &  &  & $\mathcal{I}^{\left(b\right)}$ &  & \tabularnewline
\hline 
\hline 
$\left(k\cdot p\right)\,\left(k\cdot q\right)\,k_{\alpha\beta}$ &  & $p_{\alpha\beta}\,I_{d}$ &  & $C_{\alpha\gamma}\,C_{\beta\delta}\,\left(k\cdot q\right)\,p_{\epsilon\zeta}\,k^{\gamma\zeta}\,k^{\delta\epsilon}$ &  & $-p_{\alpha\beta}\,I_{d}$\tabularnewline
$C_{\alpha\gamma}\,C_{\beta\delta}\,\left(k\cdot p\right)\,q_{\epsilon\zeta}\,k^{\gamma\epsilon}\,k^{\delta\zeta}$ &  & $p_{\alpha\beta}\,I_{d}$ &  & $C_{\alpha\gamma}\,C_{\beta\delta}\,p_{\epsilon\zeta}\,q_{\eta\theta}\,k^{\gamma\theta}\,k^{\delta\zeta}\,k^{\epsilon\eta}$ &  & $-p_{\alpha\beta}\,I_{d}$\tabularnewline
$\left(k\cdot q\right)\,k_{\alpha\gamma}\,k_{\beta\delta}\,p^{\gamma\delta}$ &  & $-p_{\alpha\beta}\,I_{d}$ &  & $C_{\alpha\gamma}\,C^{\delta\epsilon}\,k_{\beta\zeta}\,k_{\epsilon\eta}\,q_{\delta\theta}\,k^{\gamma\theta}\,p^{\eta\zeta}$ &  & $p_{\alpha\beta}\,I_{d}$\tabularnewline
$k_{\alpha\beta}\,k_{\gamma\delta}\,k_{\epsilon\zeta}\,p^{\delta\zeta}\,q^{\gamma\epsilon}$ &  & $2\,p_{\alpha\beta}\,I_{d}$ &  & $\left(k\cdot p\right)\,k_{\alpha\gamma}\,k_{\beta\delta}\,q^{\delta\gamma}$ &  & $p_{\alpha\beta}\,I_{d}$\tabularnewline
$k_{\alpha\gamma}\,k_{\beta\delta}\,k_{\epsilon\zeta}\,p^{\delta\epsilon}\,q^{\zeta\gamma}$ &  & $-p_{\alpha\beta}\,I_{d}$ &  & $\left(k\cdot p\right)\,k_{\alpha\beta}\,k^{2}$ &  & $-2\,p_{\alpha\beta}\,I_{d}$\tabularnewline
$\left(k\cdot q\right)\,k_{\alpha\beta}\,k^{2}$ &  & $-2\,p_{\alpha\beta}\,I_{d}$ &  & $\left(k\cdot q\right)\,p_{\alpha\beta}\,k^{2}$ &  & $3\,p_{\alpha\beta}\,I_{d}$\tabularnewline
$\left(k\cdot p\right)\,q_{\alpha\beta}\,k^{2}$ &  & $p_{\alpha\beta}\,I_{d}$ &  & $C_{\alpha\gamma}\,C_{\beta\delta}\,p_{\zeta\epsilon}\,k^{\gamma\zeta}\,k^{\delta\epsilon}\,k^{2}$ &  & $2\,p_{\alpha\beta}\,I_{d}$\tabularnewline
$C_{\alpha\gamma}\,C_{\beta\delta}\,q_{\epsilon\zeta}\,k^{\gamma\zeta}\,k^{\delta\epsilon}\,k^{2}$ &  & $-2\,p_{\alpha\beta}\,I_{d}$ &  & $C_{\alpha\gamma}\,C_{\beta\delta}\,q_{\epsilon\zeta}\,k^{\gamma\epsilon}\,k^{\delta\zeta}\,k^{2}$ &  & $-2\,p_{\alpha\beta}\,I_{d}$\tabularnewline
$k_{\alpha\gamma}\,k_{\beta\delta}\,p^{\delta\gamma}\,k^{2}$ &  & $2\,p_{\alpha\beta}\,I_{d}$ &  & $k_{\beta\gamma}\,q_{\alpha\delta}\,p^{\gamma\delta}\,k^{2}$ &  & $-p_{\alpha\beta}\,I_{d}$\tabularnewline
$C_{\alpha\gamma}\,C_{\beta\delta}\,q_{\epsilon\zeta}\,k^{\delta\epsilon}\,p^{\gamma\zeta}\,k^{2}$ &  & $3\,p_{\alpha\beta}\,I_{d}$ &  & $k_{\beta\gamma}\,p_{\alpha\delta}\,q^{\gamma\delta}\,k^{2}$ &  & $3\,p_{\alpha\beta}\,I_{d}$\tabularnewline
$C_{\alpha\gamma}\,C_{\beta\delta}\,k_{\epsilon\zeta}\,p^{\delta\zeta}\,q^{\gamma\epsilon}\,k^{2}$ &  & $3\,p_{\alpha\beta}\,I_{d}$ &  & $C_{\alpha\gamma}\,C_{\beta\delta}\,p_{\epsilon\zeta}\,k^{\delta\epsilon}\,q^{\gamma\zeta}\,k^{2}$ &  & $-p_{\alpha\beta}\,I_{d}$\tabularnewline
$k_{\alpha\gamma}\,k_{\beta\delta}\,q^{\delta\gamma}\,k^{2}$ &  & $-2\,p_{\alpha\beta}\,I_{d}$ &  & $C_{\alpha\gamma}\,C_{\beta\delta}\,k_{\epsilon\zeta}\,p^{\gamma\zeta}\,q^{\delta\epsilon}\,k^{2}$ &  & $3\,p_{\alpha\beta}\,I_{d}$\tabularnewline
$k_{\alpha\beta}\,k^{4}$ &  & $4\,p_{\alpha\beta}\,I_{d}$ &  & $p_{\alpha\beta}\,k^{4}$ &  & $-6\,p_{\alpha\beta}\,I_{d}$\tabularnewline
$q_{\alpha\beta}\,k^{4}$ &  & $-2\,p_{\alpha\beta}\,I_{d}$ &  & $C_{\beta\alpha}\left(k\cdot q\right)\,k^{2}$ &  & $3\,C_{\beta\alpha}\,I_{d}$\tabularnewline
$C_{\beta\gamma}\,q_{\alpha\delta}\,k^{\gamma\delta}\,k^{2}$ &  & $3\,C_{\beta\alpha}\,I_{d}$ &  & $C_{\alpha\gamma}\,q_{\beta\delta}\,k^{\gamma\delta}\,k^{2}$ &  & $3\,C_{\alpha\beta}\,I_{d}$\tabularnewline
$C_{\beta\gamma}\,k_{\alpha\delta}\,q^{\gamma\delta}\,k^{2}$ &  & $3\,C_{\beta\alpha}\,I_{d}$ &  & $C_{\alpha\gamma}\,k_{\beta\delta}\,q^{\gamma\delta}\,k^{2}$ &  & $3\,C_{\alpha\beta}\,I_{d}$\tabularnewline
$C_{\beta\alpha}\,k^{4}$ &  & $-6\,C_{\beta\alpha}\,I_{d}$ &  &  &  & \tabularnewline
\hline 
\hline 
 &  &  &  &  &  & \tabularnewline
\end{tabular}\caption{\label{tab:Integrals-of-diagram-S-c}Integrals of diagram $\mathcal{S}_{\Gamma\Gamma}^{\left(c\right)}$
in the two-point vertex function of gauge superfield, with $I_{d}=\frac{1}{192\pi^{2}\epsilon}$
and $\mathcal{I}^{\left(b\right)}\equiv\int\frac{d^{D}kd^{D}q}{\left(2\pi\right)^{2D}}\frac{1}{\left(k+p\right)^{2}\left(k+q\right)^{2}q^{2}\left(k^{2}\right)^{2}}$.}
\end{table}
\par\end{center}

\begin{center}
\begin{table}[h]
\centering{}%
\begin{tabular}{ccccccc}
$\mathcal{I}^{\left(c\right)}$ &  &  &  & $\mathcal{I}^{\left(c\right)}$ &  & \tabularnewline
\hline 
\hline 
$\left(k\cdot p\right)\,\left(k\cdot q\right)\,q_{\alpha\beta}$ &  & $-p_{\alpha\beta}\,I_{d}$ &  & $C_{\alpha\gamma}\,C_{\beta\delta}\,\left(p\cdot q\right)\,q_{\epsilon\zeta}\,k^{\gamma\zeta}\,k^{\delta\epsilon}$ &  & $0$\tabularnewline
$C_{\alpha\gamma}\,C^{\delta\epsilon}\,p_{\epsilon\zeta}\,q_{\beta\eta}\,q_{\delta\theta}\,k^{\gamma\theta}\,k^{\zeta\eta}$ &  & $4\,p_{\alpha\beta}\,I_{d}$ &  & $C_{\alpha\gamma}\,C^{\delta\epsilon}\,k_{\epsilon\zeta}\,q_{\beta\eta}\,q_{\delta\theta}\,k^{\gamma\theta}\,p^{\zeta\eta}$ &  & $0$\tabularnewline
$C_{\beta\gamma}\,C^{\delta\epsilon}\,k_{\alpha\zeta}\,p_{\epsilon\eta}\,q_{\delta\theta}\,k^{\eta\theta}\,q^{\gamma\zeta}$ &  & $2\,p_{\alpha\beta}\,I_{d}$ &  & $C_{\beta\gamma}\,C^{\delta\epsilon}\,k_{\alpha\zeta}\,k_{\epsilon\eta}\,q_{\delta\theta}\,p^{\eta\theta}\,q^{\gamma\zeta}$ &  & $-2\,p_{\alpha\beta}\,I_{d}$\tabularnewline
$C_{\beta\gamma}\,C^{\delta\epsilon}\,k_{\alpha\zeta}\,k_{\delta\eta}\,q_{\epsilon\theta}\,p^{\eta\theta}\,q^{\gamma\zeta}$ &  & $2\,p_{\alpha\beta}\,I_{d}$ &  & $C_{\beta\gamma}\,C^{\delta\epsilon}\,k_{\alpha\zeta}\,k_{\delta\eta}\,q_{\epsilon\theta}\,p^{\zeta\theta}\,q^{\gamma\eta}$ &  & $0$\tabularnewline
$k_{\alpha\gamma}\,k_{\delta\epsilon}\,q_{\beta\zeta}\,p^{\epsilon\zeta}\,q^{\delta\gamma}$ &  & $0$ &  & $C_{\alpha\gamma}\,C_{\beta\delta}\,\left(k\cdot q\right)\,p_{\epsilon\zeta}\,k^{\gamma\zeta}\,q^{\delta\epsilon}$ &  & $p_{\alpha\beta}\,I_{d}$\tabularnewline
$C_{\alpha\gamma}\,C_{\beta\delta}\,p_{\epsilon\zeta}\,q_{\eta\theta}\,k^{\gamma\eta}\,k^{\zeta\theta}\,q^{\delta\epsilon}$ &  & $0$ &  & $C_{\alpha\gamma}\,C_{\beta\delta}\,k_{\epsilon\zeta}\,q_{\eta\theta}\,k^{\gamma\theta}\,p^{\zeta\eta}\,q^{\delta\epsilon}$ &  & $-4\,p_{\alpha\beta}\,I_{d}$\tabularnewline
$C_{\alpha\gamma}\,C_{\beta\delta}\,k_{\epsilon\zeta}\,q_{\eta\theta}\,k^{\gamma\eta}\,p^{\epsilon\theta}\,q^{\delta\zeta}$ &  & $-4\,p_{\alpha\beta}\,I_{d}$ &  & $C_{\alpha\gamma}\,C_{\beta\delta}\,k_{\epsilon\zeta}\,p_{\eta\theta}\,k^{\gamma\theta}\,q^{\delta\zeta}\,q^{\epsilon\eta}$ &  & $0$\tabularnewline
$C_{\beta\gamma}\,C^{\delta\epsilon}\,k_{\alpha\zeta}\,k_{\delta\eta}\,p_{\epsilon\theta}\,q^{\gamma\theta}\,q^{\zeta\eta}$ &  & $0$ &  & $C_{\beta\gamma}\,C^{\delta\epsilon}\,k_{\alpha\zeta}\,k_{\delta\eta}\,p_{\epsilon\theta}\,q^{\gamma\eta}\,q^{\zeta\theta}$ &  & $4\,p_{\alpha\beta}\,I_{d}$\tabularnewline
$\left(p\cdot q\right)\,k_{\alpha\beta}\,k^{2}$ &  & $-p_{\alpha\beta}\,I_{d}$ &  & $\left(k\cdot q\right)\,p_{\alpha\beta}\,k^{2}$ &  & $-3\,p_{\alpha\beta}\,I_{d}$\tabularnewline
$\left(k\cdot p\right)\,q_{\alpha\beta}\,k^{2}$ &  & $-p_{\alpha\beta}\,I_{d}$ &  & $\left(k\cdot q\right)\,q_{\alpha\beta}\,k^{2}$ &  & $4\,p_{\alpha\beta}\,I_{d}$\tabularnewline
$\left(p\cdot q\right)\,q_{\alpha\beta}\,k^{2}$ &  & $-2\,p_{\alpha\beta}\,I_{d}$ &  & $p_{\alpha\gamma}\,q_{\beta\delta}\,k^{\gamma\delta}\,k^{2}$ &  & $-3\,p_{\alpha\beta}\,I_{d}$\tabularnewline
$k_{\alpha\gamma}\,q_{\beta\delta}\,p^{\gamma\delta}\,k^{2}$ &  & $p_{\alpha\beta}\,I_{d}$ &  & $q_{\alpha\gamma}\,q_{\beta\delta}\,p^{\delta\gamma}\,k^{2}$ &  & $2\,p_{\alpha\beta}\,I_{d}$\tabularnewline
$C_{\alpha\gamma}\,C_{\beta\delta}\,q_{\epsilon\zeta}\,k^{\gamma\epsilon}\,p^{\delta\zeta}\,k^{2}$ &  & $-3\,p_{\alpha\beta}\,I_{d}$ &  & $C_{\alpha\gamma}\,C_{\beta\delta}\,p_{\epsilon\zeta}\,k^{\gamma\zeta}\,q^{\delta\epsilon}\,k^{2}$ &  & $p_{\alpha\beta}\,I_{d}$\tabularnewline
$C_{\alpha\gamma}\,C_{\beta\delta}\,k_{\epsilon\zeta}\,q^{\gamma\zeta}\,q^{\delta\epsilon}\,k^{2}$ &  & $2\,p_{\alpha\beta}\,I_{d}$ &  & $C_{\alpha\gamma}\,C_{\beta\delta}\,p_{\epsilon\zeta}\,k^{\gamma\epsilon}\,q^{\delta\zeta}\,k^{2}$ &  & $p_{\alpha\beta}\,I_{d}$\tabularnewline
$C_{\alpha\gamma}\,C_{\beta\delta}\,k_{\epsilon\zeta}\,q^{\gamma\epsilon}\,q^{\delta\zeta}\,k^{2}$ &  & $2\,p_{\alpha\beta}\,I_{d}$ &  & $C_{\alpha\gamma}\,C_{\beta\delta}\,k_{\epsilon\zeta}\,p^{\gamma\zeta}\,q^{\delta\epsilon}\,k^{2}$ &  & $-3\,p_{\alpha\beta}\,I_{d}$\tabularnewline
$C_{\alpha\gamma}\,C_{\beta\delta}\,p_{\epsilon\zeta}\,q^{\gamma\epsilon}\,q^{\delta\zeta}\,k^{2}$ &  & $2\,p_{\alpha\beta}\,I_{d}$ &  & $p_{\alpha\beta}\,k^{4}$ &  & $0$\tabularnewline
$q_{\alpha\beta}\,k^{4}$ &  & $0$ &  & $\left(k\cdot p\right)\,k_{\alpha\beta}\,q^{2}$ &  & $-2\,p_{\alpha\beta}\,I_{d}$\tabularnewline
$\left(k\cdot p\right)\,q_{\alpha\beta}\,q^{2}$ &  & $-p_{\alpha\beta}\,I_{d}$ &  & $p_{\alpha\gamma}\,q_{\beta\delta}\,k^{\gamma\delta}\,q^{2}$ &  & $-3\,p_{\alpha\beta}\,I_{d}$\tabularnewline
$C_{\alpha\gamma}\,C_{\beta\delta}\,p_{\epsilon\zeta}\,k^{\gamma\zeta}\,k^{\delta\epsilon}\,q^{2}$ &  & $2\,p_{\alpha\beta}\,I_{d}$ &  & $C_{\alpha\gamma}\,C_{\beta\delta}\,q_{\epsilon\zeta}\,k^{\gamma\zeta}\,k^{\delta\epsilon}\,q^{2}$ &  & $2\,p_{\alpha\beta}\,I_{d}$\tabularnewline
$k_{\alpha\gamma}\,k_{\beta\delta}\,p^{\delta\gamma}\,q^{2}$ &  & $2\,p_{\alpha\beta}\,I_{d}$ &  & $k_{\alpha\gamma}\,q_{\beta\delta}\,p^{\gamma\delta}\,q^{2}$ &  & $p_{\alpha\beta}\,I_{d}$\tabularnewline
$C_{\alpha\gamma}\,C_{\beta\delta}\,q_{\epsilon\zeta}\,k^{\delta\epsilon}\,p^{\gamma\zeta}\,q^{2}$ &  & $-3\,p_{\alpha\beta}\,I_{d}$ &  & $C_{\alpha\gamma}\,C_{\beta\delta}\,p_{\epsilon\zeta}\,k^{\gamma\zeta}\,q^{\delta\epsilon}\,q^{2}$ &  & $p_{\alpha\beta}\,I_{d}$\tabularnewline
$C_{\alpha\gamma}\,C_{\beta\delta}\,k_{\epsilon\zeta}\,p^{\gamma\zeta}\,q^{\delta\epsilon}\,q^{2}$ &  & $-3\,p_{\alpha\beta}\,I_{d}$ &  & $C_{\alpha\gamma}\,C_{\beta\delta}\,p_{\epsilon\zeta}\,k^{\gamma\epsilon}\,q^{\delta\zeta}\,q^{2}$ &  & $p_{\alpha\beta}\,I_{d}$\tabularnewline
$C_{\alpha\gamma}\,C_{\beta\delta}\,k_{\epsilon\zeta}\,p^{\gamma\epsilon}\,q^{\delta\zeta}\,q^{2}$ &  & $-3\,p_{\alpha\beta}\,I_{d}$ &  & $k_{\alpha\beta}\,k^{2}\,q^{2}$ &  & $6\,p_{\alpha\beta}\,I_{d}$\tabularnewline
$p_{\alpha\beta}\,k^{2}\,q^{2}$ &  & $-6\,p_{\alpha\beta}\,I_{d}$ &  & $C_{\alpha\gamma}\,q_{\beta\delta}\,q_{\epsilon\zeta}\,k^{\gamma\zeta}\,k^{\delta\epsilon}$ &  & $0$\tabularnewline
$C_{\alpha\gamma}\,q_{\beta\delta}\,q_{\epsilon\zeta}\,k^{\gamma\epsilon}\,k^{\zeta\delta}$ &  & $0$ &  & $C_{\beta\alpha}\,\left(k\cdot q\right)\,k_{\alpha\delta}\,q^{\gamma\delta}$ &  & $-3\,C_{\beta\alpha}\,I_{d}$\tabularnewline
$C_{\beta\gamma}\,k_{\alpha\delta}\,k_{\epsilon\zeta}\,q^{\gamma\zeta}\,q^{\delta\epsilon}$ &  & $0$ &  & $C_{\alpha\gamma}\,C_{\beta\delta}\,C^{\epsilon\zeta}\,k_{\epsilon\eta}\,q_{\zeta\theta}\,k^{\gamma\theta}\,q^{\delta\eta}$ &  & $0$\tabularnewline
$C_{\beta\gamma}\,k_{\alpha\delta}\,k_{\epsilon\zeta}\,q^{\gamma\zeta}\,q^{\epsilon\delta}$ &  & $0$ &  & $C_{\alpha\gamma}\,q_{\beta\delta}\,k^{\gamma\delta}\,k^{2}$ &  & $-3\,C_{\alpha\beta}\,I_{d}$\tabularnewline
$C_{\beta\gamma}\,k_{\alpha\delta}\,q^{\gamma\delta}\,k^{2}$ &  & $-3\,C_{\beta\alpha}\,I_{d}$ &  & $C_{\alpha\gamma}\,q_{\beta\delta}\,k^{\gamma\delta}\,q^{2}$ &  & $-3\,C_{\alpha\beta}\,I_{d}$\tabularnewline
$C_{\beta\gamma}\,k_{\alpha\delta}\,q^{\gamma\delta}\,q^{2}$ &  & $-3\,C_{\beta\alpha}\,I_{d}$ &  & $C_{\beta\alpha}\,k^{2}\,q^{2}$ &  & $-6\,C_{\beta\alpha}\,I_{d}$\tabularnewline
\hline 
\hline 
 &  &  &  &  &  & \tabularnewline
\end{tabular}\caption{\label{tab:Integrals-of-diagram-S-d}Integrals of diagram $\mathcal{S}_{\Gamma\Gamma}^{\left(d\right)}$
in the two-point vertex function of gauge superfield, with $I_{d}=\frac{1}{192\pi^{2}\epsilon}$
and $\mathcal{I}^{\left(c\right)}\equiv\int\frac{d^{D}kd^{D}q}{\left(2\pi\right)^{2D}}\frac{1}{\left(k+p\right)^{2}\left(q+p\right)^{2}\left(k-q\right)^{2}q^{2}k^{2}}$.}
\end{table}
\par\end{center}

\myclearpage

\end{document}